\documentclass[12pt]{article}
\usepackage[margin=1in]{geometry} 
\usepackage{amsmath,amsthm,amssymb,amsfonts,mathtools,bm,mathdots}
\usepackage[makeroom]{cancel}
\usepackage{graphicx}
\usepackage{fancyhdr}
\usepackage{hyperref}
\hypersetup{allcolors=black}
\usepackage{centernot}
\usepackage{lipsum}
\usepackage{multirow}
\usepackage{float}
\usepackage{mathrsfs}
\usepackage{array}
\usepackage{tabto}
\usepackage{wrapfig}
\usepackage{caption}
\usepackage[bottom]{footmisc}
\usepackage{subcaption}
\usepackage[title]{appendix}
\usepackage{booktabs}
\usepackage{longtable}
\usepackage{lscape}
\newcommand\scalemath[2]{\scalebox{#1}{\mbox{\ensuremath{\displaystyle #2}}}} 
\newtheorem{theorem}{Theorem}[section]
\newtheorem*{theorem*}{\bf Theorem}
\newtheorem{definition}{Definition}[section]
\newtheorem*{definition*}{\bf Definition}

\newcommand{\comment}[1]{}
\newcommand{\IC}{\mathbb{C}}
\newcommand{\IP}{\mathbb{P}}
\newcommand{\IQ}{\mathbb{Q}}
\newcommand{\IR}{\mathbb{R}}
\newcommand{\cN}{\mathcal{N}}
\newcommand{\cM}{\mathcal{M}}

\begin{document}
\title{Machine-Learning Dessins d'Enfants: Explorations via Modular and Seiberg-Witten Curves}
\author{Yang-Hui He\textsuperscript{1,2,3}\footnote{hey@maths.ox.ac.uk}, Edward Hirst\textsuperscript{1}\footnote{edward.hirst@city.ac.uk}, Toby Peterken\textsuperscript{4}\footnote{toby.peterken@keble.ox.ac.uk}}
\date{\small\today \\ \textsuperscript{1} \textit{\small Department of Mathematics, City, University of London, EC1V 0HB, UK}\\
\textsuperscript{2}\textit{\small Merton College, University of Oxford, OX1 4JD, UK}\\
\textsuperscript{3}\textit{\small School of Physics, NanKai University, Tianjin, 300071, P.R. China}\\
\textsuperscript{4}\textit{\small Keble College, University of Oxford, OX1 3PG, UK}
}
\maketitle

\begin{abstract}
We apply machine-learning to the study of dessins d'enfants.
Specifically, we investigate a class of dessins which reside at the intersection of the investigations of modular subgroups, Seiberg-Witten curves and extremal elliptic K3 surfaces.
A deep feed-forward neural network with simple structure and standard activation functions without prior knowledge of the underlying mathematics is established and imposed onto the classification of extension degree over the rationals, known to be a difficult problem.
The classifications reached 0.92 accuracy with 0.03 standard error relatively quickly.
The Seiberg-Witten curves for those with rational coefficients are also tabulated.
\end{abstract}

\thispagestyle{empty}
\clearpage
\setcounter{page}{1}
\numberwithin{equation}{section}
\tableofcontents


\section{Introduction \& Summary}
Alexander Grothendieck's Dessins d'enfants, or childrens' drawings, laid a cornerstone of modern mathematics, inspiring a confluence of geometry, combinatorics, and number theory.
Having learnt of the remarkable theorem of Bely\v{\i} \cite{Belyi} which relates the existence of algebraic models of Riemann surfaces to that of analytic properties of rational functions thereon, Grothendieck launched an entire programme \cite{Esquisse} by pictorially representing\footnote{
Interestingly, F.~Klein had the vision, when considering and 11-fold cover of the Riemann sphere, to draw a diagram which he called {\it Linienz\"uge} (line-tracks) \cite{Klein}, which is really a prototype for a dessin.
} this structure as bipartite graphs (the dessin) drawn on the Riemann surface.
He hypothesised dessins d'enfants in their current form as a conceptual representation of the absolute Galois group over the rationals, one the most mysterious and least understood objects in number theory.
Subsequently, he developed a generalisation of Bely\v{\i}'s theorem which extends the surfaces considered in the mapping to more general Riemann surfaces. Properties of the mapping are identified with combinatorial invariants of the dessin d'enfant graphs \cite{Esquisse} (q.v.~several standard introductory books and lecture notes in \cite{Girondo,Guillot,Lando,Zapponi}).

Remarkably, this beautiful story has more recently found its place in theoretical physics, especially in supersymmetric gauge theories and string theory.
These have ranged from a novel way of factorizing \cite{SW_curves} Seiberg-Witten curves \cite{Seiberg:1994rs}, to realizing \cite{Jejjala:2010vb,Hanany:2011ra,He:2016yvz} brane-tilings and dimer models, which is the most general method \cite{Franco:2005sm} of understanding AdS$_5$/CFT$_4$ for toric Calabi-Yau spaces, as well as to certain elliptic K3-surfaces \cite{YMR} and their relation \cite{He:2012kw} to generalized quiver gauge theories \cite{Gaiotto:2009we}.

One downside to the {\it esquisse} is that Bely\v{\i} maps are notoriously difficult to compute explicitly: whilst the dessin is easy to draw, the underlying map to the Riemann sphere could very quickly involve algebraic numbers unimaginable from the shape of the graph.
To give an example from physics, the brane-tiling/dessin for the so-called suspended pinched point Calabi-Yau singularity had been known for a decade \cite{Franco:2005sm} and consists of a simple triangle-rectangle tessellation of the doubly-periodic plane, but it took years \cite{Vidunas:2017jei} to construct the Bely\v{\i} map as a rational function.
Indeed, as always, Grothendieck's power lies in the profundity of an abstract panorama, rather than in the mundanity of computation and untidy expressions.

Recently, a paradigm was established to see whether the latest techniques in machine-learning (ML) and data science can be used to study mathematical structures, and in particular, whether otherwise prohibitive calculations could be bypassed, at least stochastically  \cite{YHHdeep}.
Indeed, \cite{YHHdeep, Krefl:2017yox, Ruehle:2017mzq, Carifio:2017bov} brought about machine-learning to string/gauge theory (see \cite{He:2018jtw,Ruehle:2020jrk,BYHP} and references therein for reviews).
Various problems in mathematics and theoretical physics have been put to the test by neural-networks and classifiers with architectures without a priori knowledge of the underlying mathematical structures.

An intuitive hierarchy of difficulty is beginning to emerge \cite{YHHtalk}: numerical algorithms such as those finding Calabi-Yau volumes \cite{Krefl:2017yox} and Ricci-flat metrics \cite{Ashmore:2019wzb} seem most amenable to ML, slightly less behaved though oftentimes comparable in performance are problems in algebraic geometry over $\IC$ such as topological invariants and bundle cohomology \cite{YHHdeep,Ruehle:2017mzq,Carifio:2017bov,Bull:2018uow,Brodie:2019dfx}.
Luckily for string theory, these two classes of problems are the most typical ones encountered.
At the next level of difficulty are algebraic and combinatorial problems, such as determining the simplicity of finite groups \cite{He:2019nzx} and property of graphs \cite{quiver,laplace}.
The very zenith of complexity, as expected, is number theory. To have a neural network predict the next prime, for instance, would be unfathomable: indeed simple experiments in \cite{YHHdeep} showed that the AI is doing no better than randomly guessing.
Yet, problems in number theory are still within the grasp of algebraic geometry and analysis seems to be slightly better than random guesses: the archetypal problem, the Birch-Swinnerton-Dyer conjecture, was subject to various ML algorithms in \cite{Alessandretti:2019jbs} and found to be so.

It is therefore natural to question, uniting our above two strands of thought, how the discipline of dessins d'enfants, overarching algebraic geometry, graph theory and algebraic number theory, responds to ML.
Given the paucity of available data - in part due to the aforementioned difficulty of computing Bely\v{\i} maps - it is a challenge to find a suitable dataset on which we could try supervised machine-learning.
Fortunately, a convenient set of dessins has been established in the literature, with the added bonus of involving some further elegant mathematics and physics.
Simultaneously exploiting the method of \cite{SW_curves} which related dessins to Seiberg-Witten curves, that of \cite{Miranda_Persson,MirandaPersson_table} in recognizing that the j-invariant of elliptic K3-surfaces are Bely\v{\i}, as well as that of \cite{Sebbar,j_McKay} in representing the Cayley coset graph of congruence subgroups of the modular group $PSL(2,\mathbb{Z})$ as bipartite graphs (and hence to $\cN=2$ generalized quiver theories \cite{He:2012kw,He:2015vua}), \cite{YMR} focused on a set of 112 distinguished K3 surfaces, the so-called extremal ones.

These K3 surfaces of concern are elliptically fibred over $\IP^1$ so the $j$-invariant of the fibering elliptic curves are naturally a rational function $j(z)$ of the base projective coordinate of $\IP^1$. 
Interestingly, $j(z)$ is a Bely\v{\i} map onto the base Riemann sphere and gives a dessin d'enfant.
Most of these have been determined explicitly \cite{MirandaPersson_table}, and at least the extension-degree over $\IQ$ for all of them are computed.
Meanwhile, the monodromy group of the covering of $\IP^1$ (equivalently, the cartographic group of the dessin) were computed in \cite{YMR}.
We will therefore take a two-pronged approach in this paper.
First, we complete the task of \cite{He:2012kw} (wherein 33 special cases of the 112, corresponding to torsion-free congruence subgroups of $PSL(2,\mathbb{Z})$ were translated to Seiberg-Witten curves) and provide the full correspondence of modular-subgroup/elliptic K3 surface/dessins/Seiberg-Witten curves for the 112 extreme K3 surfaces.

Second, and this is a much more curious and potentially very useful direction, we ask the question:
{\em can aspects of dessins be machine-learnt?
}
Specially, we can ask the following simple classification problem to a neural network (NN) with no knowledge of mathematics: given the dessin, drawn as a bipartite graph on a plane, there is an intrinsic integer, viz., the transcendence degree over $\IQ$, which is crucial to the underlying number theory and which is difficult to obtain; can this degree be learned by training the NN in a supervised way?

The purpose of this paper is to address these two directions.
The Seiberg-Witten curves can be found algorithmically and are tabulated in appendix \ref{SW_appendix}.
The focus on machine-learnable qualities studies two methods of representing the dessins, each learnt on relatively simple deep-feed-forward NNs. The first takes the adjacency matrix of the dessin as input, whilst the second a list of cycles around each node (which we call cyclic edge lists); both are classified according to the transcendence degree as output, and learning accuracy reaches 0.53 and 0.92 respectively in the best cases run.
The high accuracy for the cyclic edge list learning suggests the possibility of a way to directly understand the Galois orbits of dessins, and thence the absolute Galois group via tensor representations of the dessins' graph information, at least stochastically.

\comment{
The specific subset of dessins examined in this study relate to certain subgroups of the modular group, $PSL(2,\mathbb{Z})$. They relate to transformations on the Riemann sphere, and in turn are also related to Seiberg-Witten curves used to examine supersymmetric gauge theories \cite{SW_curves}. The focus was on those dessins corresponding to specific forms of elliptically fibred K3 surfaces as examined in \cite{YMR}, whose corresponding Seiberg-Witten curves have been derived and listed in appendix \ref{SW_appendix}. To classify these dessins according to the degree of the field extension they are defined over, and hence equivalently the size of the their orbits, a deep feed-forward neural network was used working with the dessin adjacency matrices as the dataset.
}

The paper is structured to provide an introduction to the dessin d'enfant graphs in sections \S\ref{dessin_section} and \S\ref{dessinCY_section}; explaining their importance in the Galois theory context, and their relation to Calabi-Yau manifolds in the context of algebraic geometry and string theory. The dessin representations in tensor format, and respective subtleties, such that a NN can process them is discussed in \S\ref{s:data}. Next, the mechanism of producing Seiberg-Witten curves, motivating the study of dessins for considering physical theories in the IR limit, is presented in section \S\ref{SW_section}. Following this the analysis of the machine-learning methods to classify the dataset are given in section \S\ref{ML_results}. The results show successes in identifying a dessin's extension degree from its adjacency matrix, and particularly from the cyclic representation format. The appendices provide further discussion of the Bely\v{\i} pair - dessin equivalence; list the corresponding Seiberg-Witten curves for the dessins defined over rational fields; and provide the full dataset of dessins analysed, with their matrix and cyclic edge list representations.

\section{Dramatis Personae}
We  begin  with  a  rapid  introduction of  the  various ingredients  from  the mathematics  and the physics, highlighting in  a  database  of  dessins d’enfants which  will  be  central  to  our analysis.

\subsection{Dessins d'Enfants}\label{dessin_section}
\subsubsection{Bely\v{\i}'s Theorem and Graph Theory}\label{Belyi_pair_label}
Let $X$ be a smooth, compact, connected Riemann surface.
A remarkable theorem \cite{Belyi} states that
\begin{theorem}[Belyi]
$X$ has an algebraic model over $\overline{\mathbb{Q}}$  IFF there exists a (surjective) map $\beta : X \rightarrow \mathbb{P}^1$ which is ramified at exactly 3 points. 
\end{theorem}
The covering map $\beta$ has come to be known as a {\bf Bely\v{\i} Map}. The depth of this theorem stems from the following. We know that a Riemann surface of genus $g$, as an algebraic variety, can be written, for example, as a hyper-elliptic curve\footnote{We should, of course, include the point at infinity by projectivising $\mathbb{C}^2$ to $\mathbb{P}^2$ but for convenience, we will work in the affine patch with ordinary coordinates $(x,y)$ and remember to include $(\infty, \infty)$ explicitly.
}, i.e., as a polynomial of degree $2g+1$ (or $2g+2$) in $\mathbb{C}^2$ with coordinates $(x,y)$. The coefficients in this polynomial dictate the complex structure and in principle could be arbitrary complex numbers. Bely\v{\i}'s theorem tells us that IFF one can find a rational map $\zeta = \beta(x,y)$ from $X$ to $\mathbb{P}^1$ with coordinate $\zeta$ which is ramified (not 1-1) at precisely 3 points, then the complex coefficients are algebraic numbers \cite{Girondo}.

Now, by M\"obius transformation 
$\zeta \longmapsto \frac{a\zeta+b}{c\zeta+d}\,,\quad \text{for} \; a,b,c,d \in \mathbb{C}\,$
on the coordinate $\zeta$ of $\mathbb{P}^1$, any 3 generic points can be taken to $(0, 1, \infty)$. In other words, Bely\v{\i}'s theorem dictates that IFF we can find a rational function $\beta(x,y)$ on $X$ at whose pre-images for 0, 1 and $\infty$ the Taylor series after the constant term (which is, of course, 0, 1, or $\infty$ respectively) starts at an order $n > 1$, then $X$ has complex structure over $\overline{\mathbb{Q}}$. The positive integer $n$ is called the {\bf ramification index}.

We emphasize that $\beta$ is a covering map and thus highly surjective in that there could be an arbitrary number of pre-images of $(0,1,\infty)$ and we need to perform Taylor series at all of these pre-images.
We can collect all the ramification indices into a so-called {\bf passport}, which is the list of ramification indices for each of these pre-images of 0, 1, and $\infty$.
Grothendieck's insight was to realize that the Bely\v{\i} map gives an embedded graph on $X$ as follows:
\begin{definition}
  Consider $\beta^{-1}(0)$, which is a set of points on $\Sigma$ that can be coloured as white, and likewise $\beta^{-1}(1)$, black.
  The pre-image of any simple curve with endpoints 0 and 1 on $\IP^1$ is a bipartite graph embedded in $\Sigma$ whose valency at a point is given by the ramification index (i.e., order of vanishing of Taylor series) on $\beta$.
  This is the {\bf dessin d'enfant}.
\end{definition}
Restricted by Riemann-Hurwitz, $\beta^{-1}(\infty)$ is not an independent degree of freedom, but is rather taken 1-to-1 to the faces in the bipartite graph, with the number of sides of the polygonal face being twice the ramification index.
The passport can then be written as
\begin{equation}
\left[
  r_0(1), r_0(2), \ldots, r_0(W) \ \big| \
  r_1(1), r_1(2), \ldots, r_1(B) \ \big| \
  r_\infty(1), r_\infty(2), \ldots, r_\infty(I)
  \right] \ , 
  \end{equation}
where $r_0(j)$ is the valency (ramification index) of the $j$-th white node, likewise $r_1(j)$, that for the $j$-th black node, and $r_\infty(j)$, half the number of sides to the $j$-th face.
The passport does not uniquely determine it since one further needs the connectivity between the white/black nodes, but it is nevertheless an important quantity.
The {\em degree} of the Bely\v{\i} map is the row-sum $d = \sum\limits_j r_0(j) = \sum\limits_j r_1(j) = \sum\limits_j r_\infty(j)$ and is the degree of $\beta$ as a rational function.
The collection of $(X,\beta)$ gives the {\em Bely\v{\i} pair} which can be used to represent a dessin.

Throughout this paper we will only be concerned with genus zero or {\bf planar dessins}.
That is, the Bely\v{\i} maps will be from $X  =\mathbb{P}^1$ (with coordinate $z$) to $\mathbb{P}^1$ (with coordinate $\zeta$) and will be a rational function in a single complex variable.
That is, 
\begin{equation}
\zeta  = \beta(z) = P(z) / Q(z) \quad : \mathbb{P}^1_z \to \mathbb{P}^1_\zeta
\end{equation}
with polynomials $P(z)$ and $Q(z)$. 

The dessins considered in this study are also {\bf clean} as well as planar, such that the degree of all black nodes is 2, these are the pre-images of 1. Of course, which colour we choose for pre-images of 0 or 1 is purely by convention and we adhere to that of colouring pre-images of 0 and 1 as white and black respectively. Moreover, 0, 1 and $\infty$ can be transformed freely amongst themselves by $SL(2; \IC)$ so we need to fix a convention for $\beta$ from the start as well.

\begin{wrapfigure}[16]{r}{0.35\textwidth}
    \vspace{-60pt}
    \begin{center}
    \includegraphics[width=60mm]{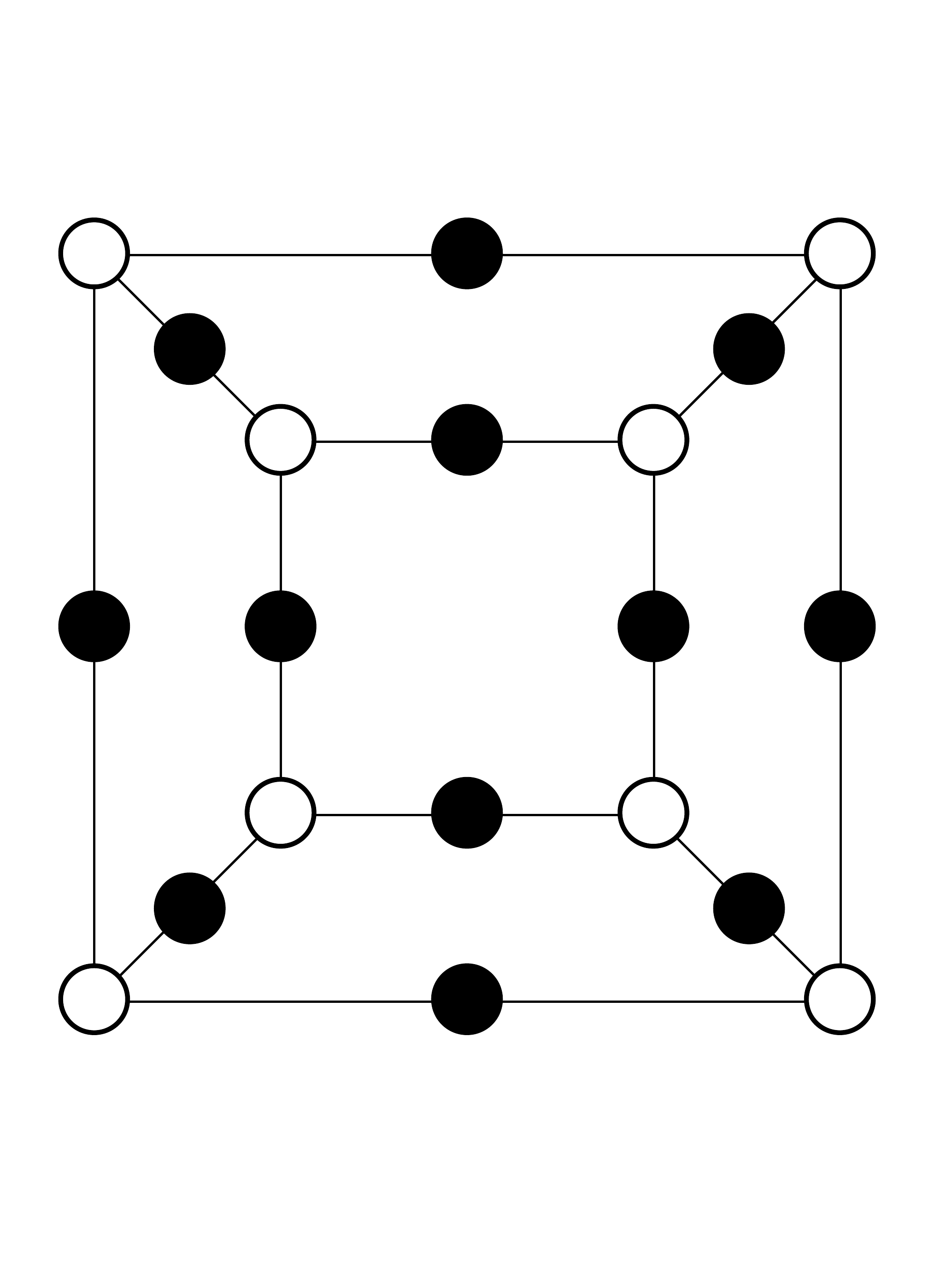} 
    \vspace{-35pt}
    \caption{An example of a dessin d'enfant. This dessin is of the type examined in this paper, it is a planar dessin, and also clean - as all same colour nodes have the same valency.}
    \label{example_dessin}
    \end{center}
\end{wrapfigure}

An example of a clean planar dessin is given in figure \ref{example_dessin}; this will be our running example in this section and the various mathematical and physical structures described in this paper will be explicitly shown as we proceed with this introduction.

\comment{
In the embedding process of a dessin, the white and black nodes are associated with the preimages 
of the points 0 and 1 respectively under the action of the Bely\v{\i} function. In addition the edges are associated to the preimage of the unit interval [0,\,1]. The ramification of the $\infty$ points are then inferred from the graph structure, as described in \S\ref{dessin->X label}. Since the ramification points represent degeneracy in the Bely\v{\i} function, the graph-theoretic degree of the nodes hence corresponds to the ramification degree of the general Riemann surface cover, $X$ \cite{Esquisse}.
}
The passport is $\begin{Bmatrix}
3^8 \\ 2^{12} \\ 4^6
\end{Bmatrix}$, where $3^8$ means there are 8 white nodes of valency 3, $2^{12}$ means there are 12 black nodes of valency 2 (this means that the dessin is clean in that all of one colour have valency 2). Finally, the $4^6$ means that there are 6 faces bounded by 8 edges - note that we are drawing on $\IP^1 \simeq S^2$ so there is a point at infinity for which there is an exterior face in addition to the 5 visible ones.
The triangulation of the dessin faces leads to a list of number of triangles for each face which is exactly twice each entry of the passport.
Of course, this example dessin is precisely the cube, drawn on a sphere.

The Bely\v{\i} pair corresponding to this dessin can be reproduced, and the mapping to $\IP^1$ gives the Bely\v{\i} function up to biholomorphism. Choosing a version equivalent to the Bely\v{\i} map provided in \cite{Beuke_Belyi_list}, we see that this dessin's Bely\v{\i} map is:
\begin{equation}\label{betaeg}
    \beta(z) = \frac{(z^8-14z^4+1)^3}{(-108z^4(z^4+1)^4)}\,,
\end{equation}
noting that this Bely\v{\i} map differs to the map provided in \cite{j_McKay} by a birational transformation. Calculating the preimages of $\{0,1,\infty\}$ under \eqref{betaeg}, the dessin in figure \ref{example_dessin} can be reproduced. The preimages of 0 correspond to roots of the Bely\v{\i} map numerator. The preimages of 1 correspond to roots of the polynomial formed from the difference of the Bely\v{\i} map numerator and denominator. Finally the preimages of $\infty$ are given by the limit $z\mapsto\infty$, or the roots of the Bely\v{\i} map denominator.
Furthermore, the ramification at preimage $z_0$ is the leading order of the Taylor series of $\beta(z)$ at $z_0$.
Summarised in table \ref{example_dessin_ramification_table}.

\begin{table}[h!]
\centering
\begin{tabular}{|l|l|l|l|}
\hline
\multicolumn{1}{|c|}{\begin{tabular}[c]{@{}c@{}}Images\\ $\beta(z)$\end{tabular}} & \multicolumn{1}{c|}{\begin{tabular}[c]{@{}c@{}}Preimages\\ $z$\end{tabular}} & \multicolumn{1}{c|}{\begin{tabular}[c]{@{}c@{}}No.~of\\ Preimages\end{tabular}} & \multicolumn{1}{c|}{\begin{tabular}[c]{@{}c@{}}Ram.~\\ Ind.\end{tabular}} \\ \hline
\multicolumn{1}{|c|}{0} & \multicolumn{1}{c|}{$\sqrt{2\pm\sqrt{3}}\,,-\sqrt{2\pm\sqrt{3}}\,,i\sqrt{2\pm\sqrt{3}}\,,-i\sqrt{2\pm\sqrt{3}}$} & \multicolumn{1}{c|}{8} & \multicolumn{1}{c|}{3} \\ \hline
\multicolumn{1}{|c|}{1} & \multicolumn{1}{c|}{\begin{tabular}[c]{@{}c@{}}$\pm 1\,,\pm i\,,$\\ $\pm\bigg(1+\frac{1}{\sqrt{2}}+i\sqrt{\frac{1}{2}(3+ 2\sqrt{2}}\bigg)\,, \pm\bigg(1+\frac{1}{\sqrt{2}}-i\sqrt{\frac{1}{2}(3+ 2\sqrt{2}}\bigg)\,,$\\ $\pm\bigg(1-\frac{1}{\sqrt{2}}+i\sqrt{\frac{1}{2}(3- 2\sqrt{2}}\bigg) \,, \pm\bigg(1-\frac{1}{\sqrt{2}}+i\sqrt{\frac{1}{2}(3- 2\sqrt{2}}\bigg)$\end{tabular}} & \multicolumn{1}{c|}{12} & \multicolumn{1}{c|}{2} \\ \hline
\multicolumn{1}{|c|}{$\infty$} & \multicolumn{1}{c|}{$ 0\,, \pm \frac{1}{\sqrt{2}}\big(1+i)\,, \pm \frac{1}{\sqrt{2}}\big(1-i)\,,\infty$} & \multicolumn{1}{c|}{6} & \multicolumn{1}{c|}{4} \\ \hline
\end{tabular}
\caption{The preimages of $\{0,1,\infty\}$, along with their multiplicities, for a Bely\v{\i} map corresponding to congruence modular subgroup $\Gamma(4)$.}
\label{example_dessin_ramification_table}
\end{table}
In each case the number of preimages and the multiplicities agree with this example dessin's passport. Therefore producing the 8 trivalent white nodes, 12 bivalent black nodes, and 6 faces (including the outer face) of the dessin in this example.

\subsubsection{Dessin Transformations and Galois Representation} \label{dessin->X label}
It is through Bely\v{\i}'s theorem that dessins provide the surface they are embedded in with an obvious Riemann structure, since through interpretation of the dessin, the surface $X$ can be described via algebraic curves with coefficients as algebraic numbers. 
The key to Grothendieck's {\it Esquisse} is that the combinatorial interpretation of dessins d'enfants is related to the absolute Galois group over the algebraic rational numbers. This group describes how these algebraic curves (and hence dessins also) transform into each other \cite{Lando}. This interpretation of dessins d'enfants is as an object to represent the combinatorial invariants associated with the action of the {\bf absolute Galois group} 
\footnote{
In defining the Galois group, \textit{Galois extensions} of a base field are considered. These are algebraic field extensions which are both normal, and separable \cite{Edwards, Cox}. A Galois group is then the group of all automorphisms of a Galois extension which fix the base field. Beyond these the \textit{absolute} Galois group of a field, requires a specific extension of the base field known as the separable closure of the field. Note that since the base field considered, $\mathbb{Q}$, is perfect, the separable closure is equal to its algebraic closure denoted $\overline{\mathbb{Q}}$.
}
of the rational numbers, $Gal\big(\,\overline{\mathbb{Q}}\,/\,\mathbb{Q}\,\big)$.
The group  $Gal\big(\,\overline{\mathbb{Q}}\,/\,\mathbb{Q}\,\big)$, has no direct description, it is hence through Bely\v{\i}'s theorem that dessin d'enfant graphs become of interest, to study this group \cite{Zapponi, Cox}.

The important relation which the theory of dessins capitalises on is an equivalence between several categories. The equivalence relates the category of embedded dessin graphs with: finite sets under permutation; field extensions; and certain types of algebraic curves. Since field extensions are directly related to the definition of the absolute Galois group, it is through this categorical equivalence that representations of the Galois group act {\em faithfully} on dessins d'enfants. This is what makes dessins so useful in studying the elusive $Gal\big(\,\overline{\mathbb{Q}}\,/\,\mathbb{Q}\,\big)$ group.

Triangulation of the dessin allows the original Riemann surface's algebraic curve from the Bely\v{\i} pair to be reproduced up to homeomorphism. This process is further explained in appendix \ref{dessin->riemann}. This triangulation process also directly demonstrates the category equivalence to finite sets under permutation, by defining reflection operators over the triangles produced.
This category equivalence in part inspired the second representation method for the dessins used in these investigations. Under each dessin's triangulation, action of a permutation in effect changes the orientation of a constituent triangle based on the edge positions. Thus one representation method used for the dessins (introduced in \S\ref{cyclic_rep}), reformulates the information of the constituent triangles' orientations, in terms of cyclic edge lists around the nodes.

Reconsidering the absolute Galois group, action of its elements may change the subfields of the algebraic completion which the Bely\v{\i} functions are defined over, this maps Bely\v{\i} pairs to each other, and hence corresponds to moving around orbits of dessins. The orbit of a dessin is all dessins that the absolute Galois group can transform the original dessin into. The largest orbit in the dataset we consider corresponds to a quartic extension of the rationals, as shown in appendix \ref{dataset_appendix}, corresponding to passport: 12-5-3-2-1-1. The minimal polynomial associated with the quartic root used in the extension hence has 4 distinct roots (since the extension is also separable). Therefore each dessin in this Galois orbit corresponds to an extension of the rationals by one of these 4 roots. The root corresponding to each of these dessins is used in defining the coefficients of its Bely\v{\i} map.

The dessins therefore act as tools to identify invariants and other properties of the rational absolute Galois group. Invariants such as degree sequence (the list of node degrees), and the stabiliser of the orbit act as important insights from the theory of dessins into study of this group \cite{Lando}. Importantly the size of the orbit of a dessin corresponds to the degree of the field extension. This degree of field extension is an invariant under Seiberg duality, and hence acts as a useful test for QFT duality, the physical interpretation of this degree is further examined in \cite{Hanany:2011bs}. It is classification according to this degree which machine-learning is used for in this study.

\subsection{Modular K3 Surfaces}\label{dessinCY_section}
Having introduced dessins, let us move on to two seeming unrelated subjects: subgroups of $PSL(2,\mathbb{Z})$ and K3 surfaces.
It is remarkable that they both give rise to dessins.

\subsubsection{Subgroups of the Modular Group $PSL(2,\mathbb{Z})$}\label{congruence_section}
\comment{
The dessins considered in this paper, arise through analysis of the modular group, $PSL(2,\mathbb{Z})$. This group represents fractional linear transformations on the upper half complex plane, $\mathscr{H}$, acting alike M\"obius transformations previously discussed in \S\ref{dessin_section}.
}
Consider the modular group $\Gamma = PSL(2,\mathbb{Z})$, which is an infinite group generated by two elements $\Gamma := \langle S, T | S^2 = (ST)^3 = I \rangle$.
The most familiar subgroup is the principal congruence subgroup\footnote{
More technically, a principle congruence subgroup is the kernel of a reduction modulo $n$ morphism, i.e., it is what maps to the identity under $\pi_n: SL(2,\mathbb{Z}) \longmapsto SL(2,\mathbb{Z}/n\mathbb{Z})$. A congruence subgroup is then any subgroup which contains a principle congruence subgroup for some $n \geq 1$; additionally the smallest $n$ such that the $n$th principle congruence subgroup is contained is known as the level of the subgroup \cite{Shimura}. 
}
\begin{equation}\label{gammam}
\Gamma(m) := \{A \in \Gamma : A_{ij} = I_{ij} \mod m \} \ .
\end{equation}

The Cayley graph (where elements of the groups are nodes and there is a link between two nodes if there exists a group multiplication taking one to another) for $\Gamma$ is a freely generated trivalent tree, with each trivalent node replaced by a triangle.  The triangles correspond to $S^3 = I$ and the edges, to $T^2 = I$.
For any finite index subgroup $\Gamma'$ of $\Gamma$ we can let nodes be cosets of $\Gamma'$ within $\Gamma$ and likewise create a graph, this time finite (number of nodes is equal to the index / the number of cosets). This is the Schreier-Cayley coset graph for $\Gamma'$ and is a finite trivalent graph.  In fact, every finite trivalent graph can be generated this way: one could find some appropriate $\Gamma' \subset \Gamma$.
The point is that all Schreier-Cayley coset graphs are {\it clean dessins}, by placing a black node on every edge.
For instance, our running example\footnote{There is a peculiar ADE story here, the coset graphs for $\Gamma(m)$, $m=3,4,5$, are precisely the tetrahedron, the cube/octahedron and the dodecahedron/icosahedron.
} of figure \ref{example_dessin} is the coset graph for the congruence subgroup $\Gamma(4)$.

In \cite{Sebbar}, all torsion-free, genus zero\footnote{
The genus here means the genus of modular curve formed from the quotient of the subgroup's action on the upper half plane $\mathscr{H}$. Genus zero means that $\mathscr{H}$ quotiented by the subgroup is topologically equivalent to a Riemann sphere.
}, congruence subgroups were classified.
There are only 33 of them. 
The conjugacy class decomposition of the modular group using these subgroups leads to indexes: $\{6,12,24,36,48,60\}$, where the restriction to a multiple of 6 is shown through use of the Riemann-Hurwitz formula (see \cite{YMR} for a detailed review). 

The action of the modular subgroups can be extended from $\mathscr{H}$ to $\mathscr{H} \times \mathbb{C}$ such that:
\begin{equation}
    (\tau,z) \longmapsto \bigg(\gamma \tau, \frac{z+m\tau +n}{c\tau + d}\bigg)\,, \quad \text{for} \; \gamma = \begin{pmatrix} a & b\\ c & d \end{pmatrix},\; m,n \in \mathbb{Z}\,,
\end{equation}
where $\gamma$ denotes action of an element of one of the subgroups. Now taking the quotient of $\mathscr{H} \times \mathbb{C}$ by this extended automorphism gives the modular curve from before the extension with an elliptic fibration to form the unique 'modular surface' for that subgroup, called the Shioda surface. 
The index of the subgroup used gives the Euler number of the corresponding modular surface \cite{YMR}.

The modular surfaces formed from the index 24 torsion-free genus zero congruence subgroups are K3 surfaces, whilst those from index 36 subgroups are Calabi-Yau 3-folds (although these are instead fibrations over $\mathbb{P}^2_\mathbb{C}$) \cite{YMR}. 
Higher dimensional Calabi-Yau surfaces have more complicated $j$-invariants, and any Bely\v{\i} property is not as obvious, hence in this study we focus on these K3's.
As an aside however, from here dessins may also be considered to nicely arise as the structure of brane tilings formed from these Calabi-Yau manifolds when used in Type IIB superstring theory \cite{BYHP}. Whilst also the extremal elliptic K3 (modular) surfaces arise in F-theory \cite{Kimura_Ftheory,Kimura_F_HetII}.
The point is that whilst only 9 of the 33 torsion-free congruence subgroups are index 24 and give K3 surfaces, it was realized in \cite{YMR} that if you relax the condition that the subgroup has to be defined by some congruence (mod) relation, then you can obtain all extremal K3 surfaces, to which we now turn.

\subsubsection{Dessins from \textit{j}-maps}\label{j-maps_section}
Now, a surface elliptically fibred over $\IP^1$ takes the form of a Weierstra{\ss} equation:
\begin{equation}\label{weierstrass_eqn}
    y^2 = 4x^3 - g_2(z)x - g_3(z)\,,
\end{equation}
where the coefficients $g_2$ and $g_3$ are functions of the base Riemann sphere's complex coordinate $z$.
The modular discriminant is defined
\begin{equation}
    \Delta \vcentcolon = g_2(z)^3 - 27g_3(z)^2\,,
\end{equation}
and as a discriminant it indicates the degeneracy of roots of the cubic part of the Weierstra{\ss} equation. 
Modular forms are particularly useful in creating Galois representations, as well as appearing in many other useful areas of mathematics also. In general they are holomorphic functions on $\mathscr{H}$, and satisfy the condition
\begin{equation}
    f\bigg(\frac{az+b}{cz+d} \bigg) = (cz+d)^k f(z)\,,
\end{equation}
for weight $k$. This condition shows the response of a modular form to action of the modular group on its input. Using the modular discriminant, a specific weight zero modular form can be defined, this is known as the \textbf{j-invariant}:
\begin{equation}\label{jmap}
    j(z) \vcentcolon = \frac{g_2(z)^3}{\Delta} = \bigg(\frac{g_2(z)^3}{g_2(z)^3 - 27g_3(z)^2}\bigg)\,,
\end{equation}
which is a modular function (in fact the unique one that generates the field of modular-invariants).
Since the form is weight zero it is hence invariant under action of the modular group. These \textit{j}-invariants for the K3 modular surfaces in consideration can be considered as endomorphisms of the Riemann sphere, which are ramified at $\{0,1,\infty\}$.  Note that some definitions of $j(z)$ may include an additional factor of 1728, requiring a further M\"obius transformations to convert the second ramified point from $1728 \mapsto 1$ \cite{j_McKay}.

The key observation is that due to the ramification structure of \eqref{jmap}, these \textit{j}-invariants are Bely\v{\i}! This striking property is discussed and proved in \cite{Miranda_Persson}.
To get a sense for this structure consider for $j(z) = 0$, this requires $g_2(z) = 0$, and the cubic dependence on it in the invariant function makes each root into a three-fold ramification point; these are the white nodes of the dessin. Equivalently $j(z) = 1$ requires $g_3(z) = 0$, and the square dependence here makes each root into a two-fold ramification point; the black nodes. Finally the $j(z) = \infty$ singularities require the modular discriminant $\Delta = 0$ (or $z\mapsto\infty$).
Thus, each modular surface gives a dessin via its j-invariant.

In particular, the degree of the \textit{j}-invariant's numerator equals the surface's index (here 24). So that the $z\mapsto\infty$ limit remains a singularity we require the denominator to have a lower leading order, hence the $g_2^3$ and $g_3^2$ factors must be of the same degree (both being 24), so the leading power of $g_2^3$ can be negated. This makes the invariants $g_2$ and $g_3$ of degree 8 and 12 respectively, leading to 8 trivalent white nodes, and 12 bivalent black nodes in these dessins. 

Henceforth, we will focus on {\bf extremal K3 surfaces} where all singular fibres are of Kodaira type $I_n$ and there are exactly 6 of these fibres (q.v.~\cite{Miranda_Persson} and \cite{YMR} for further discussions).
These 6 Kodaira singularities act as a 6-part partition of the mapping's degree (which is 24), giving the passport information. These 6 ramifications induce the 6 faces of the dessins (including the outside face, since the dessin is truly drawn on a sphere). 
There are 112 elliptic modular K3 surfaces which are extremal (indeed, there are 199 6-partitions of 24 but \cite{Miranda_Persson} showed only 112 are allowed) and we will focus on these in this paper.
Note that if dessins are in the same Galois orbit they necessarily have the same passport (but not vice versa).

For instance, our example in figure \ref{example_dessin} corresponds to 
\begin{equation}
\begin{split}
    g_2(z) & = z^8-14z^4+1\,,\\
    g_3(z) & = \frac{1}{3\sqrt{3}}\big(z^{12}+33z^8-33z^4-1\big)\,.
\end{split}
\end{equation}
Using these functions in \eqref{weierstrass_eqn}, the Weierstra{\ss} equation for the modular surface is:
\begin{equation}
    y^2 = 4x^3 - \big(z^8-14z^4+1\big) x - \frac{1}{3\sqrt{3}}\big(z^{12}+33z^8-33z^4-1\big)\,,
\end{equation}
which then gives the defining equation of the K3 surface as a hypersurface in the anticanonical bundle of $\IP^2_{[x:y:z]}$ via the Weierstrass equation \eqref{weierstrass_eqn}.
This is the modular surface which corresponds to the extended quotient action of the congruence modular subgroup $\Gamma (4)$ on $\mathscr{H}$.

\subsection{The Dessin Database} \label{s:data}
In summary, we study 112 extremal semi-stable K3 elliptic fibrations, in which case the modular K3 surfaces have index 24. The j-invariant of each is a Bely\v{\i} map with passport
\begin{equation}
\begin{Bmatrix}
3^W \\ 
2^B \\ 
n_1^{a_1},n_2^{a_2},...,n_k^{a_k}
\end{Bmatrix}\;,\quad
\begin{matrix}
W & = & \text{number of preimages of 0} &=& 8\\ 
B & = & \text{number of preimages of 1} &= &12 \\ 
\{n_i^{a_i}\} & = & \text{cusp widths of elliptic modular K3 surface} &&
\end{matrix}
\end{equation}
such that
\begin{equation}
    \sum_i a_i = 6\,,\quad \sum_i a_i \cdot n_i = 24\,.
\end{equation}
Each dessin is planar, trivalent, and clean and corresponds to a Schreier-Cayley coset graph of a particular subgroup of the modular group $PSL(2; \mathbb{Z})$.
Of these, 9 are congruence subgroups, including the principal ones $\Gamma(m)$, $m=3,4,5$.

All such K3 surfaces and associated dessins were classified in \cite{Miranda_Persson} and \cite{MirandaPersson_table}, the cartographic and modular subgroups were computed in \cite{YMR}.
The reader is also referred to the website \cite{Beuke_Belyi_list} for the hand-drawn dessins.
For reference, we include all dessins (along with their field extensions, adjacency matrices, and cyclic edge lists) in appendix \ref{dataset_appendix}. Of these surfaces only 9 correspond to congruence subgroups, the remainder correspond to more general subgroups \cite{YMR, Miranda_Persson}.

One might note that there are more than 112 diagrams and there are, in fact, 191 dessins in the dataset. There are 112 distinct passports in this dataset, however the dessins sort themselves into 125 orbits of varying degree extension, where the degree of extension is equal to the size of the orbit. In some cases, for example 8-8-3-3-1-1 in appendix \ref{dataset_appendix}, there are multiple orbits per passport (here one orbit of size one - with no degree extension, and one orbit of size two - with quadratic degree extension); and this accounts for the extra 13 orbits. 

Within orbits some graphs are isomorphic, this is where some roots of the minimal polynomial are complex. Since the minimal polynomial is defined over the field of (real) rational numbers, any root with non-zero imaginary part also has its complex conjugate as a root. These pairs of conjugate roots of the polynomial correspond to dessins which are isomorphic, but chirality flipped. For example in appendix \ref{dataset_appendix}, considering passport: 7-6-5-3-2-1, there is one orbit associated to this passport which is cubic, it hence has 3 dessins belonging to it, corresponding to the 3 roots of a cubic polynomial (the minimal polynomial). As the polynomial has only real rational coefficients, its roots must be decomposed into: one real root, and one pair of complex conjugate roots (note there could have been 3 distinct real roots, but here this is not the case). The conjugate pair gives the isomorphic B $\&$ C dessins, which share the same adjacency matrix. To further exemplify this, note that in some cases (7-7-3-3-2-2) the polynomial has distinct real roots giving non-isomorphic dessins, but in others (7-7-4-4-1-1) the roots are a complex conjugate pair giving isomorphic dessins, but flipped. As explained in \S\ref{dessin->X label}, dessins are embedded graphs so how they are drawn on the Riemann sphere makes a difference.

In addition, isomorphisms may also occur between dessins in different orbits. For example consider in appendix \ref{dataset_appendix}, the dessins labelled (via their passports) 10-8-3-1-1-1 and 11-6-4-1-1-1 A, these dessins are isomorphic in a graph-theoretic context as they are drawn the same except with the edge connecting nodes 3 \& 7 in a different position. In terms of graph theory these dessins are hence isomorphic, and although these isomorphisms out of orbit are rare, this nevertheless causes a subtlety in using the adjacency matrix representation method for the dessins. Both subtleties associated to graph isomorphism, in and out of orbit, are rectified with the cyclic representation method which is sensitive to the edge placement.

Removing repeated matrices due to isomorphisms leads to 152 distinct matrices in the dataset, corresponding to the 152 dessins unique up to this isomorphism within their own orbit (where isomorphisms occur out of orbit both matrices are removed). Whereas for the cyclic edge list representation all 191 lists are independent and can all be used.

Since these dessin objects are drawn as graphs, it is logical to first try learning this representation method for them based on the graph-theory standard of using adjacency matrices. In the rare cases where dessins in different orbits had the same adjacency matrix (including up to node relabellings), these dessins were be removed from the dataset. Therefore under this representation method, since our dessins are clean, it suffices to only draw the white nodes and the graph is, at least combinatorically, completely captured by $8 \times 8$ adjacency matrices, which we will define shortly.
These are shown next to the dessins in appendix \ref{dataset_appendix}, and will constitute the first data input style to be analysed in \S\ref{ML_results}. The second data input style represents dessins by cyclic lists of the edges surrounding each node, as discussed further in section \S\ref{cyclic_rep}. This is believed to be a faithful representation method, and hence all 191 dessins are included in this dataset.

As mentioned, one key feature of a dessin d'enfant is the algebraic field in which the algebraic model for the underlying Riemann surface $X$ (as well as the Bely\v{\i} map) lives.
Oftentimes, the minimal polynomial defining the extension, even for relatively innocuous looking dessins, could be enormous and because dessins are number-theoretic, the coefficients in these polynomials are completely rigid and are very precise integers (giving the roots as specific algebraic numbers).
For example, for the relatively simple brane-tiling of the so-called $X^{3,0}$ toric gauge theory, the defining polynomial's coefficients were found \cite{Hanany:2011bs} to be integers on the order of $10^{200}$.

Importantly, for our dataset, the dessin Bely\v{\i} maps are defined over $\mathbb{Q}$, or some extension of $\mathbb{Q}$ involving a square, cubic, or quartic root.
This means that the extension degree of all our dessins are 1, 2, 3, or 4 (where $1$ means $\mathbb{Q}$ itself).
Where the field is an extension of $\mathbb{Q}$, the Galois action maps between the roots of the polynomial defining the extension and in each case this corresponds to a different dessin. 
Hence the number of dessins corresponding to each passport equals the sum over the orbits' field extension degrees for all orbits with this passport ramification data. For example, in appendix \ref{dataset_appendix}, passport: 10-10-1-1-1-1, has 3 dessins; it also has two orbits: one degree 2 from the quadratic extension, and one degree 1 where there is no extension, satisfying the $3=2+1$.

\subsection{Seiberg-Witten Theory}\label{SW_section}
In this final subsection of setting the stage, we come to the physics of our dessins.
\subsubsection{Sieberg-Witten Curves}
Seiberg-Witten theory \cite{Seiberg:1994rs} considers the low energy IR limit of $\mathcal{N}=2$ supersymmetric gauge theories. More specifically it allows for exact description of the coulomb branch vacuum manifold by defining it in terms of a hyperelliptic curve - the 'Seiberg-Witten curve'.
Now, points of the $\mathcal{N}=2$ vacuum manifold defined by the hyperelliptic curve can be lifted to $\mathcal{N}=1$ vacua through deformation with a tree level superpotential. At these special points, which are singularities of the manifold, dessins naturally arise as combinatoric representations of the root structure of the Seiberg-Witten hyperelliptic curve. A correspondence can hence be considered between phases of the $\mathcal{N}=1$ vacua and classes of dessins in Galois orbits \cite{SW_curves}.

For a general $U(N)$ gauge theory with $L$ massive scalars (masses $m_i$ respectively), the Seiberg-Witten curve takes the form:\\
\begin{equation}\label{SW_eqn}
    y^2 =\; \langle \det(z I-\Phi) \rangle ^2 - \;4\Lambda^{2N-L}\prod_{i=1}^L (z+m_i)\,,
\end{equation}
with $\Phi$ the adjoint scalar in the coulomb branches vector multiplet, $\Lambda$, a cut-off measure which controls the energy limit of the theory \cite{SW_curves, Tachikawa}, and the deformation term (second on the RHS) is prescribed by the $\cN=1$ superpotential.

The insight of \cite{SW_curves} is that \eqref{SW_eqn} can be identified with Bely\v{\i} maps of clean dessins by first making the association:
\begin{equation}
\begin{split}
    P_N(z) &\vcentcolon = \; \langle \det(z I -\Phi) \rangle \,,\\
    B(z) &\vcentcolon = - \;4\Lambda^{2N-L}\prod_{i=1}^L (z-m_i)\,.
\end{split}
\end{equation}
In addition another function $A(z)$ is defined, such that it has an equivalent form to the function $B(z)$, which can more generally be written
\begin{equation}\label{AB_eqn}
    A(z) = \prod_i J^i_{u_i}(z) \,,\quad B(z) = \prod_i G^i_{v_i}(z) \,,
\end{equation}
for $J_{u_i}$ and $G_{v_i}$ as sets of polynomials of degrees $u_i$ and $v_i$ respectively. 
These functions are designed so as to satisfy the relation: $A-B=P_N^2$. Then, the map defined by
    \begin{equation}\label{Belyi_map_form}
    \beta(z) \vcentcolon = \frac{A(z)}{B(z)} = 1+\frac{P_N^2(z)}{B(z)}\,,
\end{equation}
is Bely\v{\i}.
The preimages of 0 correspond to the roots of $A(z)$, due to the form of $A(z)$ defined in \eqref{AB_eqn} each of the $J_{u_i}$ polynomials of degree $u_i$ will lead to $u_i$ roots, and as the whole polynomial is raised to the power of $i$ in $A(z)$'s definition, each root/ramification point will have valency $i$ also. Correspondingly the roots of $P_N(z)$ give the Bely\v{\i} map preimages of 1, and since it is squared in the Bely\v{\i} map definition these points will have valency 2 - giving clean dessins exclusively. Finally the preimages of $\infty$, the final ramification point of the Bely\v{\i} map, are given by the roots of $B(z)$ (as well as the $A(z)\mapsto \infty$ limit).

A general Seiberg-Witten (SW) curve does not factorise as simply as represented by \eqref{AB_eqn}. However tuning the superpotential parameters can cause the roots of the functions to coincide, giving this polynomials raised to powers structure. It is at these isolated singularities along the $\mathcal{N}=1$ branches caused by the superpotential tuning that the dessins are formed from the usual branch-less structure of the roots, since the roots now have higher valencies. 

Specifically, for the dessins we are considering, $A(z)$ and $B(z)$ take the form:
\begin{equation}
    A(z) = J_8^3\,,\quad B(z) = G_{12}^2\,,
\end{equation}
to match the form of the passport given in \S\ref{Belyi_pair_label}.

In the physical theory at these points the superpotential has triggered a magnetic Higgs mechanism. Here the symmetry breaking of the full gauge group for the $\mathcal{N}=2$ Seiberg-Witten curve leaves a residual symmetry group preserving a submanifold of the curve which corresponds to one of the $\mathcal{N}=1$ vacua. Thus, there are number-theoretic special points in $\mathcal{N}=1$ vacua. Indeed, \cite{He:2012kw} further associates elliptic K3 surfaces and modular congruence subgroups to these distinguished points.

A primary aim in the field of Galois theory is to find invariants which can distinguish between different orbits of $Gal\big(\,\overline{\mathbb{Q}}\,/\,\mathbb{Q}\,\big)$. Since these orbits correspond to orbits of dessins, they thus also provide tools for distinguishing between different branches of the $\mathcal{N}=1$ vacua. The tools they relate to are holonomy measures for the gauge connections of the residual symmetry groups occurring at these $\mathcal{N}=1$ vacua. Through this connection if there is an exact correspondence between all invariants distinguishing Galois orbits, and order parameters which distinguish vacua branches, there is potential to draw physical insight from the use of dessins more explicitly in the physical theories \cite{SW_curves, CSW}.

\subsubsection{Dessins to Seiberg-Witten Curves}\label{dessin_to_SW}
Thus armed, we can now put our foregoing discussions together and algorithmically follow the process laid out in Cachazo et al. \cite{SW_curves} to convert the Bely\v{\i} maps of the relevant dessins, some of whose Bely\v{\i} maps have already been computed by F.~Beukers \cite{Beuke_Belyi_list}, into Seiberg-Witten curves.
 
We reverse the procedure in \eqref{AB_eqn} and \eqref{Belyi_map_form}, and for convenience set
$\alpha := -4\Lambda^{2N-L}$.
Then starting with \eqref{Belyi_map_form}, where $A(z)$ and $P_N(z)$ are monic polynomials and $B(z)=\alpha \prod_i^L (z-m_i)$ with $m_i$ as the complex roots of $B(z)$.
The SW curve is then given by:
\begin{equation}
    y^2=P_N(z)^2+\alpha\prod_i^L (z+m_i)\,.
\end{equation}
The polynomials for the Bely\v{\i} maps of dessins with rational coefficients are provided in reference \cite{Beuke_Belyi_list} with the following correspondence to the polynomials listed above:
\begin{equation}
\begin{split}
    A(z)&=x^3(z)/a\,,\\
    B(z)&=k(z)/a\,,\\
    P_N^2(z)&=y^2(z)/a\,,\\
\end{split}
\end{equation}
where $a$ is a constant that may be required to make the polynomials $A(z)$ and $P_N(z)$ monic. The algorithm for this computation is given by Algorithm 1.
\newpage
\noindent\rule{\textwidth}{1pt}
\textbf{Algorithm 1:} Algorithm taking Bely\v{\i} maps and calculating the equivalent SW curve\\
\rule{\textwidth}{1pt}
$\quad$\textbf{Input:} $x(z)$ and $k(z)$\\
\textbf{Output:} $y^2$, the SW curve corresponding to the Bely\v{\i} map $\frac{x^3}{k}$\\
$q(z)=x^3(z)-k(z)$\\
$a=\text{coefficient in front of the highest order term in $q(z)$}$\\
$P_N^2(z)=\frac{q(z)}{a}$ // Now $P_N$ is a monic polynomial\\
$B(z)=\frac{k(z)}{a}$ // Now the polynomials are in the form required by the correspondence above\\
$\alpha=\text{coefficient in front of the highest order term in $B(z)$}$\\
$B'(z)=\frac{B(-z)}{\alpha}$\\
$B'(z)=\pm B'(z)$ // Needed to keep the leading order term positive\\
$y^2=P_N^2(z)+\alpha \cdot B'(z)$\\
\noindent\rule{\textwidth}{1pt}\\

For our running example of figure \ref{example_dessin},
we make use of \eqref{betaeg} and \eqref{SW_eqn} and substitute the relation: $P_N^2(z) = A(z)-B(z)$. Then noticing that since the roots of $B(z)$ (given by the preimages of $\infty$ in table \ref{example_dessin_ramification_table}, excluding $\infty$) are either 0 or occur in $\pm$ pairs, then the form swap
\begin{equation}
    B(z) = \alpha \prod_{i=1}^L (z-m_i)\longmapsto \alpha \prod_{i=1}^L (z+m_i) = B(z)\,,
\end{equation}
reproduces $B(z)$. This makes the Seiberg-Witten curve
\begin{equation}
    y^2 = (A(z)-B(z))+B(z) = A(z) = (z^8-14^4+1)^3\,,
\end{equation}
as given in appendix \ref{SW_appendix}. The Bely\v{\i} maps have been computed by \cite{Beuke_Belyi_list} for the cases where the field extension is degree 1 (i.e., all coefficients in $\beta$ are rationals and scalable to integers).
We start from these and obtain the list of SW curves, as presented in appendix \ref{SW_appendix}.

\section{Classifying the Dessin Extensions}\label{ML_results}
We now come to the discussion of our computations, where simple fully connected Neural Networks (NNs), with no knowledge of the underlying mathematics, were created that had some success in predicting a dessin's extension degree.
\newpage
\subsection{Data Engineering}
\subsubsection{Adjacency Matrices}
\begin{wrapfigure}[15]{r}{0.5\textwidth}
    \vspace{-130pt}
    \begin{center}
    \includegraphics[width=60mm]{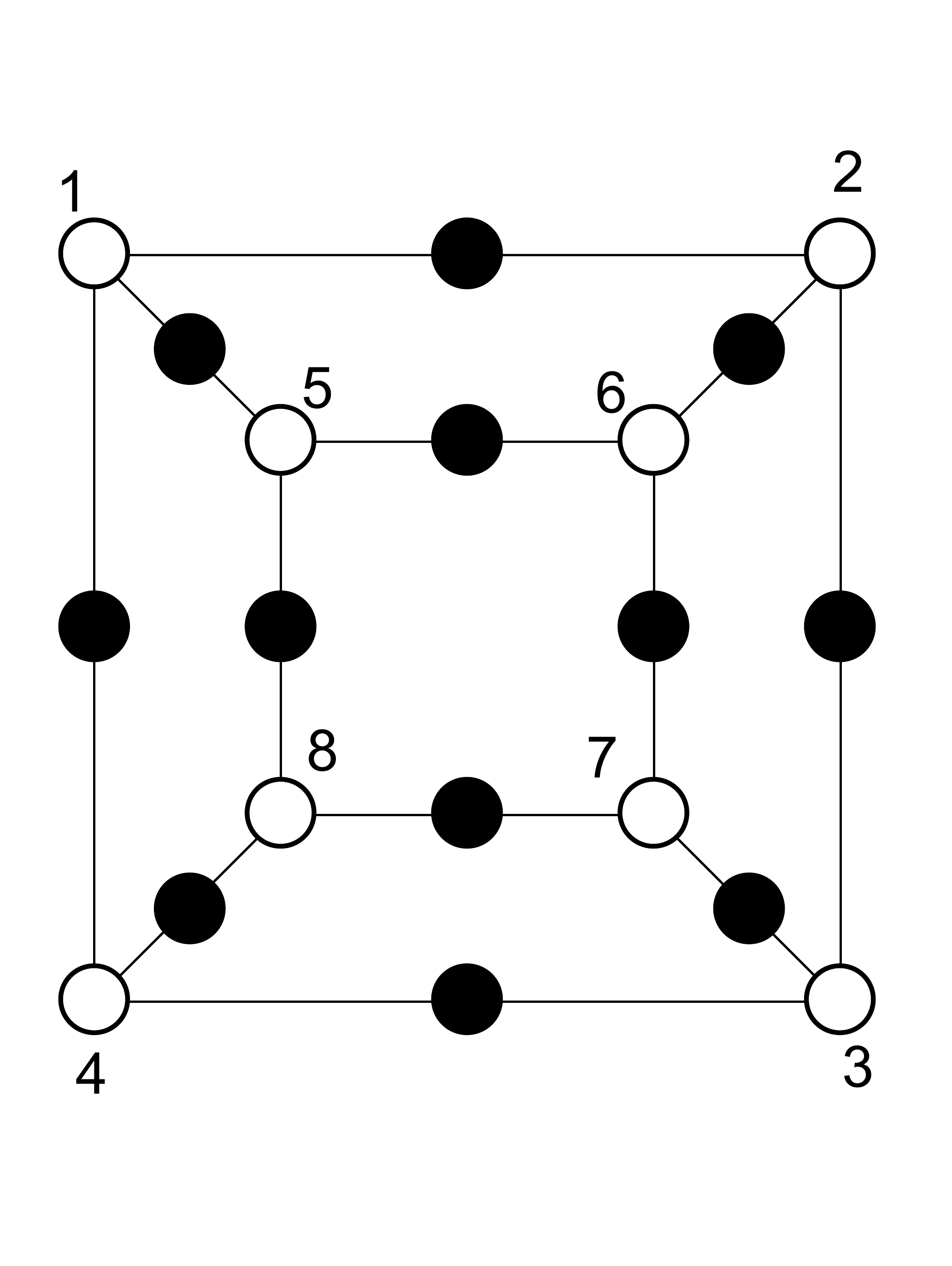}
    \vspace{-40pt}
    \caption*{$\begin{pmatrix}
        0 & 1 & 0 & 1 & 1 & 0 & 0 & 0\\ 
        1 & 0 & 1 & 0 & 0 & 1 & 0 & 0\\
        0 & 1 & 0 & 1 & 0 & 0 & 1 & 0\\
        1 & 0 & 1 & 0 & 0 & 0 & 0 & 1\\
        1 & 0 & 0 & 0 & 0 & 1 & 0 & 1\\
        0 & 1 & 0 & 0 & 1 & 0 & 1 & 0\\
        0 & 0 & 1 & 0 & 0 & 1 & 0 & 1\\
        0 & 0 & 0 & 1 & 1 & 0 & 1 & 0
    \end{pmatrix}$\\}
    \caption{The matrix representation of our example dessin, with white nodes labelled, and the corresponding adjacency matrix shown.}
    \label{dessin_example_mat}
    \end{center}
\end{wrapfigure}

We begin by setting up the data. In our first investigation we use adjacency matrices to represent the dessins, since NNs require a tensor input and this graph-theoretic format is the most obvious (despite subtle drawbacks associated to isomorphisms).
Because the dessins are all clean, the information from the black nodes can be suppressed (each black node and its two adjacent edges are contracted into a single edge) and we only need to know how the white nodes are linked combinatorically (in fact, each white node has valency 3 by construction).
Thus, graph-theoretically, each dessin may be described by adjacency matrices $\mathcal{M}$, defined such that:

 \begin{itemize}
     \item Each white node for a given dessin is labelled between $1$ and $8$.
     \item If a single edge connected nodes $i$ and $j$ then a $1$ was entered in the elements\\
     $\mathcal{M}_{ij}$ and $\mathcal{M}_{ji}$.
     \item If there were 2 edges connecting these nodes then a 2 would be entered into the corresponding element. 
     Also if an edge connected a node to itself then this would count as 2 connections and hence a 2 was entered in the element $\mathcal{M}_{ii}$.  
 \end{itemize}
As the dessins are not directed, the adjacency matrices are symmetric. An example of this adjacency matrix representation is given in figure \S\ref{dessin_example_mat} for our example dessin. The full dessin dataset, along with the corresponding matrices are further shown in appendix \ref{dataset_appendix}.

We make a few remarks.
First, $\cM$ is defined only up to permutation via relabelling the nodes, so we need to take this into account.
Next, we see from the appendix that multiple dessins could have the same $\cM$.
In the majority of cases these are precisely Galois conjugates so they have the same extension degree.
Therefore, there is only minor ambiguity in this form of dessin representation and our classification problem. In the cases of duplicate matrices, these occurrences are removed from the dataset.

As is well established, the effectiveness of a NN is almost always improved by increasing the size of the dataset used to train it \cite{data_size_1,data_size_2}. To artificially increase the size of the dataset from 152 distinct dessins matrices to $\sim$200,000 we precisely exploit the freedom in choice of relabelling the nodes. In the process of increasing the dataset size in this manner, for each additional matrix a random permutation of the numbers 1 to 8 was generated, and accordingly the matrix row \& columns were swapped to reflect this new ordering. This was repeated for $\sim$1000 of the 8! permutations possible for each of the 152 dessin matrices in the dataset. Where a permutation reproduced the same matrix (due to some symmetry in the dessin), this permutation was ignored and another generated to maintain the same number of permuted matrices per dessin.

The proportion of the dataset corresponding to a particular extension was as follows: rational 48.7$\%$, quadratic 30.9$\%$, cubic 19.1$\%$, quartic 1.3$\%$. Which remained approximately the same after data extension. These proportions were inverted and used as class weights in the training, encouraging the NN to pay proportionately more attention to classification of lower frequency classes, improving the true effectiveness of the classification.

In summary, we now have a labelled dataset of size 199,405 of the form
\begin{equation}
\cM_{8 \times 8} \longrightarrow \mbox{Extension Degree over $\mathbb{Q}$} \ ,
\end{equation}
where the degree is 1, 2, 3 or 4.
This constitutes a perfect 4-category classification problem for machine-learning.

\subsubsection{Cyclic Edge Lists}\label{cyclic_rep}
\begin{wrapfigure}[20]{r}{0.5\textwidth}
    \vspace{-60pt}
    \begin{center}
    \includegraphics[width=60mm]{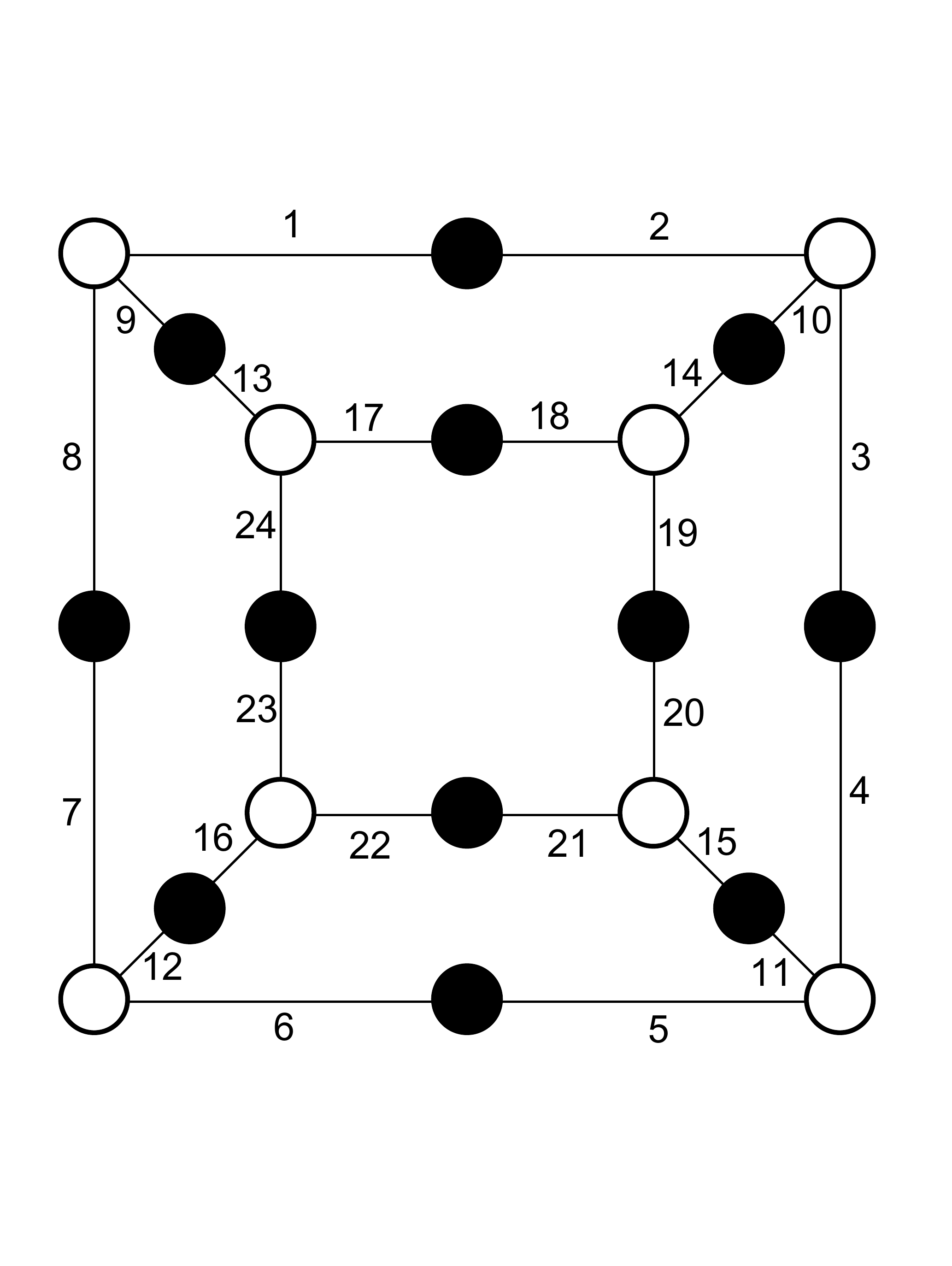} 
    \vspace{-27pt}
    \caption*{\{\{\{8,1,9\},\{2,3,10\},\{4,5,11\},\{12,6,7\},\{18,14,19\},
    \{22,16,23\},\{13,24,17\},\{21,20,15\}\}, \\ \{\{1,2\},\{9,13\},\{10,14\},\{3,4\},\{15,11\},\{24,23\},\{17,18\},
    \{8,7\},\{5,6\},\{12,16\},\{21,22\},\{20,19\}\}\}}
    \caption{The cyclic edge list representation of our example dessin. The dessin edges are first randomly labelled from 1 to 24 (although here in a systematic manner to aid understanding). Then for each white node a clockwise list of edges is given, and respectively for each black node an anticlockwise list. This dessin is hence represented in this form as shown.}
    \label{dessin_example_cyclic}
    \end{center}
\end{wrapfigure}

In the second investigation, a different representation method for the dessins was trialed. In this method the category equivalence to finite sets under permutation was used, in that the dessin information was represented by cyclic lists of edges surrounding each node.

To define each dessin, all edges were numbered from 1 to 24, then about each white node a list was made with the 3 edge numbers of the 3 edges surrounding it in a clockwise manner. This list was cyclic in that the only the order of the 3 edge numbers was really important. In addition for each black node an equivalent anticlockwise list was produced, although for these clean dessins (with bivalent black nodes) the cyclic ordering was irrelevant.

Since the NN takes a tensor object as its data input these lists have redundancy in 3 distinct forms. These are: ordering of the node lists in the full list for each dessin; cyclic ordering of the edges within each node list, and also edge relabellings of the dessin. In an equivalent manner to the adjacency matrix data format, these redundancies were used to inflate the dataset size for improved learning.

To demonstrate this dessin representation method, we return again to the running example shown in figure \S\ref{example_dessin}, now with edges labelled to reflect this data format and shown in figure \S\ref{dessin_example_cyclic}. The full dataset is shown alongside the dessins in appendix \ref{dataset_appendix}.

All 3 redundancy types were used to inflate the dataset size, each individually, and also altogether, as shown in the following results. The number of permutations was kept consistent across all investigations, and in particular the learning was most successful with the cyclic permutations of edges within each node list, suggesting it is a more important property in this method of dessin representation. In each case, the final list of node lists for each of the dessins (/each dessin permutations) were flattened to input into the NN.

The class proportions were: 39.8\%, 34.6\%, 23.6\%, 2.1\% respectively, which remained the same after dataset extension. Again these proportions where used as class weights in NN training. This produced a similar 4-category classification of the form
\begin{equation}
    \mathcal{V}_{48} \longrightarrow \mbox{Extension Degree over $\mathbb{Q}$} ,
\end{equation}
where $\mathcal{V}_{48}$ represents the 48-component flattened vector giving the the cyclic edge lists around white, then black, nodes.

\subsection{Setup of the Machine-Learning Model}
\begin{figure}[H]
    \centering
    \includegraphics[width=\textwidth]{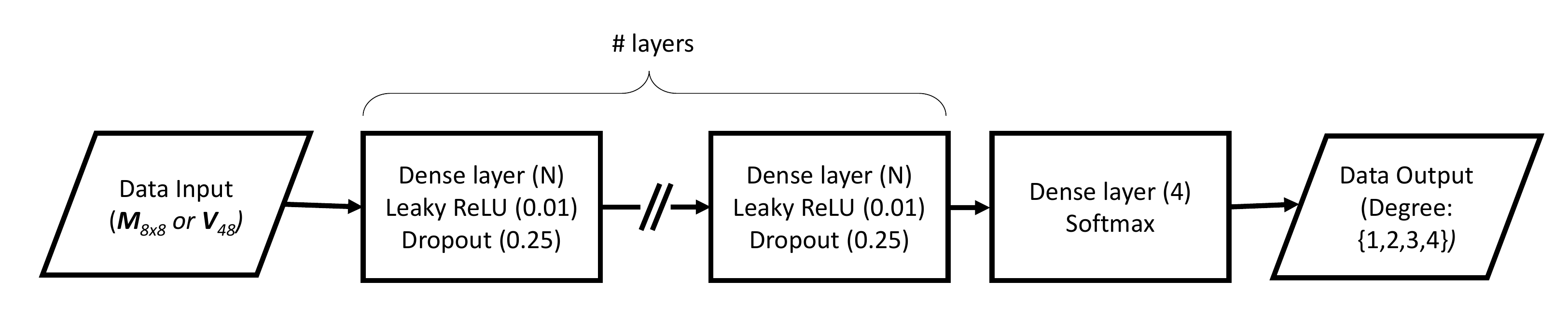} 
    \caption{Schematic showing the layers of the neural network. The dessins are inputted in a tensor format (either adjacency matrix or cyclic edge list vector), which go through as many dense layers of N neurons as specified, each layer with a Leaky ReLU activation (with $\alpha = 0.01$) and Dropout layer (with factor 0.25). The information processed then passes through another dense layer with softmax activation for classification, outputting 1 of the 4 degree extensions.}
    \label{fig:NNdiagram}
\end{figure}

A deep feed-forward  neural network (a multi-layer perceptron) was created using the TensorFlow Keras Library \cite{tensorflow}, consisting of several fully connected (known as dense in the Keras documentation) layers  \cite{Fundamentals_of_deep_learning, Machine_learning_python}.  
The activation function for each of these fully connected layers was the Leaky ReLU function\footnote{
This is the Leaky Re(ctified) L(inear) U(nit), defined as $f(x) = \max(-\alpha x,x)$ for $x \in \mathbb{R}$. This investigation used $\alpha = 0.01$.
}. In addition a Dropout layer, with a 0.25 dropout factor, was added after each activation. Dropout layers randomly ignores a fraction of the previous layer's neurons as each batch of data passes through the NN, this reduces the probability of overfitting the NN to the training data.
Finally, the output layer was a dense layer with 4 neurons with the softmax activation function\footnote{A softmax layer returns \textit{n} floating point values which sum to unity. Each value represents the probability that input is of a particular category.
That is, $f(z_i) = \exp(z_i) \left( \sum_{j=1}^k \exp(z_j) \right)^{-1} : \IR^k \to \IR^k$.
}. The output of each of these 4 neurons represents the probability that the dessin had a particular extension, taking the most probable extension as the predicted classification. A schematic of this model is found in figure \ref{fig:NNdiagram}.

The model was complied with the following specifications \cite{tensorflow}:
\begin{itemize}
    \item Loss function: {\fontfamily{qcr}\selectfont sparse$\_$categorical$\_$cross-entropy}.   When training the model, the parameters of the NN get adjusted to minimise the loss function. Cross-entropy is the required loss function for discrete categories of data \cite{Machine_learning_python}.
   Specifically, the cross-entropy used is
   $-\frac{1}{n} \sum\limits_{j=1}^n \sum\limits_{i=1}^C y_i^j \log ( \hat{y}_i^j )$ where $n$ is the number of samples in the training, indexed by $j$ and $C$ is the number of categories (here 4), indexed by $i$. Moreover, $y$ is the sample label (here the extension degree) and $\hat{y}$ is the predicted value for the degree.

    \item Optimiser:  {\fontfamily{qcr}\selectfont adam} with the default Keras hyper-parameters.  Adam is one of the many modern improvements on the stochastic gradient descent algorithm.  Comparisons with other, similar, optimisers suggest {\fontfamily{qcr}\selectfont adam} to be the best generic choice \cite{optimisers}.

    \item Metric: {\fontfamily{qcr}\selectfont accuracy}.  Metrics do not influence the training of the model, they specify what data to output at the end to judge the effectiveness of the model.  {\fontfamily{qcr}\selectfont accuracy} means the output is the fraction of predictions the model got correct \cite{tensorflow}.
\end{itemize}

To measure the bias we introduce a bias loss function defined as:
\begin{equation}
   L=\bigg(\sum_i\bigg(\frac{\frac{\tilde{n_i}}{N_{\text{test}}}-\frac{n_i}{N_{\text{total}}}}{\frac{n_i}{N_{\text{total}}}}\bigg)^2\bigg)/N \,,
\end{equation}
the summation is over all the different categories, $\tilde{n}_i/N_{\text{test}}$ is the predicted fraction of that category in the test data,  $n_i/N_{\text{total}}$ is the actual fraction of that category in the full dataset, and $N$ is the number of categories.
This measure gives the average fractional difference between the predicted and actual fraction of each category.
Note also that during training the number of occurrences in each class is inverted to give a class weight which is fed into the NNs to encourage them to pay more proportionate attention to less common classes. This reduces bias of the NN classifier, and should produce an overall better classifier.

\subsection{NN Results}
The style of machine-learning used in this study is: \textit{supervised} learning. In this style the initial dataset is partitioned into a training set and a test set; and both sets are composed of defined inputs (here adjacency matrices or cyclic edge lists) and outputs (here degrees of the field extension). The network is fed the training set in batches of a given size, as many times as the number of epochs specified, during the fitting process. Then after the network has been trained, it is tested on the test data which it has not seen before. The proportion of correct classifications on the unseen test data gives the accuracy measure of the network's success.

The other measure is Matthew's Correlation Coefficient (MCC), it is defined in reference to the confusion matrix which is a 4-entry matrix measuring probabilities of correct (positive) and incorrect (negative) predictions of True and False hypotheses respectively. Using this, MCC is defined as the square root of the normalised $\chi^2$ statistical measure:
\begin{equation}
    \phi \vcentcolon = \sqrt{\frac{\chi^2}{N}} = \frac{tp\cdot tn - fp\cdot fn}{\sqrt{(tp+fp)(tp+fn)(tn+fp)(tn+fn)}}\,,
\end{equation}
for $\{tp,tn,fp,fn\}$ representing \{true positive, true negative, false positive, false negative\} respectively \cite{He:2018jtw}.

From the dataset $20\%$ of the dessins were selected, at random, to act as the test data, and the rest were used to train the model. A different random set was selected every time the NN was trained and tested, removing systematic errors that could have occurred if the test data was not representative of the whole population. This was performed as a 5-fold cross-validation. Here a dataset is partitioned into 5, each partition acts as a test dataset for 1 of 5 independently trained NNs (trained on the remaining $80\%$ of the data). For each of the 5 NNs, all the learning measures are calculated, providing a set of 5 measures for each investigation which are then averaged over and standard error calculated. This procedure produces the results displayed in table \ref{ml_results}. 

\begin{longtable}[c]{|c|c|c|c|c|}
\hline
\multicolumn{2}{|c|}{\multirow{2}{*}{Data Input Type}} & \multicolumn{3}{c|}{Average Measures}  \\ 
\cline{3-5} \multicolumn{2}{|c|}{} & Accuracy  & MCC  & Bias \\ \hline
\endfirsthead
\endhead
\multicolumn{2}{|c|}{\begin{tabular}[c]{@{}c@{}} $\mathcal{M}_{8x8}$ \\ (adjacency matrix)\end{tabular}} & \begin{tabular}[c]{@{}c@{}} 0.536 \\ $\pm$ 0.002 \end{tabular} & \begin{tabular}[c]{@{}c@{}} 0.180 \\ $\pm$ 0.008 \end{tabular} & \begin{tabular}[c]{@{}c@{}} 0.54 \\ $\pm$ 0.03 \end{tabular} \\ \hline
 & 1000:0:0  & \begin{tabular}[c]{@{}c@{}} 0.92 \\ $\pm$ 0.03 \end{tabular} & \begin{tabular}[c]{@{}c@{}} 0.88 \\ $\pm$ 0.04 \end{tabular} & \begin{tabular}[c]{@{}c@{}} 0.47 \\ $\pm$ 0.13  \end{tabular} \\ \cline{2-5} 
$\mathcal{V}_{48}$  & 0:1000:0  & \begin{tabular}[c]{@{}c@{}} 0.38 \\ $\pm$ 0.02 \end{tabular} & \begin{tabular}[c]{@{}c@{}} nan \\ $\pm$ nan \end{tabular} & \begin{tabular}[c]{@{}c@{}} 1.32 \\ $\pm$ 0.00 \end{tabular} \\ \cline{2-5} 
\begin{tabular}[c]{@{}c@{}}(cyclic\\ edge list)\end{tabular}   & 0:0:1000  & \begin{tabular}[c]{@{}c@{}}0.553\\ $\pm$ 0.008 \end{tabular} & \begin{tabular}[c]{@{}c@{}} 0.28 \\ $\pm$ 0.02 \end{tabular} & \begin{tabular}[c]{@{}c@{}} 0.42 \\ $\pm$ 0.04 \end{tabular} \\ \cline{2-5} 
 & 10:10:10  & \begin{tabular}[c]{@{}c@{}} 0.42 \\ $\pm$ 0.02 \end{tabular} & \begin{tabular}[c]{@{}c@{}} nan \\ $\pm$ nan \end{tabular} &  \begin{tabular}[c]{@{}c@{}} 1.32 \\ $\pm$ 0.00 \end{tabular} \\ \hline
\caption{Results table showing the learning measures of Accuracy, MCC, and Bias, each averaged with standard error over the 5 NNs trained in each investigation for the 5-fold cross-validation performed. The 5 investigations constitute 2 data formats, the first the adjacency matrix format, the second the cyclic edge list format. For the cyclic edge list format the number of permutations generated under each redundancy are also given in the form \{Edge Cycles : Node Reordering : Edge Relabelling\}.}
\label{ml_results}\\
\end{longtable}

The results here show mediocre success for learning the adjacency matrix representations, since an accuracy of 0.25, and MCC of 0, denotes true random guessing amongst the 4 categories. This lesser performance here may be perhaps attributed to the subtleties in using this representation method for dessins associated to isomorphisms occasionally occurring outside of Galois orbits. For the cyclic edge list representation, the latter 3 experiments were not especially successful, for the NN's with high Bias (reaching the upper bound for this dataset of 1.32), classes were completely ignored in classification, leading to MCC being incalculable (shown by the value 'nan'). However the cyclic edge list representation, with dataset size inflated using the cyclic reordering of edges within each node list was highly successful, exceeding 90\% accuracy. This result shows the importance of the cyclic property in this representation, and suggests a promisingly strong link between this cyclic edge representation of dessins and their Galois orbit size.

The NN used in each case trained with batches of 32 inputs over 20 epochs, they all had 4 dense layers of 512 neurons before the final 4-neuron dense layer. Compiling and testing the models took on the order of about two hours each using a standard personal computer (64 bit, i5 2.3Ghz i5 dual core, 8GB RAM).

\section{Conclusion \& Outlook}
In this paper we have initiated the study of machine-learning applied to dessins d'enfants, as part of the recently advocated programme to see whether AI can deep-learn structures of mathematics \cite{YHHdeep,YHHtalk,He:2018jtw}.
The key aspect of the programme is the following:  given the rapid increase of available data in pure mathematics and mathematical physics, gathered over the last decade or so, ranging from algebraic geometry, to group theory and combinatorics, to number theory, can neural networks and classifiers - often without any prior knowledge of the underlying mathematics - uncover more efficient methods of computation and raise unexpected conjectures.
As mentioned in the introduction, various experiments by a host of collaborations have shown this to be a fruitful venture.

As a concrete play-ground, we have taken 191 dessins which reside at the cusp of several intertwining fields of investigation: subgroups of the modular group, extremal elliptic K3 surfaces as Shioda modular surfaces, and $\cN=1$ Seiberg-Witten curves.
We have first completed the mapping amongst these quantities in the present context.
Then, we have launched the examination of the question: {\it can one ``look'' at a dessin and classify its degree of extension over $\IQ$?}
Indeed, a central theme of Grothendieck's {\it Esquisse} is to understand how dessins furnish a representation of the absolute Galois group which governs all algebraic extensions over $\IQ$.

Rather surprisingly, we find that a simple multi-layer perceptron with standard Leaky ReLU activation function and dense layers can classify this problem within our dataset, enhanced by permutation equivalence to a size of the order $10^5$, to 92\% accuracy.
This is quite contrary to the fact how difficult it is to compute explicit Bely\v{\i} maps and to the intuition built over the last couple of years how number-theoretical problems respond poorly to machine-learning.
Perhaps, dessins are closer to complex analysis than to number theory, as opposed to BSD \cite{Alessandretti:2019jbs}, which seems more resilient to modern data-scientific methods.

Of course, we have only skimmed the surface. There is an infinitude of dessins of varying complexity and it is known that $Gal\big(\,\overline{\mathbb{Q}}\,/\,\mathbb{Q}\,\big)$ acts even faithfully on genus 0 and genus 1 families thereof individually.
Can our neural network extrapolate to these?
Indeed, as a part of more ambitious plan, can we establish machine-learning models which help compute actual Bely\v{\i} maps? Indeed, what about other aspects of dessins, such as cartographic groups and permutation triples, many of which have also found their place in theoretical physics; can these be machine-learnt?
There are undoubtedly many other questions we leave to future investigations.

\comment{
This study showed the success of machine-learning methods in predicting the degree of the field extension over the rationals which a dessin d'enfant graph's Bely\v{\i} map is defined over. The simplicity of the networks used indicates that perhaps there is a simpler description undermining the dessin structure which allows the extension degree to be extracted. Since the field extension is closely related to the orbit structure of the dessins, perhaps this indicates a direction for further research, and supports the use of dessins in examining the Galois group structure of $Gal\big(\,\overline{\mathbb{Q}}\,/\,\mathbb{Q}\,\big)$.

Specifically the dessins associated with semi-stable elliptically fibred K3 Calabi-Yau surfaces were considered, and this dataset was documented in an appendix, along with the list of corresponding Seiberg-Witten curves which was calculated. The paper also serves as an introduction to the dessin d'enfant graphs, as well as the machine-learning methods used.}

\section*{Acknowledgements}
YHH would like to thank STFC for grant ST/J00037X/1. EH
would like to thank STFC for the PhD studentship.
The authors would also like to thank Matthias Hutzler, Prof. Maxim Smirnov and the remainder of the group at University of Augsburg for their identification of a data analysis error, and discussion that lead authors to identify the dessin representation subtlety and improve this analysis.

\newpage
\begin{appendices}
\section{Reproducing the Riemann Surface}\label{dessin->riemann}
To further motivate the study of dessins, importantly one can reproduce the Riemann surface's algebraic curve, $X$, used in the Bely\v{\i} pair definition of a dessin, as described in \S\ref{Belyi_pair_label}, up to a homeomorphism. In this process the dessin is triangulated, and this triangulation also proves important in linking the dessins to the finite set category.

Starting with a dessin, first nodes are introduced into each isolated part of the dessin (i.e., one on each face, including the face outside the dessin); these are associated to the points of the Bely\v{\i} function ramified at infinity. Then edges from each of these `infinity' nodes are drawn to all nodes of the dessin along the boundary of the isolated section of the dessin.  This is all nodes which edges can be drawn to without having to cross another edge, noting there may be duplicate nodes where the nodes appear more than once in the region's boundary.

\begin{figure}[h!]
    \centering
    \includegraphics[width=\textwidth]{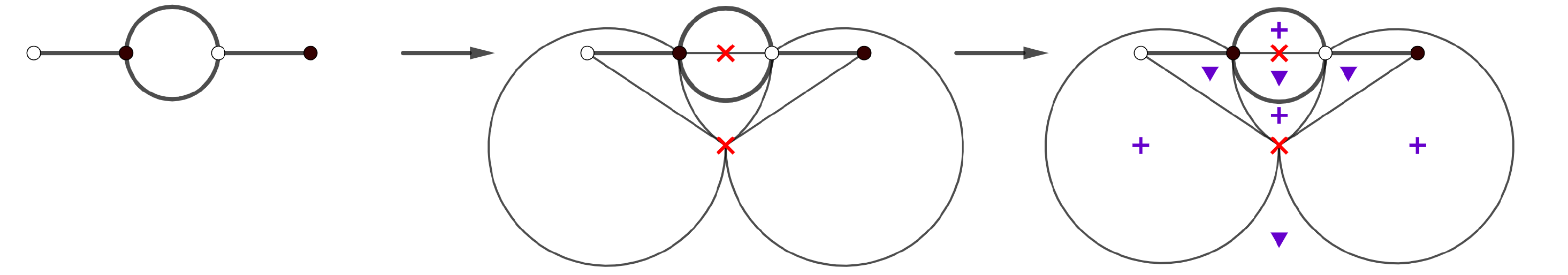} 
    \caption{The process of converting a dessin into the corresponding Riemann surface. The first graph shows an arbitrary dessin. The second shows the inclusion of the 'infinity' ramification nodes (red crosses) in each isolated region of the dessin; and the additional edges drawn to all nodes bounding the region - the triangulation process. The final graph shows each triangulated region associated to either the lower (blue triangle) or upper (blue plus) half planes according to the boundary  orientations. These planes are then glued together along each boundary to give the Riemann surface.}
    \label{Dessin_to_X}
\end{figure}

This process triangulates the dessin, such that all the isolated sections are now bound by 3 nodes (one of each of \{white $\circ$, black $\bullet$, infinity $\infty$\}). This is shown in figure \ref{Dessin_to_X}. If the triangle is bounded in a clockwise/anticlockwise direction by the node order: white $\mapsto$ black $\mapsto \infty$; then the section is associated to a lower/upper half plane. The half planes are then glued along the appropriate boundaries \cite{Girondo}. Importantly since the dessin is truly embedded on the Riemann sphere in our case, the outside 'face' is identified with the opposite half plane to the usual node ordering of the boundary as seen from the representation in the figure.

This produces a Riemann surface homeomorphic to the original $X$ in the Bely\v{\i} pair used to define the dessin. Note since the graph-theoretic degree of a node leads to twice as many half planes around it, which are correspondingly glued, it becomes clear this degree represents the ramification degree of the surface cover. This Riemann surface produced may then be mapped to the Riemann sphere, and this mapping is then equivalent to the Bely\v{\i} function from the original Bely\v{\i} pair, up to a biholomorphism. Therefore an equivalent Bely\v{\i} pair is produced, and the relation between these pairs is appropriately defined within the Galois-theoretic structure.

After triangulation there now exists three order 2 permutations on the triangles, which are reflections in each of the triangles edges. The full cartographic group is then the group of permutations of these triangles generated by these 3 reflections. Combining the reflection operators (denoted a, b, and c), three permutation operators can be defined:
$\{\sigma,\alpha,\phi\} \vcentcolon = \{ab,bc,ca\}$ respectively; note that also due to the order two nilpotence of reflection operators 
\begin{equation}\label{permutation_identity}
    \sigma\cdot\alpha\cdot\phi = 1\,.
\end{equation}
Since each triangle is associated to one edge from the original dessin, these permutation operators may be thought of as generating permutations of the dessin. This defines the cartographic group, as the edge set permutations generated by these permutation operators. However due to the property from \eqref{permutation_identity}, only two generators are required. This then defines the equivalence, there is a bijection between isomorphism classes of dessins (under permutation) and conjugacy classes of finite index subgroups of the free group generated by $\{\sigma,\alpha\}$. This gives another practical use for dessins in examining free groups \cite{Guillot}. In part, this equivalence inspired the cyclic edge list representation method of the dessins used in the investigations.

\section{Generated Seiberg-Witten Curves}\label{SW_appendix}
Here we include the Seiberg-Witten curves for all dessins with rational coefficients, and hence defined over the field $\mathbb{Q}$, as listed in appendix \ref{dataset_appendix}, and \cite{MirandaPersson_table}. Note those defined over fields which are extensions of the rationals are not included.

{\small
\begin{landscape}
\begin{longtable}{ll}

Dessin             & SW curve $y^2=$   \\
19, 1, 1, 1, 1, 1  & $-\frac{135 z^4}{2}-\frac{81 z^2}{2}+(z^8-2 z^7+7 z^6-6 z^5+11 z^4+4 z^3+12 z+1))^3+\frac{837}{2}$     \\
18, 2, 1, 1, 1, 1  & $\frac{1}{729} ((9 z^8+36 z^6+54 z^4+28 z^2+1))^3$     \\
17, 3, 1, 1, 1, 1  & $\frac{((9 z^4-14 z^3-9 z^2+6 z+21)) (4 z+1)^3+((54 z^{12}+162 z^{11}-270 z^9+162 z^8+432 z^7-252 z^6-216 z^5+270 z^4-63 z^2+21 z-5))^2}{2916}$        \\
16, 4, 1, 1, 1, 1  & $((z^8+16 z^4+16))^3$    \\
16, 3, 2, 1, 1, 1  & $\frac{1}{729} ((-165888 z^7-442368 z^3+((9 z^8+48 z^7+64 z^6+48 z^4+128 z^3+16))^3))$   \\
16, 2, 2, 2, 1, 1  & $(z^8-8 z^6+20 z^4-16 z^2+1)^3$  \\
15, 3, 3, 1, 1, 1  & $\frac{1}{729} (-64 (z-1)^3 (z^3-3 z-3) (3 z-4)^3+64 (z+1)^3 (3 z+4)^3 (z^3-3 z+3)+$\\
& $(z^2-z-1)^3 (9 z^6-27 z^5+45 z^3-51 z+23)^3)$        \\
15, 3, 2, 2, 1, 1  & $-1728 (z^2-3 z+1)^2 (z^2+6 z+10) (2 z-5)^3+1728 (2 z+5)^3 (z^2-6 z+10) (z^2+3 z+1)^2+$\\
& $(z^2-5)^3 (z^6-15 z^4+75 z^2-48 z-5)^3$   \\
14, 5, 2, 1, 1, 1  & $\frac{207360 z^9+912384 z^7+387072 z^5+(25 z^8-40 z^7+56 z^6-112 z^5+32 z+16)^3}{15625}$   \\
14, 4, 3, 1, 1, 1  & $-8707129344 (z-5)^3 (5 z^3-3 z^2-264 z-5980) (1-3 z)^4+8707129344 (z+5)^3 (3 z+1)^4 (5 z^3+3 z^2-264 z+5980)+$\\
& $(z^8-10 z^7-35 z^6-700 z^5+10780 z^4-12544 z^3+12880 z^2-677440 z+2843200)^3$   \\
14, 3, 3, 2, 1, 1  & $30233088 z^9-120932352 z^7+6348948480 z^5-19863138816 z^3+(z^8-4 z^7-14 z^6+315 z^4-1134 z^2-1620 z+1377)^3+$\\
& $40814668800z$  \\
14, 3, 3, 2, 1, 1  & $-1259712 (z+8)^2 (5 z^2+21 z-441) (17 z^2-71 z+59)^3+1259712 (z-8)^2 (5 z^2-21 z-441) (17 z^2+71 z+59)^3+$\\
& $(z^8+12 z^7-102 z^6-1096 z^5+4275 z^4+22632 z^3-90566 z^2+42756 z+122913)^3$      \\
14, 3, 2, 2, 2, 1  & $\frac{27 (z-4) (12 z+1)^2 (3 z^2-14 z+14)^2 z^3-27 (1-12 z)^2 (z+4) (3 z^2+14 z+14)^2 z^3+4 (9 z^8+120 z^7+616 z^6+1512 z^5+1764 z^4+784 z^3-12 z+1)^3}{2916}$   \\
13, 7, 1, 1, 1, 1  & $8460288 z^{10}+132865920 z^8+135543240 z^6+\frac{26168373 z^4}{2}+\frac{290655 z^2}{2}+$\\
& $(z^8+6 z^7+21 z^6+98 z^5+182 z^4+210 z^3+301 z^2-214 z+4)^3-\frac{135}{2}$      \\
13, 5, 3, 1, 1, 1  & $\frac{1}{729} (64 (4 z+5)^3 (5 z^3+15 z^2-9 z-355) (9 z-31)^5-64 (4 z-5)^3 (9 z+31)^5 (5 z^3-15 z^2-9 z+355)+$\\
& $(9 z^8-180 z^6-720 z^5+2790 z^4+7008 z^3-7540 z^2-38640 z+14825)^3)$    \\
13, 4, 3, 2, 1, 1  & $-918330048 (109-5 z)^2 (8 z-73)^3 (7 z^2+730 z+21543) (303-17 z)^4+$\\
& $918330048 (5 z+109)^2 (8 z+73)^3 (17 z+303)^4 (7 z^2-730 z+21543)+$\\
& $(z^8-72 z^7-1092 z^6+135016 z^5+1687350 z^4-76683768 z^3-1742916644 z^2-9658896744 z+6717775617)^3$        \\
13, 3, 3, 2, 2, 1  & $322486272 (z+10) (5 z^2-24 z+612)^2 (4 z^2-17 z+430)^3-322486272 (z-10) (4 z^2+17 z+430)^3 (5 z^2+24 z+612)^2+$\\
& $(z^8+240 z^6-1600 z^5+14880 z^4-274944 z^3+490240 z^2-10544640 z+29606400)^3$           \\
12, 6, 3, 1, 1, 1  & $34012224 (z-3)^3 (z^3+9 z^2-216) (z-4)^6-34012224 (z+3)^3 (z+4)^6 (z^3-9 z^2+216)+$\\
& $(z^2-18)^3 (z^6-54 z^4+1620 z^2-5184 z+4536)^3$       \\
12, 6, 2, 2, 1, 1  & $(z^8+8 z^6-16 z^2+16)^3$   \\
12, 6, 2, 2, 1, 1  & $\frac{-13226976 z^{11}+5668704 z^9+419904 z^7+(5184 z^8+5184 z^7+3024 z^6+21 z^2-6 z+1)^3}{139314069504}$    \\
12, 5, 2, 2, 2, 1  & $\frac{-432 (z+4) (5 z+14)^2 (70 z^2-24 z+9)^2 z^5+432 (14-5 z)^2 (z-4) (70 z^2+24 z+9)^2 z^5+64 (25 z^8+240 z^7+756 z^6+784 z^5+70 z^2-24 z+9)^3}{1000000}$    \\
12, 4, 4, 2, 1, 1  & $\frac{108 (7-4 z)^4 (z-2)^2 (z^2+2) (z-1)^4-108 (z+1)^4 (z+2)^2 (4 z+7)^4 (z^2+2)+(16 z^8+128 z^7+448 z^6+960 z^5+1488 z^4+1728 z^3+1392 z^2+696 z+177)^3}{4096}$           \\
12, 4, 3, 3, 1, 1  & $(z^8-12 z^6+30 z^4-12 z^2+9)^3$  \\
12, 4, 3, 3, 1, 1  & $110592 (z^2+12 z+45) (5 z^2-36 z+65)^3 (z-3)^4-110592 (z+3)^4 (z^2-12 z+45) (5 z^2+36 z+65)^3+$\\
& $(z^2-15)^3 (z^6-45 z^4+1155 z^2-3456 z+2865)^3$    \\
12, 4, 3, 2, 2, 1  & $-1296 z^{11}+6156 z^9-11340 z^7-8100 z^5-216 z^3+(z^8+8 z^7+16 z^6-8 z^5-28 z^4+16 z^3+4 z^2-4 z+1)^3$  \\
12, 3, 3, 3, 2, 1  & $1728 (z-4)^2 (z+1)^3 (z+5) (5 z^2-8 z-40)^3-1728 (z-5) (z-1)^3 (z+4)^2 (5 z^2+8 z-40)^3+$\\
& $(z^2-10)^3 (z^6-30 z^4+180 z^2-192 z-40)^3$               \\
12, 3, 3, 2, 2, 2  & $(z^8+12 z^6+48 z^4+66 z^2+9)^3$ \\
11, 9, 1, 1, 1, 1  & $-\frac{27}{4} (z^4+14 z^3+47 z^2-102 z+45) (4 z-3)^9+\frac{27}{4} (4 z+3)^9 (z^4-14 z^3+47 z^2+102 z+45)+$\\
& $(z^8+18 z^7+95 z^6-18 z^5-701 z^4+1068 z^3-432 z^2-108 z+81)^3$        \\
11, 5, 5, 1, 1, 1  & $1728 (z^3+3 z^2+11 z+121) (4 z^2-9 z+91)^5-1728 (4 z^2+9 z+91)^5 (z^3-3 z^2+11 z-121)+$\\
& $(z^8+44 z^6-176 z^5+374 z^4-5984 z^3+2156 z^2-44176 z+64801)^3$  \\
11, 5, 3, 2, 2, 1  & $\frac{27}{4} (3 z-53) (7 z+43)^3 (243-5 z^2)^2 (8 z-141)^5-\frac{27}{4} (3 z+53) (7 z-43)^3 (8 z+141)^5 (243-5 z^2)^2$+\\
& $(z^8-48 z^7+588 z^6+3304 z^5-85470 z^4-9408 z^3+3745084 z^2-923724 z-45225702)^3$   \\
11, 4, 4, 3, 1, 1  & $-1728 (z-2)^3 (3 z^2+z+111) (5 z^2-9 z+9)^4+1728 (z+2)^3 (3 z^2-z+111) (5 z^2+9 z+9)^4+$\\
&$ (z^8+6 z^7+57 z^6+278 z^5+900 z^4+2190 z^3+3145 z^2+1110 z-855)^3$   \\
11, 4, 3, 3, 2, 1  & $-1259712 (z-26)^2 (4 z-153) (7 z^2+4 z+4)^3 (9-5 z)^4+1259712 (z+26)^2 (4 z+153) (5 z+9)^4 (7 z^2-4 z+4)^3+$\\
& $(z^8+96 z^7+3192 z^6+41888 z^5+168000 z^4+97440 z^3+244720 z^2-728640 z-200880)^3$   \\
10, 10, 1, 1, 1, 1 & $(z^8+12 z^6+14 z^4-12 z^2+1)^3$ \\
10, 8, 3, 1, 1, 1  & $\frac{1}{729} (-64 (z+1)^3 (11 z^3-105 z^2+1008 z+124) (2-9 z)^8+64 (z-1)^3 (9 z+2)^8 (11 z^3+105 z^2+1008 z-124)+$\\
& $(9 z^8-108 z^7+1062 z^6-2196 z^5+2385 z^4+20568 z^3+42104 z^2+19104 z+1808)^3)$    \\
10, 8, 3, 1, 1, 1  & $-442368 (7 z+15)^4 (11 z^3+3 z^2-2864 z+1600) (z-1)^7+442368 (15-7 z)^4 (z+1)^7 (11 z^3-3 z^2-2864 z-1600)+$\\
& $(z^8+2 z^7-287 z^6-336 z^5+9520 z^4+3136 z^3+108192 z^2-52992 z-4736)^3$  \\
10, 7, 2, 2, 2, 1  & $-5292 z^{13}-23814 z^{11}-45900 z^9-33750 z^7+(z^8+4 z^7-14 z^3+14 z^2-20 z+25)^3$      \\
10, 6, 5, 1, 1, 1  & $-\frac{335544320 z^{13}}{9}-1006632960 z^{11}-\frac{45164265472 z^9}{27}-\frac{1677721600 z^7}{9}+$\\
& $\frac{1}{729} (9 z^8+120 z^7+940 z^6+3240 z^5+5670 z^4+1288 z^3+2700 z^2+600 z+25)^3$   \\
10, 6, 4, 2, 1, 1  & $1259712 (2 z+1)^2 (7 z+8)^4 (z^2-18 z+297) (z-1)^6-1259712 (8-7 z)^4 (1-2 z)^2 (z+1)^6 (z^2+18 z+297)+$\\
& $(z^8+24 z^7+420 z^6+2072 z^5+4830 z^4-36120 z^3+122500 z^2-101880 z+27945)^3$   \\
10, 6, 4, 2, 1, 1  & $\frac{108 z^4 (z+3)^2 (9 z^2+42 z-5) (1-8 z)^6-108 (z-3)^2 z^4 (8 z+1)^6 (9 z^2-42 z-5)+(144 z^8-1536 z^7+5248 z^6-5568 z^5-720 z^4+512 z^3+192 z^2+24 z+1)^3}{2985984}$  \\
10, 6, 3, 2, 2, 1  & $-\frac{27}{4} (z+2)^3 (z+6) (z^2-2 z+2)^2 (1-4 z)^6+\frac{27}{4} (z-6) (z-2)^3 (4 z+1)^6 (z^2+2 z+2)^2+$\\
& $(z^8+8 z^7+8 z^6-24 z^5+20 z^4+48 z^3-16 z^2-76 z+193)^3$    \\
10, 5, 2, 2, 2, 2  & $\frac{1}{64} (432 (z-4)^3 (-6 z^3+22 z^2+12 z+9)^2 z^5-432 (z+4)^3 (6 z^3+22 z^2-12 z+9)^2 z^5+$\\
& $64 (z^8+12 z^7+48 z^6+64 z^5+6 z^3+22 z^2-12 z+9)^3)$    \\
10, 4, 4, 3, 2, 1  & $(z^8-4 z^7-98 z^6+28 z^5+2345 z^4+3136 z^3-6944 z^2+8192 z+80128)^3+$\\
& $1728 (4-7 z)^4 (16-3 z)^2 (z+3)^3 (z+11) (z-4)^4-1728 (z-11) (z-3)^3 (z+4)^4 (3 z+16)^2 (7 z+4)^4$  \\
10, 4, 3, 3, 2, 2  & $1259712 (25-2 z)^2 (z+1)^2 (z^2-14 z+25)^3 z^4-1259712 (z-1)^2 (2 z+25)^2 (z^2+14 z+25)^3 z^4+$\\
& $(z^8+40 z^7+580 z^6+3560 z^5+7550 z^4-1192 z^3-9500 z^2-5000 z+15625)^3$   \\
9, 9, 2, 2, 1, 1   & $\frac{1}{729} (-16777216 (z^2+3)^2 (z^2-6 z-3) z^9+16777216 (z^2+3)^2 (z^2+6 z-3) z^9+$\\
& $(z-1)^3 (z+3)^3 (9 z^6+54 z^5+27 z^4+260 z^3-81 z^2+486 z-243)^3)$    \\
9, 7, 5, 1, 1, 1   & $\frac{1}{729} (-64 (7 z-3)^5 (5 z^3-51 z^2+879 z-369) (4 z-3)^7+64 (4 z+3)^7 (7 z+3)^5 (5 z^3+51 z^2+879 z+369)+$ \\
& $(9 z^8-144 z^7+2268 z^6-11872 z^5+42070 z^4-16464 z^3-15876 z^2+11232 z-1863)^3)$   \\
9, 6, 4, 3, 1, 1   & $\frac{1}{729} (64 (z-1)^4 (4 z+5)^3 (z^2+4 z-20) (14-5 z)^6-64 (z+1)^4 (4 z-5)^3 (5 z+14)^6 (z^2-4 z-20)-$ \\
& $(z^2-10)^3 (-9 z^6+270 z^4+800 z^3+780 z^2+672 z+1160)^3)$  \\
9, 5, 5, 2, 2, 1   & $\frac{1}{729} (64 (2 z-7) (z^2+5 z+7)^2 (7 z^2-10 z-50)^5-64 (2 z+7) (z^2-5 z+7)^2 (7 z^2+10 z-50)^5+$\\
& $(9 z^8-252 z^6-224 z^5+2478 z^4+3920 z^3-7700 z^2-18000 z-4375)^3)$  \\
9, 5, 3, 3, 3, 1   & $\frac{1}{729} (-64 (z-3) (5 z^3-3 z-1)^3 z^5+64 (z+3) (5 z^3-3 z+1)^3 z^5+(z^2+z-1)^3 (9 z^6+27 z^5-5 z^3+3 z-1)^3)$                 \\
9, 4, 3, 3, 3, 2   & $\frac{1}{729} (-64 (z-3)^2 (4 z^3-9 z^2-6 z-1)^3 z^4+64 (z+3)^2 (4 z^3+9 z^2-6 z+1)^3 z^4+$\\
& $(z^2+2 z-1)^3 (9 z^6+54 z^5+81 z^4-4 z^3-9 z^2+6 z-1)^3)$   \\
8, 8, 4, 2, 1, 1   & $(z^8-4 z^6-10 z^4+28 z^2+1)^3$  \\
8, 8, 3, 3, 1, 1   & $\frac{-4 (3 z^2-1)^3 (9 z^2+24 z-11) (1-3 z)^8+4 (3 z+1)^8 (3 z^2-1)^3 (9 z^2-24 z-11)+(9 z^4+12 z^3-1)^3 (9 z^4+12 z^3-36 z^2+24 z-5)^3}{531441}$   \\
8, 8, 2, 2, 2, 2   & $(z^8+z^4+1)^3$ \\
8, 7, 3, 3, 2, 1   & $\frac{1}{4096}(108 (z-17)^2 (3 z-2) (7 z^2-112 z+16)^3 (5 z-4)^7-108 (z+17)^2 (3 z+2) (5 z+4)^7 (7 z^2+112 z+16)^3+$\\
& $(16 z^8+832 z^7+15568 z^6+123424 z^5+375025 z^4+432880 z^3+323680 z^2+167680 z+37120)^3)$   \\
8, 6, 5, 2, 2, 1   & $\frac{1}{2985984}(-108 (8-15 z)^2 (7-3 z)^2 z^5 (z+6) (8-7 z)^6+108 (z-6) z^5 (3 z+7)^2 (7 z+8)^6 (15 z+8)^2+$\\
& $(144 z^8+192 z^7-3248 z^6+4704 z^5+5145 z^4-20384 z^3+29568 z^2-18432 z+4096)^3)$   \\
8, 6, 4, 2, 2, 2   & $\frac{432 (4-3 z)^2 (z-1)^4 (3 z^2+2 z+1)^2 z^6-432 (z+1)^4 (9 z^3+6 z^2-5 z+4)^2 z^6+64 (9 z^8+24 z^7+16 z^6+3 z^4+4 z^3+1)^3}{46656}$                 \\
8, 5, 4, 3, 2, 2   & $\frac{1}{64} (-432 (z-8)^3 (3 z+1)^4 (-5 z^2+42 z+9)^2 z^5+432 (1-3 z)^4 (z+8)^3 (5 z^2+42 z-9)^2 z^5+$\\
& $64 (z^8-24 z^7+192 z^6-512 z^5-45 z^4+348 z^3+328 z^2+96 z+9)^3)$   \\
8, 4, 4, 4, 3, 1   & $108 (z-7) (z-1)^4 (z+1)^3 (3 z^2+2 z+1)^4-108 (z-1)^3 (z+7) (3 z^3+z^2-z+1)^4+$ \\
& $(z^8+8 z^7+4 z^6-24 z^5+18 z^4-24 z^3+60 z^2-24 z-3)^3$                         \\
8, 4, 4, 4, 2, 2   & $(z^8-4 z^6+5 z^4-2 z^2+1)^3$ \\
7, 7, 7, 1, 1, 1   & $51840 z^{16}+819072 z^{14}+1572480 z^{12}+438912 z^{10}-6912 z^8+(z^8-12 z^7+42 z^6-56 z^5+35 z^4-14 z^2+4 z+1)^3$ \\
7, 6, 6, 2, 2, 1   & $46656 (z-4) (z^2+5 z+13)^2 (z^2-3 z-3)^6-46656 (z+4) (z^2-5 z+13)^2 (z^2+3 z-3)^6+$\\
& $(z^8+4 z^7-14 z^6+315 z^4-1134 z^2+8100 z+27297)^3$   \\
7, 6, 5, 4, 1, 1   & $\frac{1}{729} (-1728 (799-33 z)^4 (4 z+163)^5 (3 z^2+40 z-3416) (z-28)^6+1728 (z+28)^6 (4 z-163)^5 (33 z+799)^4 (3 z^2-40 z-3416)$\\
& $+ (9 z^8+48 z^7-35000 z^6-481152 z^5+44788800 z^4+967680672 z^3-15708981456 z^2-549509702400 z-3524652257136)^3)$ \\
7, 6, 4, 4, 2, 1   & $\frac{1}{729} (108 (z+2)^2 (12 z-25) (3 z^2+2 z-2)^4 (z-2)^6-108 (z+2)^6 (12 z+25) (-3 z^2+2 z+2)^4 (z-2)^2+$\\
& $(9 z^8+24 z^7-56 z^6-168 z^5+84 z^4+336 z^3+336)^3)$  \\
& \\
7, 5, 5, 3, 2, 2   & $1728 (2 z+7)^3 (z^2-3 z+63)^2 (3 z^2-2 z+42)^5-1728 (2 z-7)^3 (z^2+3 z+63)^2 (3 z^2+2 z+42)^5+$\\
& $(z^8+84 z^6-224 z^5+1806 z^4-8400 z^3+15484 z^2+411600 z-497007)^3$  \\
7, 5, 4, 4, 3, 1   & $-108 z^3 (4 z-125) (5 z^2+10 z+14)^4 (3 z-8)^5+108 z^3 (3 z+8)^5 (4 z+125) (5 z^2-10 z+14)^4+$ \\
& $(z^8+40 z^7+280 z^6-280 z^5+2660 z^4-22288 z^3+1120 z^2-4160 z+10000)^3$    \\
7, 5, 4, 3, 3, 2   & $\frac{1}{729} (-64 (106-9 z)^2 (z-9)^4 (5 z^2+12 z+12)^3 (4 z-9)^5+64 (z+9)^4 (4 z+9)^5 (9 z+106)^2 (5 z^2-12 z+12)^3+$\\
& $(9 z^8+288 z^7+3192 z^6+13216 z^5+6720 z^4-39648 z^3+33264 z^2-1041984 z+431568)^3)$   \\
6, 6, 6, 4, 1, 1   & $1728 (1-2 z)^6 z^4 (z^2-18 z+9) (z-1)^6-1728 z^4 (z+1)^6 (2 z+1)^6 (z^2+18 z+9)+$\\
& $(z^2+6 z+3)^3 (z^6+18 z^5+21 z^4+36 z^3+39 z^2+18 z+3)^3$     \\
6, 6, 6, 2, 2, 2   & $(z^8+12 z^6+30 z^4-228 z^2+9)^3$\\
6, 6, 5, 4, 2, 1   & $\frac{729}{4} (z+3)^4 (2 z+5)^2 (4 z-15) (z^2-2 z-5)^6-\frac{729}{4} (5-2 z)^2 (z-3)^4 (4 z+15) (z^2+2 z-5)^6+$\\
& $(z^8-30 z^6-20 z^5+300 z^4-24 z^3-1430 z^2+3300 z+9000)^3$ \\
6, 6, 5, 3, 3, 1   & $-1728 (2 z-5) (z^2+6 z+10)^3 (z^2-3 z+1)^6+1728 (2 z+5) (z^2-6 z+10)^3 (z^2+3 z+1)^6+$\\
& $(z^2-5)^3 (z^6-15 z^4+75 z^2+432 z-1205)^3$ \\
6, 6, 4, 4, 2, 2   & $-108 (3-2 z)^2 (z+3)^2 (z^2-3)^4 (z-1)^6+108 (z-3)^2 (z+1)^6 (2 z+3)^2 (z^2-3)^4+$\\
& $(z^8-12 z^6+8 z^5+42 z^4-48 z^3+28 z^2-120 z+117)^3$\\
6, 6, 4, 3, 3, 2   & $(z^8+4 z^6+60 z^4+58 z^2+1)^3$\\
6, 6, 3, 3, 3, 3   & $(z^8-4 z^6+6 z^4-12 z^2+9)^3$\\
6, 5, 4, 4, 3, 2   & $\frac{1}{729} (-4 (7-2 z)^2 (3 z+2)^3 (5 z^2-2 z+2)^4 (z-2)^5+4 (z+2)^5 (2 z+7)^2 (3 z-2)^3 (5 z^2+2 z+2)^4+$\\
& $(9 z^8-72 z^7+192 z^6-184 z^5+20 z^4+48 z^3+416 z^2-256 z-48)^3)$\\
5, 5, 5, 5, 2, 2   & $1728 (2 z+5)^5 (z^2+5 z-25)^2 (z^2-6 z+10)^5-1728 (2 z-5)^5 (z^2-5 z-25)^2 (z^2+6 z+10)^5+$\\
& $(z^8-60 z^6+160 z^5+670 z^4-10800 z^3+59500 z^2-130000 z+90625)^3$\\
5, 5, 4, 4, 3, 3   & $\frac{1}{729} (-4 (4 z^2-5 z+25)^3 (9 z^2+20 z+25)^4 (z-1)^5+4 (z+1)^5 (4 z^2+5 z+25)^3 (9 z^2-20 z+25)^4+$\\
& $(9 z^8+60 z^6-40 z^5+130 z^4-2800 z^3+500 z^2-5000 z+3125)^3)$\\
4, 4, 4, 4, 4, 4   & $(z^8-14 z^4+1)^3$ 
\end{longtable}
\end{landscape}
}

{

\section{The Dessin d'Enfant Dataset} \label{dataset_appendix}
Below the 191 dessins d'enfants, as shown in \cite{YMR, MirandaPersson_table}, are presented. Here the white nodes have been arbitrarily labelled from 1 to 8, and the edges arbitrarily labelled 1 to 24. The respective adjacency matrix, and cyclic edge list, for each dessin's labelling is given. The cyclic edge list representations follow the convention outlined in section \S\ref{cyclic_rep}, and demonstrated in figure \S\ref{dessin_example_cyclic}, with edges listed in a clockwise manner about white nodes. 

Analysis using the cyclic edge list representation used all 191 independent dessins' edge lists. However, analysis using the matrix representation used the 152 independent matrices (less due to dessin isomorphism). The true datasets analysed inflated these dataset sizes by permuting the labellings of the edges/nodes respectively. 

The dessins are organised with respect to their 6-tuple ramification data, and those with identical 6-tuples are sub-denoted with alphabetic denominations. If these are also isomorphic they are drawn together with the same adjacency matrix. Isomorphisms outside of orbits (such as: 10-6-5-1-1-1 A \& 144-3-1-1-1, 10-8-3-1-1-1 \& 11-6-4-1-1-1 A) are drawn separately, and these isomorphisms highlight a subtlety in the effectiveness of using matrices to represent dessins. Note there was an error in drawing of the dessin for passport 8-6-5-3-1-1 B in \cite{YMR}, which has been corrected here.

In addition to the 6-tuple partition of 24 (the dessin passport) which represents the 6 singular Kodaira fibre types of the corresponding semi-stable elliptically fibred K3 surface, the field which the modular surface's Weierstra{\ss} equation is defined over is given in brackets. Where there is no extension $\mathbb{Q}$ is written; where the extension is by roots of a quadratic polynomial one of the roots is given (i.e. $\sqrt{a}$); and where the extension is by a root of a cubic or quartic polynomial, 'cubic' or 'quartic' are written respectively.

Note that since dessins of the same orbit have the same 6-tuple partition whilst there may be multiple orbits with the same 6-tuple partition, this causes the number of dessins per partition to equal the sum of the degrees of all orbits which share the partition in consideration.

\begin{figure}[H]
    \begin{subfigure}{0.5\textwidth}
        \centering \captionsetup{justification=centering}
        $\scalemath{0.75}{
        \displaystyle \begin{pmatrix}
            0 & 1 & 0 & 1 & 1 & 0 & 0 & 0\\ 
            1 & 0 & 1 & 0 & 0 & 1 & 0 & 0\\
            0 & 1 & 0 & 1 & 0 & 0 & 1 & 0\\
            1 & 0 & 1 & 0 & 0 & 0 & 0 & 1\\
            1 & 0 & 0 & 0 & 0 & 1 & 0 & 1\\
            0 & 1 & 0 & 0 & 1 & 0 & 1 & 0\\
            0 & 0 & 1 & 0 & 0 & 1 & 0 & 1\\
            0 & 0 & 0 & 1 & 1 & 0 & 1 & 0
        \end{pmatrix}}$
        $\vcenter{\hbox{\includegraphics[width=0.5\textwidth]{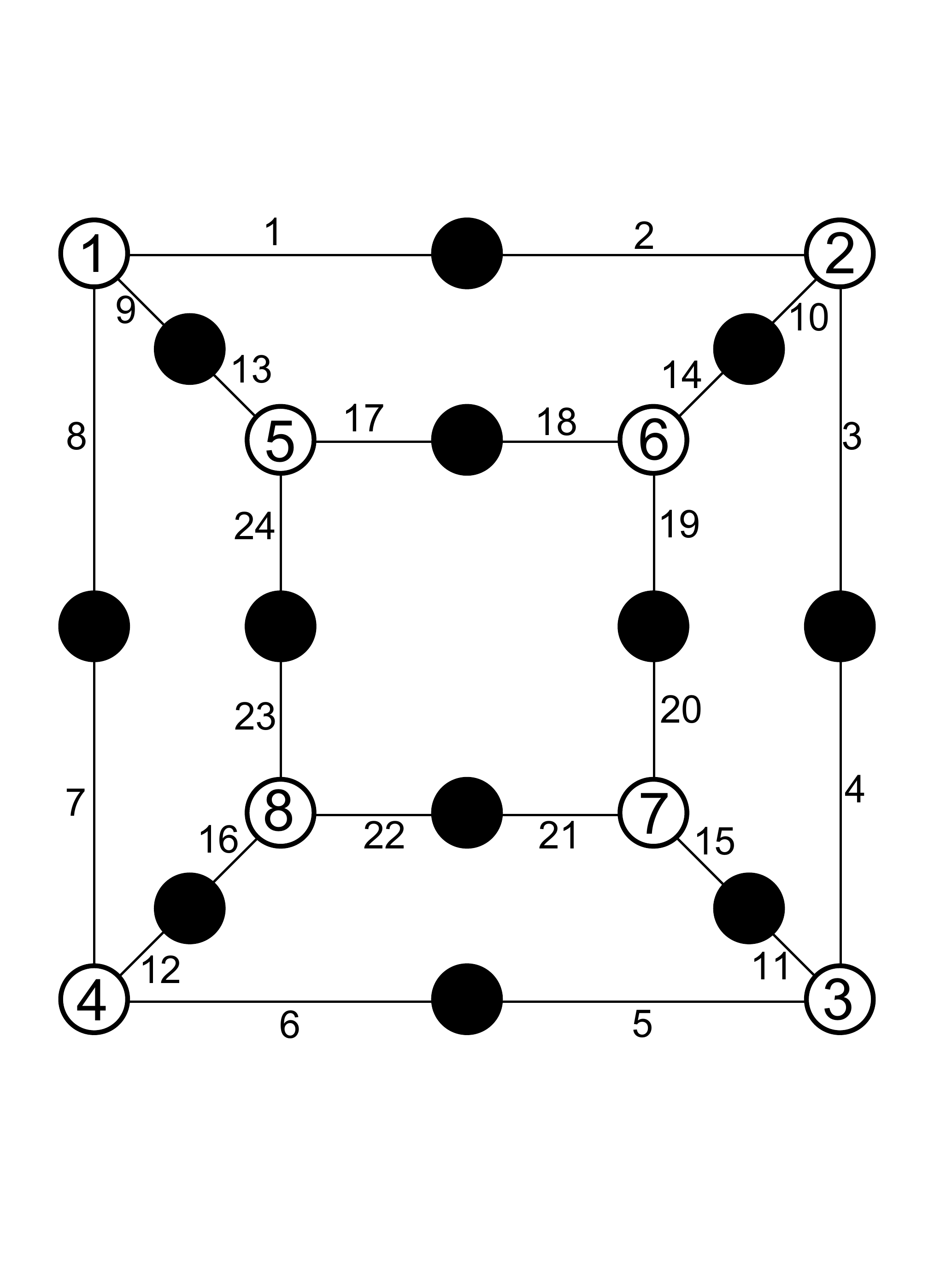}}}$
        \vspace*{-0.5cm}
        \caption{ \{\{\{8,1,9\},\{2,3,10\},
        \{4,5,11\},\{12,6,7\},\{18,14,19\},
        \{22,16,23\},\{13,24,17\},\{21,20,15\}\}, \\ \{\{1,2\},\{9,13\},\{10,14\},\{3,4\},
        \{15,11\},\{24,23\},\{17,18\},
        \{8,7\},\{5,6\},\{12,16\},
        \{21,22\},\{20,19\}\}\}}
        \caption{4-4-4-4-4-4 $(\mathbb{Q})$}
        \label{Dessin}
    \end{subfigure} \hfill
    \begin{subfigure}{0.5\textwidth}
        \centering \captionsetup{justification=centering}
        $\scalemath{0.75}{
        \displaystyle \begin{pmatrix}
            0 & 1 & 0 & 1 & 1 & 0 & 0 & 0\\ 
            1 & 0 & 1 & 0 & 0 & 1 & 0 & 0\\
            0 & 1 & 0 & 0 & 0 & 0 & 1 & 1\\
            1 & 0 & 0 & 0 & 1 & 0 & 0 & 1\\
            1 & 0 & 0 & 1 & 0 & 1 & 0 & 0\\
            0 & 1 & 0 & 0 & 1 & 0 & 1 & 0\\
            0 & 0 & 1 & 0 & 0 & 1 & 0 & 1\\
            0 & 0 & 1 & 1 & 0 & 0 & 1 & 0
        \end{pmatrix}}$
        $\vcenter{\hbox{\includegraphics[width=0.5\textwidth]{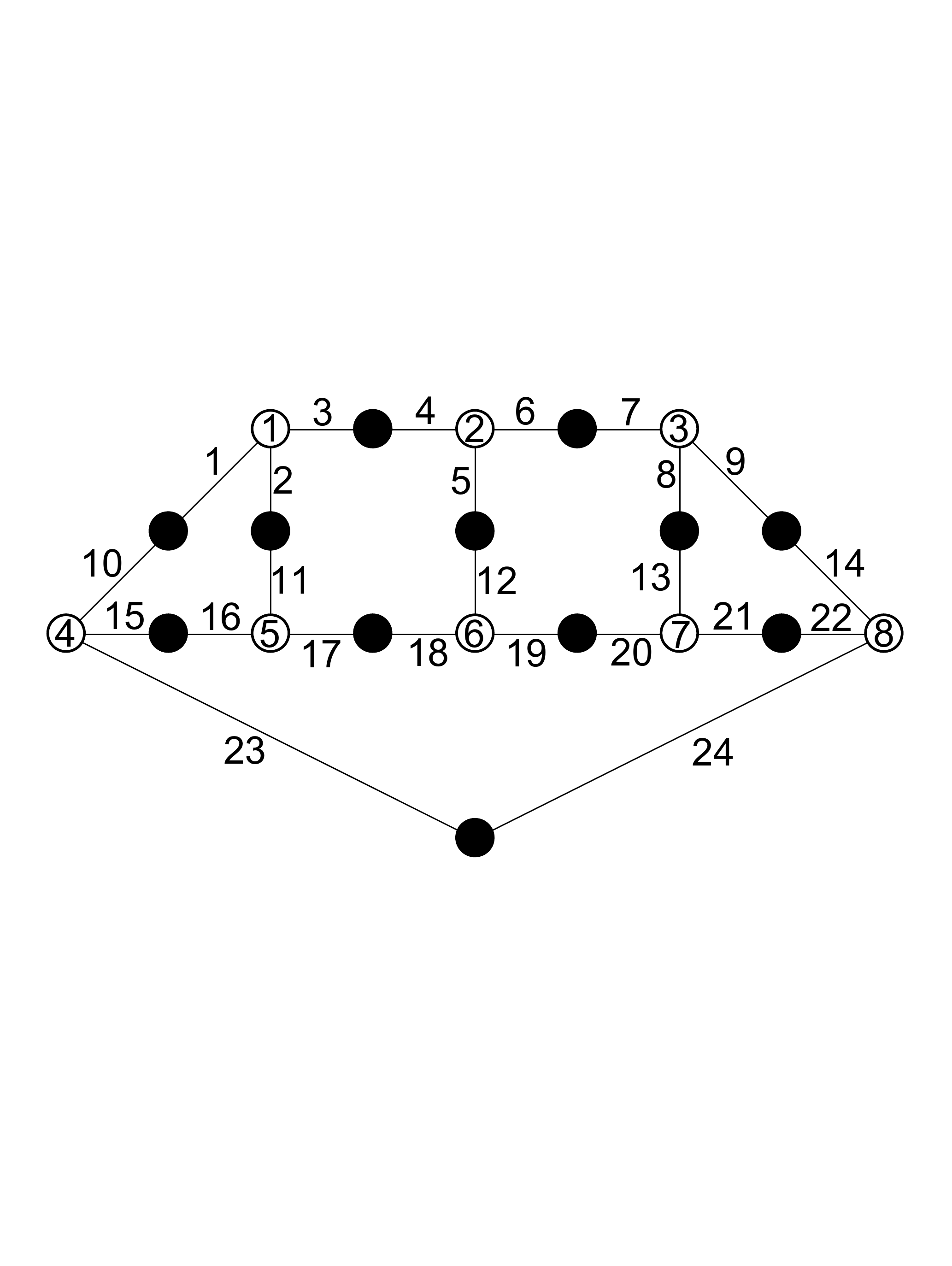}}}$
        \vspace*{-0.5cm}
        \caption{ \{\{\{10,15,23\},\{21,20,13\},
        \{4,6,5\},\{1,3,2\},\{11,17,16\},
        \{22,14,24\},\{7,9,8\},\{12,19,18\}\}, \\ 
        \{\{24,23\},\{14,9\},\{8,13\},
        \{16,15\},\{1,10\},\{6,7\},
        \{3,4\},\{11,2\},\{20,19\},
        \{21,22\},\{17,18\},\{12,5\}\}\}}
        \caption{5-5-4-4-3-3 $(\mathbb{Q})$}
        \label{Dessin}
    \end{subfigure}\hfill
\end{figure}

\begin{figure}[H]
    \begin{subfigure}{0.5\textwidth}
        \centering \captionsetup{justification=centering}
        $\scalemath{0.75}{
        \displaystyle \begin{pmatrix}
            0 & 1 & 1 & 1 & 0 & 0 & 0 & 0\\ 
            1 & 0 & 0 & 0 & 2 & 0 & 0 & 0\\
            1 & 0 & 0 & 0 & 0 & 1 & 1 & 0\\
            1 & 0 & 0 & 0 & 0 & 1 & 0 & 1\\
            0 & 2 & 0 & 0 & 0 & 1 & 0 & 0\\
            0 & 0 & 1 & 1 & 1 & 0 & 0 & 0\\
            0 & 0 & 1 & 0 & 0 & 0 & 0 & 2\\
            0 & 0 & 0 & 1 & 0 & 0 & 2 & 0
        \end{pmatrix}}$
        $\vcenter{\hbox{\includegraphics[width=0.35\textwidth]{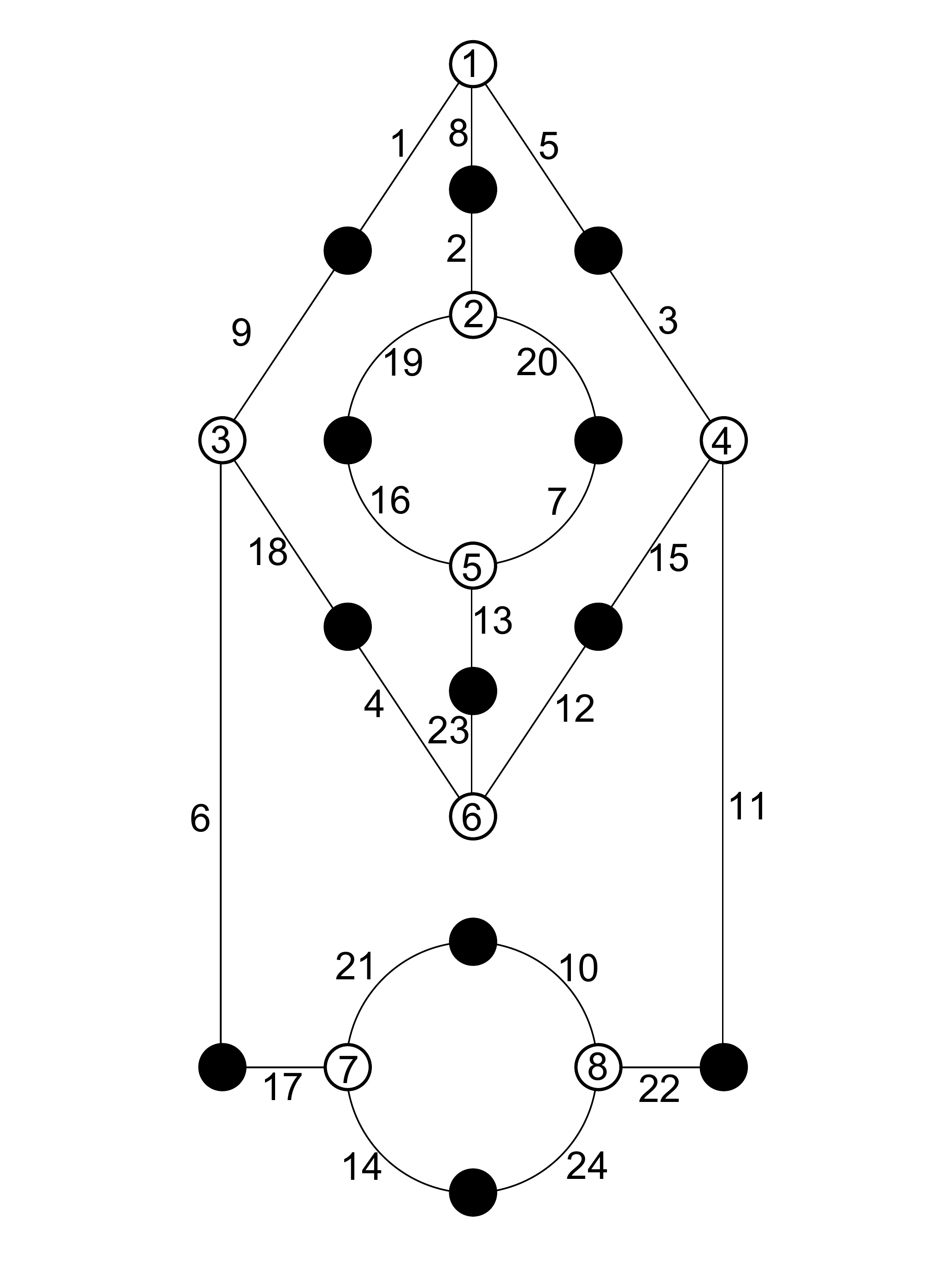}}}$
        \caption{ \{\{\{1,5,8\},\{2,20,19\},
        \{9,18,6\},\{3,11,15\},\{16,7,13\},
        \{23,12,4\},\{21,14,17\},\{10,22,24\}\}, \\ 
        \{\{9,1\},\{5,3\},\{20,7\},
        \{21,10\},\{14,24\},\{17,6\},
        \{22,11\},\{12,15\},\{4,18\},
        \{23,13\},\{16,19\},\{8,2\}\}\}}
        \caption{5-5-5-5-2-2 $(\mathbb{Q})$}
        \label{Dessin}
    \end{subfigure} \hfill
    \begin{subfigure}{0.5\textwidth}
        \centering \captionsetup{justification=centering}
        $\scalemath{0.75}{
        \displaystyle \begin{pmatrix}
            0 & 2 & 1 & 0 & 0 & 0 & 0 & 0\\ 
            2 & 0 & 0 & 0 & 1 & 0 & 0 & 0\\
            1 & 0 & 0 & 1 & 0 & 1 & 0 & 0\\
            0 & 0 & 1 & 0 & 1 & 0 & 1 & 0\\
            0 & 1 & 0 & 1 & 0 & 0 & 0 & 1\\
            0 & 0 & 1 & 0 & 0 & 0 & 1 & 1\\
            0 & 0 & 0 & 1 & 0 & 1 & 0 & 1\\
            0 & 0 & 0 & 0 & 1 & 1 & 1 & 0
        \end{pmatrix}}$
        $\vcenter{\hbox{\includegraphics[width=0.35\textwidth]{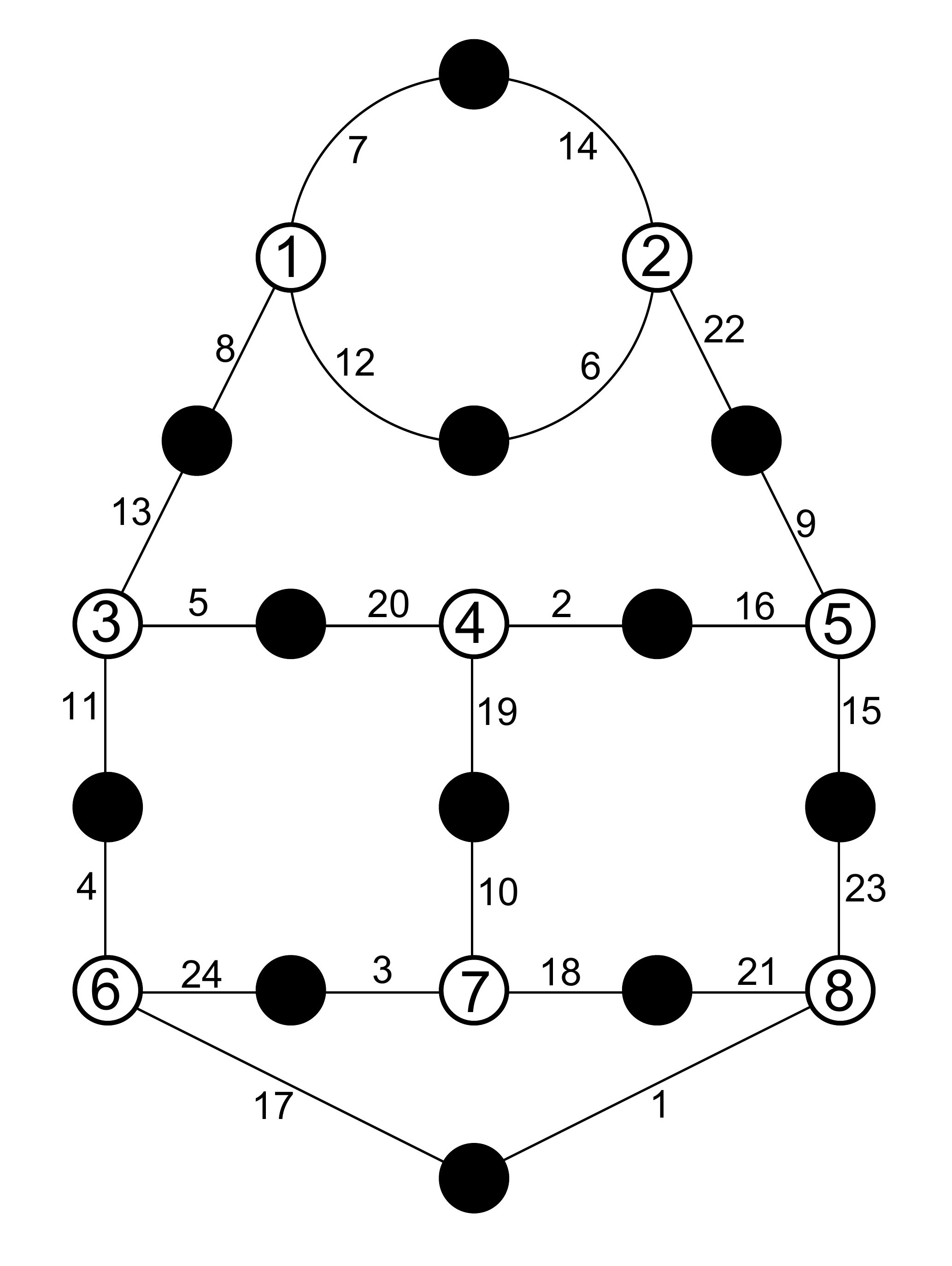}}}$
        \caption{ \{\{\{7,12,8\},\{13,5,11\},
        \{3,10,18\},\{23,1,21\},\{24,17,4\},
        \{14,22,6\},\{2,19,20\},\{15,16,9\}\}, \\ 
        \{\{1,17\},\{21,18\},\{3,24\},
        \{10,19\},\{4,11\},\{15,23\},
        \{2,16\},\{20,5\},\{13,8\},
        \{22,9\},\{12,6\},\{7,14\}\}\}}
        \caption{6-5-4-4-3-2 $(\mathbb{Q})$}
        \label{Dessin}
    \end{subfigure}\hfill
\end{figure}

\begin{figure}[H]
    \begin{subfigure}{0.5\textwidth}
        \centering \captionsetup{justification=centering}
        $\scalemath{0.75}{
        \displaystyle \begin{pmatrix}
            0 & 1 & 1 & 1 & 0 & 0 & 0 & 0\\ 
            1 & 0 & 0 & 1 & 0 & 1 & 0 & 0\\
            1 & 0 & 0 & 1 & 0 & 0 & 1 & 0\\
            1 & 1 & 1 & 0 & 0 & 0 & 0 & 0\\
            0 & 0 & 0 & 0 & 0 & 1 & 1 & 1\\
            0 & 1 & 0 & 0 & 1 & 0 & 0 & 1\\
            0 & 0 & 1 & 0 & 1 & 0 & 0 & 1\\
            0 & 0 & 0 & 0 & 1 & 1 & 1 & 0
        \end{pmatrix}}$
        $\vcenter{\hbox{\includegraphics[width=0.35\textwidth]{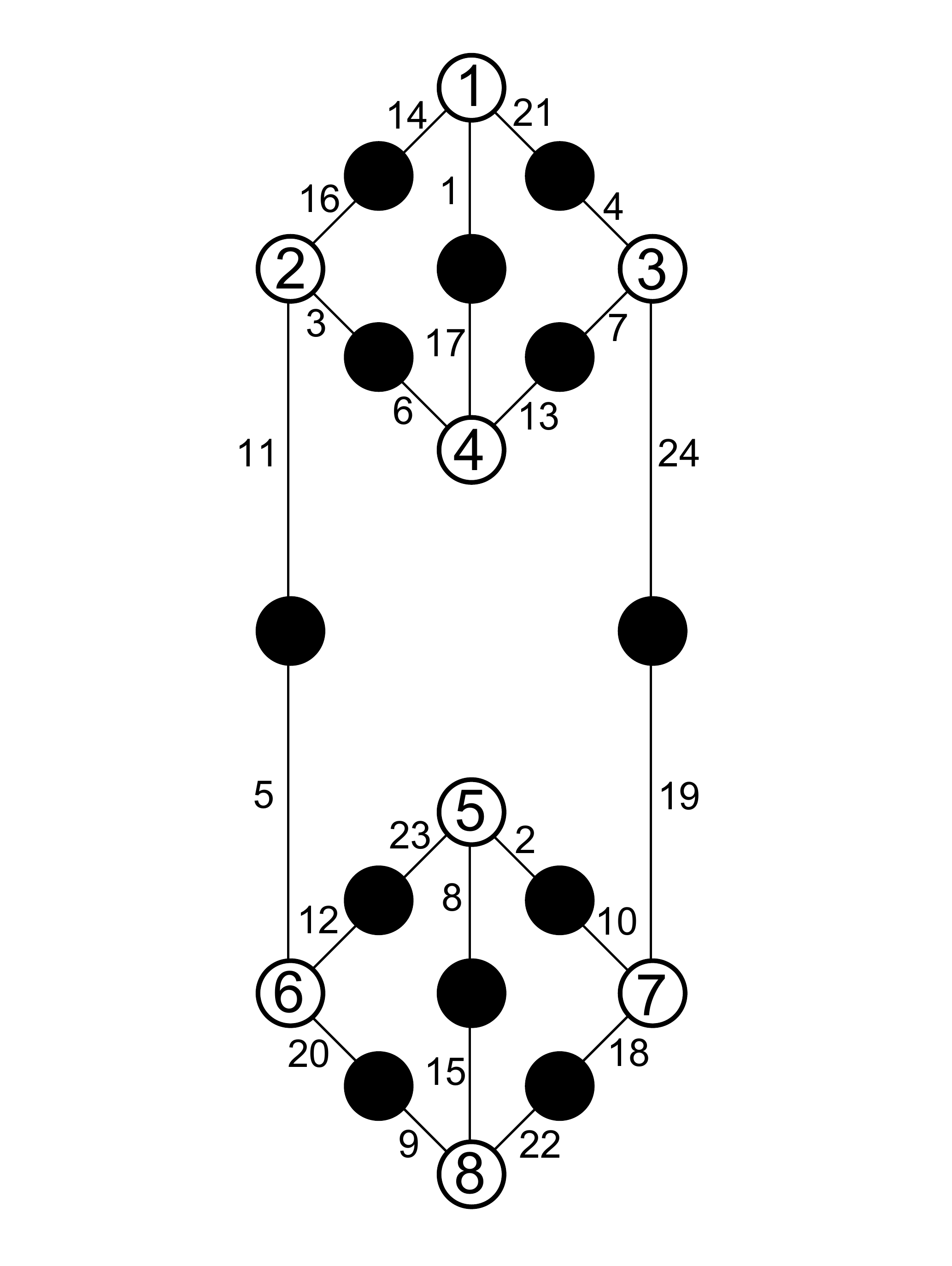}}}$
        \caption{ \{\{\{23,2,8\},\{10,19,18\},
        \{6,17,13\},\{21,1,14\},\{12,20,5\},
        \{16,3,11\},\{10,19,18\},\{4,24,7\}\}, \\ 
        \{\{3,6\},\{4,21\},\{14,16\},
        \{9,20\},\{12,23\},\{24,19\},
        \{5,11\},\{8,15\},\{18,22\},
        \{2,10\},\{17,1\},\{13,7\}\}\}}
        \caption{6-6-3-3-3-3 $(\mathbb{Q})$}
        \label{Dessin}
    \end{subfigure} \hfill
    \begin{subfigure}{0.5\textwidth}
        \centering \captionsetup{justification=centering}
        $\scalemath{0.75}{
        \displaystyle \begin{pmatrix}
            0 & 2 & 0 & 0 & 1 & 0 & 0 & 0\\ 
            2 & 0 & 0 & 0 & 0 & 1 & 0 & 0\\
            0 & 0 & 0 & 1 & 1 & 0 & 1 & 0\\
            0 & 0 & 1 & 0 & 0 & 1 & 0 & 1\\
            1 & 0 & 1 & 0 & 0 & 0 & 1 & 0\\
            0 & 1 & 0 & 1 & 0 & 0 & 0 & 1\\
            0 & 0 & 1 & 0 & 1 & 0 & 0 & 1\\
            0 & 0 & 0 & 1 & 0 & 1 & 1 & 0
        \end{pmatrix}}$
        $\vcenter{\hbox{\includegraphics[width=0.35\textwidth]{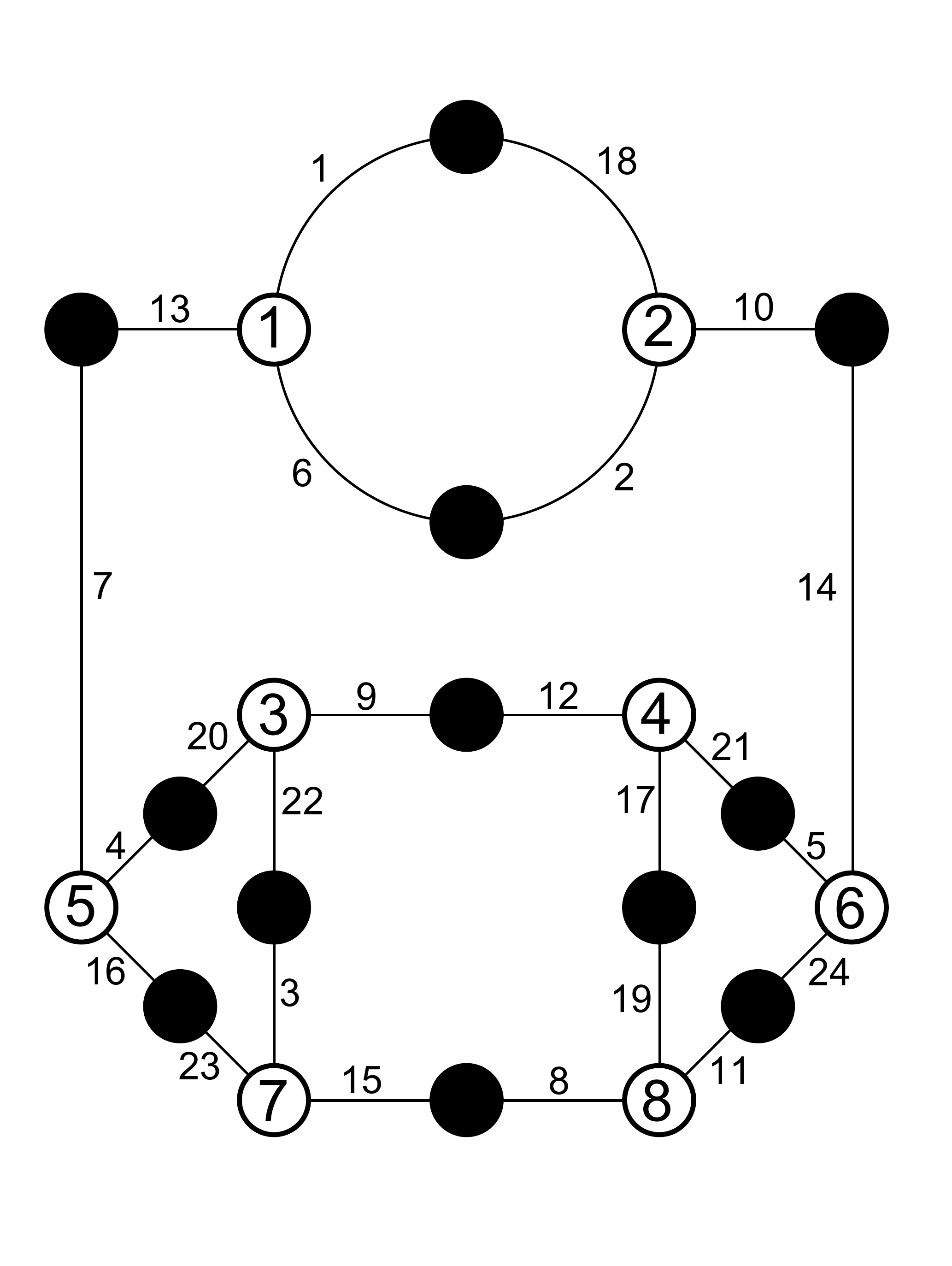}}}$
        \caption{ \{\{\{1,6,13\},\{18,10,2\},
        \{5,14,24\},\{20,9,22\},\{4,16,7\},
        \{11,8,19\},\{23,3,15\},\{12,21,17\}\}, \\ 
        \{\{7,13\},\{3,22\},\{11,24\},
        \{17,19\},\{23,16\},\{15,8\},
        \{1,18\},\{10,14\},\{6,2\},
        \{4,20\},\{5,21\},\{9,12\}\}\}}
        \caption{6-6-4-3-3-2 $(\mathbb{Q})$}
        \label{Dessin}
    \end{subfigure}\hfill
\end{figure}

\begin{figure}[H]
    \begin{subfigure}{0.5\textwidth}
        \centering \captionsetup{justification=centering}
        $\scalemath{0.75}{
        \displaystyle \begin{pmatrix}
            0 & 2 & 1 & 0 & 0 & 0 & 0 & 0\\ 
            2 & 0 & 0 & 1 & 0 & 0 & 0 & 0\\
            1 & 0 & 0 & 1 & 1 & 0 & 0 & 0\\
            0 & 1 & 1 & 0 & 0 & 0 & 0 & 1\\
            0 & 0 & 1 & 0 & 0 & 1 & 0 & 1\\
            0 & 0 & 0 & 0 & 1 & 0 & 2 & 0\\
            0 & 0 & 0 & 0 & 0 & 2 & 0 & 1\\
            0 & 0 & 0 & 1 & 1 & 0 & 1 & 0
        \end{pmatrix}}$
        $\vcenter{\hbox{\includegraphics[width=0.35\textwidth]{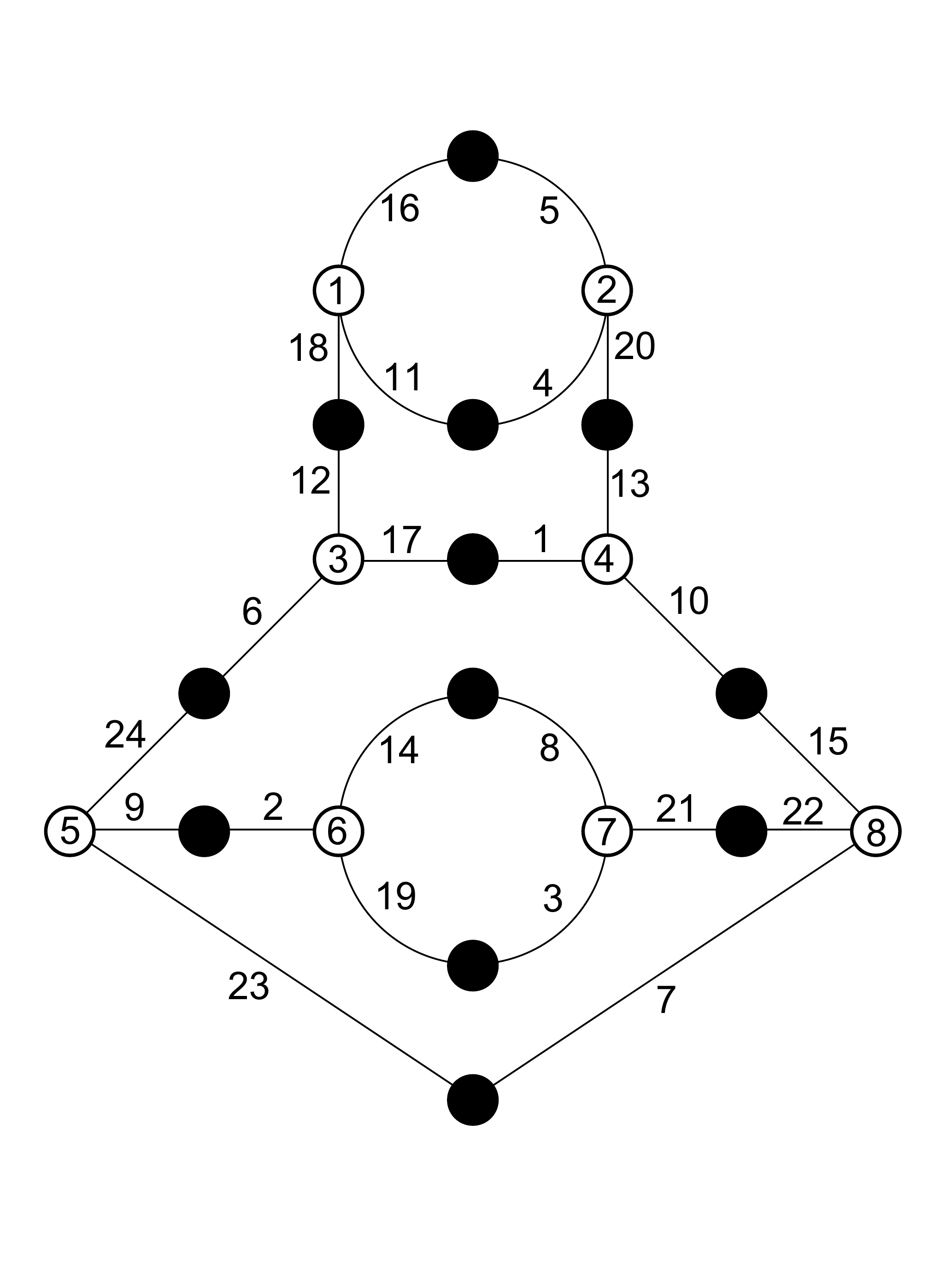}}}$
        \caption{ \{\{\{9,23,24\},\{2,14,19\},
        \{8,21,3\},\{22,15,7\},\{6,12,17\},
        \{1,13,10\},\{18,16,11\},\{5,20,4\}\}, \\ 
        \{\{16,5\},\{11,4\},\{12,18\},
        \{13,20\},\{1,17\},\{6,24\},
        \{10,15\},\{9,2\},\{8,14\},
        \{3,19\},\{21,22\},\{7,23\}\}\}}
        \caption{6-6-4-4-2-2 $(\mathbb{Q})$}
        \label{Dessin}
    \end{subfigure} \hfill
    \begin{subfigure}{0.5\textwidth}
        \centering \captionsetup{justification=centering}
        $\scalemath{0.75}{
        \displaystyle \begin{pmatrix}
            0 & 1 & 1 & 1 & 0 & 0 & 0 & 0\\ 
            1 & 0 & 0 & 1 & 0 & 1 & 0 & 0\\
            1 & 0 & 2 & 0 & 0 & 0 & 0 & 0\\
            1 & 1 & 0 & 0 & 0 & 0 & 1 & 0\\
            0 & 0 & 0 & 0 & 0 & 1 & 1 & 1\\
            0 & 1 & 0 & 0 & 1 & 0 & 0 & 1\\
            0 & 0 & 0 & 1 & 1 & 0 & 0 & 1\\
            0 & 0 & 0 & 0 & 1 & 1 & 1 & 0
        \end{pmatrix}}$
        $\vcenter{\hbox{\includegraphics[width=0.35\textwidth]{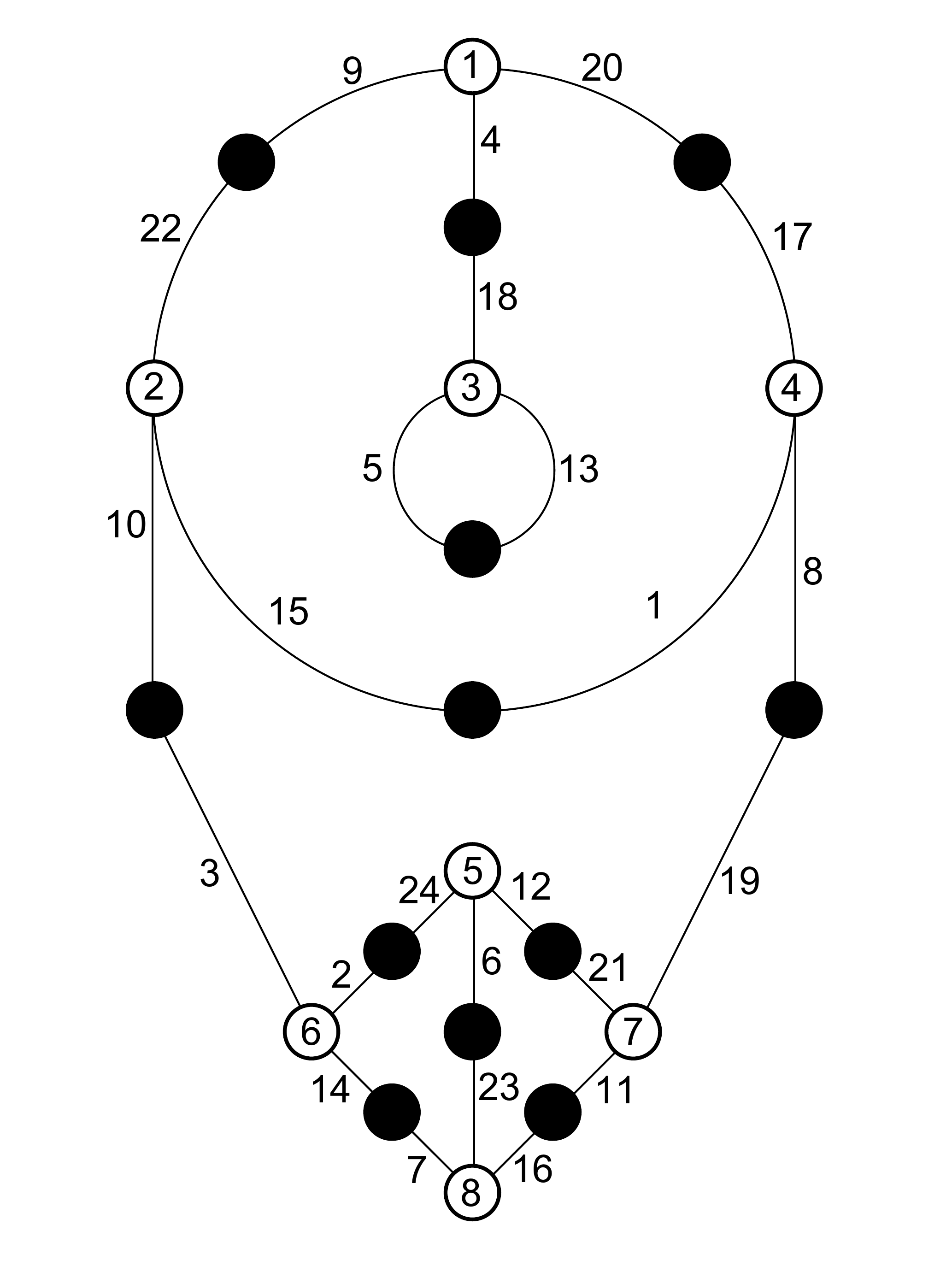}}}$
        \caption{ \{\{\{22,15,10\},\{9,20,4\},
        \{17,8,1\},\{18,13,5\},\{24,12,6\},
        \{2,14,3\},\{21,19,11\},\{23,16,7\}\}, \\ 
        \{\{7,14\},\{16,11\},\{23,6\},
        \{12,21\},\{2,24\},\{3,10\},
        \{19,8\},\{1,15\},\{5,13\},
        \{18,4\},\{9,22\},\{17,20\}\}\}}
        \caption{6-6-5-3-3-1 $(\mathbb{Q})$}
        \label{Dessin}
    \end{subfigure}\hfill
\end{figure}

\begin{figure}[H]
    \begin{subfigure}{0.5\textwidth}
        \centering \captionsetup{justification=centering}
        $\scalemath{0.75}{
        \displaystyle \begin{pmatrix}
            0 & 2 & 1 & 0 & 0 & 0 & 0 & 0\\ 
            2 & 0 & 0 & 1 & 0 & 0 & 0 & 0\\
            1 & 0 & 0 & 1 & 1 & 0 & 0 & 0\\
            0 & 1 & 1 & 0 & 0 & 0 & 1 & 0\\
            0 & 0 & 1 & 0 & 0 & 0 & 1 & 1\\
            0 & 0 & 0 & 0 & 0 & 2 & 0 & 1\\
            0 & 0 & 0 & 1 & 1 & 0 & 0 & 1\\
            0 & 0 & 0 & 0 & 1 & 1 & 1 & 0
        \end{pmatrix}}$
        $\vcenter{\hbox{\includegraphics[width=0.35\textwidth]{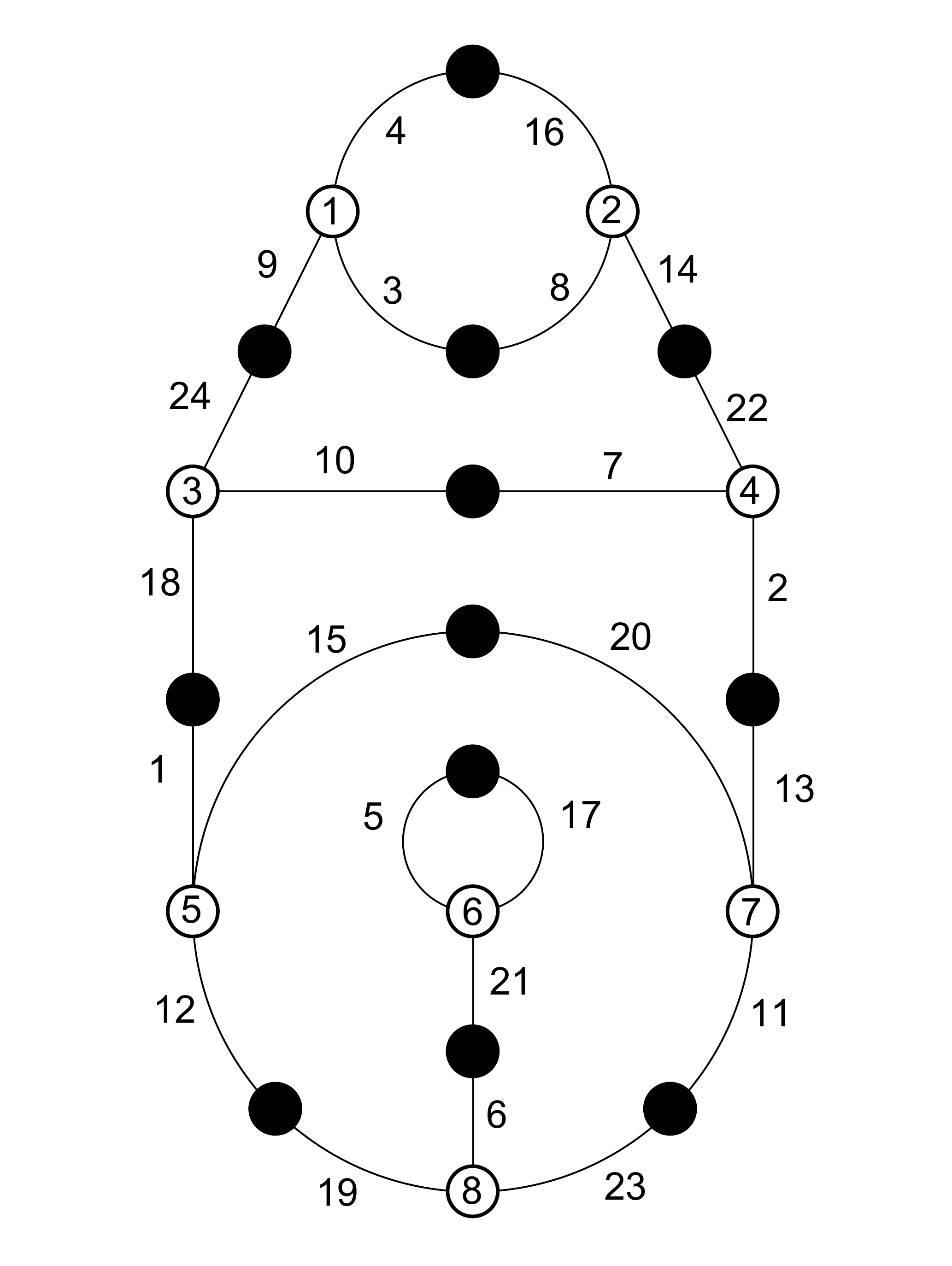}}}$
        \caption{ \{\{\{24,10,18\},\{4,3,9\},
        \{16,14,8\},\{22,2,7\},\{1,15,12\},
        \{13,11,20\},\{5,17,21\},\{6,23,19\}\}, \\ 
        \{\{12,19\},\{6,21\},\{23,11\},
        \{5,17\},\{1,18\},\{2,23\},
        \{20,15\},\{10,7\},\{14,22\},
        \{9,24\},\{3,8\},\{16,4\}\}\}}
        \caption{6-6-5-4-2-1 $(\mathbb{Q})$}
        \label{Dessin}
    \end{subfigure} \hfill
    \begin{subfigure}{0.5\textwidth}
        \centering \captionsetup{justification=centering}
        $\scalemath{0.75}{
        \displaystyle \begin{pmatrix}
            0 & 1 & 2 & 0 & 0 & 0 & 0 & 0\\ 
            1 & 2 & 0 & 0 & 0 & 0 & 0 & 0\\
            2 & 0 & 0 & 1 & 0 & 0 & 0 & 0\\
            0 & 0 & 1 & 0 & 0 & 0 & 0 & 2\\
            0 & 0 & 0 & 0 & 0 & 1 & 2 & 0\\
            0 & 0 & 0 & 0 & 1 & 2 & 0 & 0\\
            0 & 0 & 0 & 0 & 2 & 0 & 0 & 1\\
            0 & 0 & 0 & 2 & 0 & 0 & 1 & 0
        \end{pmatrix}}$
        $\vcenter{\hbox{\includegraphics[width=0.35\textwidth]{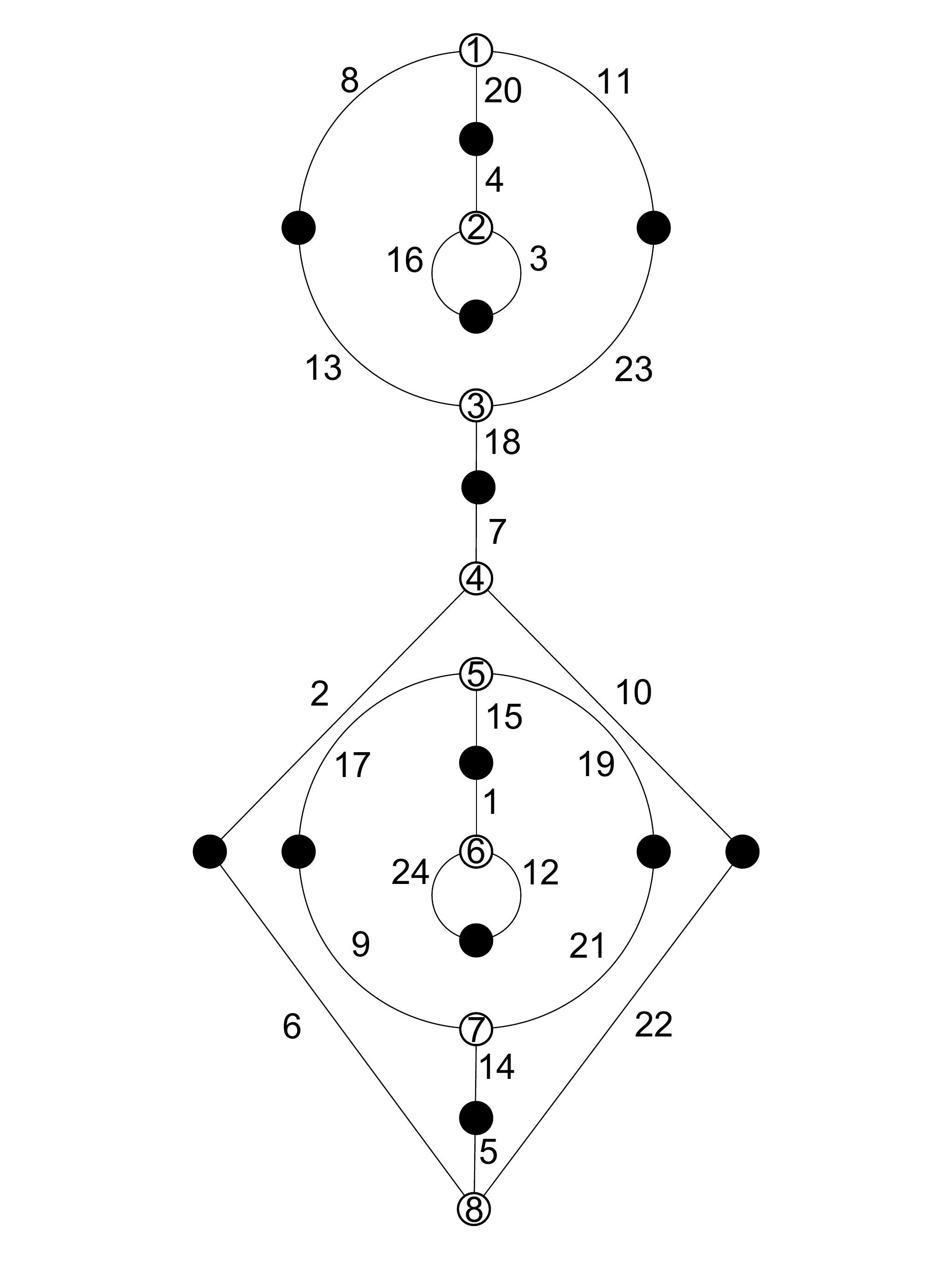}}}$
        \caption{ \{\{\{11,20,8\},\{3,16,4\},
        \{13,23,18\},\{7,10,2\},\{15,17,19\},
        \{12,24,1\},\{9,21,14\},\{5,22,6\}\}, \\ 
        \{\{6,2\},\{9,17\},\{15,1\},
        \{24,12\},\{5,14\},\{21,19\},
        \{22,10\},\{7,18\},\{8,13\},
        \{3,16\},\{4,20\},\{11,23\}\}\}}
        \caption{6-6-5-5-1-1 A $(\sqrt{3})$}
        \label{Dessin}
    \end{subfigure}\hfill
\end{figure}

\begin{figure}[H]
    \begin{subfigure}{0.5\textwidth}
        \centering \captionsetup{justification=centering}
        $\scalemath{0.75}{
        \displaystyle \begin{pmatrix}
            2 & 1 & 0 & 0 & 0 & 0 & 0 & 0\\ 
            1 & 0 & 1 & 0 & 0 & 0 & 1 & 0\\
            0 & 1 & 0 & 1 & 0 & 0 & 1 & 0\\
            0 & 0 & 1 & 0 & 0 & 1 & 0 & 1\\
            0 & 0 & 0 & 0 & 2 & 0 & 0 & 1\\
            0 & 0 & 0 & 1 & 0 & 0 & 1 & 1\\
            0 & 1 & 1 & 0 & 0 & 1 & 0 & 0\\
            0 & 0 & 0 & 1 & 1 & 1 & 0 & 0
        \end{pmatrix}}$
        $\vcenter{\hbox{\includegraphics[width=0.35\textwidth]{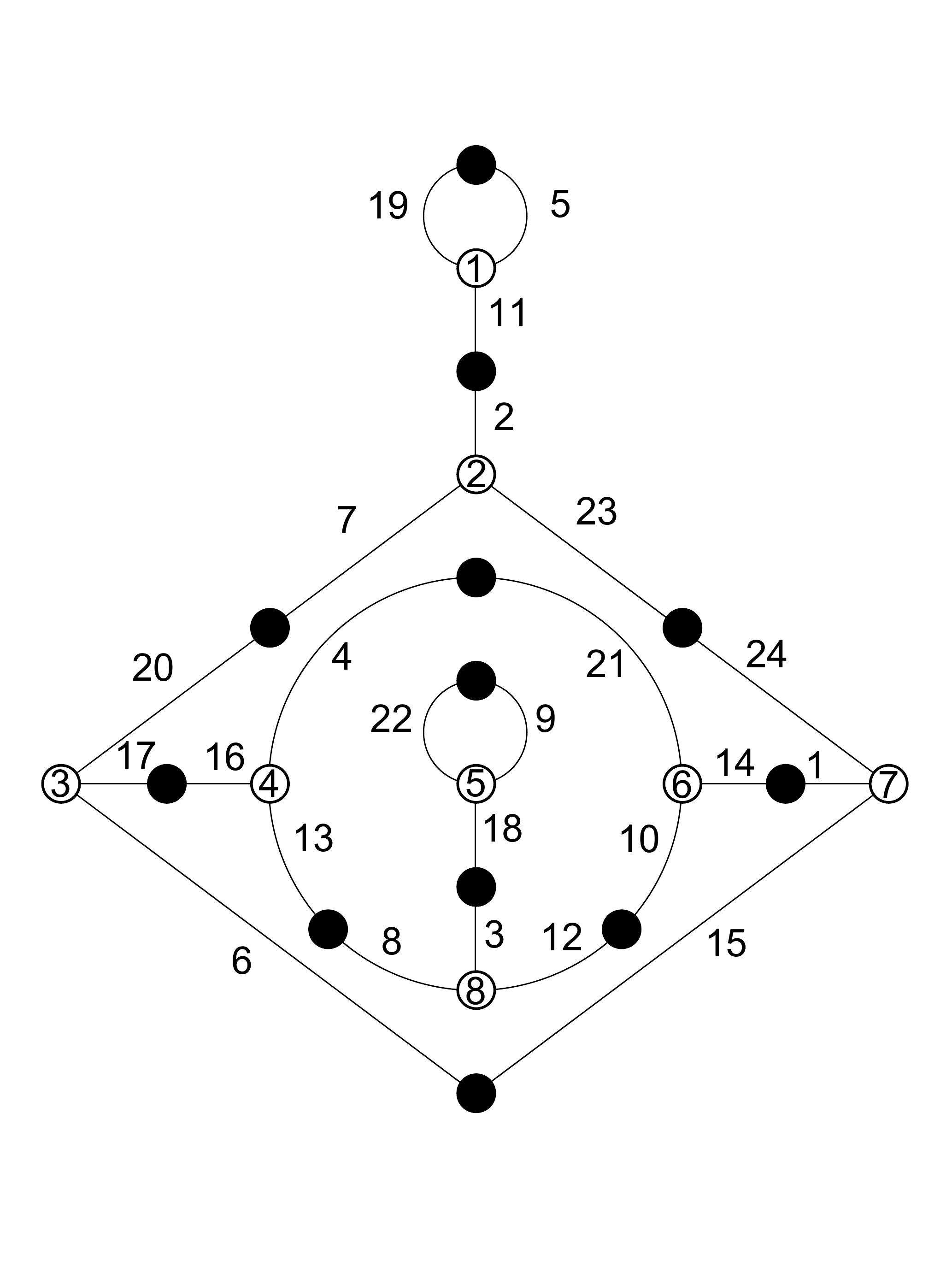}}}$
        \caption{ \{\{\{1,24,15\},\{14,10,21\},
        \{9,18,22\},\{3,12,8\},\{16,4,13\},
        \{17,6,20\},\{23,7,2\},\{11,19,5\}\}, \\ 
        \{\{5,19\},\{2,11\},\{7,20\},
        \{23,24\},\{1,14\},\{10,12\},
        \{6,15\},\{8,13\},\{9,22\},
        \{21,4\},\{16,17\},\{18,3\}\}\}}
        \caption{6-6-5-5-1-1 B $(\sqrt{3})$}
        \label{Dessin}
    \end{subfigure} \hfill
    \begin{subfigure}{0.5\textwidth}
        \centering \captionsetup{justification=centering}
        $\scalemath{0.75}{
        \displaystyle \begin{pmatrix}
            0 & 2 & 1 & 0 & 0 & 0 & 0 & 0\\ 
            2 & 0 & 0 & 0 & 0 & 1 & 0 & 0\\
            1 & 0 & 0 & 1 & 0 & 0 & 1 & 0\\
            0 & 0 & 1 & 0 & 2 & 0 & 0 & 0\\
            0 & 0 & 0 & 2 & 0 & 1 & 0 & 0\\
            0 & 1 & 0 & 0 & 1 & 0 & 0 & 1\\
            0 & 0 & 1 & 0 & 0 & 0 & 0 & 2\\
            0 & 0 & 0 & 0 & 0 & 1 & 2 & 0
        \end{pmatrix}}$
        $\vcenter{\hbox{\includegraphics[width=0.35\textwidth]{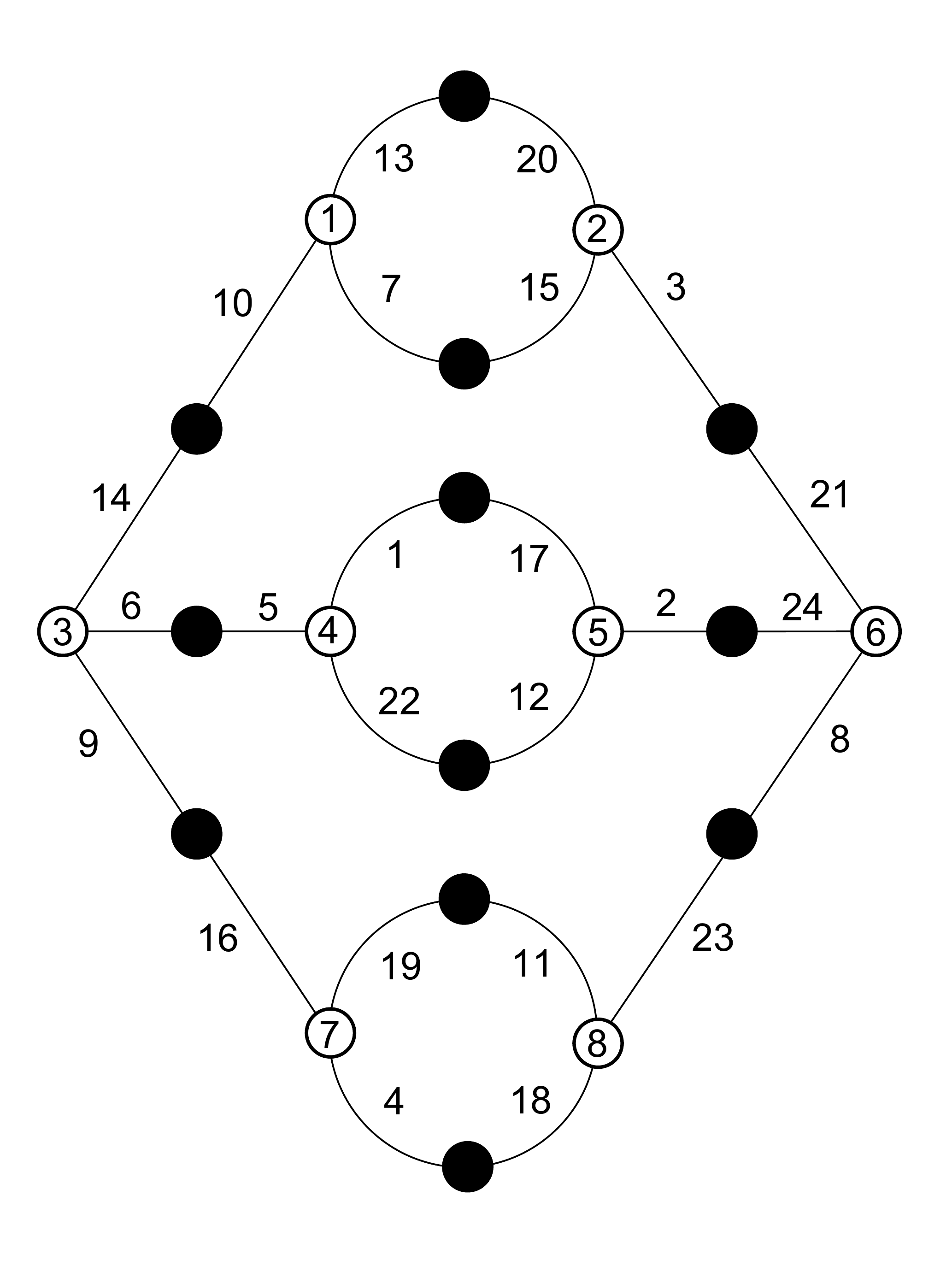}}}$
        \caption{ \{\{\{7,10,13\},\{3,15,20\},
        \{1,22,5\},\{17,2,12\},\{24,21,8\},
        \{6,9,14\},\{4,16,19\},\{11,23,18\}\}, \\ 
        \{\{4,18\},\{19,11\},\{16,9\},
        \{8,23\},\{22,12\},\{17,1\},
        \{6,5\},\{2,24\},\{21,3\},
        \{10,14\},\{15,7\},\{20,13\}\}\}}
        \caption{6-6-6-2-2-2 $(\mathbb{Q})$}
        \label{Dessin}
    \end{subfigure}\hfill
\end{figure}

\begin{figure}[H]
    \begin{subfigure}{0.5\textwidth}
        \centering \captionsetup{justification=centering}
        $\scalemath{0.75}{
        \displaystyle \begin{pmatrix}
            0 & 1 & 1 & 1 & 0 & 0 & 0 & 0\\ 
            1 & 0 & 0 & 1 & 1 & 0 & 0 & 0\\
            1 & 0 & 2 & 0 & 0 & 0 & 0 & 0\\
            1 & 1 & 0 & 0 & 0 & 0 & 1 & 0\\
            0 & 1 & 0 & 0 & 0 & 0 & 1 & 1\\
            0 & 0 & 0 & 0 & 0 & 2 & 0 & 1\\
            0 & 0 & 0 & 1 & 1 & 0 & 0 & 1\\
            0 & 0 & 0 & 0 & 1 & 1 & 1 & 0
        \end{pmatrix}}$
        $\vcenter{\hbox{\includegraphics[width=0.35\textwidth]{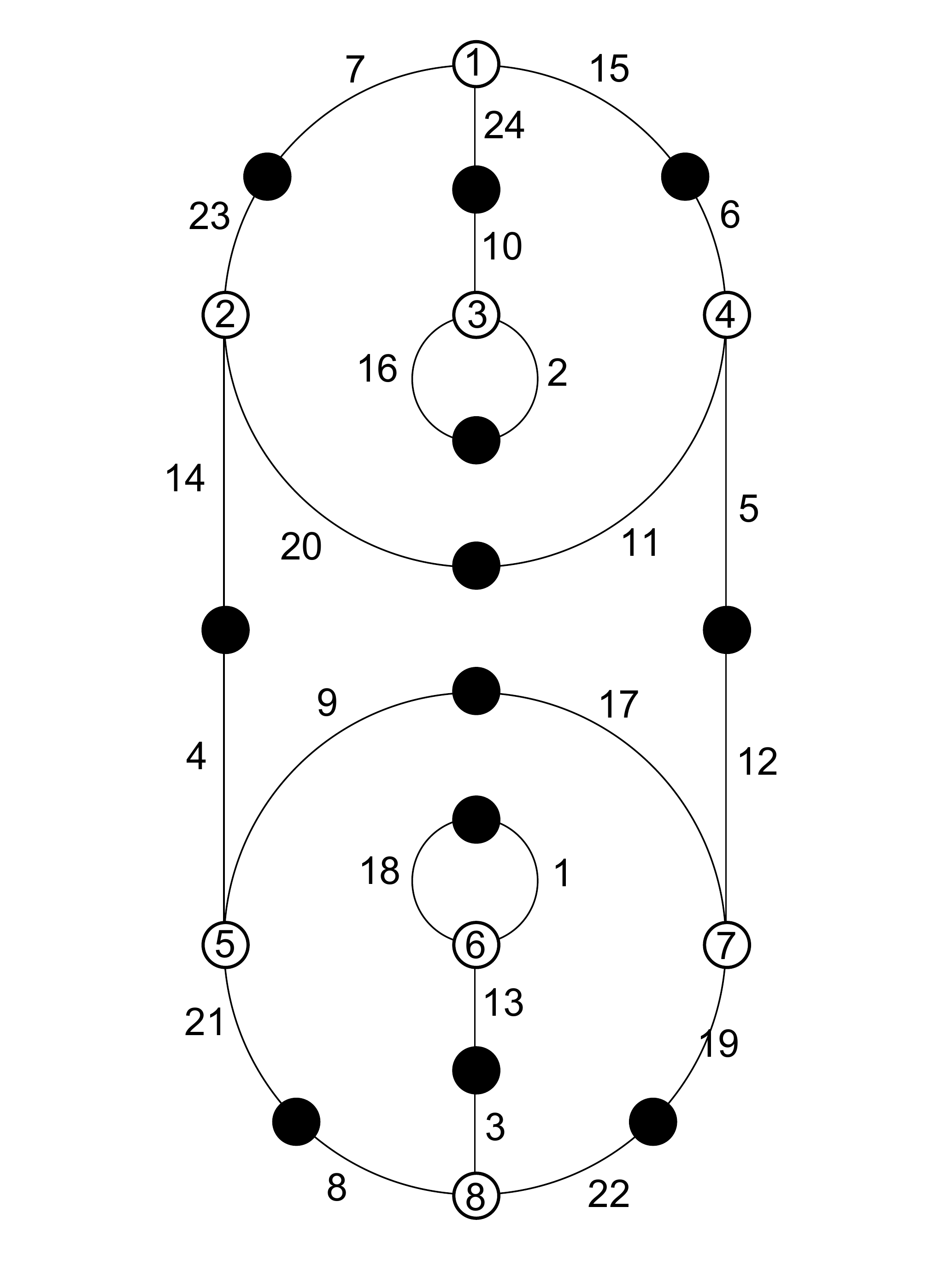}}}$
        \caption{ \{\{\{23,23,14\},\{7,15,24\},
        \{6,5,11\},\{2,16,10\},\{19,17,12\},
        \{1,13,18\},\{4,9,21\},\{3,22,8\}\}, \\ 
        \{\{1,18\},\{3,13\},\{9,17\},
        \{22,19\},\{8,21\},\{14,4\},
        \{12,5\},\{10,24\},\{2,16\},
        \{7,23\},\{15,6\},\{11,20\}\}\}}
        \caption{6-6-6-4-1-1 $(\mathbb{Q})$}
        \label{Dessin}
    \end{subfigure} \hfill
    \begin{subfigure}{0.5\textwidth}
        \centering \captionsetup{justification=centering}
        $\scalemath{0.75}{
        \displaystyle \begin{pmatrix}
            0 & 2 & 1 & 0 & 0 & 0 & 0 & 0\\ 
            2 & 0 & 0 & 1 & 0 & 0 & 0 & 0\\
            1 & 0 & 0 & 1 & 0 & 1 & 0 & 0\\
            0 & 1 & 1 & 0 & 0 & 0 & 1 & 0\\
            0 & 0 & 0 & 0 & 0 & 1 & 1 & 1\\
            0 & 0 & 1 & 0 & 1 & 0 & 0 & 1\\
            0 & 0 & 0 & 1 & 1 & 0 & 0 & 1\\
            0 & 0 & 0 & 0 & 1 & 1 & 1 & 0
        \end{pmatrix}}$
        $\vcenter{\hbox{\includegraphics[width=0.35\textwidth]{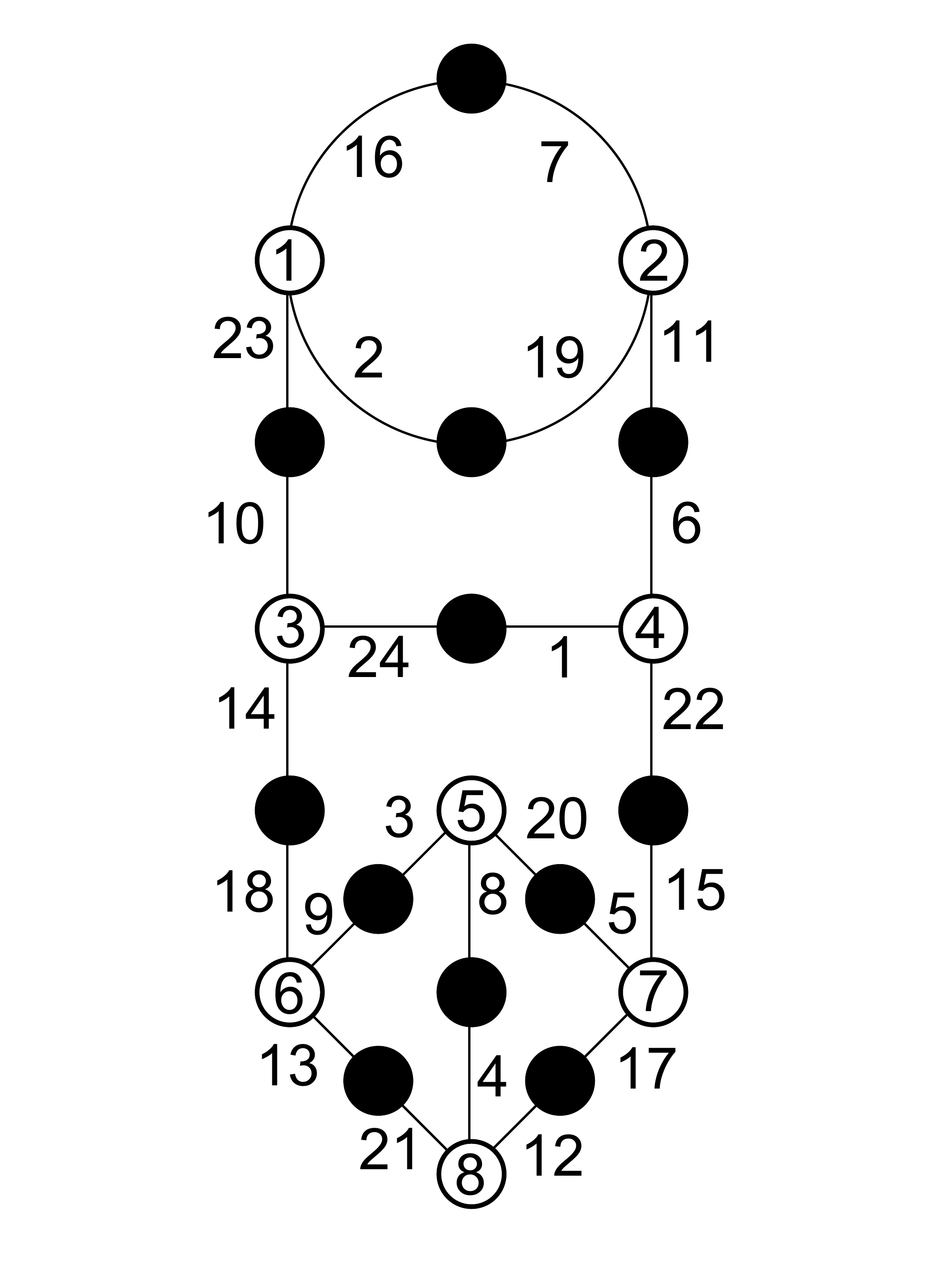}}}$
        \caption{ \{\{\{2,23,16\},\{11,19,7\},
        \{6,22,1\},\{24,14,10\},\{3,20,8\},
        \{9,13,18\},\{5,15,17\},\{12,21,4\}\}, \\ 
        \{\{13,21\},\{12,17\},\{8,4\},
        \{9,3\},\{22,15\},\{5,20\},
        \{18,14\},\{1,24\},\{10,23\},
        \{6,11\},\{2,19\},\{16,7\}\}\}}
        \caption{7-5-4-3-3-2 $(\mathbb{Q})$}
        \label{Dessin}
    \end{subfigure}\hfill
\end{figure}

\begin{figure}[H]
    \begin{subfigure}{0.5\textwidth}
        \centering \captionsetup{justification=centering}
        $\scalemath{0.75}{
        \displaystyle \begin{pmatrix}
            2 & 1 & 0 & 0 & 0 & 0 & 0 & 0\\ 
            1 & 0 & 1 & 0 & 0 & 1 & 0 & 0\\
            0 & 1 & 0 & 1 & 0 & 0 & 0 & 1\\
            0 & 0 & 1 & 0 & 1 & 0 & 1 & 0\\
            0 & 0 & 0 & 1 & 0 & 1 & 1 & 0\\
            0 & 1 & 0 & 0 & 1 & 0 & 0 & 1\\
            0 & 0 & 0 & 1 & 1 & 0 & 0 & 1\\
            0 & 0 & 1 & 0 & 0 & 1 & 1 & 0
        \end{pmatrix}}$
        $\vcenter{\hbox{\includegraphics[width=0.35\textwidth]{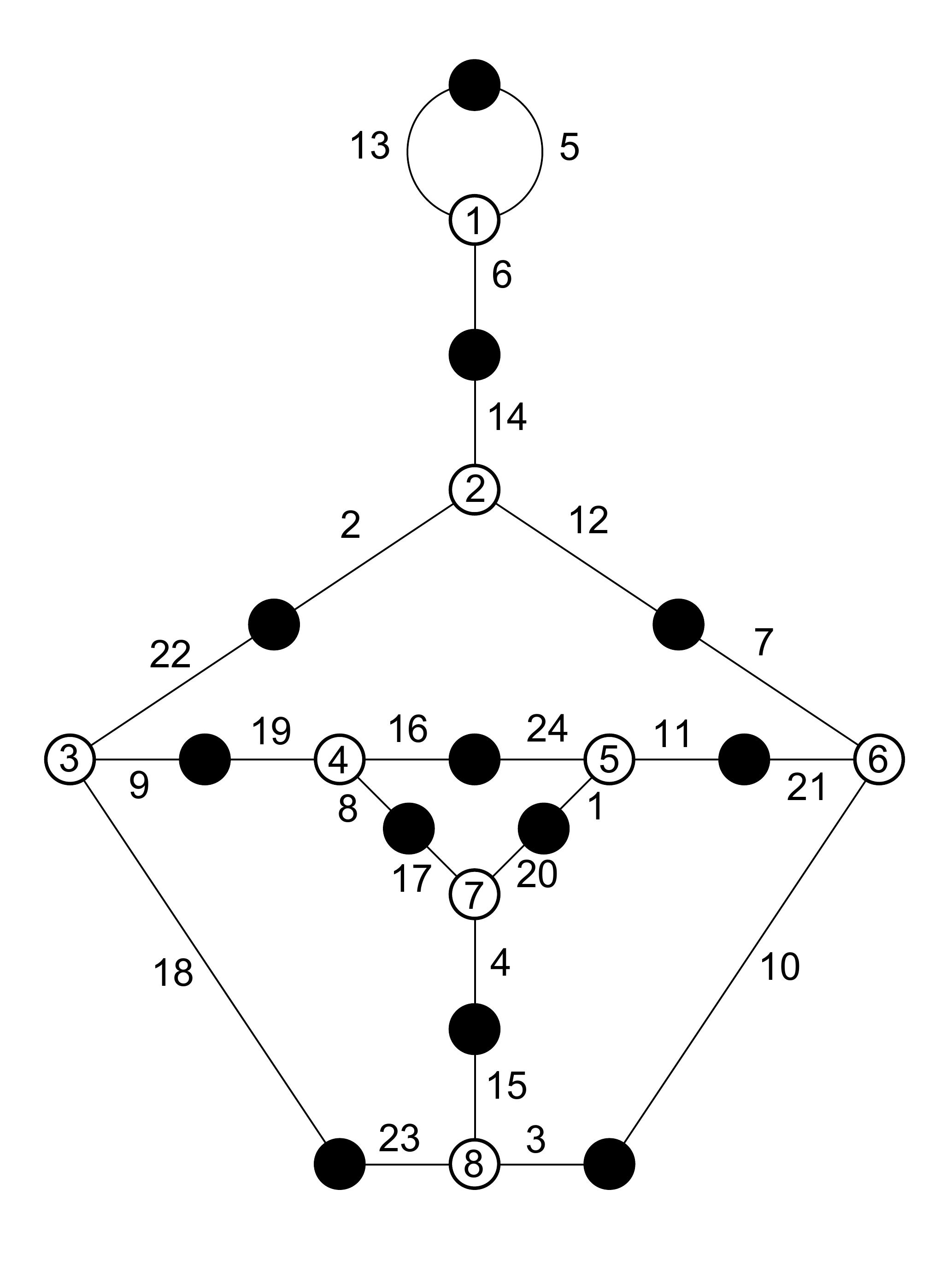}}}$
        \caption{ \{\{\{5,6,13\},\{2,14,12\},
        \{10,21,7\},\{11,1,24\},\{16,8,19\},
        \{20,4,17\},\{22,9,18\},\{23,15,3\}\}, \\ 
        \{\{3,10\},\{4,15\},\{23,18\},
        \{8,17\},\{1,20\},\{16,24\},
        \{9,19\},\{11,21\},\{2,22\},
        \{7,12\},\{14,6\},\{13,5\}\}\}}
        \caption{7-5-4-4-3-1 $(\mathbb{Q})$}
        \label{Dessin}
    \end{subfigure} \hfill
    \begin{subfigure}{0.5\textwidth}
        \centering \captionsetup{justification=centering}
        $\scalemath{0.75}{
        \displaystyle \begin{pmatrix}
            0 & 1 & 0 & 1 & 0 & 1 & 0 & 0\\ 
            1 & 0 & 2 & 0 & 0 & 0 & 0 & 0\\
            0 & 2 & 0 & 0 & 1 & 0 & 0 & 0\\
            1 & 0 & 0 & 0 & 1 & 1 & 0 & 0\\
            0 & 0 & 1 & 1 & 0 & 0 & 0 & 1\\
            1 & 0 & 0 & 1 & 0 & 0 & 1 & 0\\
            0 & 0 & 0 & 0 & 0 & 1 & 0 & 2\\
            0 & 0 & 0 & 0 & 1 & 0 & 2 & 0
        \end{pmatrix}}$
        $\vcenter{\hbox{\includegraphics[width=0.35\textwidth]{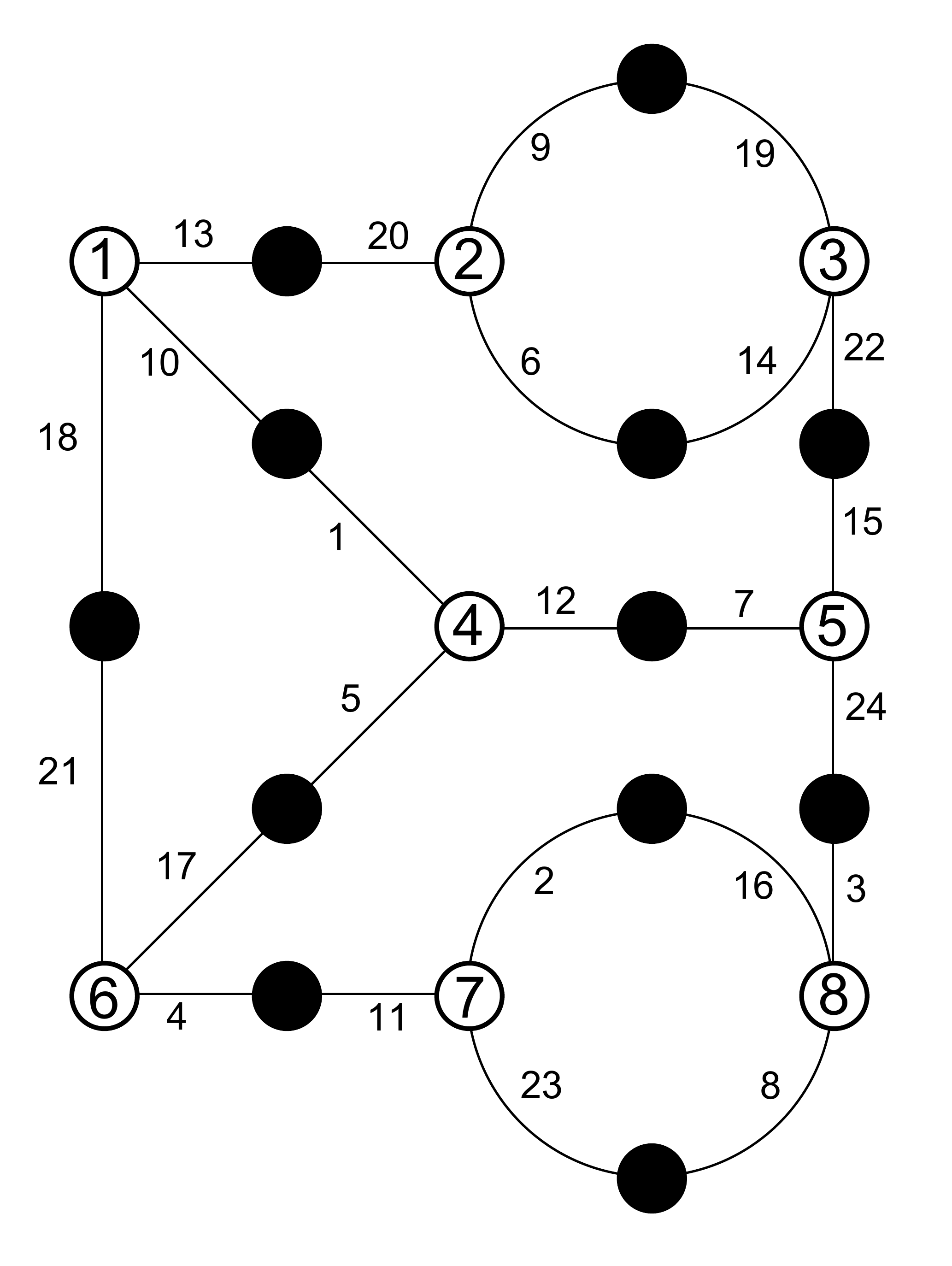}}}$
        \caption{ \{\{\{9,6,20\},\{19,22,14\},
        \{15,24,7\},\{12,5,1\},\{13,10,18\},
        \{21,17,4\},\{11,2,23\},\{3,8,16\}\}, \\ 
        \{\{13,20\},\{19,9\},\{14,6\},
        \{1,10\},\{18,21\},\{5,17\},
        \{7,12\},\{3,24\},\{4,11\},
        \{2,16\},\{8,23\},\{3,24\}\}\}}
        \caption{7-5-5-3-2-2 $(\mathbb{Q})$}
        \label{Dessin}
    \end{subfigure}\hfill
\end{figure}

\begin{figure}[H]
    \begin{subfigure}{0.5\textwidth}
        \centering \captionsetup{justification=centering}
        $\scalemath{0.75}{
        \displaystyle \begin{pmatrix}
            0 & 1 & 2 & 0 & 0 & 0 & 0 & 0\\ 
            1 & 2 & 0 & 0 & 0 & 0 & 0 & 0\\
            2 & 0 & 0 & 1 & 0 & 0 & 0 & 0\\
            0 & 0 & 1 & 0 & 1 & 0 & 0 & 1\\
            0 & 0 & 0 & 1 & 0 & 1 & 0 & 1\\
            0 & 0 & 0 & 0 & 1 & 0 & 2 & 0\\
            0 & 0 & 0 & 0 & 0 & 2 & 0 & 1\\
            0 & 0 & 0 & 1 & 1 & 0 & 1 & 0
        \end{pmatrix}}$
        $\vcenter{\hbox{\includegraphics[width=0.35\textwidth]{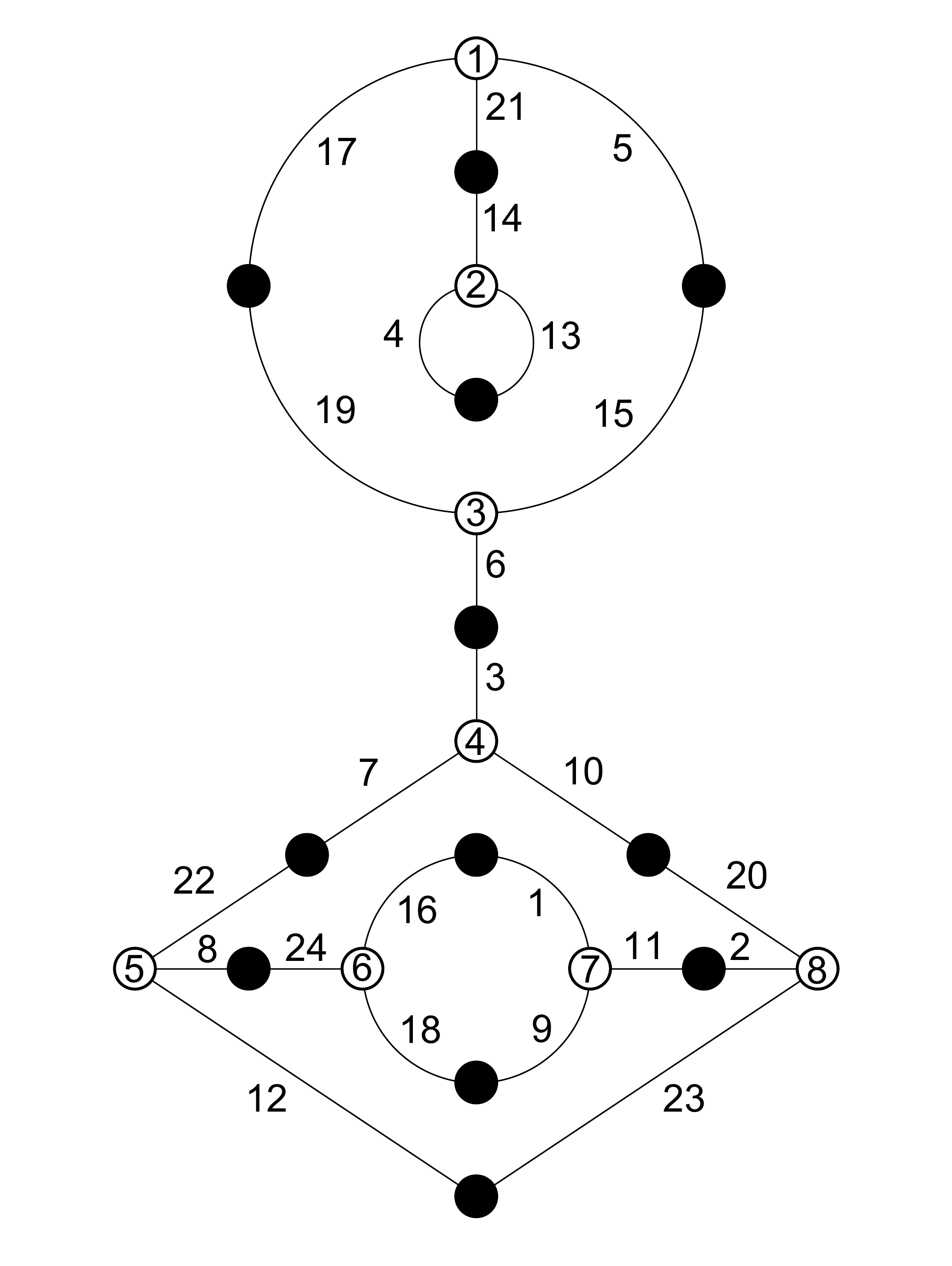}}}$
        \caption{ \{\{\{21,17,5\},\{13,4,14\},
        \{19,15,6\},\{3,10,7\},\{20,23,2\},
        \{11,9,1\},\{18,24,16\},\{22,8,12\}\}, \\ 
        \{\{19,17\},\{21,14\},\{4,13\},
        \{15,5\},\{6,3\},\{22,7\},
        \{10,20\},\{1,16\},\{12,23\},
        \{8,24\},\{2,11\},\{18,9\}\}\}}
        \caption{7-5-5-4-2-1 A $(\sqrt{2})$}
        \label{Dessin}
    \end{subfigure} \hfill
    \begin{subfigure}{0.5\textwidth}
        \centering \captionsetup{justification=centering}
        $\scalemath{0.75}{
        \displaystyle \begin{pmatrix}
            2 & 1 & 0 & 0 & 0 & 0 & 0 & 0\\ 
            1 & 0 & 1 & 0 & 1 & 0 & 0 & 0\\
            0 & 1 & 0 & 1 & 0 & 0 & 0 & 1\\
            0 & 0 & 1 & 0 & 1 & 1 & 0 & 0\\
            0 & 1 & 0 & 1 & 0 & 0 & 0 & 1\\
            0 & 0 & 0 & 1 & 0 & 0 & 2 & 0\\
            0 & 0 & 0 & 0 & 0 & 2 & 0 & 1\\
            0 & 0 & 1 & 0 & 1 & 0 & 1 & 0
        \end{pmatrix}}$
        $\vcenter{\hbox{\includegraphics[width=0.35\textwidth]{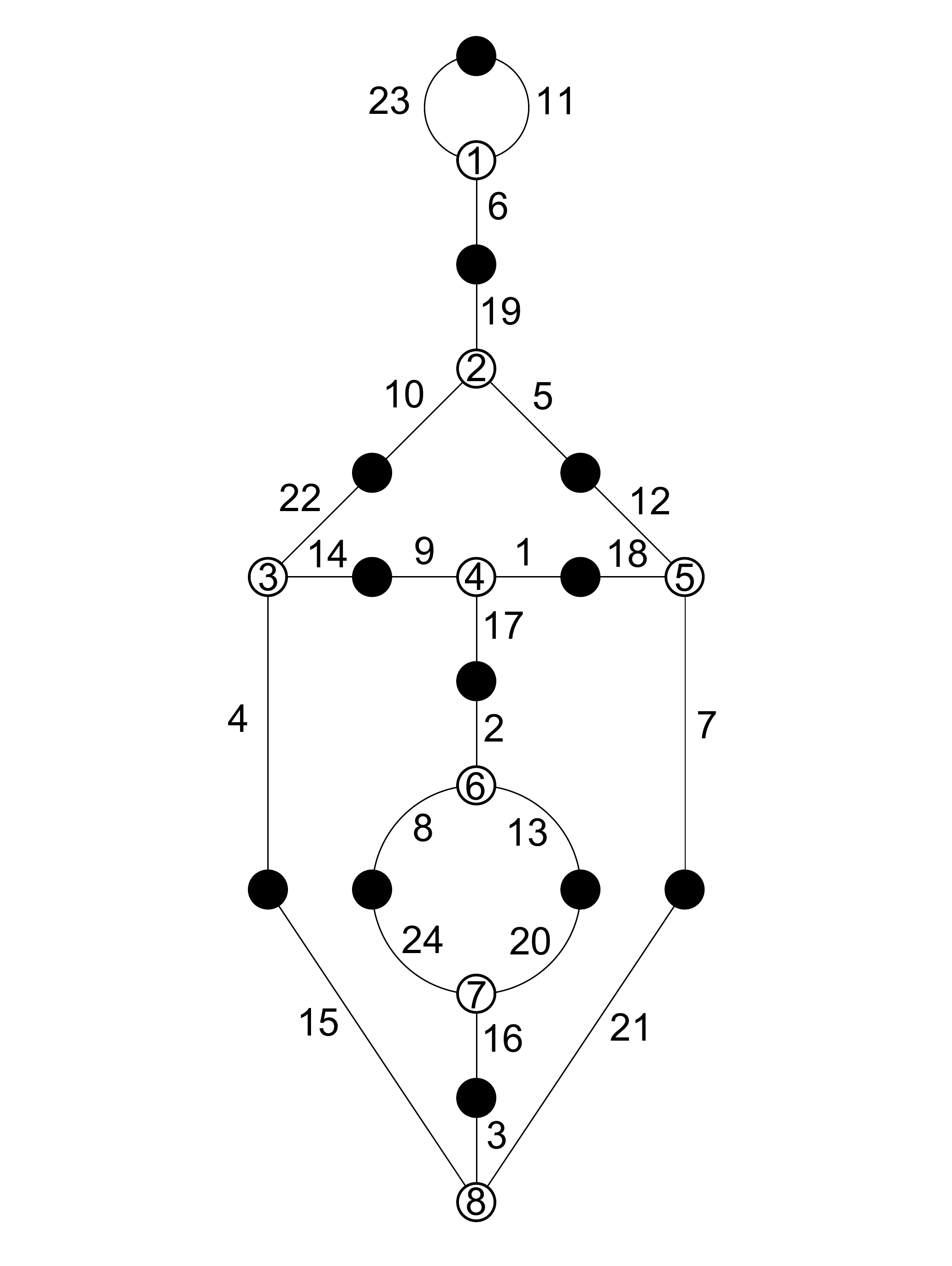}}}$
        \caption{ \{\{\{23,11,6\},\{19,5,10\},
        \{22,14,4\},\{9,1,17\},\{18,12,7\},
        \{2,13,8\},\{24,20,16\},\{3,21,15\}\}, \\ 
        \{\{15,4\},\{3,16\},\{21,7\},
        \{8,24\},\{20,13\},\{2,17\},
        \{1,18\},\{9,14\},\{10,22\},
        \{12,5\},\{19,6\},\{23,11\}\}\}}
        \caption{7-5-5-4-2-1 B $(\sqrt{2})$}
        \label{Dessin}
    \end{subfigure}\hfill
\end{figure}

\begin{figure}[H]
    \begin{subfigure}{0.5\textwidth}
        \centering \captionsetup{justification=centering}
        $\scalemath{0.75}{
        \displaystyle \begin{pmatrix}
            0 & 1 & 1 & 1 & 0 & 0 & 0 & 0\\ 
            1 & 0 & 0 & 1 & 1 & 0 & 0 & 0\\
            1 & 0 & 2 & 0 & 0 & 0 & 0 & 0\\
            1 & 1 & 0 & 0 & 0 & 1 & 0 & 0\\
            0 & 1 & 0 & 0 & 0 & 1 & 1 & 0\\
            0 & 0 & 0 & 1 & 1 & 0 & 0 & 1\\
            0 & 0 & 0 & 0 & 1 & 0 & 0 & 2\\
            0 & 0 & 0 & 0 & 0 & 1 & 2 & 0
        \end{pmatrix}}$
        $\vcenter{\hbox{\includegraphics[width=0.35\textwidth]{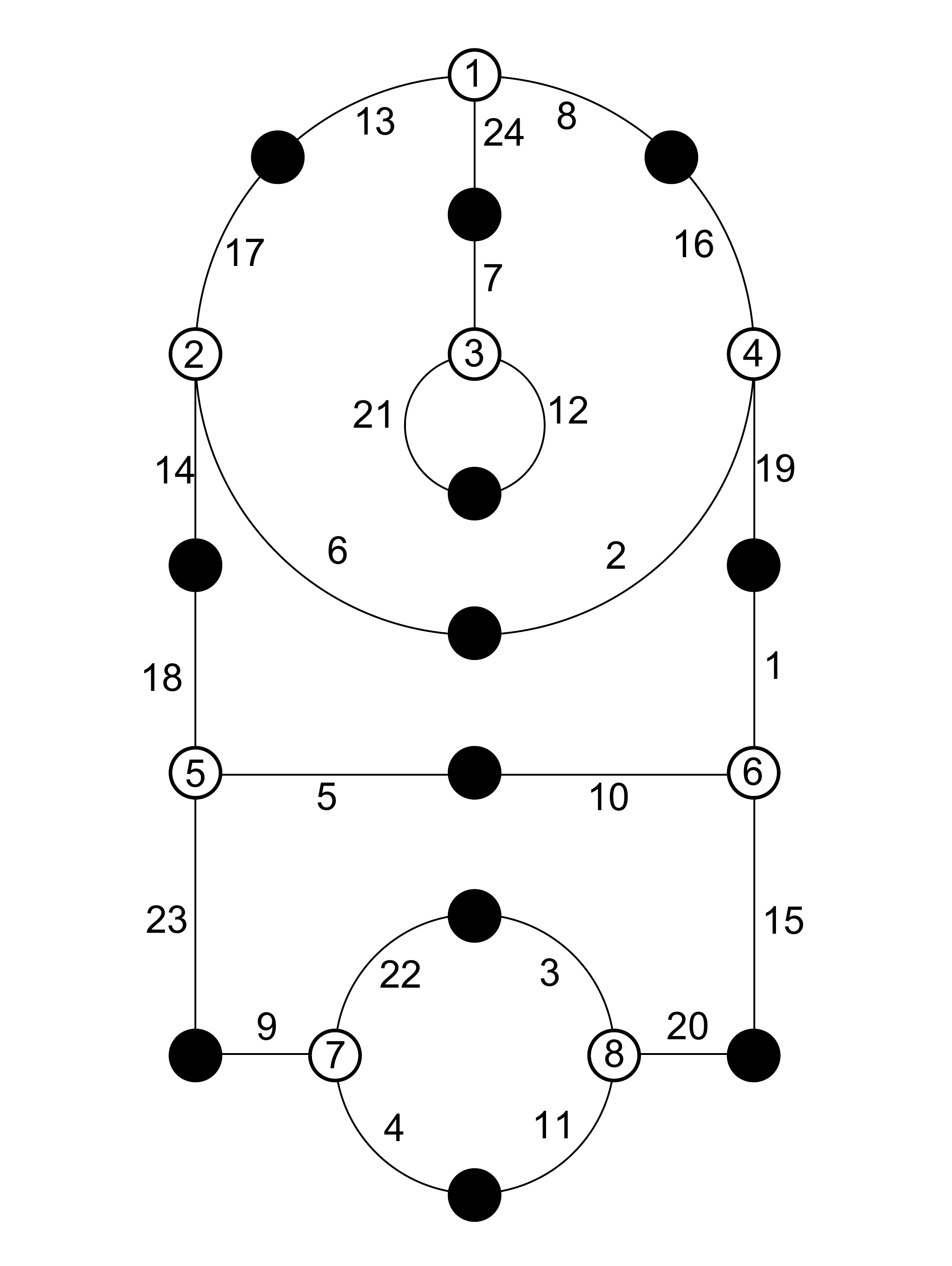}}}$
        \caption{ \{\{\{8,24,13\},\{16,19,2\},
        \{7,12,21\},\{17,6,14\},\{18,5,23\},
        \{10,1,15\},\{9,22,4\},\{3,20,11\}\}, \\ 
        \{\{4,11\},\{3,22\},\{15,20\},
        \{9,23\},\{10,5\},\{14,18\},
        \{1,19\},\{2,6\},\{21,12\},
        \{7,24\},\{17,13\},\{8,16\}\}\}}
        \caption{7-6-4-4-2-1 $(\mathbb{Q})$}
        \label{Dessin}
    \end{subfigure} \hfill
    \begin{subfigure}{0.5\textwidth}
        \centering \captionsetup{justification=centering}
        $\scalemath{0.75}{
        \displaystyle \begin{pmatrix}
            0 & 2 & 1 & 0 & 0 & 0 & 0 & 0\\ 
            2 & 0 & 1 & 0 & 0 & 0 & 0 & 0\\
            1 & 1 & 0 & 1 & 0 & 0 & 0 & 0\\
            0 & 0 & 1 & 0 & 0 & 0 & 0 & 2\\
            0 & 0 & 0 & 0 & 0 & 1 & 2 & 0\\
            0 & 0 & 0 & 0 & 1 & 2 & 0 & 0\\
            0 & 0 & 0 & 0 & 2 & 0 & 0 & 1\\
            0 & 0 & 0 & 2 & 0 & 0 & 1 & 0
        \end{pmatrix}}$
        $\vcenter{\hbox{\includegraphics[width=0.35\textwidth]{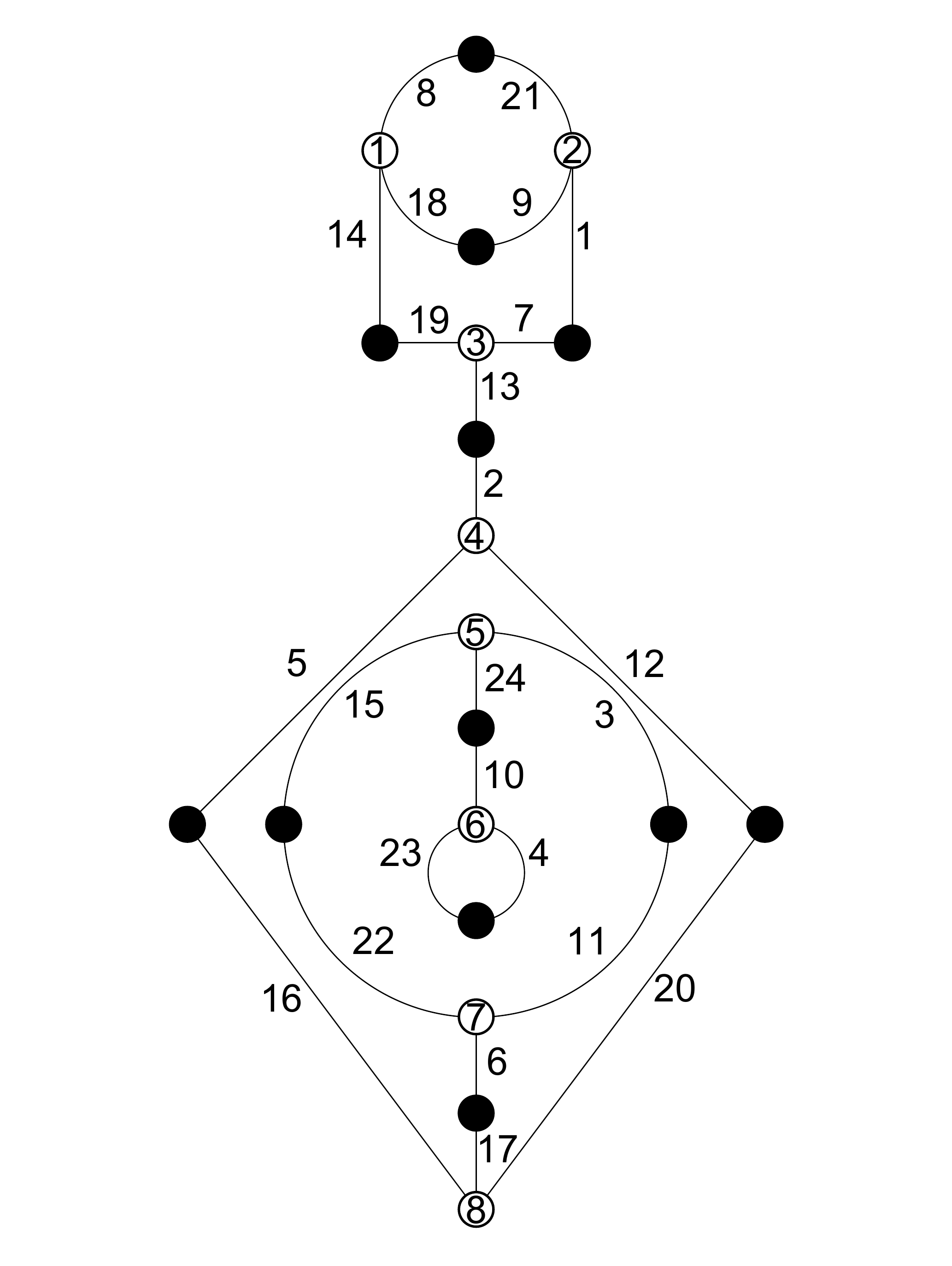}}}$
        \caption{ \{\{\{8,18,14\},\{21,1,9\},
        \{19,7,13\},\{2,12,5\},\{24,15,3\},
        \{10,4,23\},\{22,11,6\},\{17,20,16\}\}, \\ 
        \{\{5,16\},\{22,15\},\{10,24\},
        \{12,20\},\{23,4\},\{11,3\},
        \{2,13\},\{1,7\},\{14,19\},
        \{8,21\},\{9,18\},\{6,17\}\}\}}
        \caption{7-6-5-3-2-1 A (cubic)}
        \label{Dessin}
    \end{subfigure}\hfill
\end{figure}

\begin{figure}[H]
    \begin{subfigure}{0.6\textwidth}
        \centering \captionsetup{justification=centering}
        $\scalemath{0.75}{
        \displaystyle \begin{pmatrix}
            0 & 1 & 1 & 0 & 0 & 1 & 0 & 0\\ 
            1 & 0 & 1 & 0 & 0 & 0 & 1 & 0\\
            1 & 1 & 0 & 1 & 0 & 0 & 0 & 0\\
            0 & 0 & 1 & 0 & 2 & 0 & 0 & 0\\
            0 & 0 & 0 & 2 & 0 & 1 & 0 & 0\\
            1 & 0 & 0 & 0 & 1 & 0 & 1 & 0\\
            0 & 1 & 0 & 0 & 0 & 1 & 0 & 1\\
            0 & 0 & 0 & 0 & 0 & 0 & 1 & 2
        \end{pmatrix}}$
        $\vcenter{\hbox{\includegraphics[width=0.25\textwidth]{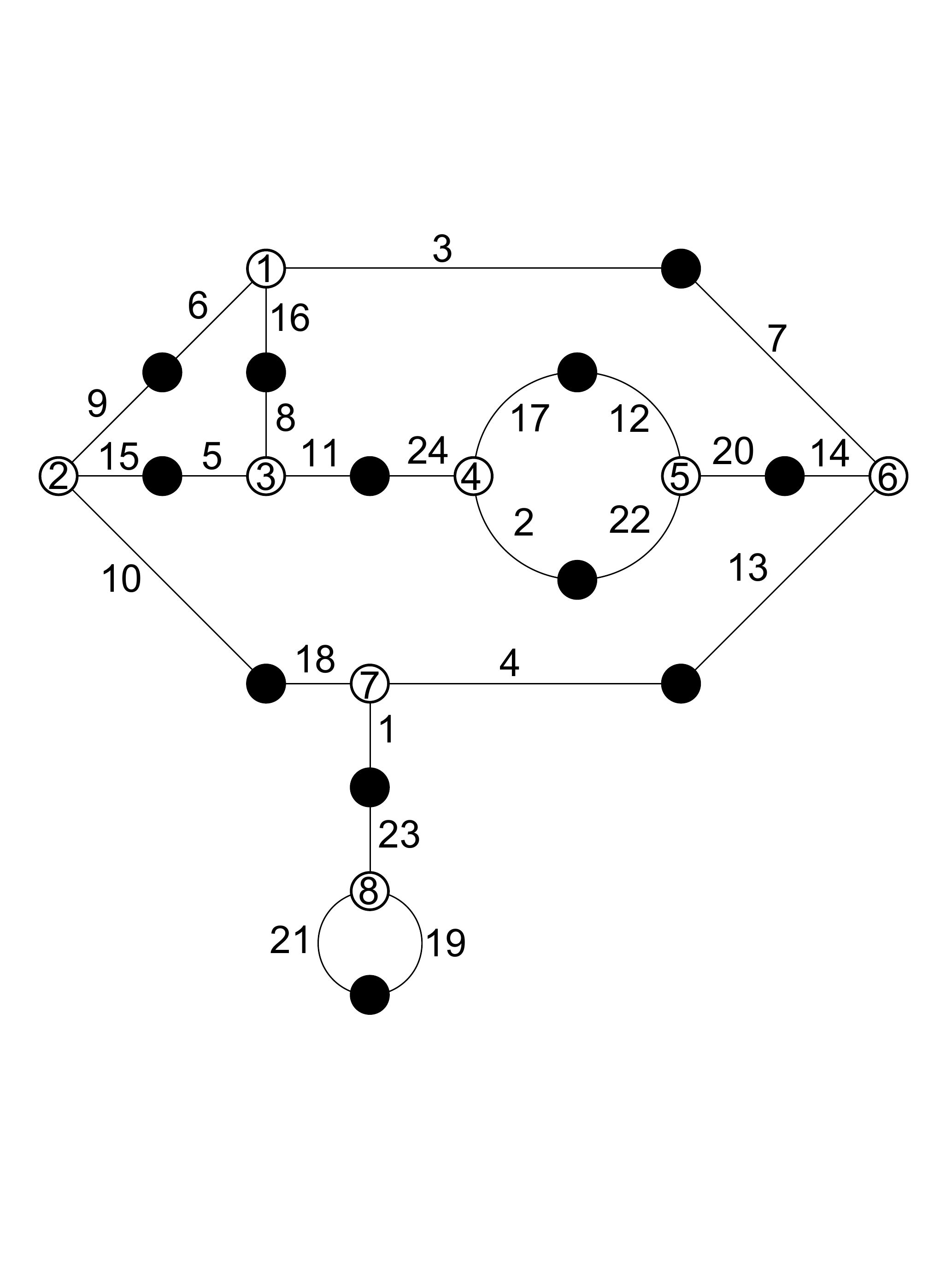}}}$
        $\vcenter{\hbox{\includegraphics[width=0.25\textwidth]{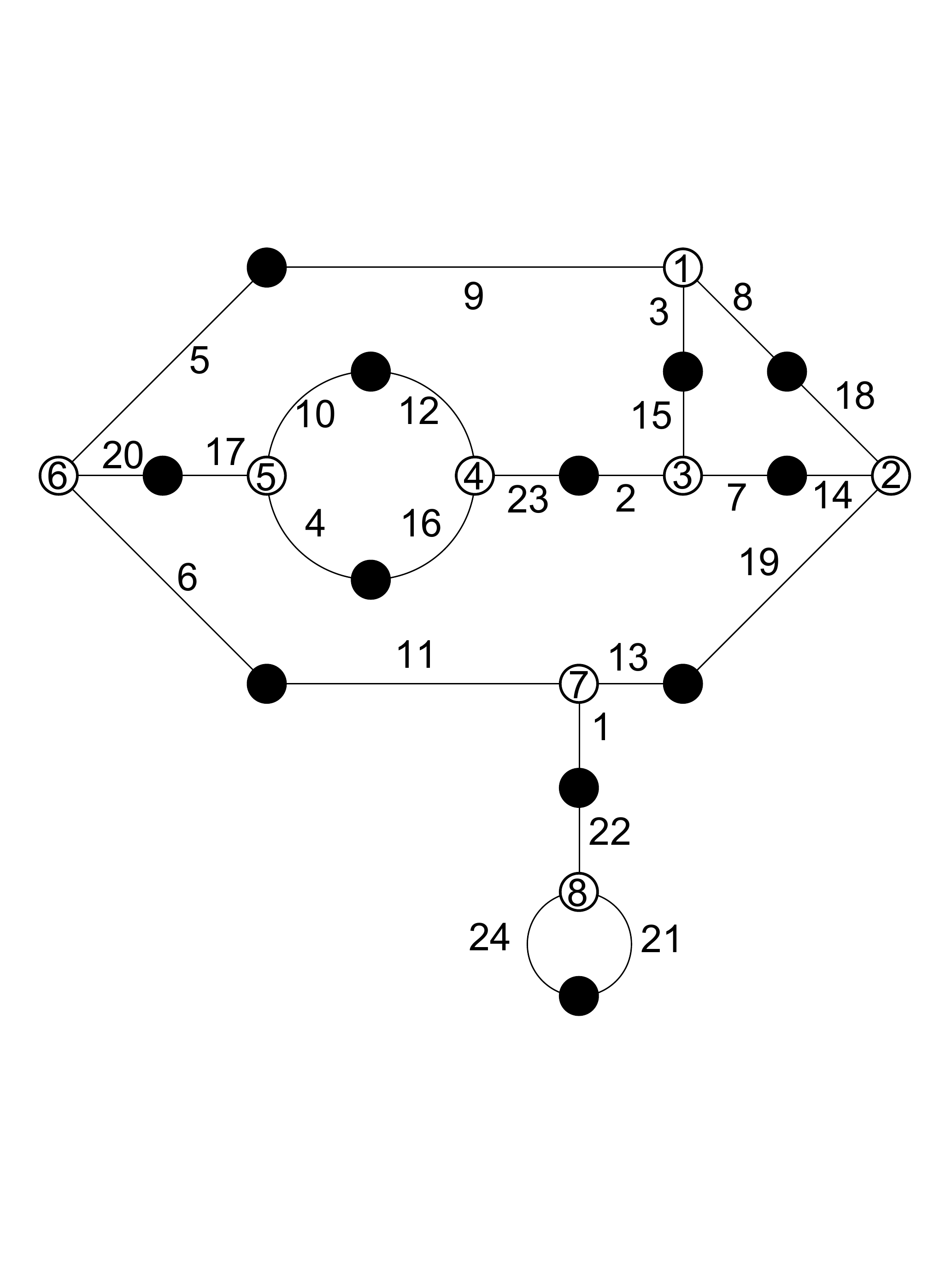}}}$
        \caption{ B: \{\{\{15,10,9\},\{5,8,11\},
        \{6,3,16\},\{7,13,14\},\{20,22,12\},
        \{2,24,17\},\{18,4,1\},\{23,19,21\}\}, \\ 
        \{\{19,21\},\{23,1\},\{10,18\},
        \{4,13\},\{14,20\},\{3,7\},
        \{8,16\},\{6,9\},\{5,15\},
        \{12,17\},\{11,24\},\{2,22\}\}\}\\
        C: \{\{\{5,20,6\},\{17,10,4\},
        \{12,23,16\},\{3,9,8\},\{7,2,15\},
        \{18,19,14\},\{13,1,11\},\{21,224,22\}\}, \\ 
        \{\{8,18\},\{15,3\},\{5,9\},
        \{20,17\},\{6,11\},\{13,19\},
        \{1,22\},\{24,21\},\{12,10\},
        \{4,16\},\{2,23\},\{7,14\}\}\}}
        \caption{7-6-5-3-2-1 B \& C (cubic)}
        \label{Dessin}
    \end{subfigure} \hfill
    \begin{subfigure}{0.4\textwidth}
        \centering \captionsetup{justification=centering}
        $\scalemath{0.75}{
        \displaystyle \begin{pmatrix}
            0 & 1 & 2 & 0 & 0 & 0 & 0 & 0\\ 
            1 & 2 & 0 & 0 & 0 & 0 & 0 & 0\\
            2 & 0 & 0 & 1 & 0 & 0 & 0 & 0\\
            0 & 0 & 1 & 0 & 0 & 1 & 0 & 1\\
            0 & 0 & 0 & 0 & 0 & 1 & 1 & 1\\
            0 & 0 & 0 & 1 & 1 & 0 & 0 & 1\\
            0 & 0 & 0 & 0 & 1 & 0 & 2 & 0\\
            0 & 0 & 0 & 1 & 1 & 1 & 0 & 0
        \end{pmatrix}}$
        $\vcenter{\hbox{\includegraphics[width=0.35\textwidth]{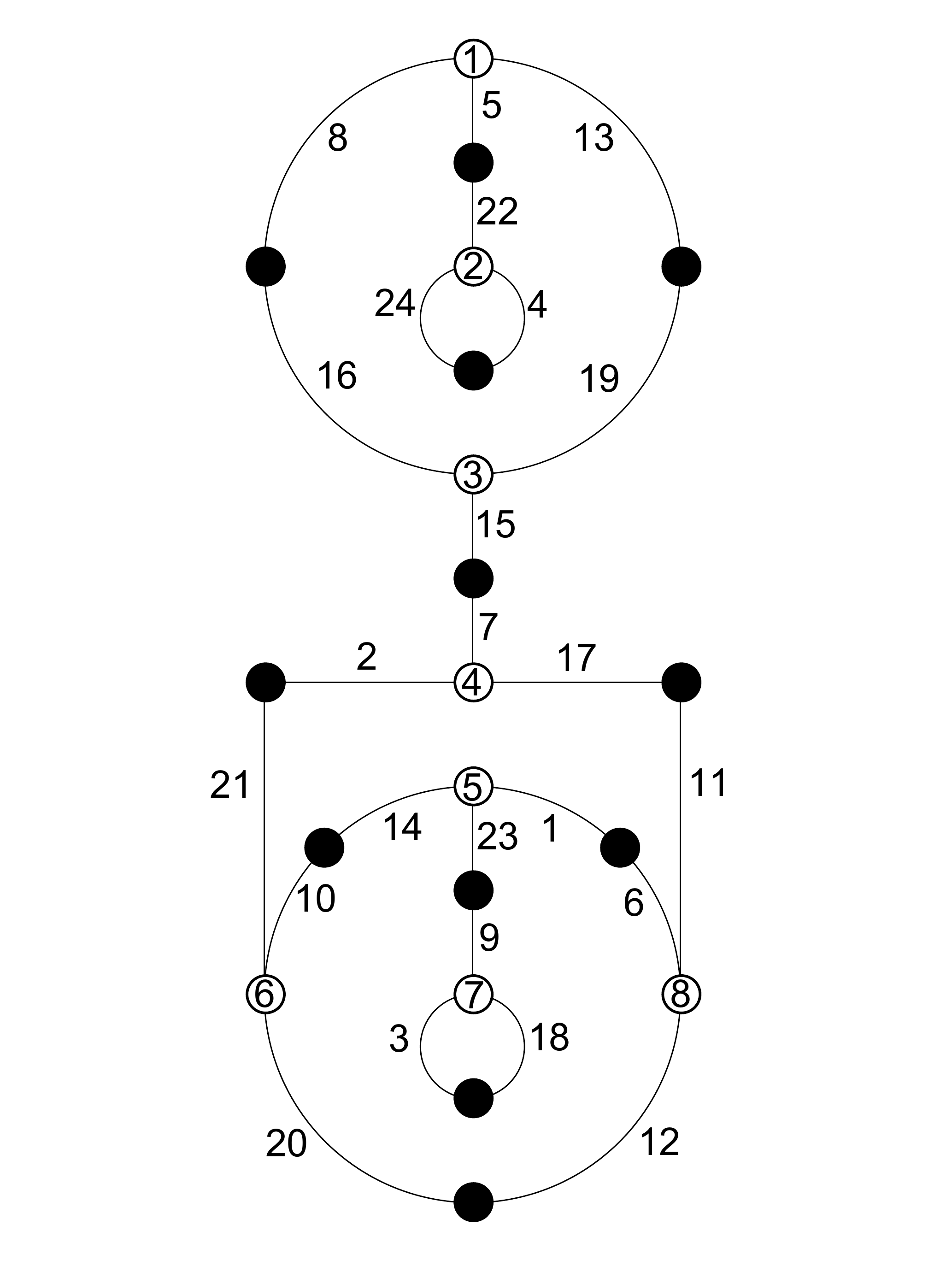}}}$
        \caption{ \{\{\{5,8,13\},\{22,4,24\},
        \{16,19,15\},\{7,17,2\},\{21,10,20\},
        \{3,9,18\},\{6,11,12\},\{14,1,23\}\}, \\ 
        \{\{21,2\},\{17,11\},\{10,14\},
        \{1,6\},\{9,23\},\{3,18\},
        \{20,12\},\{7,15\},\{8,16\},
        \{5,22\},\{4,24\},\{19,13\}\}\}}
        \caption{7-6-5-4-1-1 $(\mathbb{Q})$}
        \label{Dessin}
    \end{subfigure}\hfill
\end{figure}

\begin{figure}[H]
    \begin{subfigure}{0.5\textwidth}
        \centering \captionsetup{justification=centering}
        $\scalemath{0.75}{
        \displaystyle \begin{pmatrix}
            0 & 2 & 0 & 0 & 1 & 0 & 0 & 0\\ 
            2 & 0 & 1 & 0 & 0 & 0 & 0 & 0\\
            0 & 1 & 0 & 2 & 0 & 0 & 0 & 0\\
            0 & 0 & 2 & 0 & 0 & 0 & 1 & 0\\
            1 & 0 & 0 & 0 & 0 & 0 & 1 & 1\\
            0 & 0 & 0 & 0 & 0 & 2 & 0 & 1\\
            0 & 0 & 0 & 1 & 1 & 0 & 0 & 1\\
            0 & 0 & 0 & 0 & 1 & 1 & 1 & 0
        \end{pmatrix}}$
        $\vcenter{\hbox{\includegraphics[width=0.35\textwidth]{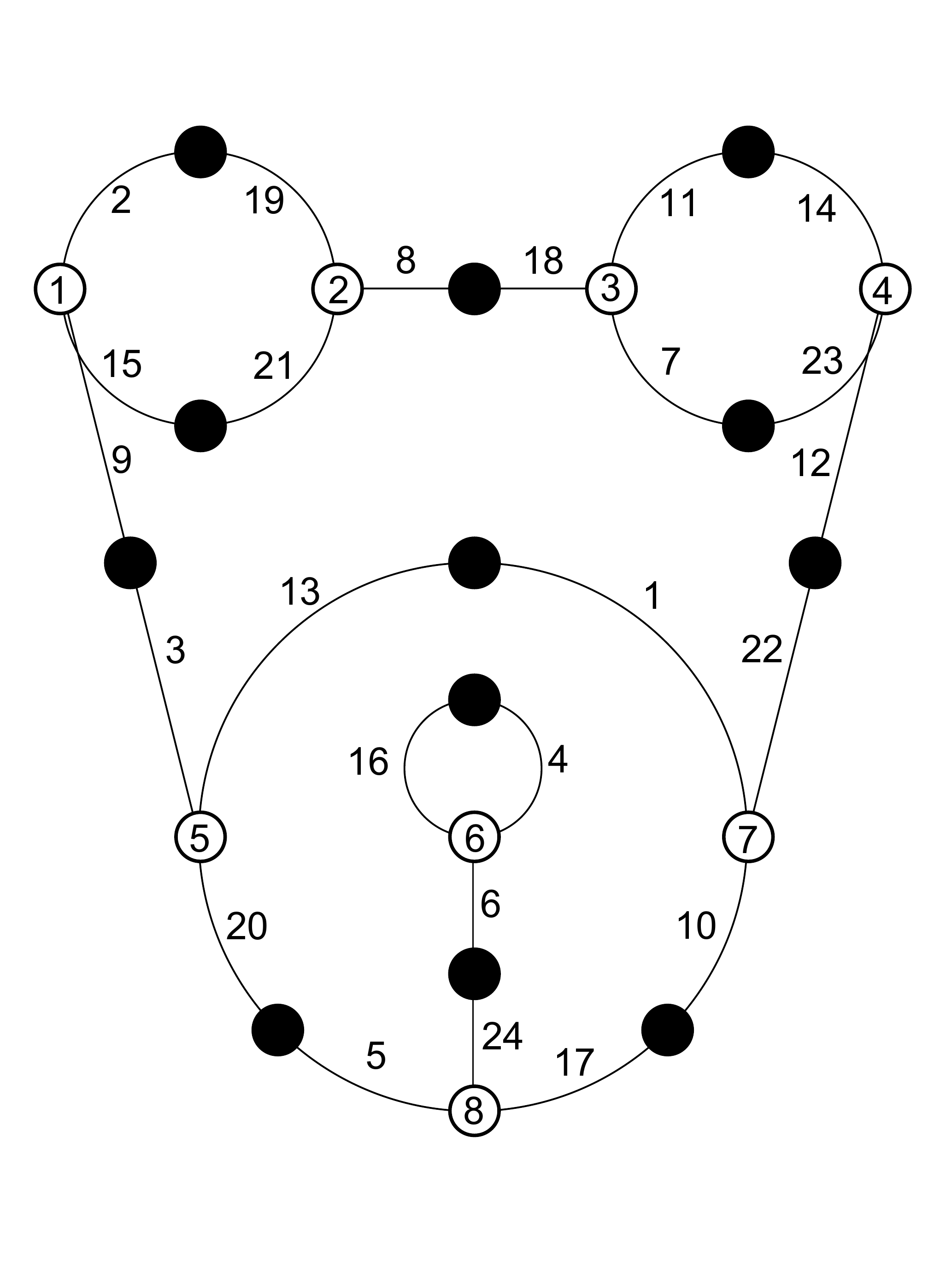}}}$
        \caption{ \{\{\{2,15,9\},\{19,8,21\},
        \{18,11,7\},\{14,12,23\},\{4,6,16\},
        \{20,3,13\},\{22,10,1\},\{24,17,5\}\}, \\ 
        \{\{2,19\},\{8,18\},\{23,7\},
        \{21,15\},\{11,14\},\{22,12\},
        \{13,1\},\{6,24\},\{20,5\},
        \{17,10\},\{4,16\},\{9,3\}\}\}}
        \caption{7-6-6-2-2-1 $(\mathbb{Q})$}
        \label{Dessin}
    \end{subfigure} \hfill
    \begin{subfigure}{0.5\textwidth}
        \centering \captionsetup{justification=centering}
        $\scalemath{0.75}{
        \displaystyle \begin{pmatrix}
            0 & 2 & 0 & 0 & 0 & 1 & 0 & 0\\ 
            2 & 0 & 1 & 0 & 0 & 0 & 0 & 0\\
            0 & 1 & 0 & 2 & 0 & 0 & 0 & 0\\
            0 & 0 & 2 & 0 & 0 & 0 & 1 & 0\\
            0 & 0 & 0 & 0 & 0 & 1 & 1 & 1\\
            1 & 0 & 0 & 0 & 1 & 0 & 0 & 1\\
            0 & 0 & 0 & 1 & 1 & 0 & 0 & 1\\
            0 & 0 & 0 & 0 & 1 & 1 & 1 & 0
        \end{pmatrix}}$
        $\vcenter{\hbox{\includegraphics[width=0.35\textwidth]{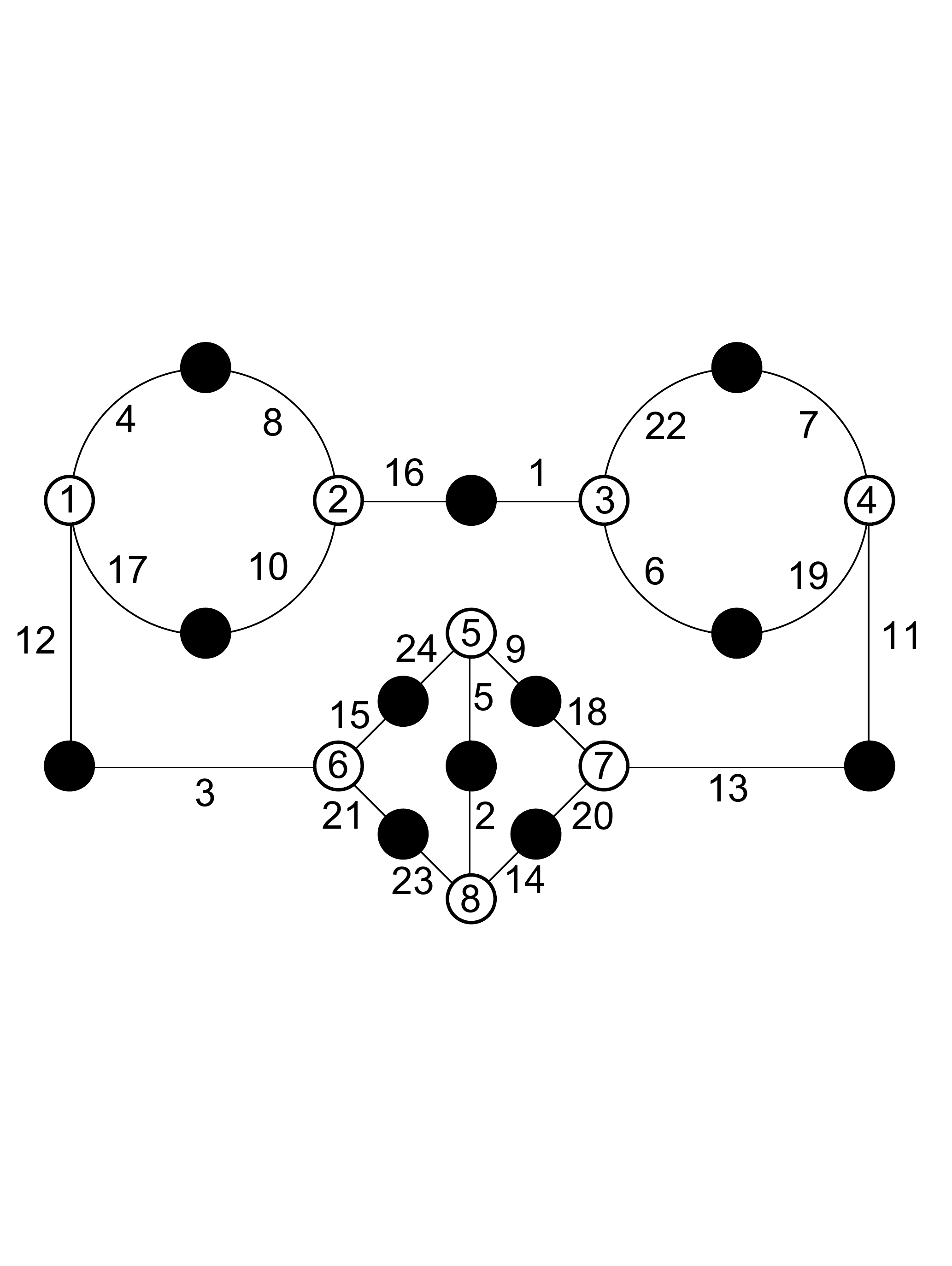}}}$
        \caption{ \{\{\{4,17,12\},\{8,16,10\},
        \{1,22,6\},\{7,11,19\},\{13,20,18\},
        \{2,14,23\},\{9,5,24\},\{15,21,3\}\}, \\ 
        \{\{15,24\},\{20,14\},\{16,1\},
        \{2,5\},\{23,21\},\{18,9\},
        \{12,3\},\{13,11\},\{6,19\},
        \{4,8\},\{17,10\},\{22,7\}\}\}}
        \caption{7-7-3-3-2-2 A $(\sqrt{7})$}
        \label{Dessin}
    \end{subfigure}\hfill
\end{figure}

\begin{figure}[H]
    \begin{subfigure}{0.4\textwidth}
        \centering \captionsetup{justification=centering}
        $\scalemath{0.75}{
        \displaystyle \begin{pmatrix}
            0 & 2 & 1 & 0 & 0 & 0 & 0 & 0\\ 
            2 & 0 & 1 & 0 & 0 & 0 & 0 & 0\\
            1 & 1 & 0 & 1 & 0 & 0 & 0 & 0\\
            0 & 0 & 1 & 0 & 0 & 0 & 0 & 2\\
            0 & 0 & 0 & 0 & 0 & 2 & 1 & 0\\
            0 & 0 & 0 & 0 & 2 & 0 & 1 & 0\\
            0 & 0 & 0 & 0 & 1 & 1 & 0 & 1\\
            0 & 0 & 0 & 2 & 0 & 0 & 1 & 0
        \end{pmatrix}}$
        $\vcenter{\hbox{\includegraphics[width=0.35\textwidth]{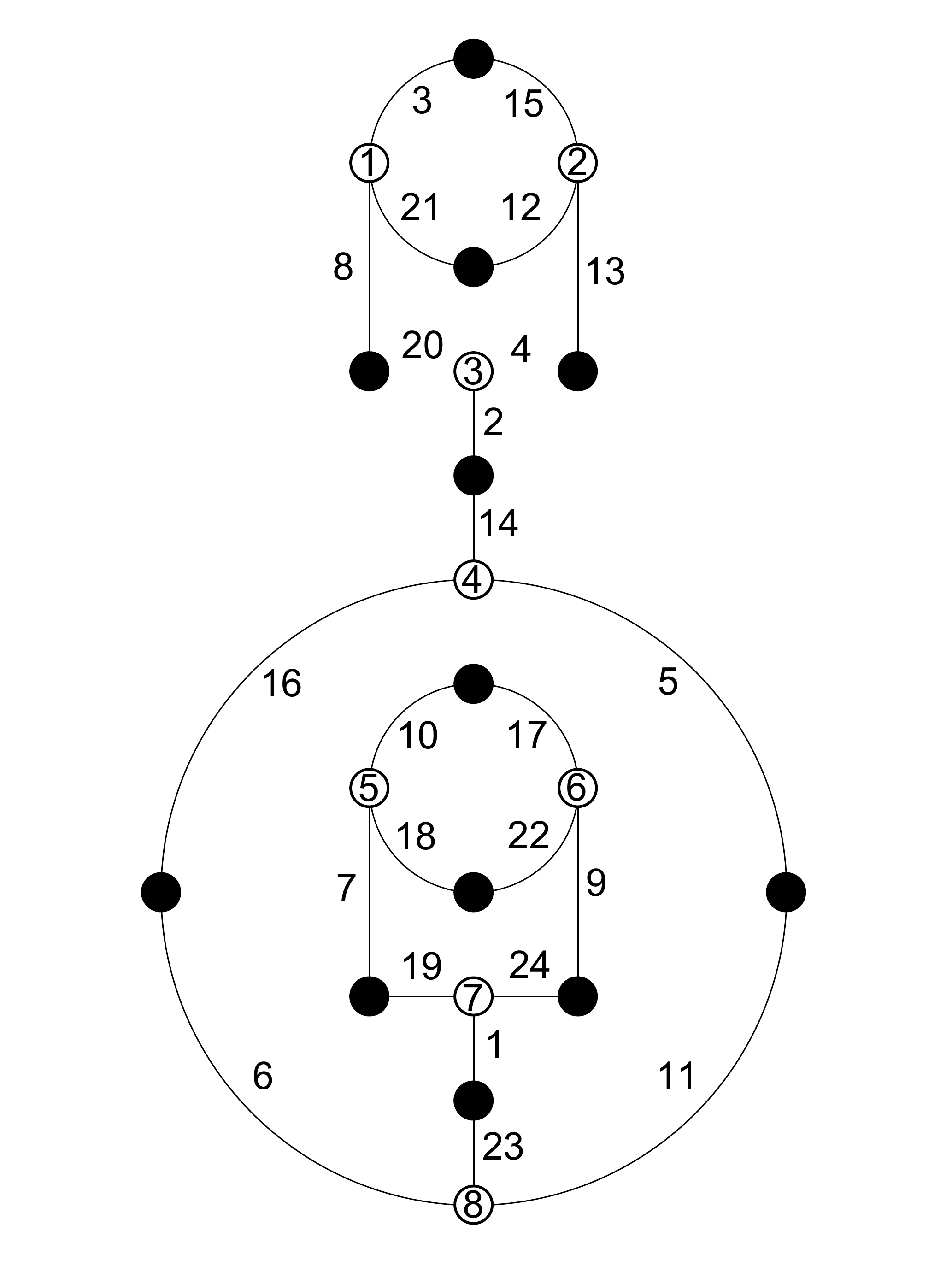}}}$
        \caption{ \{\{\{3,21,8\},\{15,13,12\},
        \{4,2,20\},\{14,5,16\},\{23,11,6\},
        \{19,24,1\},\{17,9,22\},\{10,18,7\}\}, \\ 
        \{\{3,15\},\{20,8\},\{21,12\},
        \{10,17\},\{2,14\},\{4,13\},
        \{16,6\},\{11,5\},\{1,23\},
        \{18,22\},\{7,19\},\{9,24\}\}\}}
        \caption{7-7-3-3-2-2 B $(\sqrt{7})$}
        \label{Dessin}
    \end{subfigure} \hfill
    \begin{subfigure}{0.6\textwidth}
        \centering \captionsetup{justification=centering}
        $\scalemath{0.75}{
        \displaystyle \begin{pmatrix}
            0 & 0 & 1 & 0 & 1 & 0 & 1 & 0\\ 
            0 & 2 & 0 & 1 & 0 & 0 & 0 & 0\\
            1 & 0 & 0 & 1 & 0 & 0 & 0 & 1\\
            0 & 1 & 1 & 0 & 1 & 0 & 0 & 0\\
            1 & 0 & 0 & 1 & 0 & 0 & 0 & 1\\
            0 & 0 & 0 & 0 & 0 & 2 & 1 & 0\\
            1 & 0 & 0 & 0 & 0 & 1 & 0 & 1\\
            0 & 0 & 1 & 0 & 1 & 0 & 1 & 0
        \end{pmatrix}}$
        $\vcenter{\hbox{\includegraphics[width=0.25\textwidth]{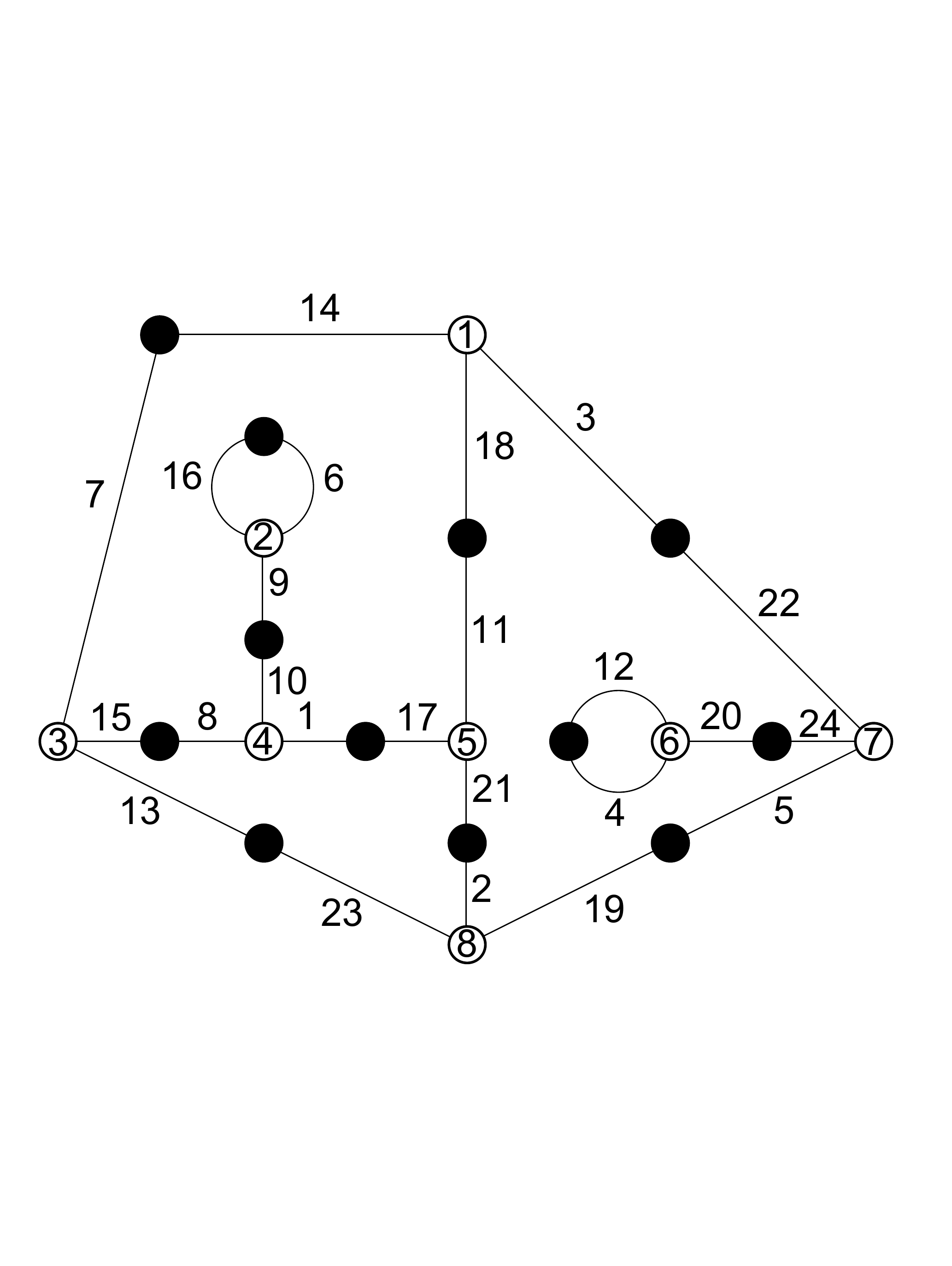}}}$
        $\vcenter{\hbox{\includegraphics[width=0.25\textwidth]{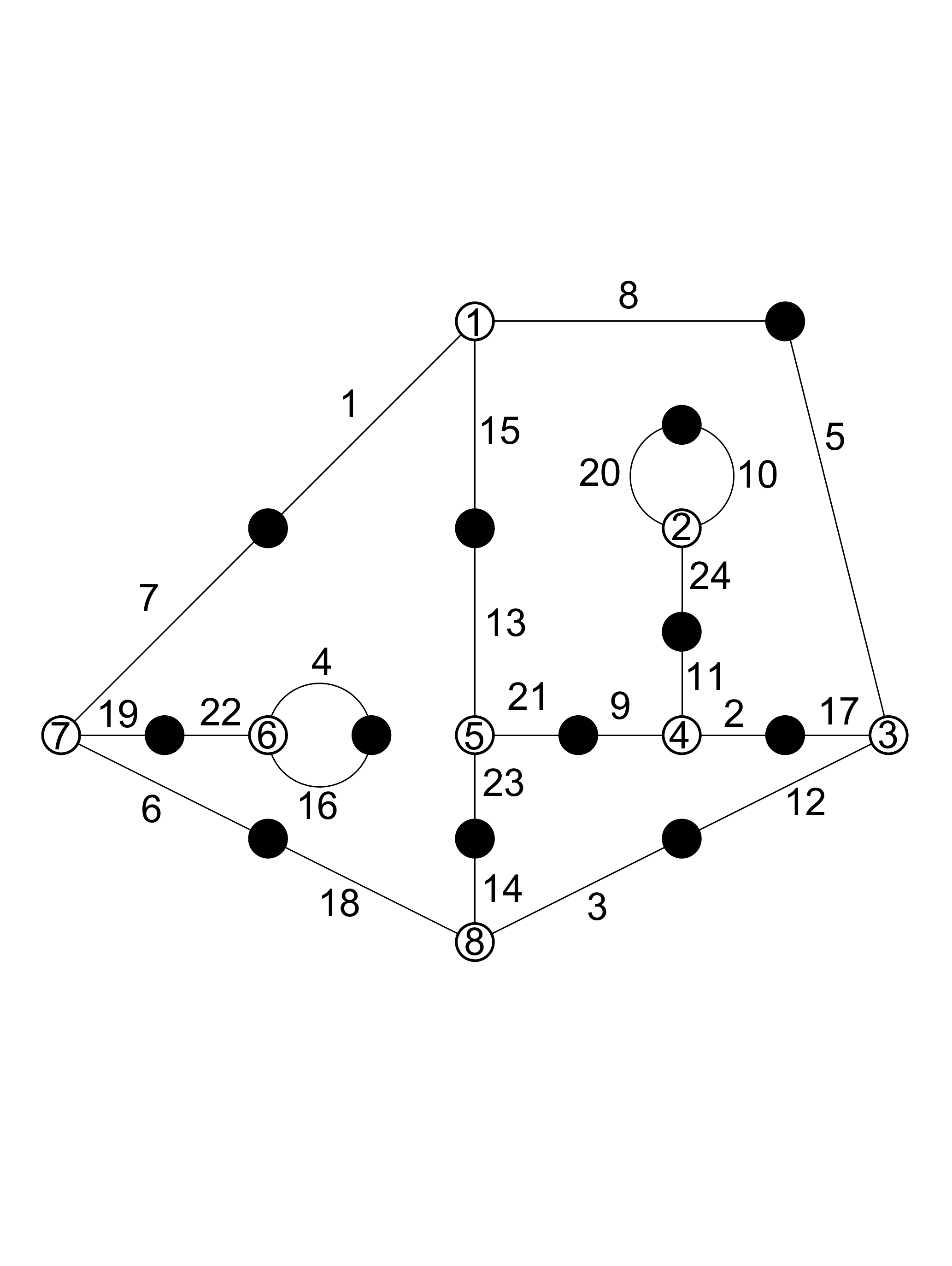}}}$
        \caption{ A: \{\{\{7,15,13\},\{8,10,1\},
        \{16,6,9\},\{17,11,21\},\{2,19,23\},
        \{3,18,14\},\{12,20,4\},\{22,5,24\}\}, \\ 
        \{\{7,14\},\{3,22\},\{18,11\},
        \{6,16\},\{9,10\},\{15,8\},
        \{1,17\},\{23,13\},\{2,21\},
        \{5,19\},\{24,20\},\{4,12\}\}\} \\
        B: \{\{\{8,15,1\},\{10,24,20\},
        \{2,9,11\},\{5,12,17\},\{3,18,14\},
        \{7,19,6\},\{16,22,4\},\{21,23,13\}\}, \\ 
        \{\{1,7\},\{5,8\},\{15,13\},
        \{20,10\},\{11,24\},\{9,21\},
        \{17,2\},\{3,12\},\{14,23\},
        \{6,18\},\{4,16\},\{22,19\}\}\}}
        \caption{7-7-4-4-1-1 A \& B $(\sqrt{-7})$}
        \label{Dessin}
    \end{subfigure}\hfill
\end{figure}

\begin{figure}[H]
    \begin{subfigure}{0.5\textwidth}
        \centering \captionsetup{justification=centering}
        $\scalemath{0.75}{
        \displaystyle \begin{pmatrix}
            0 & 1 & 1 & 1 & 0 & 0 & 0 & 0\\ 
            1 & 0 & 1 & 0 & 0 & 0 & 1 & 0\\
            1 & 1 & 0 & 0 & 0 & 0 & 0 & 1\\
            1 & 0 & 0 & 0 & 1 & 0 & 0 & 1\\
            0 & 0 & 0 & 1 & 2 & 0 & 0 & 0\\
            0 & 0 & 0 & 0 & 0 & 2 & 1 & 0\\
            0 & 1 & 0 & 0 & 0 & 1 & 0 & 1\\
            0 & 0 & 1 & 1 & 0 & 0 & 1 & 0
        \end{pmatrix}}$
        $\vcenter{\hbox{\includegraphics[width=0.35\textwidth]{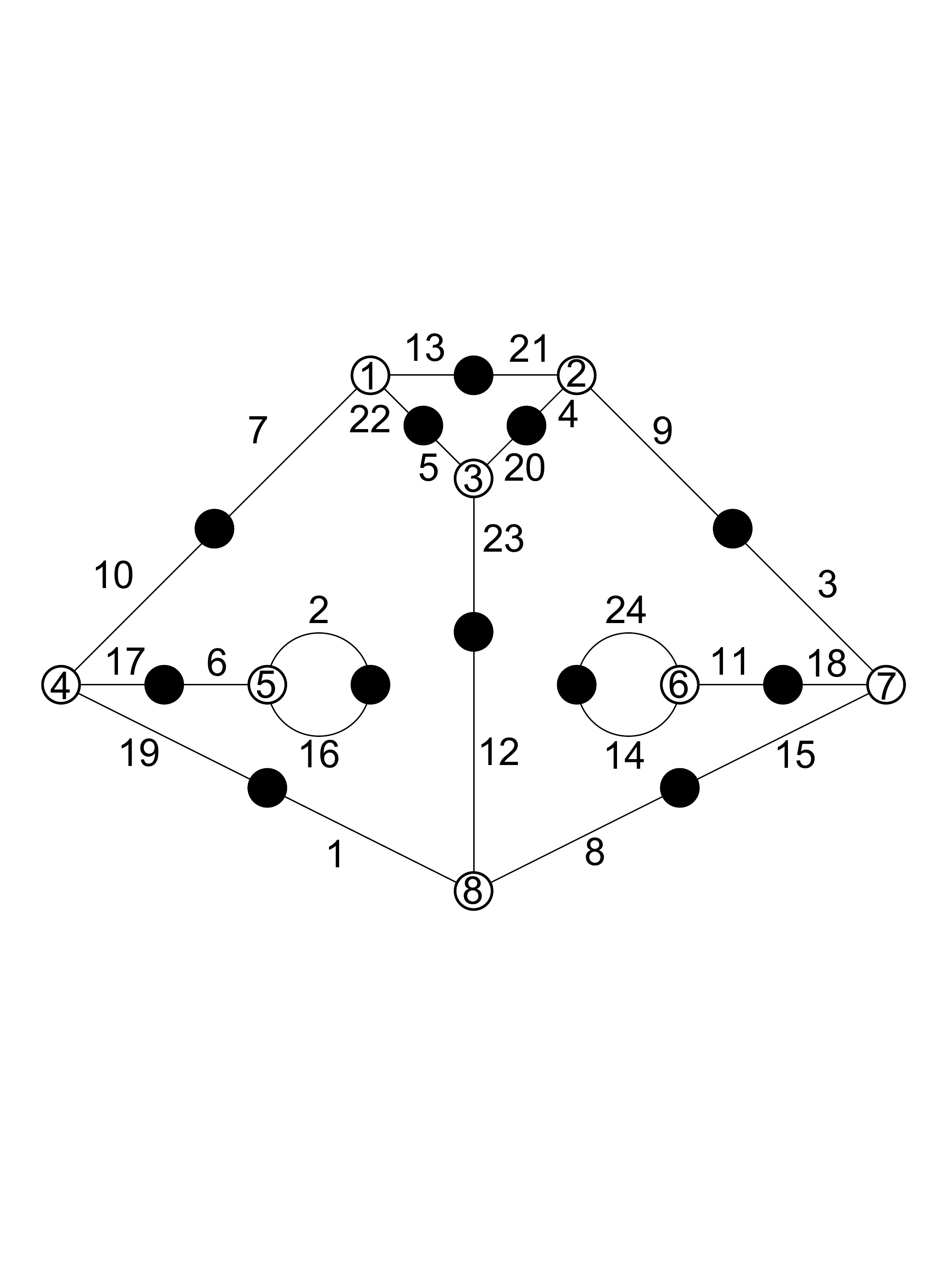}}}$
        \caption{ \{\{\{13,22,7\},\{20,23,5\},
        \{24,11,14\},\{8,1,12\},\{2,16,6\},
        \{10,17,19\},\{18,3,15\},\{21,9,4\}\}, \\ 
        \{\{21,13\},\{5,22\},\{10,7\},
        \{6,17\},\{2,16\},\{1,19\},
        \{15,8\},\{24,14\},\{18,11\},
        \{12,23\},\{4,20\},\{3,9\}\}\}}
        \caption{7-7-5-3-1-1 A  $(\sqrt{21})$}
        \label{Dessin}
    \end{subfigure} \hfill
    \begin{subfigure}{0.5\textwidth}
        \centering \captionsetup{justification=centering}
        $\scalemath{0.75}{
        \displaystyle \begin{pmatrix}
            2 & 1 & 0 & 0 & 0 & 0 & 0 & 0\\ 
            1 & 0 & 1 & 1 & 0 & 0 & 0 & 0\\
            0 & 1 & 0 & 1 & 0 & 0 & 0 & 1\\
            0 & 1 & 1 & 0 & 0 & 0 & 0 & 1\\
            0 & 0 & 0 & 0 & 0 & 1 & 2 & 0\\
            0 & 0 & 0 & 0 & 1 & 2 & 0 & 0\\
            0 & 0 & 0 & 0 & 2 & 0 & 0 & 1\\
            0 & 0 & 1 & 1 & 0 & 0 & 1 & 0
        \end{pmatrix}}$
        $\vcenter{\hbox{\includegraphics[width=0.35\textwidth]{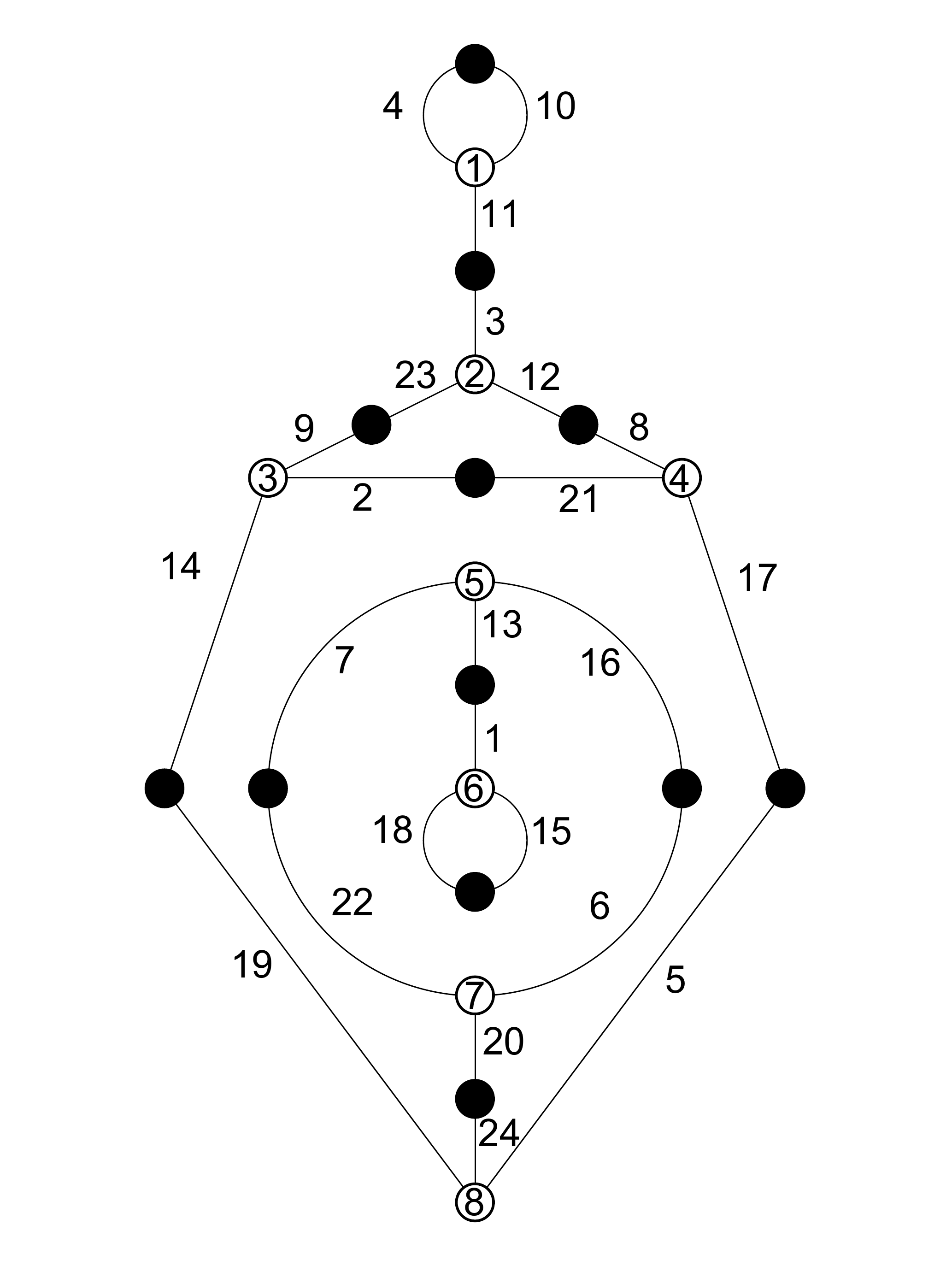}}}$
        \caption{ \{\{\{4,10,11\},\{3,12,23\},
        \{8,17,21\},\{2,14,9\},\{24,5,19\},
        \{15,18,1\},\{13,7,16\},\{8,17,21\}\}, \\ 
        \{\{4,10\},\{11,3\},\{9,23\},
        \{8,12\},\{2,21\},\{5,17\},
        \{19,14\},\{24,20\},\{18,15\},
        \{1,13\},\{7,22\},\{16,6\}\}\}}
        \caption{7-7-5-3-1-1 B  $(\sqrt{21})$}
        \label{Dessin}
    \end{subfigure}\hfill
\end{figure}

\begin{figure}[H]
    \begin{subfigure}{0.5\textwidth}
        \centering \captionsetup{justification=centering}
        $\scalemath{0.75}{
        \displaystyle \begin{pmatrix}
            2 & 1 & 0 & 0 & 0 & 0 & 0 & 0\\ 
            1 & 0 & 0 & 0 & 1 & 0 & 1 & 0\\
            0 & 0 & 2 & 1 & 0 & 0 & 0 & 0\\
            0 & 0 & 1 & 0 & 1 & 0 & 1 & 0\\
            0 & 1 & 0 & 1 & 0 & 0 & 0 & 1\\
            0 & 0 & 0 & 0 & 0 & 2 & 0 & 1\\
            0 & 1 & 0 & 1 & 0 & 0 & 0 & 1\\
            0 & 0 & 0 & 0 & 1 & 1 & 1 & 0
        \end{pmatrix}}$
        $\vcenter{\hbox{\includegraphics[width=0.35\textwidth]{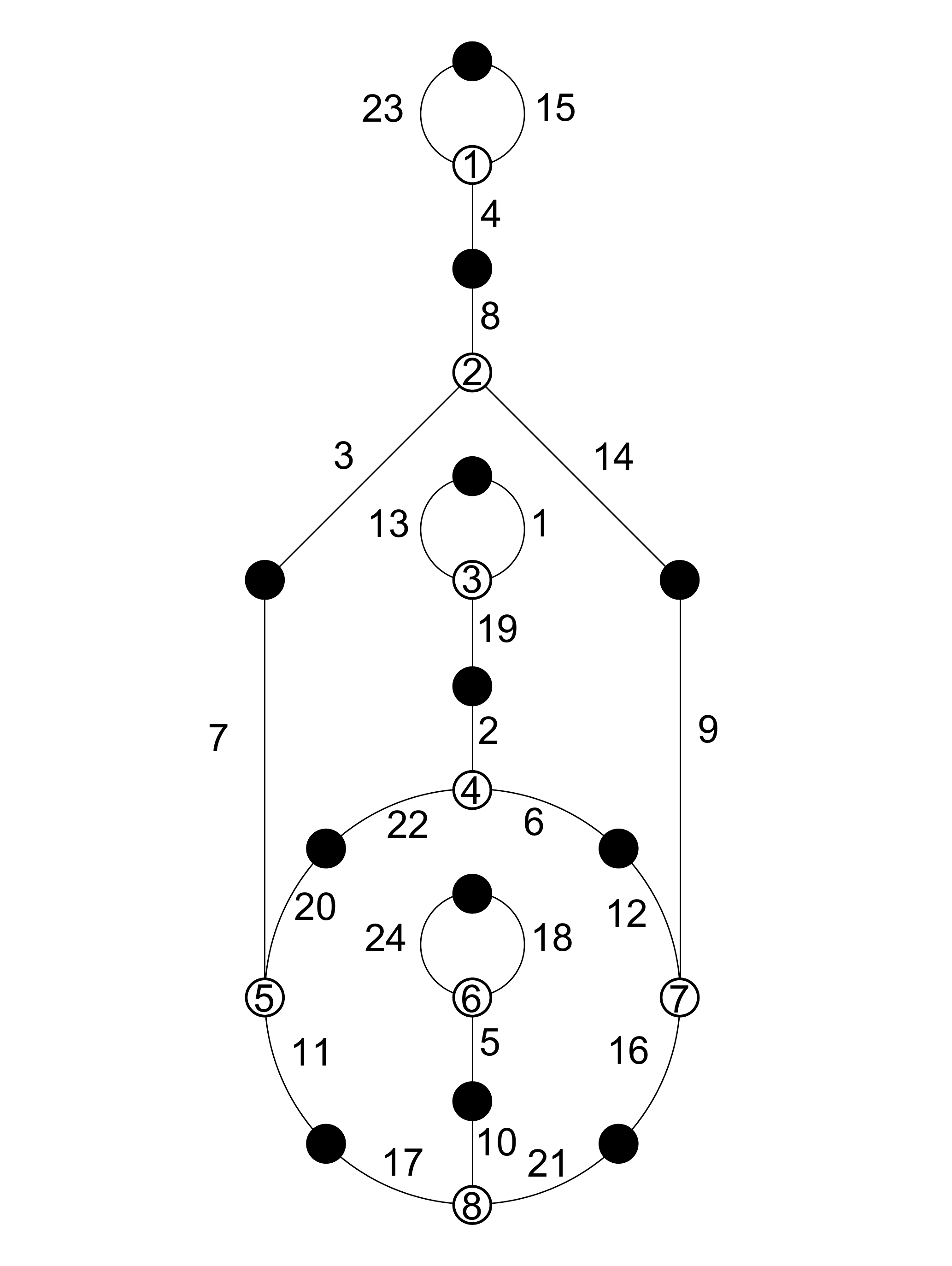}}}$
        \caption{ \{\{\{23,15,4\},\{8,14,3\},
        \{13,1,19\},\{2,6,22\},\{12,9,16\},
        \{10,21,17\},\{24,18,5\},\{20,11,7\}\}, \\ 
        \{\{23,15\},\{4,8\},\{7,3\},
        \{14,9\},\{2,19\},\{1,13\},
        \{20,22\},\{6,12\},\{24,18\},
        \{5,10\},\{11,17\},\{16,21\}\}\}}
        \caption{7-7-7-1-1-1 $(\mathbb{Q})$}
        \label{Dessin}
    \end{subfigure} \hfill
    \begin{subfigure}{0.5\textwidth}
        \centering \captionsetup{justification=centering}
        $\scalemath{0.75}{
        \displaystyle \begin{pmatrix}
            0 & 2 & 1 & 0 & 0 & 0 & 0 & 0\\ 
            2 & 0 & 0 & 1 & 0 & 0 & 0 & 0\\
            1 & 0 & 0 & 1 & 1 & 0 & 0 & 0\\
            0 & 1 & 1 & 0 & 0 & 1 & 0 & 0\\
            0 & 0 & 1 & 0 & 0 & 1 & 1 & 0\\
            0 & 0 & 0 & 1 & 1 & 0 & 0 & 1\\
            0 & 0 & 0 & 0 & 1 & 0 & 0 & 2\\
            0 & 0 & 0 & 0 & 0 & 1 & 2 & 0
        \end{pmatrix}}$
        $\vcenter{\hbox{\includegraphics[width=0.35\textwidth]{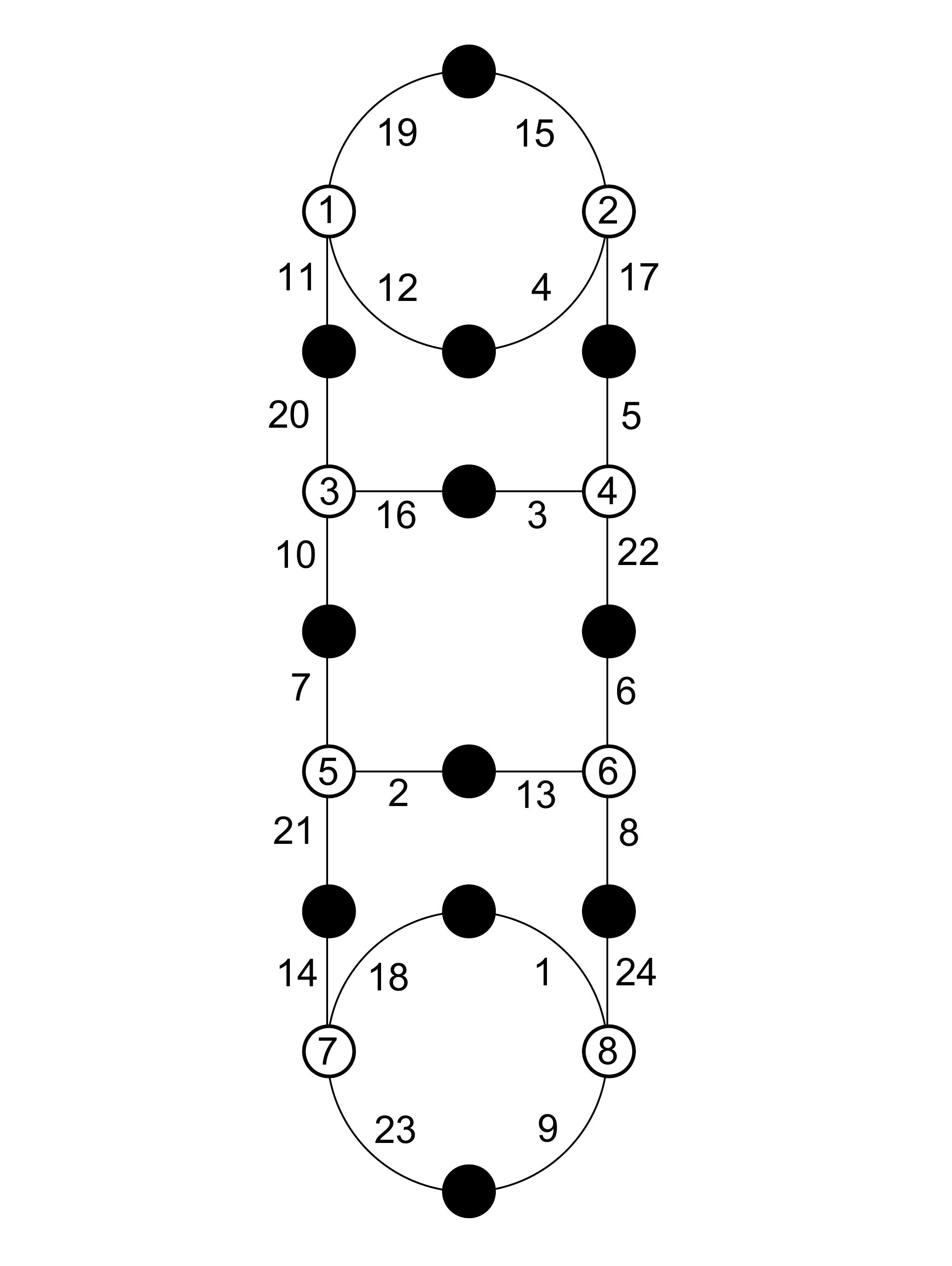}}}$
        \caption{ \{\{\{19,12,11\},\{15,17,4\},
        \{20,16,10\},\{5,22,3\},\{7,2,21\},
        \{6,8,13\},\{14,18,23\},\{1,24,9\}\}, \\ 
        \{\{19,15\},\{12,4\},\{11,20\},
        \{17,5\},\{3,16\},\{10,7\},
        \{22,6\},\{2,13\},\{21,14\},
        \{8,24\},\{18,1\},\{23,9\}\}\}}
        \caption{8-4-4-4-2-2 $(\mathbb{Q})$}
        \label{Dessin}
    \end{subfigure}\hfill
\end{figure}

\begin{figure}[H]
    \begin{subfigure}{0.5\textwidth}
        \centering \captionsetup{justification=centering}
        $\scalemath{0.75}{
        \displaystyle \begin{pmatrix}
            2 & 1 & 0 & 0 & 0 & 0 & 0 & 0\\ 
            1 & 0 & 1 & 0 & 1 & 0 & 0 & 0\\
            0 & 1 & 0 & 1 & 0 & 1 & 0 & 0\\
            0 & 0 & 1 & 0 & 1 & 0 & 1 & 0\\
            0 & 1 & 0 & 1 & 0 & 0 & 0 & 1\\
            0 & 0 & 1 & 0 & 0 & 0 & 1 & 1\\
            0 & 0 & 0 & 1 & 0 & 1 & 0 & 1\\
            0 & 0 & 0 & 0 & 1 & 1 & 1
            & 0
        \end{pmatrix}}$
        $\vcenter{\hbox{\includegraphics[width=0.35\textwidth]{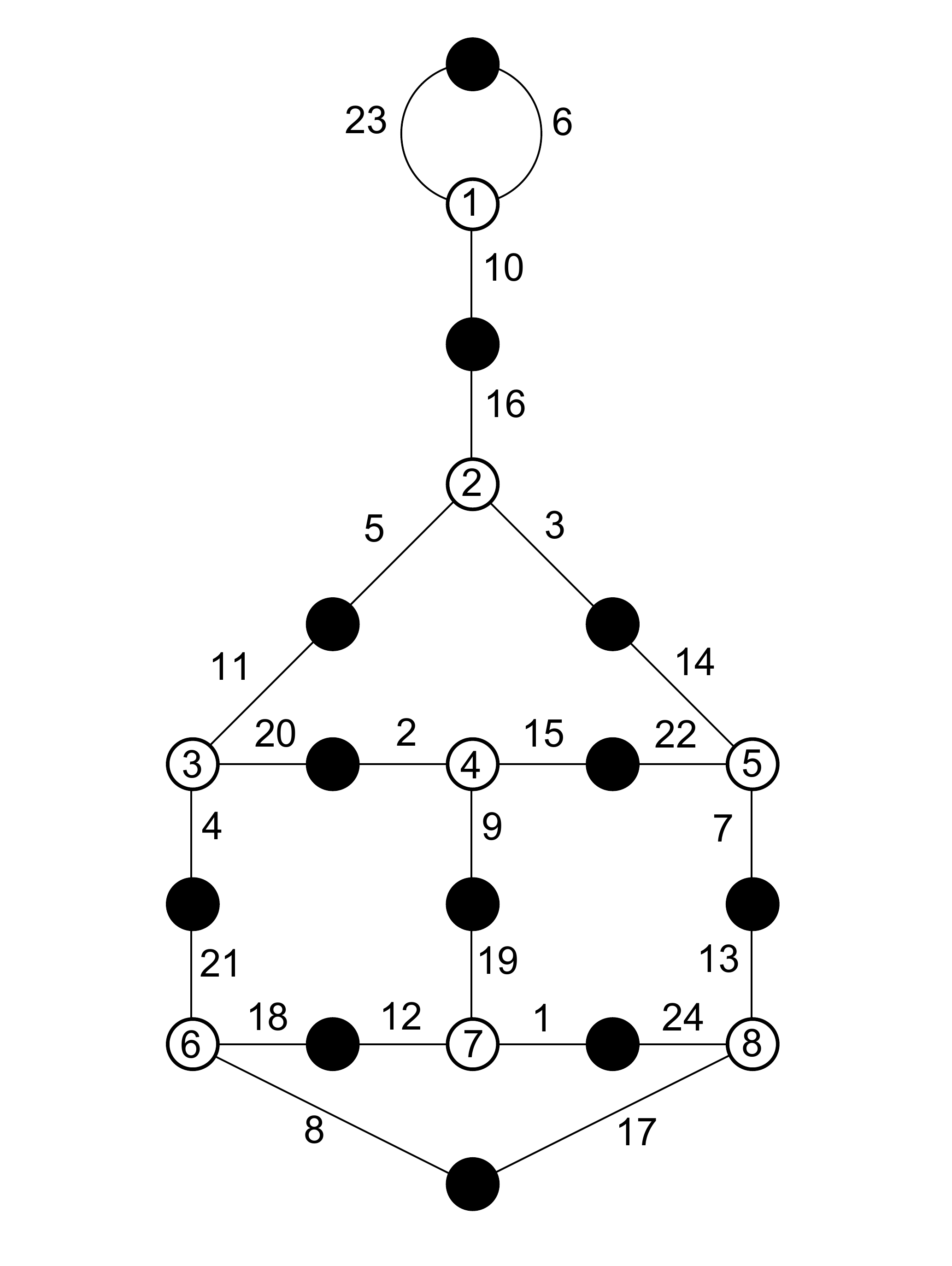}}}$
        \caption{ \{\{\{6,10,23\},\{16,3,5\},
        \{14,7,22\},\{15,9,2\},\{20,4,11\},
        \{21,18,8\},\{12,19,1\},\{24,13,17\}\}, \\ 
        \{\{8,17\},\{1,24\},\{19,9\},
        \{18,12\},\{4,21\},\{13,7\},
        \{2,20\},\{15,22\},\{5,11\},
        \{3,14\},\{16,10\},\{6,23\}\}\}}
        \caption{8-4-4-4-3-1 $(\mathbb{Q})$}
        \label{Dessin}
    \end{subfigure} \hfill
    \begin{subfigure}{0.5\textwidth}
        \centering \captionsetup{justification=centering}
        $\scalemath{0.75}{
        \displaystyle \begin{pmatrix}
            0 & 2 & 1 & 0 & 0 & 0 & 0 & 0\\ 
            2 & 0 & 1 & 0 & 0 & 0 & 0 & 0\\
            1 & 1 & 0 & 1 & 0 & 0 & 0 & 0\\
            0 & 0 & 1 & 0 & 1 & 0 & 0 & 1\\
            0 & 0 & 0 & 1 & 0 & 1 & 0 & 1\\
            0 & 0 & 0 & 0 & 1 & 0 & 2 & 0\\
            0 & 0 & 0 & 0 & 0 & 2 & 0 & 1\\
            0 & 0 & 0 & 1 & 1 & 0 & 1 & 0
        \end{pmatrix}}$
        $\vcenter{\hbox{\includegraphics[width=0.35\textwidth]{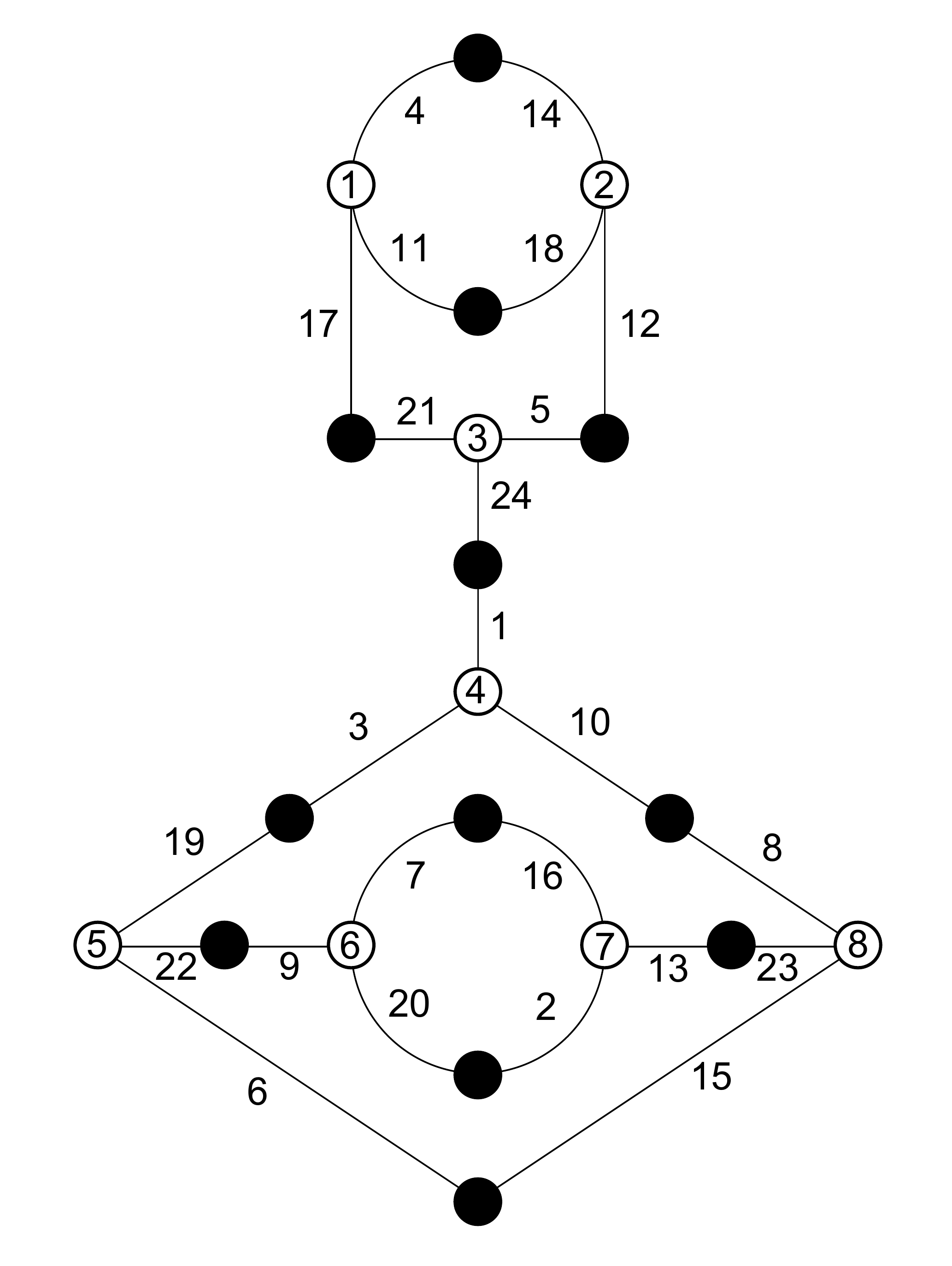}}}$
        \caption{ \{\{\{4,11,17\},\{14,12,18\},
        \{21,5,24\},\{1,10,3\},\{8,15,23\},
        \{16,13,2\},\{7,20,9\},\{19,22,6\}\}, \\ 
        \{\{4,14\},\{11,18\},\{17,21\},
        \{5,12\},\{24,1\},\{3,19\},
        \{10,8\},\{22,9\},\{13,23\},
        \{7,16\},\{2,20\},\{6,15\}\}\}}
        \caption{8-5-4-3-2-2 $(\mathbb{Q})$}
        \label{Dessin}
    \end{subfigure}\hfill
\end{figure}

\begin{figure}[H]
    \begin{subfigure}{0.5\textwidth}
        \centering \captionsetup{justification=centering}
        $\scalemath{0.75}{
        \displaystyle \begin{pmatrix}
            2 & 1 & 0 & 0 & 0 & 0 & 0 & 0\\ 
            1 & 0 & 0 & 0 & 1 & 1 & 0 & 0\\
            0 & 0 & 0 & 1 & 1 & 0 & 1 & 0\\
            0 & 0 & 1 & 0 & 0 & 1 & 0 & 1\\
            0 & 1 & 1 & 0 & 0 & 0 & 1 & 0\\
            0 & 1 & 0 & 1 & 0 & 0 & 0 & 1\\
            0 & 0 & 1 & 0 & 1 & 0 & 0 & 1\\
            0 & 0 & 0 & 1 & 0 & 1 & 1 & 0
        \end{pmatrix}}$
        $\vcenter{\hbox{\includegraphics[width=0.35\textwidth]{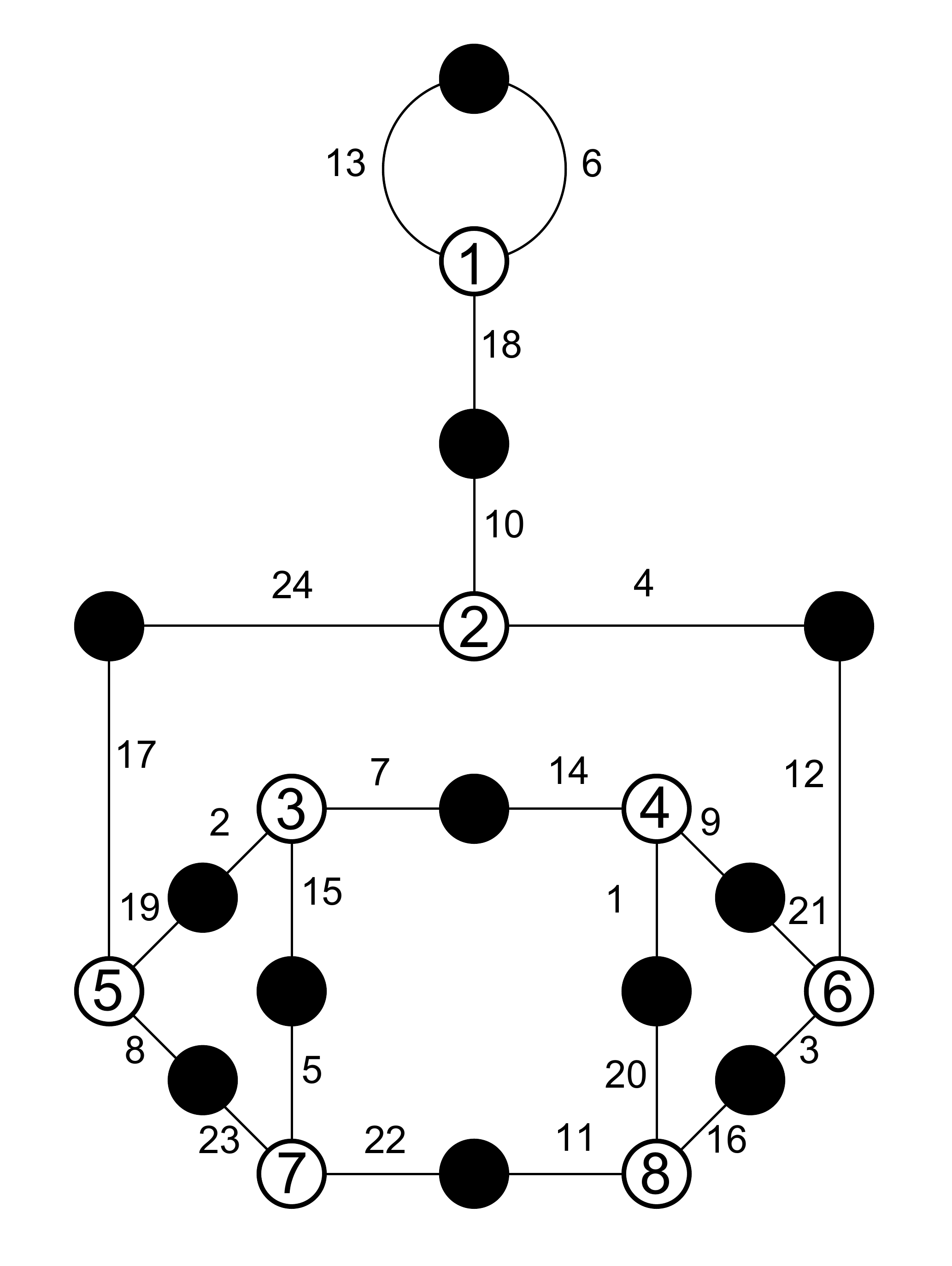}}}$
        \caption{ \{\{\{13,6,18\},\{10,4,24\},
        \{17,19,8\},\{2,7,15\},\{14,9,1\},
        \{21,12,3\},\{20,16,11\},\{5,22,23\}\}, \\ 
        \{\{13,6\},\{18,10\},\{17,24\},
        \{4,12\},\{7,14\},\{19,2\},
        \{8,23\},\{5,15\},\{11,22\},
        \{1,20\},\{9,21\},\{3,16\}\}\}}
        \caption{8-5-4-3-3-1 A $(\sqrt{10})$}
        \label{Dessin}
    \end{subfigure} \hfill
    \begin{subfigure}{0.5\textwidth}
        \centering \captionsetup{justification=centering}
        $\scalemath{0.75}{
        \displaystyle \begin{pmatrix}
            0 & 1 & 2 & 0 & 0 & 0 & 0 & 0\\ 
            1 & 2 & 0 & 0 & 0 & 0 & 0 & 0\\
            2 & 0 & 0 & 1 & 0 & 0 & 0 & 0\\
            0 & 0 & 1 & 0 & 0 & 1 & 1 & 0\\
            0 & 0 & 0 & 0 & 0 & 1 & 1 & 1\\
            0 & 0 & 0 & 1 & 1 & 0 & 0 & 1\\
            0 & 0 & 0 & 1 & 1 & 0 & 0 & 1\\
            0 & 0 & 0 & 0 & 1 & 1 & 1 & 0
        \end{pmatrix}}$
        $\vcenter{\hbox{\includegraphics[width=0.35\textwidth]{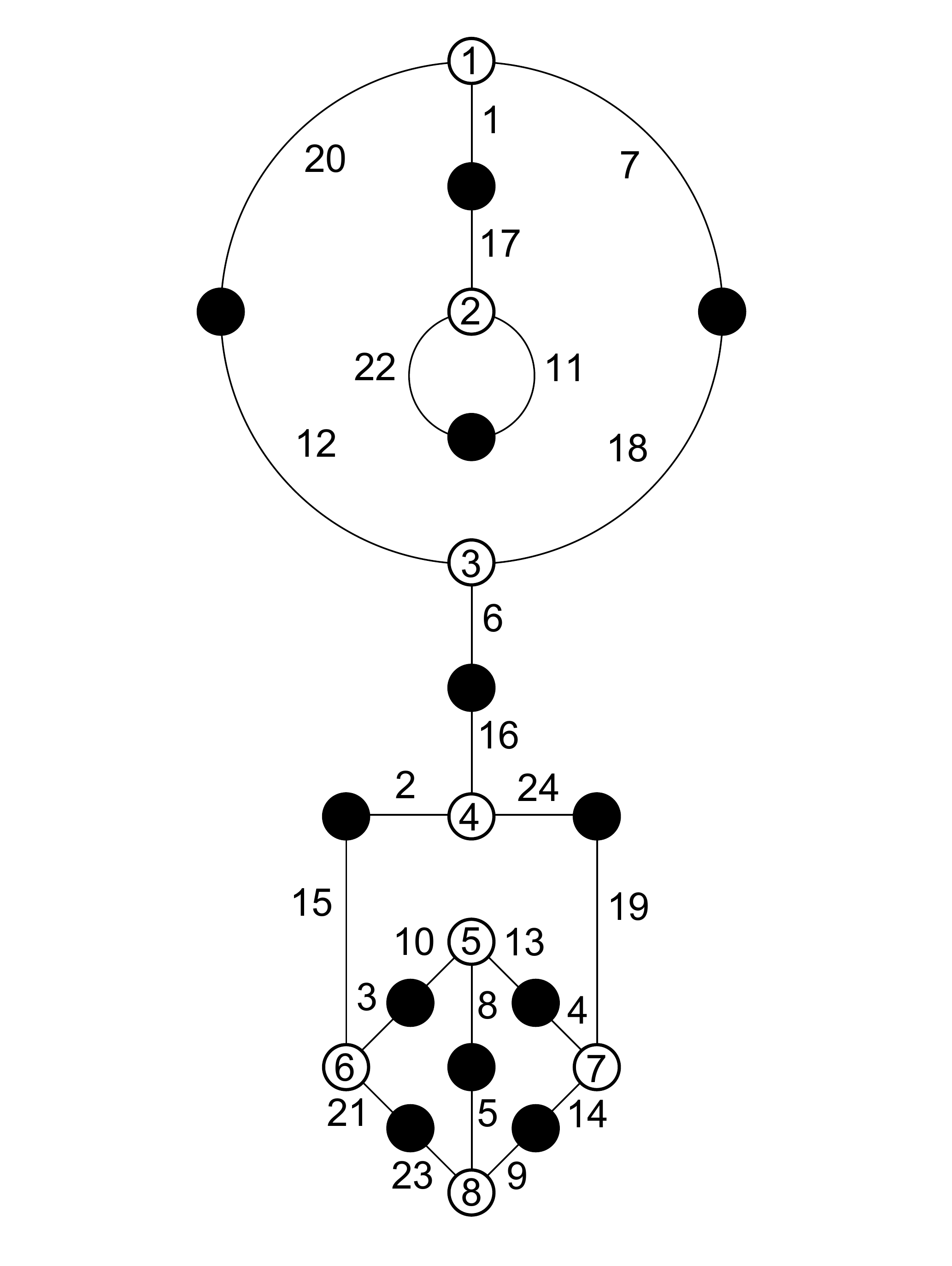}}}$
        \caption{ \{\{\{1,20,7\},\{17,11,22\},
        \{12,18,6\},\{16,24,2\},\{15,3,21\},
        \{10,13,8\},\{4,19,14\},\{5,9,23\}\}, \\ 
        \{\{20,12\},\{7,18\},\{22,11\},
        \{17,1\},\{16,6\},\{2,15\},
        \{24,19\},\{3,10\},\{4,13\},
        \{5,8\},\{21,23\},\{9,14\}\}\}}
        \caption{8-5-4-3-3-1 B $(\sqrt{10})$}
        \label{Dessin}
    \end{subfigure}\hfill
\end{figure}

\begin{figure}[H]
    \begin{subfigure}{0.5\textwidth}
        \centering \captionsetup{justification=centering}
        $\scalemath{0.75}{
        \displaystyle \begin{pmatrix}
            0 & 2 & 1 & 0 & 0 & 0 & 0 & 0\\ 
            2 & 0 & 0 & 1 & 0 & 0 & 0 & 0\\
            1 & 0 & 0 & 1 & 1 & 0 & 0 & 0\\
            0 & 1 & 1 & 0 & 0 & 1 & 0 & 0\\
            0 & 0 & 1 & 0 & 0 & 0 & 2 & 0\\
            0 & 0 & 0 & 1 & 0 & 0 & 0 & 2\\
            0 & 0 & 0 & 0 & 2 & 0 & 0 & 1\\
            0 & 0 & 0 & 0 & 0 & 2 & 1 & 0
        \end{pmatrix}}$
        $\vcenter{\hbox{\includegraphics[width=0.35\textwidth]{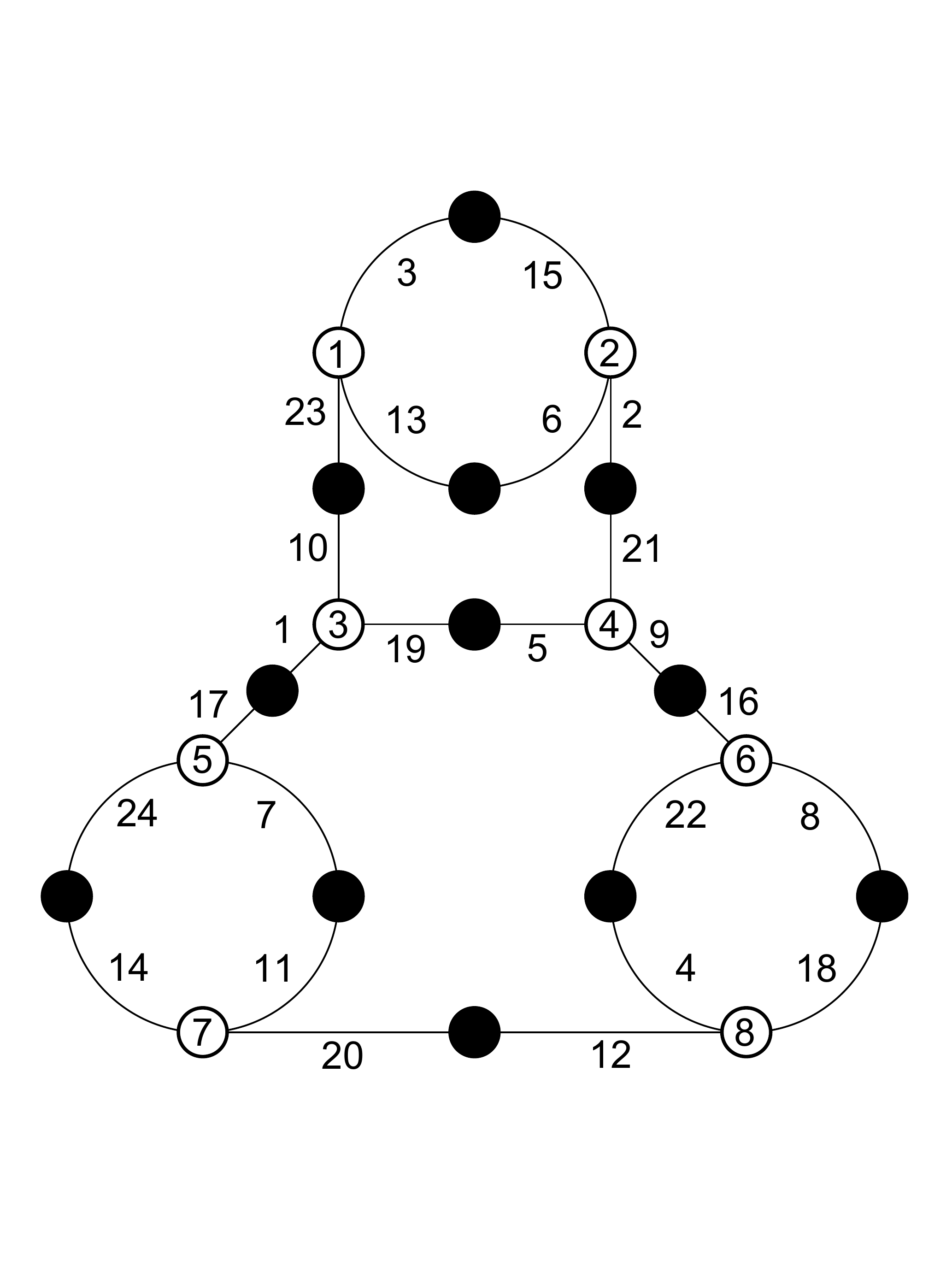}}}$
        \caption{ \{\{\{3,13,23\},\{15,2,6\},
        \{10,19,1\},\{21,9,5\},\{16,8,22\},
        \{4,18,12\},\{20,14,11\},\{24,17,7\}\}, \\ 
        \{\{20,12\},\{11,7\},\{14,24\},
        \{1,17\},\{5,19\},\{10,23\},
        \{6,13\},\{3,15\},\{2,21\},
        \{9,16\},\{8,18\},\{22,4\}\}\}}
        \caption{8-6-4-2-2-2 $(\mathbb{Q})$}
        \label{Dessin}
    \end{subfigure} \hfill
    \begin{subfigure}{0.5\textwidth}
        \centering \captionsetup{justification=centering}
        $\scalemath{0.75}{
        \displaystyle \begin{pmatrix}
            2 & 1 & 0 & 0 & 0 & 0 & 0 & 0\\ 
            1 & 0 & 1 & 1 & 0 & 0 & 0 & 0\\
            0 & 1 & 0 & 1 & 0 & 0 & 1 & 0\\
            0 & 1 & 1 & 0 & 0 & 0 & 0 & 1\\
            0 & 0 & 0 & 0 & 0 & 2 & 1 & 0\\
            0 & 0 & 0 & 0 & 2 & 0 & 0 & 1\\
            0 & 0 & 1 & 0 & 1 & 0 & 0 & 1\\
            0 & 0 & 0 & 1 & 0 & 1 & 1 & 0
        \end{pmatrix}}$
        $\vcenter{\hbox{\includegraphics[width=0.35\textwidth]{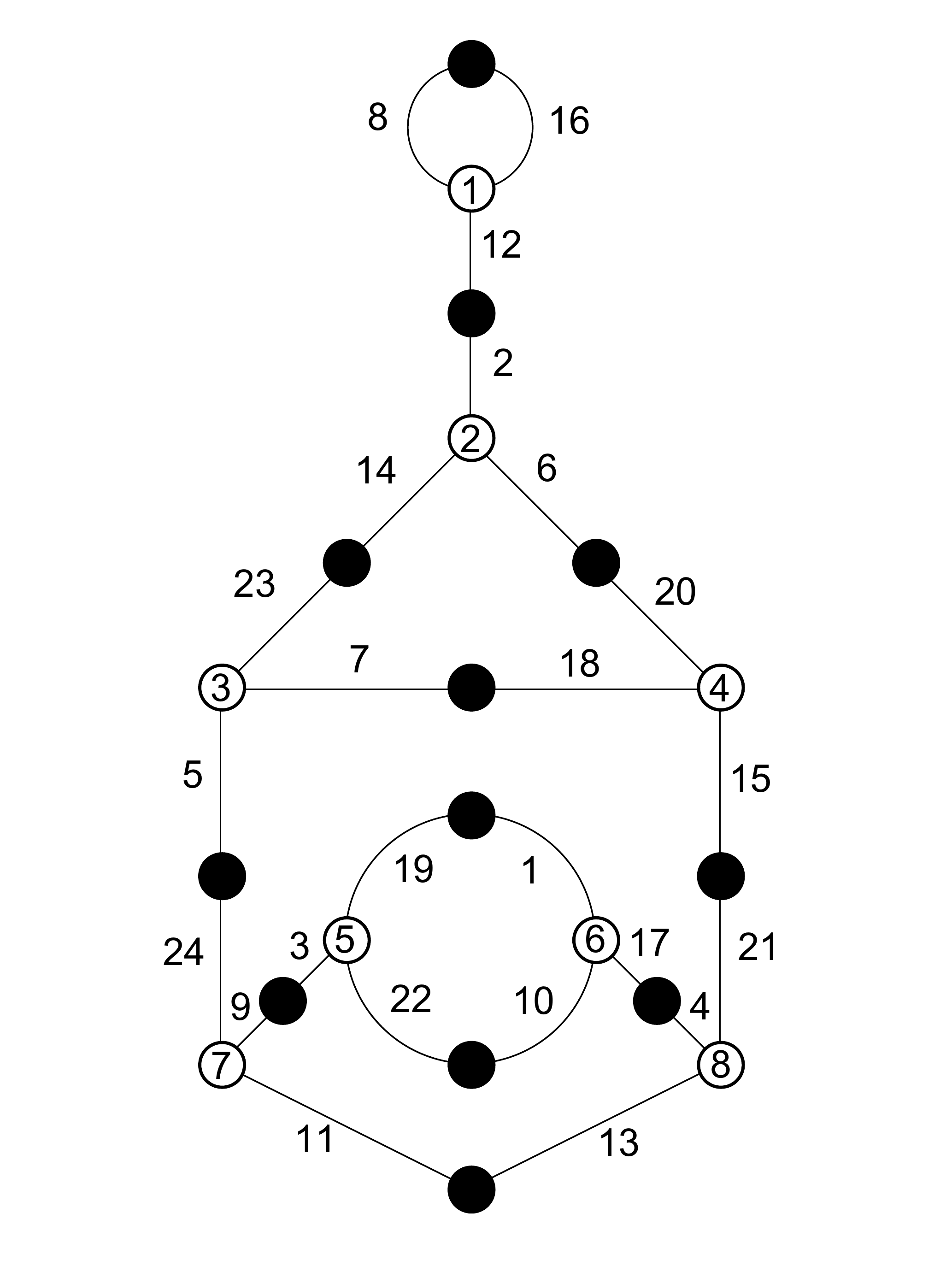}}}$
        \caption{ \{\{\{12,8,16\},\{2,6,14\},
        \{23,7,5\},\{18,20,15\},\{9,11,24\},
        \{3,19,22\},\{1,17,10\},\{4,21,13\}\}, \\ 
        \{\{11,13\},\{10,22\},\{1,19\},
        \{9,3\},\{4,17\},\{21,15\},
        \{24,5\},\{7,18\},\{23,14\},
        \{20,6\},\{2,12\},\{8,16\}\}\}}
        \caption{8-6-4-3-2-1 A $(\sqrt{2})$}
        \label{Dessin}
    \end{subfigure}\hfill
\end{figure}

\begin{figure}[H]
    \begin{subfigure}{0.5\textwidth}
        \centering \captionsetup{justification=centering}
        $\scalemath{0.75}{
        \displaystyle \begin{pmatrix}
            0 & 2 & 1 & 0 & 0 & 0 & 0 & 0\\ 
            2 & 0 & 1 & 0 & 0 & 0 & 0 & 0\\
            1 & 1 & 0 & 1 & 0 & 0 & 0 & 0\\
            0 & 0 & 1 & 0 & 0 & 1 & 0 & 1\\
            0 & 0 & 0 & 0 & 0 & 1 & 1 & 1\\
            0 & 0 & 0 & 1 & 1 & 0 & 0 & 1\\
            0 & 0 & 0 & 0 & 1 & 0 & 2 & 0\\
            0 & 0 & 0 & 1 & 1 & 1 & 0 & 0
        \end{pmatrix}}$
        $\vcenter{\hbox{\includegraphics[width=0.35\textwidth]{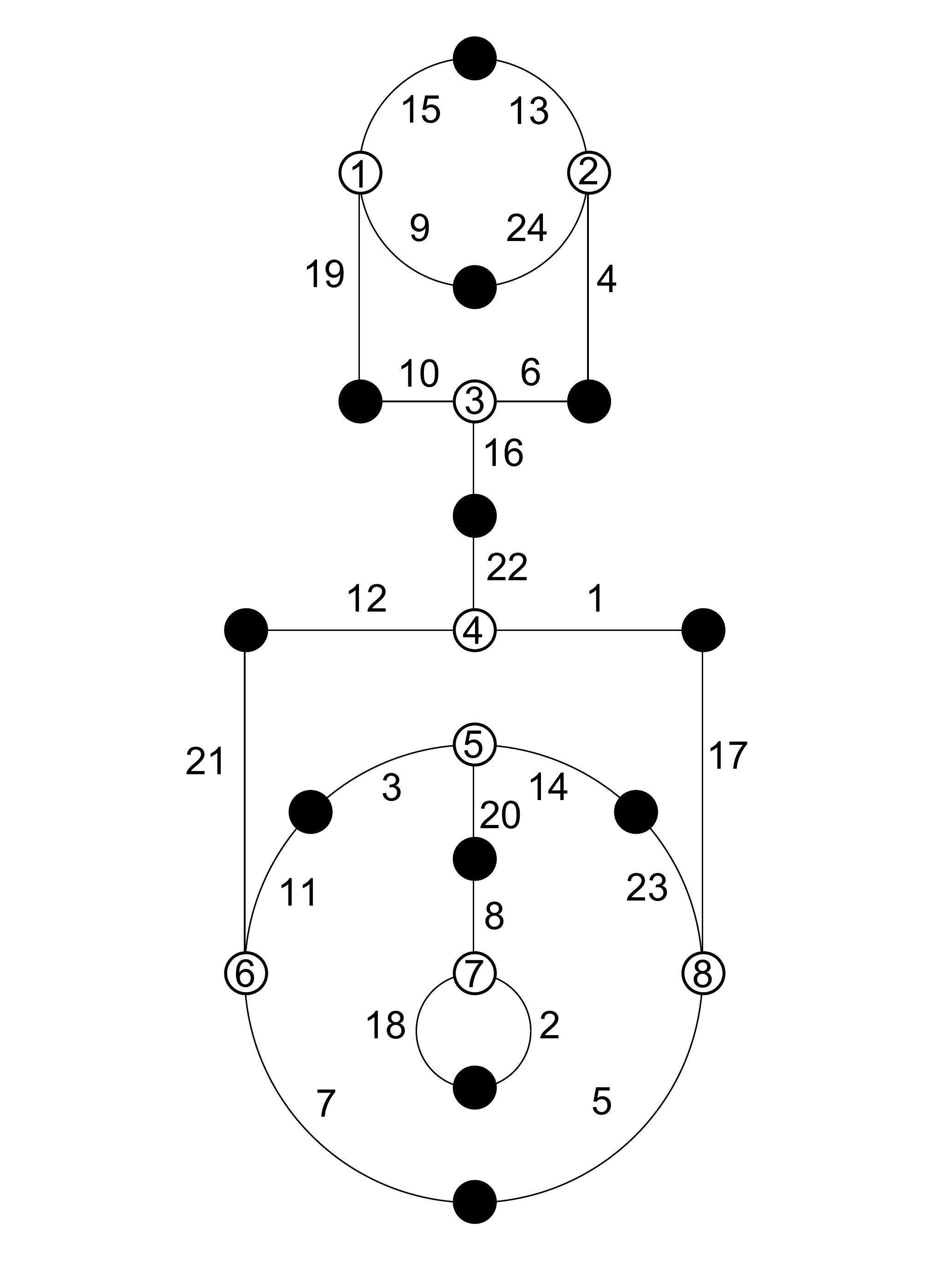}}}$
        \caption{ \{\{\{15,9,19\},\{24,13,4\},
        \{10,6,16\},\{22,1,12\},\{21,11,7\},
        \{3,14,20\},\{23,17,5\},\{8,2,18\}\}, \\ 
        \{\{5,7\},\{2,18\},\{3,11\},
        \{14,23\},\{8,20\},\{12,21\},
        \{1,17\},\{22,16\},\{4,6\},
        \{10,19\},\{9,24\},\{15,13\}\}\}}
        \caption{8-6-4-3-2-1 B $(\sqrt{2})$}
        \label{Dessin}
    \end{subfigure} \hfill
    \begin{subfigure}{0.5\textwidth}
        \centering \captionsetup{justification=centering}
        $\scalemath{0.75}{
        \displaystyle \begin{pmatrix}
            2 & 1 & 0 & 0 & 0 & 0 & 0 & 0\\ 
            1 & 0 & 1 & 0 & 0 & 1 & 0 & 0\\
            0 & 1 & 0 & 1 & 0 & 0 & 1 & 0\\
            0 & 0 & 1 & 0 & 2 & 0 & 0 & 0\\
            0 & 0 & 0 & 2 & 0 & 1 & 0 & 0\\
            0 & 1 & 0 & 0 & 1 & 0 & 0 & 1\\
            0 & 0 & 1 & 0 & 0 & 0 & 0 & 2\\
            0 & 0 & 0 & 0 & 0 & 1 & 2 & 0
        \end{pmatrix}}$
        $\vcenter{\hbox{\includegraphics[width=0.35\textwidth]{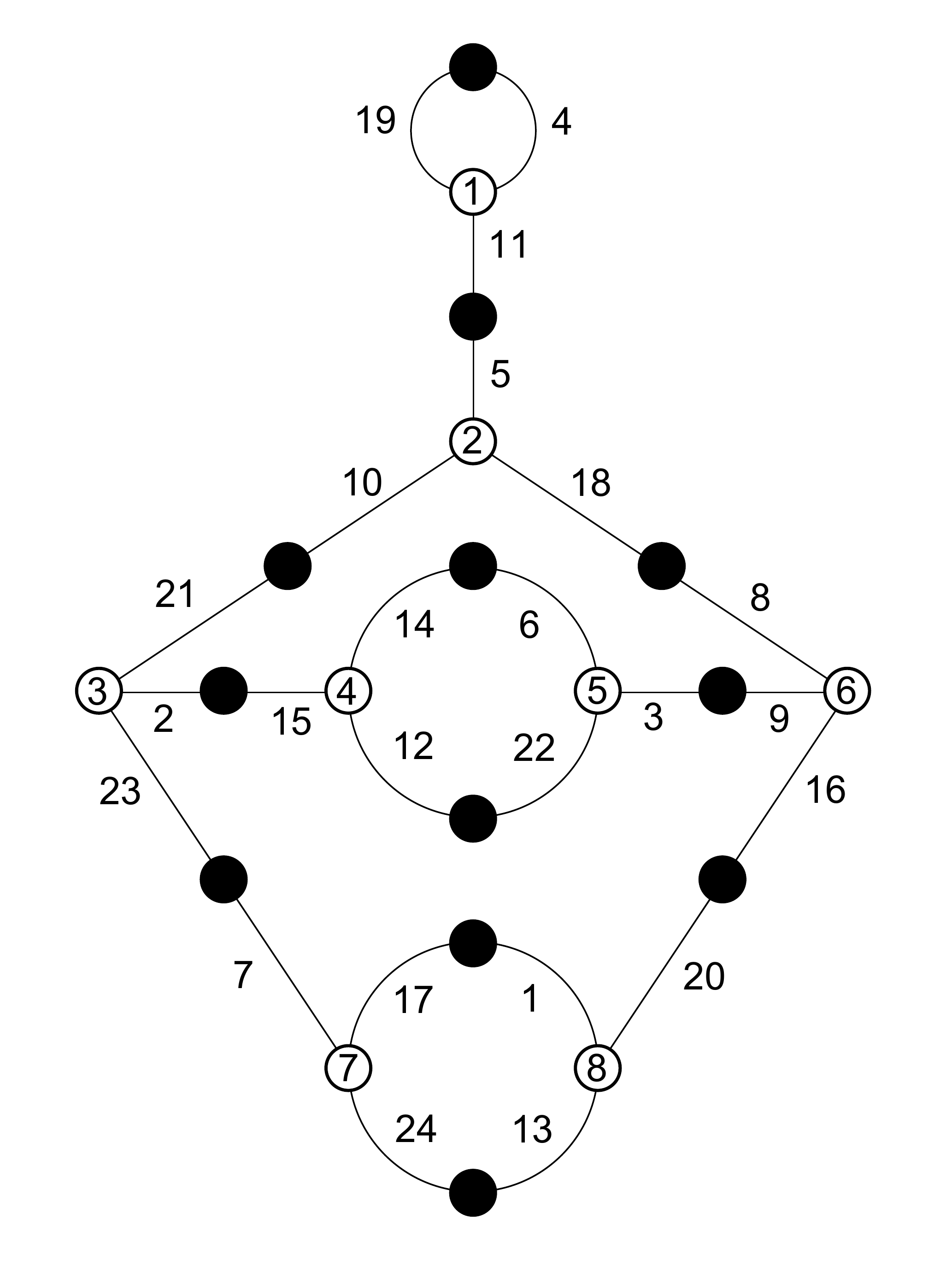}}}$
        \caption{ \{\{\{19,4,11\},\{5,18,10\},
        \{21,2,23\},\{15,14,12\},\{6,3,22\},
        \{8,16,9\},\{7,17,24\},\{1,20,13\}\}, \\ 
        \{\{13,24\},\{17,1\},\{7,23\},
        \{16,20\},\{9,3\},\{22,12\},
        \{6,14\},\{2,15\},\{10,21\},
        \{8,18\},\{11,5\},\{4,19\}\}\}}
        \caption{8-6-5-2-2-1 $(\mathbb{Q})$}
        \label{Dessin}
    \end{subfigure}\hfill
\end{figure}

\begin{figure}[H]
    \begin{subfigure}{0.5\textwidth}
        \centering \captionsetup{justification=centering}
        $\scalemath{0.75}{
        \displaystyle \begin{pmatrix}
            2 & 1 & 0 & 0 & 0 & 0 & 0 & 0\\ 
            1 & 0 & 1 & 1 & 0 & 0 & 0 & 0\\
            0 & 1 & 0 & 1 & 0 & 1 & 0 & 0\\
            0 & 1 & 1 & 0 & 0 & 0 & 0 & 1\\
            0 & 0 & 0 & 0 & 0 & 1 & 1 & 1\\
            0 & 0 & 1 & 0 & 1 & 0 & 0 & 1\\
            0 & 0 & 0 & 0 & 1 & 0 & 2 & 0\\
            0 & 0 & 0 & 1 & 1 & 1 & 0 & 0
        \end{pmatrix}}$
        $\vcenter{\hbox{\includegraphics[width=0.35\textwidth]{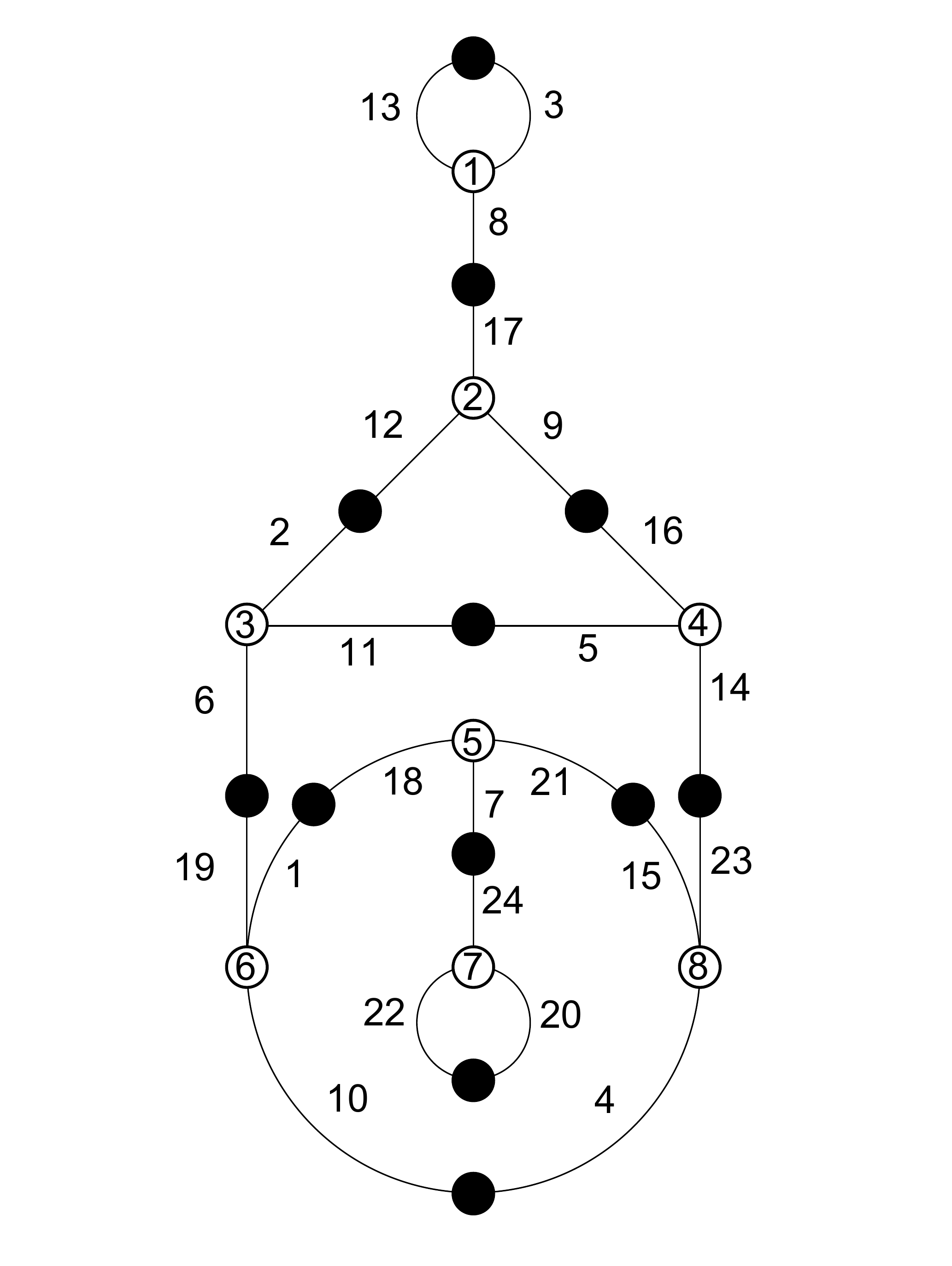}}}$
        \caption{ \{\{\{13,3,8\},\{17,9,12\},
        \{16,14,5\},\{2,11,6\},\{19,1,10\},
        \{18,21,7\},\{15,23,4\},\{24,20,22\}\}, \\ 
        \{\{20,22\},\{24,7\},\{15,21\},
        \{18,1\},\{19,6\},\{23,14\},
        \{11,5\},\{2,12\},\{9,16\},
        \{17,8\},\{13,3\},\{10,4\}\}\}}
        \caption{8-6-5-3-1-1 A $(\sqrt{5})$}
        \label{Dessin}
    \end{subfigure} \hfill
    \begin{subfigure}{0.5\textwidth}
        \centering \captionsetup{justification=centering}
        $\scalemath{0.75}{
        \displaystyle \begin{pmatrix}
            0 & 1 & 2 & 0 & 0 & 0 & 0 & 0\\
            1 & 2 & 0 & 0 & 0 & 0 & 0 & 0\\
            2 & 0 & 0 & 1 & 0 & 0 & 0 & 0\\
            0 & 0 & 1 & 0 & 1 & 1 & 0 & 0\\
            0 & 0 & 0 & 1 & 0 & 1 & 0 & 1\\
            0 & 0 & 0 & 1 & 1 & 0 & 0 & 1\\
            0 & 0 & 0 & 0 & 0 & 0 & 2 & 1\\
            0 & 0 & 0 & 0 & 1 & 1 & 1 & 0
        \end{pmatrix}}$
        $\vcenter{\hbox{\includegraphics[width=0.25\textwidth]{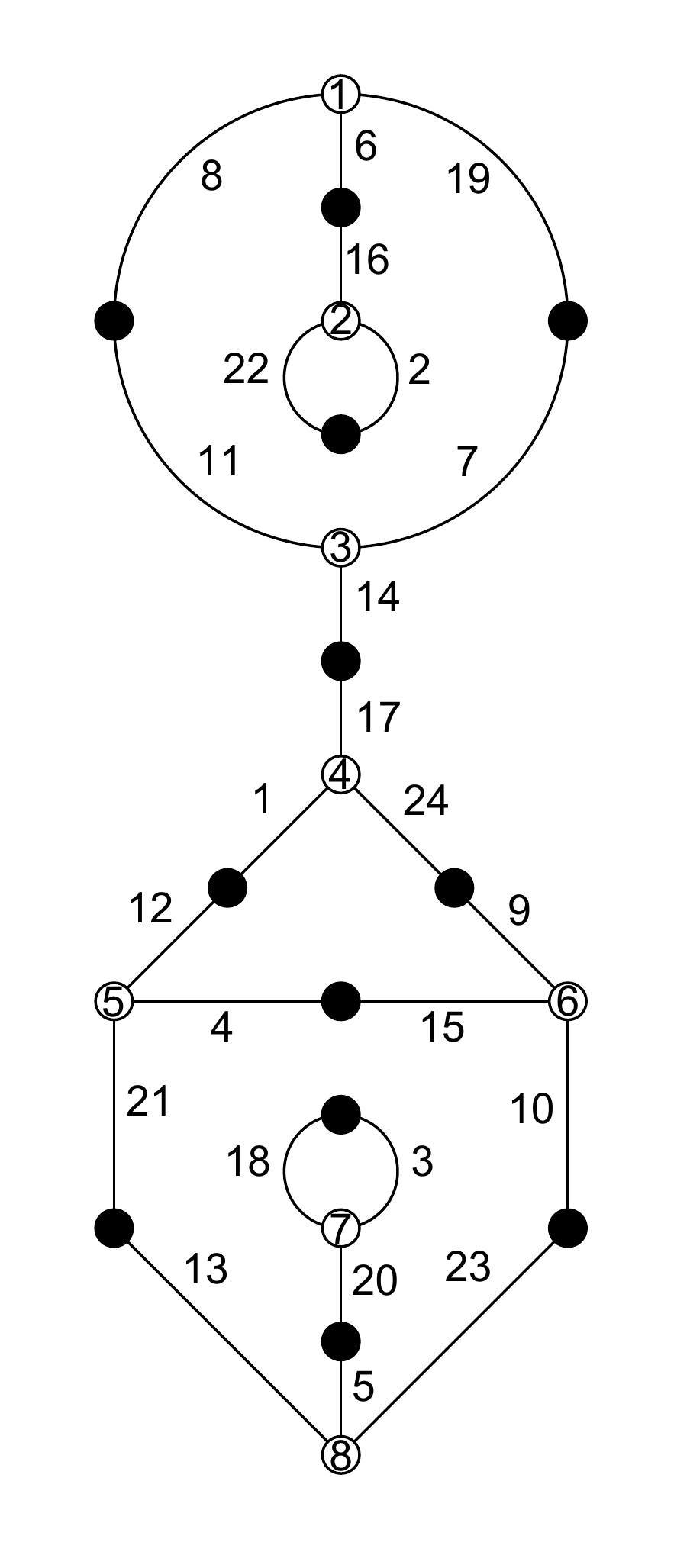}}}$
        \caption{ \{\{\{6,8,19\},\{16,2,22\},
        \{11,7,14\},\{17,24,1\},\{12,4,21\},
        \{9,10,15\},\{3,20,18\},\{5,23,13\}\}, \\ 
        \{\{5,20\},\{13,21\},\{4,15\},
        \{10,23\},\{18,3\},\{1,12\},
        \{9,24\},\{17,14\},\{7,19\},
        \{11,8\},\{2,22\},\{16,6\}\}\}}
        \caption{8-6-5-3-1-1 B $(\sqrt{5})$}
        \label{Dessin}
    \end{subfigure}\hfill
\end{figure}

\begin{figure}[H]
    \begin{subfigure}{0.6\textwidth}
        \centering \captionsetup{justification=centering}
        $\scalemath{0.75}{
        \displaystyle \begin{pmatrix}
            0 & 1 & 1 & 1 & 0 & 0 & 0 & 0\\ 
            1 & 0 & 0 & 1 & 0 & 1 & 0 & 0\\
            1 & 0 & 2 & 0 & 0 & 0 & 0 & 0\\
            1 & 1 & 0 & 0 & 1 & 0 & 0 & 0\\
            0 & 0 & 0 & 1 & 0 & 0 & 2 & 0\\
            0 & 1 & 0 & 0 & 0 & 0 & 1 & 1\\
            0 & 0 & 0 & 0 & 2 & 1 & 0 & 0\\
            0 & 0 & 0 & 0 & 0 & 1 & 0 & 2
        \end{pmatrix}}$
        $\vcenter{\hbox{\includegraphics[width=0.25\textwidth]{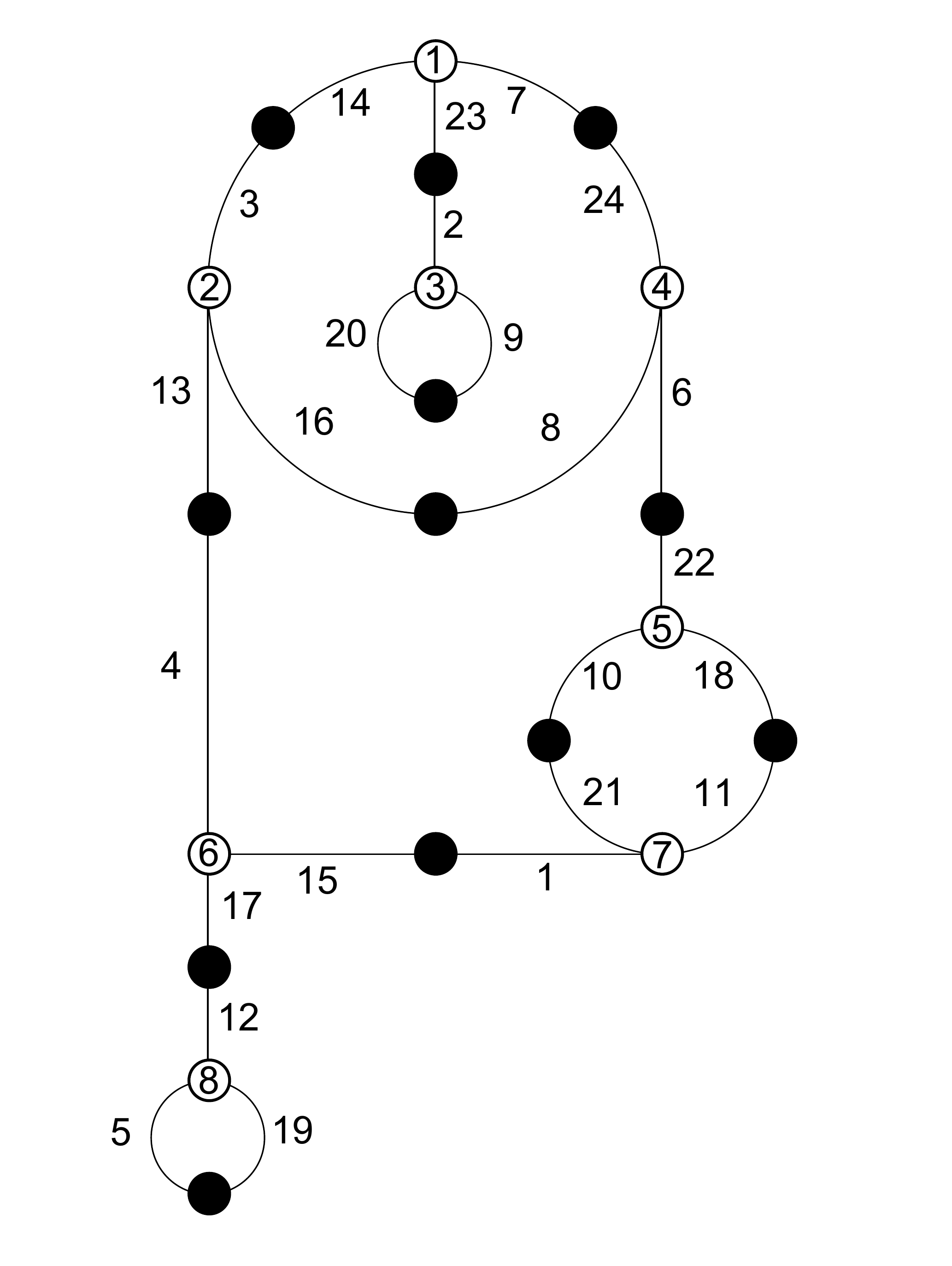}}}$
        $\vcenter{\hbox{\includegraphics[width=0.25\textwidth]{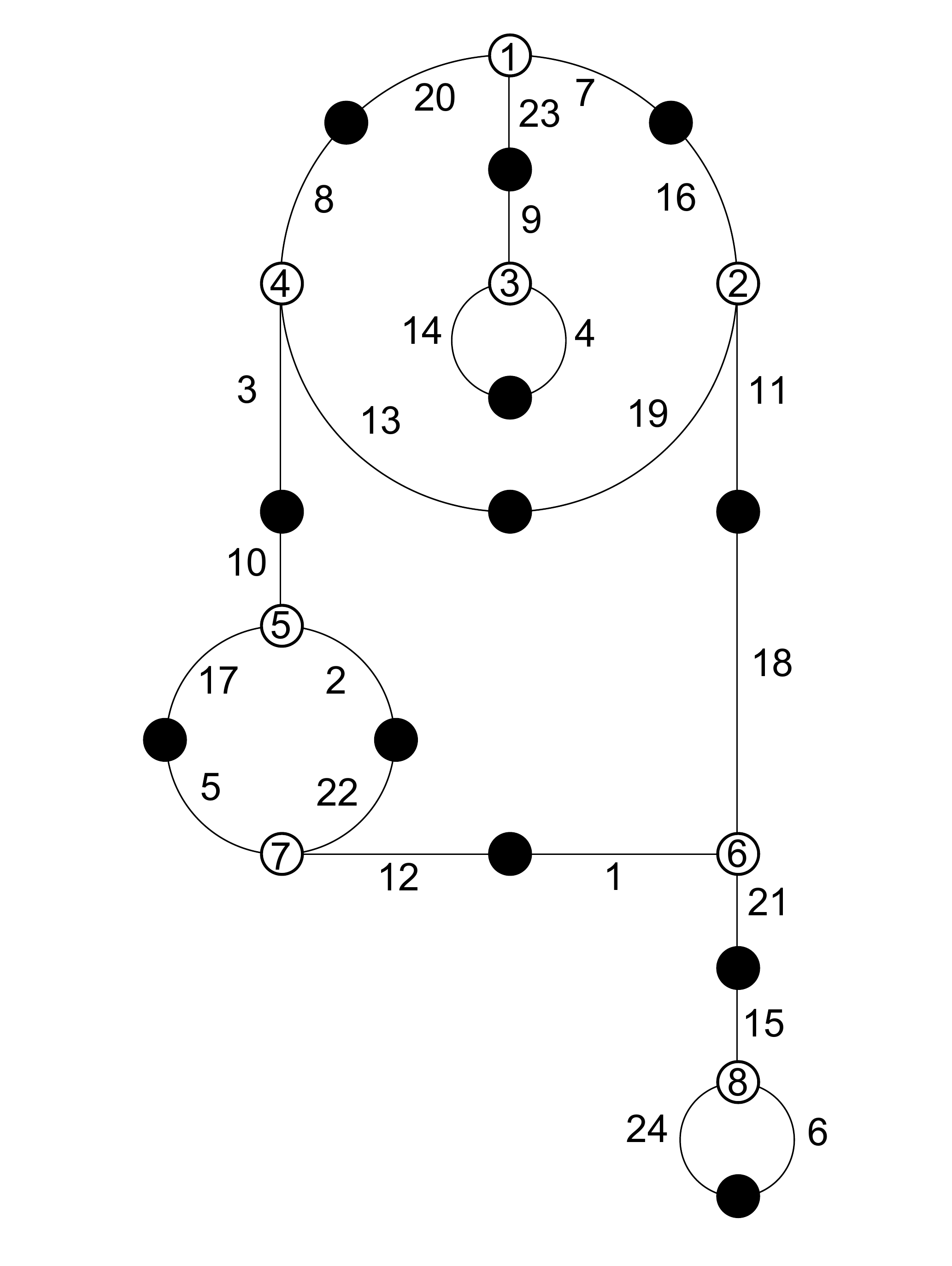}}}$
        \caption{ A: \{\{\{14,7,23\},\{2,9,20\},
        \{24,6,8\},\{16,13,3\},\{4,15,17\},
        \{10,22,18\},\{11,1,21\},\{19,5,12\}\}, \\ 
        \{\{5,19\},\{12,17\},\{15,1\},
        \{4,13\},\{11,18\},\{10,21\},
        \{22,6\},\{24,7\},\{23,2\},
        \{9,20\},\{3,14\},\{8,16\}\}\} \\
        B: \{\{\{20,7,23\},\{9,4,14\},
        \{8,13,3\},\{11,19,16\},\{10,2,17\},
        \{22,12,5\},\{1,18,21\},\{15,6,24\}\}, \\ 
        \{\{6,24\},\{15,21\},\{1,12\},
        \{22,2\},\{5,17\},\{10,3\},
        \{18,11\},\{13,19\},\{8,20\},
        \{7,16\},\{23,9\},\{4,14\}\}\}}
        \caption{8-6-6-2-1-1 A \& B  $(\sqrt{-3})$}
        \label{Dessin}
    \end{subfigure} \hfill
    \begin{subfigure}{0.4\textwidth}
        \centering \captionsetup{justification=centering}
        $\scalemath{0.75}{
        \displaystyle \begin{pmatrix}
            0 & 2 & 1 & 0 & 0 & 0 & 0 & 0\\ 
            2 & 0 & 1 & 0 & 0 & 0 & 0 & 0\\
            1 & 1 & 0 & 1 & 0 & 0 & 0 & 0\\
            0 & 0 & 1 & 0 & 0 & 0 & 1 & 1\\
            0 & 0 & 0 & 0 & 2 & 1 & 0 & 0\\
            0 & 0 & 0 & 0 & 1 & 0 & 1 & 1\\
            0 & 0 & 0 & 1 & 0 & 1 & 0 & 1\\
            0 & 0 & 0 & 1 & 0 & 1 & 1 & 0
        \end{pmatrix}}$
        $\vcenter{\hbox{\includegraphics[width=0.35\textwidth]{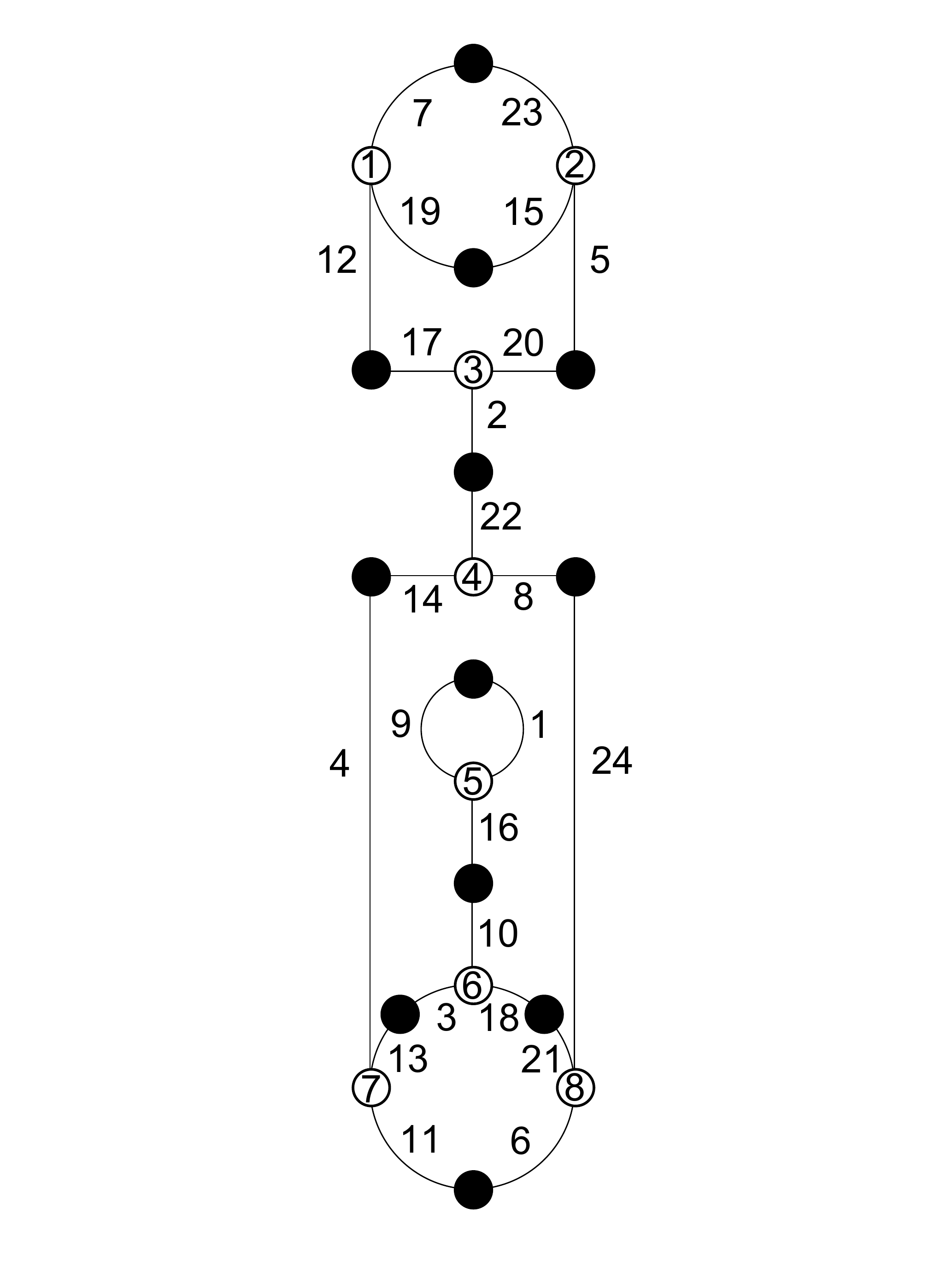}}}$
        \caption{ \{\{\{7,19,12\},\{23,5,15\},
        \{17,20,2\},\{22,8,14\},\{1,16,9\},
        \{10,18,3\},\{21,24,6\},\{13,11,4\}\}, \\ 
        \{\{6,11\},\{3,13\},\{18,21\},
        \{10,16\},\{1,9\},\{4,14\},
        \{8,24\},\{22,2\},\{12,17\},
        \{5,20\},\{19,15\},\{7,23\}\}\}}
        \caption{8-7-3-3-2-1 $(\mathbb{Q})$}
        \label{Dessin}
    \end{subfigure}\hfill
\end{figure}

\begin{figure}[H]
    \begin{subfigure}{0.5\textwidth}
        \centering \captionsetup{justification=centering}
        $\scalemath{0.75}{
        \displaystyle \begin{pmatrix}
            0 & 1 & 1 & 0 & 0 & 0 & 1 & 0\\ 
            1 & 0 & 0 & 2 & 0 & 0 & 0 & 0\\
            1 & 0 & 0 & 0 & 2 & 0 & 0 & 0\\
            0 & 2 & 0 & 0 & 0 & 1 & 0 & 0\\
            0 & 0 & 2 & 0 & 0 & 0 & 1 & 0\\
            0 & 0 & 0 & 1 & 0 & 0 & 1 & 1\\
            1 & 0 & 0 & 0 & 1 & 1 & 0 & 0\\
            0 & 0 & 0 & 0 & 0 & 1 & 0 & 2
        \end{pmatrix}}$
        $\vcenter{\hbox{\includegraphics[width=0.25\textwidth]{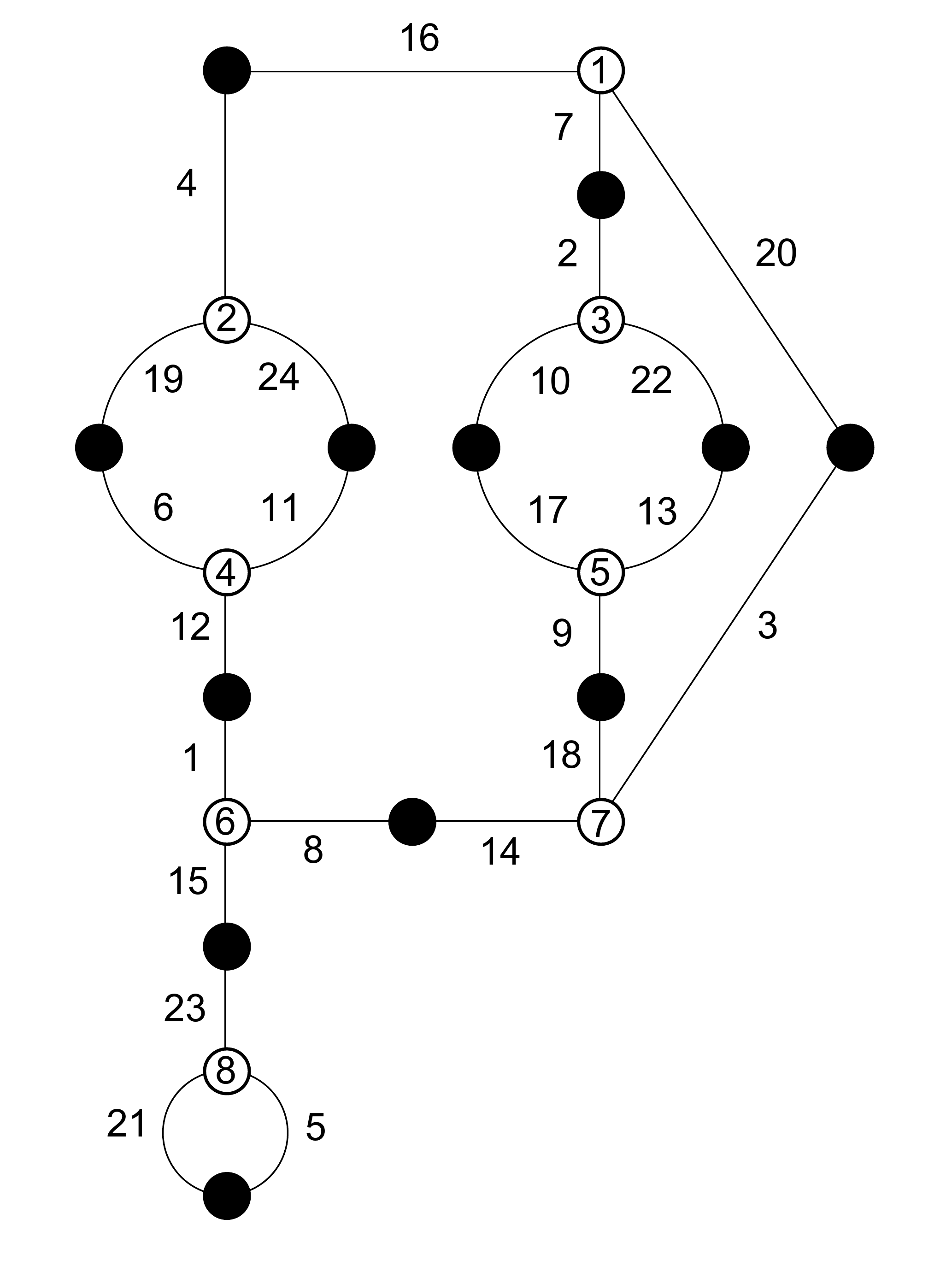}}}$
        $\vcenter{\hbox{\includegraphics[width=0.25\textwidth]{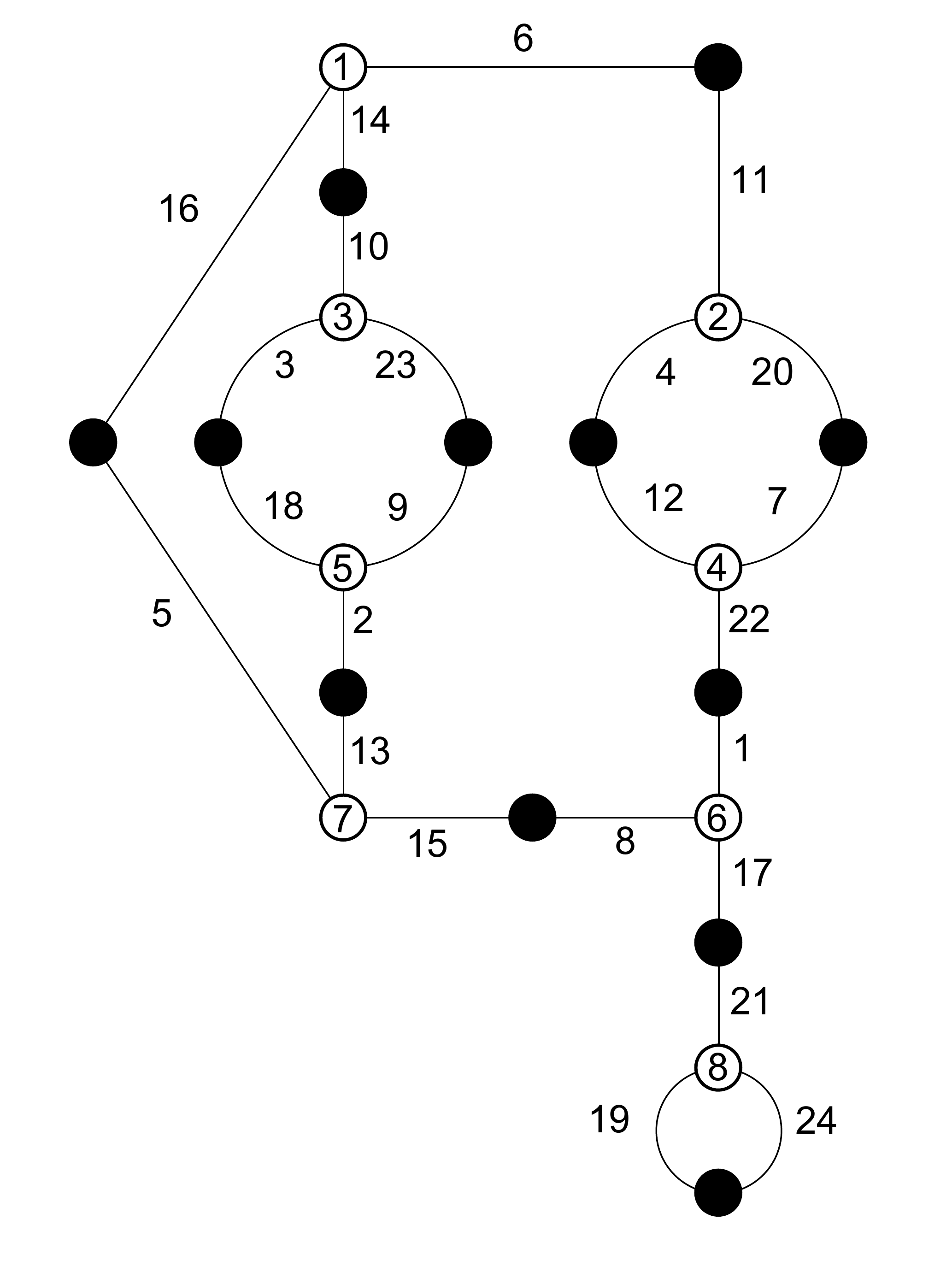}}}$
        \caption{ A: \{\{\{4,24,19\},\{6,11,12\},
        \{16,20,7\},\{2,22,10\},\{17,13,9\},
        \{18,3,14\},\{1,8,15\},\{23,5,21\}\}, \\ 
        \{\{5,21\},\{23,15\},\{1,12\},
        \{8,14\},\{9,18\},\{3,20\},
        \{17,10\},\{13,22\},\{2,7\},
        \{16,4\},\{19,6\},\{11,24\}\}\} \\
        B: \{\{\{16,6,14\},\{10,23,3\},
        \{18,9,2\},\{23,15,5\},\{8,1,17\},
        \{21,24,19\},\{22,12,7\},\{4,11,20\}\}, \\ 
        \{\{6,11\},\{4,12\},\{20,7\},
        \{22,1\},\{17,21\},\{8,15\},
        \{13,2\},\{5,16\},\{18,3\},
        \{9,23\},\{10,14\},\{19,24\}\}\}}
        \caption{8-7-4-2-2-1 A \& B $(\sqrt{-7})$}
        \label{Dessin}
    \end{subfigure} \hfill
    \begin{subfigure}{0.5\textwidth}
        \centering \captionsetup{justification=centering}
        $\scalemath{0.75}{
        \displaystyle \begin{pmatrix}
            2 & 0 & 1 & 0 & 0 & 0 & 0 & 0\\ 
            0 & 0 & 1 & 1 & 1 & 0 & 0 & 0\\
            1 & 1 & 0 & 0 & 0 & 1 & 0 & 0\\
            0 & 1 & 0 & 0 & 1 & 0 & 0 & 1\\
            0 & 1 & 0 & 1 & 0 & 1 & 0 & 0\\
            0 & 0 & 1 & 0 & 1 & 0 & 0 & 1\\
            0 & 0 & 0 & 0 & 0 & 0 & 2 & 1\\
            0 & 0 & 0 & 1 & 0 & 1 & 1 & 0
        \end{pmatrix}}$
        $\vcenter{\hbox{\includegraphics[width=0.25\textwidth]{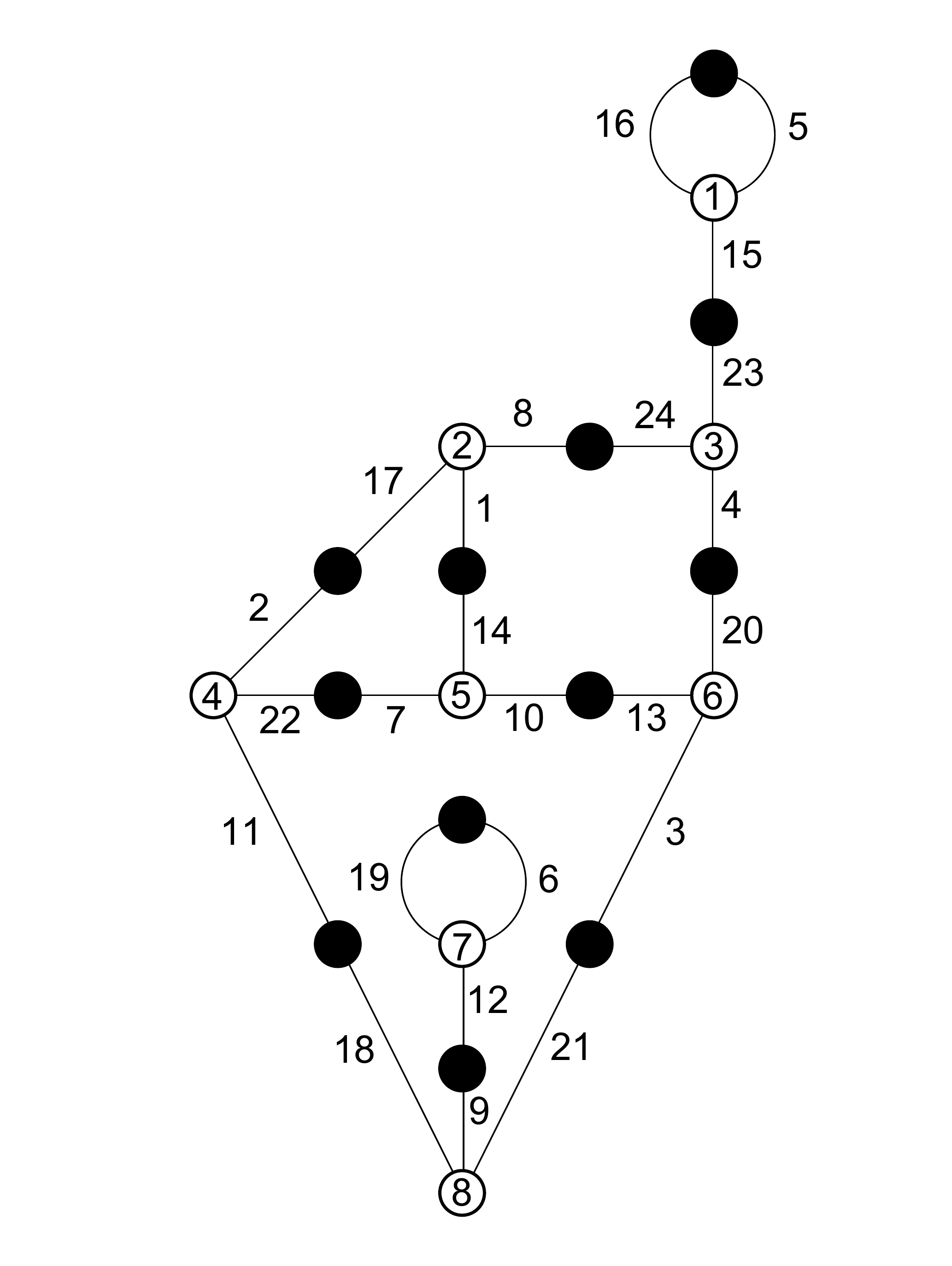}}}$
        $\vcenter{\hbox{\includegraphics[width=0.25\textwidth]{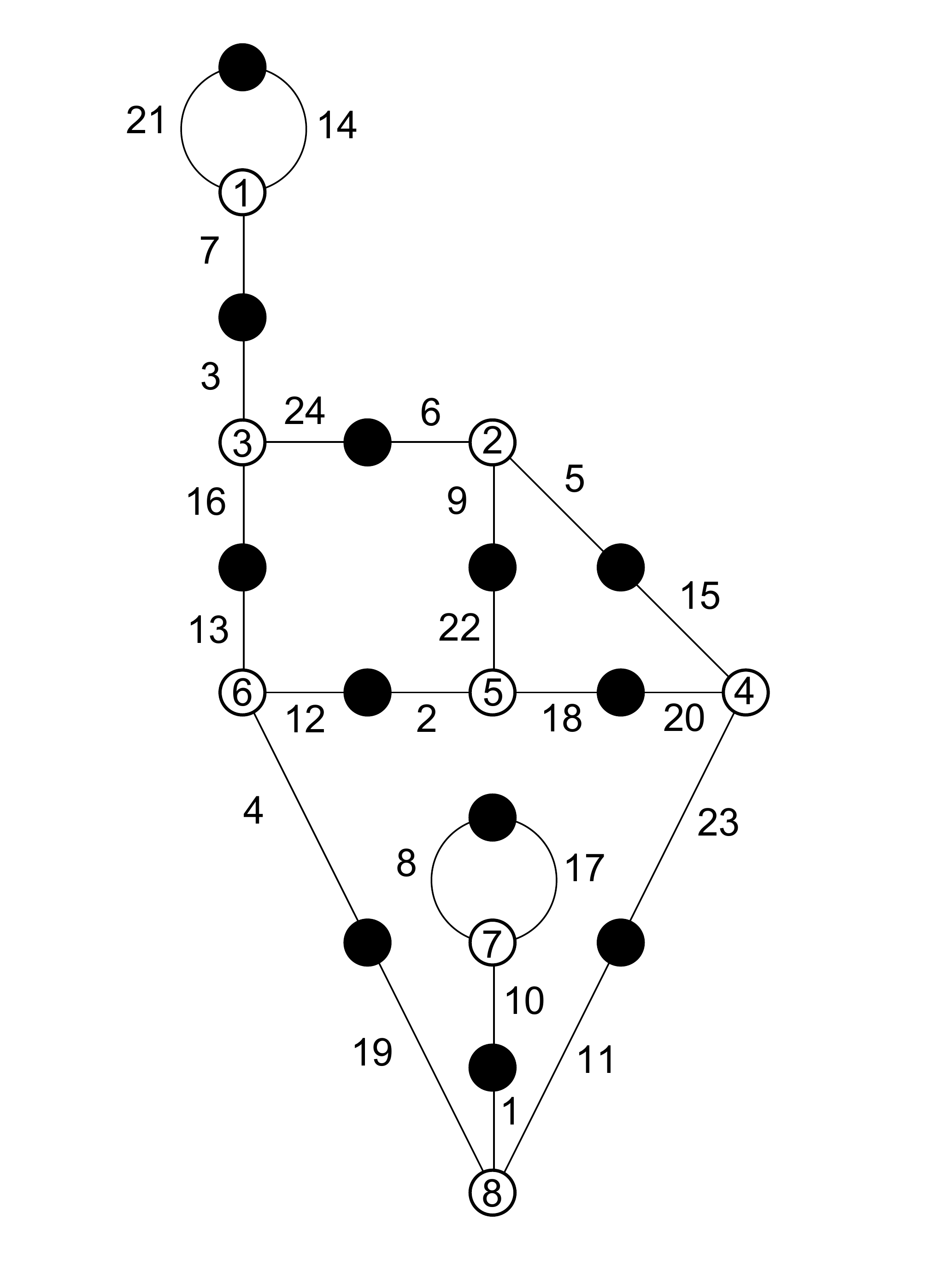}}}$
        \caption{ A: \{\{\{16,5,15\},\{23,4,24\},
        \{20,3,13\},\{10,7,14\},\{8,1,17\},
        \{2,22,11\},\{9,21,18\},\{12,19,6\}\}, \\ 
        \{\{19,6\},\{9,12\},\{18,11\},
        \{3,21\},\{10,13\},\{7,22\},
        \{2,17\},\{8,24\},\{1,14\},
        \{20,4\},\{23,15\},\{5,16\}\}\} \\
        B: \{\{\{21,14,7\},\{3,24,16\},
        \{6,5,9\},\{15,23,20\},\{18,2,22\},
        \{13,12,4\},\{8,17,10\},\{19,1,11\}\}, \\ 
        \{\{11,23\},\{10,1\},\{8,17\},
        \{19,4\},\{12,2\},\{20,18\},
        \{22,9\},\{15,5\},\{6,24\},
        \{13,16\},\{3,7\},\{14,21\}\}\}}
        \caption{8-7-4-3-1-1 A \& B  $(\sqrt{-6})$}
        \label{Dessin}
    \end{subfigure}\hfill
\end{figure}

\begin{figure}[H]
    \begin{subfigure}{0.4\textwidth}
        \centering \captionsetup{justification=centering}
        $\scalemath{0.75}{
        \displaystyle \begin{pmatrix}
            2 & 1 & 0 & 0 & 0 & 0 & 0 & 0\\ 
            1 & 0 & 1 & 0 & 1 & 0 & 0 & 0\\
            0 & 1 & 0 & 0 & 0 & 1 & 1 & 0\\
            0 & 0 & 0 & 2 & 0 & 1 & 0 & 0\\
            0 & 1 & 0 & 0 & 0 & 1 & 0 & 1\\
            0 & 0 & 1 & 1 & 1 & 0 & 0 & 0\\
            0 & 0 & 1 & 0 & 0 & 0 & 0 & 2\\
            0 & 0 & 0 & 0 & 1 & 0 & 2 & 0
        \end{pmatrix}}$
        $\vcenter{\hbox{\includegraphics[width=0.35\textwidth]{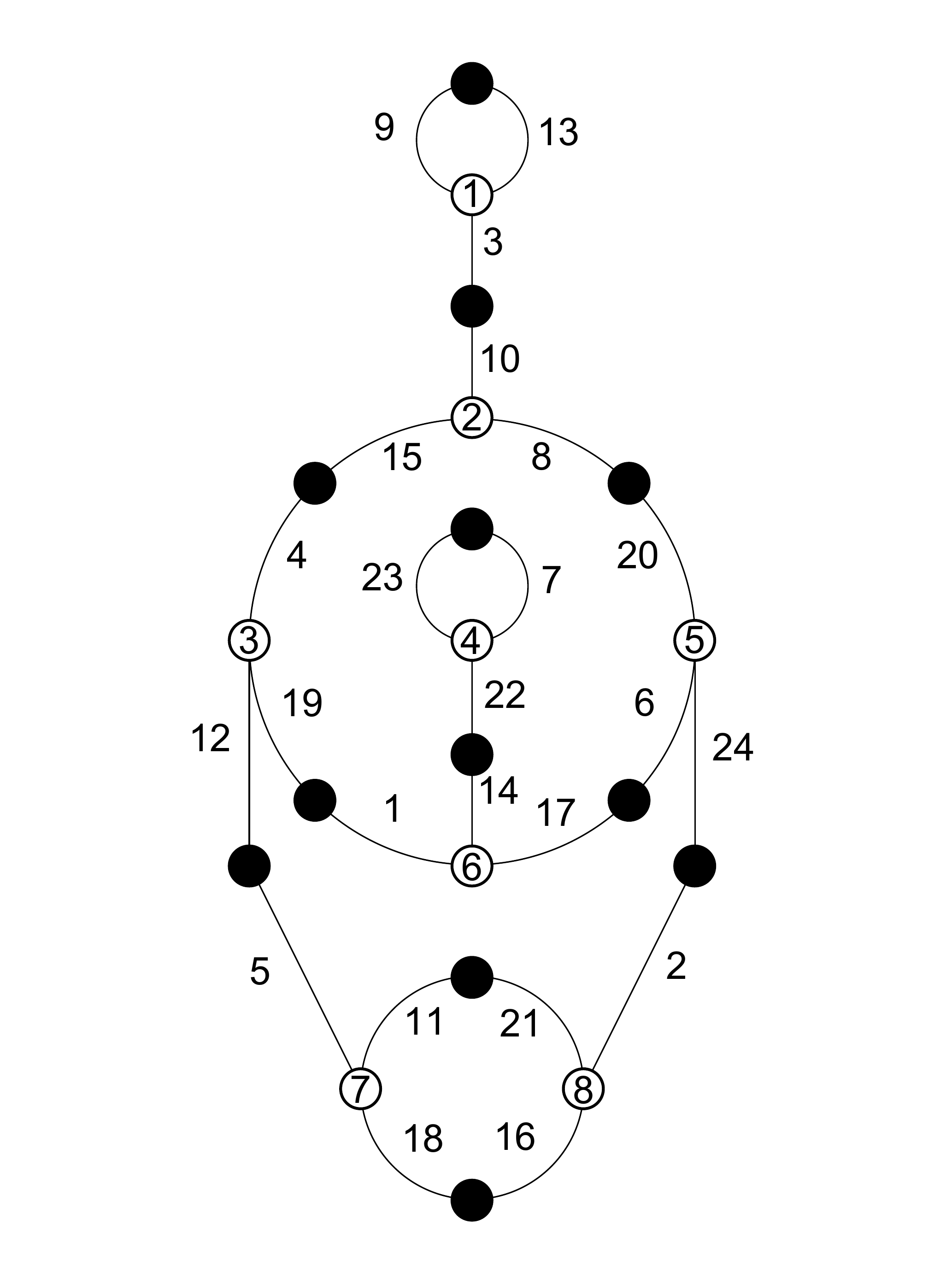}}}$
        \caption{ \{\{\{9,13,3\},\{10,8,15\},
        \{20,24,6\},\{14,17,1\},\{7,22,23\},
        \{19,12,4\},\{5,11,18\},\{21,2,16\}\}, \\ 
        \{\{11,21\},\{16,18\},\{5,12\},
        \{2,24\},\{1,19\},\{17,6\},
        \{14,22\},\{7,23\},\{4,15\},
        \{8,20\},\{10,3\},\{9,13\}\}\}}
        \caption{8-7-5-2-1-1 A (cubic)}
        \label{Dessin}
    \end{subfigure} \hfill
    \begin{subfigure}{0.6\textwidth}
        \centering \captionsetup{justification=centering}
        $\scalemath{0.75}{
        \displaystyle \begin{pmatrix}
            0 & 1 & 2 & 0 & 0 & 0 & 0 & 0\\ 
            1 & 2 & 0 & 0 & 0 & 0 & 0 & 0\\
            2 & 0 & 0 & 1 & 0 & 0 & 0 & 0\\
            0 & 0 & 1 & 0 & 1 & 0 & 0 & 1\\
            0 & 0 & 0 & 1 & 0 & 1 & 1 & 0\\
            0 & 0 & 0 & 0 & 1 & 2 & 0 & 0\\
            0 & 0 & 0 & 0 & 1 & 0 & 0 & 2\\
            0 & 0 & 0 & 1 & 0 & 0 & 2 & 0
        \end{pmatrix}}$
        $\vcenter{\hbox{\includegraphics[width=0.25\textwidth]{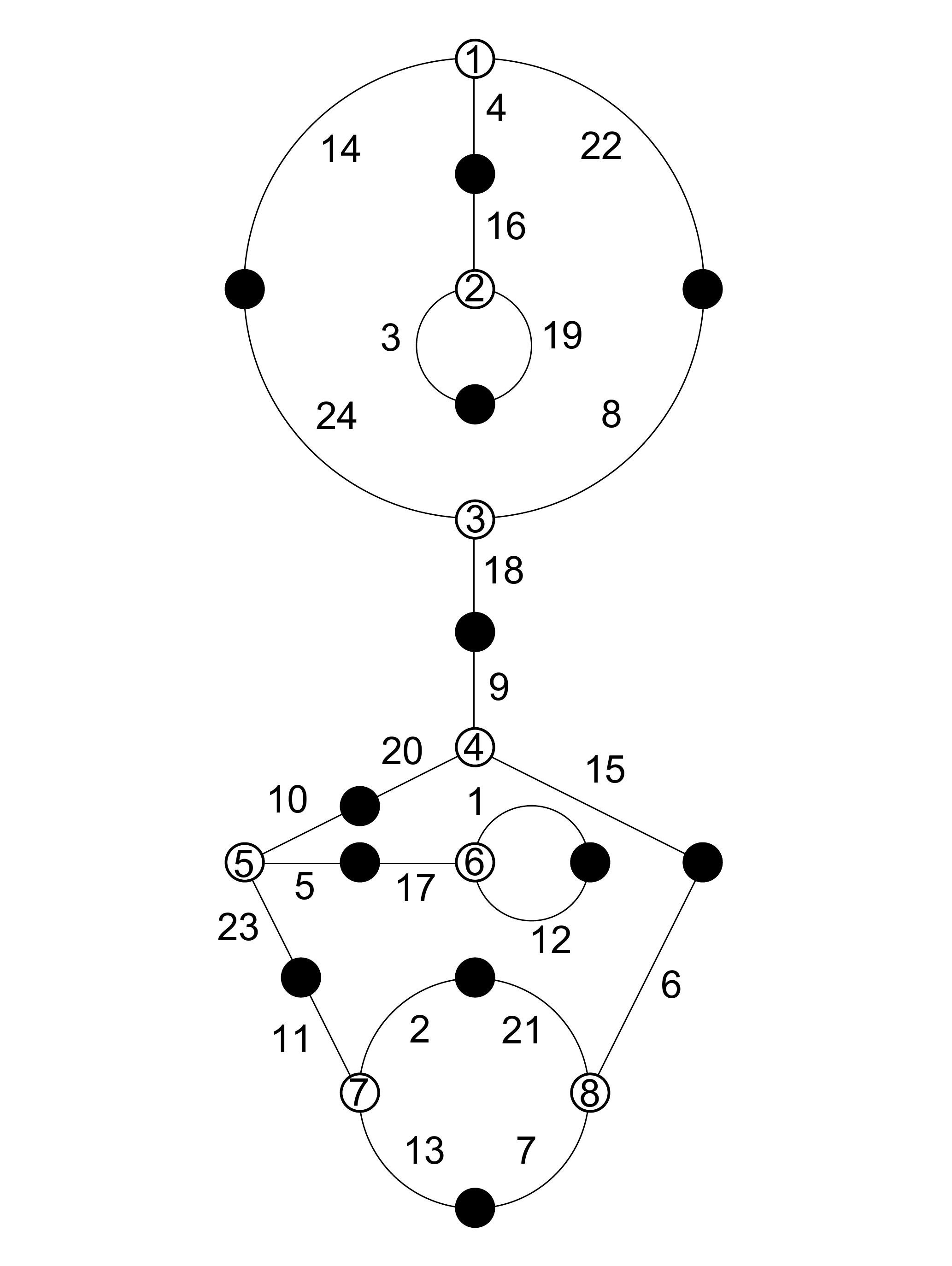}}}$
        $\vcenter{\hbox{\includegraphics[width=0.25\textwidth]{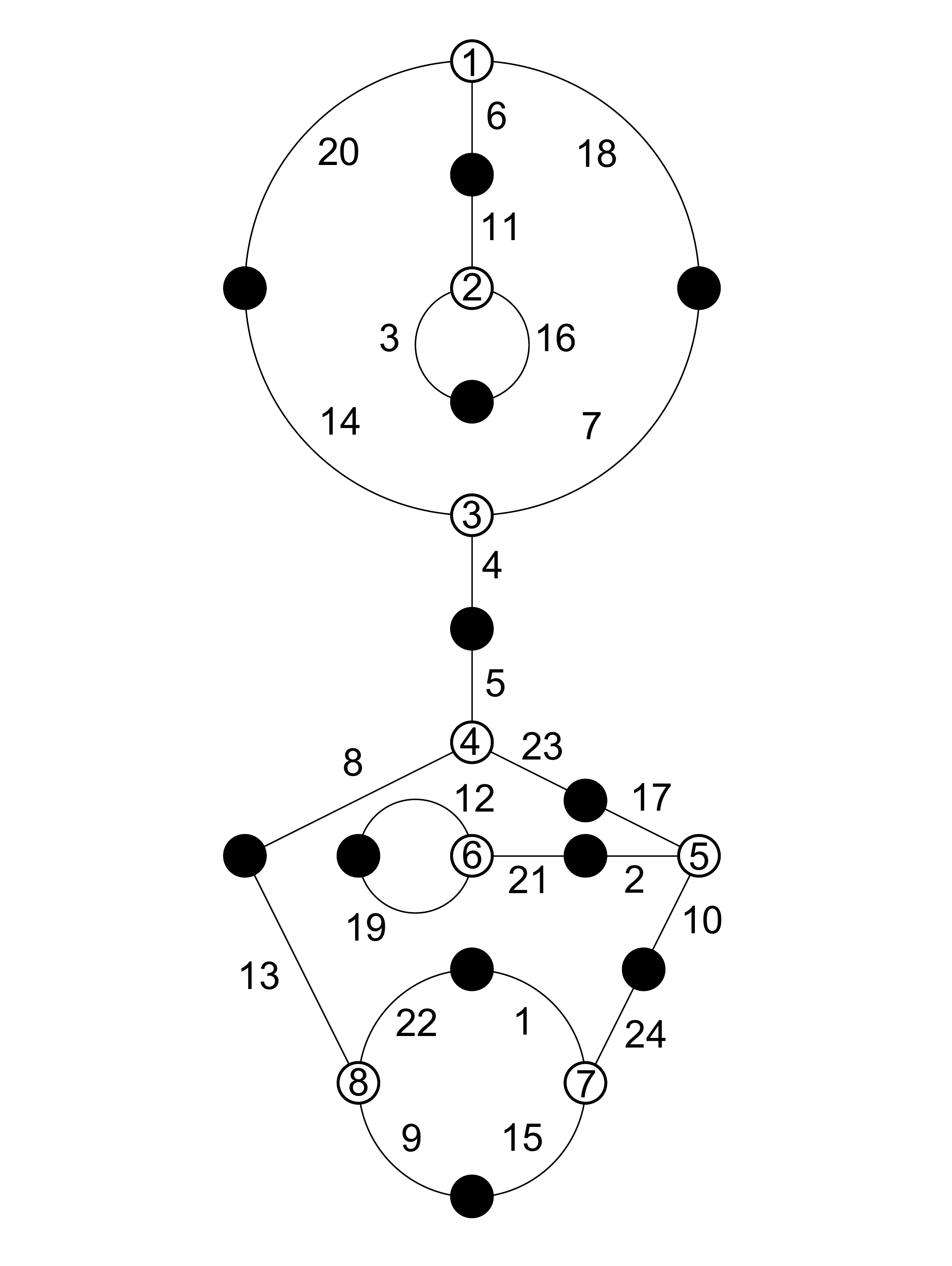}}}$
        \caption{ B: \{\{\{14,22,4\},\{16,19,3\},
        \{8,18,24\},\{9,15,20\},\{10,5,23\},
        \{17,1,12\},\{21,6,7\},\{11,2,13\}\}, \\ 
        \{\{2,21\},\{13,7\},\{11,23\},
        \{6,15\},\{10,20\},\{5,17\},
        \{1,12\},\{9,18\},\{24,14\},
        \{4,16\},\{22,8\},\{3,19\}\}\} \\
        C: \{\{\{6,20,18\},\{11,16,3\},
        \{14,7,4\},\{8,5,23\},\{13,22,9\},
        \{1,24,15\},\{17,10,2\},\{12,21,19\}\}, \\ 
        \{\{13,8\},\{22,1\},\{9,15\},
        \{10,24\},\{2,21\},\{12,19\},
        \{17,23\},\{5,4\},\{14,20\},
        \{6,11\},\{18,7\},\{3,16\}\}\}}
        \caption{8-7-5-2-1-1 B \& C (cubic)}
        \label{Dessin}
    \end{subfigure}\hfill
\end{figure}

\begin{figure}[H]
    \begin{subfigure}{0.6\textwidth}
        \centering \captionsetup{justification=centering}
        $\scalemath{0.75}{
        \displaystyle \begin{pmatrix}
            0 & 1 & 1 & 1 & 0 & 0 & 0 & 0\\ 
            1 & 0 & 0 & 1 & 1 & 0 & 0 & 0\\
            1 & 0 & 2 & 0 & 0 & 0 & 0 & 0\\
            1 & 1 & 0 & 0 & 0 & 0 & 1 & 0\\
            0 & 1 & 0 & 0 & 0 & 1 & 1 & 0\\
            0 & 0 & 0 & 0 & 1 & 2 & 0 & 0\\
            0 & 0 & 0 & 1 & 1 & 0 & 0 & 1\\
            0 & 0 & 0 & 0 & 0 & 0 & 1 & 2
        \end{pmatrix}}$
        $\vcenter{\hbox{\includegraphics[width=0.25\textwidth]{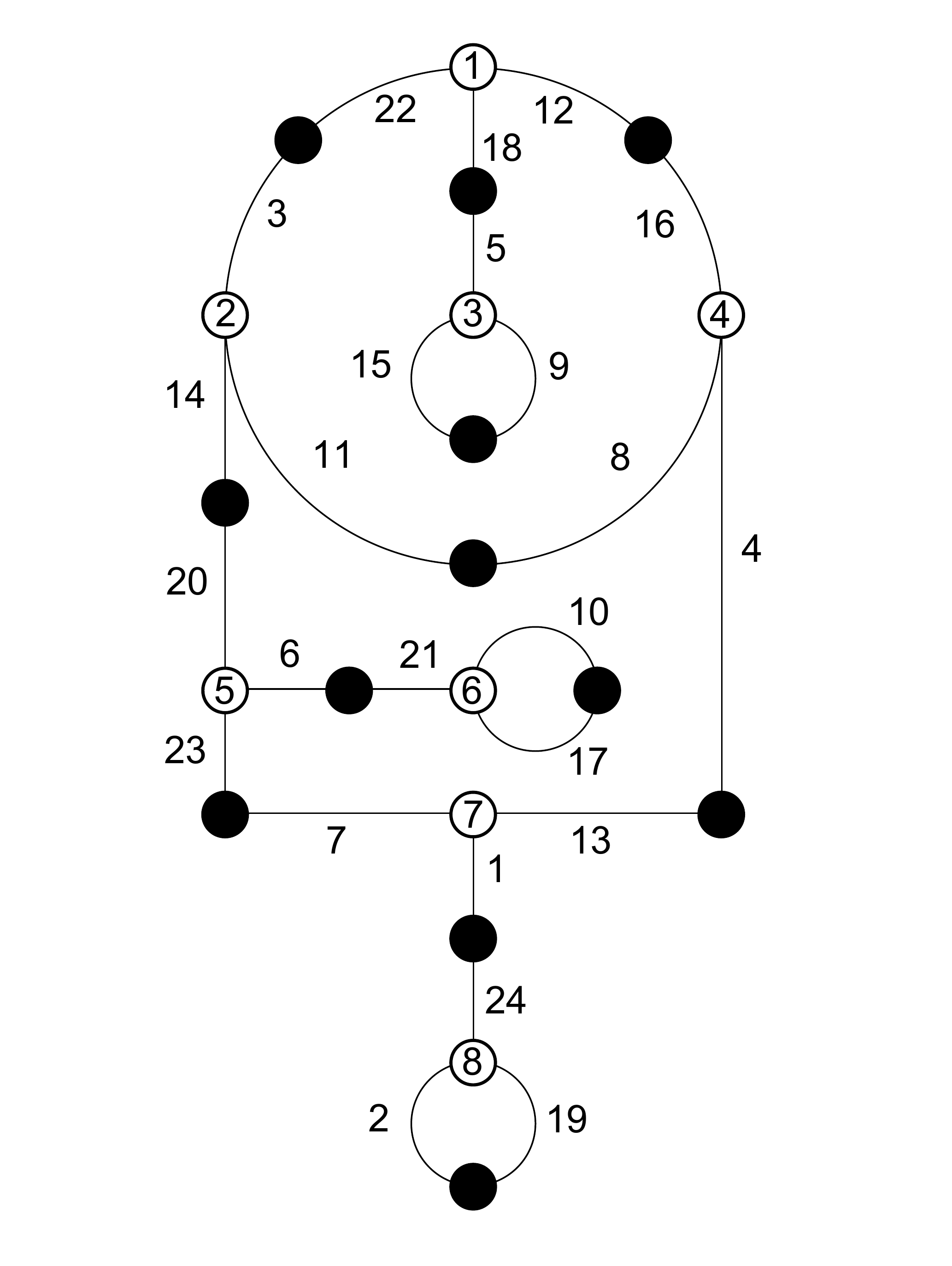}}}$
        $\vcenter{\hbox{\includegraphics[width=0.25\textwidth]{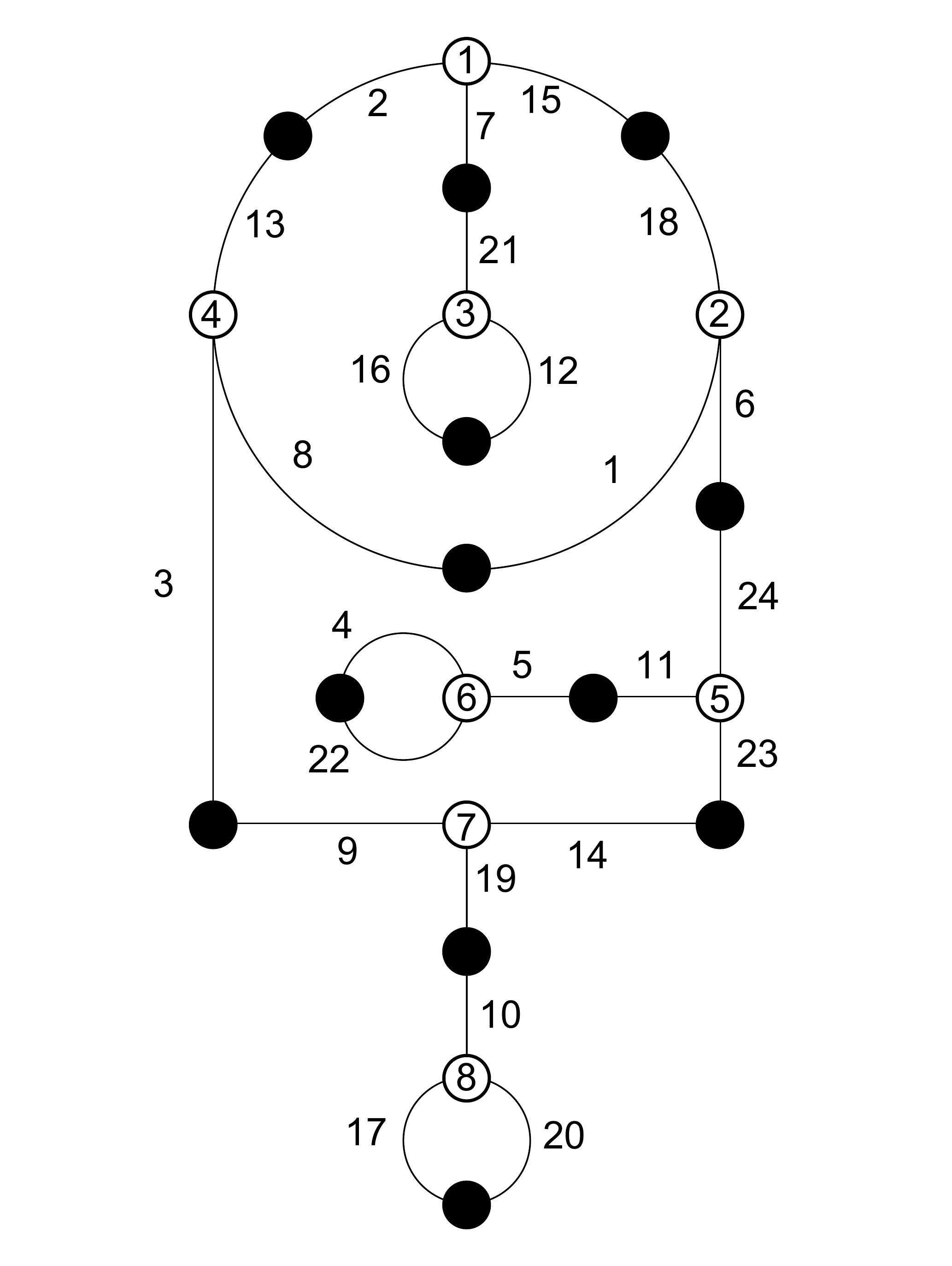}}}$
        \caption{ A: \{\{\{22,12,18\},\{16,4,8\},
        \{15,5,9\},\{14,3,11\},\{20,6,23\},
        \{21,10,17\},\{7,13,1\},\{2,24,19\}\}, \\ 
        \{\{2,19\},\{24,1\},\{7,23\},
        \{13,4\},\{8,11\},\{6,21\},
        \{10,17\},\{20,14\},\{15,9\},
        \{5,18\},\{3,22\},\{12,16\}\}\} \\
        B: \{\{\{2,15,7\},\{12,16,21\},
        \{6,1,18\},\{24,23,11\},\{22,4,5\},
        \{9,14,19\},\{10,20,17\},\{8,3,13\}\}, \\ 
        \{\{20,17\},\{10,19\},\{9,3\},
        \{14,23\},\{5,11\},\{4,22\},
        \{24,6\},\{1,8\},\{16,12\},
        \{21,7\},\{2,13\},\{15,18\}\}\}}
        \caption{8-7-6-1-1-1 A \& B  $(\sqrt{-3})$}
        \label{Dessin}
    \end{subfigure} \hfill
    \begin{subfigure}{0.4\textwidth}
        \centering \captionsetup{justification=centering}
        $\scalemath{0.75}{
        \displaystyle \begin{pmatrix}
            0 & 1 & 2 & 0 & 0 & 0 & 0 & 0\\ 
            1 & 0 & 0 & 2 & 0 & 0 & 0 & 0\\
            2 & 0 & 0 & 0 & 1 & 0 & 0 & 0\\
            0 & 2 & 0 & 0 & 0 & 1 & 0 & 0\\
            0 & 0 & 1 & 0 & 0 & 0 & 2 & 0\\
            0 & 0 & 0 & 1 & 0 & 0 & 0 & 2\\
            0 & 0 & 0 & 0 & 2 & 0 & 0 & 1\\
            0 & 0 & 0 & 0 & 0 & 2 & 1 & 0
        \end{pmatrix}}$
        $\vcenter{\hbox{\includegraphics[width=0.35\textwidth]{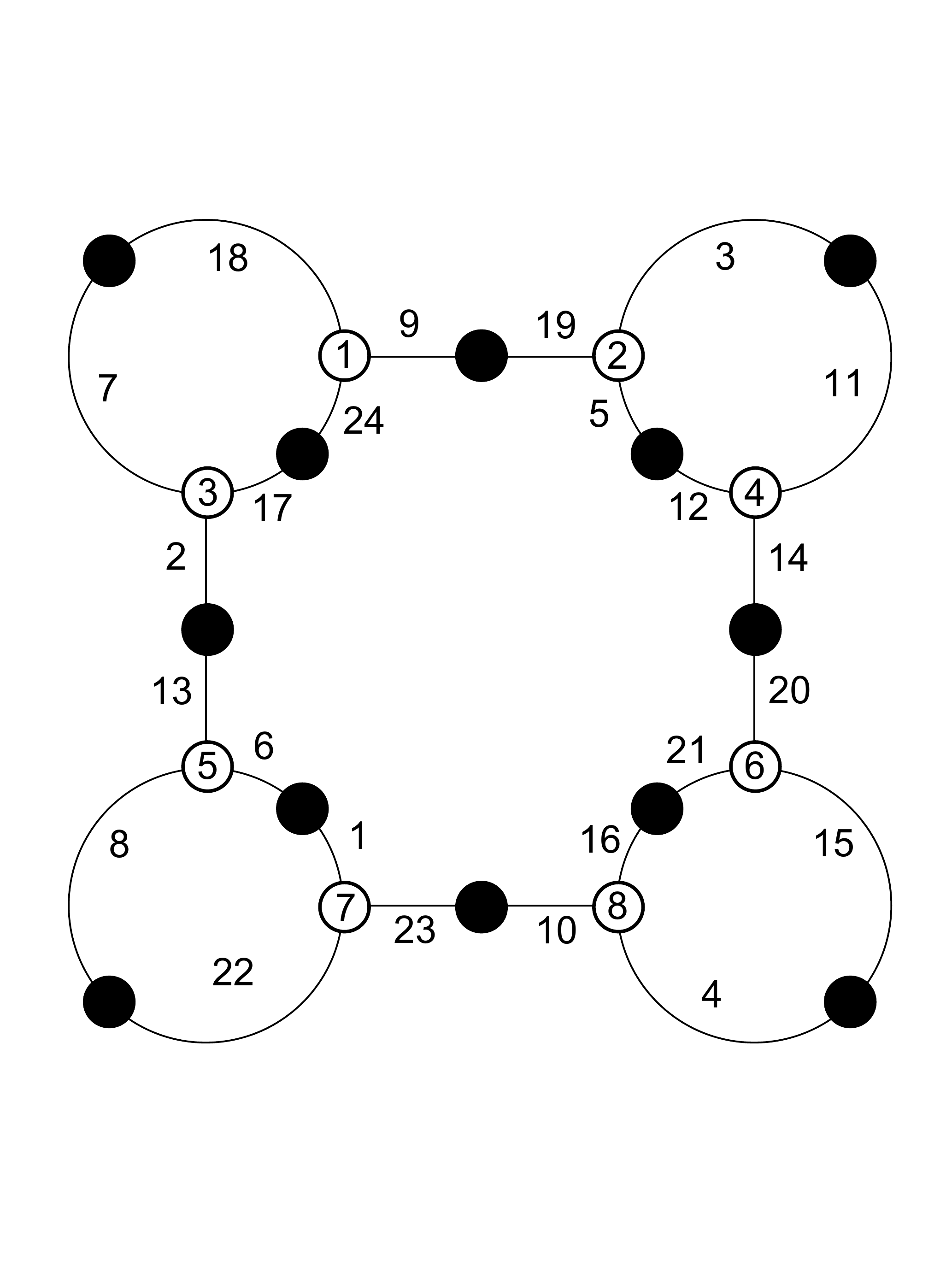}}}$
        \caption{ \{\{\{18,9,24\},\{7,17,2\},
        \{6,8,13\},\{1,23,22\},\{10,16,4\},
        \{21,20,15\},\{12,11,14\},\{19,3,5\}\}, \\ 
        \{\{7,18\},\{9,19\},\{17,24\},
        \{2,13\},\{6,1\},\{8,22\},
        \{23,10\},\{16,21\},\{4,15\},
        \{20,14\},\{5,12\},\{3,11\}\}\}}
        \caption{8-8-2-2-2-2 $(\mathbb{Q})$}
        \label{Dessin}
    \end{subfigure}\hfill
\end{figure}

\begin{figure}[H]
    \begin{subfigure}{0.4\textwidth}
        \centering \captionsetup{justification=centering}
        $\scalemath{0.75}{
        \displaystyle \begin{pmatrix}
            2 & 1 & 0 & 0 & 0 & 0 & 0 & 0\\ 
            1 & 0 & 1 & 1 & 0 & 0 & 0 & 0\\
            0 & 1 & 0 & 1 & 0 & 1 & 0 & 0\\
            0 & 1 & 1 & 0 & 0 & 0 & 0 & 1\\
            0 & 0 & 0 & 0 & 2 & 0 & 1 & 0\\
            0 & 0 & 1 & 0 & 0 & 0 & 1 & 1\\
            0 & 0 & 0 & 0 & 1 & 1 & 0 & 1\\
            0 & 0 & 0 & 1 & 0 & 1 & 1 & 0
        \end{pmatrix}}$
        $\vcenter{\hbox{\includegraphics[width=0.35\textwidth]{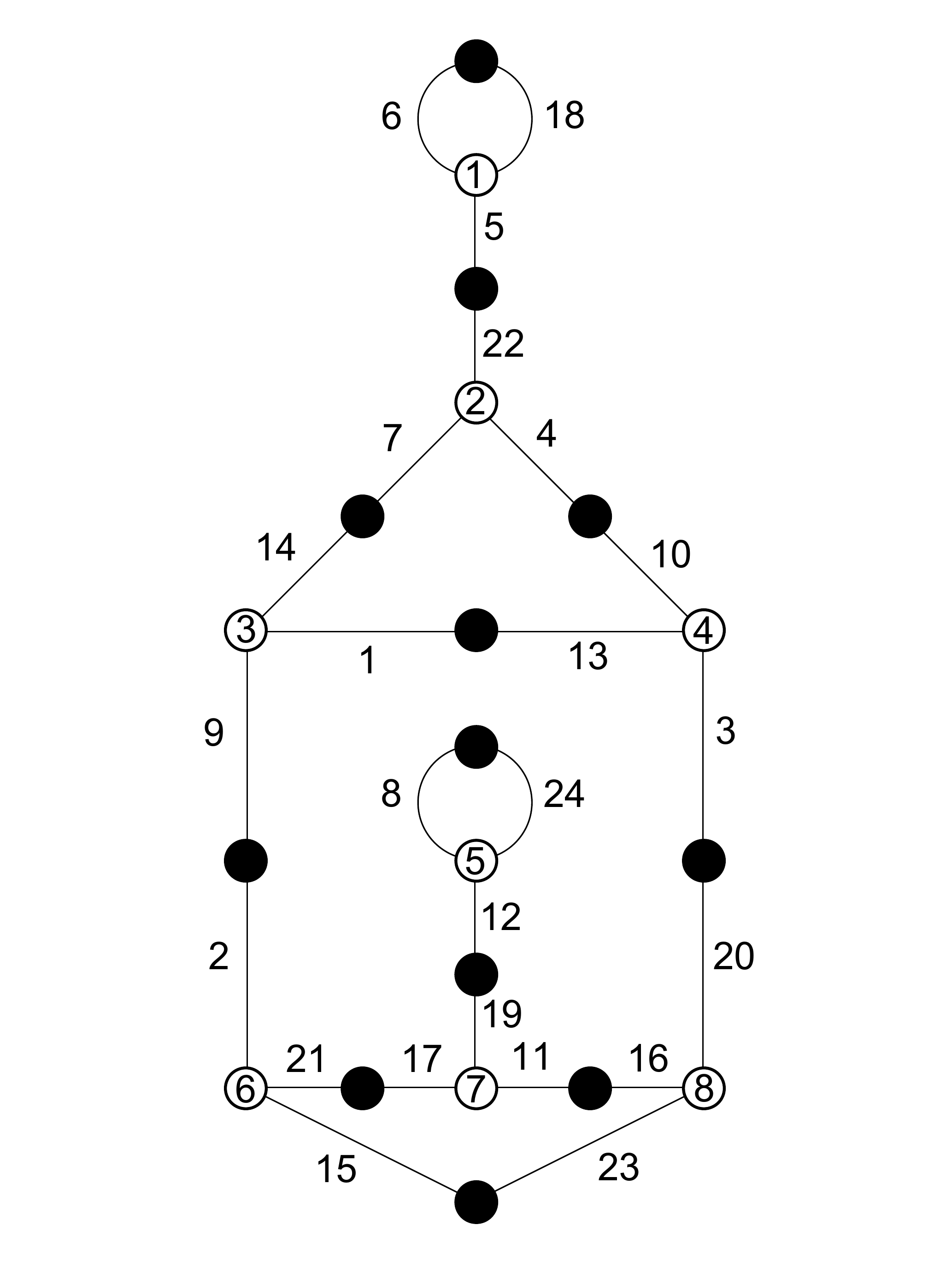}}}$
        \caption{ \{\{\{6,18,5\},\{22,4,7\},
        \{10,3,13\},\{14,1,9\},\{2,21,15\},
        \{17,19,11\},\{16,20,23\},\{12,8,24\}\}, \\ 
        \{\{6,18\},\{5,22\},\{7,14\},
        \{4,10\},\{1,13\},\{9,2\},
        \{8,24\},\{12,19\},\{21,17\},
        \{15,23\},\{11,16\},\{3,20\}\}\}}
        \caption{8-8-3-3-1-1 A  $(\mathbb{Q})$}
        \label{Dessin}
    \end{subfigure} \hfill
    \begin{subfigure}{0.6\textwidth}
        \centering \captionsetup{justification=centering}
        $\scalemath{0.75}{
        \displaystyle \begin{pmatrix}
            0 & 1 & 1 & 1 & 0 & 0 & 0 & 0\\ 
            1 & 0 & 1 & 0 & 1 & 0 & 0 & 0\\
            1 & 1 & 0 & 1 & 0 & 0 & 0 & 0\\
            1 & 0 & 1 & 0 & 0 & 0 & 1 & 0\\
            0 & 1 & 0 & 0 & 0 & 1 & 1 & 0\\
            0 & 0 & 0 & 0 & 1 & 2 & 0 & 0\\
            0 & 0 & 0 & 1 & 1 & 0 & 0 & 1\\
            0 & 0 & 0 & 0 & 0 & 0 & 1 & 2
        \end{pmatrix}}$
        $\vcenter{\hbox{\includegraphics[width=0.25\textwidth]{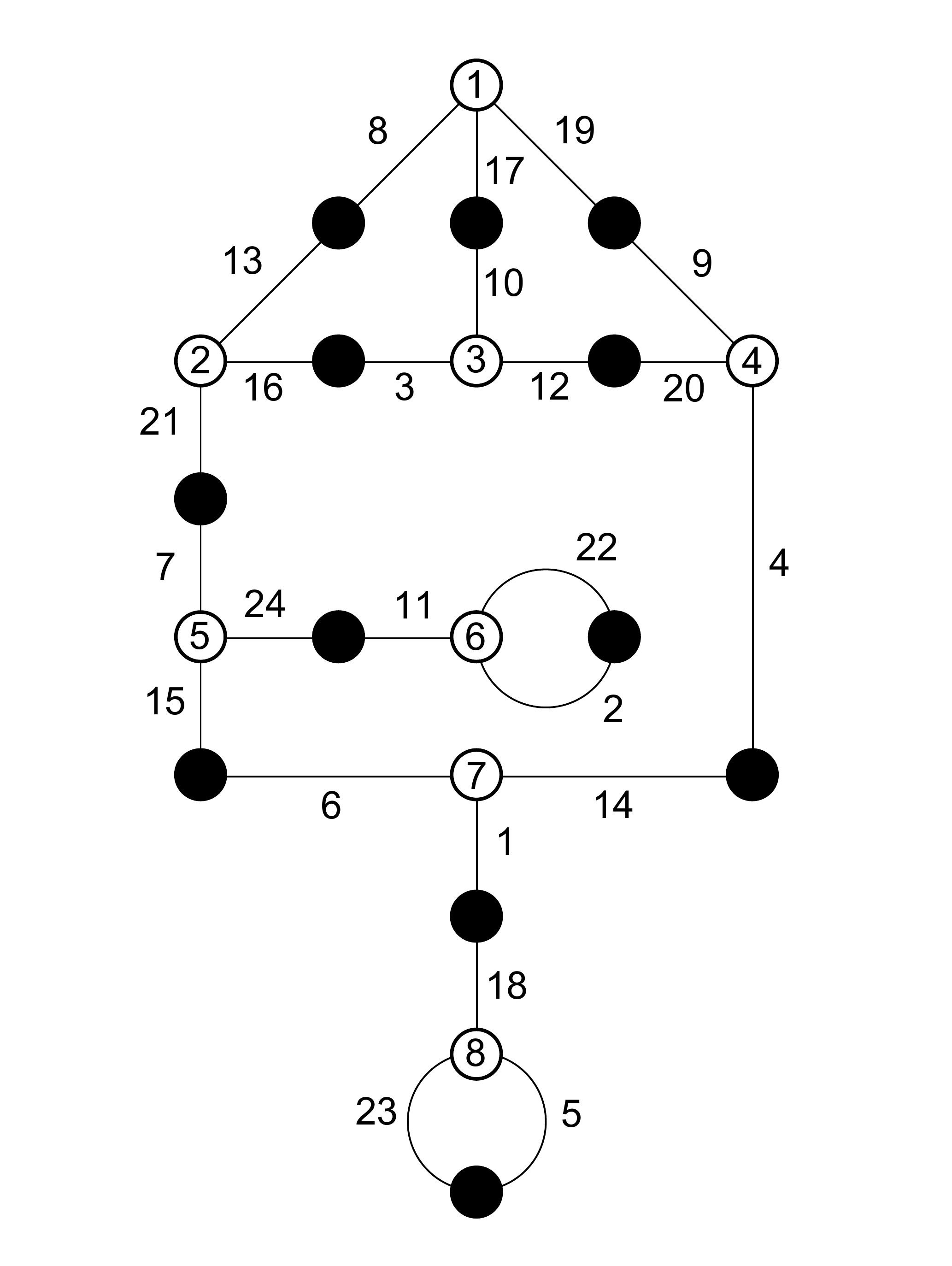}}}$
        $\vcenter{\hbox{\includegraphics[width=0.25\textwidth]{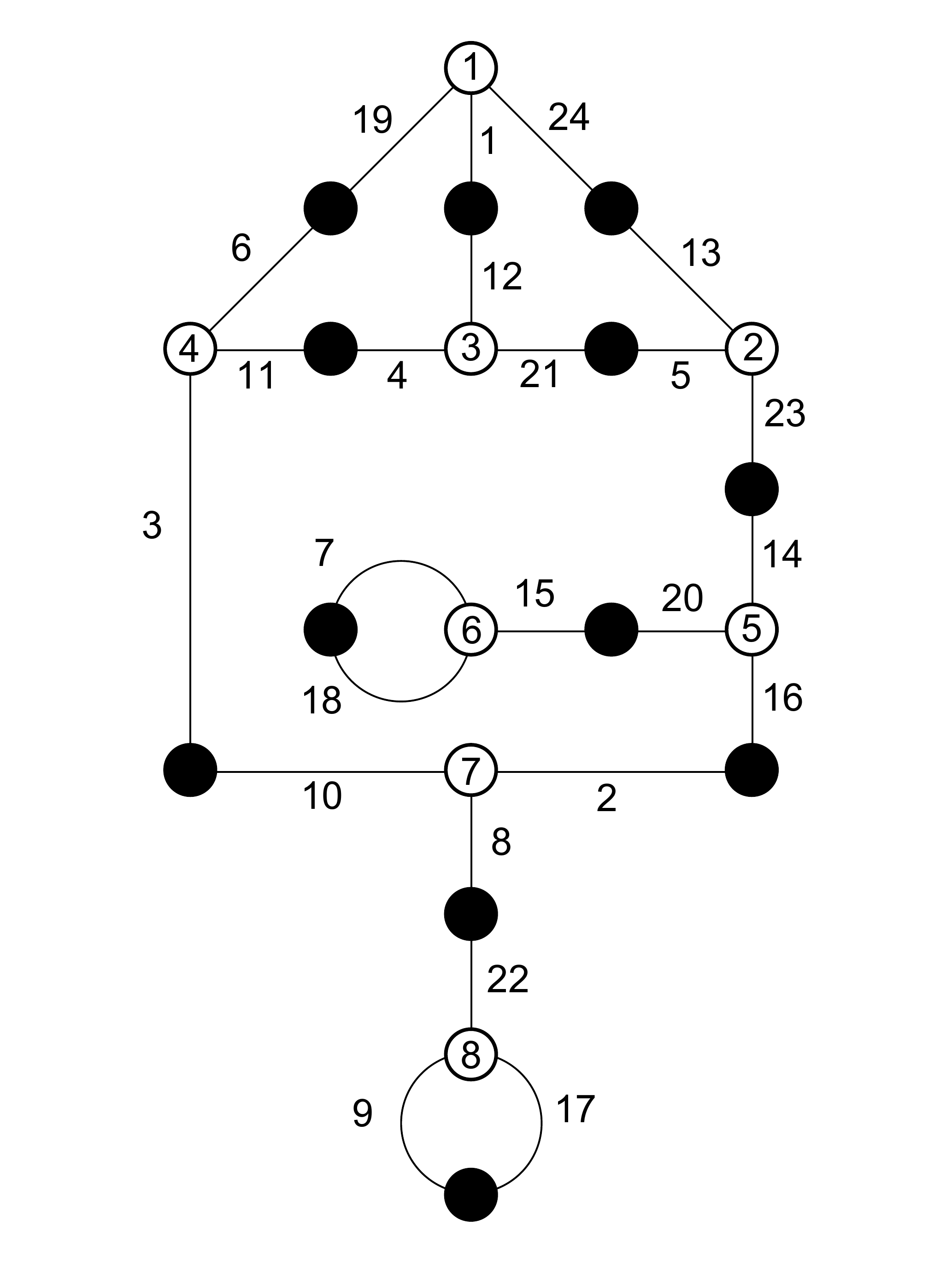}}}$
        \caption{ B: \{\{\{8,19,17\},\{10,12,3\},
        \{13,16,21\},\{9,4,20\},\{22,2,11\},
        \{1,6,14\},\{18,5,23\},\{24,15,7\}\}, \\ 
        \{\{5,23\},\{1,18\},\{6,15\},
        \{14,4\},\{22,2\},\{11,24\},
        \{7,21\},\{16,3\},\{13,8\},
        \{17,10\},\{19,9\},\{12,20\}\}\} \\
        C: \{\{\{1,19,24\},\{6,11,3\},
        \{4,12,21\},\{13,23,5\},\{7,15,18\},
        \{20,14,16\},\{2,8,10\},\{22,17,9\}\}, \\ 
        \{\{22,8\},\{2,16\},\{7,18\},
        \{9,17\},\{3,10\},\{15,20\},
        \{14,23\},\{5,21\},\{24,13\},
        \{1,12\},\{6,19\},\{11,4\}\}\}}
        \caption{8-8-3-3-1-1 B \& C  $(\sqrt{-2})$}
        \label{Dessin}
    \end{subfigure}\hfill
\end{figure}

\begin{figure}[H]
    \begin{subfigure}{0.6\textwidth}
        \centering \captionsetup{justification=centering}
        $\scalemath{0.75}{
        \displaystyle \begin{pmatrix}
            0 & 1 & 1 & 0 & 0 & 1 & 0 & 0\\ 
            1 & 2 & 0 & 0 & 0 & 0 & 0 & 0\\
            1 & 0 & 0 & 1 & 0 & 0 & 0 & 1\\
            0 & 0 & 1 & 0 & 2 & 0 & 0 & 0\\
            0 & 0 & 0 & 2 & 0 & 1 & 0 & 0\\
            1 & 0 & 0 & 0 & 1 & 0 & 0 & 1\\
            0 & 0 & 0 & 0 & 0 & 0 & 2 & 1\\
            0 & 0 & 1 & 0 & 0 & 1 & 1 & 0
        \end{pmatrix}}$
        $\vcenter{\hbox{\includegraphics[width=0.25\textwidth]{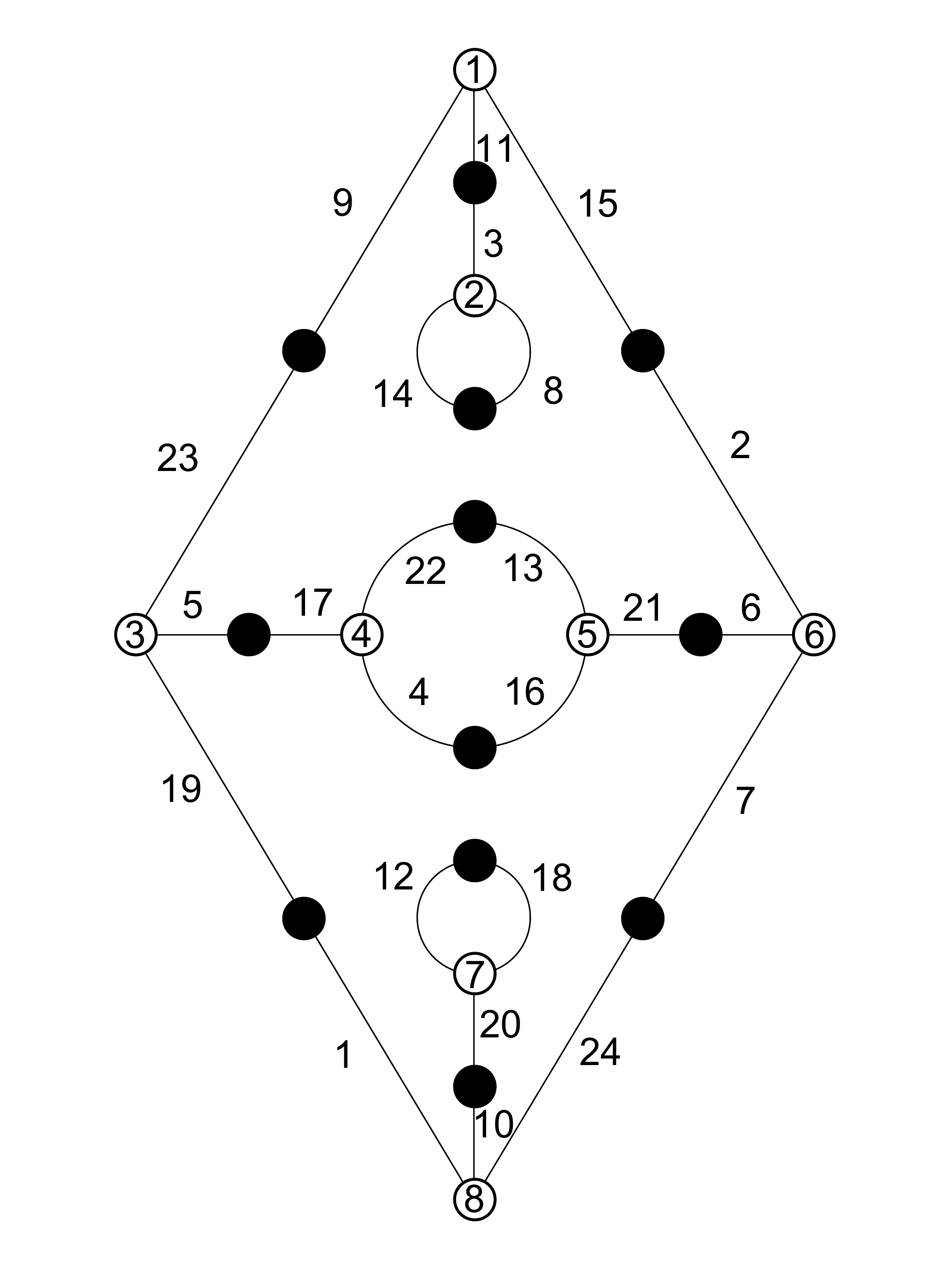}}}$
        \caption{ \{\{\{10,24,1\},\{12,18,20\},
        \{19,23,5\},\{17,22,4\},\{13,21,16\},
        \{6,2,7\},\{14,3,8\},\{9,15,11\}\}, \\ 
        \{\{9,23\},\{11,3\},\{2,15\},
        \{8,14\},\{22,13\},\{5,17\},
        \{6,21\},\{7,24\},\{1,19\},
        \{12,18\},\{10,20\},\{4,16\}\}\}}
        \caption{8-8-4-2-1-1 $(\mathbb{Q})$}
        \label{Dessin}
    \end{subfigure} \hfill
    \begin{subfigure}{0.4\textwidth}
        \centering \captionsetup{justification=centering}
        $\scalemath{0.75}{
        \displaystyle \begin{pmatrix}
            0 & 2 & 1 & 0 & 0 & 0 & 0 & 0\\ 
            2 & 0 & 1 & 0 & 0 & 0 & 0 & 0\\
            1 & 1 & 0 & 1 & 0 & 0 & 0 & 0\\
            0 & 0 & 1 & 0 & 0 & 1 & 1 & 0\\
            0 & 0 & 0 & 0 & 0 & 1 & 1 & 1\\
            0 & 0 & 0 & 1 & 1 & 0 & 0 & 1\\
            0 & 0 & 0 & 1 & 1 & 0 & 0 & 1\\
            0 & 0 & 0 & 0 & 1 & 1 & 1 & 0
        \end{pmatrix}}$
        $\vcenter{\hbox{\includegraphics[width=0.35\textwidth]{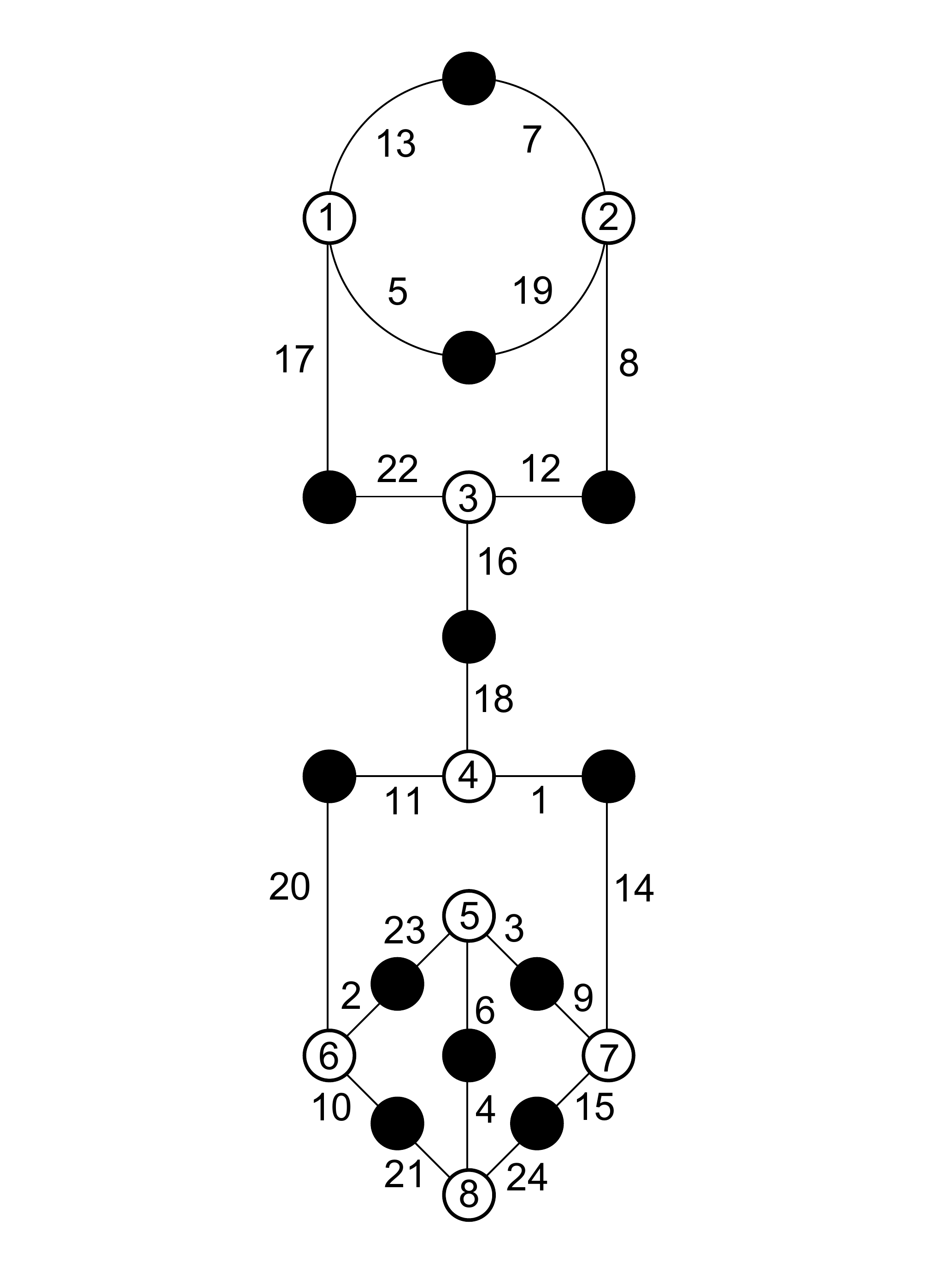}}}$
        \caption{ \{\{\{13,5,17\},\{7,8,19\},
        \{12,16,22\},\{1,11,18\},\{20,2,10\},
        \{23,3,6\},\{9,14,15\},\{4,24,21\}\}, \\ 
        \{\{10,21\},\{4,6\},\{24,15\},
        \{9,3\},\{23,2\},\{20,11\},
        \{1,14\},\{16,18\},\{22,17\},
        \{12,8\},\{5,19\},\{13,7\}\}\}}
        \caption{9-4-3-3-3-2 $(\mathbb{Q})$}
        \label{Dessin}
    \end{subfigure}\hfill
\end{figure}

\begin{figure}[H]
    \begin{subfigure}{0.5\textwidth}
        \centering \captionsetup{justification=centering}
        $\scalemath{0.75}{
        \displaystyle \begin{pmatrix}
            2 & 1 & 0 & 0 & 0 & 0 & 0 & 0\\ 
            1 & 0 & 1 & 1 & 0 & 0 & 0 & 0\\
            0 & 1 & 0 & 1 & 0 & 1 & 0 & 0\\
            0 & 1 & 1 & 0 & 0 & 0 & 1 & 0\\
            0 & 0 & 0 & 0 & 0 & 1 & 1 & 1\\
            0 & 0 & 1 & 0 & 1 & 0 & 0 & 1\\
            0 & 0 & 0 & 1 & 1 & 0 & 0 & 1\\
            0 & 0 & 0 & 0 & 1 & 1 & 1 & 0
        \end{pmatrix}}$
        $\vcenter{\hbox{\includegraphics[width=0.35\textwidth]{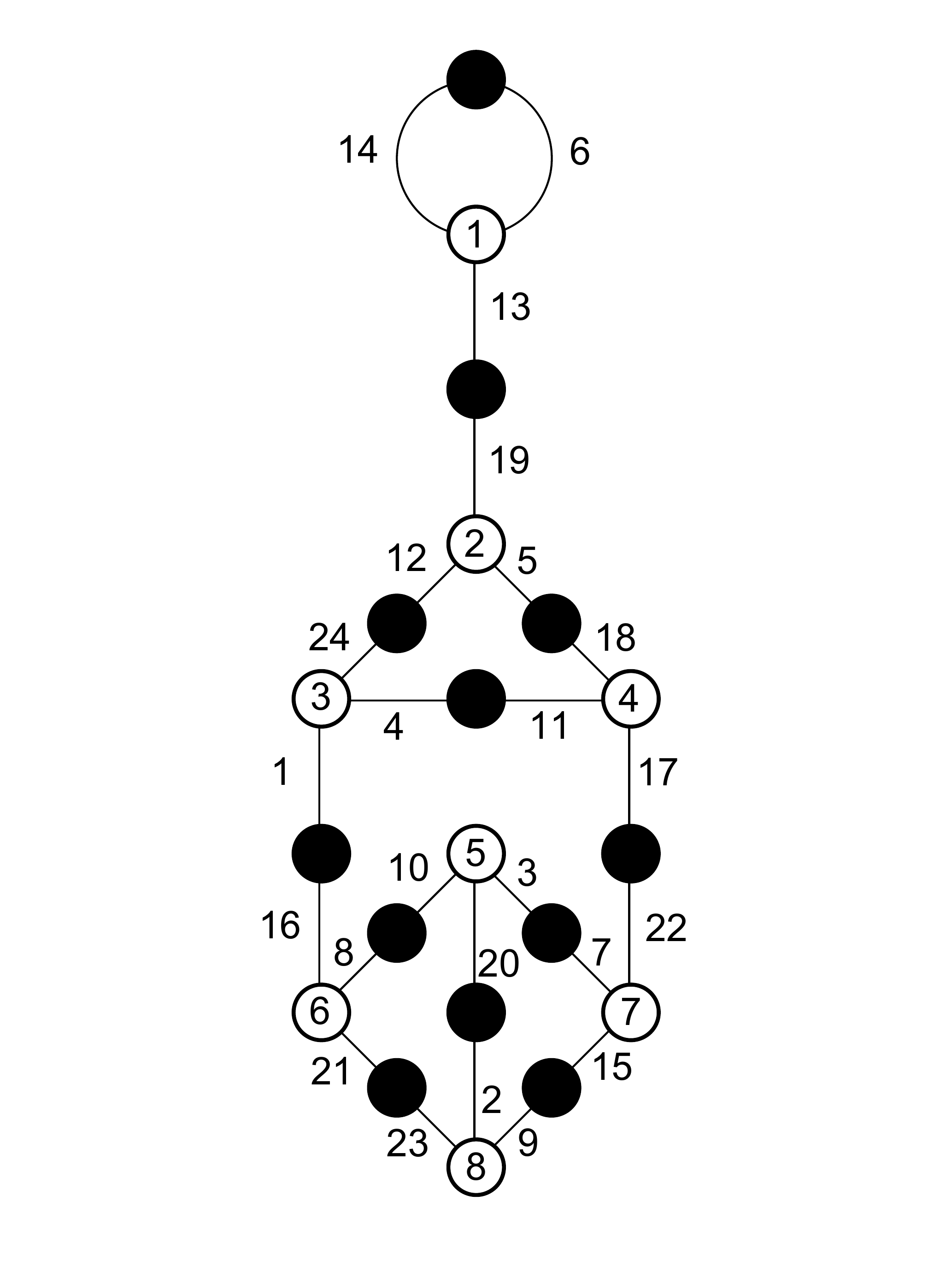}}}$
        \caption{ \{\{\{14,6,13\},\{19,5,12\},
        \{18,17,11\},\{4,1,24\},\{10,3,20\},
        \{22,15,7\},\{8,21,16\},\{2,9,23\}\}, \\ 
        \{\{21,23\},\{2,20\},\{10,8\},
        \{3,7\},\{9,15\},\{22,17\},
        \{1,16\},\{4,11\},\{24,12\},
        \{19,13\},\{14,6\},\{5,18\}\}\}}
        \caption{9-5-3-3-3-1 $(\mathbb{Q})$}
        \label{Dessin}
    \end{subfigure} \hfill
    \begin{subfigure}{0.5\textwidth}
        \centering \captionsetup{justification=centering}
        $\scalemath{0.75}{
        \displaystyle \begin{pmatrix}
            0 & 2 & 1 & 0 & 0 & 0 & 0 & 0\\ 
            2 & 0 & 0 & 1 & 0 & 0 & 0 & 0\\
            1 & 0 & 0 & 1 & 1 & 0 & 0 & 0\\
            0 & 1 & 1 & 0 & 1 & 0 & 0 & 0\\
            0 & 0 & 1 & 1 & 0 & 1 & 0 & 0\\
            0 & 0 & 0 & 0 & 1 & 0 & 0 & 2\\
            0 & 0 & 0 & 0 & 0 & 0 & 2 & 1\\
            0 & 0 & 0 & 0 & 0 & 2 & 1 & 0
        \end{pmatrix}}$
        $\vcenter{\hbox{\includegraphics[width=0.35\textwidth]{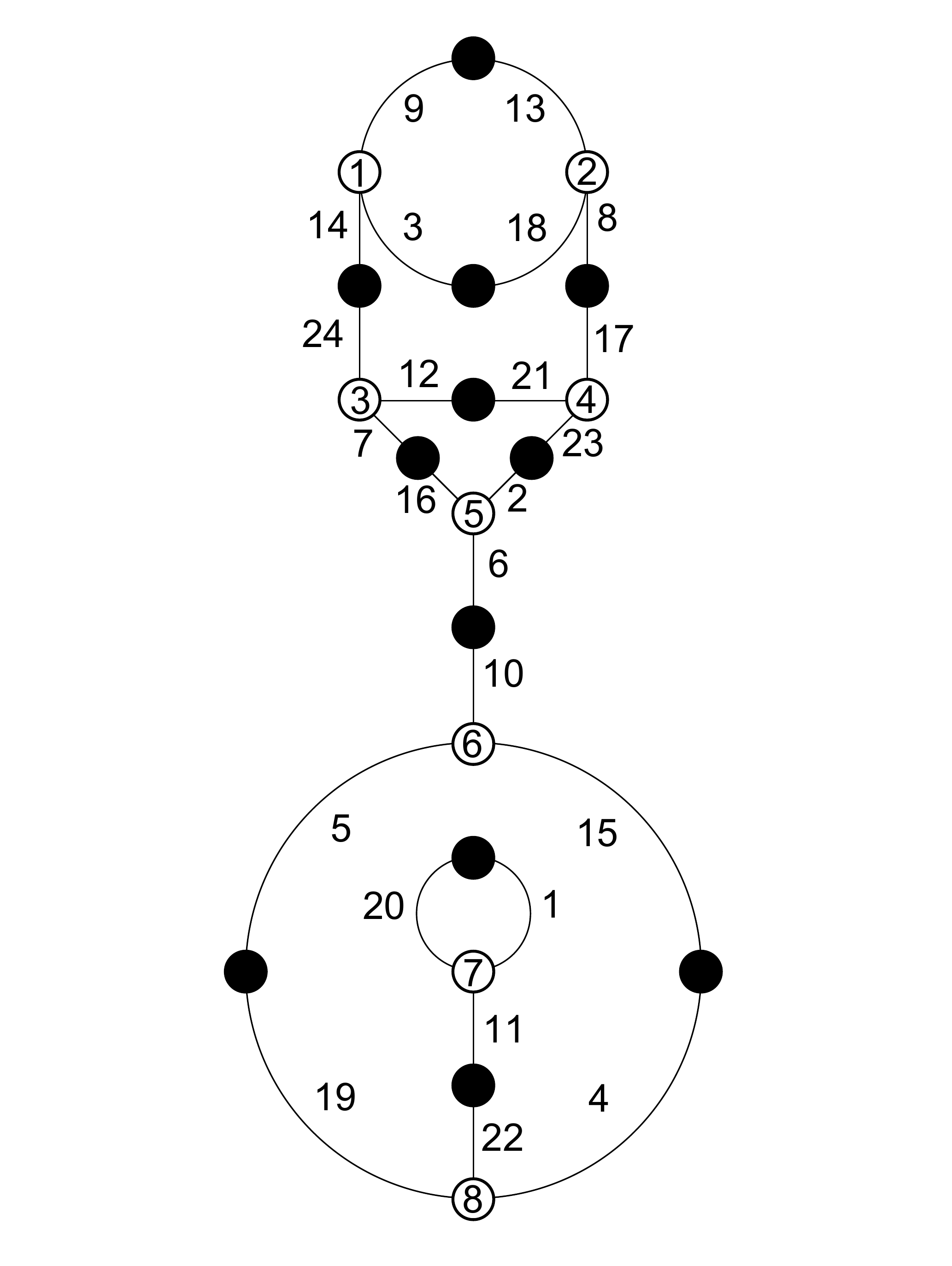}}}$
        \caption{ \{\{\{9,3,14\},\{13,8,18\},
        \{24,12,7\},\{21,17,23\},\{16,2,6\},
        \{10,15,5\},\{20,1,11\},\{22,4,19\}\}, \\ 
        \{\{20,1\},\{5,19\},\{11,22\},
        \{4,15\},\{10,6\},\{2,23\},
        \{7,16\},\{12,21\},\{14,24\},
        \{3,18\},\{8,17\},\{9,13\}\}\}}
        \caption{9-5-4-3-2-1 A (cubic)}
        \label{Dessin}
    \end{subfigure}\hfill
\end{figure}

\begin{figure}[H]
    \begin{subfigure}{0.6\textwidth}
        \centering \captionsetup{justification=centering}
        $\scalemath{0.75}{
        \displaystyle \begin{pmatrix}
            0 & 2 & 1 & 0 & 0 & 0 & 0 & 0\\ 
            2 & 0 & 0 & 0 & 1 & 0 & 0 & 0\\
            1 & 0 & 0 & 1 & 0 & 1 & 0 & 0\\
            0 & 0 & 1 & 0 & 1 & 1 & 0 & 0\\
            0 & 1 & 0 & 1 & 0 & 0 & 1 & 0\\
            0 & 0 & 1 & 1 & 0 & 0 & 1 & 0\\
            0 & 0 & 0 & 0 & 1 & 1 & 0 & 1\\
            0 & 0 & 0 & 0 & 0 & 0 & 1 & 2
        \end{pmatrix}}$
        $\vcenter{\hbox{\includegraphics[width=0.25\textwidth]{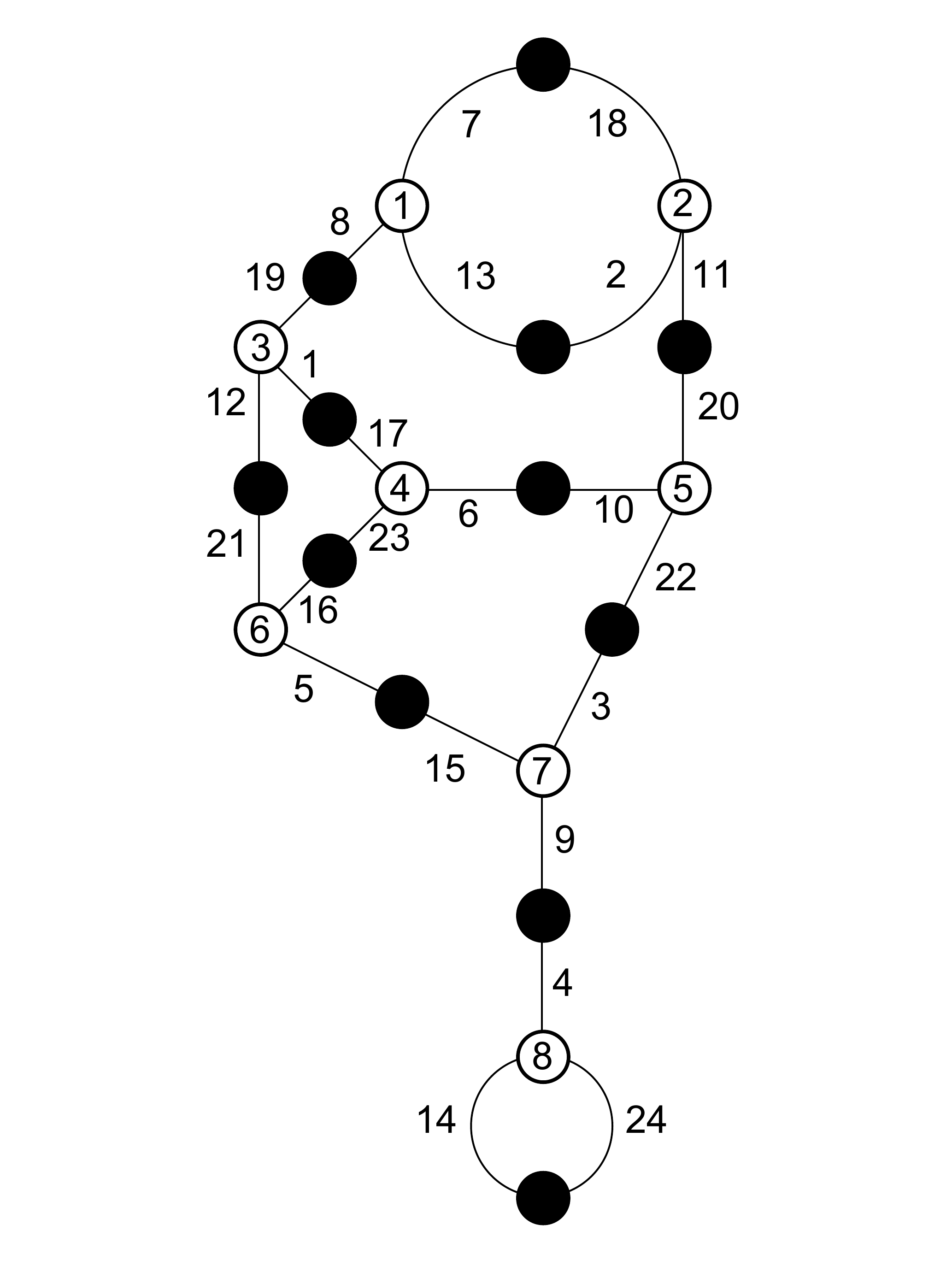}}}$
        $\vcenter{\hbox{\includegraphics[width=0.25\textwidth]{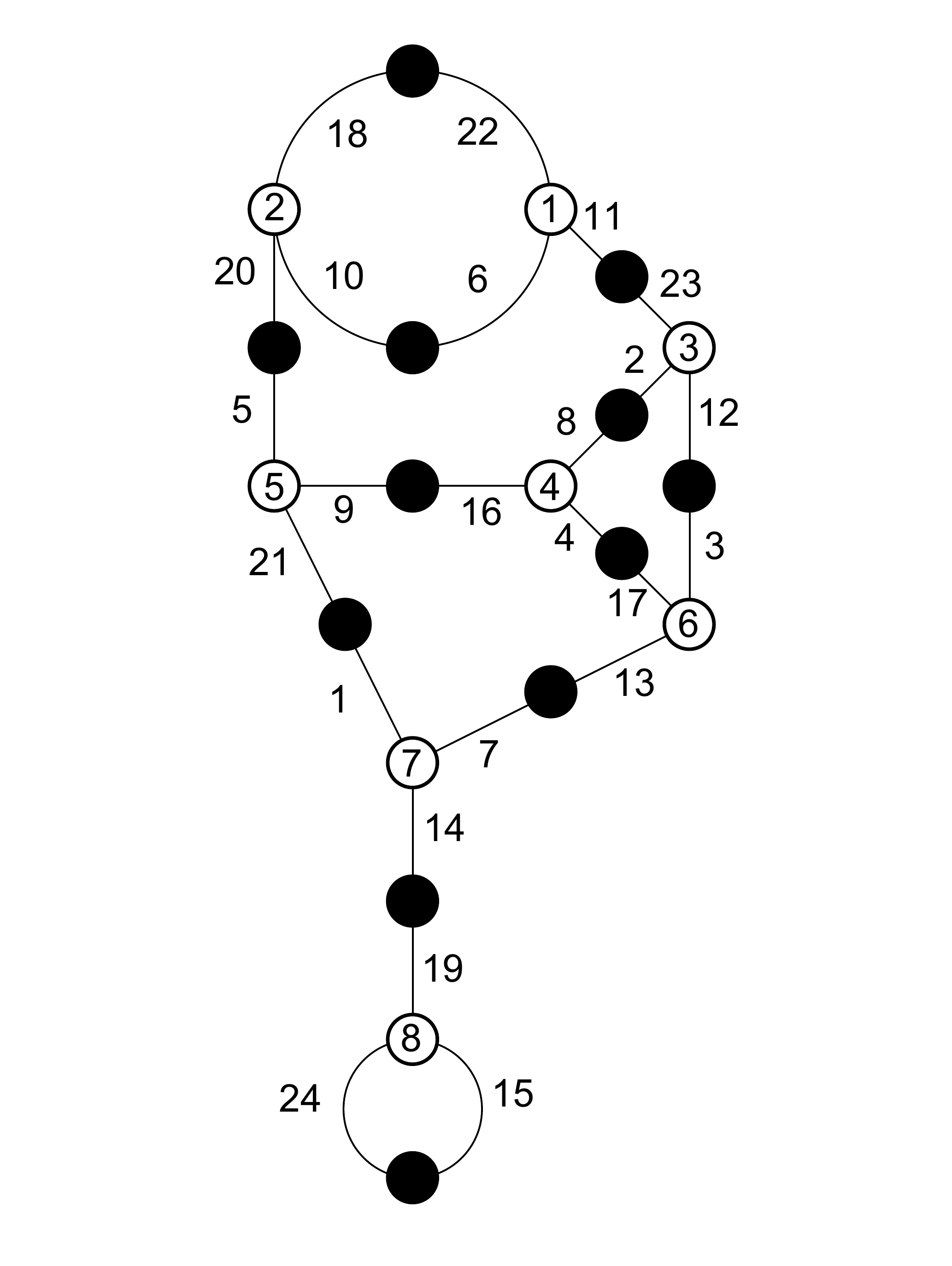}}}$
        \caption{ B: \{\{\{7,13,8\},\{18,11,2\},
        \{20,22,10\},\{6,23,17\},\{21,16,5\},
        \{1,12,19\},\{3,9,15\},\{4,24,14\}\}, \\ 
        \{\{14,24\},\{4,9\},\{15,5\},
        \{3,22\},\{6,10\},\{23,16\},
        \{1,17\},\{12,21\},\{19,8\},
        \{7,18\},\{13,2\},\{11,20\}\}\} \\
        C: \{\{\{15,24,19\},\{14,1,7\},
        \{21,5,9\},\{16,8,4\},\{17,3,13\},
        \{12,2,23\},\{11,6,22\},\{10,20,18\}\}, \\ 
        \{\{18,22\},\{11,23\},\{6,10\},
        \{20,5\},\{9,16\},\{21,1\},
        \{4,17\},\{8,2\},\{12,3\},
        \{13,7\},\{14,19\},\{24,15\}\}\}}
        \caption{9-5-4-3-2-1 B \& C (cubic)}
        \label{Dessin}
    \end{subfigure} \hfill
    \begin{subfigure}{0.4\textwidth}
        \centering \captionsetup{justification=centering}
        $\scalemath{0.75}{
        \displaystyle \begin{pmatrix}
            0 & 1 & 2 & 0 & 0 & 0 & 0 & 0\\ 
            1 & 2 & 0 & 0 & 0 & 0 & 0 & 0\\
            2 & 0 & 0 & 1 & 0 & 0 & 0 & 0\\
            0 & 0 & 1 & 0 & 1 & 0 & 1 & 0\\
            0 & 0 & 0 & 1 & 0 & 2 & 0 & 0\\
            0 & 0 & 0 & 0 & 2 & 0 & 0 & 1\\
            0 & 0 & 0 & 1 & 0 & 0 & 0 & 2\\
            0 & 0 & 0 & 0 & 0 & 1 & 2 & 0
        \end{pmatrix}}$
        $\vcenter{\hbox{\includegraphics[width=0.35\textwidth]{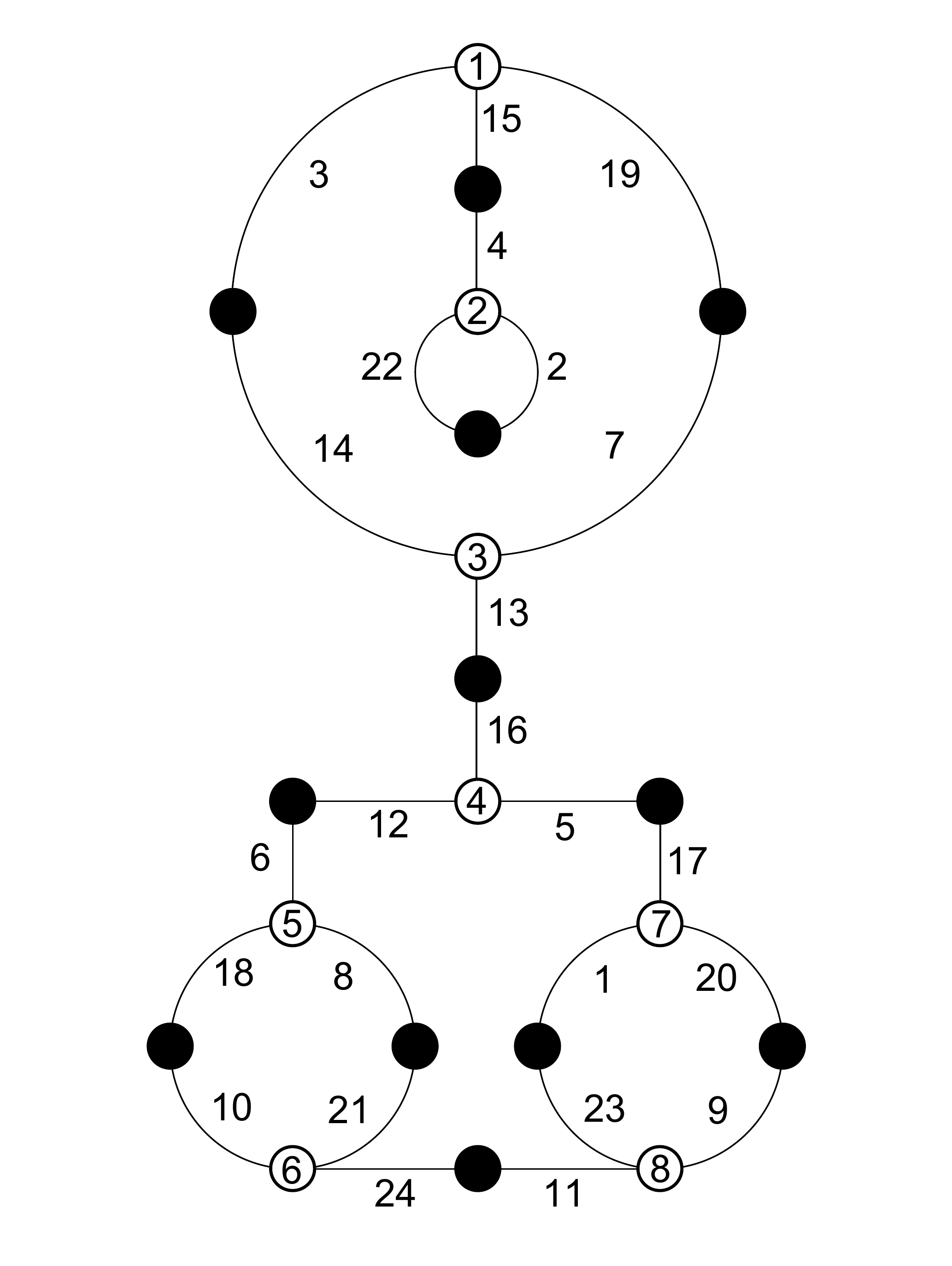}}}$
        \caption{ \{\{\{3,19,15\},\{4,2,22\},
        \{14,7,13\},\{16,5,12\},\{17,20,1\},
        \{6,8,18\},\{21,24,10\},\{23,9,11\}\}, \\ 
        \{\{24,11\},\{10,18\},\{8,21\},
        \{1,23\},\{20,9\},\{17,5\},
        \{12,6\},\{16,13\},\{14,3\},
        \{19,7\},\{15,4\},\{22,2\}\}\}}
        \caption{9-5-5-2-2-1 $(\mathbb{Q})$}
        \label{Dessin}
    \end{subfigure}\hfill
\end{figure}

\begin{figure}[H]
    \begin{subfigure}{0.4\textwidth}
        \centering \captionsetup{justification=centering}
        $\scalemath{0.75}{
        \displaystyle \begin{pmatrix}
            0 & 2 & 1 & 0 & 0 & 0 & 0 & 0\\ 
            2 & 0 & 1 & 0 & 0 & 0 & 0 & 0\\
            1 & 1 & 0 & 1 & 0 & 0 & 0 & 0\\
            0 & 0 & 1 & 0 & 1 & 0 & 1 & 0\\
            0 & 0 & 0 & 1 & 0 & 0 & 1 & 1\\
            0 & 0 & 0 & 0 & 0 & 2 & 0 & 1\\
            0 & 0 & 0 & 1 & 1 & 0 & 0 & 1\\
            0 & 0 & 0 & 0 & 1 & 1 & 1 & 0
        \end{pmatrix}}$
        $\vcenter{\hbox{\includegraphics[width=0.35\textwidth]{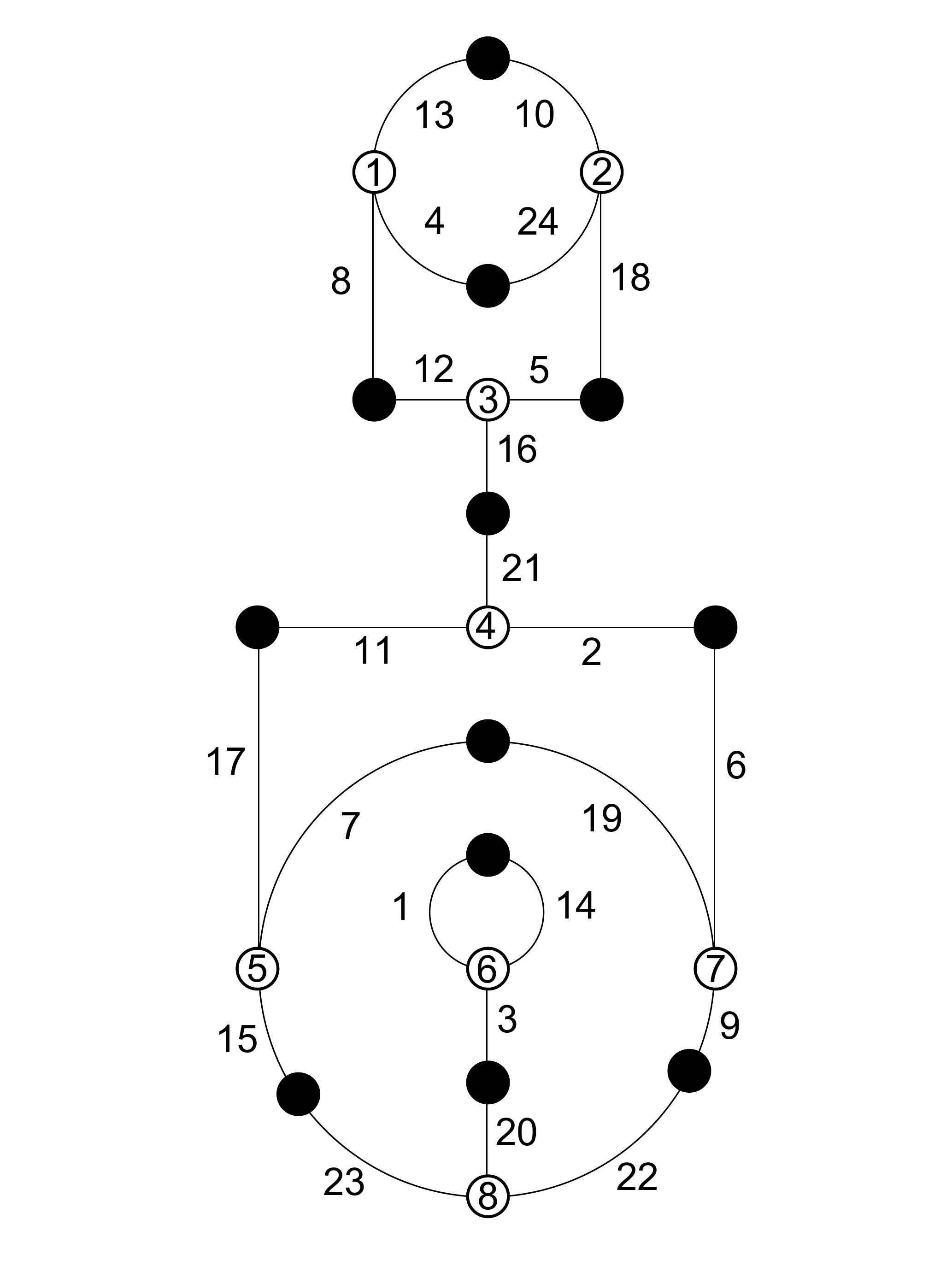}}}$
        \caption{ \{\{\{13,4,8\},\{10,18,24\},
        \{5,16,12\},\{21,2,11\},\{6,9,19\},
        \{7,15,17\},\{1,14,3\},\{20,22,23\}\}, \\ 
        \{\{7,19\},\{14,1\},\{20,3\},
        \{22,9\},\{15,23\},\{11,17\},
        \{2,6\},\{21,16\},\{12,8\},
        \{5,18\},\{4,24\},\{10,13\}\}\}}
        \caption{9-6-3-3-2-1 A (cubic)}
        \label{Dessin}
    \end{subfigure} \hfill
    \begin{subfigure}{0.6\textwidth}
        \centering \captionsetup{justification=centering}
        $\scalemath{0.75}{
        \displaystyle \begin{pmatrix}
            0 & 2 & 1 & 0 & 0 & 0 & 0 & 0\\ 
            2 & 0 & 0 & 0 & 0 & 0 & 1 & 0\\
            1 & 0 & 0 & 1 & 1 & 0 & 0 & 0\\
            0 & 0 & 1 & 0 & 1 & 1 & 0 & 0\\
            0 & 0 & 1 & 1 & 0 & 1 & 0 & 0\\
            0 & 0 & 0 & 1 & 1 & 0 & 1 & 0\\
            0 & 1 & 0 & 0 & 0 & 1 & 0 & 1\\
            0 & 0 & 0 & 0 & 0 & 0 & 1 & 2
        \end{pmatrix}}$
        $\vcenter{\hbox{\includegraphics[width=0.25\textwidth]{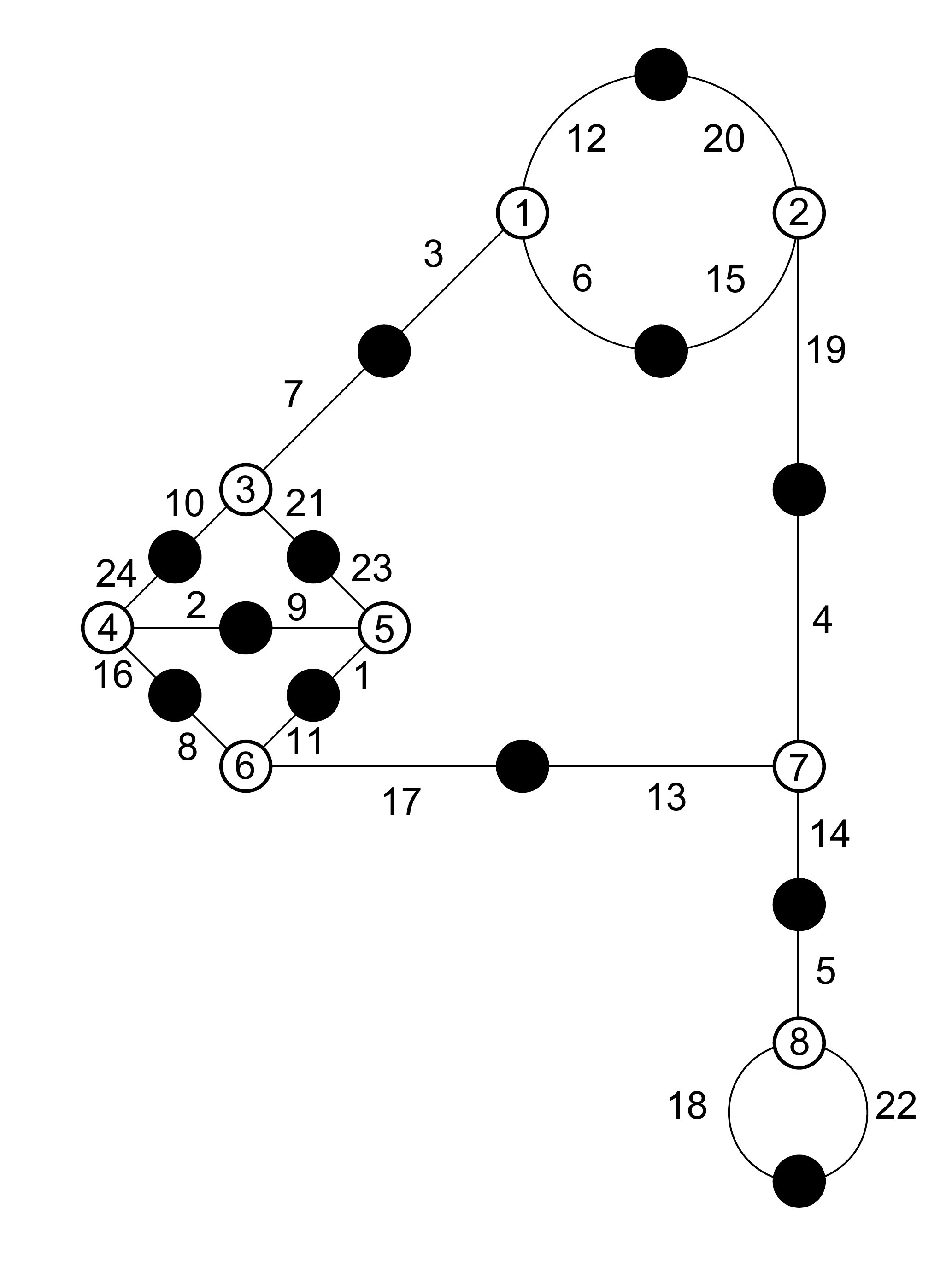}}}$
        $\vcenter{\hbox{\includegraphics[width=0.25\textwidth]{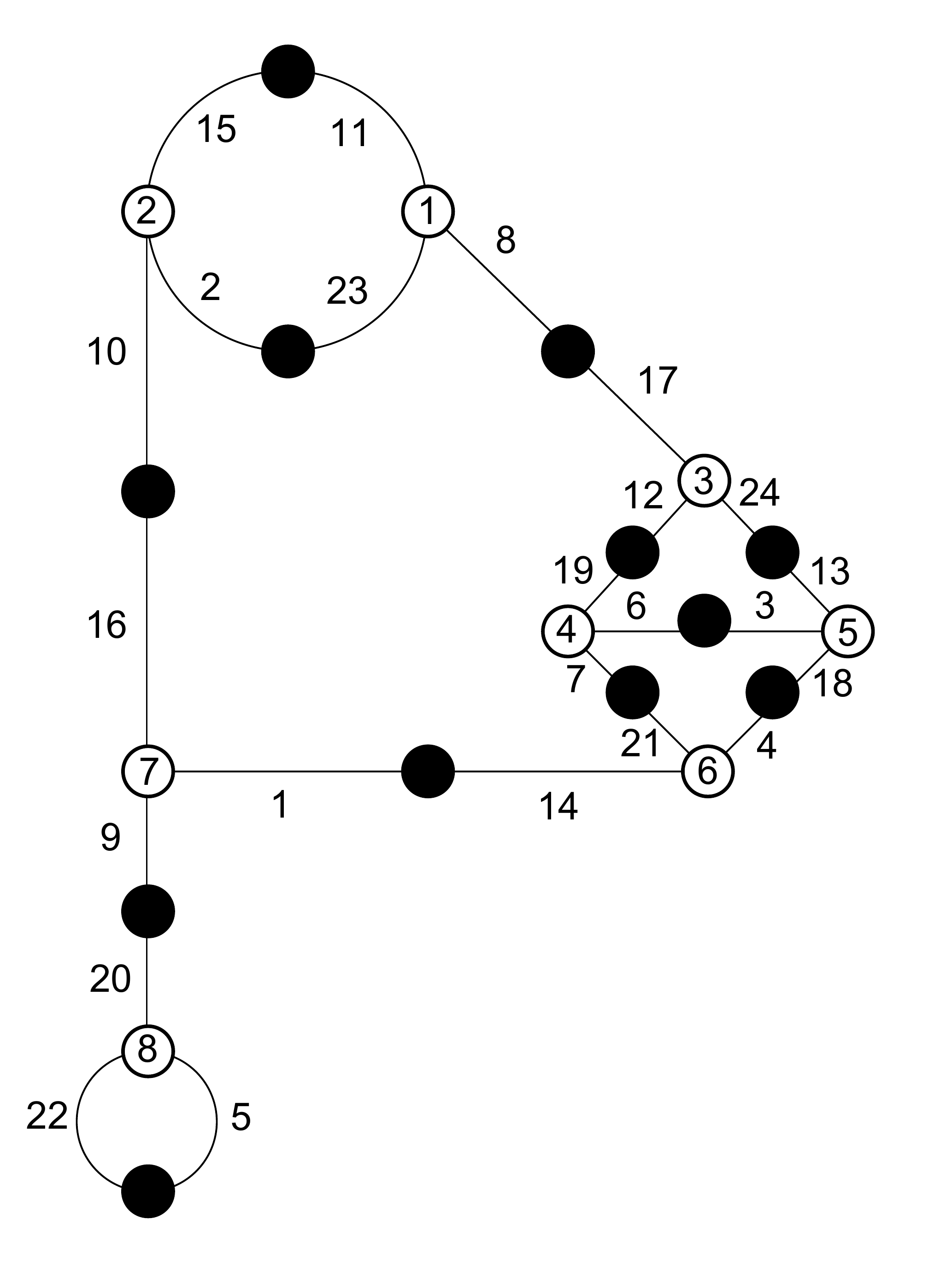}}}$
        \caption{ B: \{\{\{20,19,15\},\{6,3,12\},
        \{7,21,10\},\{24,2,16\},\{9,23,1\},
        \{11,17,8\},\{13,4,14\},\{5,22,18\}\}, \\ 
        \{\{22,18\},\{5,14\},\{13,17\},
        \{4,19\},\{6,15\},\{12,20\},
        \{3,7\},\{21,23\},\{10,24\},
        \{2,9\},\{16,8\},\{1,11\}\}\} \\
        C: \{\{\{15,2,10\},\{11,8,23\},
        \{17,24,12\},\{13,18,3\},\{6,7,19\},
        \{4,14,21\},\{1,9,16\},\{5,22,20\}\}, \\ 
        \{\{22,5\},\{9,20\},\{1,14\},
        \{16,10\},\{2,23\},\{15,11\},
        \{8,17\},\{24,13\},\{3,6\},
        \{19,12\},\{7,21\},\{4,18\}\}\}}
        \caption{9-6-3-3-2-1 B \& C (cubic)}
        \label{Dessin}
    \end{subfigure}\hfill
\end{figure}

\begin{figure}[H]
    \begin{subfigure}{0.5\textwidth}
        \centering \captionsetup{justification=centering}
        $\scalemath{0.75}{
        \displaystyle \begin{pmatrix}
            2 & 1 & 0 & 0 & 0 & 0 & 0 & 0\\ 
            1 & 0 & 1 & 1 & 0 & 0 & 0 & 0\\
            0 & 1 & 0 & 1 & 0 & 1 & 0 & 0\\
            0 & 1 & 1 & 0 & 0 & 0 & 0 & 1\\
            0 & 0 & 0 & 0 & 0 & 1 & 1 & 1\\
            0 & 0 & 1 & 0 & 1 & 0 & 0 & 1\\
            0 & 0 & 0 & 0 & 1 & 0 & 2 & 0\\
            0 & 0 & 0 & 1 & 1 & 1 & 0 & 0
        \end{pmatrix}}$
        $\vcenter{\hbox{\includegraphics[width=0.35\textwidth]{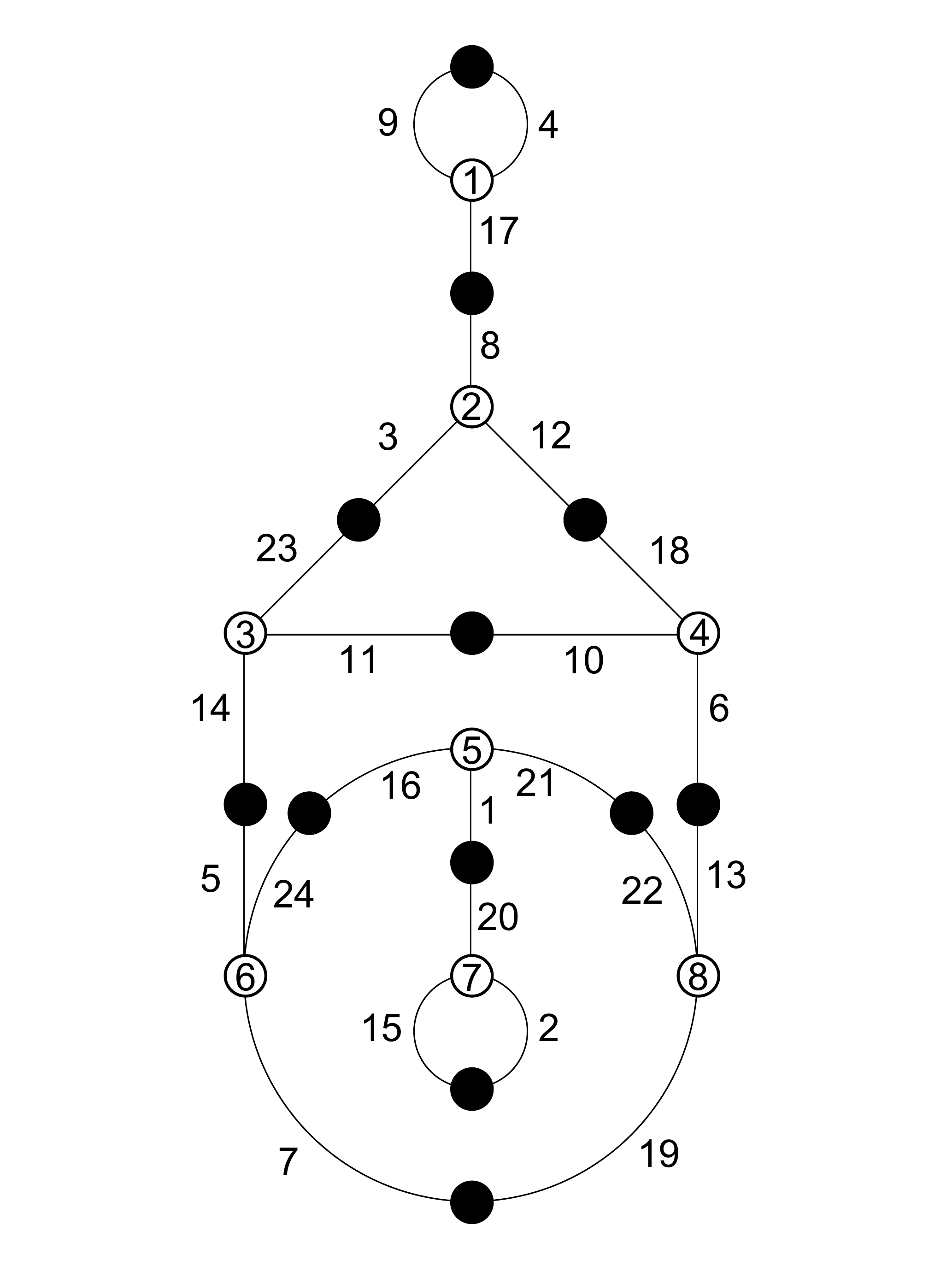}}}$
        \caption{ \{\{\{16,21,1\},\{20,2,15\},
        \{22,13,19\},\{24,7,5\},\{14,23,11\},
        \{3,8,12\},\{18,6,10\},\{17,9,4\}\}, \\
        \{\{4,9\},\{8,17\},\{3,23\},
        \{12,18\},\{11,10\},\{14,5\},
        \{16,24\},\{6,13\},\{1,20\},
        \{2,15\},\{7,19\},\{21,22\}\}\}}
        \caption{9-6-4-3-1-1 $(\mathbb{Q})$}
        \label{Dessin}
    \end{subfigure} \hfill
    \begin{subfigure}{0.5\textwidth}
        \centering \captionsetup{justification=centering}
        $\scalemath{0.75}{
        \displaystyle \begin{pmatrix}
            2 & 1 & 0 & 0 & 0 & 0 & 0 & 0\\ 
            1 & 0 & 2 & 0 & 0 & 0 & 0 & 0\\
            0 & 2 & 0 & 1 & 0 & 0 & 0 & 0\\
            0 & 0 & 1 & 0 & 0 & 0 & 0 & 2\\
            0 & 0 & 0 & 0 & 0 & 1 & 2 & 0\\
            0 & 0 & 0 & 0 & 1 & 2 & 0 & 0\\
            0 & 0 & 0 & 0 & 2 & 0 & 0 & 1\\
            0 & 0 & 0 & 2 & 0 & 0 & 1 & 0
        \end{pmatrix}}$
        $\vcenter{\hbox{\includegraphics[width=0.35\textwidth]{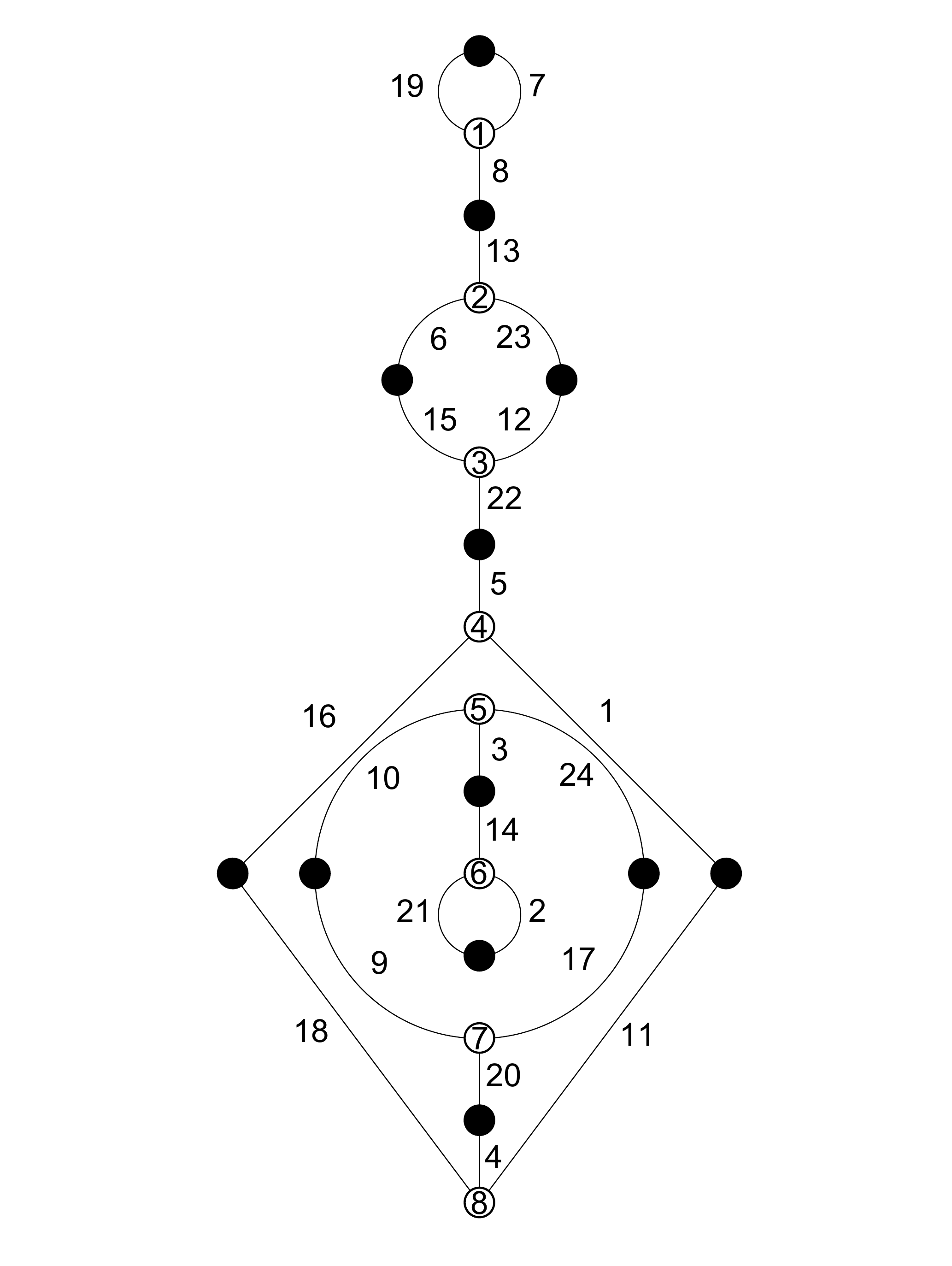}}}$
        \caption{ \{\{\{19,7,8\},\{13,23,6\},
        \{15,12,22\},\{5,1,16\},\{3,10,24\},
        \{14,2,21\},\{9,17,20\},\{4,11,18\}\}, \\ 
        \{\{18,16\},\{1,11\},\{2,21\},
        \{14,3\},\{10,9\},\{24,17\},
        \{5,22\},\{15,6\},\{23,12\},
        \{13,8\},\{19,7\},\{4,20\}\}\}}
        \caption{9-6-5-2-1-1 A (cubic)}
        \label{Dessin}
    \end{subfigure}\hfill
\end{figure}

\begin{figure}[H]
    \begin{subfigure}{0.6\textwidth}
        \centering \captionsetup{justification=centering}
        $\scalemath{0.75}{
        \displaystyle \begin{pmatrix}
            0 & 2 & 1 & 0 & 0 & 0 & 0 & 0\\ 
            2 & 0 & 0 & 0 & 0 & 0 & 1 & 0\\
            1 & 0 & 0 & 1 & 0 & 1 & 0 & 0\\
            0 & 0 & 1 & 0 & 1 & 1 & 0 & 0\\
            0 & 0 & 0 & 1 & 2 & 0 & 0 & 0\\
            0 & 0 & 1 & 1 & 0 & 0 & 1 & 0\\
            0 & 1 & 0 & 0 & 0 & 1 & 0 & 1\\
            0 & 0 & 0 & 0 & 0 & 0 & 1 & 2
        \end{pmatrix}}$
        $\vcenter{\hbox{\includegraphics[width=0.25\textwidth]{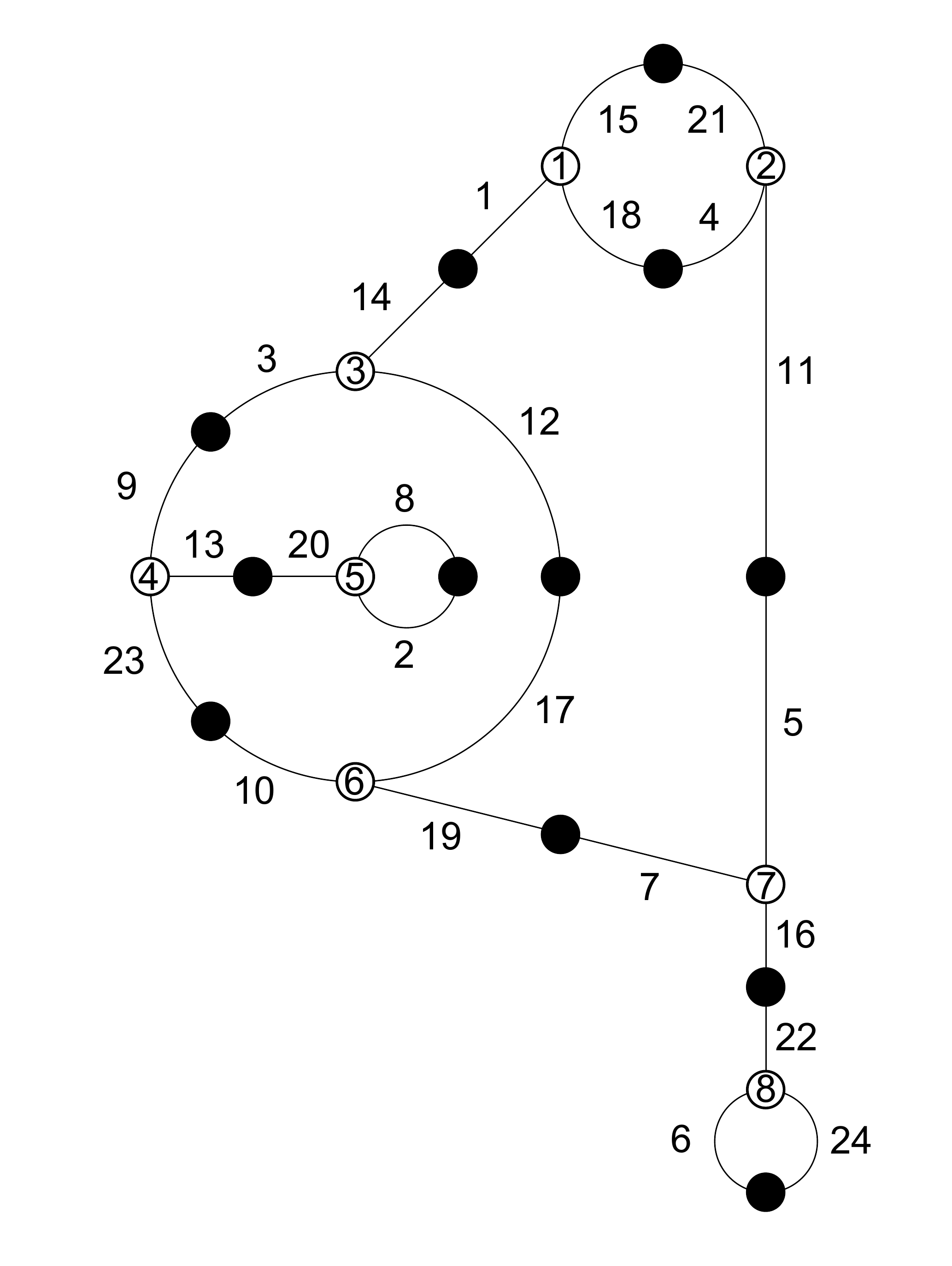}}}$
        $\vcenter{\hbox{\includegraphics[width=0.25\textwidth]{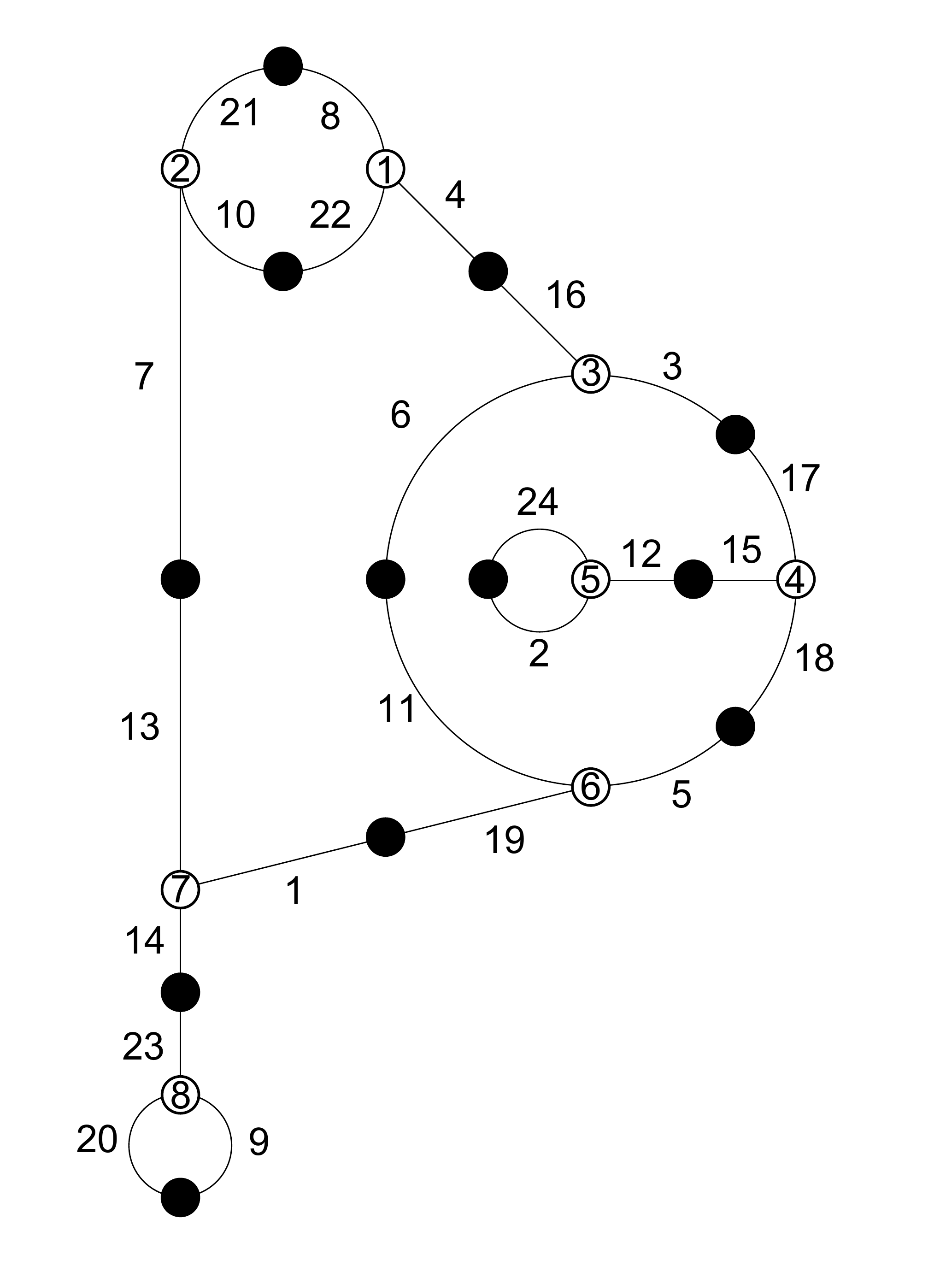}}}$
        \caption{ B: \{\{\{15,18,1\},\{21,11,4\},
        \{14,12,3\},\{20,8,2\},\{13,23,9\},
        \{10,17,19\},\{7,5,16\},\{22,24,6\}\}, \\ 
        \{\{6,24\},\{22,16\},\{5,11\},
        \{7,19\},\{10,23\},\{9,3\},
        \{13,20\},\{8,2\},\{12,17\},
        \{14,1\},\{15,21\},\{18,4\}\}\} \\
        C: \{\{\{21,10,7\},\{8,4,22\},
        \{16,3,6\},\{17,18,15\},\{24,12,2\},
        \{5,19,11\},\{13,1,14\},\{23,9,20\}\}, \\ 
        \{\{20,9\},\{14,23\},\{7,13\},
        \{19,1\},\{11,6\},\{2,24\},
        \{12,15\},\{5,18\},\{3,17\},
        \{16,4\},\{10,22\},\{21,8\}\}\}}
        \caption{9-6-5-2-1-1 B \& C  (cubic)}
        \label{Dessin}
    \end{subfigure} \hfill
    \begin{subfigure}{0.5\textwidth}
        \centering \captionsetup{justification=centering}
        $\scalemath{0.75}{
        \displaystyle \begin{pmatrix}
            2 & 1 & 0 & 0 & 0 & 0 & 0 & 0\\ 
            1 & 0 & 2 & 0 & 0 & 0 & 0 & 0\\
            0 & 2 & 0 & 1 & 0 & 0 & 0 & 0\\
            0 & 0 & 1 & 0 & 0 & 0 & 0 & 2\\
            0 & 0 & 0 & 0 & 0 & 2 & 1 & 0\\
            0 & 0 & 0 & 0 & 2 & 0 & 1 & 0\\
            0 & 0 & 0 & 0 & 1 & 1 & 0 & 1\\
            0 & 0 & 0 & 2 & 0 & 0 & 1 & 0
        \end{pmatrix}}$
        $\vcenter{\hbox{\includegraphics[width=0.35\textwidth]{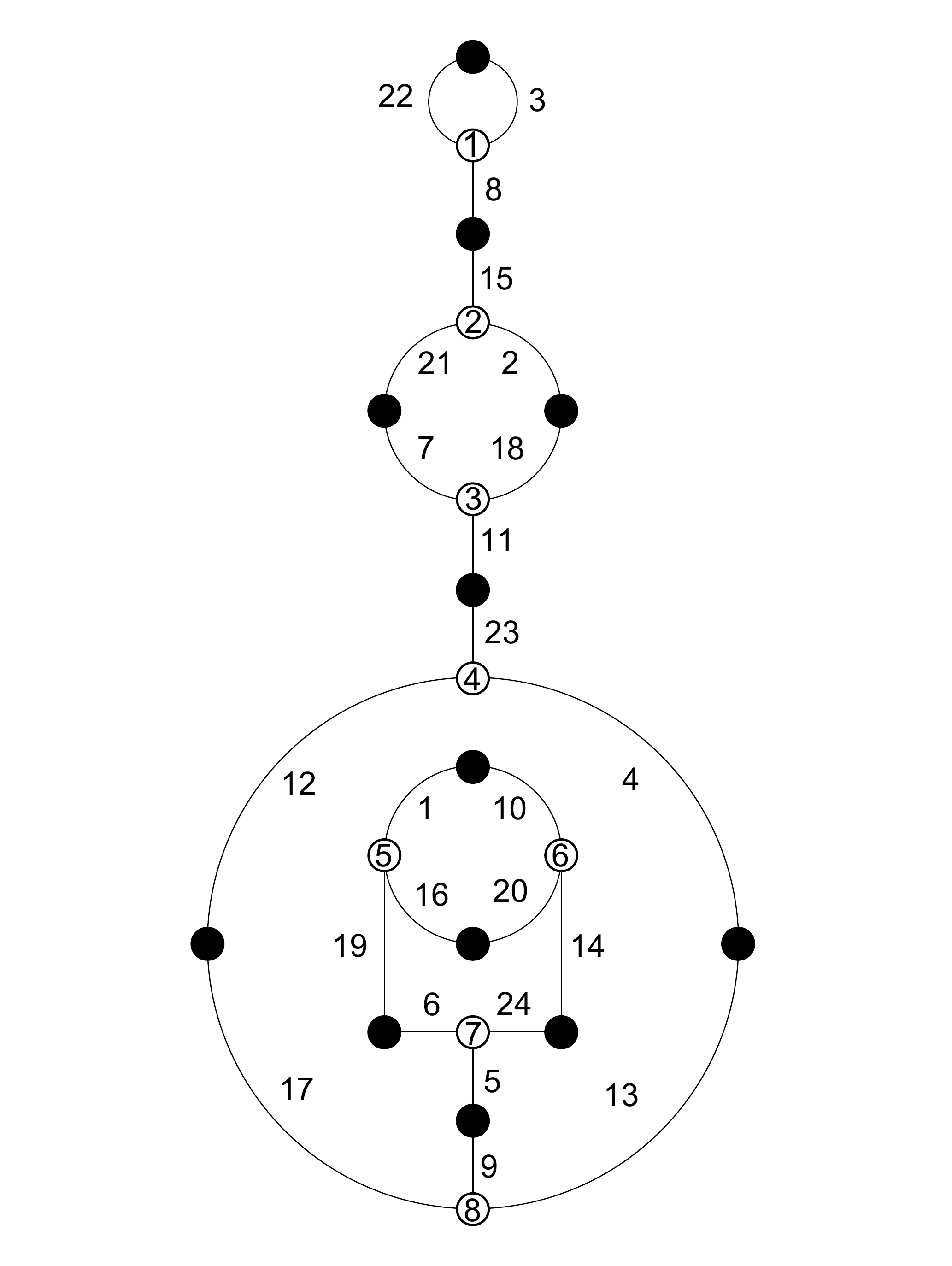}}}$
        \caption{ \{\{\{22,3,8\},\{15,2,21\},
        \{7,18,11\},\{23,4,12\},\{17,9,13\},
        \{5,6,24\},\{16,19,1\},\{20,10,14\}\}, \\ 
        \{\{1,10\},\{16,20\},\{19,6\},
        \{14,24\},\{5,9\},\{13,4\},
        \{17,12\},\{23,11\},\{7,21\},
        \{2,18\},\{15,8\},\{22,3\}\}\}}
        \caption{9-7-3-2-2-1 A (cubic)}
        \label{Dessin}
    \end{subfigure}\hfill
\end{figure}

\begin{figure}[H]
    \begin{subfigure}{0.5\textwidth}
        \centering \captionsetup{justification=centering}
        $\scalemath{0.75}{
        \displaystyle \begin{pmatrix}
            0 & 2 & 1 & 0 & 0 & 0 & 0 & 0\\ 
            2 & 0 & 1 & 0 & 0 & 0 & 0 & 0\\
            1 & 1 & 0 & 1 & 0 & 0 & 0 & 0\\
            0 & 0 & 1 & 0 & 0 & 1 & 1 & 0\\
            0 & 0 & 0 & 0 & 2 & 1 & 0 & 0\\
            0 & 0 & 0 & 1 & 1 & 0 & 0 & 1\\
            0 & 0 & 0 & 1 & 0 & 0 & 0 & 2\\
            0 & 0 & 0 & 0 & 0 & 1 & 2 & 0
        \end{pmatrix}}$
        $\vcenter{\hbox{\includegraphics[width=0.25\textwidth]{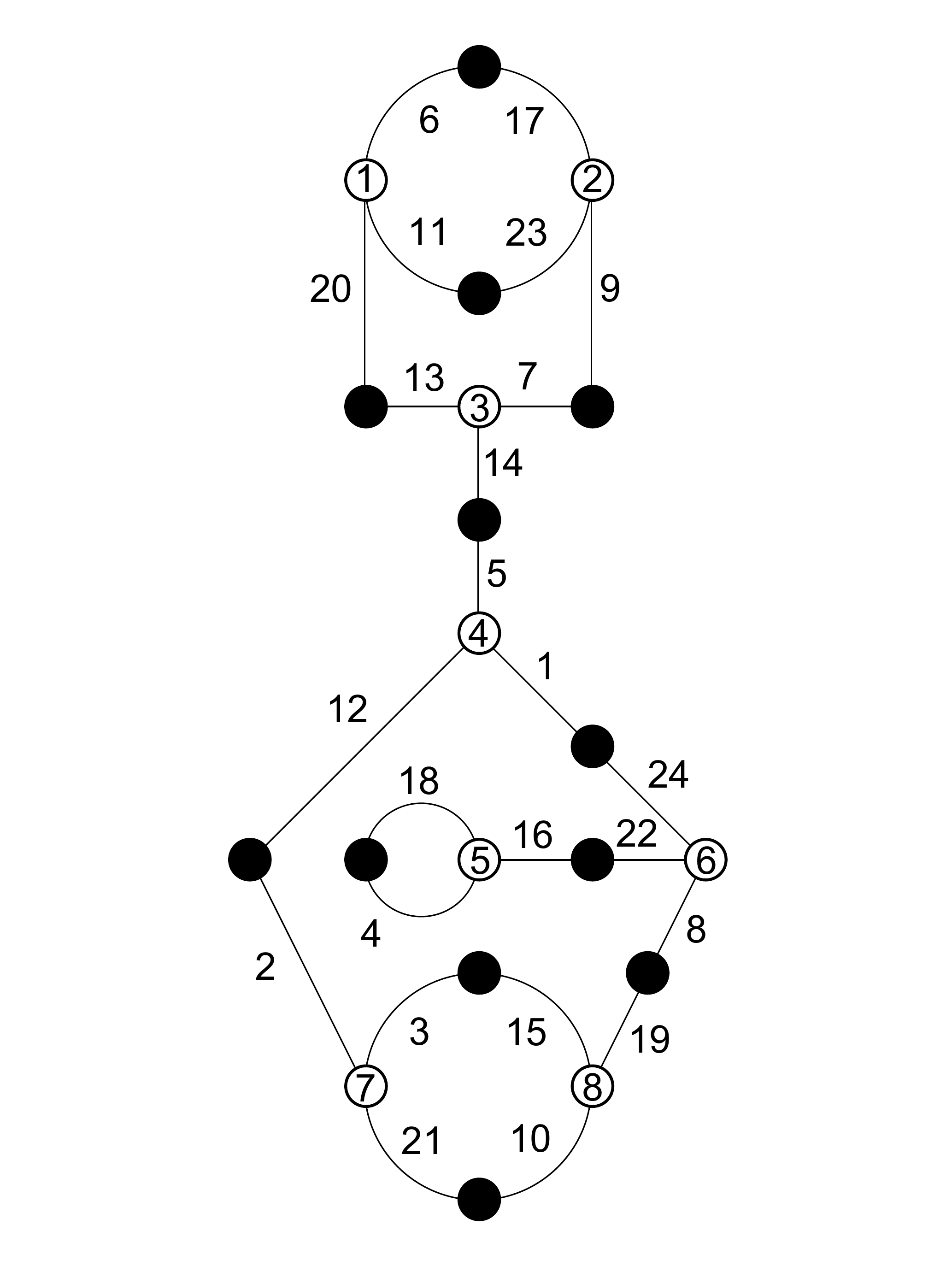}}}$
        $\vcenter{\hbox{\includegraphics[width=0.25\textwidth]{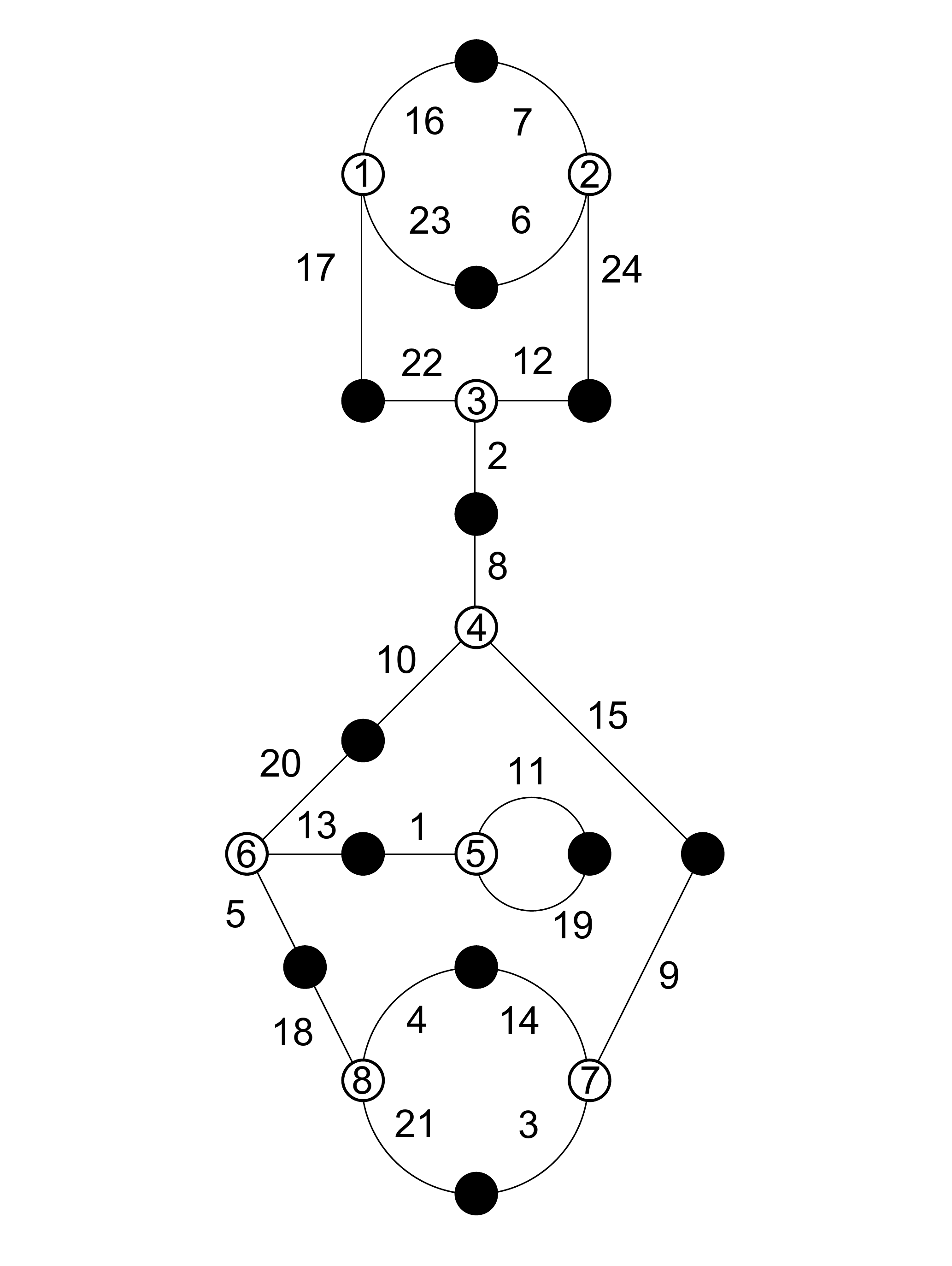}}}$
        \caption{ B: \{\{\{6,11,20\},\{17,9,23\},
        \{13,7,14\},\{5,1,12\},\{24,8,22\},
        \{16,4,18\},\{2,3,21\},\{10,15,19\}\}, \\ 
        \{\{21,10\},\{3,15\},\{8,19\},
        \{4,18\},\{16,22\},\{1,24\},
        \{2,12\},\{5,14\},\{13,20\},
        \{7,9\},\{11,23\},\{6,17\}\}\} \\
        C: \{\{\{23,17,16\},\{7,24,6\},
        \{22,12,2\},\{8,15,10\},\{20,13,5\},
        \{1,11,19\},\{9,3,14\},\{21,18,4\}\}, \\ 
        \{\{21,3\},\{4,14\},\{1,13\},
        \{11,19\},\{20,10\},\{18,5\},
        \{8,2\},\{22,17\},\{12,24\},
        \{23,6\},\{16,7\},\{15,9\}\}\}}
        \caption{9-7-3-2-2-1 B \& C (cubic)}
        \label{Dessin}
    \end{subfigure} \hfill
    \begin{subfigure}{0.5\textwidth}
        \centering \captionsetup{justification=centering}
        $\scalemath{0.75}{
        \displaystyle \begin{pmatrix}
            0 & 2 & 1 & 0 & 0 & 0 & 0 & 0\\ 
            2 & 0 & 0 & 1 & 0 & 0 & 0 & 0\\
            1 & 0 & 0 & 1 & 1 & 0 & 0 & 0\\
            0 & 1 & 1 & 0 & 0 & 0 & 1 & 0\\
            0 & 0 & 1 & 0 & 0 & 1 & 1 & 0\\
            0 & 0 & 0 & 0 & 1 & 2 & 0 & 0\\
            0 & 0 & 0 & 1 & 1 & 0 & 0 & 1\\
            0 & 0 & 0 & 0 & 0 & 0 & 1 & 2
        \end{pmatrix}}$
        $\vcenter{\hbox{\includegraphics[width=0.25\textwidth]{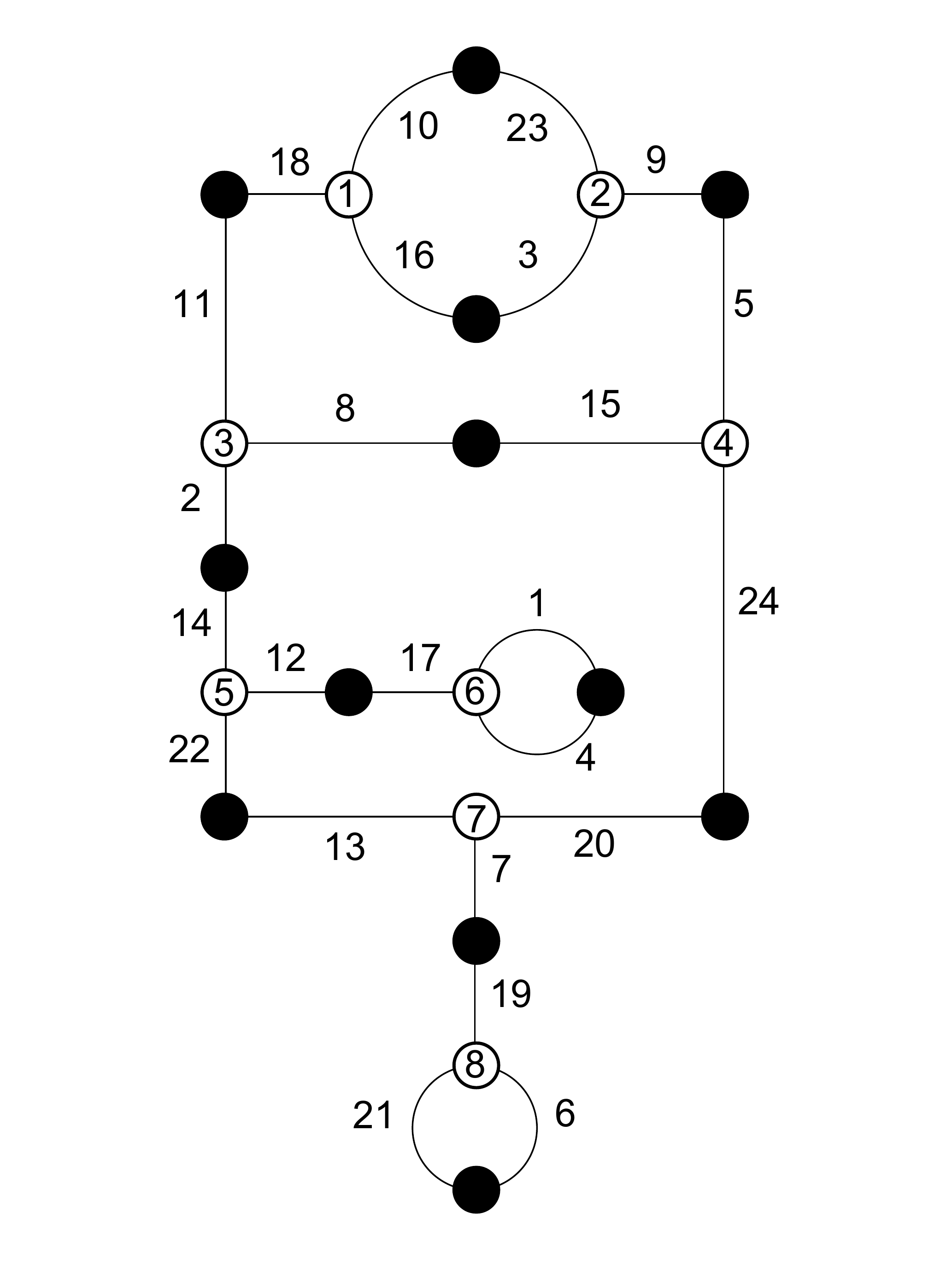}}}$
        $\vcenter{\hbox{\includegraphics[width=0.25\textwidth]{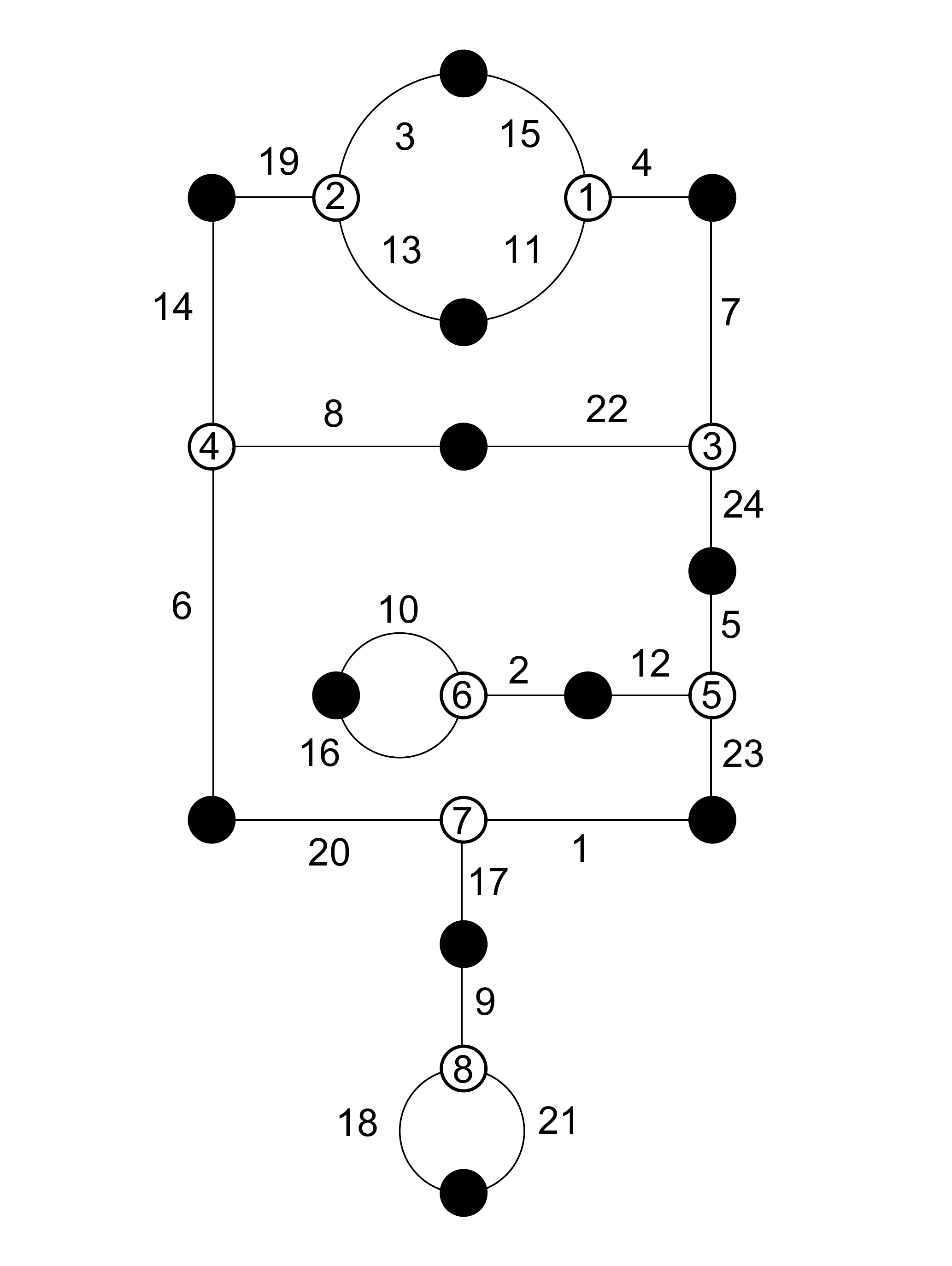}}}$
        \caption{ A: \{\{\{23,9,3\},\{10,16,18\},
        \{11,8,2\},\{15,5,24\},\{1,4,17\},
        \{12,22,14\},\{7,13,20\},\{21,19,6\}\}, \\ 
        \{\{6,21\},\{19,7\},\{20,24\},
        \{1,4\},\{12,17\},\{22,13\},
        \{2,14\},\{8,15\},\{11,18\},
        \{9,5\},\{10,23\},\{16,3\}\}\} \\
        B: \{\{\{4,11,15\},\{3,13,19\},
        \{14,8,6\},\{22,7,24\},\{5,23,12\},
        \{2,16,10\},\{1,17,20\},\{9,21,18\}\}, \\ 
        \{\{18,21\},\{9,17\},\{1,23\},
        \{20,6\},\{12,2\},\{10,16\},
        \{8,22\},\{7,4\},\{11,13\},
        \{3,15\},\{19,14\},\{5,24\}\}\}}
        \caption{9-7-4-2-1-1 A \& B $(\sqrt{-7})$}
        \label{Dessin}
    \end{subfigure}\hfill
\end{figure}

\begin{figure}[H]
    \begin{subfigure}{0.5\textwidth}
        \centering \captionsetup{justification=centering}
        $\scalemath{0.75}{
        \displaystyle \begin{pmatrix}
            2 & 1 & 0 & 0 & 0 & 0 & 0 & 0\\ 
            1 & 0 & 1 & 0 & 0 & 0 & 1 & 0\\
            0 & 1 & 0 & 1 & 0 & 0 & 1 & 0\\
            0 & 0 & 1 & 0 & 0 & 2 & 0 & 0\\
            0 & 0 & 0 & 0 & 2 & 1 & 0 & 0\\
            0 & 0 & 0 & 2 & 1 & 0 & 0 & 0\\
            0 & 1 & 1 & 0 & 0 & 0 & 0 & 1\\
            0 & 0 & 0 & 0 & 0 & 0 & 1 & 2
        \end{pmatrix}}$
        $\vcenter{\hbox{\includegraphics[width=0.35\textwidth]{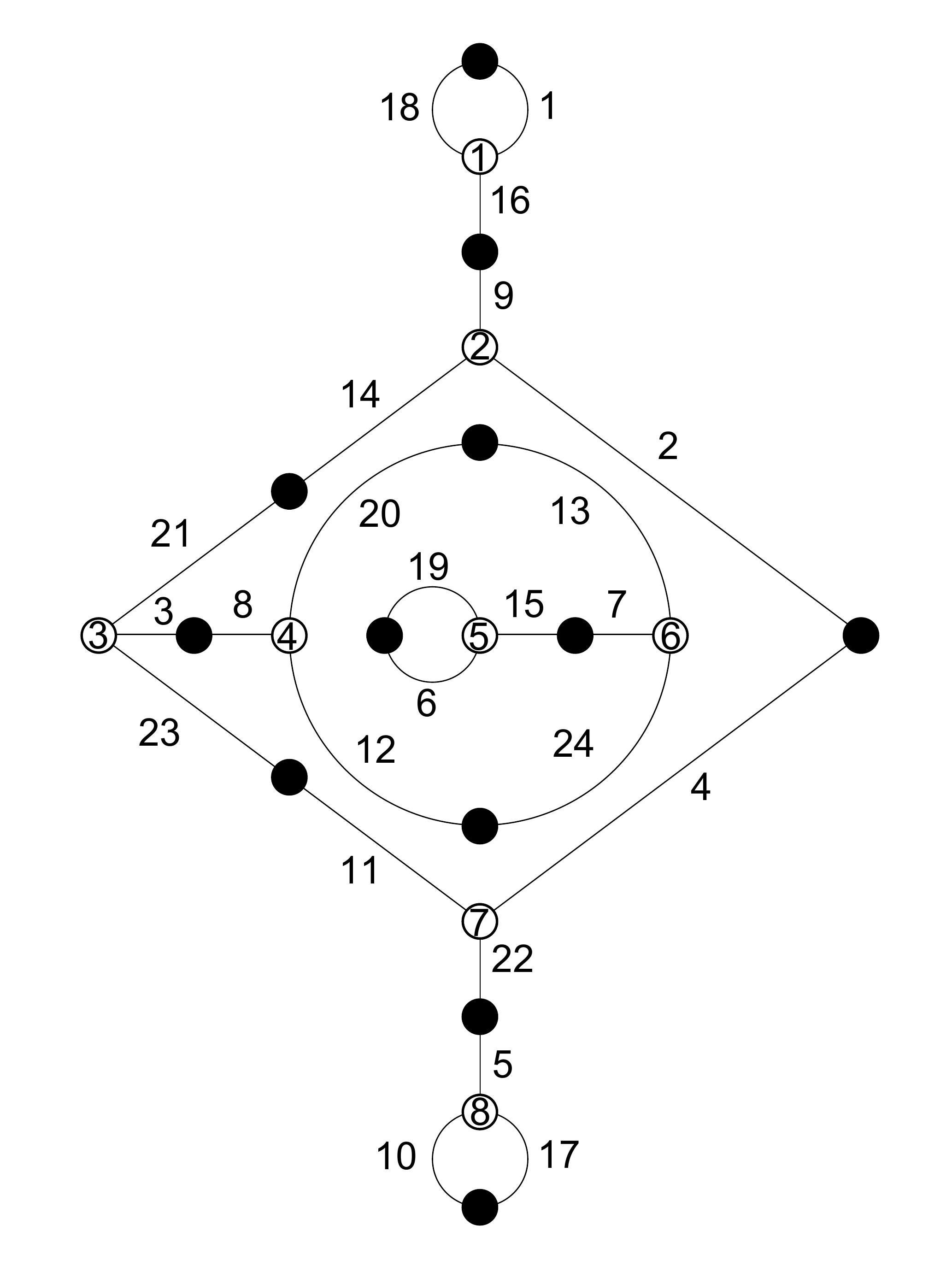}}}$
        \caption{ \{\{\{1,16,18\},\{9,2,14\},
        \{21,3,23\},\{8,20,12\},\{19,15,6\},
        \{7,13,24\},\{11,4,22\},\{5,17,10\}\}, \\ 
        \{\{10,17\},\{5,22\},\{4,2\},
        \{11,23\},\{3,8\},\{21,14\},
        \{9,16\},\{1,18\},\{20,13\},
        \{12,24\},\{7,15\},\{19,6\}\}\}}
        \caption{9-7-5-1-1-1 $(\mathbb{Q})$}
        \label{Dessin}
    \end{subfigure} \hfill
    \begin{subfigure}{0.4\textwidth}
        \centering \captionsetup{justification=centering}
        $\scalemath{0.75}{
        \displaystyle \begin{pmatrix}
            2 & 1 & 0 & 0 & 0 & 0 & 0 & 0\\ 
            1 & 0 & 0 & 1 & 0 & 0 & 1 & 0\\
            0 & 0 & 0 & 0 & 1 & 2 & 0 & 0\\
            0 & 1 & 0 & 0 & 1 & 0 & 1 & 0\\
            0 & 0 & 1 & 1 & 0 & 1 & 0 & 0\\
            0 & 0 & 2 & 0 & 1 & 0 & 0 & 0\\
            0 & 1 & 0 & 1 & 0 & 0 & 0 & 1\\
            0 & 0 & 0 & 0 & 0 & 0 & 1 & 2
        \end{pmatrix}}$
        $\vcenter{\hbox{\includegraphics[width=0.35\textwidth]{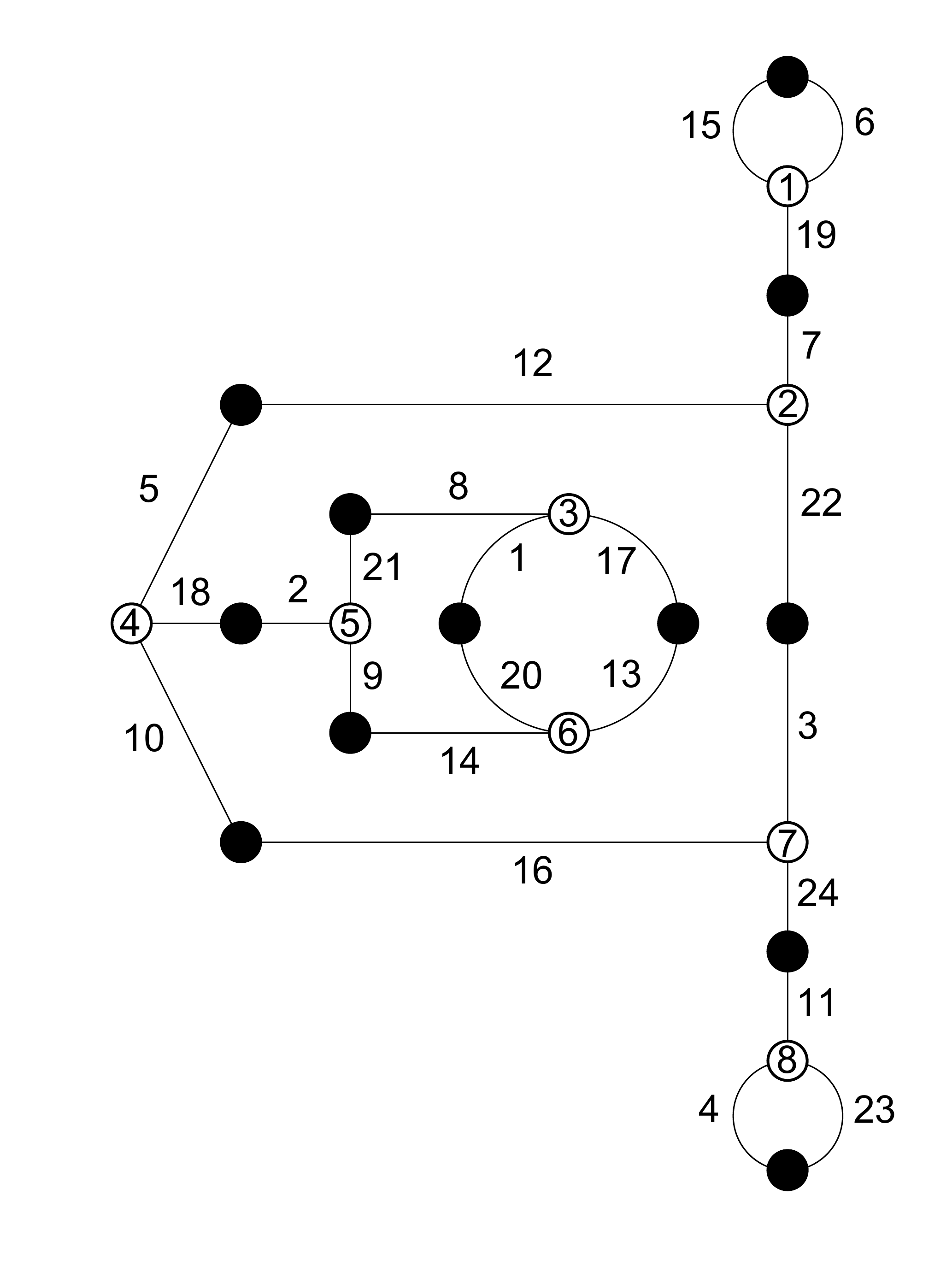}}}$
        \caption{ \{\{\{15,6,19\},\{7,22,12\},
        \{5,18,10\},\{2,21,9\},\{8,17,1\},
        \{20,13,14\},\{3,24,16\},\{4,11,23\}\}, \\ 
        \{\{4,23\},\{11,24\},\{16,10\},
        \{3,22\},\{5,12\},\{18,2\},
        \{21,8\},\{9,14\},\{1,20\},
        \{17,13\},\{7,19\},\{15,6\}\}\}}
        \caption{9-8-3-2-1-1 A (cubic)}
        \label{Dessin}
    \end{subfigure}\hfill
\end{figure}

\begin{figure}[H]
    \begin{subfigure}{0.6\textwidth}
        \centering \captionsetup{justification=centering}
        $\scalemath{0.75}{
        \displaystyle \begin{pmatrix}
            0 & 2 & 1 & 0 & 0 & 0 & 0 & 0\\ 
            2 & 0 & 0 & 0 & 0 & 1 & 0 & 0\\
            1 & 0 & 0 & 1 & 1 & 0 & 0 & 0\\
            0 & 0 & 1 & 2 & 0 & 0 & 0 & 0\\
            0 & 0 & 1 & 0 & 0 & 1 & 1 & 0\\
            0 & 1 & 0 & 0 & 1 & 0 & 1 & 0\\
            0 & 0 & 0 & 0 & 1 & 1 & 0 & 1\\
            0 & 0 & 0 & 0 & 0 & 0 & 1 & 2
        \end{pmatrix}}$
        $\vcenter{\hbox{\includegraphics[width=0.25\textwidth]{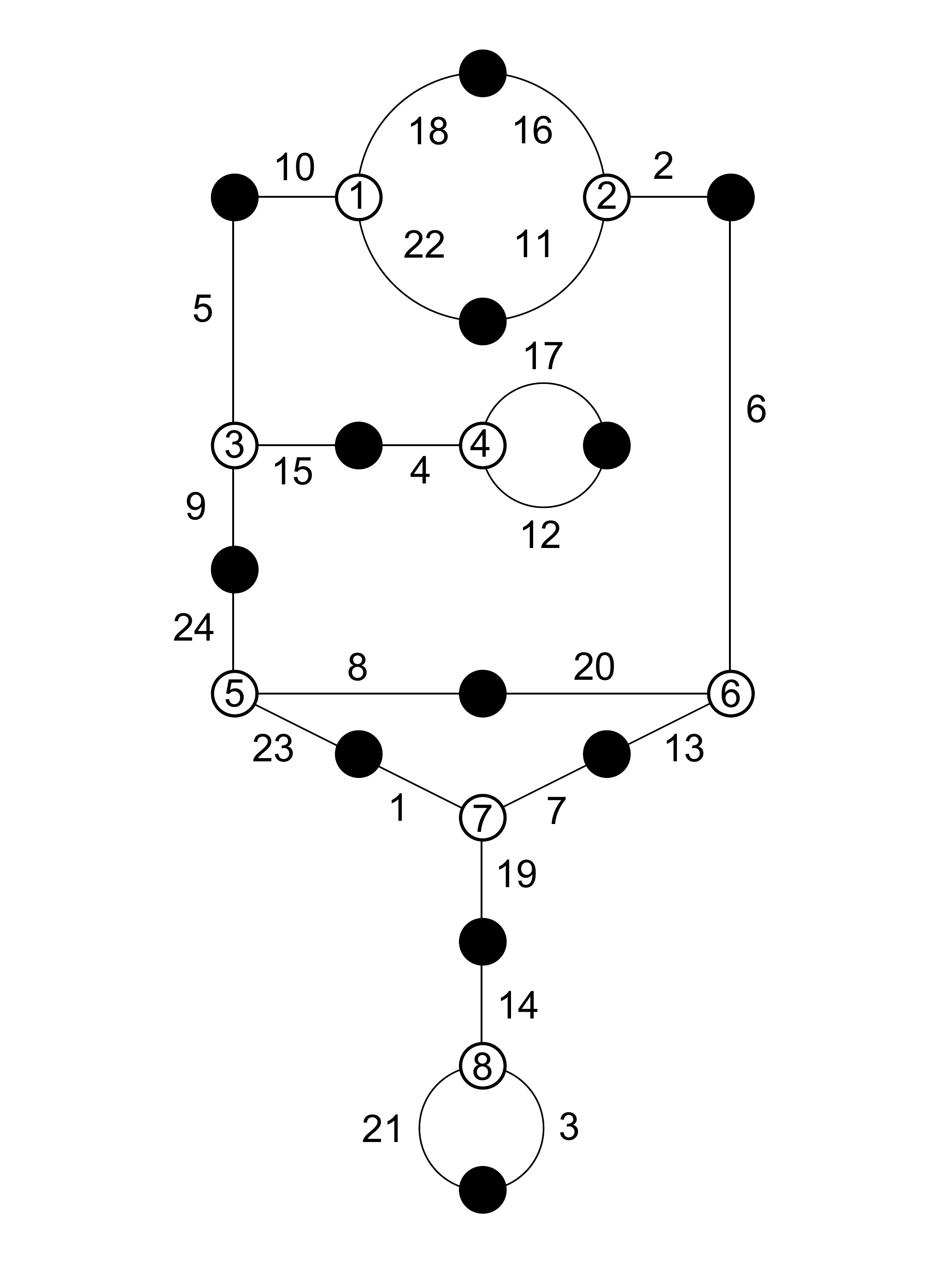}}}$
        $\vcenter{\hbox{\includegraphics[width=0.25\textwidth]{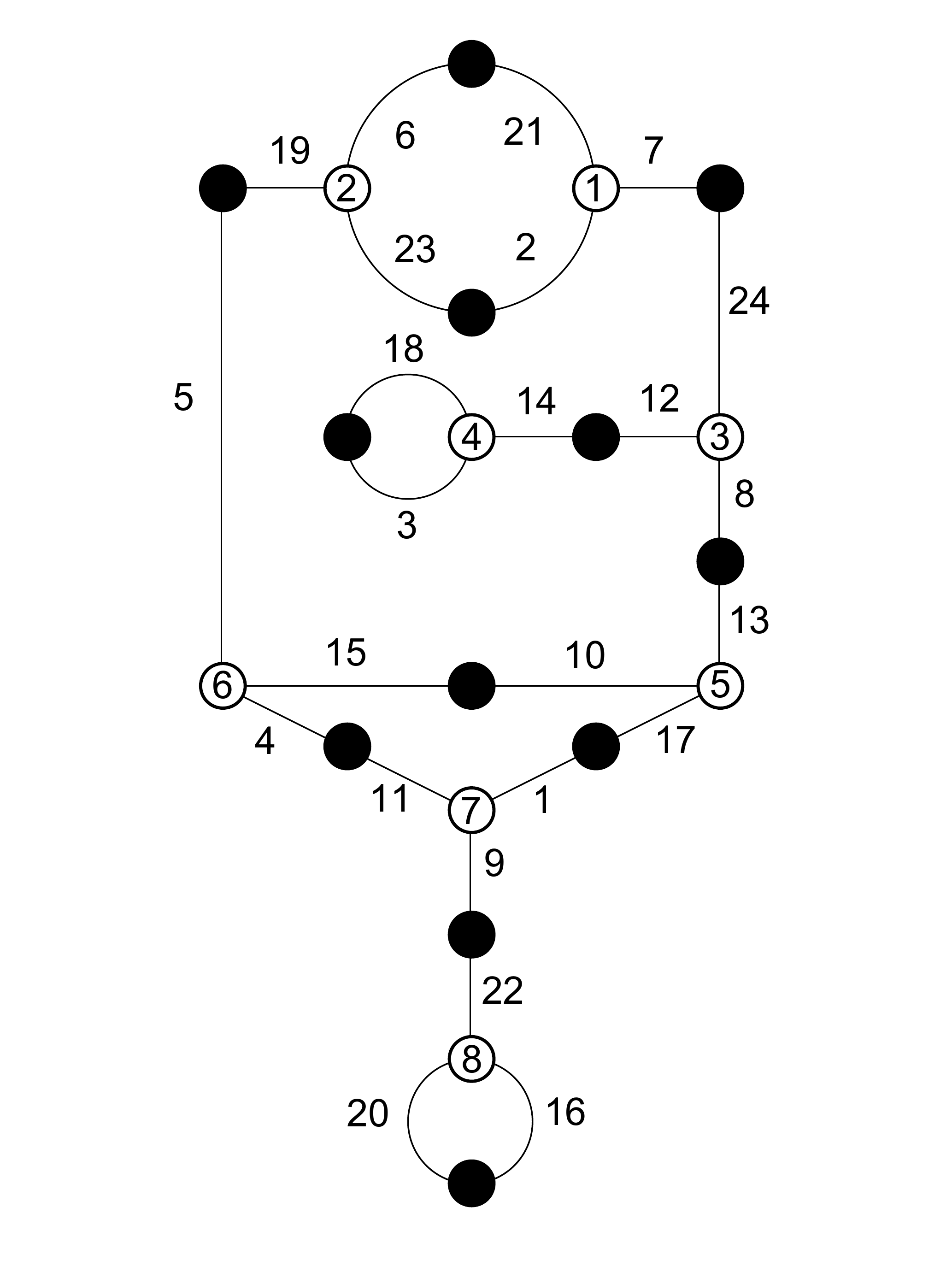}}}$
        \caption{ B: \{\{\{10,18,22\},\{16,2,11\},
        \{5,15,9\},\{4,17,12\},\{24,8,23\},
        \{20,6,13\},\{1,7,19\},\{14,3,21\}\}, \\ 
        \{\{3,21\},\{14,19\},\{1,23\},
        \{7,13\},\{8,20\},\{24,9\},
        \{6,2\},\{15,4\},\{17,12\},
        \{5,10\},\{18,16\},\{22,11\}\}\} \\
        C: \{\{\{19,6,23\},\{21,7,2\},
        \{24,8,12\},\{14,3,18\},\{13,17,10\},
        \{15,4,5\},\{1,9,11\},\{22,16,20\}\}, \\ 
        \{\{20,16\},\{22,9\},\{1,17\},
        \{11,4\},\{15,10\},\{5,19\},
        \{7,24\},\{8,13\},\{12,14\},
        \{18,3\},\{23,2\},\{6,21\}\}\}}
        \caption{9-8-3-2-1-1 B \& C (cubic)}
        \label{Dessin}
    \end{subfigure} \hfill
    \begin{subfigure}{0.4\textwidth}
        \centering \captionsetup{justification=centering}
        $\scalemath{0.75}{
        \displaystyle \begin{pmatrix}
            2 & 1 & 0 & 0 & 0 & 0 & 0 & 0\\ 
            1 & 0 & 1 & 1 & 0 & 0 & 0 & 0\\
            0 & 1 & 0 & 0 & 0 & 2 & 0 & 0\\
            0 & 1 & 0 & 0 & 0 & 0 & 2 & 0\\
            0 & 0 & 0 & 0 & 2 & 0 & 0 & 1\\
            0 & 0 & 2 & 0 & 0 & 0 & 0 & 1\\
            0 & 0 & 0 & 2 & 0 & 0 & 0 & 1\\
            0 & 0 & 0 & 0 & 1 & 1 & 1 & 0
        \end{pmatrix}}$
        $\vcenter{\hbox{\includegraphics[width=0.35\textwidth]{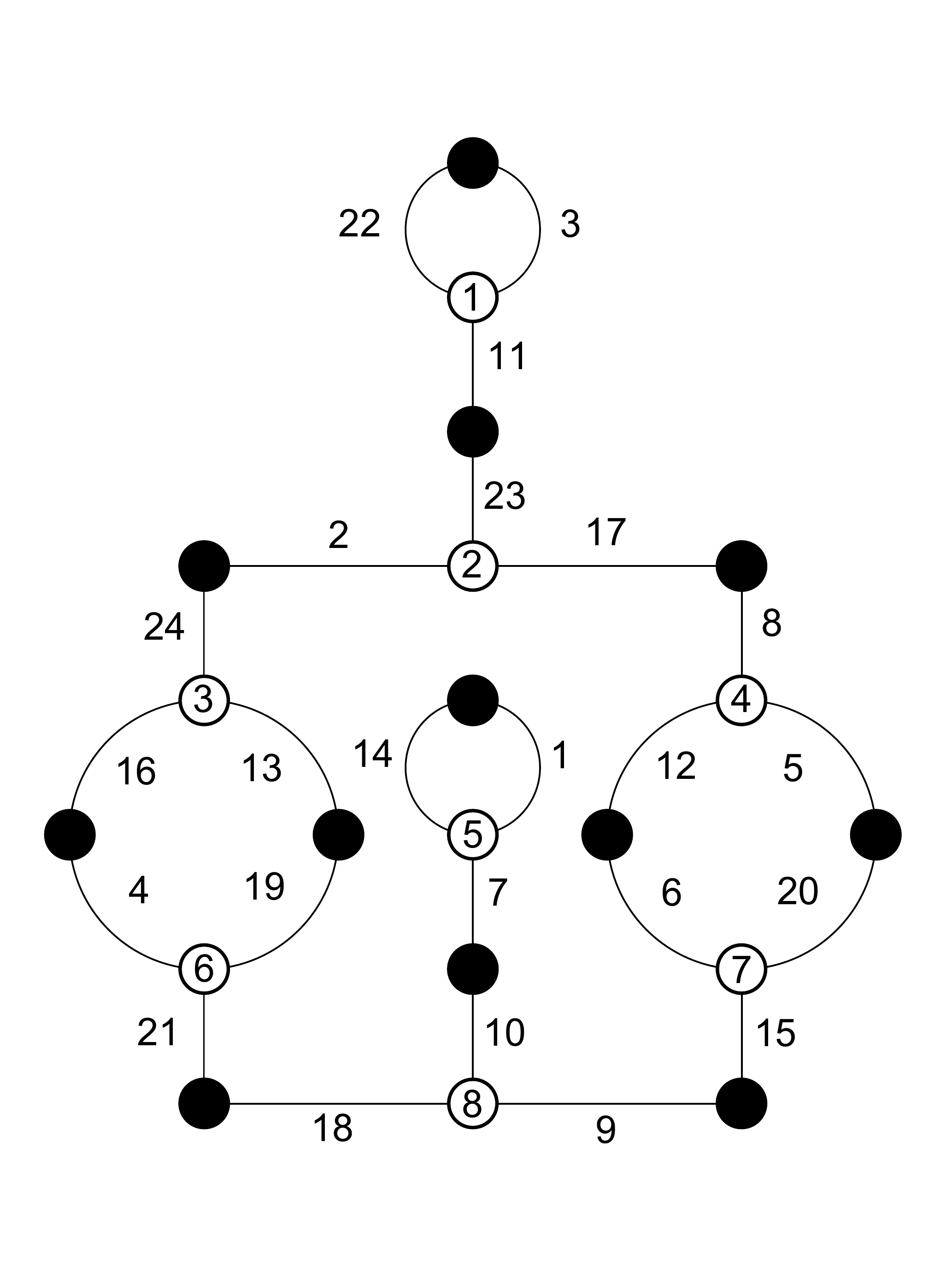}}}$
        \caption{ \{\{\{22,3,11\},\{23,17,2\},
        \{24,13,16\},\{4,19,21\},\{14,1,7\},
        \{10,9,18\},\{15,6,20\},\{8,5,12\}\}, \\ 
        \{\{14,1\},\{7,10\},\{9,15\},
        \{18,21\},\{4,16\},\{19,13\},
        \{6,12\},\{5,20\},\{8,17\},
        \{24,2\},\{23,11\},\{22,3\}\}\}}
        \caption{9-9-2-2-1-1 A $(\mathbb{Q})$}
        \label{Dessin}
    \end{subfigure}\hfill
\end{figure}

\begin{figure}[H]
    \begin{subfigure}{0.4\textwidth}
        \centering \captionsetup{justification=centering}
        $\scalemath{0.75}{
        \displaystyle \begin{pmatrix}
            2 & 1 & 0 & 0 & 0 & 0 & 0 & 0\\ 
            1 & 0 & 2 & 0 & 0 & 0 & 0 & 0\\
            0 & 2 & 0 & 1 & 0 & 0 & 0 & 0\\
            0 & 0 & 1 & 0 & 0 & 0 & 0 & 2\\
            0 & 0 & 0 & 0 & 2 & 1 & 0 & 0\\
            0 & 0 & 0 & 0 & 1 & 0 & 2 & 0\\
            0 & 0 & 0 & 0 & 0 & 2 & 0 & 1\\
            0 & 0 & 0 & 2 & 0 & 0 & 1 & 0
        \end{pmatrix}}$
        $\vcenter{\hbox{\includegraphics[width=0.35\textwidth]{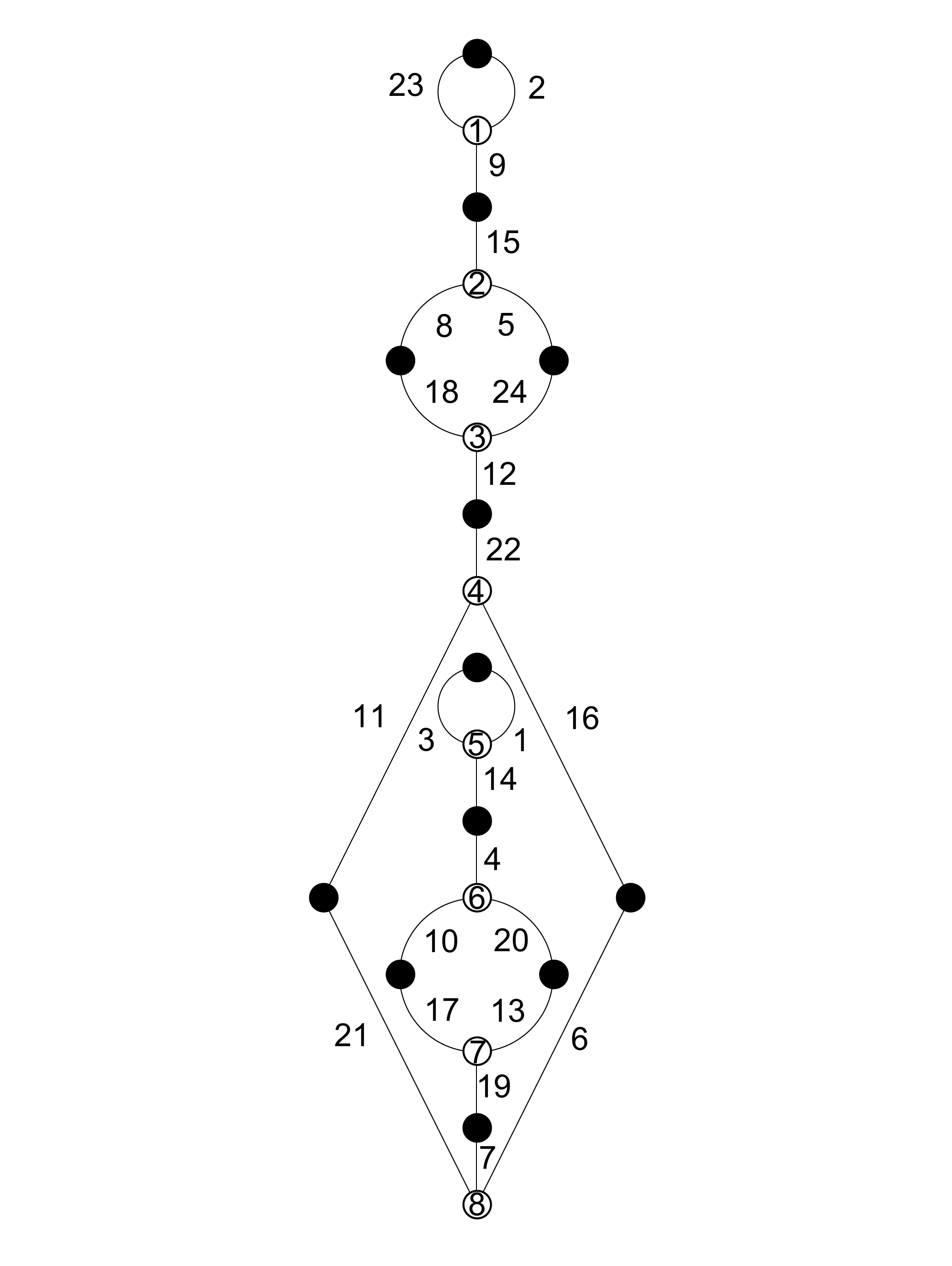}}}$
        \caption{ \{\{\{23,2,9\},\{15,5,8\},
        \{24,12,18\},\{22,16,11\},\{1,14,3\},
        \{4,20,10\},\{13,19,17\},\{7,6,21\}\}, \\ 
        \{\{7,19\},\{21,11\},\{6,16\},
        \{17,10\},\{20,13\},\{4,14\},
        \{1,3\},\{22,12\},\{18,8\},
        \{24,5\},\{15,9\},\{2,23\}\}\}}
        \caption{9-9-2-2-1-1 B (cubic)}
        \label{Dessin}
    \end{subfigure} \hfill
    \begin{subfigure}{0.6\textwidth}
        \centering \captionsetup{justification=centering}
        $\scalemath{0.75}{
        \displaystyle \begin{pmatrix}
            0 & 2 & 1 & 0 & 0 & 0 & 0 & 0\\ 
            2 & 0 & 0 & 0 & 0 & 1 & 0 & 0\\
            1 & 0 & 0 & 2 & 0 & 0 & 0 & 0\\
            0 & 0 & 2 & 0 & 0 & 0 & 1 & 0\\
            0 & 0 & 0 & 0 & 2 & 1 & 0 & 0\\
            0 & 1 & 0 & 0 & 1 & 0 & 1 & 0\\
            0 & 0 & 0 & 1 & 0 & 1 & 0 & 1\\
            0 & 0 & 0 & 0 & 0 & 0 & 1 & 2
        \end{pmatrix}}$
        $\vcenter{\hbox{\includegraphics[width=0.25\textwidth]{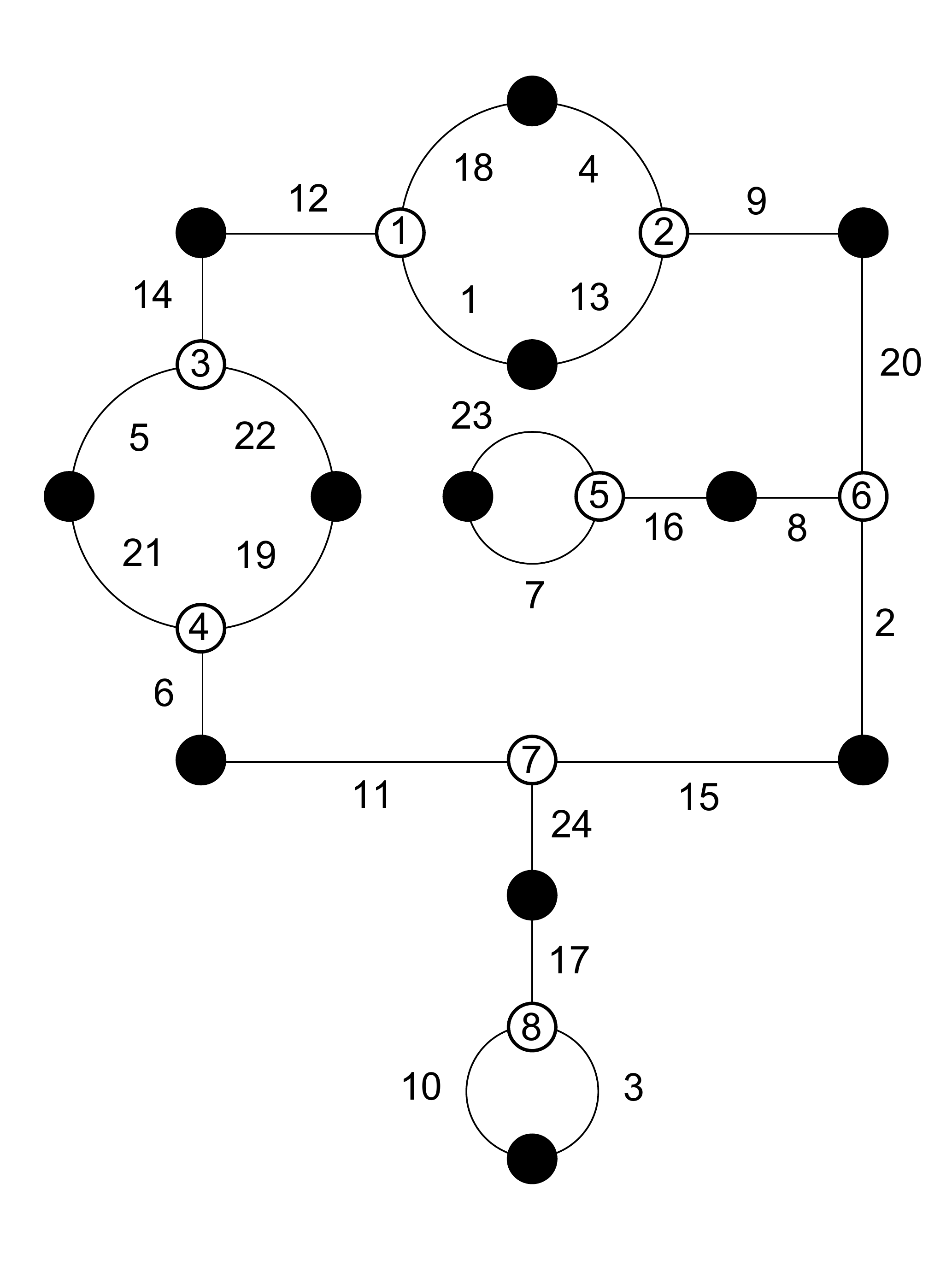}}}$
        $\vcenter{\hbox{\includegraphics[width=0.25\textwidth]{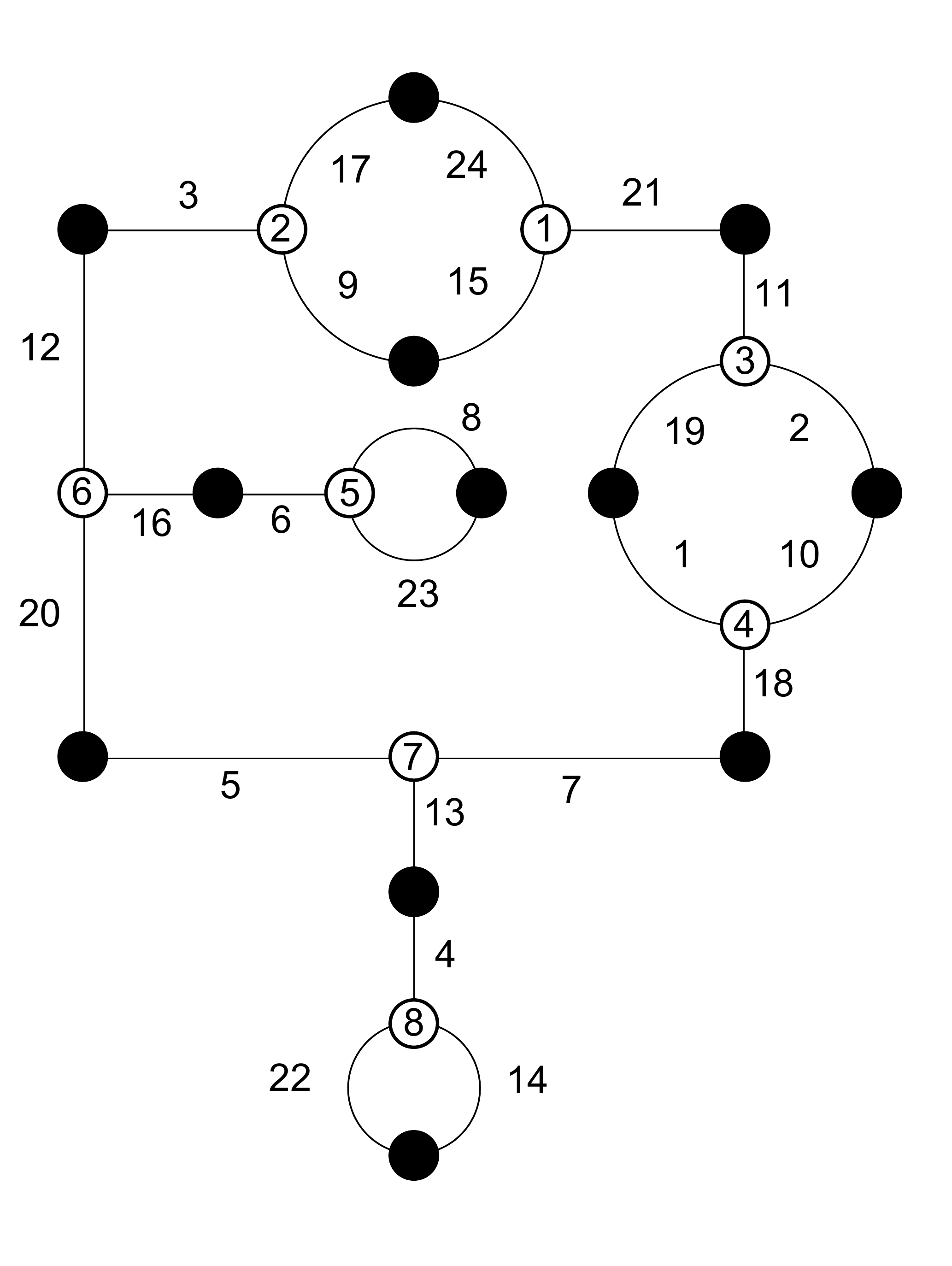}}}$
        \caption{ C: \{\{\{12,18,1\},\{4,9,13\},
        \{20,2,8\},\{16,7,23\},\{5,14,22\},
        \{21,19,6\},\{11,15,24\},\{17,3,10\}\}, \\ 
        \{\{3,10\},\{17,24\},\{15,2\},
        \{6,11\},\{21,5\},\{22,19\},
        \{14,12\},\{18,4\},\{1,13\},
        \{9,20\},\{8,16\},\{23,7\}\}\} \\
        D: \{\{\{3,17,9\},\{24,21,15\},
        \{11,2,19\},\{1,10,18\},\{7,13,5\},
        \{4,14,22\},\{20,12,16\},\{6,8,23\}\}, \\ 
        \{\{22,14\},\{4,13\},\{7,18\},
        \{5,20\},\{16,6\},\{8,23\},
        \{1,19\},\{2,10\},\{11,21\},
        \{12,3\},\{17,24\},\{9,15\}\}\}}
        \caption{9-9-2-2-1-1 C \& D (cubic)}
        \label{Dessin}
    \end{subfigure}\hfill
\end{figure}

\begin{figure}[H]
    \begin{subfigure}{0.5\textwidth}
        \centering \captionsetup{justification=centering}
        $\scalemath{0.75}{
        \displaystyle \begin{pmatrix}
            0 & 2 & 1 & 0 & 0 & 0 & 0 & 0\\ 
            2 & 0 & 1 & 0 & 0 & 0 & 0 & 0\\
            1 & 1 & 0 & 1 & 0 & 0 & 0 & 0\\
            0 & 0 & 1 & 0 & 1 & 1 & 0 & 0\\
            0 & 0 & 0 & 1 & 0 & 1 & 1 & 0\\
            0 & 0 & 0 & 1 & 1 & 0 & 0 & 1\\
            0 & 0 & 0 & 0 & 1 & 0 & 0 & 2\\
            0 & 0 & 0 & 0 & 0 & 1 & 2 & 0
        \end{pmatrix}}$
        $\vcenter{\hbox{\includegraphics[width=0.35\textwidth]{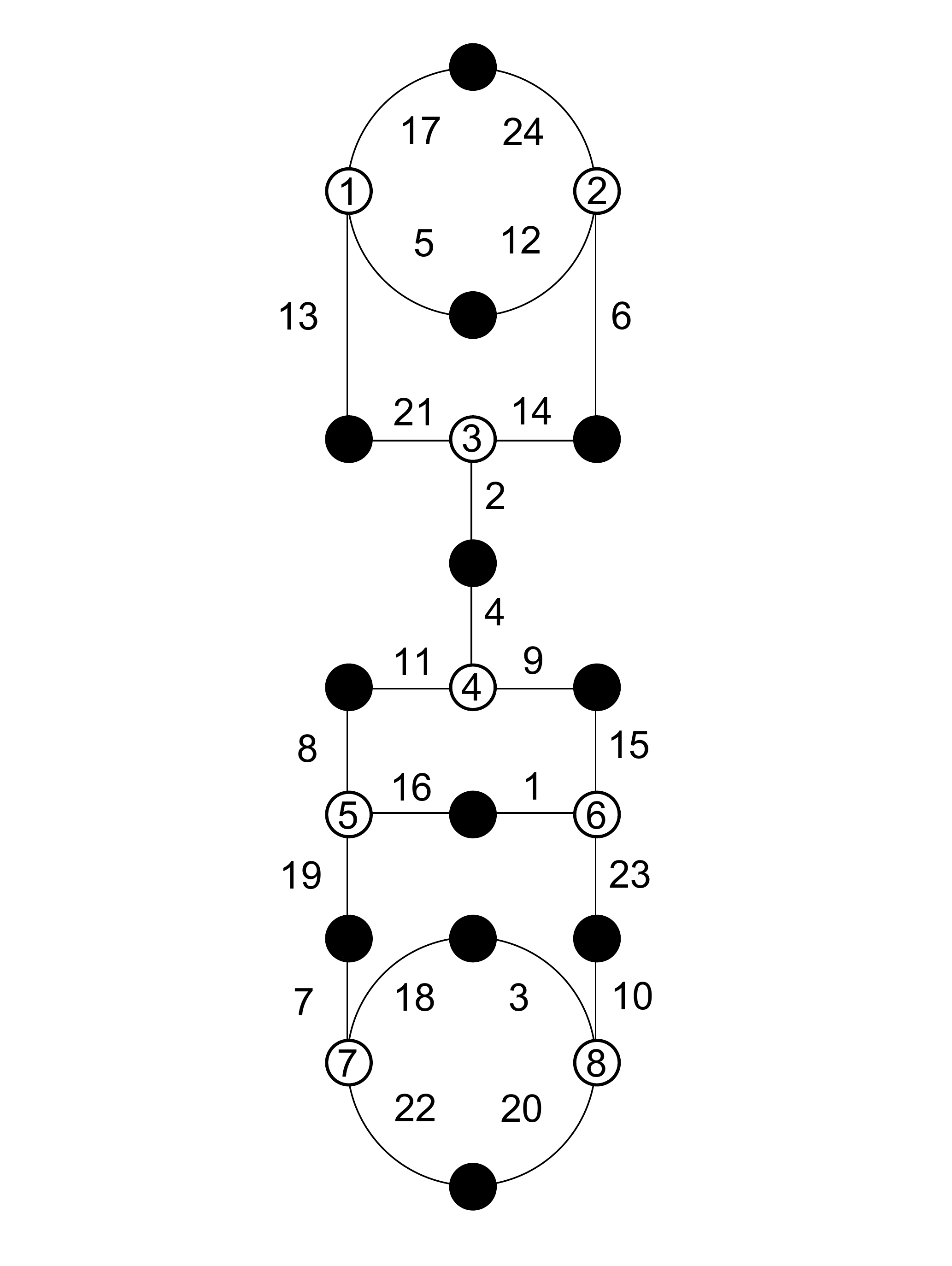}}}$
        \caption{ \{\{\{17,5,13\},\{24,6,12\},
        \{21,14,2\},\{4,9,11\},\{8,16,19\},
        \{1,15,23\},\{10,20,3\},\{22,7,18\}\}, \\ 
        \{\{22,20\},\{18,3\},\{10,23\},
        \{7,19\},\{16,1\},\{8,11\},
        \{9,15\},\{4,2\},\{14,6\},
        \{13,21\},\{5,12\},\{17,24\}\}\}}
        \caption{10-4-3-3-2-2 $(\mathbb{Q})$}
        \label{Dessin}
    \end{subfigure} \hfill
    \begin{subfigure}{0.5\textwidth}
        \centering \captionsetup{justification=centering}
        $\scalemath{0.75}{
        \displaystyle \begin{pmatrix}
            2 & 1 & 0 & 0 & 0 & 0 & 0 & 0\\ 
            1 & 0 & 1 & 1 & 0 & 0 & 0 & 0\\
            0 & 1 & 0 & 1 & 1 & 0 & 0 & 0\\
            0 & 1 & 1 & 0 & 0 & 1 & 0 & 0\\
            0 & 0 & 1 & 0 & 0 & 1 & 1 & 0\\
            0 & 0 & 0 & 1 & 1 & 0 & 0 & 1\\
            0 & 0 & 0 & 0 & 1 & 0 & 0 & 2\\
            0 & 0 & 0 & 0 & 0 & 1 & 2 & 0
        \end{pmatrix}}$
        $\vcenter{\hbox{\includegraphics[width=0.35\textwidth]{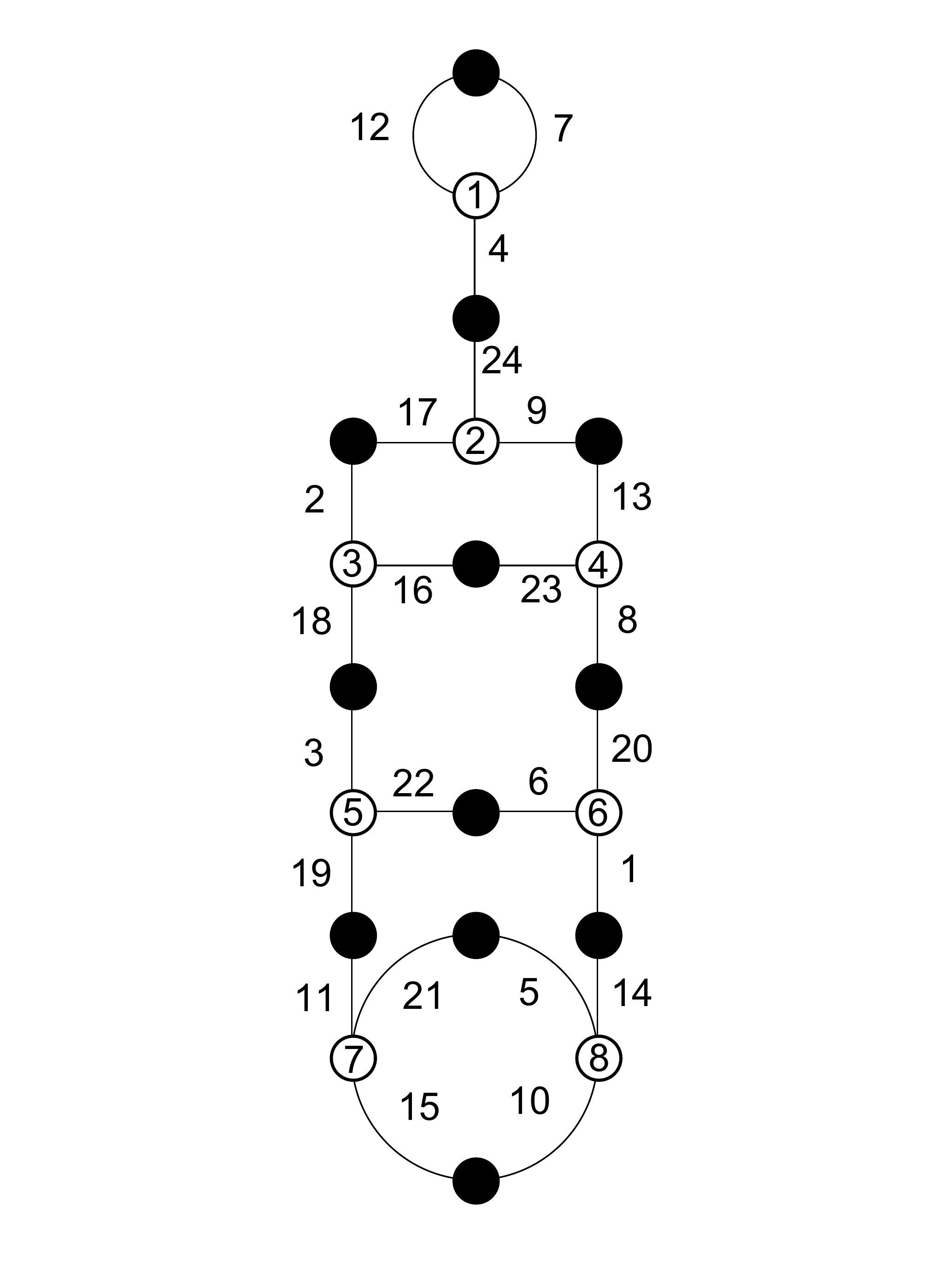}}}$
        \caption{ \{\{\{12,7,4\},\{24,9,17\},
        \{13,8,23\},\{16,18,2\},\{20,1,6\},
        \{3,22,19\},\{11,21,15\},\{14,10,5\}\}, \\ 
        \{\{15,10\},\{21,5\},\{11,19\},
        \{1,14\},\{6,22\},\{3,18\},
        \{8,20\},\{16,23\},\{2,17\},
        \{9,13\},\{24,4\},\{12,7\}\}\}}
        \caption{10-4-4-3-2-1 $(\mathbb{Q})$}
        \label{Dessin}
    \end{subfigure}\hfill
\end{figure}

\begin{figure}[H]
    \begin{subfigure}{0.5\textwidth}
        \centering \captionsetup{justification=centering}
        $\scalemath{0.75}{
        \displaystyle \begin{pmatrix}
            0 & 2 & 1 & 0 & 0 & 0 & 0 & 0\\ 
            2 & 0 & 1 & 0 & 0 & 0 & 0 & 0\\
            1 & 1 & 0 & 1 & 0 & 0 & 0 & 0\\
            0 & 0 & 1 & 0 & 1 & 1 & 0 & 0\\
            0 & 0 & 0 & 1 & 0 & 0 & 2 & 0\\
            0 & 0 & 0 & 1 & 0 & 0 & 0 & 2\\
            0 & 0 & 0 & 0 & 2 & 0 & 0 & 1\\
            0 & 0 & 0 & 0 & 0 & 2 & 1 & 0
        \end{pmatrix}}$
        $\vcenter{\hbox{\includegraphics[width=0.35\textwidth]{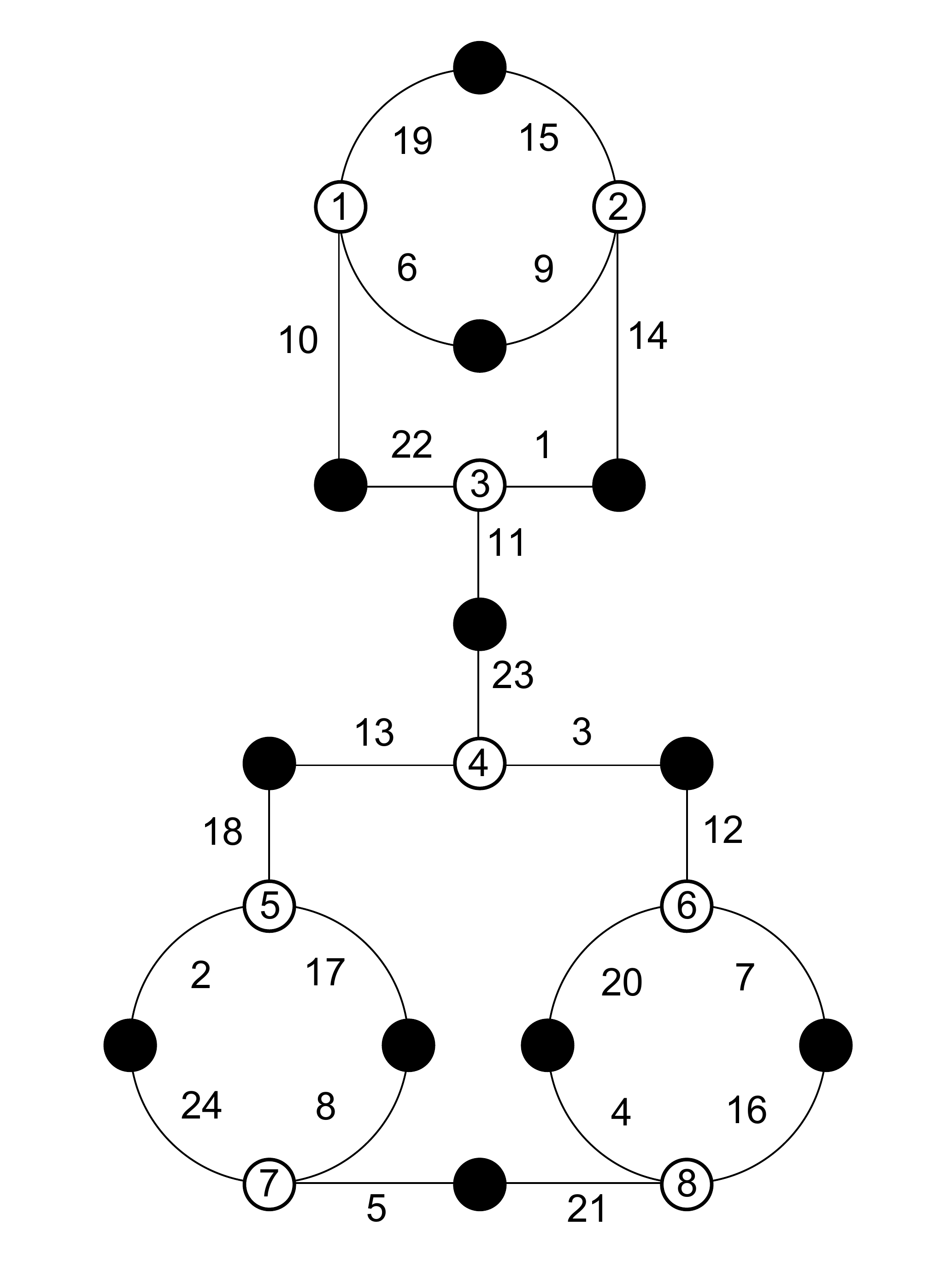}}}$
        \caption{ \{\{\{19,6,10\},\{15,14,9\},
        \{1,11,22\},\{23,3,13\},\{12,7,20\},
        \{4,16,21\},\{5,24,8\},\{2,18,17\}\}, \\ 
        \{\{5,21\},\{8,17\},\{2,24\},
        \{4,20\},\{7,16\},\{12,3\},
        \{13,18\},\{23,11\},\{1,14\},
        \{22,10\},\{6,9\},\{19,15\}\}\}}
        \caption{10-5-3-2-2-2 $(\mathbb{Q})$}
        \label{Dessin}
    \end{subfigure} \hfill
    \begin{subfigure}{0.5\textwidth}
        \centering \captionsetup{justification=centering}
        $\scalemath{0.75}{
        \displaystyle \begin{pmatrix}
            2 & 1 & 0 & 0 & 0 & 0 & 0 & 0\\ 
            1 & 0 & 2 & 0 & 0 & 0 & 0 & 0\\
            0 & 2 & 0 & 1 & 0 & 0 & 0 & 0\\
            0 & 0 & 1 & 0 & 1 & 0 & 0 & 1\\
            0 & 0 & 0 & 1 & 0 & 1 & 0 & 1\\
            0 & 0 & 0 & 0 & 1 & 0 & 2 & 0\\
            0 & 0 & 0 & 0 & 0 & 2 & 0 & 1\\
            0 & 0 & 0 & 1 & 1 & 0 & 1 & 0
        \end{pmatrix}}$
        $\vcenter{\hbox{\includegraphics[width=0.35\textwidth]{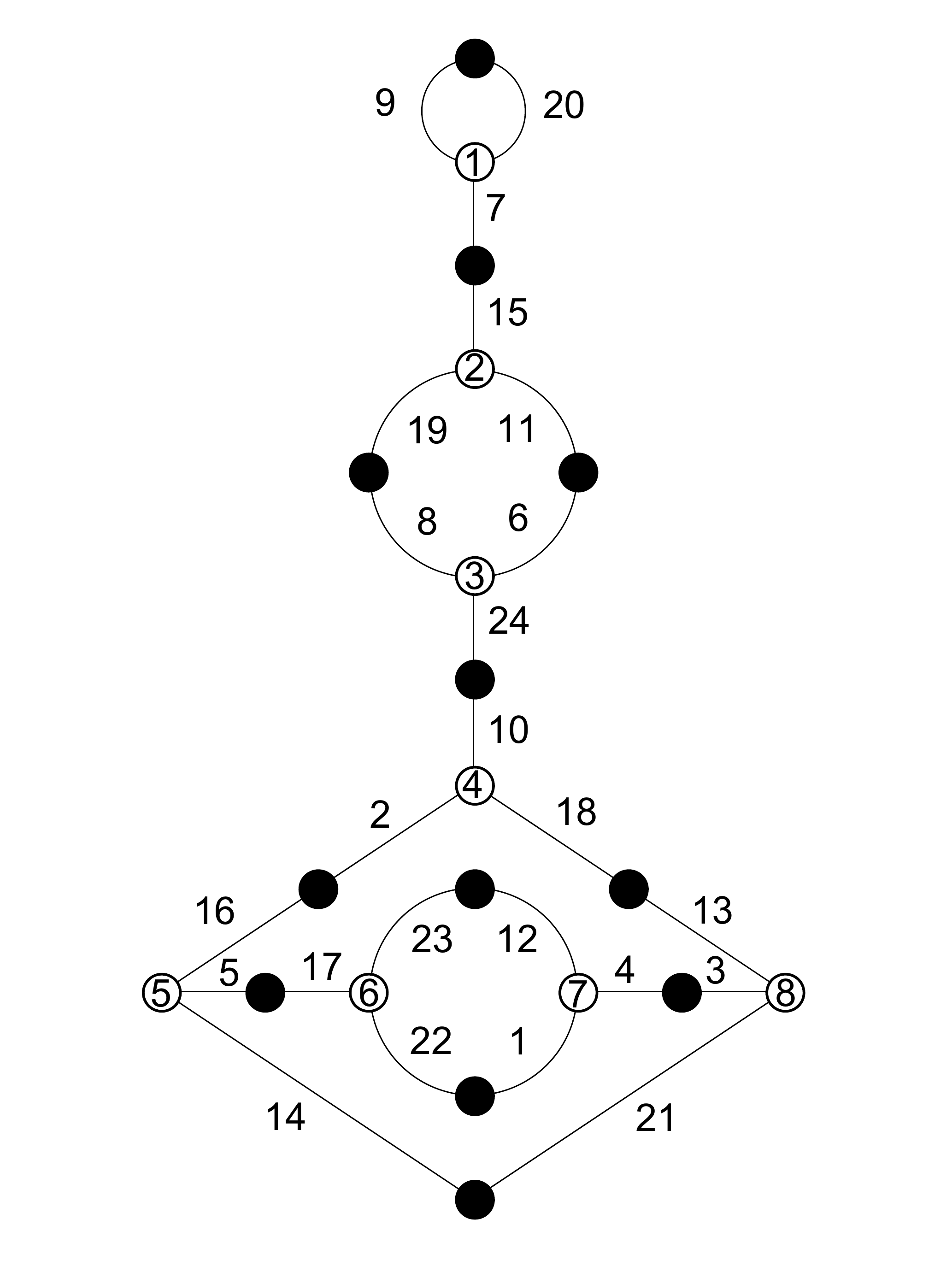}}}$
        \caption{ \{\{\{9,20,7\},\{15,11,19\},
        \{8,6,24\},\{10,18,2\},\{16,5,14\},
        \{17,23,22\},\{12,4,1\},\{3,13,21\}\}, \\ 
        \{\{14,21\},\{22,1\},\{23,12\},
        \{5,17\},\{4,3\},\{13,18\},
        \{16,2\},\{10,24\},\{6,11\},
        \{8,19\},\{15,7\},\{9,20\}\}\}}
        \caption{10-5-4-2-2-1 A (cubic)}
        \label{Dessin}
    \end{subfigure}\hfill
\end{figure}

\begin{figure}[H]
    \begin{subfigure}{0.6\textwidth}
        \centering \captionsetup{justification=centering}
        $\scalemath{0.75}{
        \displaystyle \begin{pmatrix}
            0 & 2 & 1 & 0 & 0 & 0 & 0 & 0\\ 
            2 & 0 & 0 & 1 & 0 & 0 & 0 & 0\\
            1 & 0 & 0 & 1 & 1 & 0 & 0 & 0\\
            0 & 1 & 1 & 0 & 0 & 0 & 1 & 0\\
            0 & 0 & 1 & 0 & 0 & 2 & 0 & 0\\
            0 & 0 & 0 & 0 & 2 & 0 & 1 & 0\\
            0 & 0 & 0 & 1 & 0 & 1 & 0 & 1\\
            0 & 0 & 0 & 0 & 0 & 0 & 1 & 2
        \end{pmatrix}}$
        $\vcenter{\hbox{\includegraphics[width=0.25\textwidth]{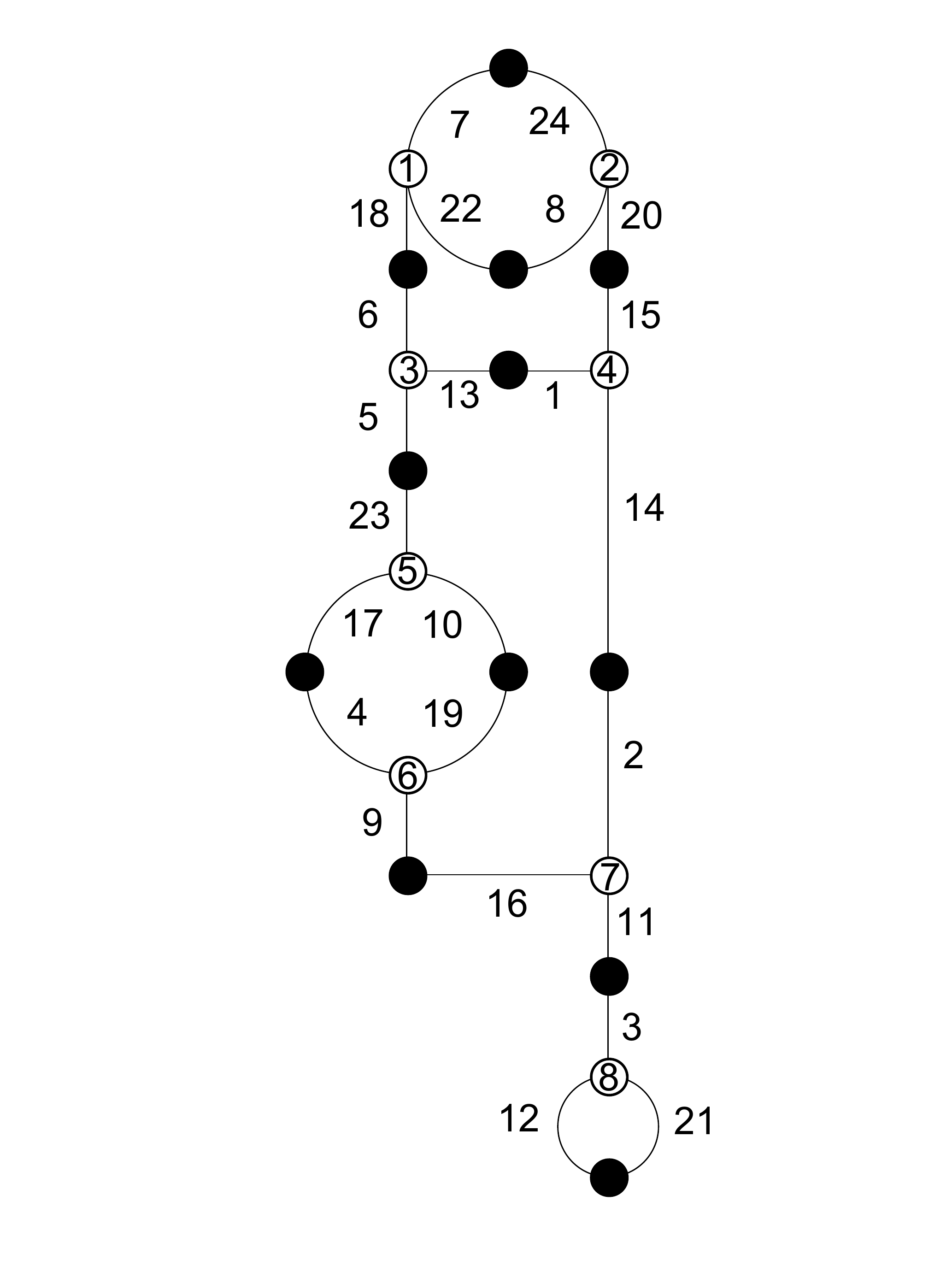}}}$
        $\vcenter{\hbox{\includegraphics[width=0.25\textwidth]{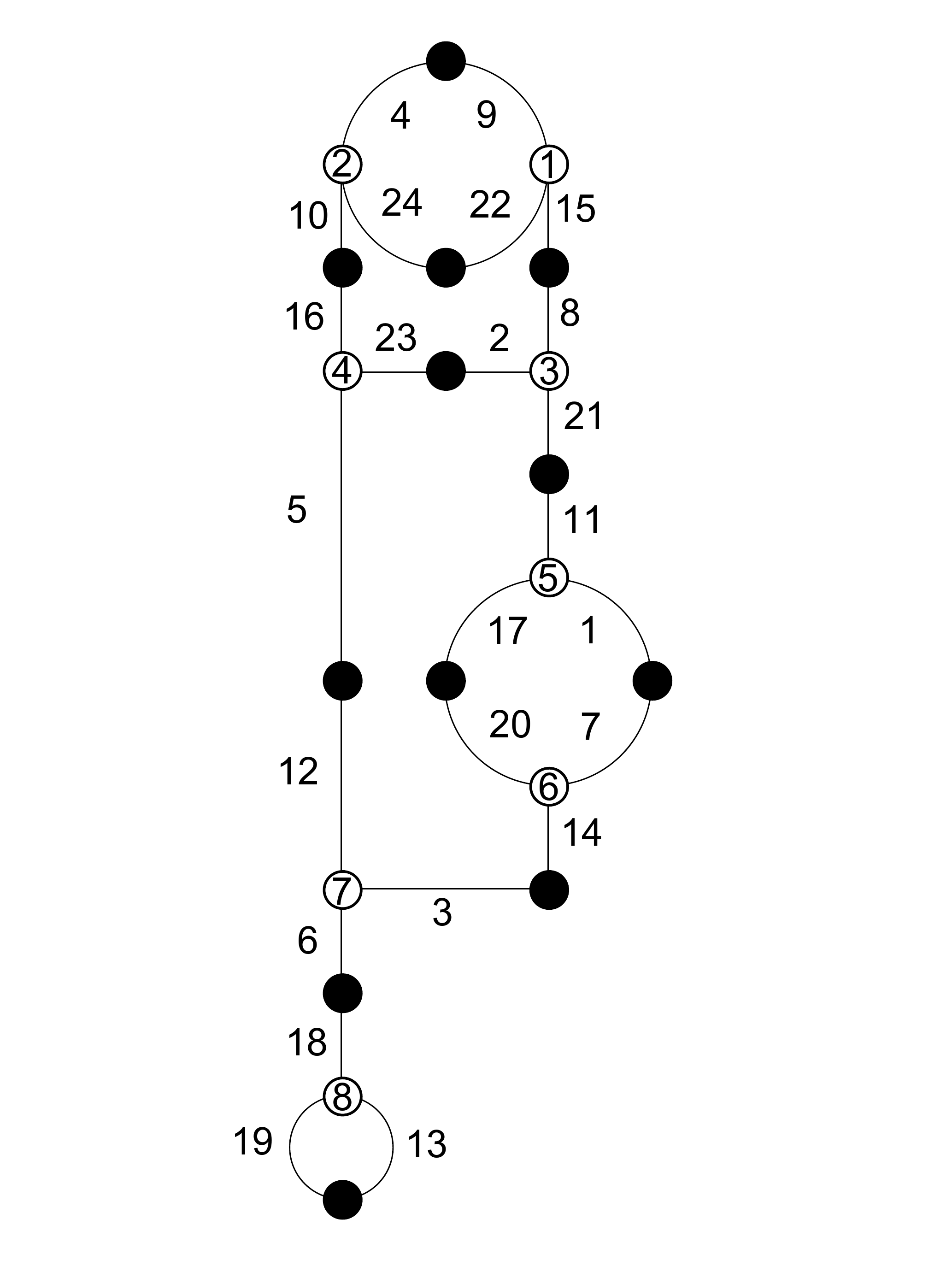}}}$
        \caption{ B: \{\{\{7,22,18\},\{24,20,8\},
        \{15,14,1\},\{6,13,5\},\{23,10,17\},
        \{19,9,4\},\{16,2,11\},\{3,21,12\}\}, \\ 
        \{\{7,24\},\{22,8\},\{18,6\},
        \{20,15\},\{1,13\},\{5,23\},
        \{17,4\},\{10,19\},\{9,16\},
        \{14,2\},\{11,3\},\{12,21\}\}\} \\
        C: \{\{\{4,24,10\},\{9,15,22\},
        \{8,21,2\},\{23,5,16\},\{11,1,17\},
        \{20,7,14\},\{3,6,12\},\{18,13,19\}\}, \\ 
        \{\{19,13\},\{18,6\},\{3,14\},
        \{7,1\},\{20,17\},\{11,21\},
        \{8,15\},\{2,23\},\{16,10\},
        \{5,12\},\{4,9\},\{24,22\}\}\}}
        \caption{10-5-4-2-2-1 B \& C (cubic)}
        \label{Dessin}
    \end{subfigure} \hfill
    \begin{subfigure}{0.4\textwidth}
        \centering \captionsetup{justification=centering}
        $\scalemath{0.75}{
        \displaystyle \begin{pmatrix}
            2 & 1 & 0 & 0 & 0 & 0 & 0 & 0\\ 
            1 & 0 & 1 & 0 & 0 & 1 & 0 & 0\\
            0 & 1 & 0 & 1 & 0 & 0 & 1 & 0\\
            0 & 0 & 1 & 0 & 2 & 0 & 0 & 0\\
            0 & 0 & 0 & 2 & 0 & 1 & 0 & 0\\
            0 & 1 & 0 & 0 & 1 & 0 & 1 & 0\\
            0 & 0 & 1 & 0 & 0 & 1 & 0 & 1\\
            0 & 0 & 0 & 0 & 0 & 0 & 1 & 2
        \end{pmatrix}}$
        $\vcenter{\hbox{\includegraphics[width=0.35\textwidth]{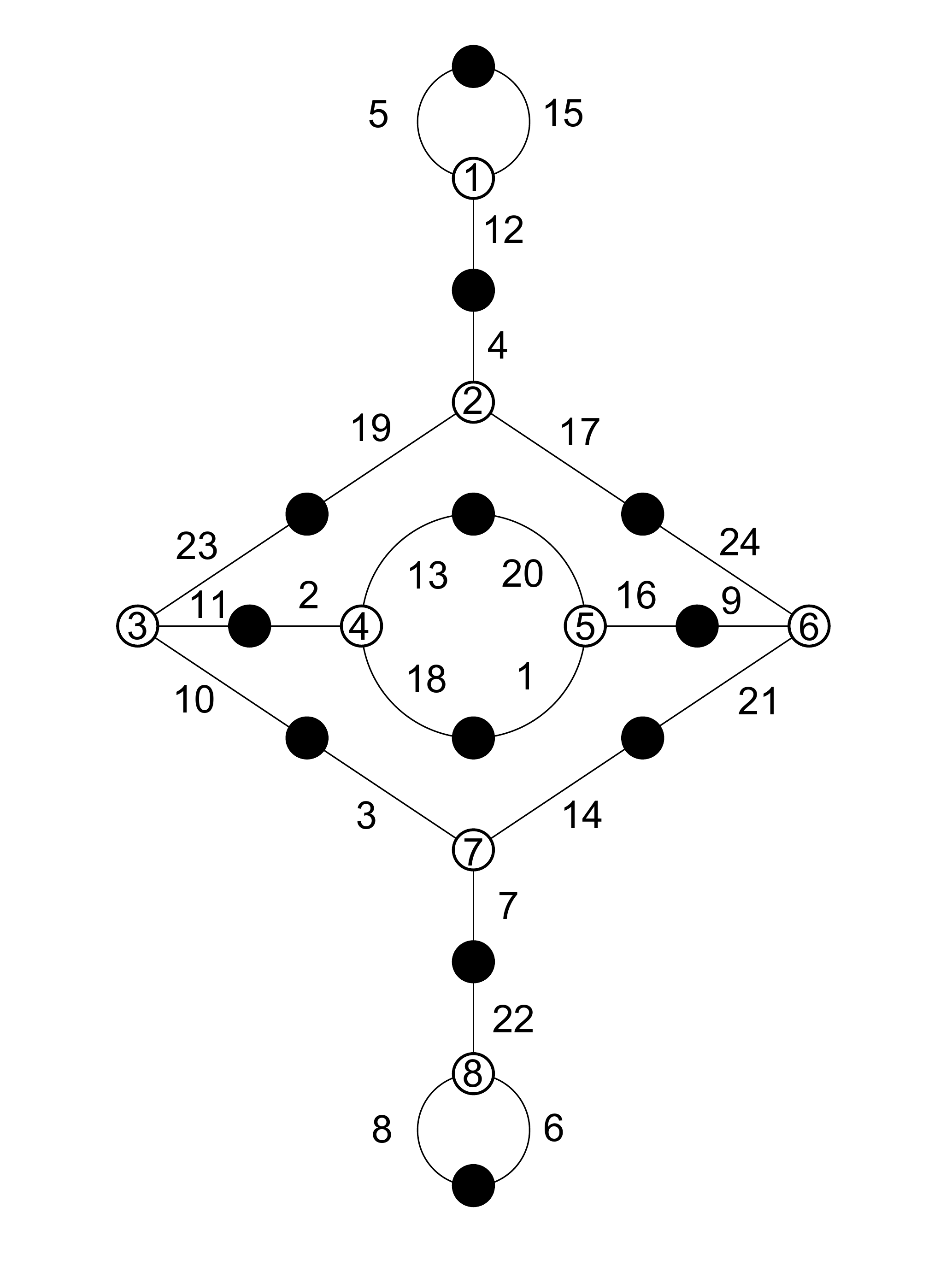}}}$
        \caption{ \{\{\{5,15,12\},\{4,17,19\},
        \{24,21,9\},\{16,1,20\},\{13,18,2\},
        \{11,10,23\},\{3,14,7\},\{22,6,8\}\}, \\ 
        \{\{6,8\},\{22,7\},\{3,10\},
        \{14,21\},\{9,16\},\{18,1\},
        \{13,20\},\{2,11\},\{23,19\},
        \{17,24\},\{4,12\},\{5,15\}\}\}}
        \caption{10-5-5-2-1-1 A $(\sqrt{5})$}
        \label{Dessin}
    \end{subfigure}\hfill
\end{figure}

\begin{figure}[H]
    \begin{subfigure}{0.5\textwidth}
        \centering \captionsetup{justification=centering}
        $\scalemath{0.75}{
        \displaystyle \begin{pmatrix}
            0 & 1 & 2 & 0 & 0 & 0 & 0 & 0\\ 
            1 & 2 & 0 & 0 & 0 & 0 & 0 & 0\\
            2 & 0 & 0 & 1 & 0 & 0 & 0 & 0\\
            0 & 0 & 1 & 0 & 2 & 0 & 0 & 0\\
            0 & 0 & 0 & 2 & 0 & 1 & 0 & 0\\
            0 & 0 & 0 & 0 & 1 & 0 & 0 & 2\\
            0 & 0 & 0 & 0 & 0 & 0 & 2 & 1\\
            0 & 0 & 0 & 0 & 0 & 2 & 1 & 0
        \end{pmatrix}}$
        $\vcenter{\hbox{\includegraphics[width=0.35\textwidth]{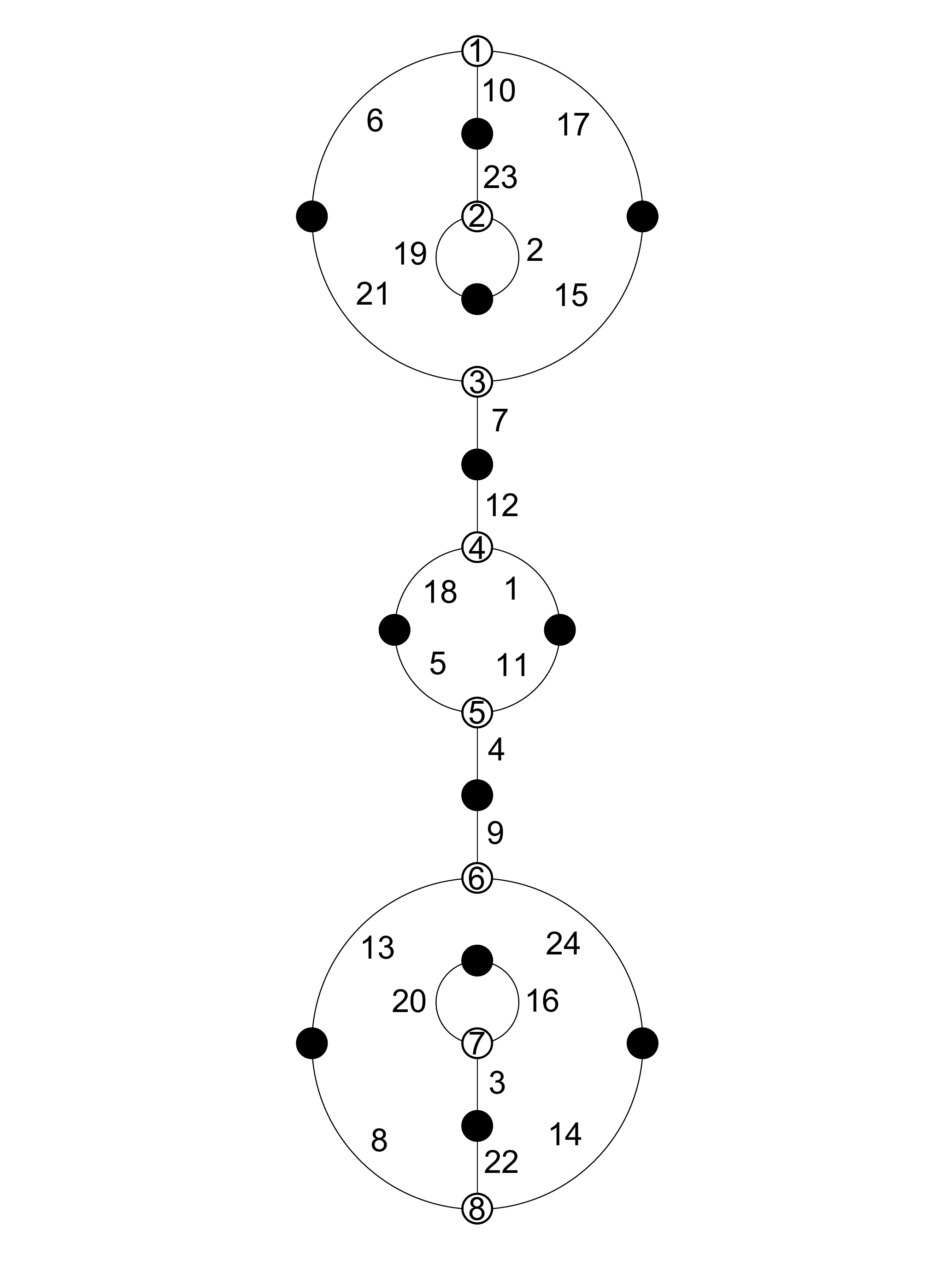}}}$
        \caption{ \{\{\{10,6,17\},\{23,2,19\},
        \{21,15,7\},\{12,1,18\},\{5,11,4\},
        \{9,24,13\},\{20,16,3\},\{22,14,8\}\}, \\ 
        \{\{14,24\},\{8,13\},\{20,16\},
        \{3,22\},\{9,4\},\{5,18\},
        \{1,11\},\{12,7\},\{21,6\},
        \{17,15\},\{23,10\},\{2,19\}\}\}}
        \caption{10-5-5-2-1-1 B $(\sqrt{5})$}
        \label{Dessin}
    \end{subfigure} \hfill
    \begin{subfigure}{0.5\textwidth}
        \centering \captionsetup{justification=centering}
        $\scalemath{0.75}{
        \displaystyle \begin{pmatrix}
            2 & 1 & 0 & 0 & 0 & 0 & 0 & 0\\ 
            1 & 0 & 1 & 1 & 0 & 0 & 0 & 0\\
            0 & 1 & 0 & 1 & 1 & 0 & 0 & 0\\
            0 & 1 & 1 & 0 & 0 & 1 & 0 & 0\\
            0 & 0 & 1 & 0 & 0 & 0 & 2 & 0\\
            0 & 0 & 0 & 1 & 0 & 0 & 0 & 2\\
            0 & 0 & 0 & 0 & 2 & 0 & 0 & 1\\
            0 & 0 & 0 & 0 & 0 & 2 & 1 & 0
        \end{pmatrix}}$
        $\vcenter{\hbox{\includegraphics[width=0.35\textwidth]{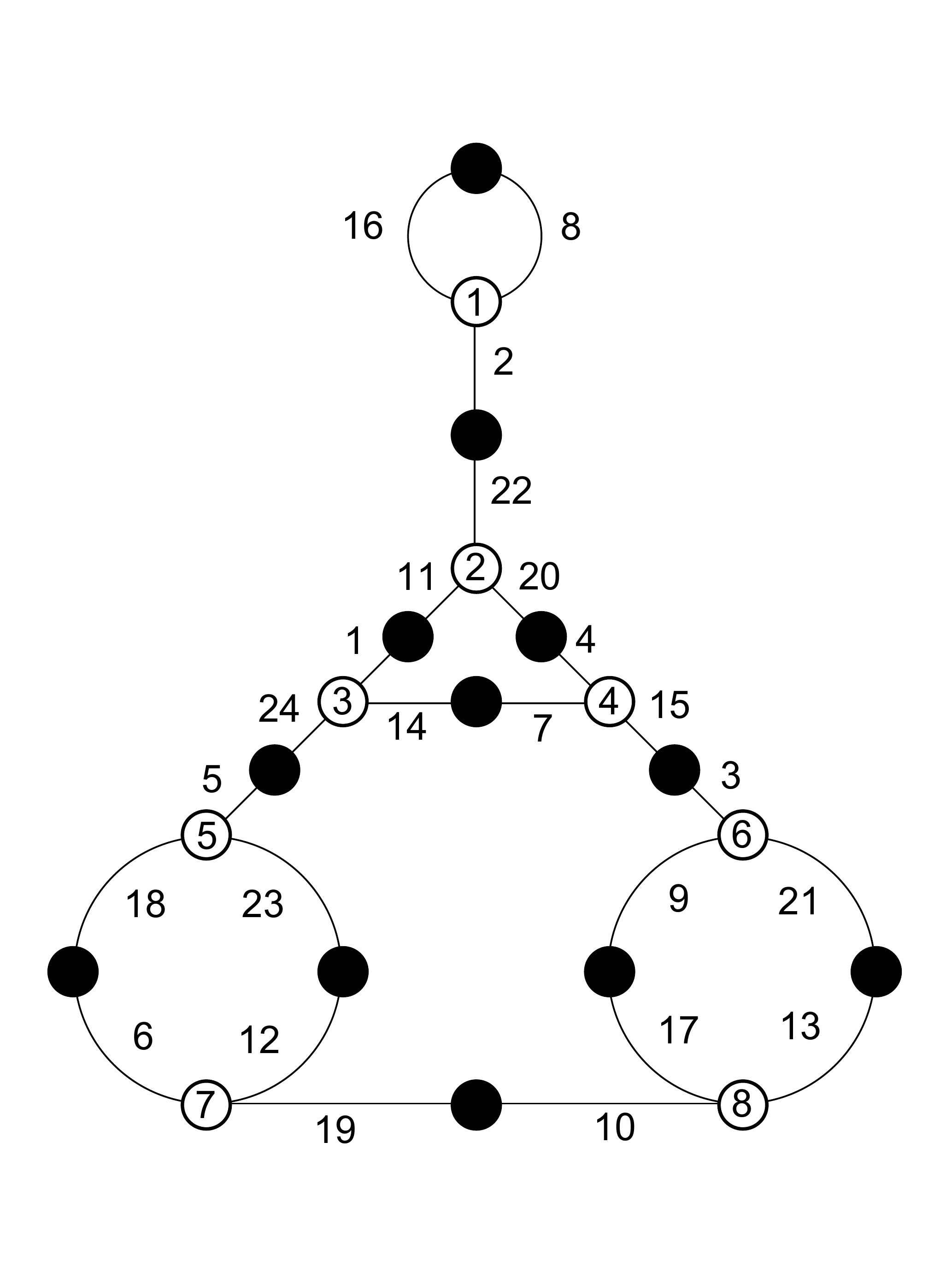}}}$
        \caption{ \{\{\{16,8,2\},\{22,20,11\},
        \{4,15,7\},\{1,14,24\},\{5,23,18\},
        \{6,12,19\},\{17,13,10\},\{9,3,21\}\}, \\ 
        \{\{19,10\},\{12,23\},\{6,18\},
        \{5,24\},\{17,9\},\{21,13\},
        \{3,15\},\{14,7\},\{1,11\},
        \{20,4\},\{22,2\},\{8,16\}\}\}}
        \caption{10-6-3-2-2-1 $(\mathbb{Q})$}
        \label{Dessin}
    \end{subfigure}\hfill
\end{figure}

\begin{figure}[H]
    \begin{subfigure}{0.5\textwidth}
        \centering \captionsetup{justification=centering}
        $\scalemath{0.75}{
        \displaystyle \begin{pmatrix}
            2 & 1 & 0 & 0 & 0 & 0 & 0 & 0\\ 
            1 & 0 & 1 & 0 & 0 & 1 & 0 & 0\\
            0 & 1 & 0 & 1 & 0 & 0 & 1 & 0\\
            0 & 0 & 1 & 0 & 2 & 0 & 0 & 0\\
            0 & 0 & 0 & 2 & 0 & 0 & 1 & 0\\
            0 & 1 & 0 & 0 & 0 & 0 & 1 & 1\\
            0 & 0 & 1 & 0 & 1 & 1 & 0 & 0\\
            0 & 0 & 0 & 0 & 0 & 1 & 0 & 2
        \end{pmatrix}}$
        $\vcenter{\hbox{\includegraphics[width=0.35\textwidth]{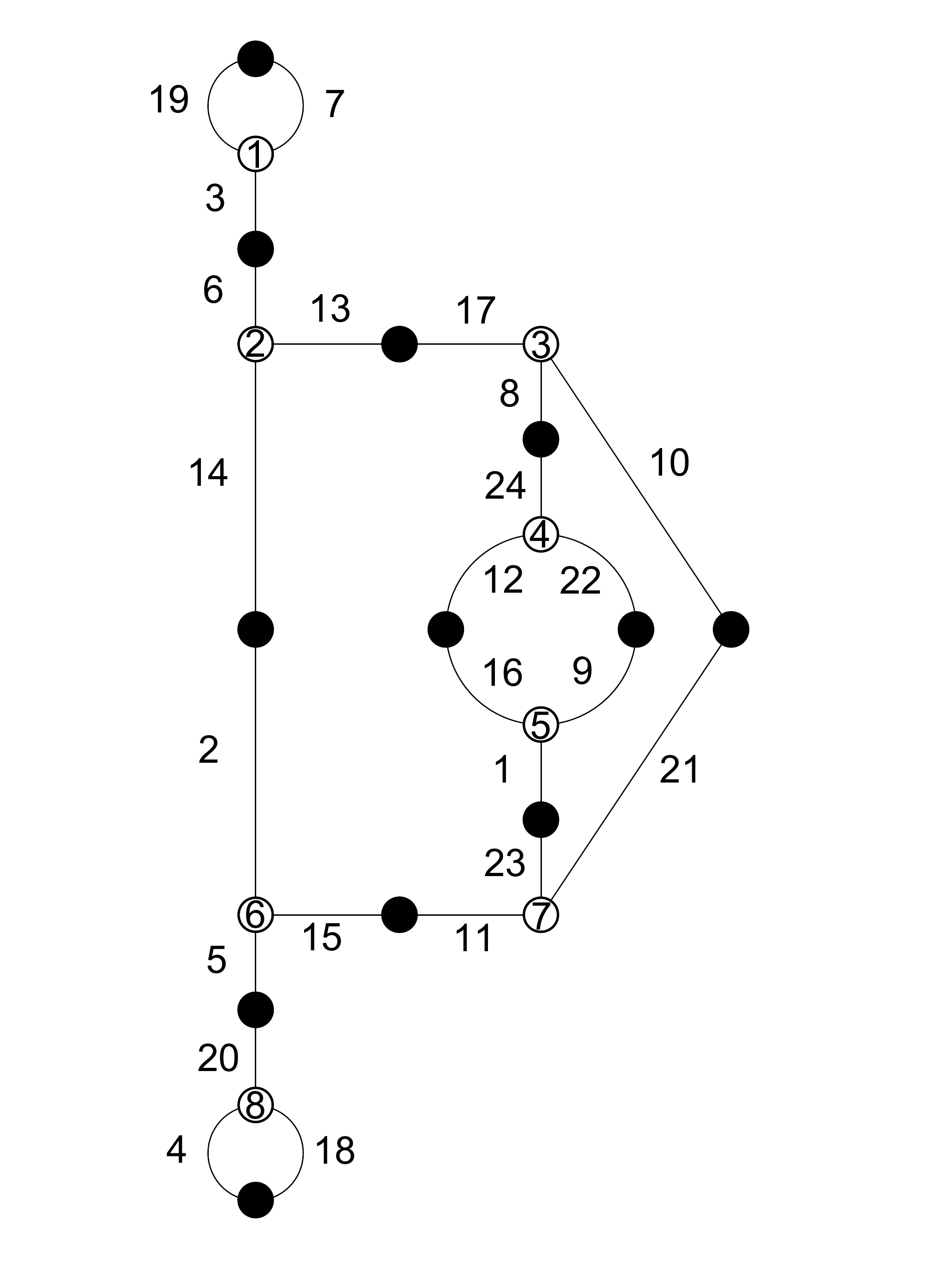}}}$
        \caption{ \{\{\{19,7,3\},\{6,13,14\},
        \{17,10,8\},\{24,22,12\},\{16,9,1\},
        \{23,21,11\},\{15,5,2\},\{20,18,4\}\}, \\ 
        \{\{4,18\},\{20,5\},\{15,11\},
        \{23,1\},\{21,10\},\{9,22\},
        \{16,12\},\{24,8\},\{17,13\},
        \{6,3\},\{19,7\},\{14,2\}\}\}}
        \caption{10-6-4-2-1-1 A $(\mathbb{Q})$}
        \label{Dessin}
    \end{subfigure} \hfill
    \begin{subfigure}{0.5\textwidth}
        \centering \captionsetup{justification=centering}
        $\scalemath{0.75}{
        \displaystyle \begin{pmatrix}
            2 & 1 & 0 & 0 & 0 & 0 & 0 & 0\\ 
            1 & 0 & 2 & 0 & 0 & 0 & 0 & 0\\
            0 & 2 & 0 & 0 & 1 & 0 & 0 & 0\\
            0 & 0 & 0 & 0 & 1 & 1 & 1 & 0\\
            0 & 0 & 1 & 1 & 0 & 1 & 0 & 0\\
            0 & 0 & 0 & 1 & 1 & 0 & 1 & 0\\
            0 & 0 & 0 & 1 & 0 & 1 & 0 & 1\\
            0 & 0 & 0 & 0 & 0 & 0 & 1 & 2
        \end{pmatrix}}$
        $\vcenter{\hbox{\includegraphics[width=0.35\textwidth]{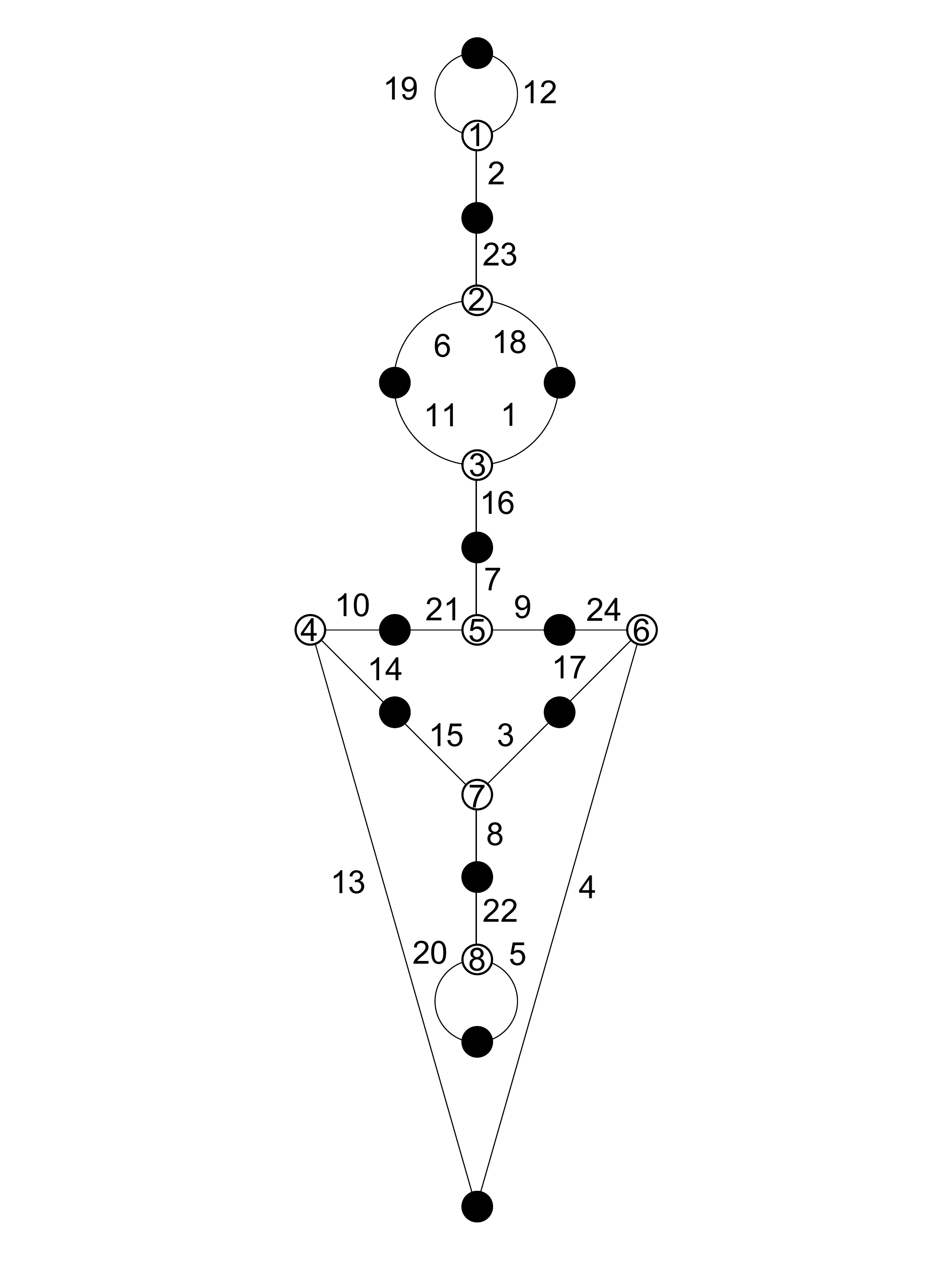}}}$
        \caption{ \{\{\{19,12,2\},\{23,18,6\},
        \{11,1,16\},\{7,9,21\},\{24,4,17\},
        \{10,14,13\},\{15,3,8\},\{22,5,20\}\}, \\ 
        \{\{5,20\},\{8,22\},\{14,15\},
        \{3,17\},\{9,24\},\{10,21\},
        \{13,4\},\{7,16\},\{1,18\},
        \{6,11\},\{23,2\},\{12,19\}\}\}}
        \caption{10-6-4-2-1-1 B $(\mathbb{Q})$}
        \label{Dessin}
    \end{subfigure}\hfill
\end{figure}

\begin{figure}[H]
    \begin{subfigure}{0.5\textwidth}
        \centering \captionsetup{justification=centering}
        $\scalemath{0.75}{
        \displaystyle \begin{pmatrix}
            2 & 1 & 0 & 0 & 0 & 0 & 0 & 0\\ 
            1 & 0 & 1 & 0 & 0 & 1 & 0 & 0\\
            0 & 1 & 0 & 1 & 0 & 0 & 1 & 0\\
            0 & 0 & 1 & 0 & 1 & 0 & 1 & 0\\
            0 & 0 & 0 & 1 & 2 & 0 & 0 & 0\\
            0 & 1 & 0 & 0 & 0 & 0 & 1 & 1\\
            0 & 0 & 1 & 1 & 0 & 1 & 0 & 0\\
            0 & 0 & 0 & 0 & 0 & 1 & 0 & 2
        \end{pmatrix}}$
        $\vcenter{\hbox{\includegraphics[width=0.35\textwidth]{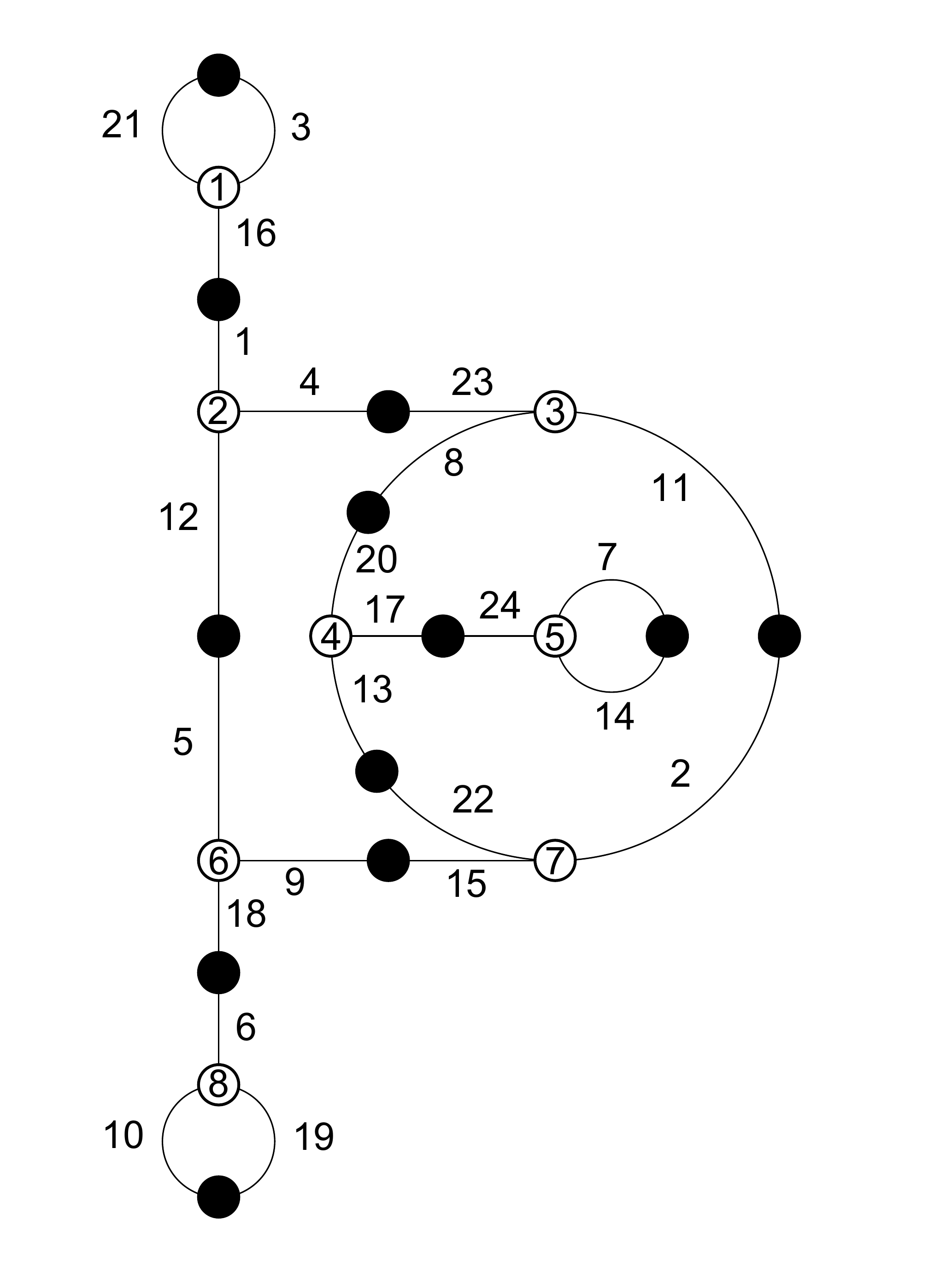}}}$
        \caption{ \{\{\{21,3,16\},\{1,4,12\},
        \{23,11,8\},\{20,17,13\},\{24,7,14\},
        \{2,15,22\},\{9,18,5\},\{6,19,10\}\}, \\ 
        \{\{10,19\},\{6,18\},\{9,15\},
        \{13,22\},\{2,11\},\{17,24\},
        \{7,14\},\{20,8\},\{4,23\},
        \{5,12\},\{1,16\},\{3,21\}\}\}}
        \caption{10-6-5-1-1-1 A $(\mathbb{Q})$}
        \label{Dessin}
    \end{subfigure} \hfill
    \begin{subfigure}{0.5\textwidth}
        \centering \captionsetup{justification=centering}
        $\scalemath{0.75}{
        \displaystyle \begin{pmatrix}
            2 & 1 & 0 & 0 & 0 & 0 & 0 & 0\\ 
            1 & 0 & 1 & 1 & 0 & 0 & 0 & 0\\
            0 & 1 & 2 & 0 & 0 & 0 & 0 & 0\\
            0 & 1 & 0 & 0 & 0 & 0 & 0 & 2\\
            0 & 0 & 0 & 0 & 0 & 1 & 2 & 0\\
            0 & 0 & 0 & 0 & 1 & 2 & 0 & 0\\
            0 & 0 & 0 & 0 & 2 & 0 & 0 & 1\\
            0 & 0 & 0 & 2 & 0 & 0 & 1 & 0
        \end{pmatrix}}$
        $\vcenter{\hbox{\includegraphics[width=0.35\textwidth]{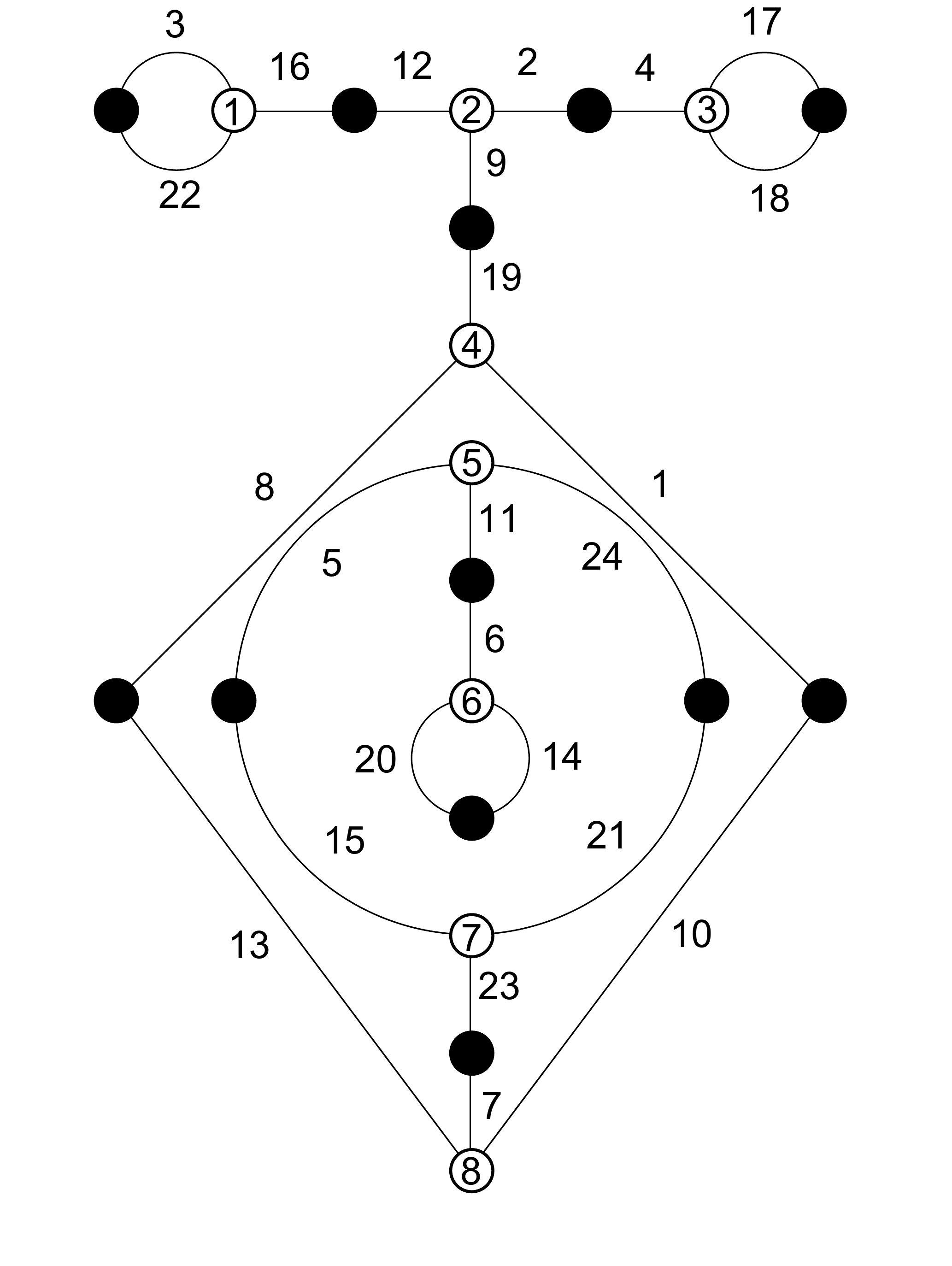}}}$
        \caption{ \{\{\{3,16,22\},\{12,2,9\},
        \{4,17,18\},\{19,1,8\},\{11,5,24\},
        \{6,14,20\},\{15,21,23\},\{7,10,13\}\}, \\ 
        \{\{7,23\},\{10,1\},\{13,8\},
        \{20,14\},\{15,5\},\{21,24\},
        \{11,6\},\{19,9\},\{2,4\},
        \{17,18\},\{16,12\},\{3,22\}\}\}}
        \caption{10-6-5-1-1-1 B (cubic)}
        \label{Dessin}
    \end{subfigure}\hfill
\end{figure}

\begin{figure}[H]
    \begin{subfigure}{0.6\textwidth}
        \centering \captionsetup{justification=centering}
        $\scalemath{0.75}{
        \displaystyle \begin{pmatrix}
            0 & 1 & 2 & 0 & 0 & 0 & 0 & 0\\ 
            1 & 2 & 0 & 0 & 0 & 0 & 0 & 0\\
            2 & 0 & 0 & 1 & 0 & 0 & 0 & 0\\
            0 & 0 & 1 & 0 & 0 & 1 & 1 & 0\\
            0 & 0 & 0 & 0 & 2 & 1 & 0 & 0\\
            0 & 0 & 0 & 1 & 1 & 0 & 1 & 0\\
            0 & 0 & 0 & 1 & 0 & 1 & 0 & 1\\
            0 & 0 & 0 & 0 & 0 & 0 & 1 & 2
        \end{pmatrix}}$
        $\vcenter{\hbox{\includegraphics[width=0.25\textwidth]{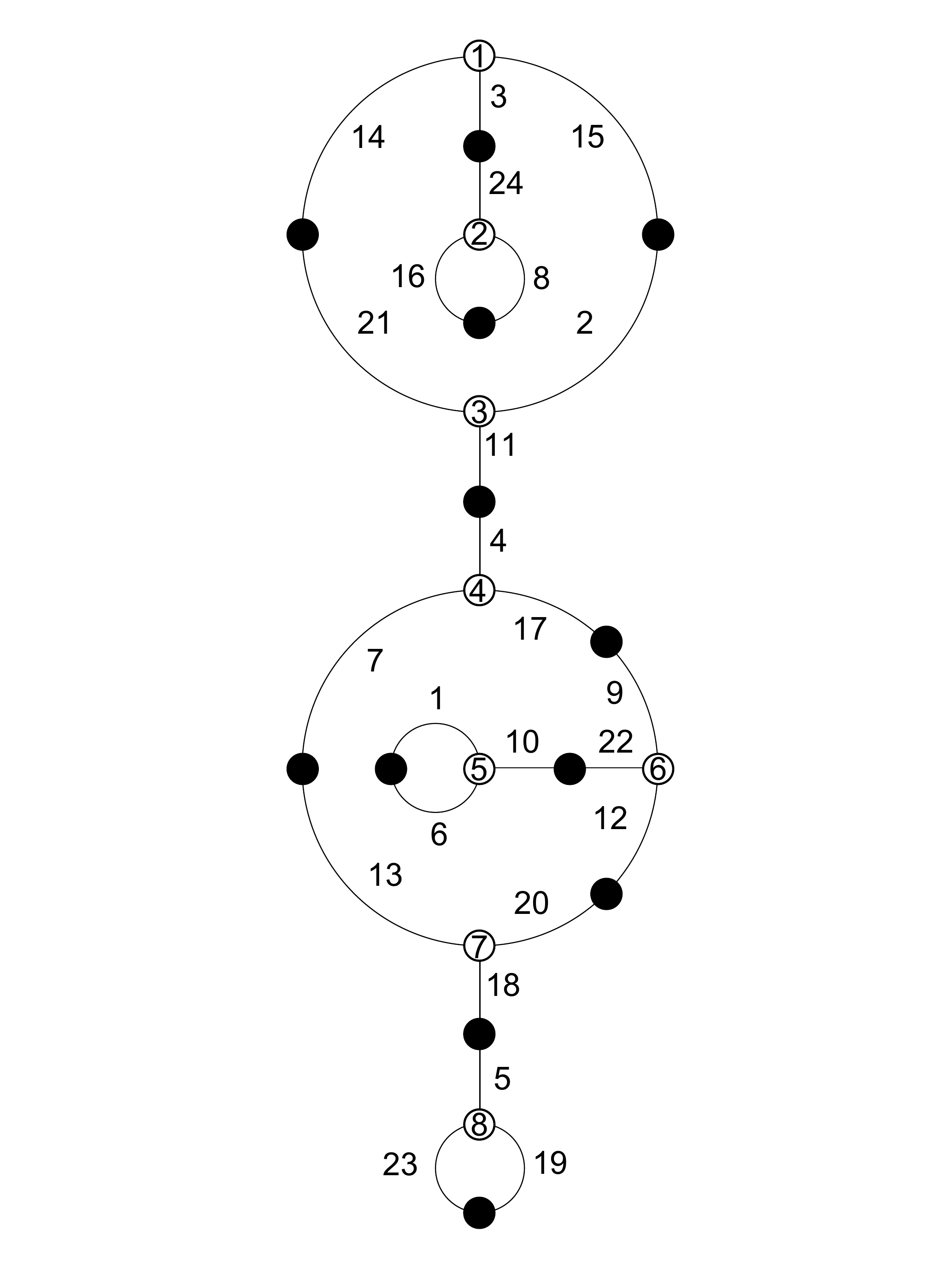}}}$
        $\vcenter{\hbox{\includegraphics[width=0.25\textwidth]{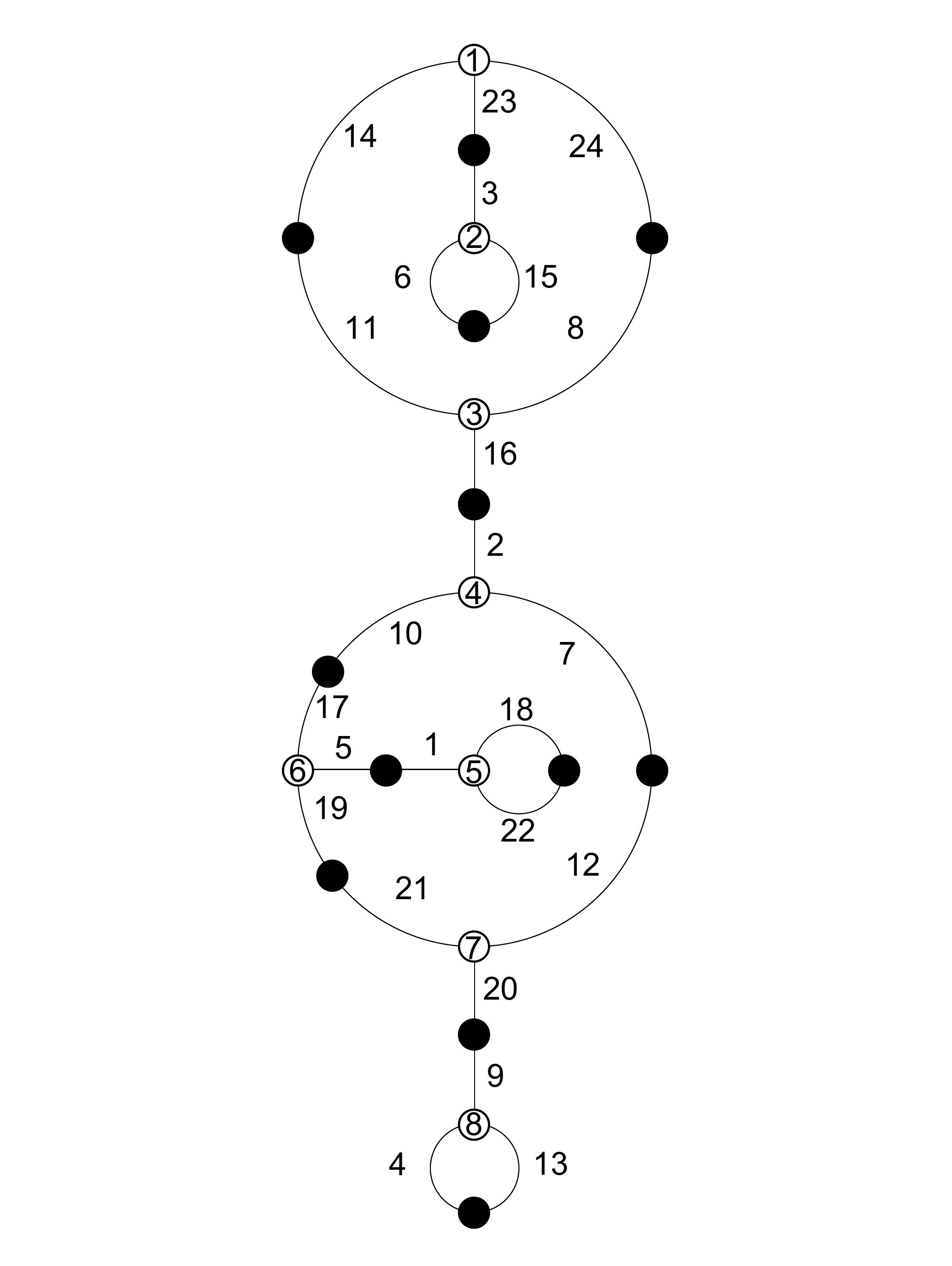}}}$
        \caption{ C: \{\{\{3,14,15\},\{24,8,16\},
        \{21,2,11\},\{4,17,7\},\{9,12,22\},
        \{10,6,1\},\{20,18,13\},\{5,19,23\}\}, \\ 
        \{\{23,19\},\{5,18\},\{20,12\},
        \{10,22\},\{1,6\},\{13,7\},
        \{17,9\},\{4,11\},\{2,15\},
        \{14,21\},\{3,24\},\{16,8\}\}\} \\
        D: \{\{\{23,14,24\},\{3,15,6\},
        \{11,8,16\},\{2,7,10\},\{17,5,19\},
        \{1,18,22\},\{21,12,20\},\{9,13,4\}\}, \\ 
        \{\{4,13\},\{9,20\},\{19,21\},
        \{12,7\},\{5,1\},\{18,22\},
        \{17,10\},\{2,16\},\{11,14\},
        \{24,8\},\{23,3\},\{6,15\}\}\}}
        \caption{10-6-5-1-1-1 C \& D (cubic)}
        \label{Dessin}
    \end{subfigure} \hfill
    \begin{subfigure}{0.4\textwidth}
        \centering \captionsetup{justification=centering}
        $\scalemath{0.75}{
        \displaystyle \begin{pmatrix}
            2 & 1 & 0 & 0 & 0 & 0 & 0 & 0\\ 
            1 & 0 & 1 & 1 & 0 & 0 & 0 & 0\\
            0 & 1 & 0 & 0 & 2 & 0 & 0 & 0\\
            0 & 1 & 0 & 0 & 0 & 2 & 0 & 0\\
            0 & 0 & 2 & 0 & 0 & 0 & 1 & 0\\
            0 & 0 & 0 & 2 & 0 & 0 & 0 & 1\\
            0 & 0 & 0 & 0 & 1 & 0 & 0 & 2\\
            0 & 0 & 0 & 0 & 0 & 1 & 2 & 0
        \end{pmatrix}}$
        $\vcenter{\hbox{\includegraphics[width=0.35\textwidth]{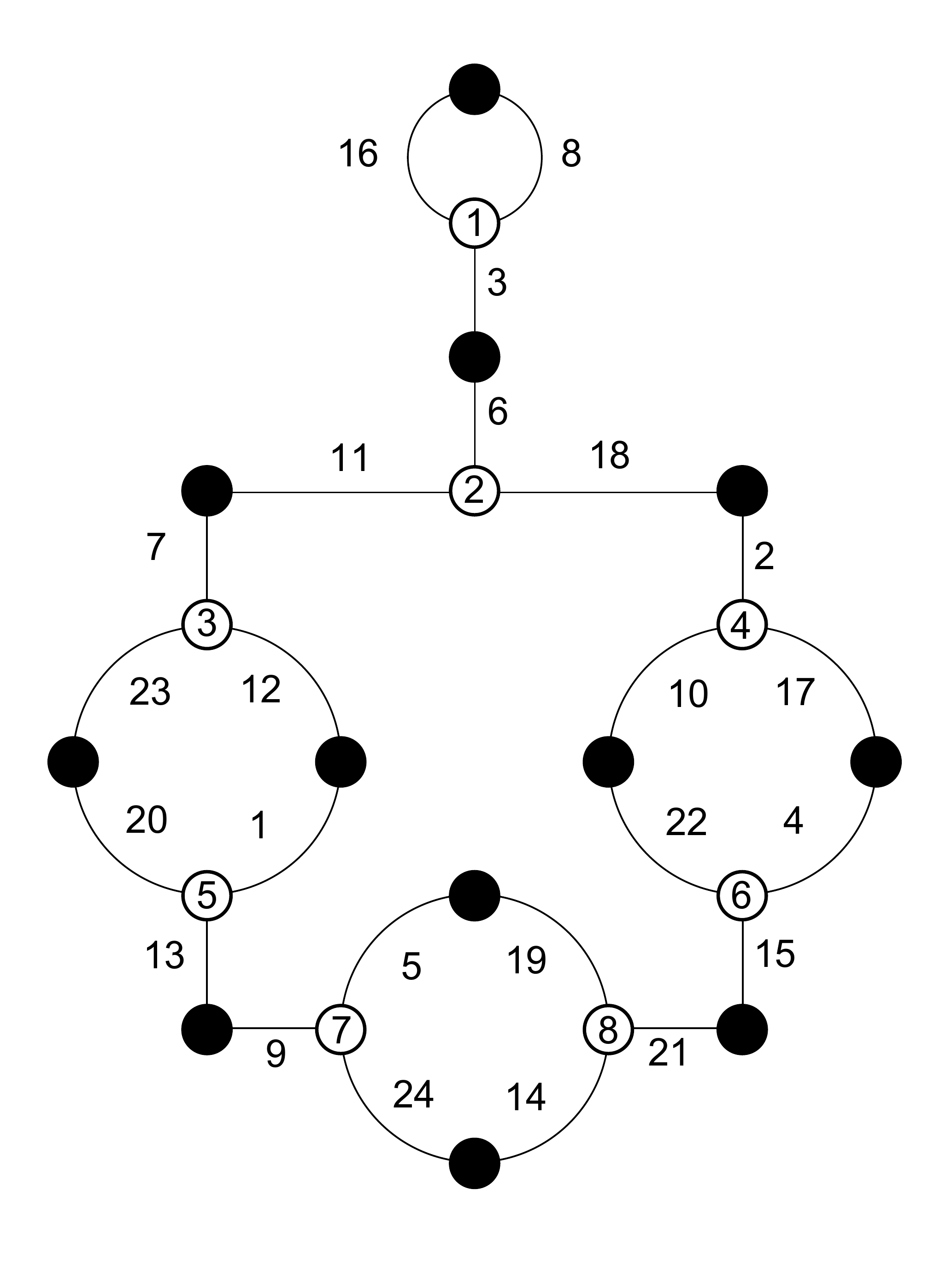}}}$
        \caption{ \{\{\{16,8,3\},\{6,18,11\},
        \{7,12,23\},\{2,17,10\},\{22,4,15\},
        \{21,14,19\},\{24,9,5\},\{13,20,1\}\}, \\ 
        \{\{24,14\},\{5,19\},\{9,13\},
        \{21,15\},\{1,12\},\{20,23\},
        \{7,11\},\{22,10\},\{17,4\},
        \{2,18\},\{6,3\},\{8,16\}\}\}}
        \caption{10-7-2-2-2-1 $(\mathbb{Q})$}
        \label{Dessin}
    \end{subfigure}\hfill
\end{figure}

\begin{figure}[H]
    \begin{subfigure}{0.5\textwidth}
        \centering \captionsetup{justification=centering}
        $\scalemath{0.75}{
        \displaystyle \begin{pmatrix}
            2 & 1 & 0 & 0 & 0 & 0 & 0 & 0\\ 
            1 & 0 & 1 & 1 & 0 & 0 & 0 & 0\\
            0 & 1 & 2 & 0 & 0 & 0 & 0 & 0\\
            0 & 1 & 0 & 0 & 0 & 0 & 0 & 2\\
            0 & 0 & 0 & 0 & 0 & 2 & 1 & 0\\
            0 & 0 & 0 & 0 & 2 & 0 & 1 & 0\\
            0 & 0 & 0 & 0 & 1 & 1 & 0 & 1\\
            0 & 0 & 0 & 2 & 0 & 0 & 1 & 0
        \end{pmatrix}}$
        $\vcenter{\hbox{\includegraphics[width=0.35\textwidth]{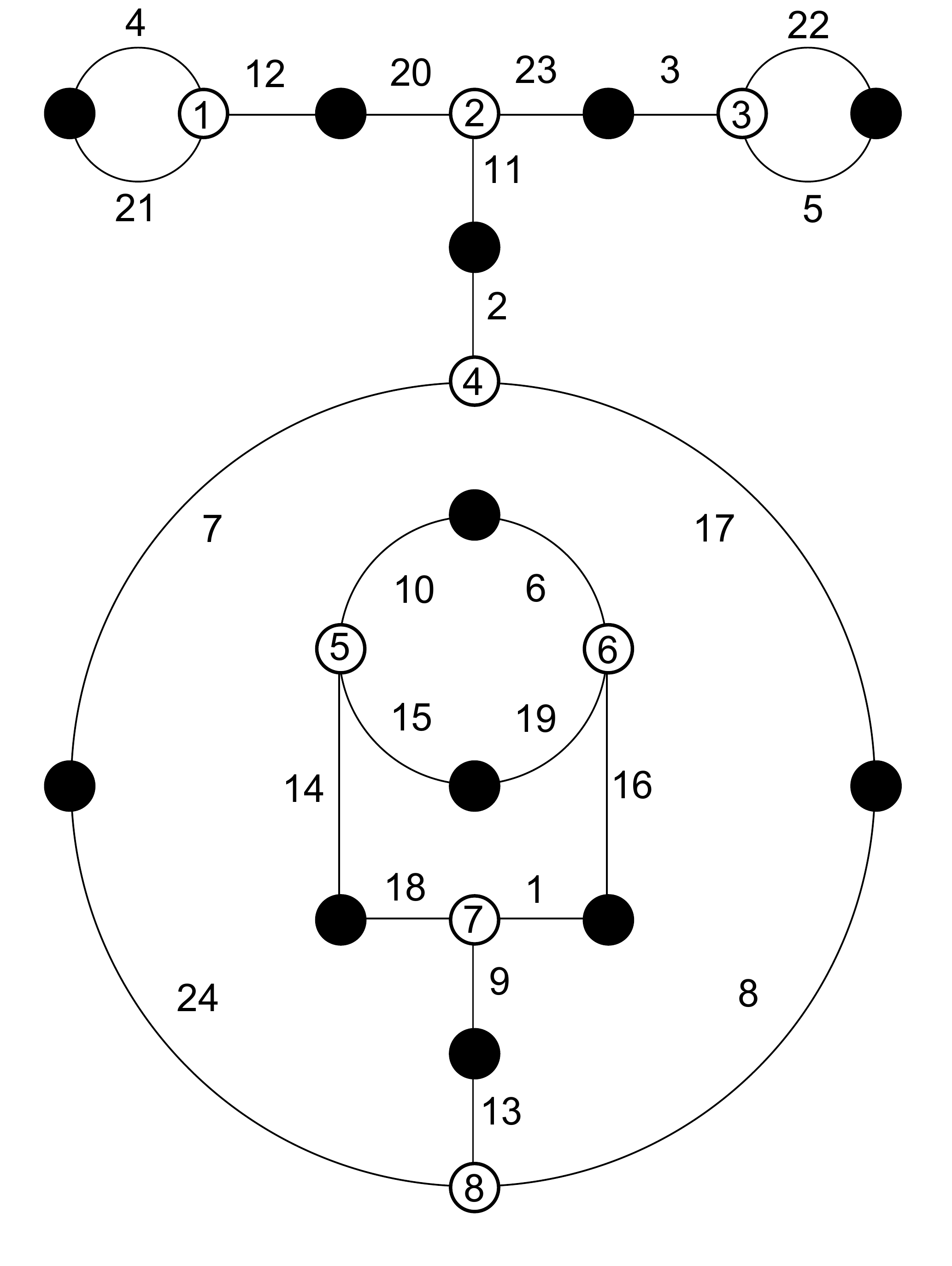}}}$
        \caption{ \{\{\{4,12,21\},\{20,23,11\},
        \{3,22,5\},\{2,17,7\},\{24,13,8\},
        \{9,18,1\},\{14,10,15\},\{6,16,19\}\}, \\ 
        \{\{4,21\},\{12,20\},\{23,3\},
        \{22,5\},\{11,2\},\{7,24\},
        \{17,8\},\{13,9\},\{1,16\},
        \{18,14\},\{10,6\},\{15,19\}\}\}}
        \caption{10-7-3-2-1-1 A $(\sqrt{21})$}
        \label{Dessin}
    \end{subfigure} \hfill
    \begin{subfigure}{0.5\textwidth}
        \centering \captionsetup{justification=centering}
        $\scalemath{0.75}{
        \displaystyle \begin{pmatrix}
            0 & 1 & 1 & 0 & 1 & 0 & 0 & 0\\ 
            1 & 0 & 1 & 1 & 0 & 0 & 0 & 0\\
            1 & 1 & 0 & 0 & 1 & 0 & 0 & 0\\
            0 & 1 & 0 & 2 & 0 & 0 & 0 & 0\\
            1 & 0 & 1 & 0 & 0 & 1 & 0 & 0\\
            0 & 0 & 0 & 0 & 1 & 0 & 2 & 0\\
            0 & 0 & 0 & 0 & 0 & 2 & 0 & 1\\
            0 & 0 & 0 & 0 & 0 & 0 & 1 & 2
        \end{pmatrix}}$
        $\vcenter{\hbox{\includegraphics[width=0.35\textwidth]{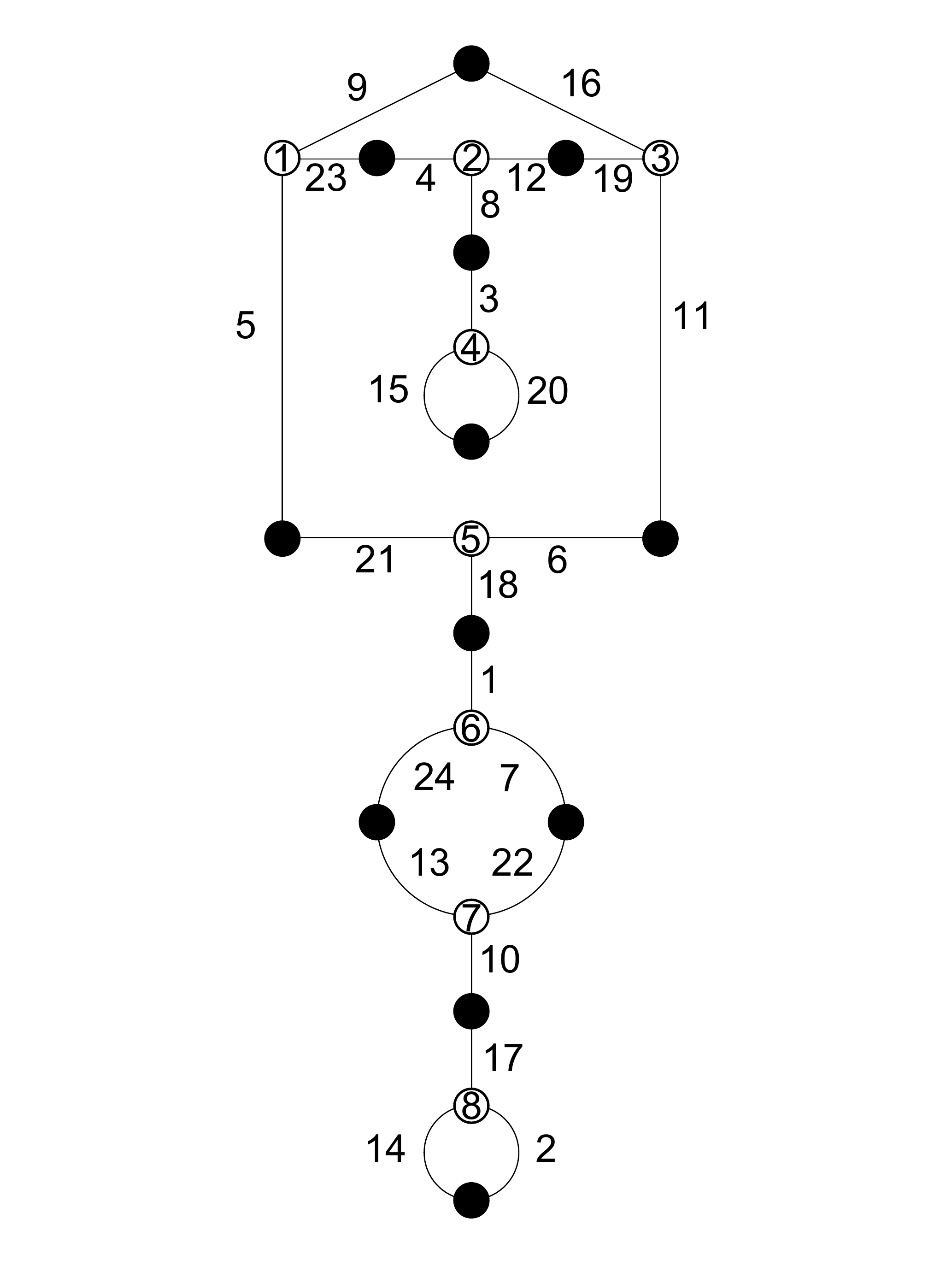}}}$
        \caption{ \{\{\{9,23,5\},\{4,12,8\},
        \{3,20,15\},\{19,16,11\},\{21,6,18\},
        \{1,7,24\},\{13,22,10\},\{17,2,14\}\}, \\ 
        \{\{14,2\},\{17,10\},\{22,7\},
        \{13,24\},\{1,18\},\{21,5\},
        \{6,11\},\{15,20\},\{3,8\},
        \{23,4\},\{12,19\},\{9,16\}\}\}}
        \caption{10-7-3-2-1-1 B $(\sqrt{21})$}
        \label{Dessin}
    \end{subfigure}\hfill
\end{figure}

\begin{figure}[H]
    \begin{subfigure}{0.5\textwidth}
        \centering \captionsetup{justification=centering}
        $\scalemath{0.75}{
        \displaystyle \begin{pmatrix}
            2 & 1 & 0 & 0 & 0 & 0 & 0 & 0\\ 
            1 & 0 & 1 & 0 & 1 & 0 & 0 & 0\\
            0 & 1 & 0 & 0 & 0 & 1 & 1 & 0\\
            0 & 0 & 0 & 2 & 0 & 1 & 0 & 0\\
            0 & 1 & 0 & 0 & 0 & 1 & 1 & 0\\
            0 & 0 & 1 & 1 & 1 & 0 & 0 & 0\\
            0 & 0 & 1 & 0 & 1 & 0 & 0 & 1\\
            0 & 0 & 0 & 0 & 0 & 0 & 1 & 2
        \end{pmatrix}}$
        $\vcenter{\hbox{\includegraphics[width=0.35\textwidth]{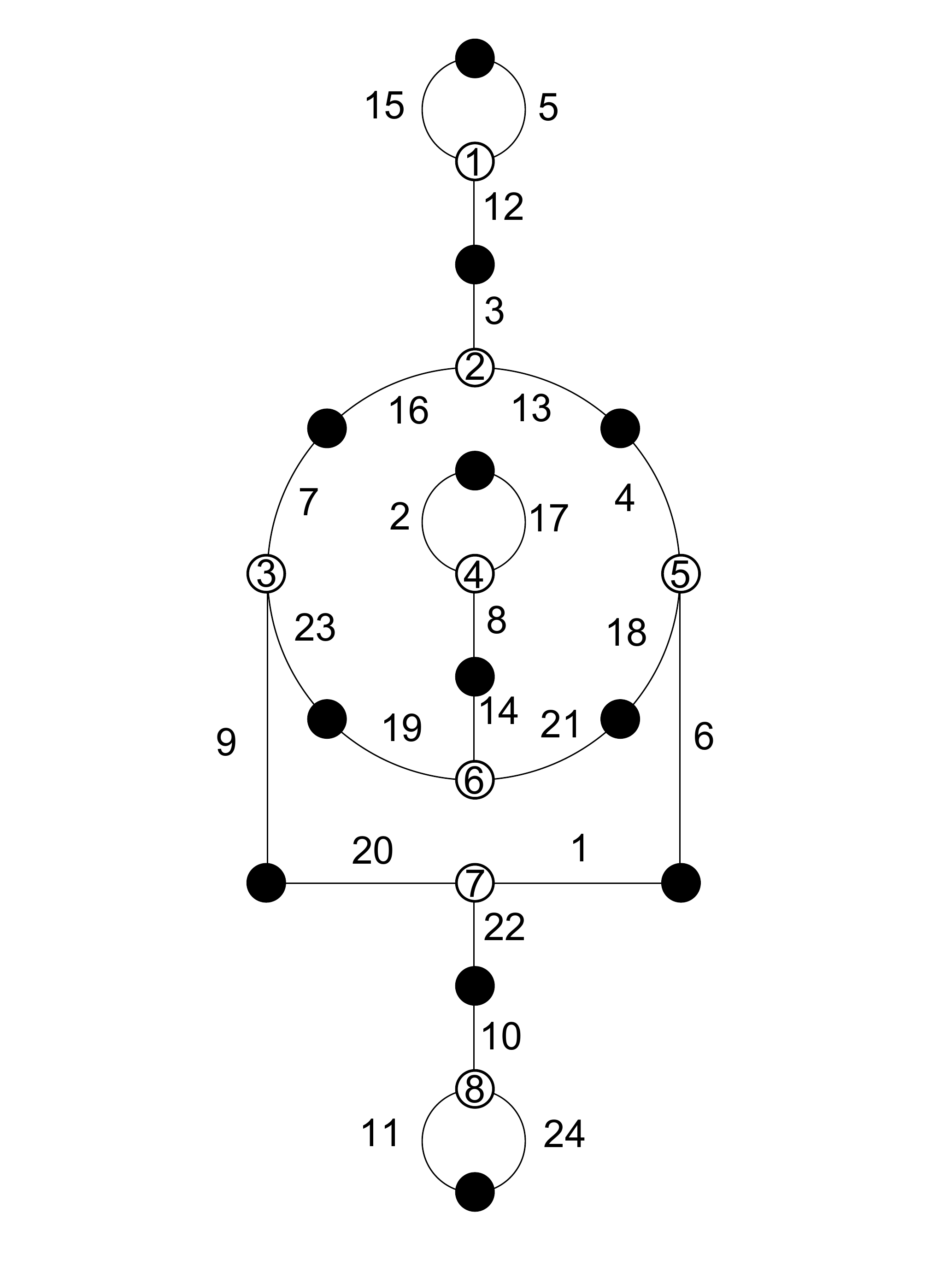}}}$
        \caption{ \{\{\{15,5,12\},\{3,13,16\},
        \{4,6,18\},\{2,17,8\},\{14,21,19\},
        \{23,9,7\},\{20,1,22\},\{10,24,11\}\}, \\ 
        \{\{11,24\},\{10,22\},\{1,6\},
        \{20,9\},\{23,19\},\{14,8\},
        \{21,18\},\{2,17\},\{4,13\},
        \{7,16\},\{3,12\},\{5,15\}\}\}}
        \caption{10-7-4-1-1-1 $(\mathbb{Q})$}
        \label{Dessin}
    \end{subfigure} \hfill
    \begin{subfigure}{0.5\textwidth}
        \centering \captionsetup{justification=centering}
        $\scalemath{0.75}{
        \displaystyle \begin{pmatrix}
            2 & 1 & 0 & 0 & 0 & 0 & 0 & 0\\ 
            1 & 0 & 1 & 0 & 0 & 1 & 0 & 0\\
            0 & 1 & 0 & 0 & 1 & 0 & 1 & 0\\
            0 & 0 & 0 & 2 & 1 & 0 & 0 & 0\\
            0 & 0 & 1 & 1 & 0 & 0 & 1 & 0\\
            0 & 1 & 0 & 0 & 0 & 0 & 1 & 1\\
            0 & 0 & 1 & 0 & 1 & 1 & 0 & 0\\
            0 & 0 & 0 & 0 & 0 & 1 & 0 & 2
        \end{pmatrix}}$
        $\vcenter{\hbox{\includegraphics[width=0.35\textwidth]{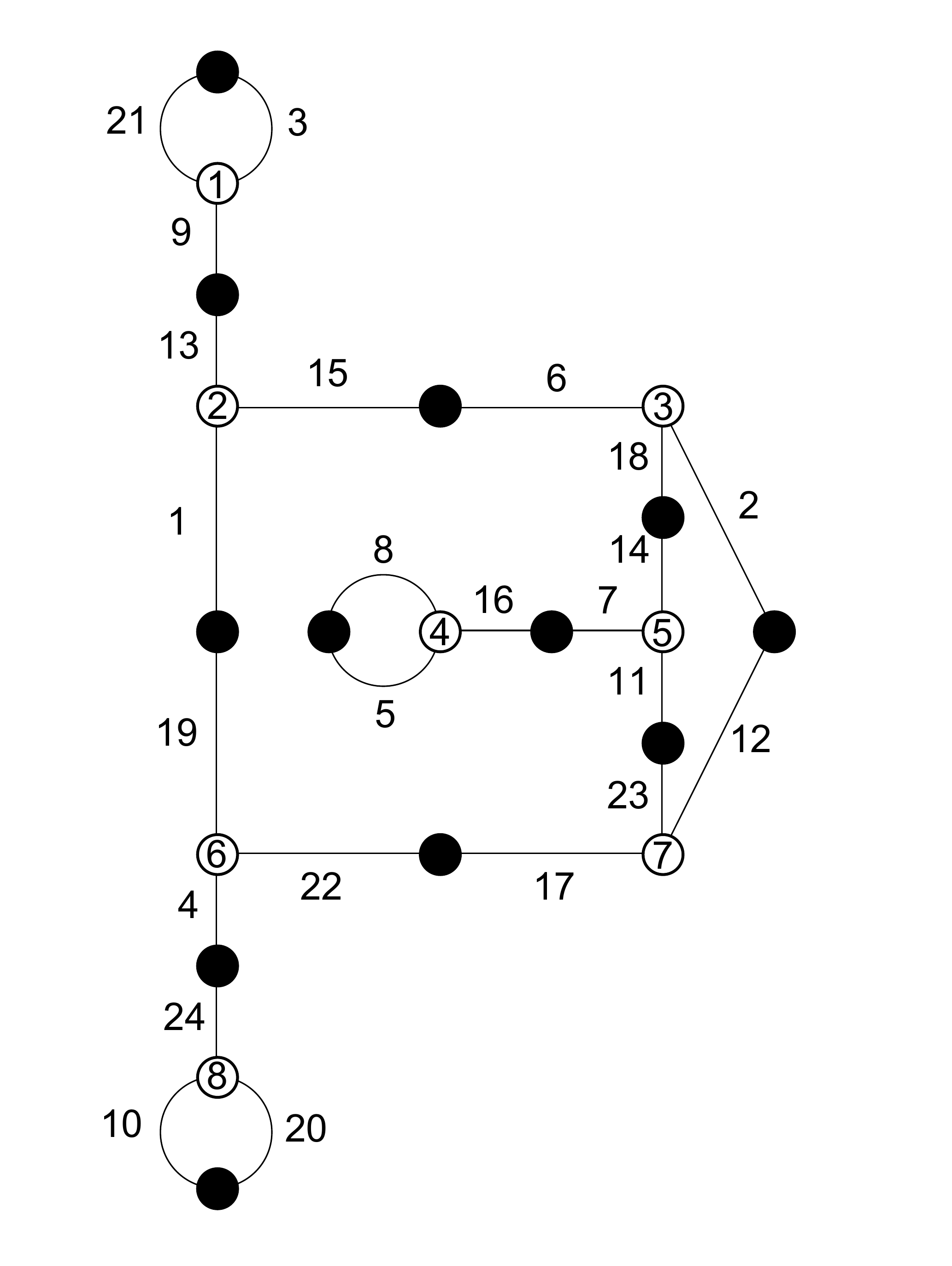}}}$
        \caption{ \{\{\{21,3,9\},\{13,15,1\},
        \{6,2,18\},\{14,11,7\},\{16,5,8\},
        \{23,12,17\},\{22,4,19\},\{24,20,10\}\}, \\ 
        \{\{10,20\},\{24,4\},\{22,17\},
        \{12,2\},\{19,1\},\{8,5\},
        \{16,7\},\{11,23\},\{14,18\},
        \{6,15\},\{13,9\},\{21,3\}\}\}}
        \caption{10-8-3-1-1-1 $(\mathbb{Q})$}
        \label{Dessin}
    \end{subfigure}\hfill
\end{figure}

\begin{figure}[H]
    \begin{subfigure}{0.5\textwidth}
        \centering \captionsetup{justification=centering}
        $\scalemath{0.75}{
        \displaystyle \begin{pmatrix}
            2 & 1 & 0 & 0 & 0 & 0 & 0 & 0\\ 
            1 & 0 & 1 & 1 & 0 & 0 & 0 & 0\\
            0 & 1 & 2 & 0 & 0 & 0 & 0 & 0\\
            0 & 1 & 0 & 0 & 0 & 0 & 0 & 2\\
            0 & 0 & 0 & 0 & 2 & 1 & 0 & 0\\
            0 & 0 & 0 & 0 & 1 & 0 & 2 & 0\\
            0 & 0 & 0 & 0 & 0 & 2 & 0 & 1\\
            0 & 0 & 0 & 2 & 0 & 0 & 1 & 0
        \end{pmatrix}}$
        $\vcenter{\hbox{\includegraphics[width=0.35\textwidth]{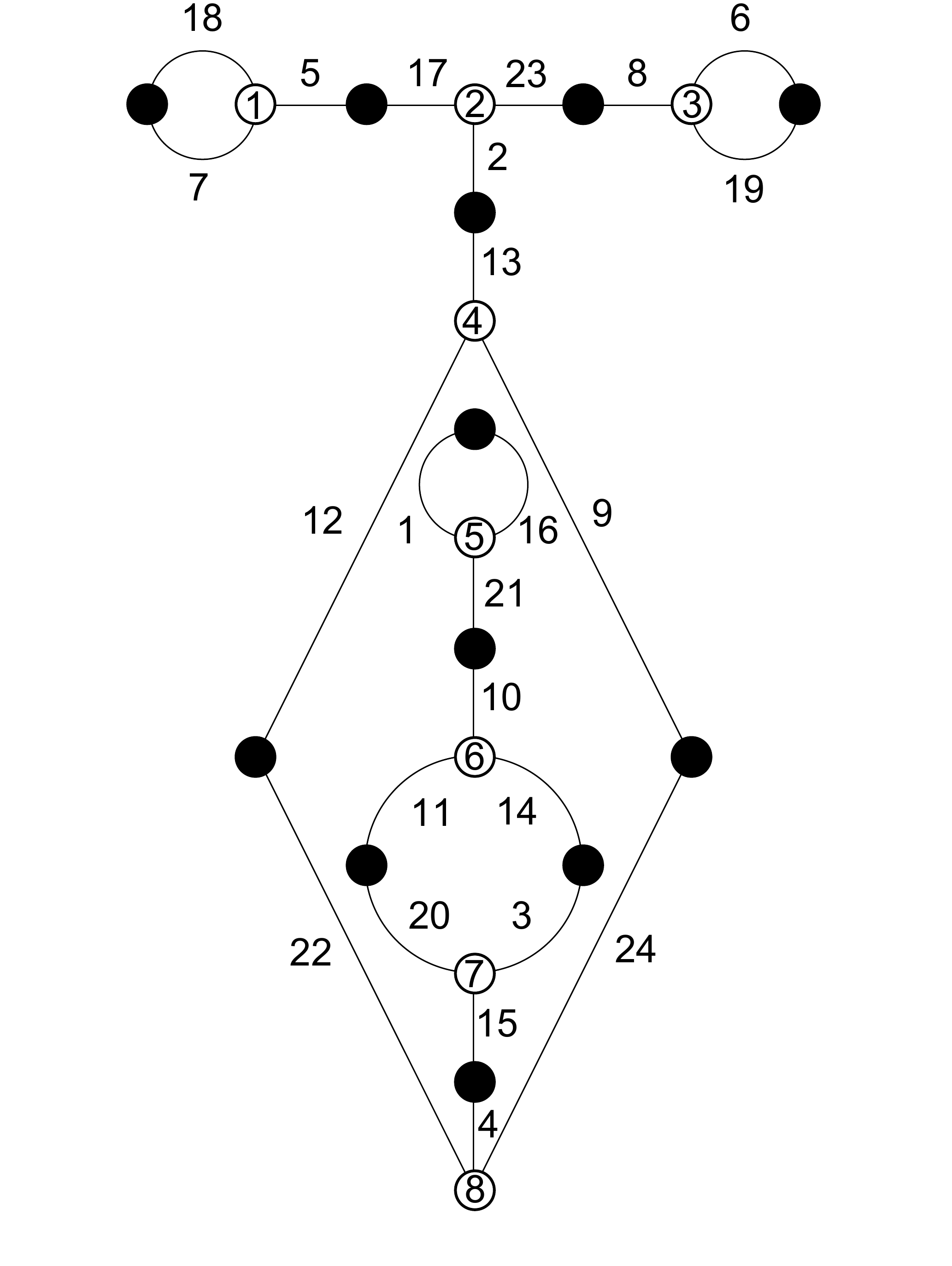}}}$
        \caption{ \{\{\{18,5,7\},\{17,23,2\},
        \{8,6,19\},\{13,9,12\},\{1,16,21\},
        \{10,14,11\},\{20,3,15\},\{4,24,22\}\}, \\ 
        \{\{4,15\},\{22,12\},\{24,9\},
        \{20,11\},\{14,3\},\{10,21\},
        \{1,16\},\{13,2\},\{17,5\},
        \{23,8\},\{6,19\},\{18,7\}\}\}}
        \caption{10-9-2-1-1-1 A $(\sqrt{5})$}
        \label{Dessin}
    \end{subfigure} \hfill
    \begin{subfigure}{0.5\textwidth}
        \centering \captionsetup{justification=centering}
        $\scalemath{0.75}{
        \displaystyle \begin{pmatrix}
            2 & 1 & 0 & 0 & 0 & 0 & 0 & 0\\ 
            1 & 0 & 2 & 0 & 0 & 0 & 0 & 0\\
            0 & 2 & 0 & 1 & 0 & 0 & 0 & 0\\
            0 & 0 & 1 & 0 & 0 & 0 & 1 & 1\\
            0 & 0 & 0 & 0 & 2 & 0 & 1 & 0\\
            0 & 0 & 0 & 0 & 0 & 2 & 0 & 1\\
            0 & 0 & 0 & 1 & 1 & 0 & 0 & 1\\
            0 & 0 & 0 & 1 & 0 & 1 & 1 & 0
        \end{pmatrix}}$
        $\vcenter{\hbox{\includegraphics[width=0.35\textwidth]{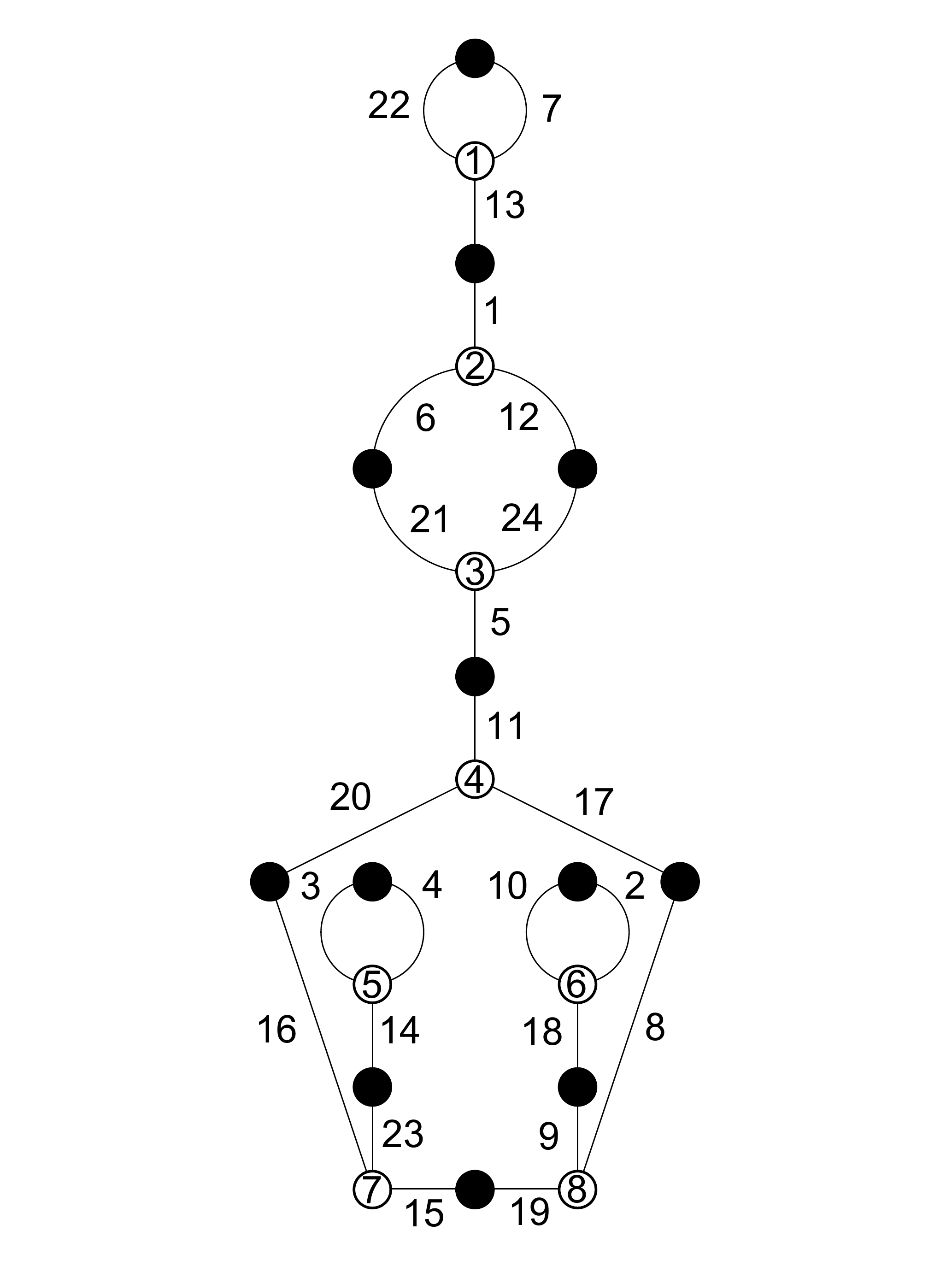}}}$
        \caption{ \{\{\{22,7,13\},\{1,12,6\},
        \{21,24,5\},\{11,17,20\},\{8,19,9\},
        \{15,16,23\},\{14,3,4\},\{2,18,10\}\}, \\ 
        \{\{3,4\},\{2,10\},\{18,9\},
        \{14,23\},\{15,19\},\{8,17\},
        \{16,20\},\{11,5\},\{24,12\},
        \{6,21\},\{1,13\},\{22,7\}\}\}}
        \caption{10-9-2-1-1-1 B $(\sqrt{5})$}
        \label{Dessin}
    \end{subfigure}\hfill
\end{figure}

\begin{figure}[H]
    \begin{subfigure}{0.6\textwidth}
        \centering \captionsetup{justification=centering}
        $\scalemath{0.75}{
        \displaystyle \begin{pmatrix}
            2 & 1 & 0 & 0 & 0 & 0 & 0 & 0\\ 
            1 & 0 & 1 & 0 & 0 & 1 & 0 & 0\\
            0 & 1 & 0 & 1 & 0 & 0 & 1 & 0\\
            0 & 0 & 1 & 2 & 0 & 0 & 0 & 0\\
            0 & 0 & 0 & 0 & 2 & 1 & 0 & 0\\
            0 & 1 & 0 & 0 & 1 & 0 & 1 & 0\\
            0 & 0 & 1 & 0 & 0 & 1 & 0 & 1\\
            0 & 0 & 0 & 0 & 0 & 0 & 1 & 2
        \end{pmatrix}}$
        $\vcenter{\hbox{\includegraphics[width=0.25\textwidth]{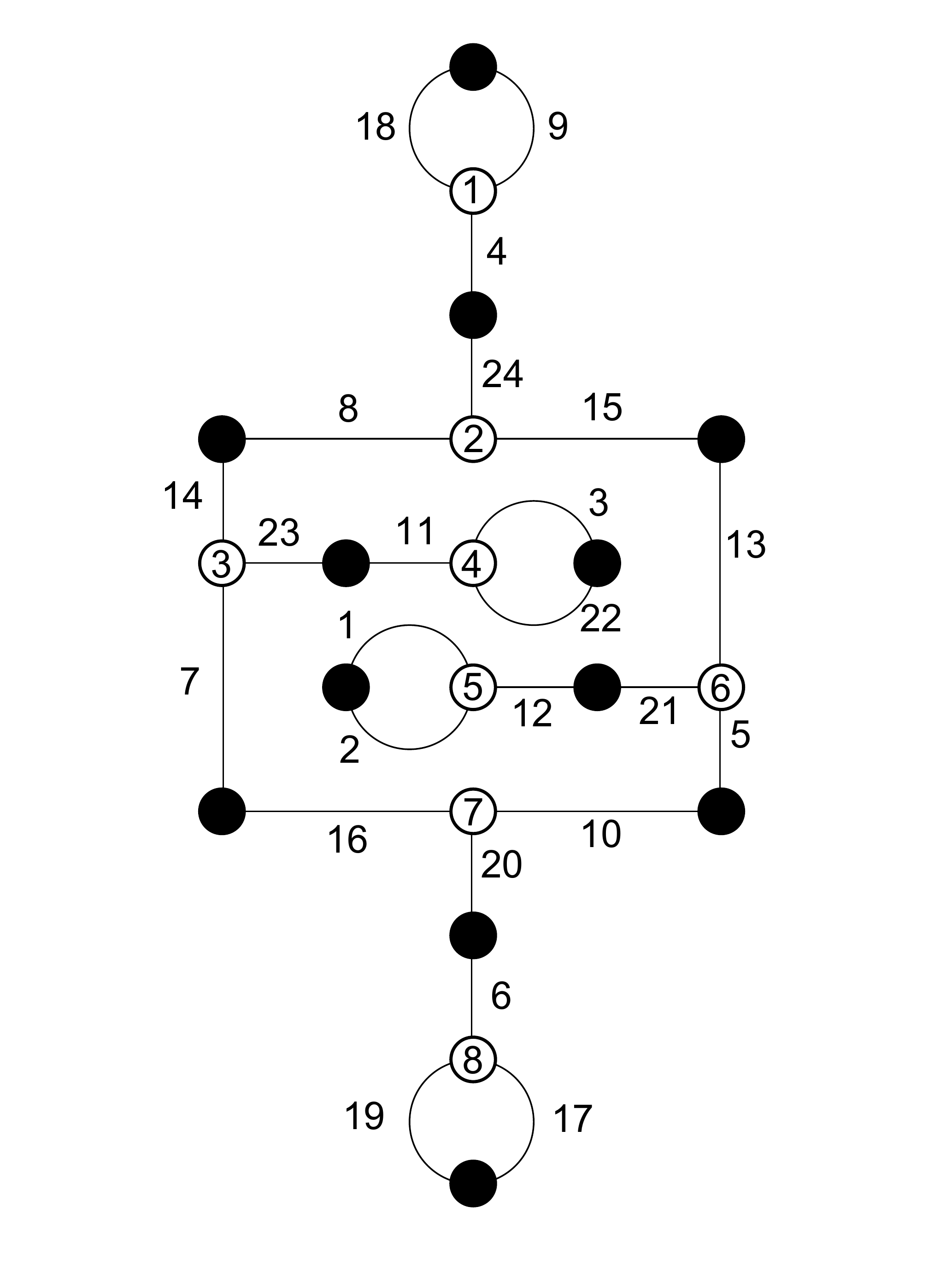}}}$
        $\vcenter{\hbox{\includegraphics[width=0.25\textwidth]{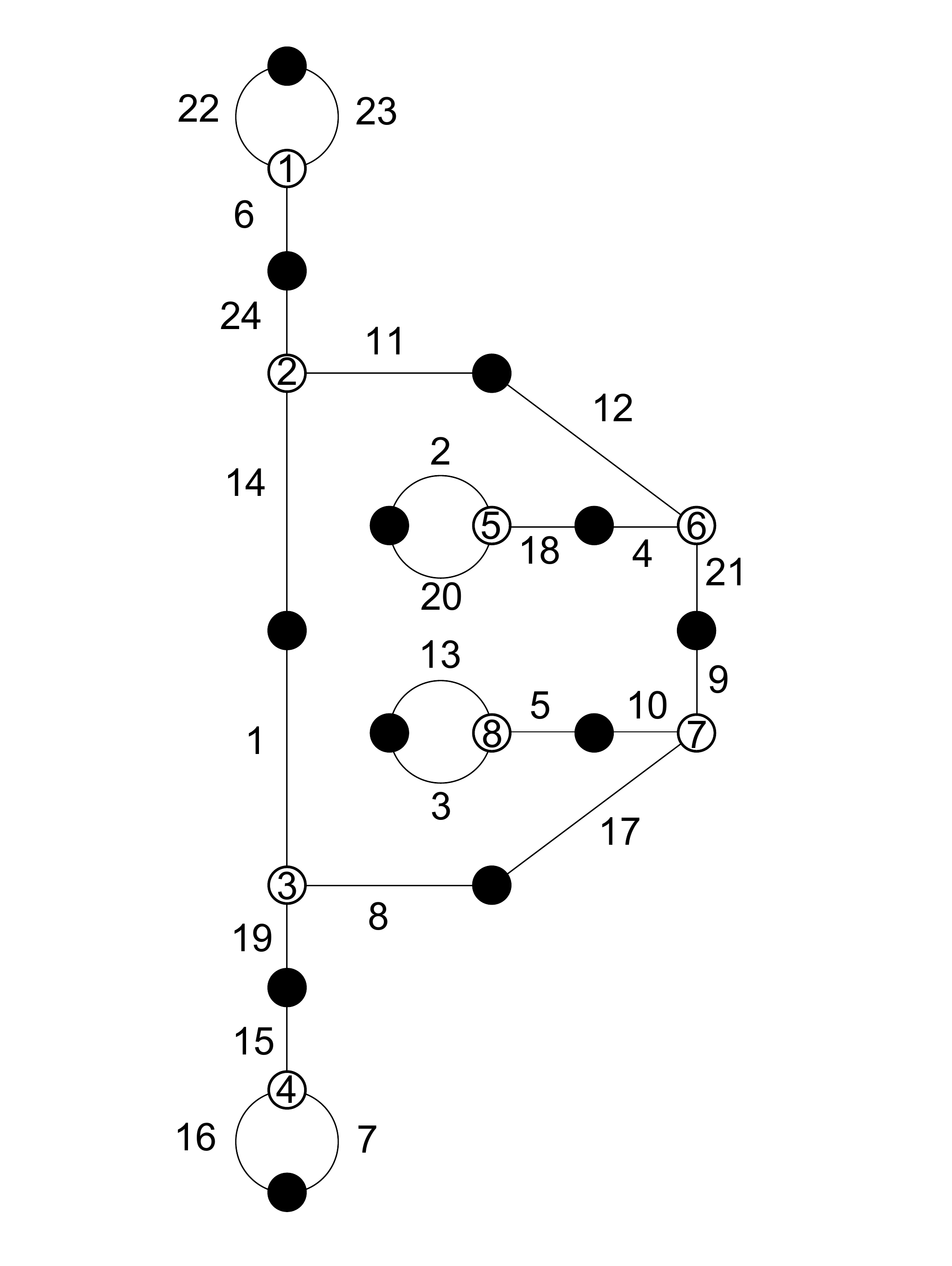}}}$
        \caption{ A: \{\{\{18,9,4\},\{24,15,8\},
        \{3,22,11\},\{14,23,7\},\{1,12,2\},
        \{21,13,5\},\{10,20,16\},\{17,19,6\}\}, \\ 
        \{\{17,19\},\{6,20\},\{16,7\},
        \{10,5\},\{21,12\},\{1,2\},
        \{3,22\},\{11,23\},\{14,8\},
        \{13,15\},\{24,4\},\{9,18\}\}\} \\
        B: \{\{\{22,23,6\},\{24,11,14\},
        \{12,21,4\},\{2,18,20\},\{9,17,10\},
        \{5,3,13\},\{8,19,1\},\{15,7,16\}\}, \\
        \{\{16,7\},\{15,19\},\{1,14\},
        \{13,3\},\{5,10\},\{8,17\},
        \{9,21\},\{4,18\},\{2,20\},
        \{12,11\},\{24,6\},\{22,23\}\}\}}
        \caption{10-10-1-1-1-1 A \& B $(\sqrt{5})$}
        \label{Dessin}
    \end{subfigure} \hfill
    \begin{subfigure}{0.4\textwidth}
        \centering \captionsetup{justification=centering}
        $\scalemath{0.75}{
        \displaystyle \begin{pmatrix}
            2 & 1 & 0 & 0 & 0 & 0 & 0 & 0\\ 
            1 & 0 & 1 & 1 & 0 & 0 & 0 & 0\\
            0 & 1 & 2 & 0 & 0 & 0 & 0 & 0\\
            0 & 1 & 0 & 0 & 0 & 0 & 0 & 2\\
            0 & 0 & 0 & 0 & 2 & 0 & 1 & 0\\
            0 & 0 & 0 & 0 & 0 & 2 & 1 & 0\\
            0 & 0 & 0 & 0 & 1 & 1 & 0 & 1\\
            0 & 0 & 0 & 2 & 0 & 0 & 1 & 0
        \end{pmatrix}}$
        $\vcenter{\hbox{\includegraphics[width=0.35\textwidth]{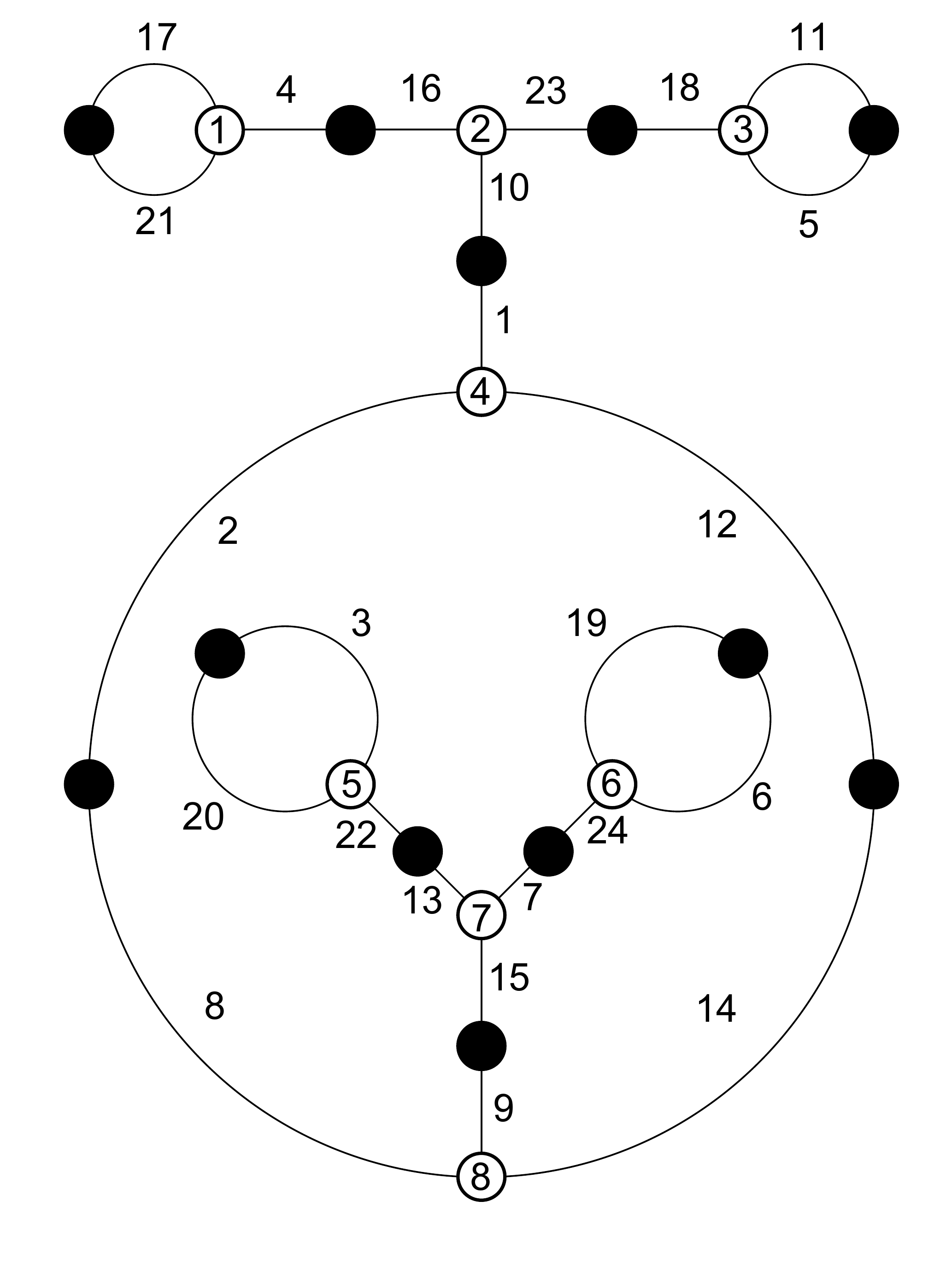}}}$
        \caption{ \{\{\{17,4,21\},\{16,23,10\},
        \{18,11,5\},\{1,12,2\},\{9,14,8\},
        \{15,13,7\},\{22,20,3\},\{19,6,24\}\}, \\ 
        \{\{9,15\},\{8,2\},\{14,12\},
        \{7,24\},\{13,22\},\{20,3\},
        \{19,6\},\{1,10\},\{23,18\},
        \{11,5\},\{16,4\},\{17,21\}\}\}}
        \caption{10-10-1-1-1-1 C $(\mathbb{Q})$}
        \label{Dessin}
    \end{subfigure}\hfill
\end{figure}

\begin{figure}[H]
    \begin{subfigure}{0.5\textwidth}
        \centering \captionsetup{justification=centering}
        $\scalemath{0.75}{
        \displaystyle \begin{pmatrix}
            2 & 1 & 0 & 0 & 0 & 0 & 0 & 0\\ 
            1 & 0 & 2 & 0 & 0 & 0 & 0 & 0\\
            0 & 2 & 0 & 1 & 0 & 0 & 0 & 0\\
            0 & 0 & 1 & 0 & 0 & 1 & 1 & 0\\
            0 & 0 & 0 & 0 & 0 & 1 & 1 & 1\\
            0 & 0 & 0 & 1 & 1 & 0 & 0 & 1\\
            0 & 0 & 0 & 1 & 1 & 0 & 0 & 1\\
            0 & 0 & 0 & 0 & 1 & 1 & 1 & 0
        \end{pmatrix}}$
        $\vcenter{\hbox{\includegraphics[width=0.35\textwidth]{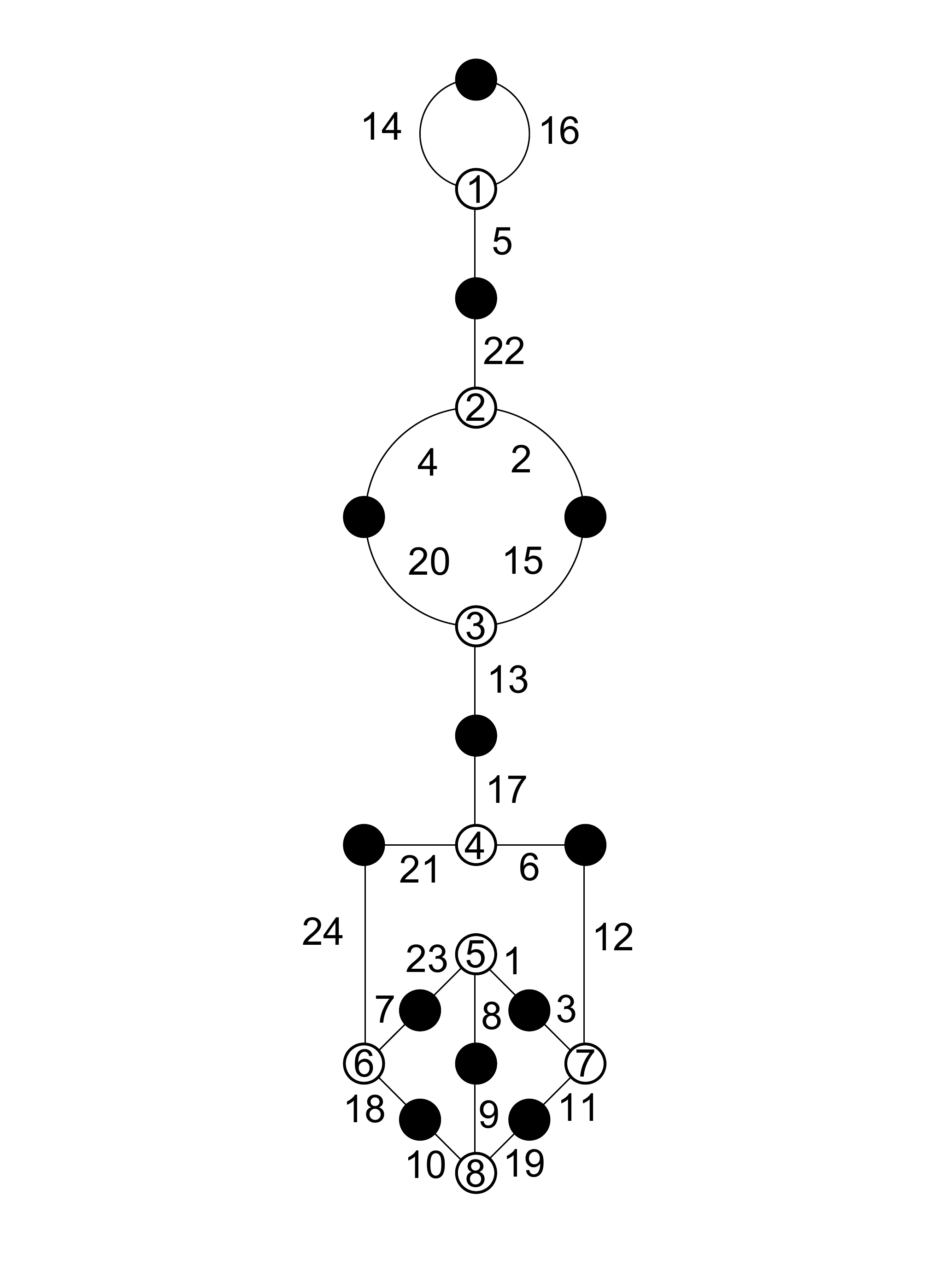}}}$
        \caption{ \{\{\{14,16,5\},\{22,2,4\},
        \{20,15,13\},\{17,6,21\},\{12,11,3\},
        \{1,8,23\},\{7,18,24\},\{9,19,10\}\}, \\ 
        \{\{18,10\},\{7,23\},\{1,3\},
        \{8,9\},\{19,11\},\{24,21\},
        \{6,12\},\{17,13\},\{20,4\},
        \{2,15\},\{22,5\},\{14,16\}\}\}}
        \caption{11-4-3-3-2-1 $(\mathbb{Q})$}
        \label{Dessin}
    \end{subfigure} \hfill
    \begin{subfigure}{0.5\textwidth}
        \centering \captionsetup{justification=centering}
        $\scalemath{0.75}{
        \displaystyle \begin{pmatrix}
            2 & 1 & 0 & 0 & 0 & 0 & 0 & 0\\ 
            1 & 0 & 1 & 1 & 0 & 0 & 0 & 0\\
            0 & 1 & 0 & 0 & 1 & 0 & 1 & 0\\
            0 & 1 & 0 & 0 & 1 & 1 & 0 & 0\\
            0 & 0 & 1 & 1 & 0 & 0 & 1 & 0\\
            0 & 0 & 0 & 1 & 0 & 0 & 1 & 1\\
            0 & 0 & 1 & 0 & 1 & 1 & 0 & 0\\
            0 & 0 & 0 & 0 & 0 & 1 & 0 & 2
        \end{pmatrix}}$
        $\vcenter{\hbox{\includegraphics[width=0.35\textwidth]{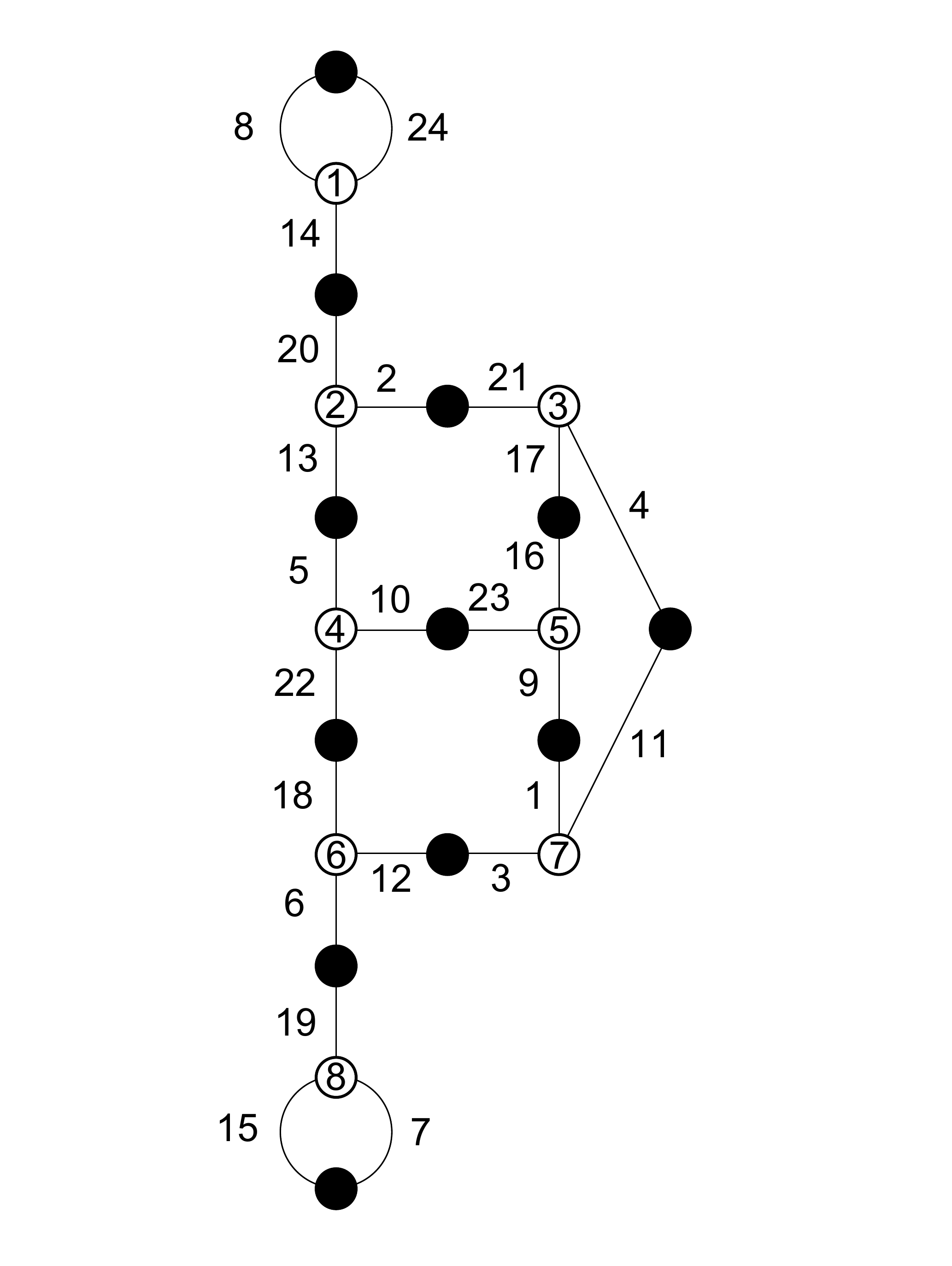}}}$
        \caption{ \{\{\{8,24,14\},\{2,13,20\},
        \{21,4,17\},\{16,9,23\},\{10,22,5\},
        \{18,12,6\},\{1,11,3\},\{19,7,15\}\}, \\ 
        \{\{15,7\},\{19,6\},\{18,22\},
        \{12,3\},\{1,9\},\{11,4\},
        \{10,23\},\{16,17\},\{2,21\},
        \{5,13\},\{20,14\},\{8,24\}\}\}}
        \caption{11-4-4-3-1-1 $(\mathbb{Q})$}
        \label{Dessin}
    \end{subfigure}\hfill
\end{figure}

\begin{figure}[H]
    \begin{subfigure}{0.5\textwidth}
        \centering \captionsetup{justification=centering}
        $\scalemath{0.75}{
        \displaystyle \begin{pmatrix}
            0 & 2 & 1 & 0 & 0 & 0 & 0 & 0\\ 
            2 & 0 & 1 & 0 & 0 & 0 & 0 & 0\\
            1 & 1 & 0 & 1 & 0 & 0 & 0 & 0\\
            0 & 0 & 1 & 0 & 2 & 0 & 0 & 0\\
            0 & 0 & 0 & 2 & 0 & 1 & 0 & 0\\
            0 & 0 & 0 & 0 & 1 & 0 & 0 & 2\\
            0 & 0 & 0 & 0 & 0 & 0 & 2 & 1\\
            0 & 0 & 0 & 0 & 0 & 2 & 1 & 0
        \end{pmatrix}}$
        $\vcenter{\hbox{\includegraphics[width=0.35\textwidth]{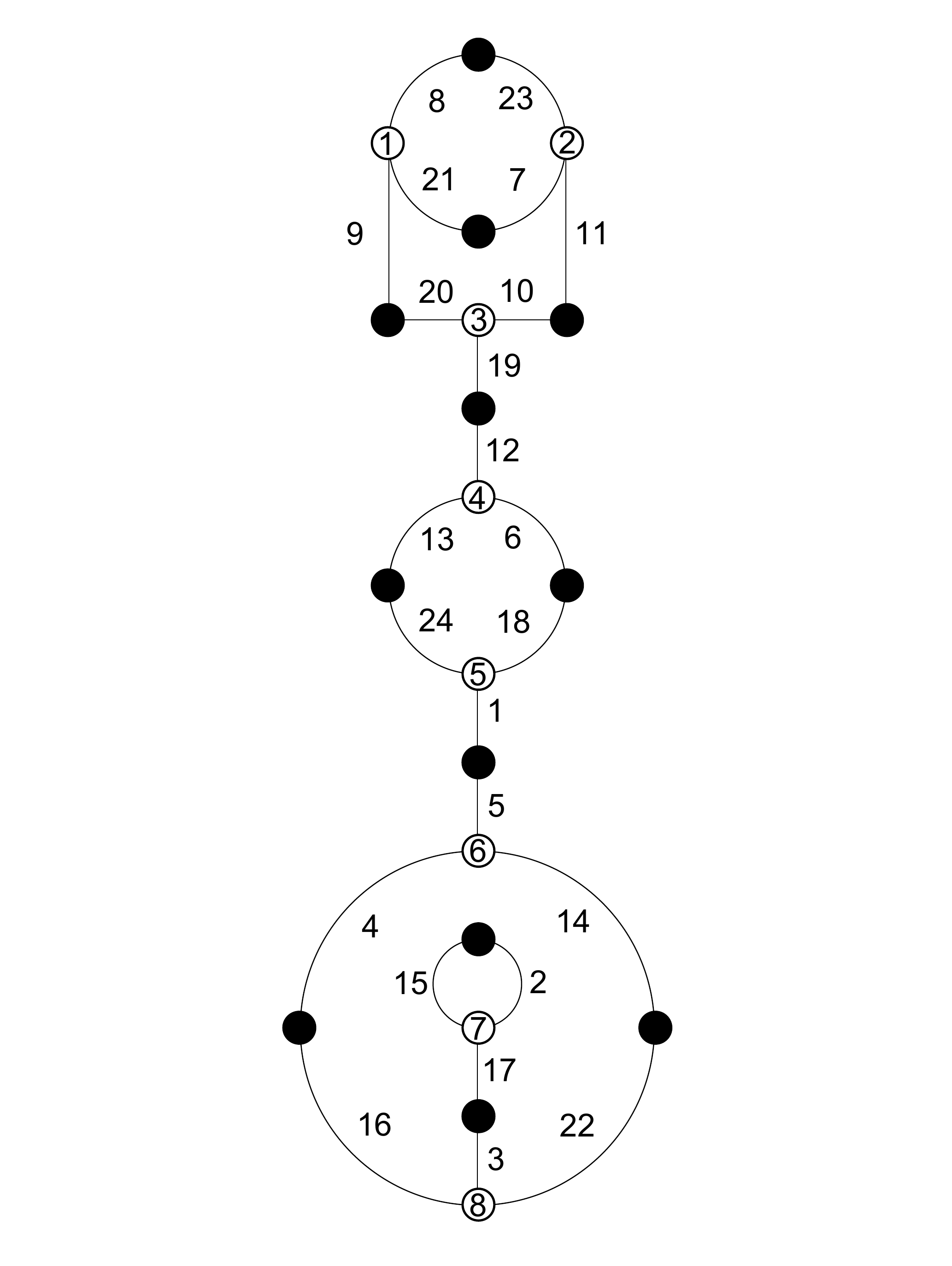}}}$
        \caption{ \{\{\{8,21,9\},\{23,11,7\},
        \{20,10,19\},\{12,6,13\},\{24,18,1\},
        \{5,14,4\},\{15,2,17\},\{22,16,3\}\}, \\ 
        \{\{3,17\},\{2,15\},\{22,14\},
        \{4,16\},\{5,1\},\{24,13\},
        \{6,18\},\{12,19\},\{20,9\},
        \{11,10\},\{7,21\},\{8,23\}\}\}}
        \caption{11-5-3-2-2-1 $(\mathbb{Q})$}
        \label{Dessin}
    \end{subfigure} \hfill
    \begin{subfigure}{0.5\textwidth}
        \centering \captionsetup{justification=centering}
        $\scalemath{0.75}{
        \displaystyle \begin{pmatrix}
            2 & 1 & 0 & 0 & 0 & 0 & 0 & 0\\ 
            1 & 0 & 1 & 1 & 0 & 0 & 0 & 0\\
            0 & 1 & 0 & 1 & 1 & 0 & 0 & 0\\
            0 & 1 & 1 & 0 & 1 & 0 & 0 & 0\\
            0 & 0 & 1 & 1 & 0 & 1 & 0 & 0\\
            0 & 0 & 0 & 0 & 1 & 0 & 0 & 2\\
            0 & 0 & 0 & 0 & 0 & 0 & 2 & 1\\
            0 & 0 & 0 & 0 & 0 & 2 & 1 & 0
        \end{pmatrix}}$
        $\vcenter{\hbox{\includegraphics[width=0.35\textwidth]{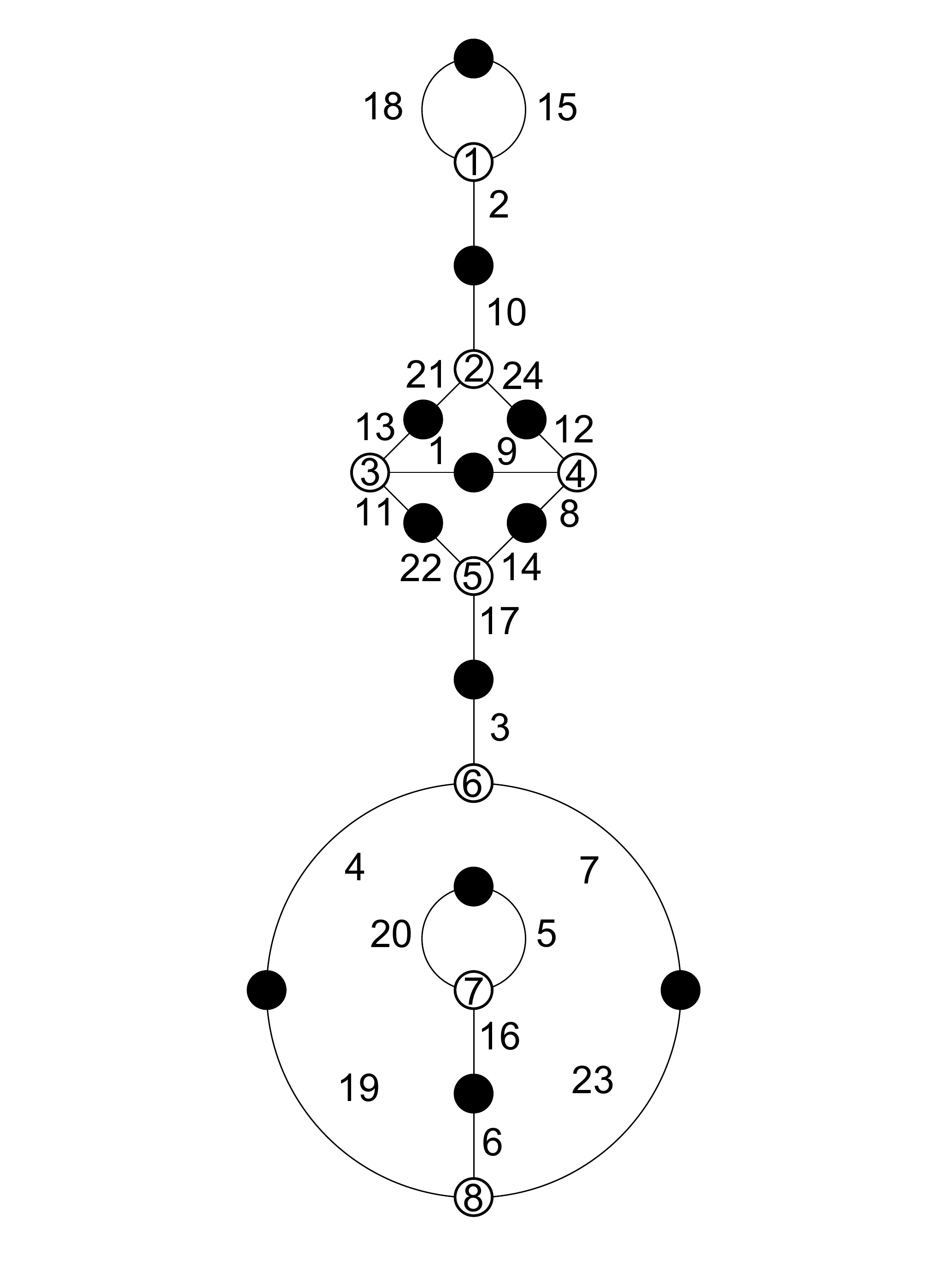}}}$
        \caption{ \{\{\{18,15,2\},\{10,24,21\},
        \{12,8,9\},\{1,11,13\},\{14,17,22\},
        \{3,7,4\},\{23,19,6\},\{16,20,5\}\}, \\
        \{\{20,5\},\{16,6\},\{19,4\},
        \{23,7\},\{3,17\},\{14,8\},
        \{22,11\},\{13,21\},\{24,12\},
        \{1,9\},\{10,2\},\{18,15\}\}\}}
        \caption{11-5-3-3-1-1 A $(\sqrt{5})$}
        \label{Dessin}
    \end{subfigure}\hfill
\end{figure}

\begin{figure}[H]
    \begin{subfigure}{0.5\textwidth}
        \centering \captionsetup{justification=centering}
        $\scalemath{0.75}{
        \displaystyle \begin{pmatrix}
            2 & 1 & 0 & 0 & 0 & 0 & 0 & 0\\ 
            1 & 0 & 1 & 0 & 0 & 1 & 0 & 0\\
            0 & 1 & 0 & 1 & 1 & 0 & 0 & 0\\
            0 & 0 & 1 & 0 & 1 & 0 & 1 & 0\\
            0 & 0 & 1 & 1 & 0 & 0 & 1 & 0\\
            0 & 1 & 0 & 0 & 0 & 0 & 1 & 1\\
            0 & 0 & 0 & 1 & 1 & 1 & 0 & 0\\
            0 & 0 & 0 & 0 & 0 & 1 & 0 & 2
        \end{pmatrix}}$
        $\vcenter{\hbox{\includegraphics[width=0.35\textwidth]{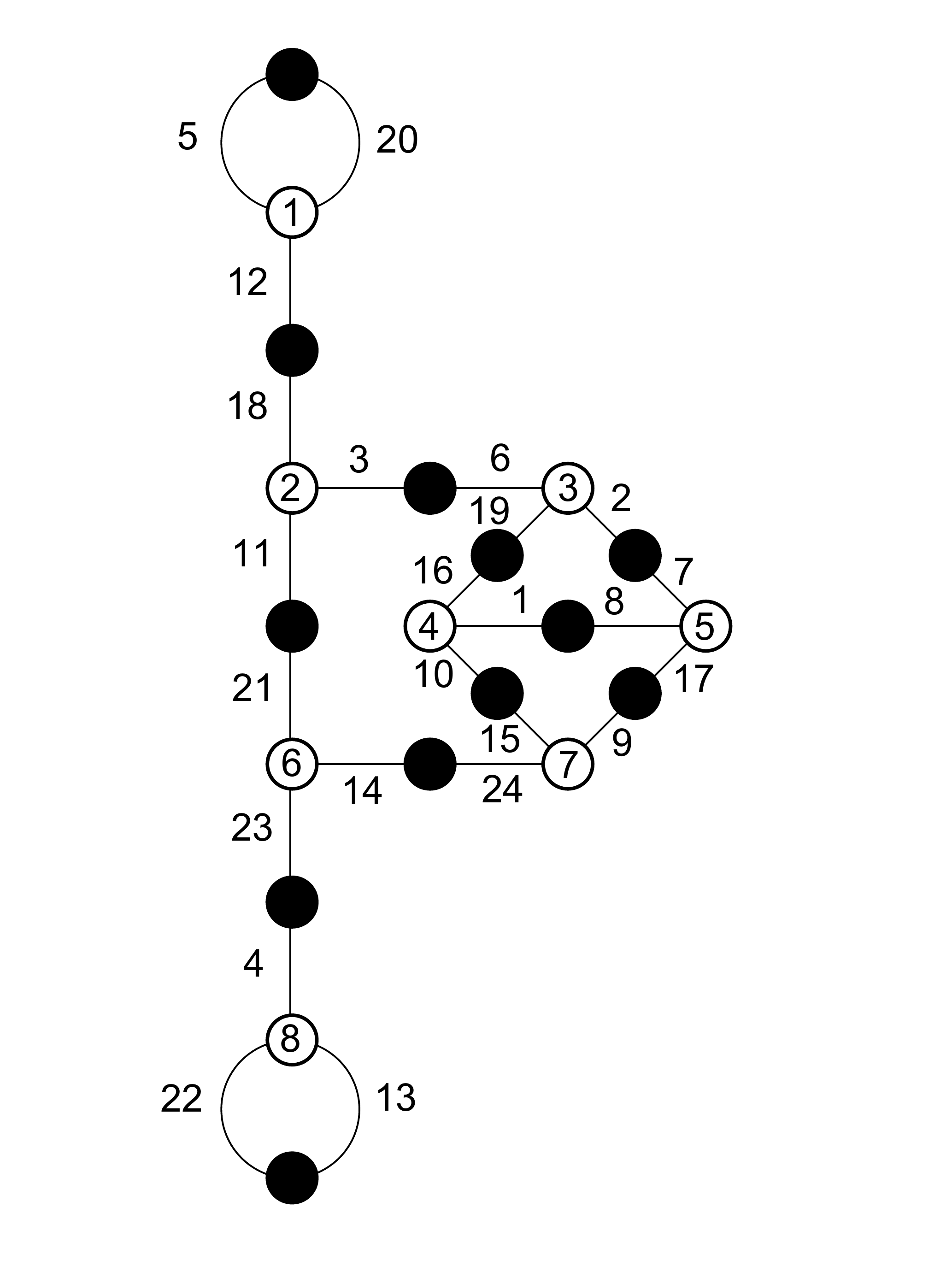}}}$
        \caption{ \{\{\{5,20,12\},\{18,3,11\},
        \{6,2,19\},\{16,1,10\},\{7,17,8\},
        \{9,24,15\},\{14,23,21\},\{4,13,22\}\}, \\ 
        \{\{22,13\},\{4,23\},\{14,24\},
        \{21,11\},\{18,12\},\{5,20\},
        \{3,6\},\{2,7\},\{16,19\},
        \{17,9\},\{1,8\},\{10,15\}\}\}}
        \caption{11-5-3-3-1-1 B $(\sqrt{5})$}
        \label{Dessin}
    \end{subfigure} \hfill
    \begin{subfigure}{0.5\textwidth}
        \centering \captionsetup{justification=centering}
        $\scalemath{0.75}{
        \displaystyle \begin{pmatrix}
            2 & 1 & 0 & 0 & 0 & 0 & 0 & 0\\ 
            1 & 0 & 1 & 1 & 0 & 0 & 0 & 0\\
            0 & 1 & 2 & 0 & 0 & 0 & 0 & 0\\
            0 & 1 & 0 & 0 & 0 & 0 & 1 & 1\\
            0 & 0 & 0 & 0 & 0 & 2 & 1 & 0\\
            0 & 0 & 0 & 0 & 2 & 0 & 0 & 1\\
            0 & 0 & 0 & 1 & 1 & 0 & 0 & 1\\
            0 & 0 & 0 & 1 & 0 & 1 & 1 & 0
        \end{pmatrix}}$
        $\vcenter{\hbox{\includegraphics[width=0.35\textwidth]{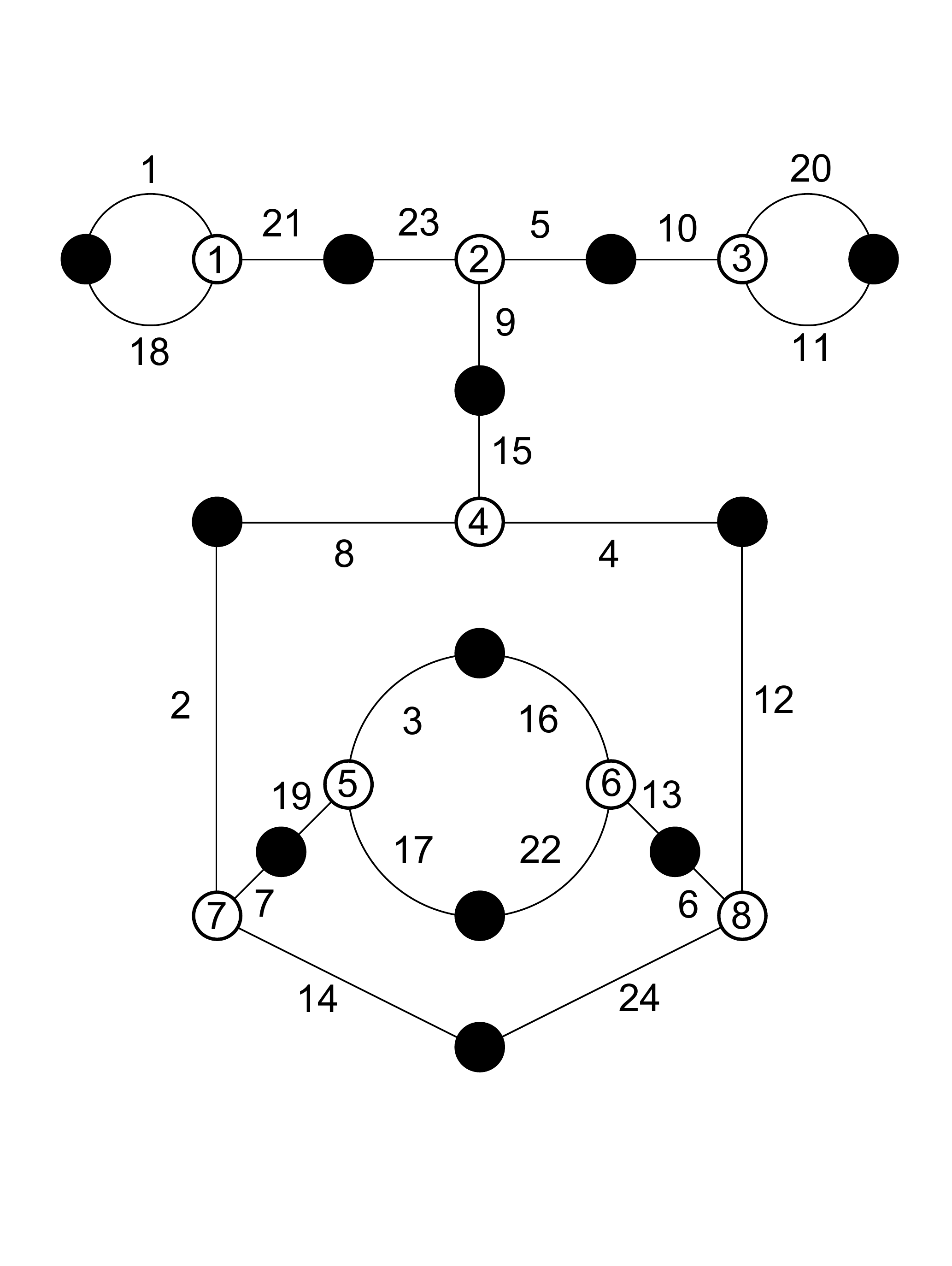}}}$
        \caption{ \{\{\{18,1,21\},\{23,5,9\},
        \{10,20,11\},\{15,4,8\},\{12,24,6\},
        \{7,14,2\},\{19,3,17\},\{16,13,22\}\}, \\ 
        \{\{1,18\},\{21,23\},\{5,10\},
        \{20,11\},\{9,15\},\{8,2\},
        \{4,12\},\{3,16\},\{17,22\},
        \{19,7\},\{14,24\},\{6,13\}\}\}}
        \caption{11-5-4-2-1-1 A (cubic)}
        \label{Dessin}
    \end{subfigure}\hfill
\end{figure}

\begin{figure}[H]
    \begin{subfigure}{0.6\textwidth}
        \centering \captionsetup{justification=centering}
        $\scalemath{0.75}{
        \displaystyle \begin{pmatrix}
            0 & 1 & 2 & 0 & 0 & 0 & 0 & 0\\ 
            1 & 2 & 0 & 0 & 0 & 0 & 0 & 0\\
            2 & 0 & 0 & 1 & 0 & 0 & 0 & 0\\
            0 & 0 & 1 & 0 & 1 & 0 & 1 & 0\\
            0 & 0 & 0 & 1 & 0 & 2 & 0 & 0\\
            0 & 0 & 0 & 0 & 2 & 0 & 1 & 0\\
            0 & 0 & 0 & 1 & 0 & 1 & 0 & 1\\
            0 & 0 & 0 & 0 & 0 & 0 & 1 & 2
        \end{pmatrix}}$
        $\vcenter{\hbox{\includegraphics[width=0.25\textwidth]{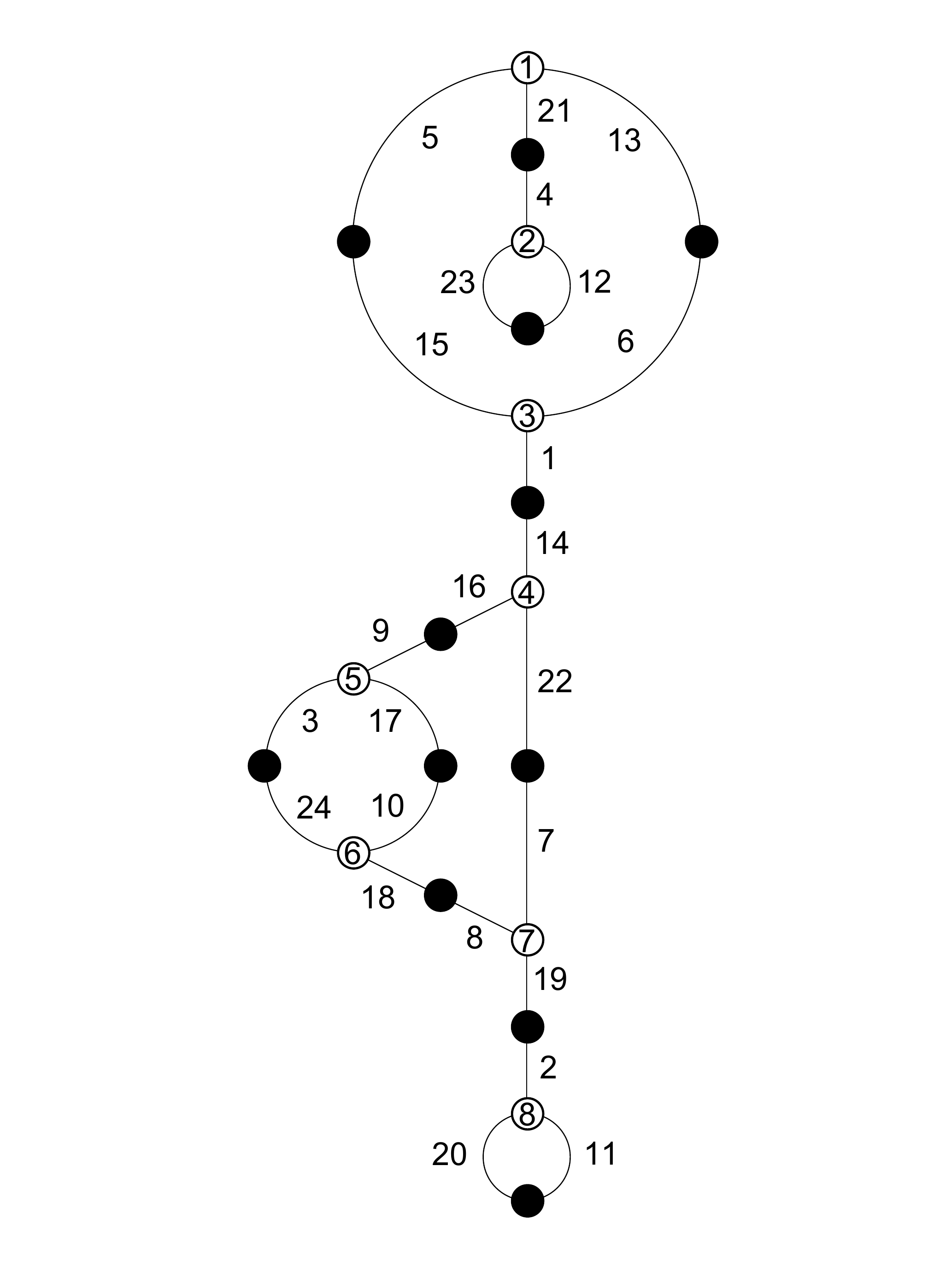}}}$
        $\vcenter{\hbox{\includegraphics[width=0.25\textwidth]{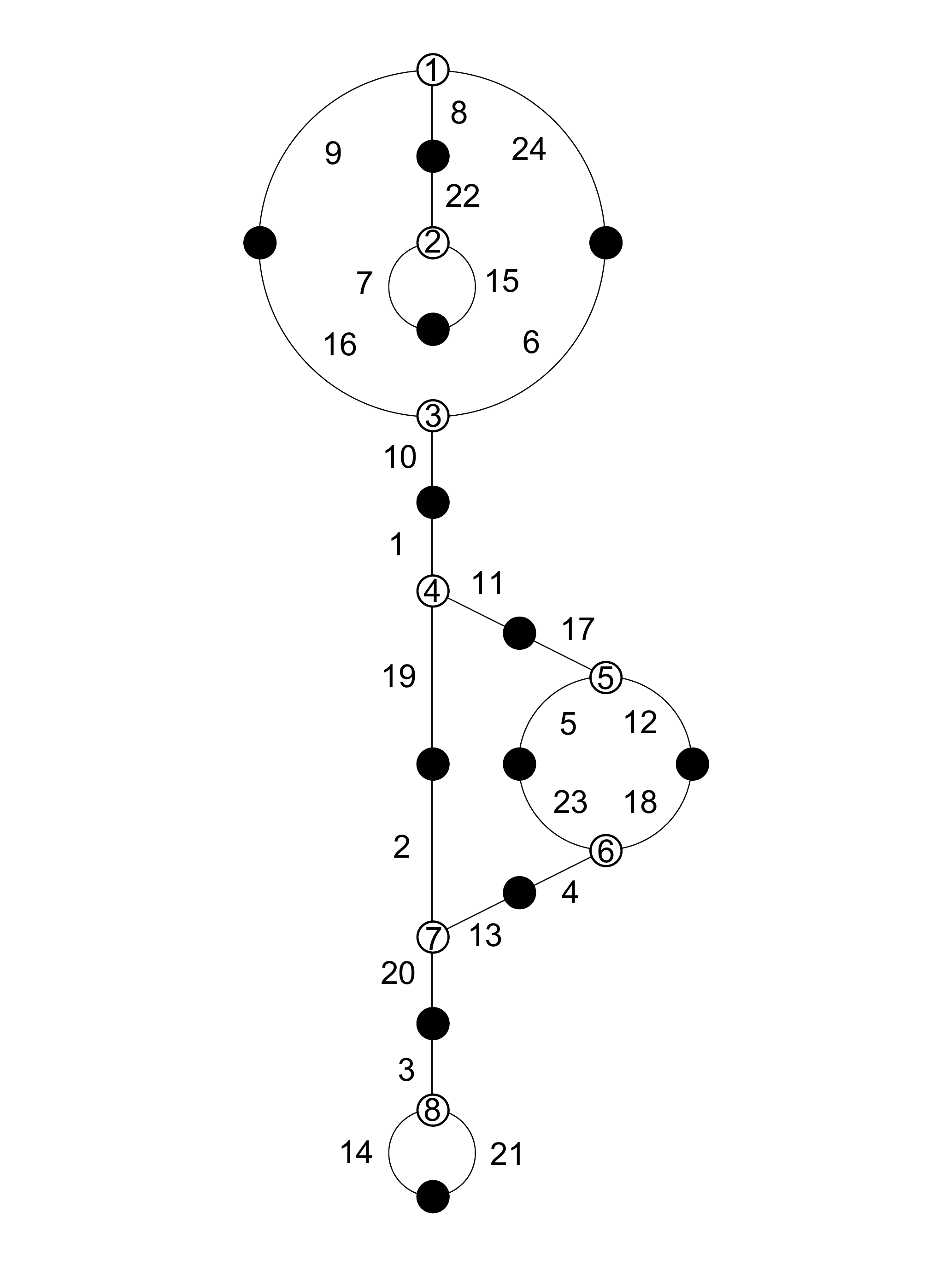}}}$
        \caption{ B: \{\{\{5,13,21\},\{4,12,23\},
        \{6,1,15\},\{14,22,16\},\{9,17,3\},
        \{24,10,18\},\{7,19,8\},\{2,11,20\}\}, \\ 
        \{\{11,20\},\{2,19\},\{8,18\},
        \{24,3\},\{17,10\},\{7,22\},
        \{9,16\},\{14,1\},\{6,13\},
        \{5,15\},\{21,4\},\{23,12\}\}\} \\
        C: \{\{\{8,9,24\},\{22,15,7\},
        \{6,10,16\},\{1,11,19\},\{17,12,5\},
        \{23,18,4\},\{2,13,20\},\{3,21,14\}\}, \\ 
        \{\{14,21\},\{20,3\},\{13,4\},
        \{2,19\},\{11,17\},\{5,23\},
        \{12,18\},\{10,1\},\{24,6\},
        \{16,9\},\{15,7\},\{8,22\}\}\}}
        \caption{11-5-4-2-1-1 B \& C (cubic)}
        \label{Dessin}
    \end{subfigure} \hfill
    \begin{subfigure}{0.4\textwidth}
        \centering \captionsetup{justification=centering}
        $\scalemath{0.75}{
        \displaystyle \begin{pmatrix}
            0 & 1 & 2 & 0 & 0 & 0 & 0 & 0\\ 
            1 & 2 & 0 & 0 & 0 & 0 & 0 & 0\\
            2 & 0 & 0 & 0 & 1 & 0 & 0 & 0\\
            0 & 0 & 0 & 2 & 1 & 0 & 0 & 0\\
            0 & 0 & 1 & 1 & 0 & 1 & 0 & 0\\
            0 & 0 & 0 & 0 & 1 & 0 & 0 & 2\\
            0 & 0 & 0 & 0 & 0 & 0 & 2 & 1\\
            0 & 0 & 0 & 0 & 0 & 2 & 1 & 0
        \end{pmatrix}}$
        $\vcenter{\hbox{\includegraphics[width=0.35\textwidth]{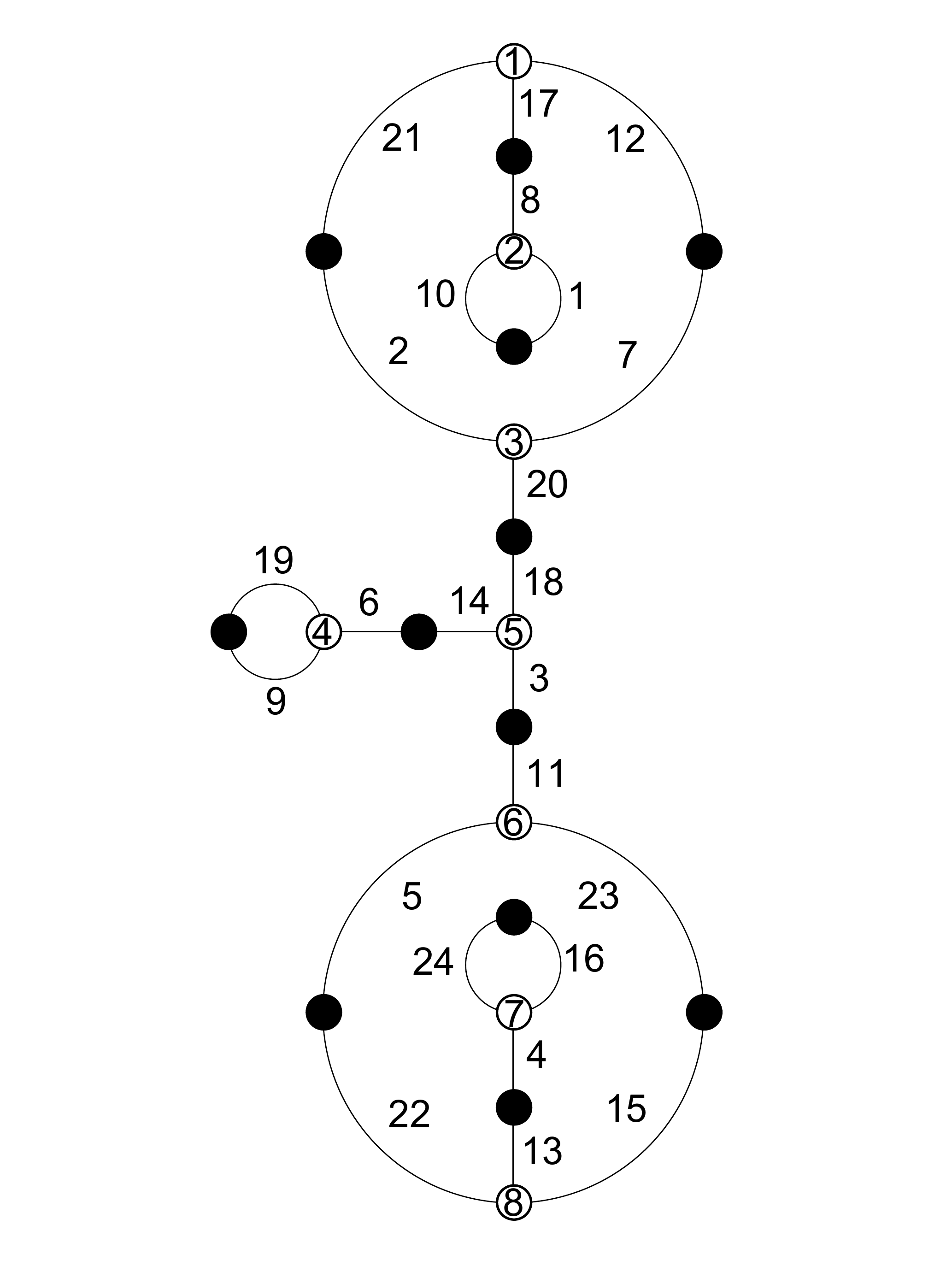}}}$
        \caption{ \{\{\{17,21,12\},\{8,1,10\},
        \{7,20,2\},\{18,3,14\},\{6,9,19\},
        \{11,23,5\},\{24,16,4\},\{13,15,22\}\}, \\ 
        \{\{13,4\},\{24,16\},\{22,5\},
        \{15,23\},\{11,3\},\{6,14\},
        \{9,19\},\{18,20\},\{2,21\},
        \{12,7\},\{17,8\},\{10,1\}\}\}}
        \caption{11-5-5-1-1-1 $(\mathbb{Q})$}
        \label{Dessin}
    \end{subfigure}\hfill
\end{figure}

\begin{figure}[H]
    \begin{subfigure}{0.4\textwidth}
        \centering \captionsetup{justification=centering}
        $\scalemath{0.75}{
        \displaystyle \begin{pmatrix}
            2 & 1 & 0 & 0 & 0 & 0 & 0 & 0\\ 
            1 & 0 & 2 & 0 & 0 & 0 & 0 & 0\\
            0 & 2 & 0 & 1 & 0 & 0 & 0 & 0\\
            0 & 0 & 1 & 0 & 1 & 0 & 1 & 0\\
            0 & 0 & 0 & 1 & 0 & 0 & 1 & 1\\
            0 & 0 & 0 & 0 & 0 & 2 & 0 & 1\\
            0 & 0 & 0 & 1 & 1 & 0 & 0 & 1\\
            0 & 0 & 0 & 0 & 1 & 1 & 1 & 0
        \end{pmatrix}}$
        $\vcenter{\hbox{\includegraphics[width=0.35\textwidth]{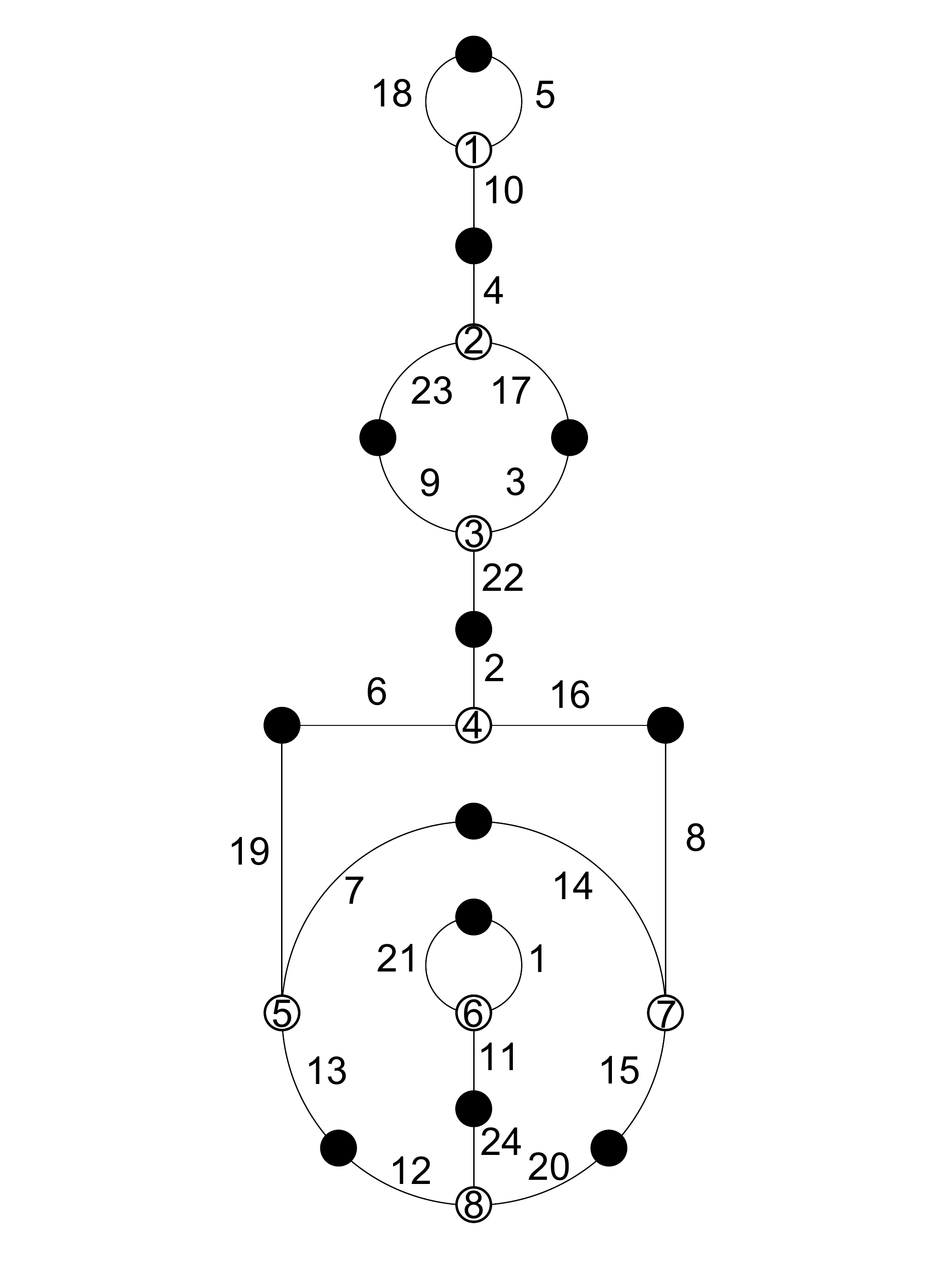}}}$
        \caption{ \{\{\{18,5,10\},\{4,17,23\},
        \{9,3,22\},\{2,16,6\},\{8,15,14\},
        \{19,7,13\},\{21,1,11\},\{24,20,12\}\}, \\ 
        \{\{21,1\},\{11,24\},\{13,12\},
        \{20,15\},\{7,14\},\{19,6\},
        \{16,8\},\{2,22\},\{9,23\},
        \{17,3\},\{4,10\},\{18,5\}\}\}}
        \caption{11-6-3-2-1-1 A (cubic)}
        \label{Dessin}
    \end{subfigure} \hfill
    \begin{subfigure}{0.6\textwidth}
        \centering \captionsetup{justification=centering}
        $\scalemath{0.75}{
        \displaystyle \begin{pmatrix}
            0 & 2 & 1 & 0 & 0 & 0 & 0 & 0\\ 
            2 & 0 & 1 & 0 & 0 & 0 & 0 & 0\\
            1 & 1 & 0 & 1 & 0 & 0 & 0 & 0\\
            0 & 0 & 1 & 0 & 1 & 0 & 1 & 0\\
            0 & 0 & 0 & 1 & 0 & 1 & 1 & 0\\
            0 & 0 & 0 & 0 & 1 & 2 & 0 & 0\\
            0 & 0 & 0 & 1 & 1 & 0 & 0 & 1\\
            0 & 0 & 0 & 0 & 0 & 0 & 1 & 2
        \end{pmatrix}}$
        $\vcenter{\hbox{\includegraphics[width=0.25\textwidth]{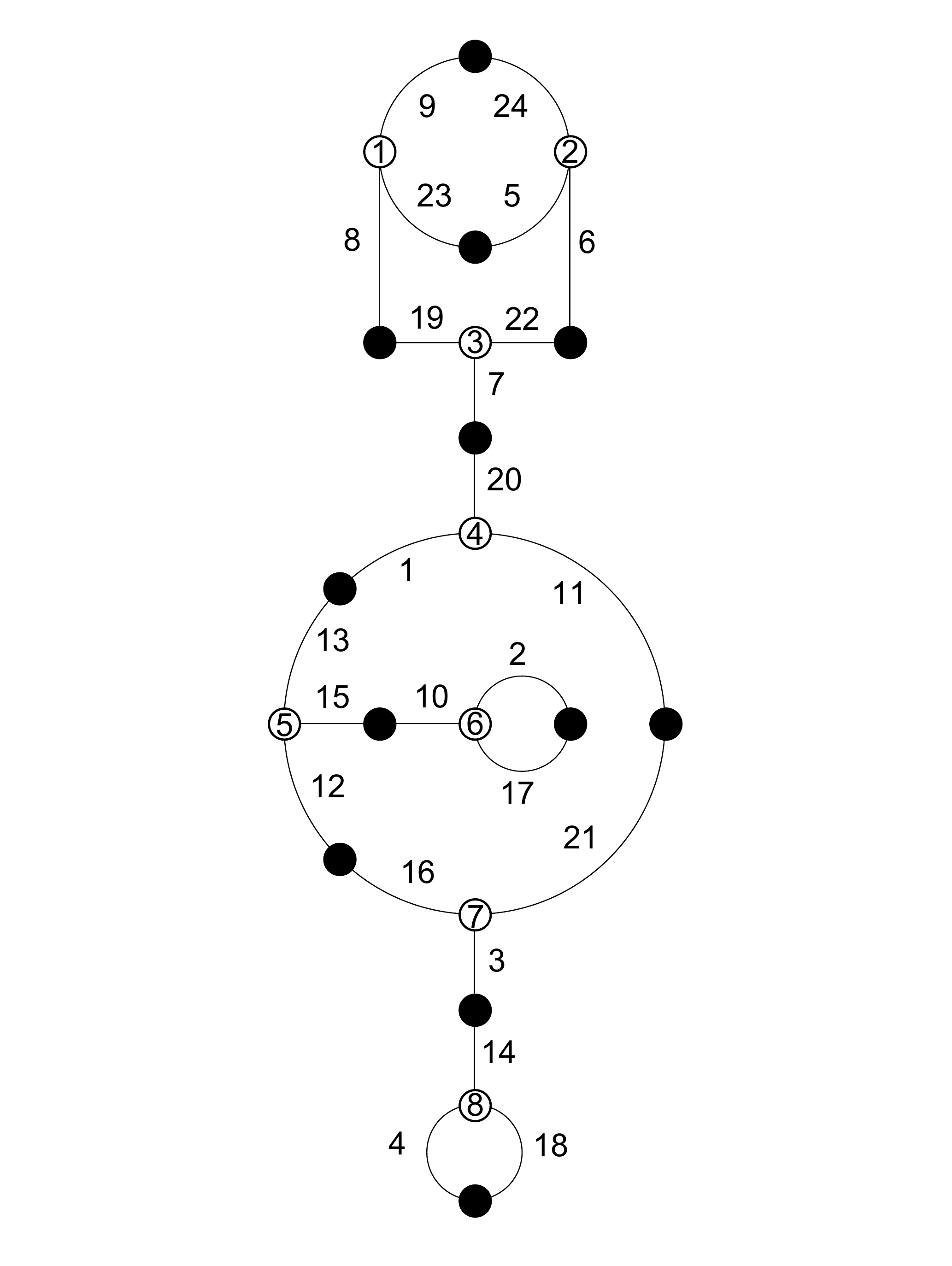}}}$
        $\vcenter{\hbox{\includegraphics[width=0.25\textwidth]{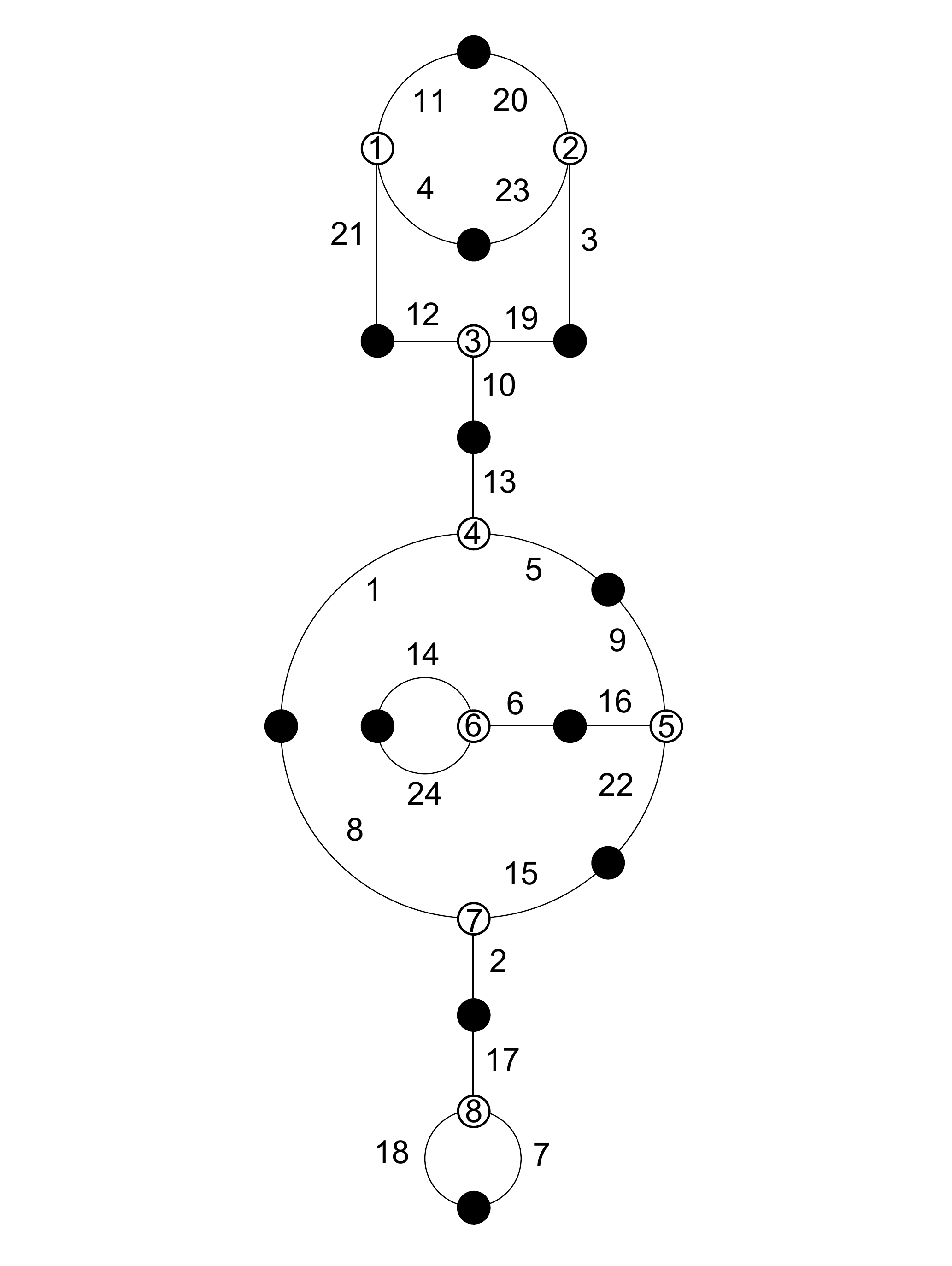}}}$
        \caption{ B: \{\{\{9,23,8\},\{24,6,5\},
        \{19,22,7\},\{20,11,1\},\{12,13,15\},
        \{10,2,17\},\{16,21,3\},\{14,18,4\}\}, \\ 
        \{\{4,18\},\{14,3\},\{12,16\},
        \{21,11\},\{15,10\},\{2,17\},
        \{13,1\},\{20,7\},\{19,8\},
        \{22,6\},\{23,5\},\{9,24\}\}\} \\
        C: \{\{\{11,4,21\},\{20,3,23\},
        \{12,19,10\},\{13,5,1\},\{9,22,16\},
        \{6,24,14\},\{8,15,2\},\{7,18,17\}\}, \\ 
        \{\{18,7\},\{17,2\},\{15,22\},
        \{8,1\},\{5,9\},\{6,16\},
        \{14,24\},\{10,13\},\{12,21\},
        \{3,19\},\{4,23\},\{11,20\}\}\}}
        \caption{11-6-3-2-1-1 B \& C (cubic)}
        \label{Dessin}
    \end{subfigure}\hfill
\end{figure}

\begin{figure}[H]
    \begin{subfigure}{0.5\textwidth}
        \centering \captionsetup{justification=centering}
        $\scalemath{0.75}{
        \displaystyle \begin{pmatrix}
            2 & 1 & 0 & 0 & 0 & 0 & 0 & 0\\ 
            1 & 0 & 1 & 0 & 0 & 1 & 0 & 0\\
            0 & 1 & 0 & 0 & 1 & 0 & 1 & 0\\
            0 & 0 & 0 & 2 & 1 & 0 & 0 & 0\\
            0 & 0 & 1 & 1 & 0 & 0 & 1 & 0\\
            0 & 1 & 0 & 0 & 0 & 0 & 1 & 1\\
            0 & 0 & 1 & 0 & 1 & 1 & 0 & 0\\
            0 & 0 & 0 & 0 & 0 & 1 & 0 & 2
        \end{pmatrix}}$
        $\vcenter{\hbox{\includegraphics[width=0.35\textwidth]{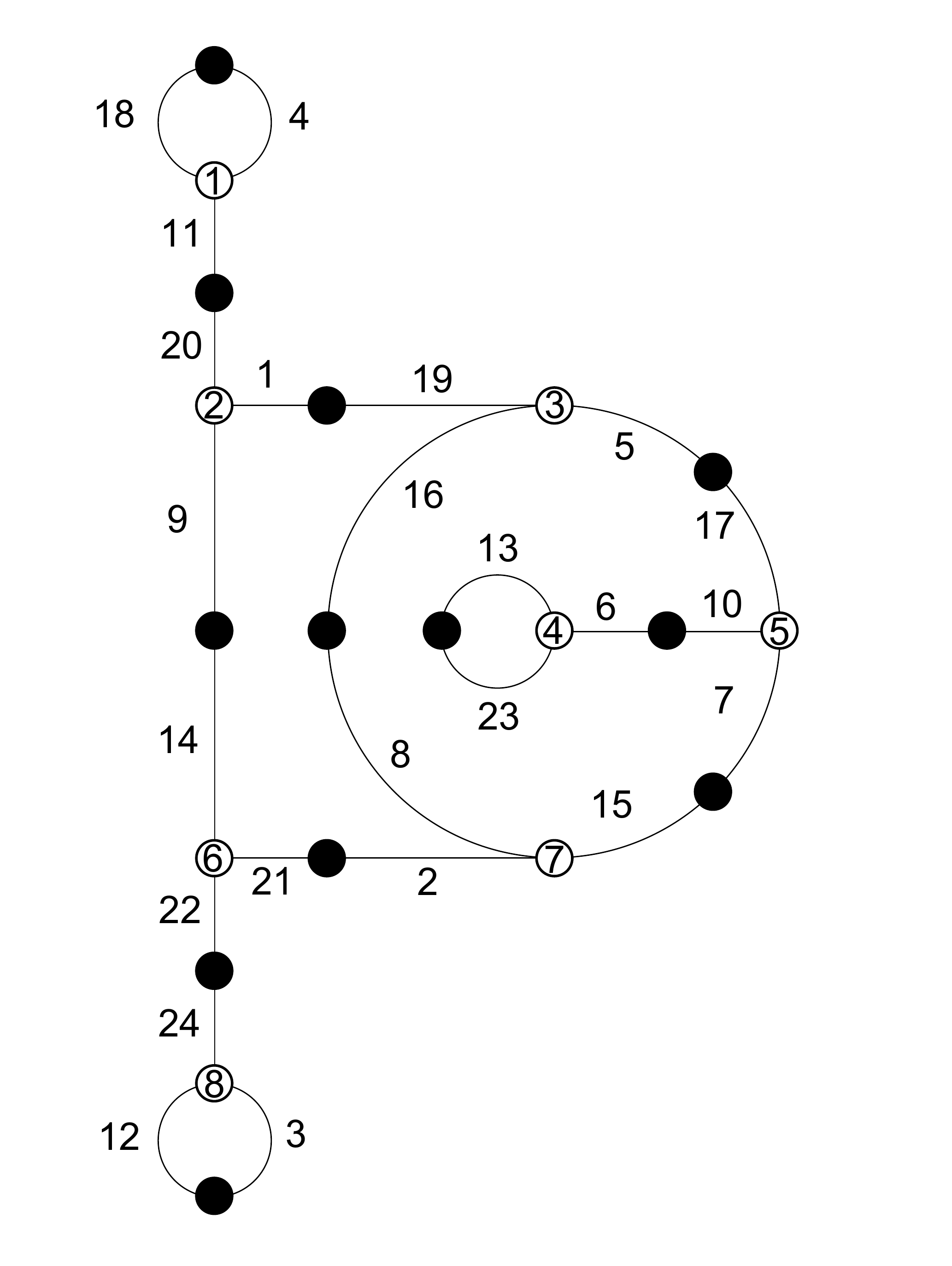}}}$
        \caption{ \{\{\{18,4,11\},\{20,1,9\},
        \{19,5,16\},\{17,7,10\},\{6,23,13\},
        \{8,15,2\},\{21,22,14\},\{24,3,12\}\}, \\ 
        \{\{3,12\},\{24,22\},\{9,14\},
        \{2,21\},\{15,7\},\{23,13\},
        \{8,16\},\{5,17\},\{6,10\},
        \{19,1\},\{20,11\},\{18,4\}\}\}}
        \caption{11-6-4-1-1-1 A $(\sqrt{33})$}
        \label{Dessin}
    \end{subfigure} \hfill
    \begin{subfigure}{0.5\textwidth}
        \centering \captionsetup{justification=centering}
        $\scalemath{0.75}{
        \displaystyle \begin{pmatrix}
            2 & 1 & 0 & 0 & 0 & 0 & 0 & 0\\ 
            1 & 0 & 1 & 1 & 0 & 0 & 0 & 0\\
            0 & 1 & 2 & 0 & 0 & 0 & 0 & 0\\
            0 & 1 & 0 & 0 & 0 & 1 & 0 & 1\\
            0 & 0 & 0 & 0 & 0 & 1 & 1 & 1\\
            0 & 0 & 0 & 1 & 1 & 0 & 0 & 1\\
            0 & 0 & 0 & 0 & 1 & 0 & 2 & 0\\
            0 & 0 & 0 & 1 & 1 & 1 & 0 & 0
        \end{pmatrix}}$
        $\vcenter{\hbox{\includegraphics[width=0.35\textwidth]{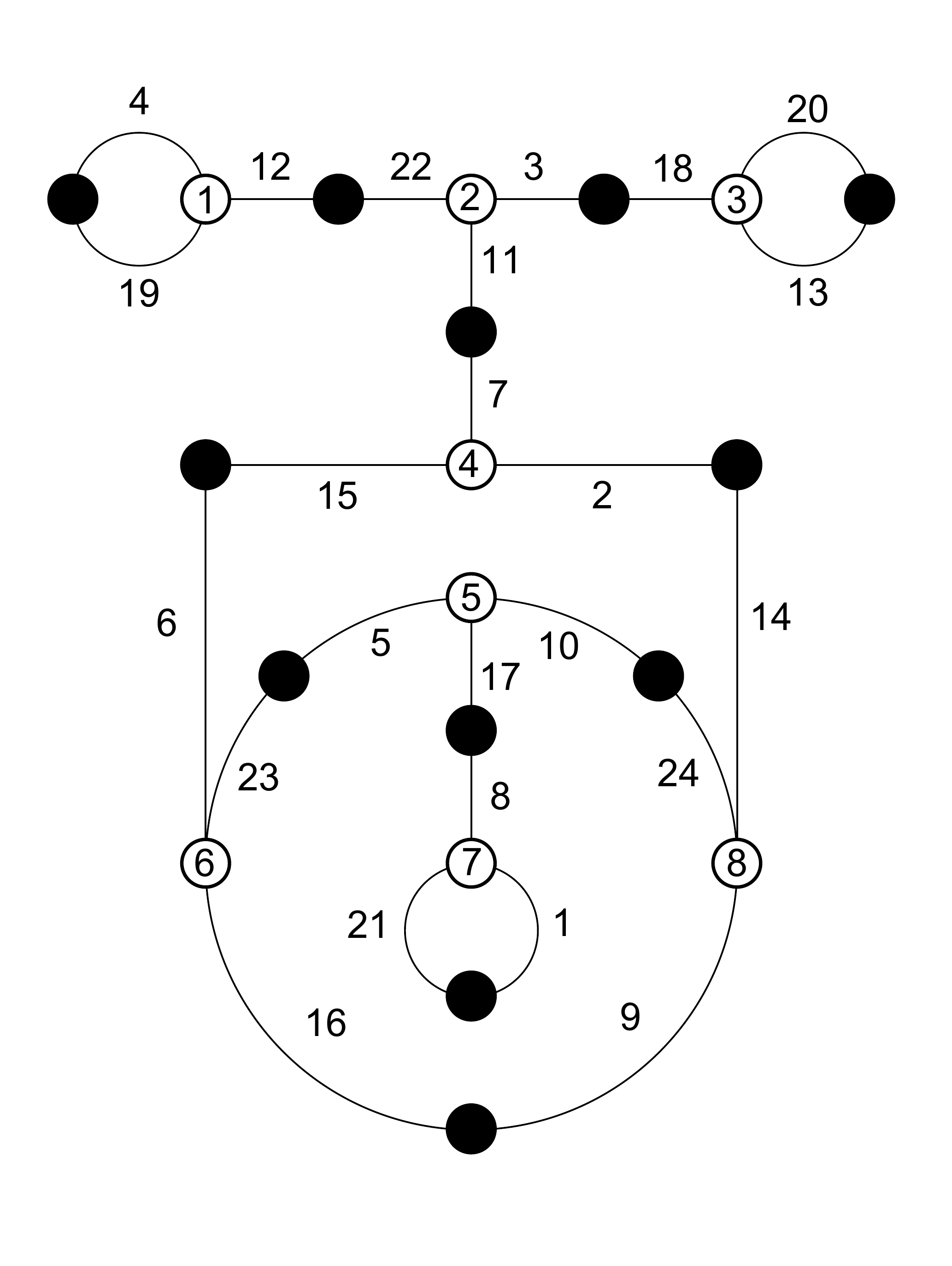}}}$
        \caption{ \{\{\{4,12,19\},\{22,3,11\},
        \{18,20,13\},\{7,2,15\},\{14,9,24\},
        \{17,5,10\},\{23,16,6\},\{8,1,21\}\}, \\ 
        \{\{21,1\},\{8,17\},\{5,23\},
        \{10,24\},\{14,2\},\{15,6\},
        \{16,9\},\{7,11\},\{22,12\},
        \{3,18\},\{20,13\},\{4,19\}\}\}}
        \caption{11-6-4-1-1-1 B $(\sqrt{33})$}
        \label{Dessin}
    \end{subfigure}\hfill
\end{figure}

\begin{figure}[H]
    \begin{subfigure}{0.6\textwidth}
        \centering \captionsetup{justification=centering}
        $\scalemath{0.75}{
        \displaystyle \begin{pmatrix}
            0 & 2 & 1 & 0 & 0 & 0 & 0 & 0\\ 
            2 & 0 & 0 & 0 & 1 & 0 & 0 & 0\\
            1 & 0 & 0 & 1 & 1 & 0 & 0 & 0\\
            0 & 0 & 1 & 2 & 0 & 0 & 0 & 0\\
            0 & 1 & 1 & 0 & 0 & 1 & 0 & 0\\
            0 & 0 & 0 & 0 & 1 & 0 & 2 & 0\\
            0 & 0 & 0 & 0 & 0 & 2 & 0 & 1\\
            0 & 0 & 0 & 0 & 0 & 0 & 1 & 2
        \end{pmatrix}}$
        $\vcenter{\hbox{\includegraphics[width=0.25\textwidth]{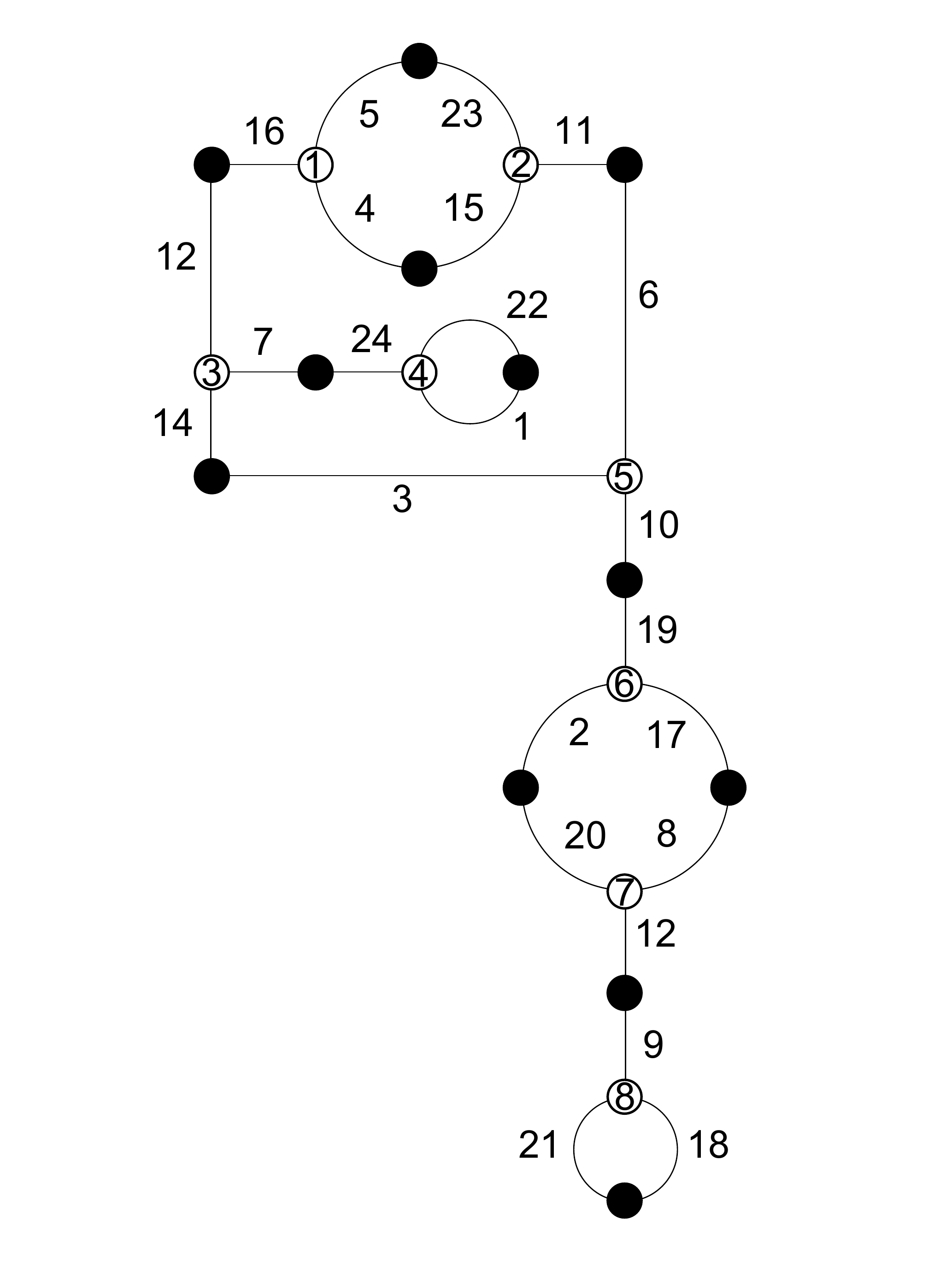}}}$
        $\vcenter{\hbox{\includegraphics[width=0.25\textwidth]{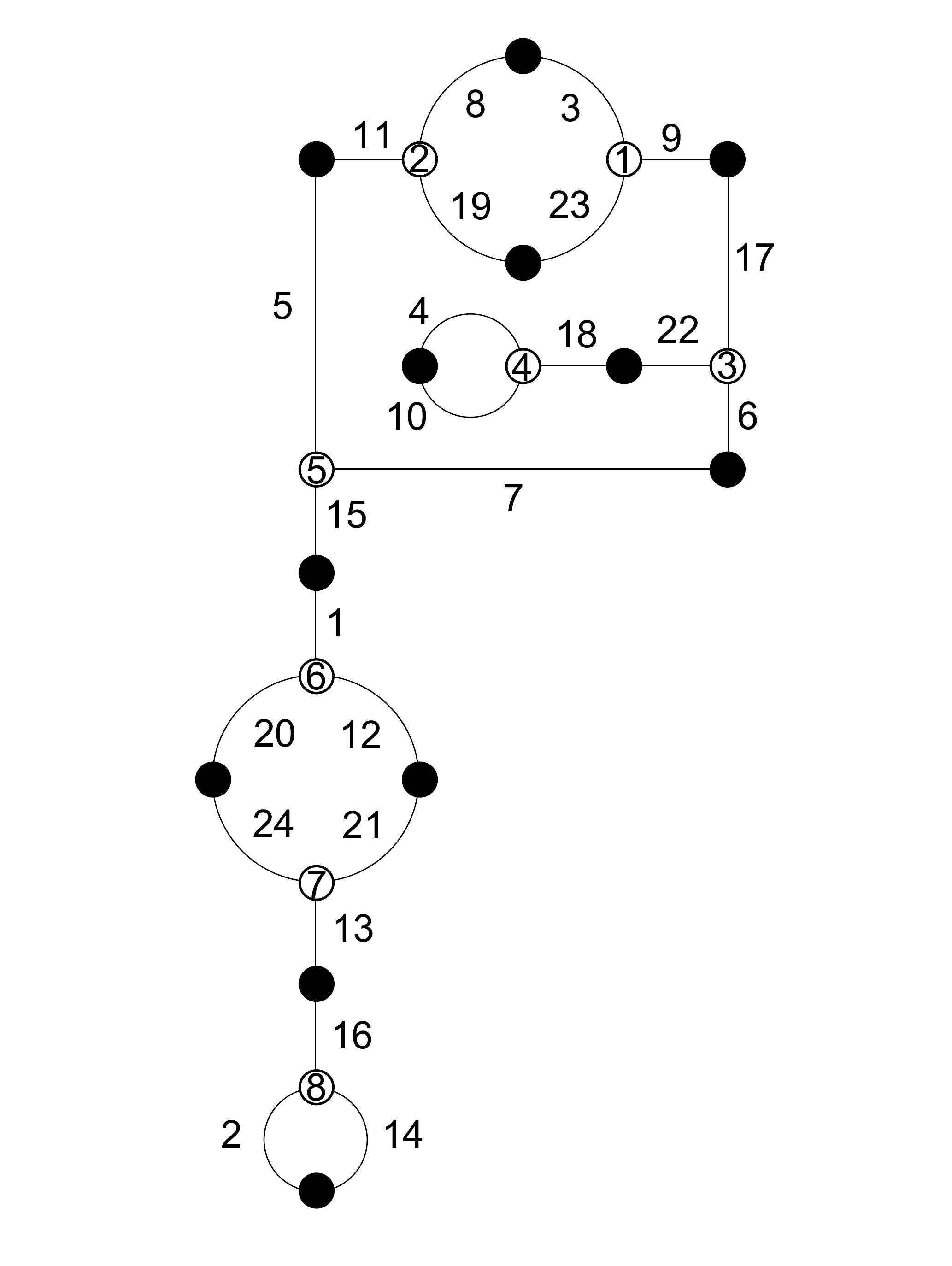}}}$
        \caption{ A: \{\{\{5,4,16\},\{23,11,15\},
        \{12,7,14\},\{24,22,1\},\{3,6,10\},
        \{19,17,2\},\{20,8,12\},\{9,18,21\}\}, \\ 
        \{\{21,18\},\{9,12\},\{8,17\},
        \{2,20\},\{19,10\},\{3,14\},
        \{6,11\},\{7,24\},\{12,16\},
        \{5,23\},\{4,15\},\{1,22\}\}\} \\
        B: \{\{\{8,19,11\},\{3,9,23\},
        \{4,18,10\},\{22,17,6\},\{5,7,15\},
        \{1,12,20\},\{24,21,13\},\{16,14,2\}\}, \\ 
        \{\{2,14\},\{16,13\},\{21,12\},
        \{20,24\},\{15,1\},\{5,11\},
        \{6,7\},\{22,18\},\{4,10\},
        \{17,9\},\{23,19\},\{8,3\}\}\}}
        \caption{11-7-2-2-1-1 A \& B $(\sqrt{-7})$}
        \label{Dessin}
    \end{subfigure} \hfill
    \begin{subfigure}{0.4\textwidth}
        \centering \captionsetup{justification=centering}
        $\scalemath{0.75}{
        \displaystyle \begin{pmatrix}
            2 & 1 & 0 & 0 & 0 & 0 & 0 & 0\\ 
            1 & 0 & 1 & 1 & 0 & 0 & 0 & 0\\
            0 & 1 & 2 & 0 & 0 & 0 & 0 & 0\\
            0 & 1 & 0 & 0 & 0 & 0 & 1 & 1\\
            0 & 0 & 0 & 0 & 2 & 1 & 0 & 0\\
            0 & 0 & 0 & 0 & 1 & 0 & 1 & 1\\
            0 & 0 & 0 & 1 & 0 & 1 & 0 & 1\\
            0 & 0 & 0 & 1 & 0 & 1 & 1 & 0
        \end{pmatrix}}$
        $\vcenter{\hbox{\includegraphics[width=0.35\textwidth]{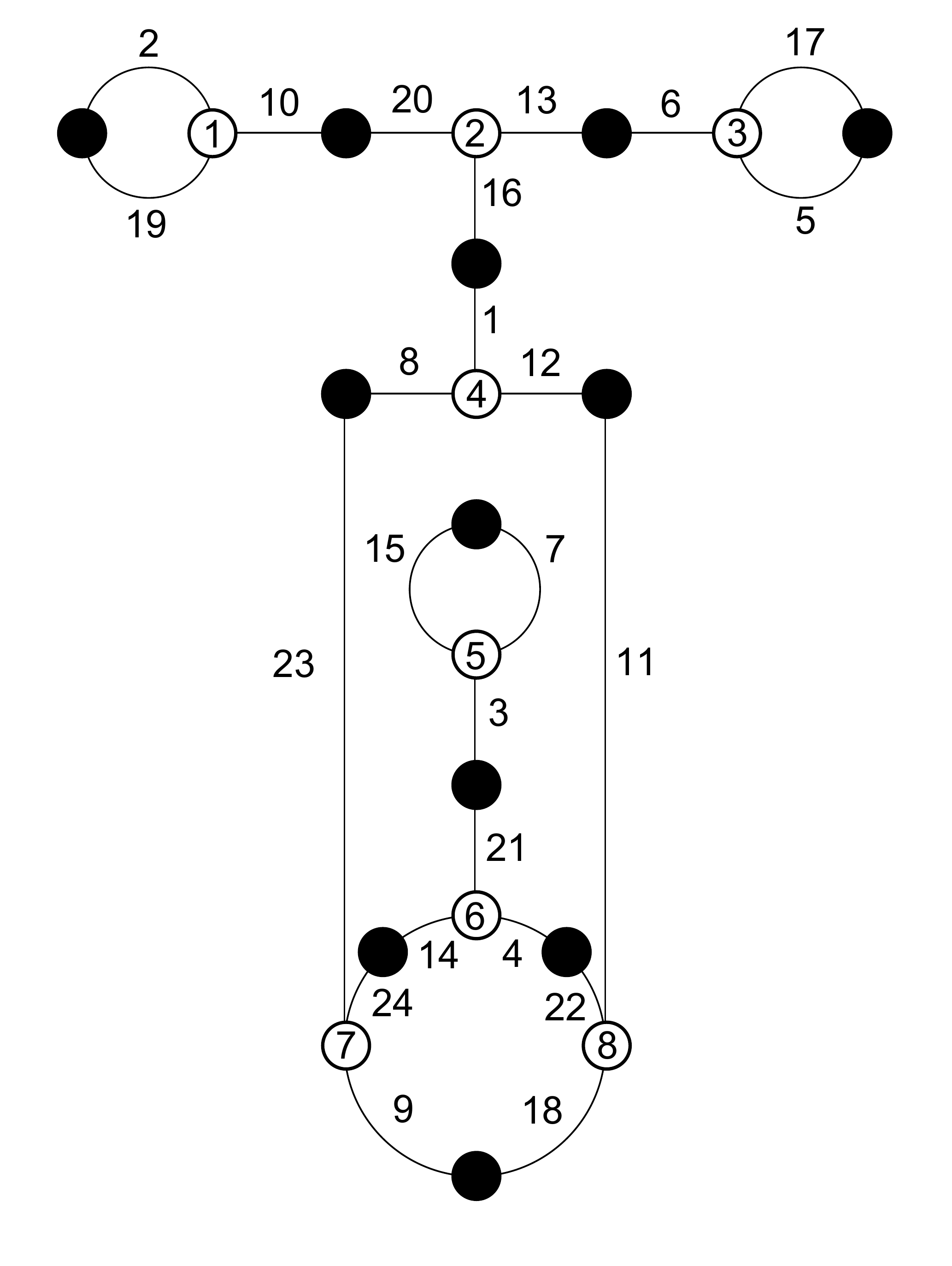}}}$
        \caption{ \{\{\{2,10,19\},\{20,13,16\},
        \{6,17,5\},\{1,12,8\},\{11,18,22\},
        \{9,23,24\},\{14,21,4\},\{3,15,7\}\}, \\ 
        \{\{9,18\},\{24,14\},\{4,22\},
        \{21,3\},\{15,7\},\{23,8\},
        \{12,11\},\{1,16\},\{13,6\},
        \{17,5\},\{2,19\},\{10,20\}\}\}}
        \caption{11-7-3-1-1-1 A (cubic)}
        \label{Dessin}
    \end{subfigure}\hfill
\end{figure}

\begin{figure}[H]
    \begin{subfigure}{0.6\textwidth}
        \centering \captionsetup{justification=centering}
        $\scalemath{0.75}{
        \displaystyle \begin{pmatrix}
            2 & 1 & 0 & 0 & 0 & 0 & 0 & 0\\ 
            1 & 0 & 1 & 0 & 0 & 1 & 0 & 0\\
            0 & 1 & 0 & 1 & 1 & 0 & 0 & 0\\
            0 & 0 & 1 & 2 & 0 & 0 & 0 & 0\\
            0 & 0 & 1 & 0 & 0 & 1 & 1 & 0\\
            0 & 1 & 0 & 0 & 1 & 0 & 1 & 0\\
            0 & 0 & 0 & 0 & 1 & 1 & 0 & 1\\
            0 & 0 & 0 & 0 & 0 & 0 & 1 & 2
        \end{pmatrix}}$
        $\vcenter{\hbox{\includegraphics[width=0.25\textwidth]{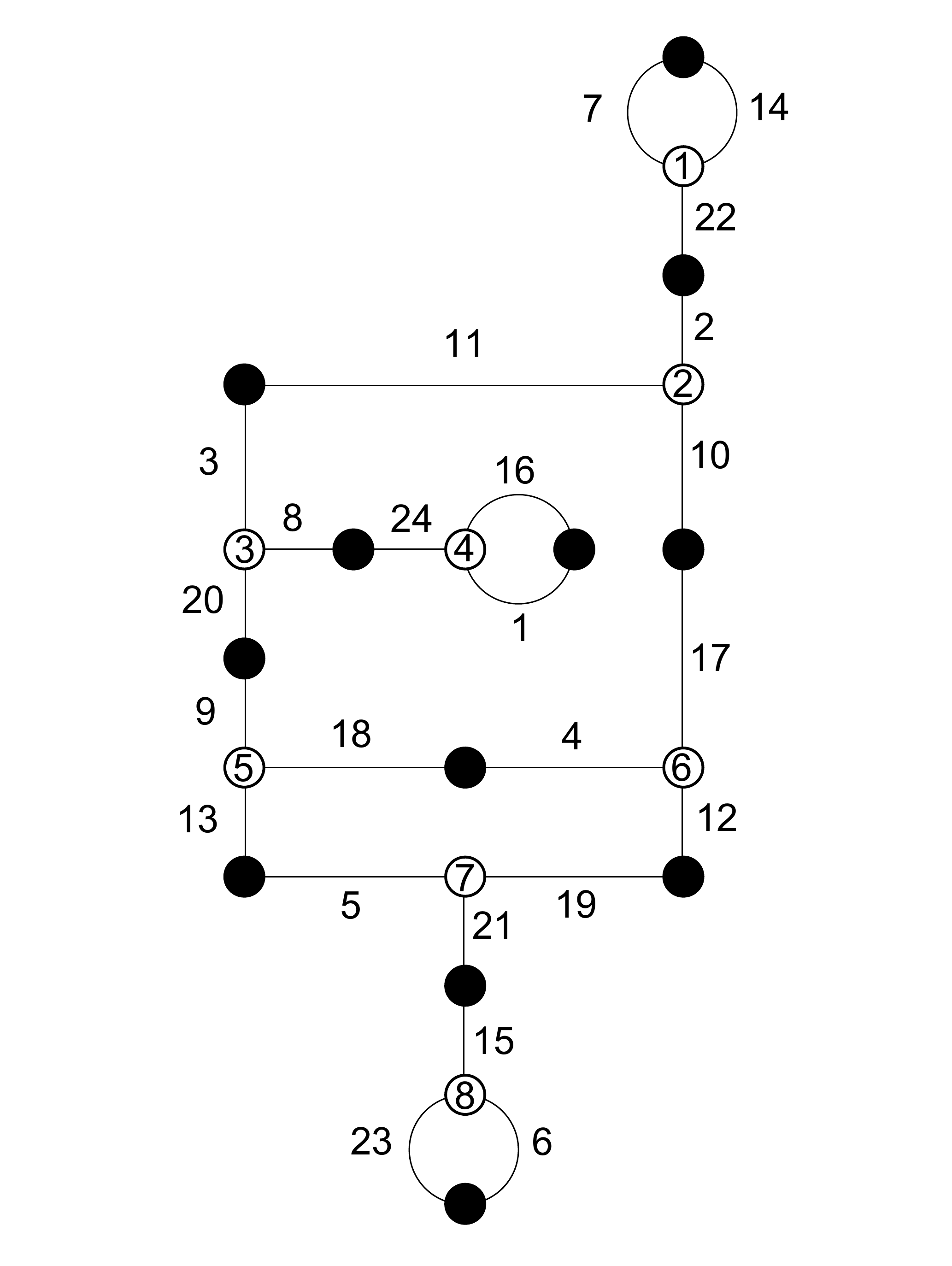}}}$
        $\vcenter{\hbox{\includegraphics[width=0.25\textwidth]{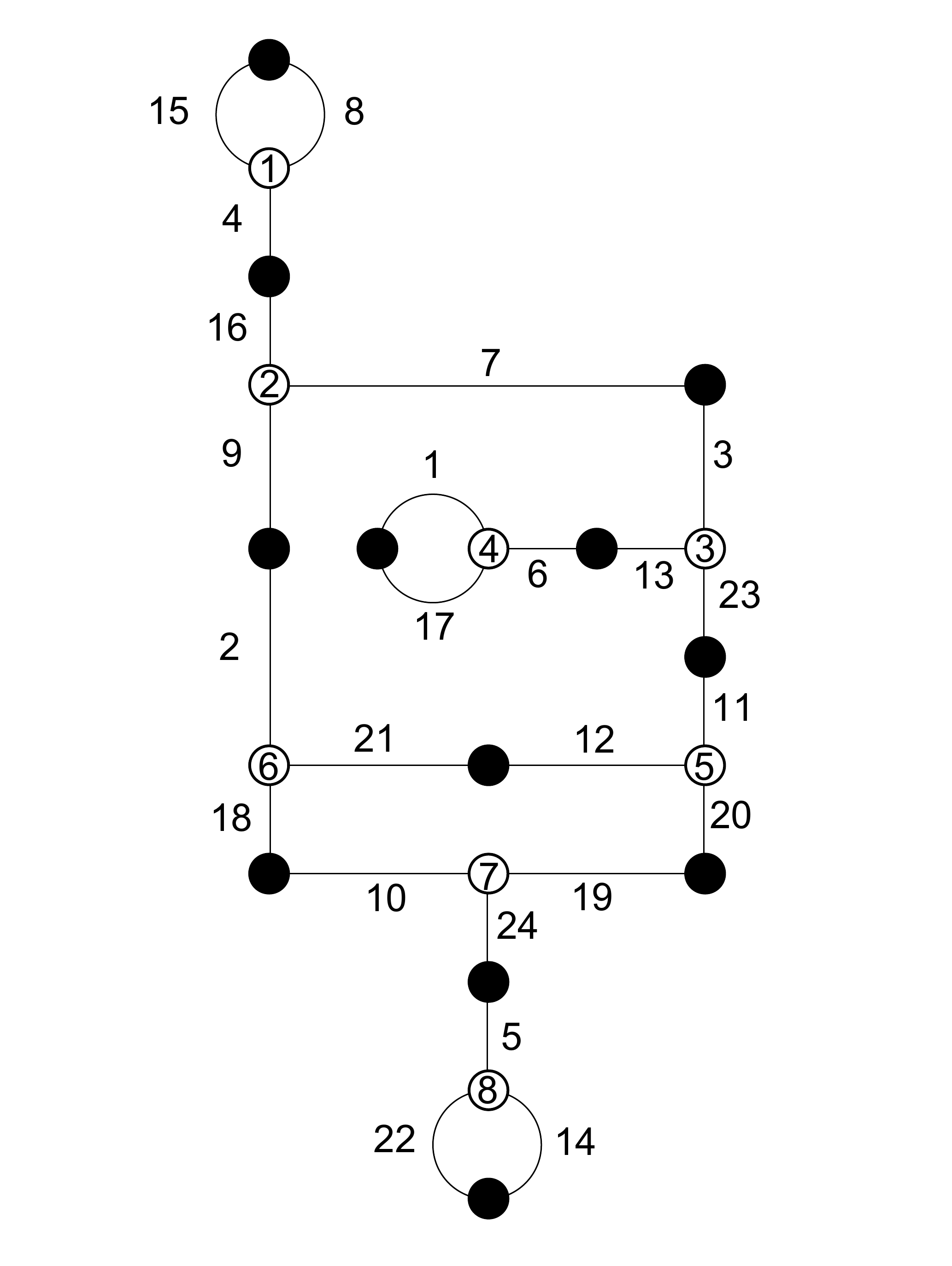}}}$
        \caption{ B: \{\{\{7,14,22\},\{2,10,11\},
        \{3,8,20\},\{24,16,1\},\{9,18,13\},
        \{4,17,12\},\{19,21,5\},\{23,15,6\}\}, \\ 
        \{\{6,23\},\{15,21\},\{19,12\},
        \{5,13\},\{18,4\},\{9,20\},
        \{3,11\},\{8,24\},\{17,10\},
        \{16,1\},\{2,22\},\{7,14\}\}\} \\
        C: \{\{\{15,8,4\},\{16,7,9\},
        \{3,23,13\},\{6,17,1\},\{11,20,12\},
        \{21,18,2\},\{19,24,10\},\{5,14,22\}\}, \\ 
        \{\{22,14\},\{5,24\},\{10,18\},
        \{19,20\},\{21,12\},\{2,9\},
        \{11,23\},\{3,7\},\{1,17\},
        \{6,13\},\{16,4\},\{8,15\}\}\}}
        \caption{11-7-3-1-1-1 B \& C (cubic)}
        \label{Dessin}
    \end{subfigure}\hfill
    \begin{subfigure}{0.4\textwidth}
        \centering \captionsetup{justification=centering}
        $\scalemath{0.75}{
        \displaystyle \begin{pmatrix}
            2 & 1 & 0 & 0 & 0 & 0 & 0 & 0\\ 
            1 & 0 & 1 & 1 & 0 & 0 & 0 & 0\\
            0 & 1 & 2 & 0 & 0 & 0 & 0 & 0\\
            0 & 1 & 0 & 0 & 1 & 0 & 0 & 1\\
            0 & 0 & 0 & 1 & 0 & 1 & 0 & 1\\
            0 & 0 & 0 & 0 & 1 & 2 & 0 & 0\\
            0 & 0 & 0 & 0 & 0 & 0 & 2 & 1\\
            0 & 0 & 0 & 1 & 1 & 0 & 1 & 0
        \end{pmatrix}}$
        $\vcenter{\hbox{\includegraphics[width=0.35\textwidth]{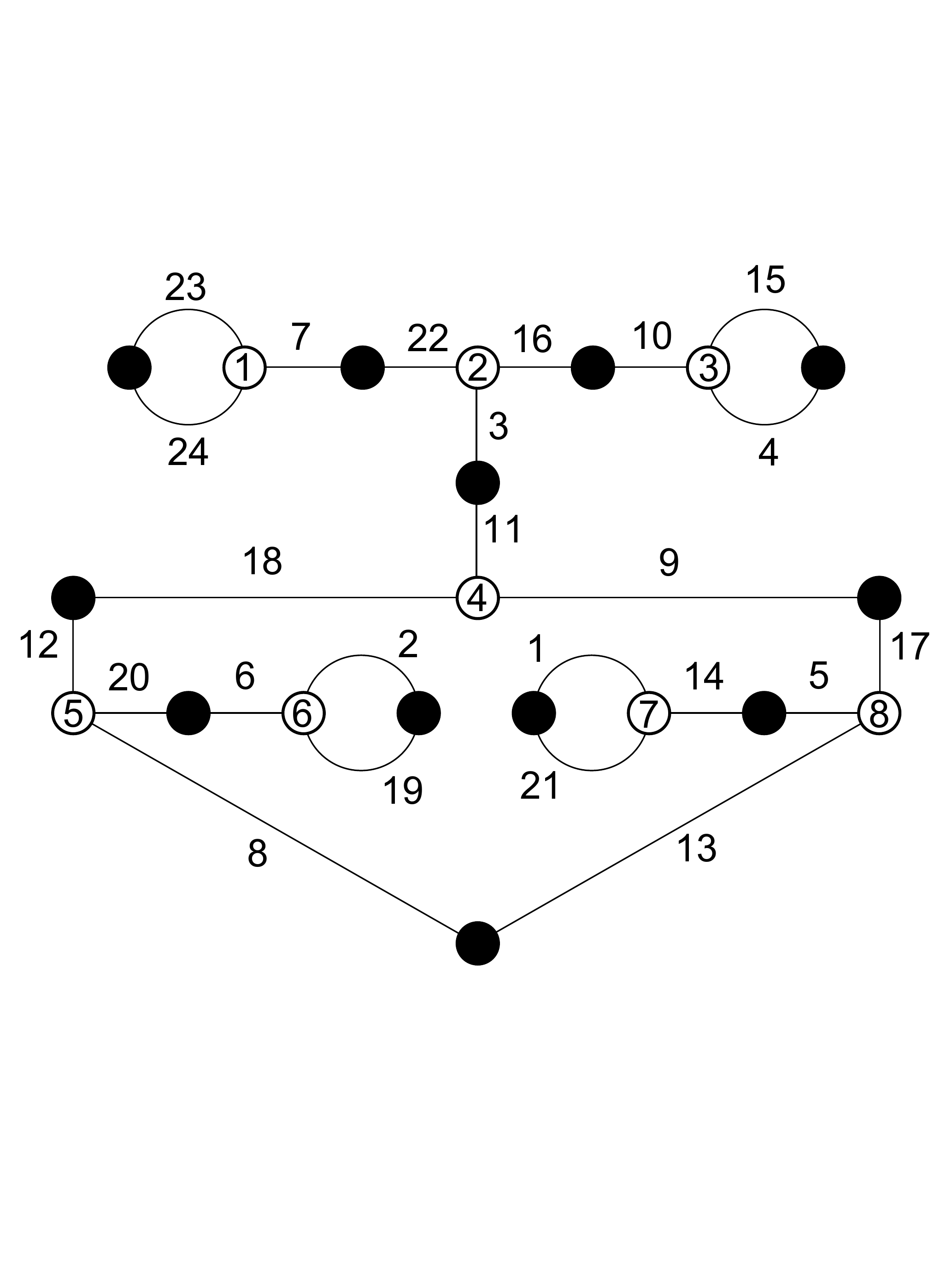}}}$
        \caption{ \{\{\{24,23,7\},\{22,16,3\},
        \{15,4,10\},\{11,9,18\},\{12,20,8\},
        \{6,2,19\},\{1,14,21\},\{5,17,13\}\}, \\ 
        \{\{8,13\},\{2,19\},\{6,20\},
        \{12,18\},\{17,9\},\{21,1\},
        \{5,14\},\{11,3\},\{22,7\},
        \{23,24\},\{15,4\},\{16,10\}\}\}}
        \caption{11-9-1-1-1-1 $(\mathbb{Q})$}
        \label{Dessin}
    \end{subfigure} \hfill
\end{figure}

\begin{figure}[H]
\begin{subfigure}{\textwidth}
    \centering \captionsetup{justification=centering}
    $\scalemath{0.75}{
    \displaystyle \begin{pmatrix}
        2 & 0 & 1 & 0 & 0 & 0 & 0 & 0\\ 
        0 & 2 & 0 & 0 & 1 & 0 & 0 & 0\\
        1 & 0 & 0 & 1 & 0 & 0 & 1 & 0\\
        0 & 0 & 1 & 0 & 1 & 1 & 0 & 0\\
        0 & 1 & 0 & 1 & 0 & 0 & 0 & 1\\
        0 & 0 & 0 & 1 & 0 & 2 & 0 & 0\\
        0 & 0 & 1 & 0 & 0 & 0 & 0 & 2\\
        0 & 0 & 0 & 0 & 1 & 0 & 2 & 0
    \end{pmatrix}}$
    $\vcenter{\hbox{\includegraphics[width=0.2\textwidth]{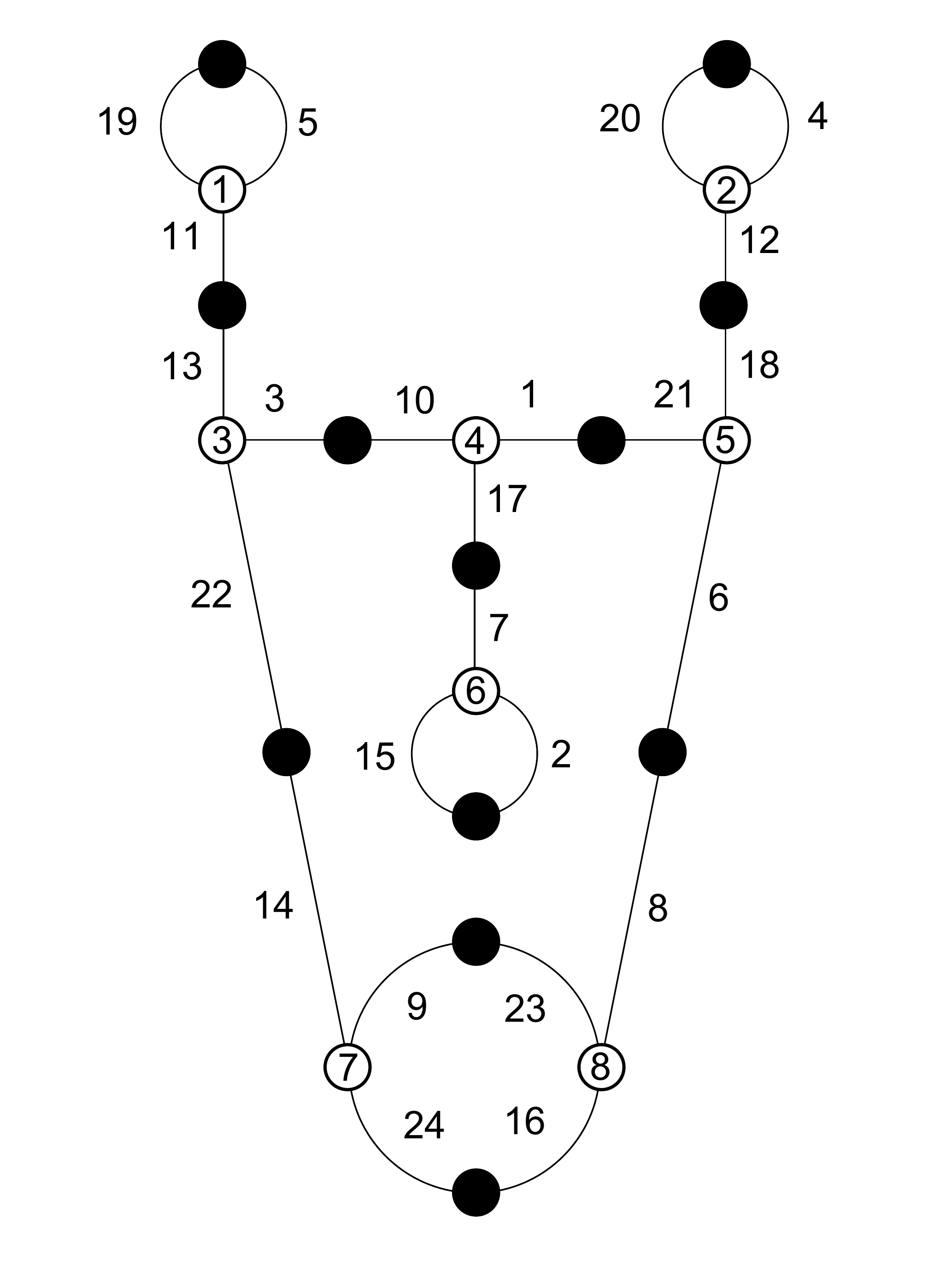}}}$
    $\vcenter{\hbox{\includegraphics[width=0.2\textwidth]{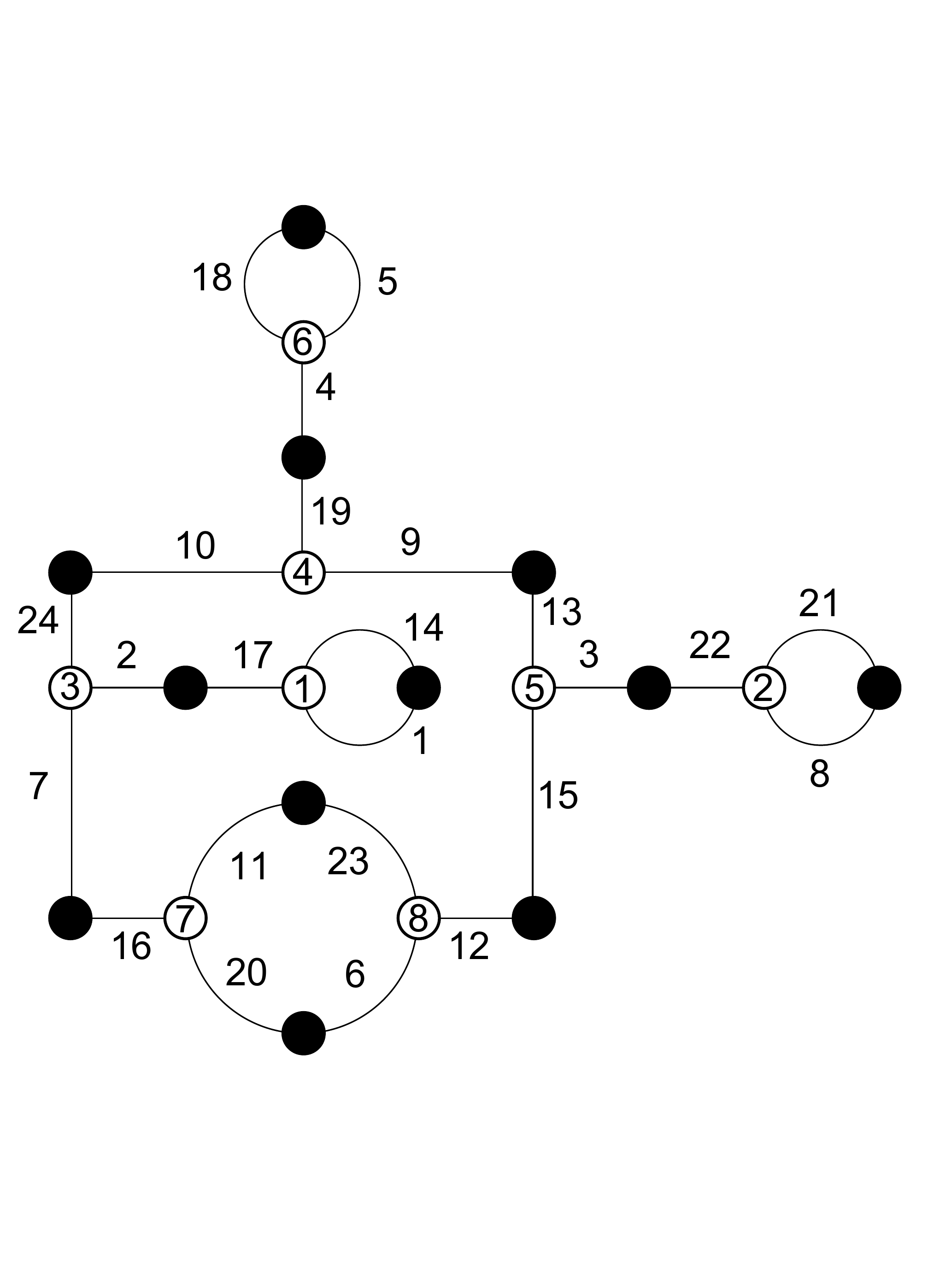}}}$
    $\vcenter{\hbox{\includegraphics[width=0.2\textwidth]{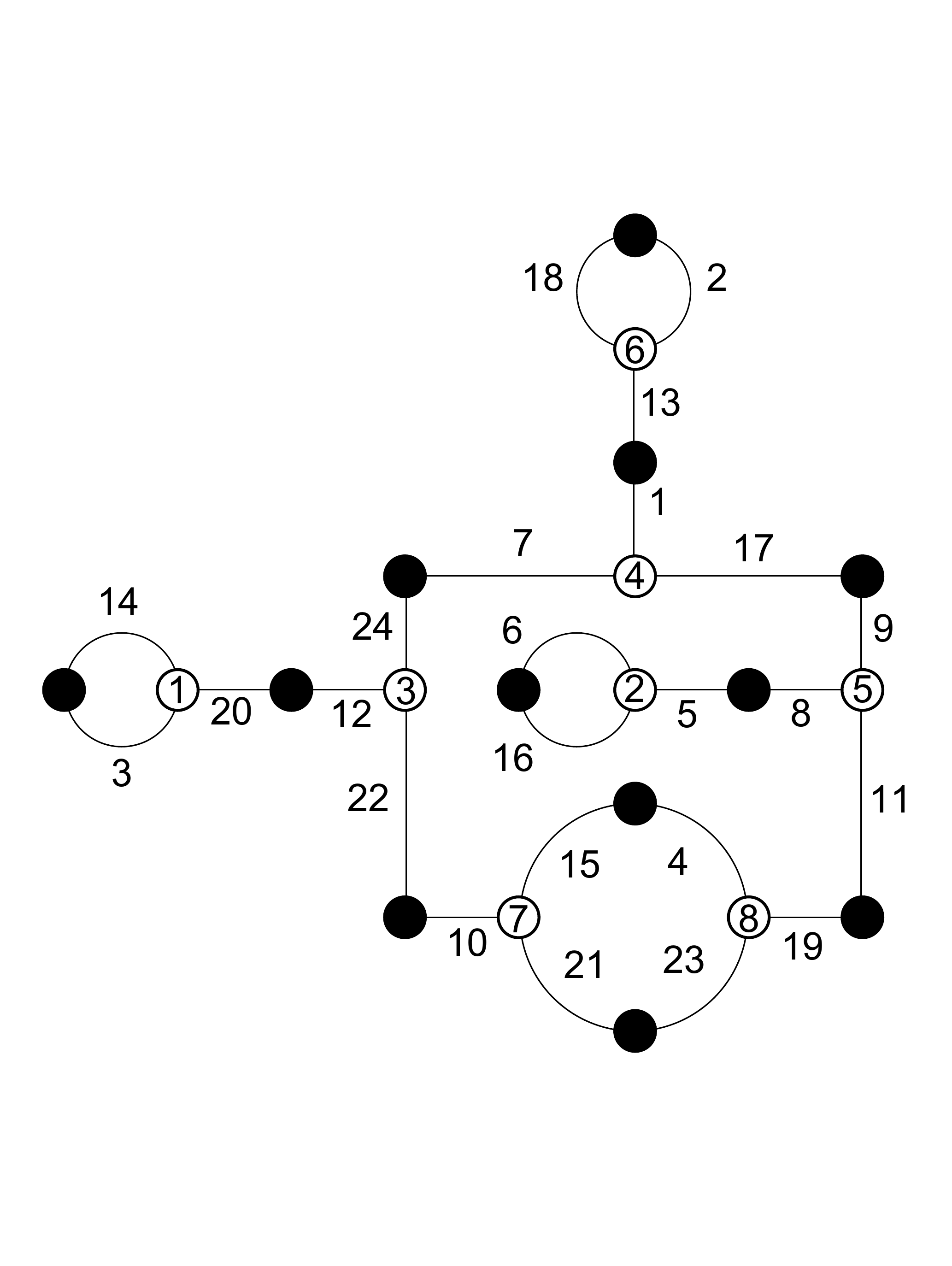}}}$
        \caption{ A: \{\{\{19,5,11\},\{13,3,22\},
        \{20,4,12\},\{18,6,21\},\{10,1,17\},
        \{7,2,15\},\{14,9,24\},\{23,8,16\}\}, \\ 
        \{\{24,16\},\{23,9\},\{14,22\},
        \{6,8\},\{7,17\},\{15,2\},
        \{10,3\},\{1,21\},\{18,12\},
        \{13,11\},\{19,5\},\{4,20\}\}\} \\
        B: \{\{\{18,5,4\},\{19,9,10\},
        \{24,2,7\},\{17,14,1\},\{16,11,20\},
        \{23,12,6\},\{15,13,3\},\{22,21,8\}\}, \\ 
        \{\{18,5\},\{4,19\},\{9,13\},
        \{10,24\},\{2,17\},\{14,1\},
        \{7,16\},\{11,23\},\{6,20\},
        \{3,22\},\{21,8\},\{16,7\}\}\} \\
        C: \{\{\{18,2,13\},\{1,17,7\},
        \{9,11,8\},\{5,16,6\},\{19,23,4\},
        \{15,21,10\},\{22,12,24\},\{20,3,14\}\}, \\ 
        \{\{3,14\},\{20,12\},\{24,7\},
        \{1,13\},\{2,18\},\{22,10\},
        \{21,23\},\{15,4\},\{19,11\},
        \{9,17\},\{5,8\},\{6,16\}\}\}}
    \caption{11-8-2-1-1-1 A, B, \& C (cubic)}
    \label{Dessin}
\end{subfigure}
\end{figure}\hfill

\begin{figure}[H]
    \begin{subfigure}{0.5\textwidth}
        \centering \captionsetup{justification=centering}
        $\scalemath{0.75}{
        \displaystyle \begin{pmatrix}
            0 & 2 & 1 & 0 & 0 & 0 & 0 & 0\\ 
            2 & 0 & 1 & 0 & 0 & 0 & 0 & 0\\
            1 & 1 & 0 & 1 & 0 & 0 & 0 & 0\\
            0 & 0 & 1 & 0 & 2 & 0 & 0 & 0\\
            0 & 0 & 0 & 2 & 0 & 1 & 0 & 0\\
            0 & 0 & 0 & 0 & 1 & 0 & 1 & 1\\
            0 & 0 & 0 & 0 & 0 & 1 & 0 & 2\\
            0 & 0 & 0 & 0 & 0 & 1 & 2 & 0
        \end{pmatrix}}$
        $\vcenter{\hbox{\includegraphics[width=0.35\textwidth]{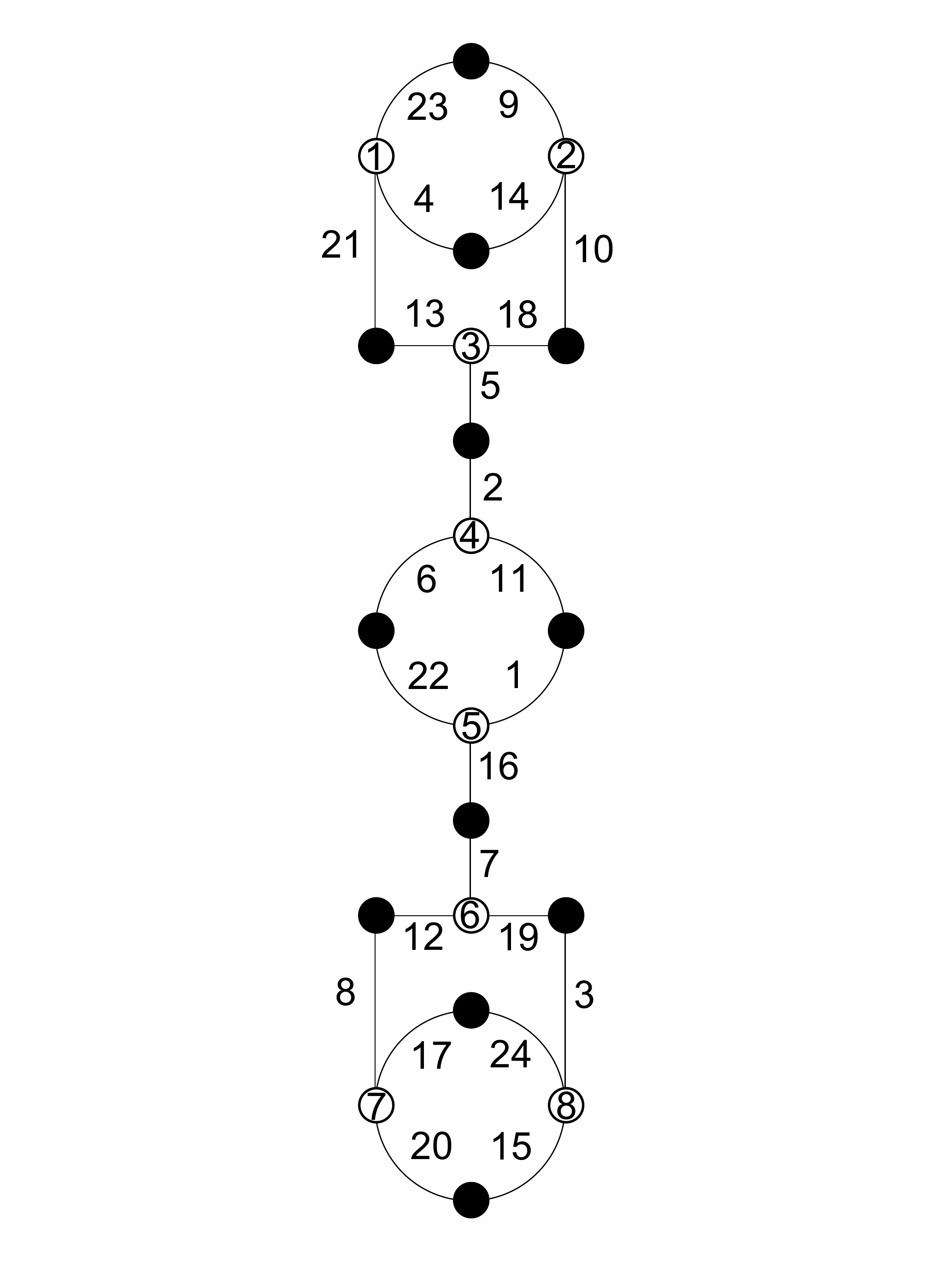}}}$
        \caption{ \{\{\{23,4,21\},\{9,10,14\},
        \{13,18,5\},\{2,11,6\},\{22,1,16\},
        \{7,19,12\},\{3,15,24\},\{8,17,20\}\}, \\ 
        \{\{20,15\},\{24,17\},\{12,8\},
        \{19,3\},\{7,16\},\{1,11\},
        \{6,22\},\{2,5\},\{13,21\},
        \{18,10\},\{14,4\},\{9,23\}\}\}}
        \caption{12-3-3-2-2-2 $(\mathbb{Q})$}
        \label{Dessin}
    \end{subfigure} \hfill
    \begin{subfigure}{0.5\textwidth}
        \centering \captionsetup{justification=centering}
        $\scalemath{0.75}{
        \displaystyle \begin{pmatrix}
            2 & 1 & 0 & 0 & 0 & 0 & 0 & 0\\ 
            1 & 0 & 1 & 1 & 0 & 0 & 0 & 0\\
            0 & 1 & 0 & 1 & 1 & 0 & 0 & 0\\
            0 & 1 & 1 & 0 & 1 & 0 & 0 & 0\\
            0 & 0 & 1 & 1 & 0 & 1 & 0 & 0\\
            0 & 0 & 0 & 0 & 1 & 0 & 1 & 1\\
            0 & 0 & 0 & 0 & 0 & 1 & 0 & 2\\
            0 & 0 & 0 & 0 & 0 & 1 & 2 & 0
        \end{pmatrix}}$
        $\vcenter{\hbox{\includegraphics[width=0.35\textwidth]{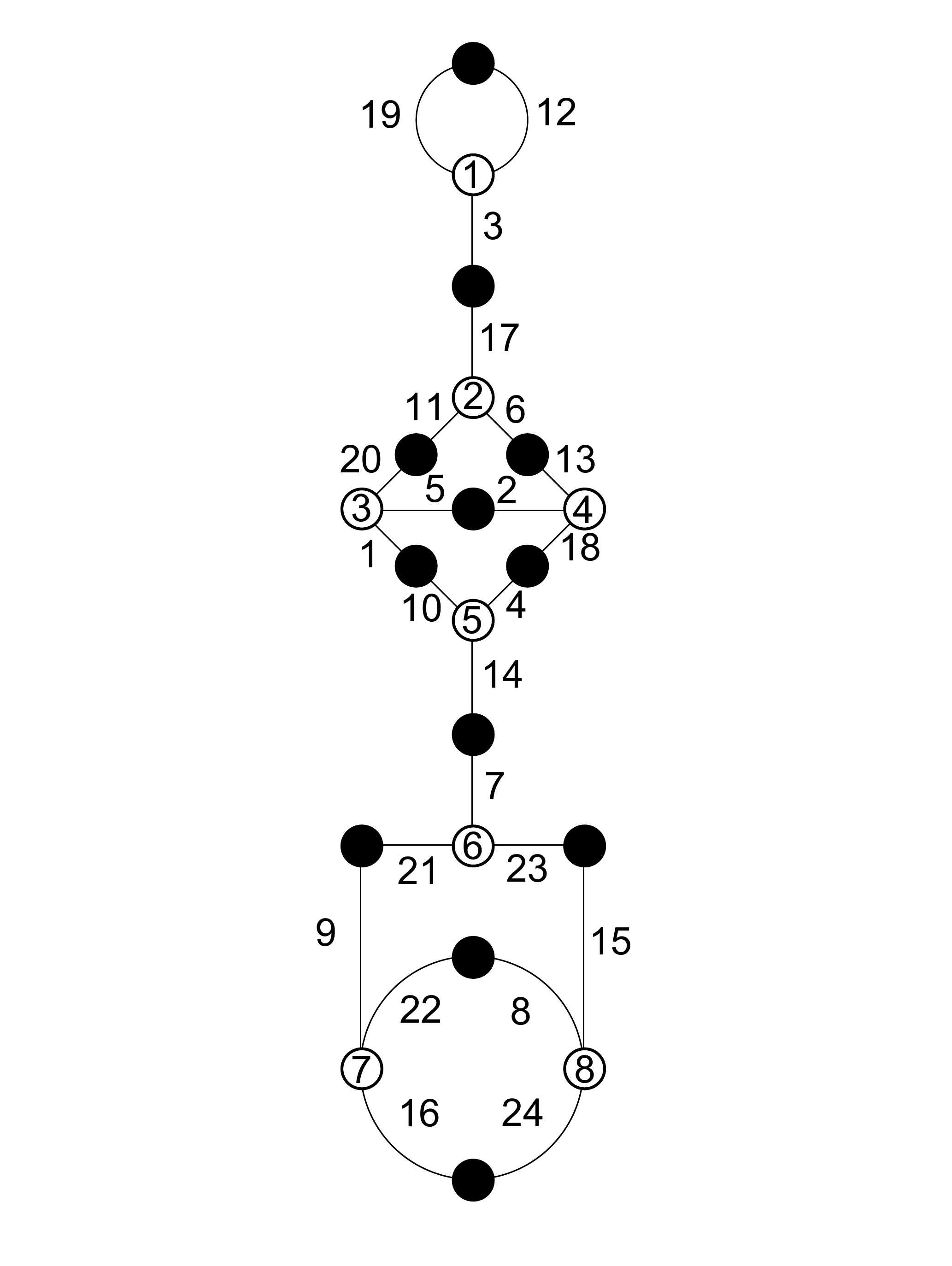}}}$
        \caption{ \{\{\{19,12,3\},\{17,6,11\},
        \{2,13,18\},\{5,1,20\},\{4,14,10\},
        \{7,23,21\},\{15,24,8\},\{22,16,9\}\}, \\ 
        \{\{16,24\},\{8,22\},\{21,9\},
        \{15,23\},\{7,14\},\{4,18\},
        \{1,10\},\{2,5\},\{11,20\},
        \{6,13\},\{17,3\},\{19,12\}\}\}}
        \caption{12-3-3-3-2-1 $(\mathbb{Q})$}
        \label{Dessin}
    \end{subfigure}\hfill
\end{figure}

\begin{figure}[H]
    \begin{subfigure}{0.4\textwidth}
        \centering \captionsetup{justification=centering}
        $\scalemath{0.75}{
        \displaystyle \begin{pmatrix}
            2 & 1 & 0 & 0 & 0 & 0 & 0 & 0\\ 
            1 & 0 & 2 & 0 & 0 & 0 & 0 & 0\\
            0 & 2 & 0 & 1 & 0 & 0 & 0 & 0\\
            0 & 0 & 1 & 0 & 1 & 1 & 0 & 0\\
            0 & 0 & 0 & 1 & 0 & 1 & 1 & 0\\
            0 & 0 & 0 & 1 & 1 & 0 & 0 & 1\\
            0 & 0 & 0 & 0 & 1 & 0 & 0 & 2\\
            0 & 0 & 0 & 0 & 0 & 1 & 2 & 0
        \end{pmatrix}}$
        $\vcenter{\hbox{\includegraphics[width=0.35\textwidth]{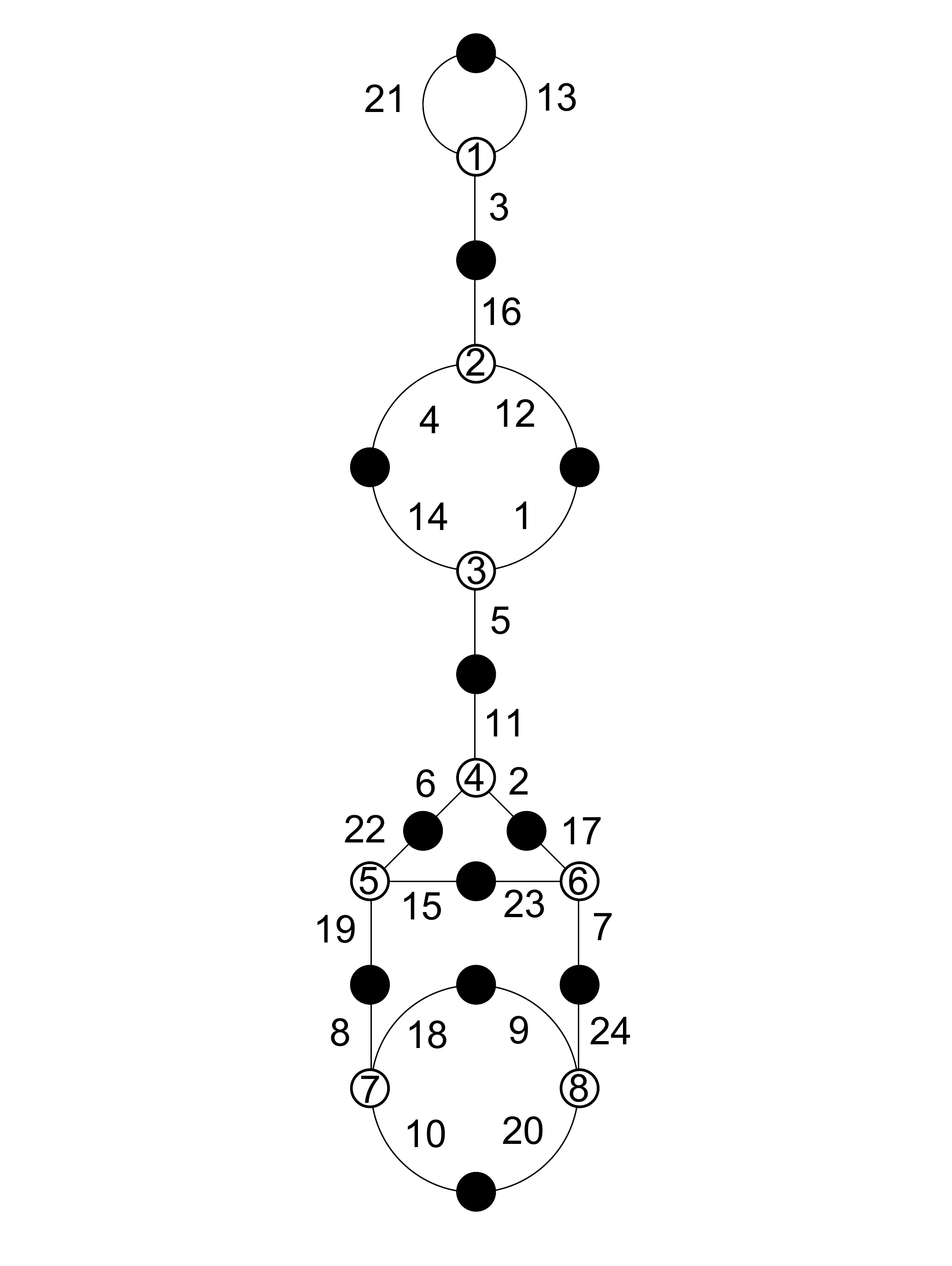}}}$
        \caption{ \{\{\{21,13,3\},\{16,12,4\},
        \{1,5,14\},\{11,2,6\},\{17,7,23\},
        \{22,15,19\},\{8,18,10\},\{24,20,9\}\}, \\ 
        \{\{10,20\},\{18,9\},\{8,19\},
        \{7,24\},\{15,23\},\{22,6\},
        \{2,17\},\{11,5\},\{1,12\},
        \{4,14\},\{16,3\},\{21,13\}\}\}}
        \caption{12-4-3-2-2-1 A $(\mathbb{Q})$}
        \label{Dessin}
    \end{subfigure} \hfill
    \begin{subfigure}{0.6\textwidth}
        \centering \captionsetup{justification=centering}
        $\scalemath{0.75}{
        \displaystyle \begin{pmatrix}
            0 & 2 & 1 & 0 & 0 & 0 & 0 & 0\\ 
            2 & 0 & 1 & 0 & 0 & 0 & 0 & 0\\
            1 & 1 & 0 & 1 & 0 & 0 & 0 & 0\\
            0 & 0 & 1 & 0 & 1 & 0 & 1 & 0\\
            0 & 0 & 0 & 1 & 0 & 2 & 0 & 0\\
            0 & 0 & 0 & 0 & 2 & 0 & 1 & 0\\
            0 & 0 & 0 & 1 & 0 & 1 & 0 & 1\\
            0 & 0 & 0 & 0 & 0 & 0 & 1 & 2
        \end{pmatrix}}$
        $\vcenter{\hbox{\includegraphics[width=0.25\textwidth]{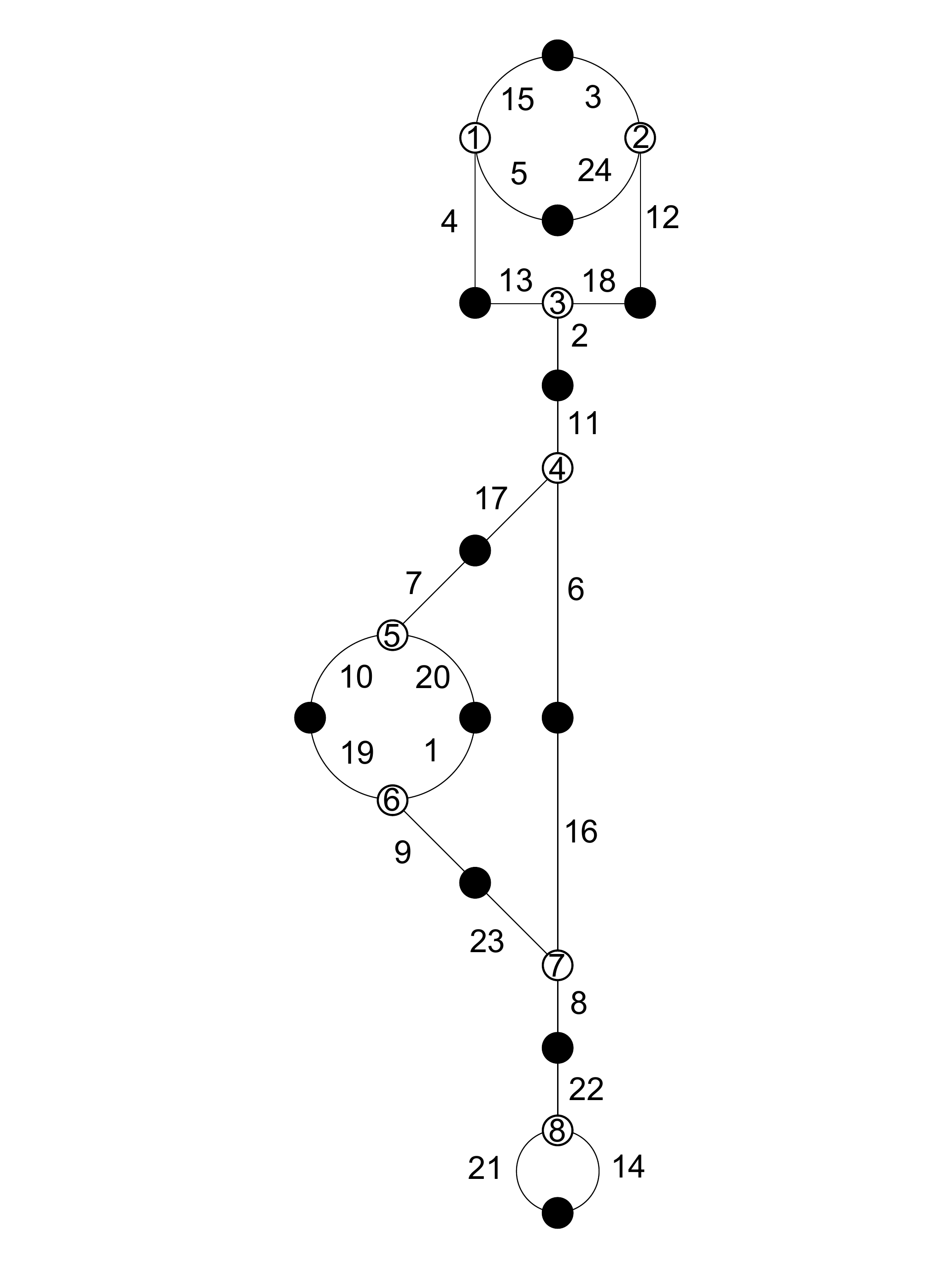}}}$
        $\vcenter{\hbox{\includegraphics[width=0.25\textwidth]{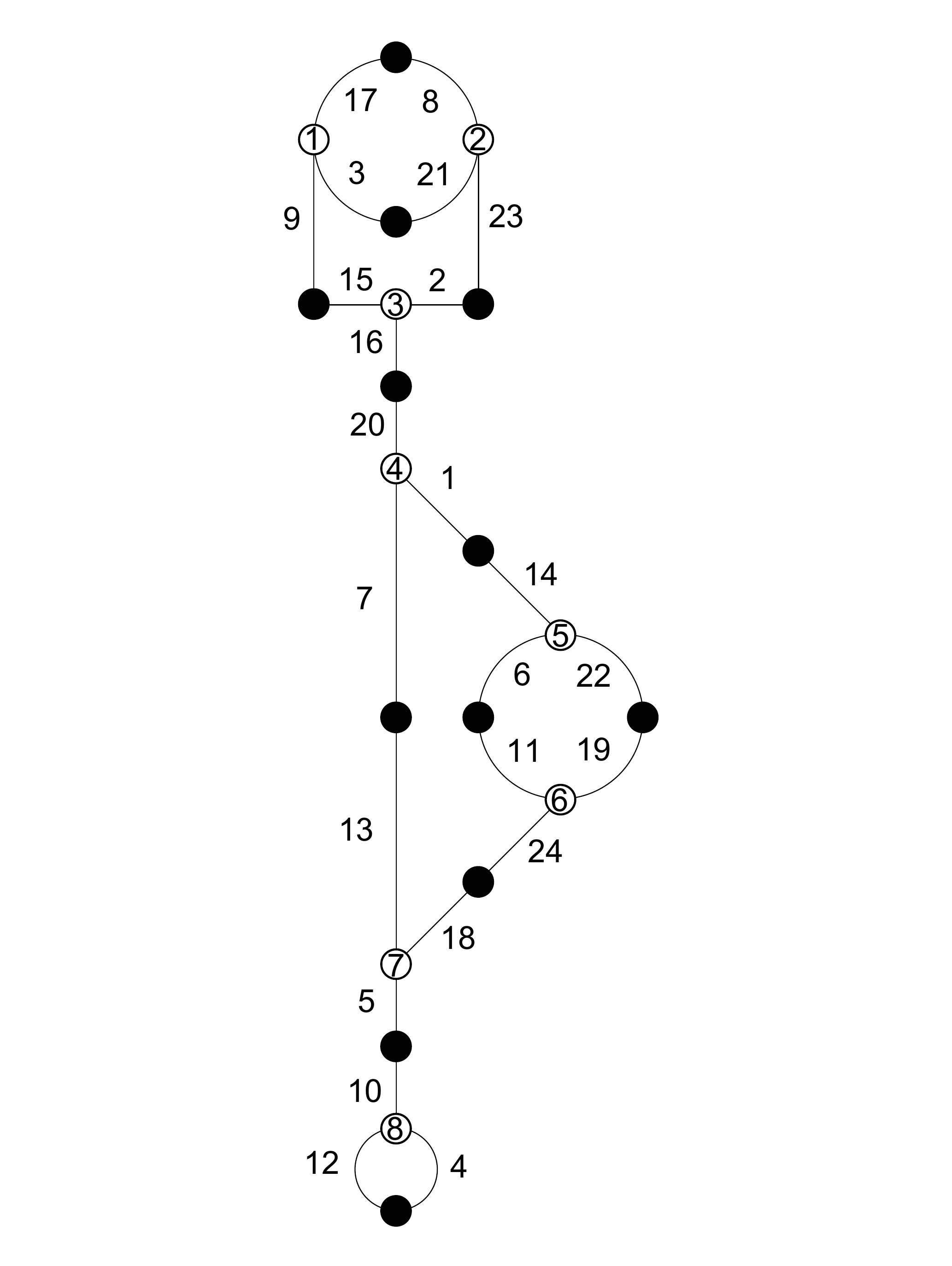}}}$
        \caption{ B: \{\{\{15,5,4\},\{24,3,12\},
        \{13,18,2\},\{11,6,17\},\{7,20,10\},
        \{1,9,19\},\{16,8,23\},\{22,14,21\}\}, \\ 
        \{\{21,14\},\{22,8\},\{23,9\},
        \{16,6\},\{1,20\},\{19,10\},
        \{7,17\},\{11,2\},\{13,4\},
        \{18,12\},\{5,24\},\{3,15\}\}\} \\
        C: \{\{\{17,3,9\},\{8,23,21\},
        \{2,16,15\},\{20,1,7\},\{14,22,6\},
        \{11,19,24\},\{18,5,13\},\{10,4,12\}\}, \\ 
        \{\{12,4\},\{5,10\},\{18,24\},
        \{13,7\},\{11,6\},\{22,19\},
        \{14,1\},\{20,16\},\{2,23\},
        \{15,9\},\{17,8\},\{3,21\}\}\}}
        \caption{12-4-3-2-2-1 B \& C $(\sqrt{-3})$}
        \label{Dessin}
    \end{subfigure}\hfill
\end{figure}

\begin{figure}[H]
    \begin{subfigure}{0.5\textwidth}
        \centering \captionsetup{justification=centering}
        $\scalemath{0.75}{
        \displaystyle \begin{pmatrix}
            2 & 1 & 0 & 0 & 0 & 0 & 0 & 0\\ 
            1 & 0 & 1 & 1 & 0 & 0 & 0 & 0\\
            0 & 1 & 0 & 1 & 1 & 0 & 0 & 0\\
            0 & 1 & 1 & 0 & 0 & 1 & 0 & 0\\
            0 & 0 & 1 & 0 & 0 & 1 & 1 & 0\\
            0 & 0 & 0 & 1 & 1 & 0 & 1 & 0\\
            0 & 0 & 0 & 0 & 1 & 1 & 0 & 1\\
            0 & 0 & 0 & 0 & 0 & 0 & 1 & 2
        \end{pmatrix}}$
        $\vcenter{\hbox{\includegraphics[width=0.35\textwidth]{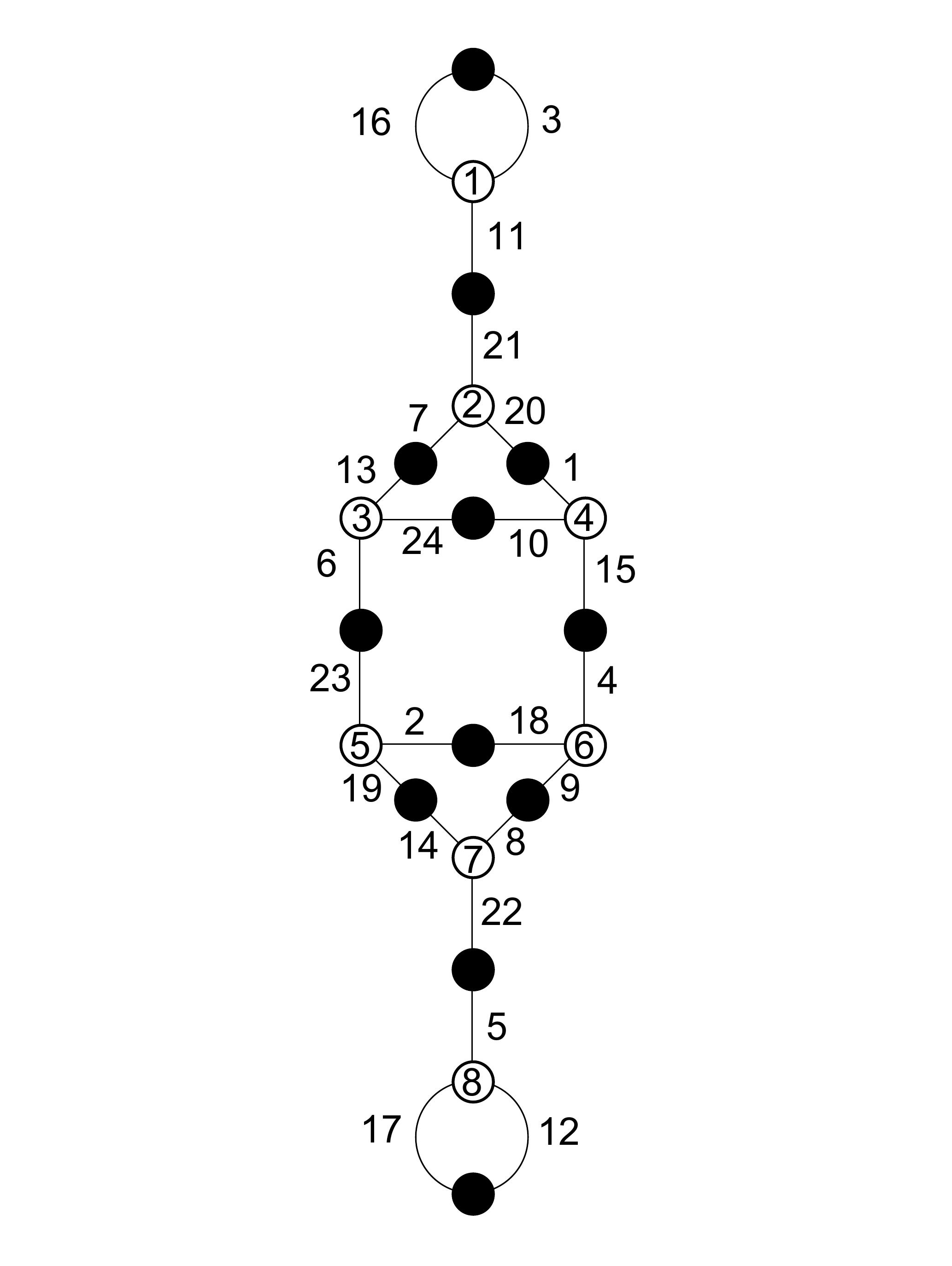}}}$
        \caption{ \{\{\{3,11,16\},\{21,20,7\},
        \{1,15,10\},\{24,6,13\},\{2,19,23\},
        \{18,4,9\},\{14,8,22\},\{5,12,17\}\}, \\ 
        \{\{12,17\},\{5,22\},\{14,19\},
        \{8,9\},\{2,18\},\{23,6\},
        \{4,15\},\{24,10\},\{1,20\},
        \{13,7\},\{21,11\},\{3,16\}\}\}}
        \caption{12-4-3-3-1-1 A $(\mathbb{Q})$}
        \label{Dessin}
    \end{subfigure} \hfill
    \begin{subfigure}{0.5\textwidth}
        \centering \captionsetup{justification=centering}
        $\scalemath{0.75}{
        \displaystyle \begin{pmatrix}
            2 & 1 & 0 & 0 & 0 & 0 & 0 & 0\\ 
            1 & 0 & 1 & 1 & 0 & 0 & 0 & 0\\
            0 & 1 & 2 & 0 & 0 & 0 & 0 & 0\\
            0 & 1 & 0 & 0 & 0 & 1 & 1 & 0\\
            0 & 0 & 0 & 0 & 0 & 1 & 1 & 1\\
            0 & 0 & 0 & 1 & 1 & 0 & 0 & 1\\
            0 & 0 & 0 & 1 & 1 & 0 & 0 & 1\\
            0 & 0 & 0 & 0 & 1 & 1 & 1 & 0
        \end{pmatrix}}$
        $\vcenter{\hbox{\includegraphics[width=0.35\textwidth]{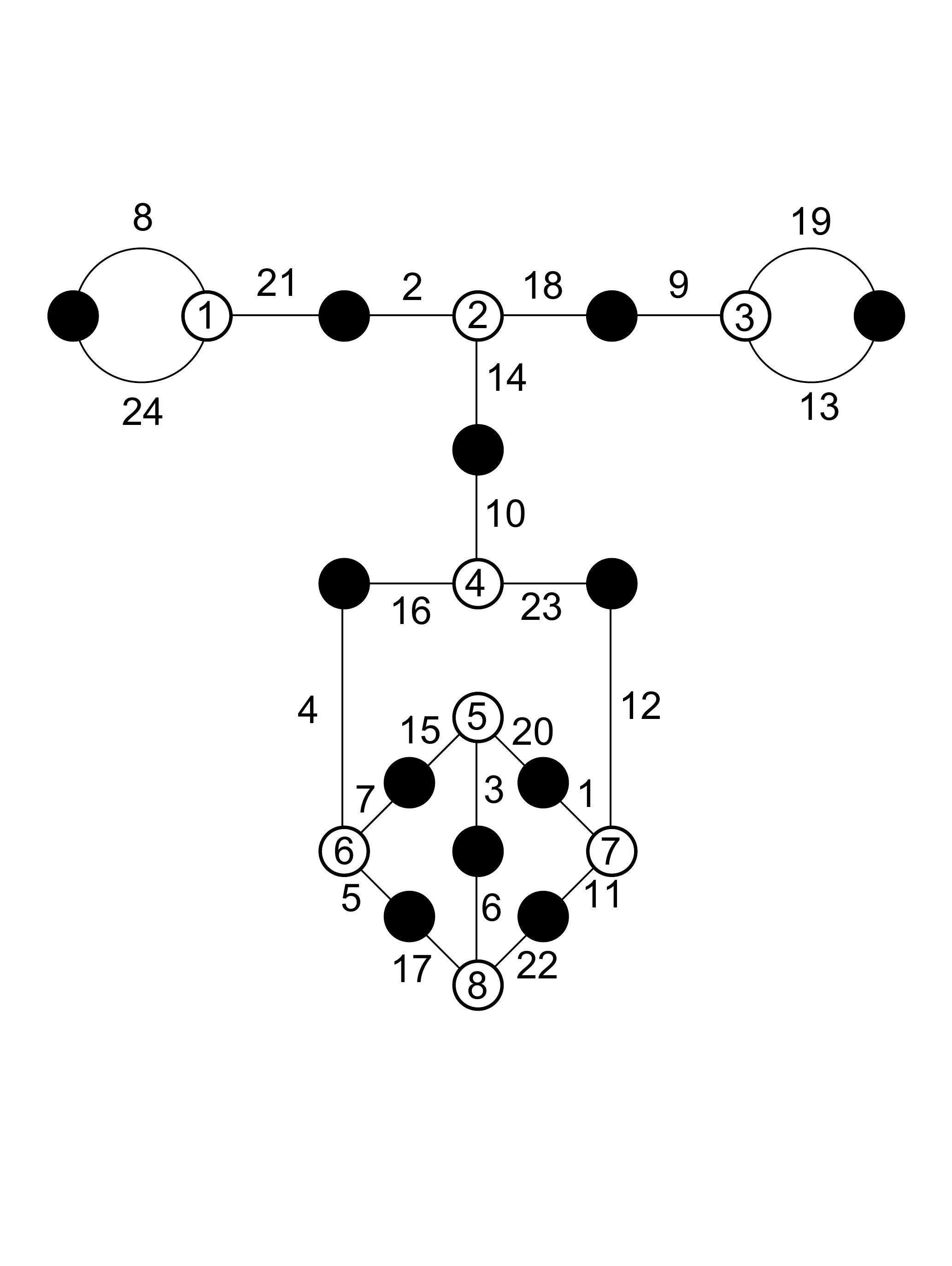}}}$
        \caption{ \{\{\{8,21,24\},\{2,18,14\},
        \{9,19,13\},\{10,23,16\},\{12,11,1\},
        \{20,3,15\},\{6,22,17\},\{5,4,7\}\}, \\
        \{\{6,3\},\{11,22\},\{17,5\},
        \{7,15\},\{20,1\},\{4,16\},
        \{23,12\},\{10,14\},\{21,2\},
        \{8,24\},\{18,9\},\{19,13\}\}\}}
        \caption{12-4-3-3-1-1 B $(\mathbb{Q})$}
        \label{Dessin}
    \end{subfigure}\hfill
\end{figure}

\begin{figure}[H]
    \begin{subfigure}{0.5\textwidth}
        \centering \captionsetup{justification=centering}
        $\scalemath{0.75}{
        \displaystyle \begin{pmatrix}
            2 & 1 & 0 & 0 & 0 & 0 & 0 & 0\\ 
            1 & 0 & 1 & 0 & 1 & 0 & 0 & 0\\
            0 & 1 & 0 & 1 & 0 & 1 & 0 & 0\\
            0 & 0 & 1 & 0 & 0 & 0 & 2 & 0\\
            0 & 1 & 0 & 0 & 0 & 1 & 0 & 1\\
            0 & 0 & 1 & 0 & 1 & 0 & 1 & 0\\
            0 & 0 & 0 & 2 & 0 & 1 & 0 & 0\\
            0 & 0 & 0 & 0 & 1 & 0 & 0 & 2
        \end{pmatrix}}$
        $\vcenter{\hbox{\includegraphics[width=0.35\textwidth]{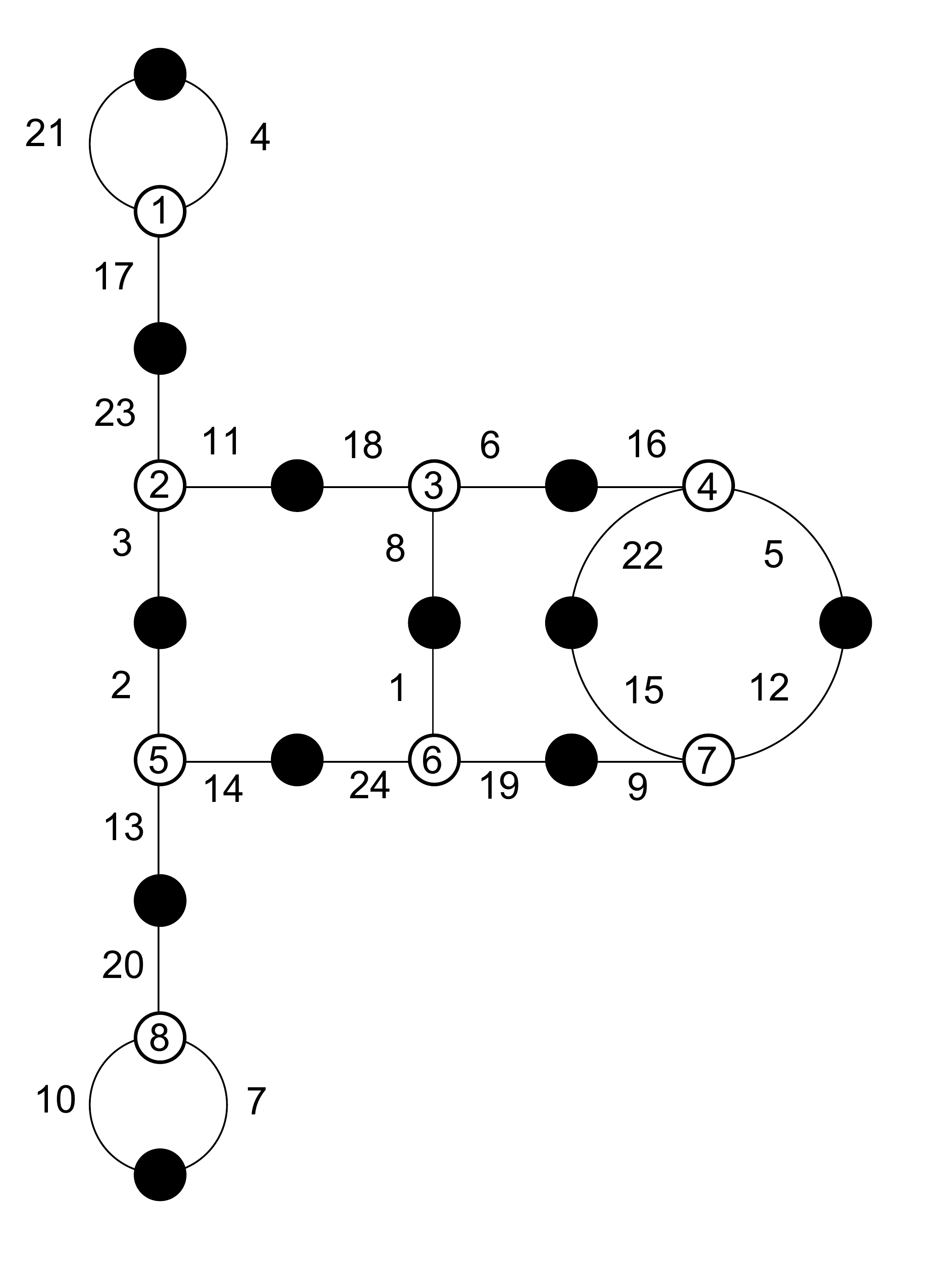}}}$
        \caption{ \{\{\{21,4,17\},\{23,11,3\},
        \{18,6,8\},\{16,5,22\},\{12,9,15\},
        \{19,24,1\},\{14,13,2\},\{20,7,10\}\}, \\ 
        \{\{10,7\},\{20,13\},\{14,24\},
        \{19,9\},\{1,8\},\{5,12\},
        \{15,22\},\{6,16\},\{11,18\},
        \{2,3\},\{23,17\},\{21,4\}\}\}}
        \caption{12-4-4-2-1-1 $(\mathbb{Q})$}
        \label{Dessin}
    \end{subfigure} \hfill
    \begin{subfigure}{0.5\textwidth}
        \centering \captionsetup{justification=centering}
        $\scalemath{0.75}{
        \displaystyle \begin{pmatrix}
            2 & 1 & 0 & 0 & 0 & 0 & 0 & 0\\ 
            1 & 0 & 2 & 0 & 0 & 0 & 0 & 0\\
            0 & 2 & 0 & 1 & 0 & 0 & 0 & 0\\
            0 & 0 & 1 & 0 & 1 & 1 & 0 & 0\\
            0 & 0 & 0 & 1 & 0 & 0 & 2 & 0\\
            0 & 0 & 0 & 1 & 0 & 0 & 0 & 2\\
            0 & 0 & 0 & 0 & 2 & 0 & 0 & 1\\
            0 & 0 & 0 & 0 & 0 & 2 & 1 & 0
        \end{pmatrix}}$
        $\vcenter{\hbox{\includegraphics[width=0.35\textwidth]{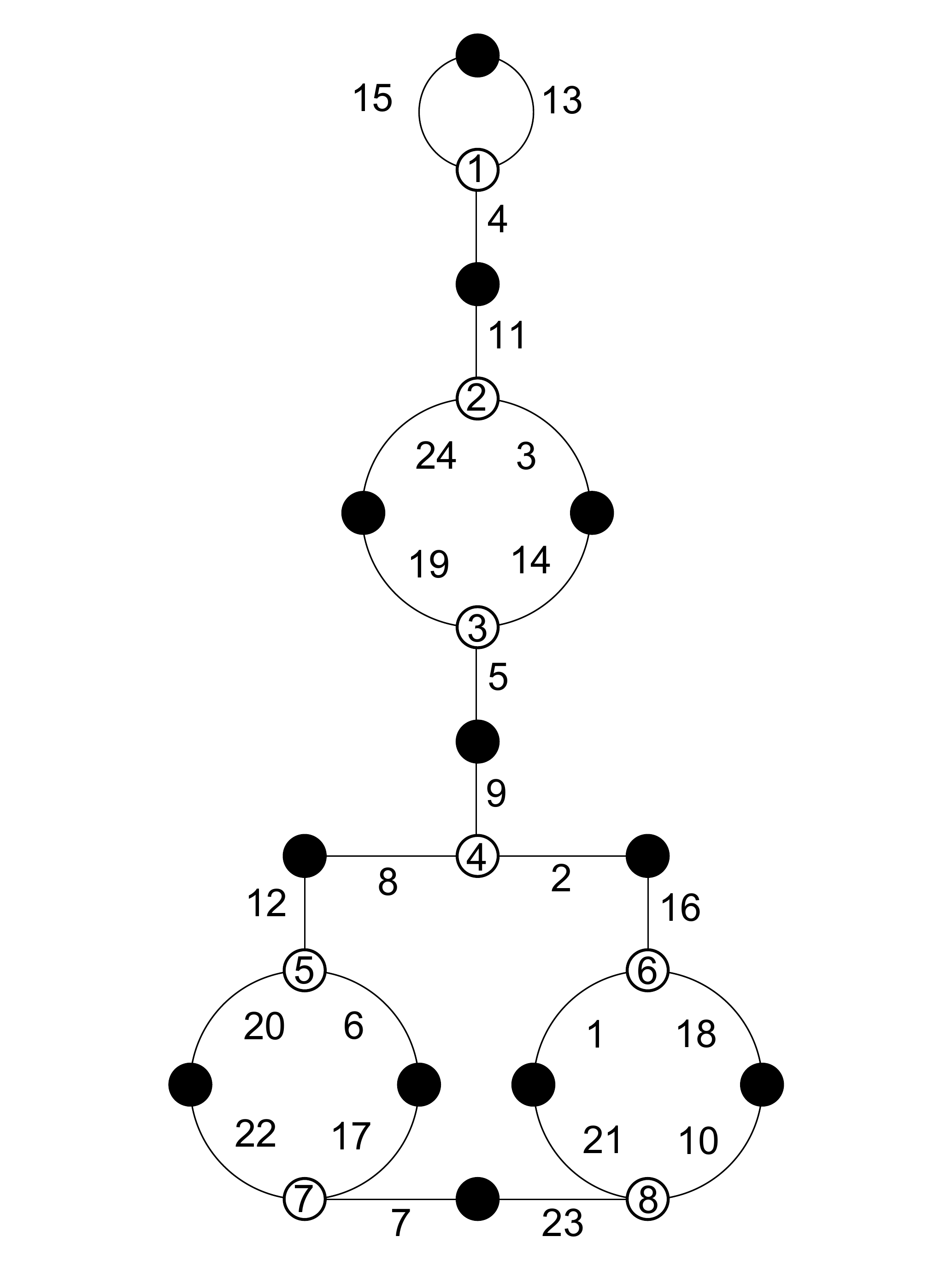}}}$
        \caption{ \{\{\{15,13,4\},\{11,3,24\},
        \{19,14,5\},\{9,2,8\},\{16,18,1\},
        \{12,6,20\},\{22,17,7\},\{23,21,10\}\}, \\ 
        \{\{7,23\},\{17,6\},\{20,22\},
        \{1,21\},\{18,10\},\{16,2\},
        \{8,12\},\{9,5\},\{14,3\},
        \{24,19\},\{11,4\},\{15,13\}\}\}}
        \caption{12-5-2-2-2-1 $(\mathbb{Q})$}
        \label{Dessin}
    \end{subfigure}\hfill
\end{figure}

\begin{figure}[H]
    \begin{subfigure}{0.5\textwidth}
        \centering \captionsetup{justification=centering}
        $\scalemath{0.75}{
        \displaystyle \begin{pmatrix}
            2 & 1 & 0 & 0 & 0 & 0 & 0 & 0\\ 
            1 & 0 & 1 & 1 & 0 & 0 & 0 & 0\\
            0 & 1 & 0 & 1 & 0 & 0 & 1 & 0\\
            0 & 1 & 1 & 0 & 1 & 0 & 0 & 0\\
            0 & 0 & 0 & 1 & 0 & 2 & 0 & 0\\
            0 & 0 & 0 & 0 & 2 & 0 & 1 & 0\\
            0 & 0 & 1 & 0 & 0 & 1 & 0 & 1\\
            0 & 0 & 0 & 0 & 0 & 0 & 1 & 2
        \end{pmatrix}}$
        $\vcenter{\hbox{\includegraphics[width=0.25\textwidth]{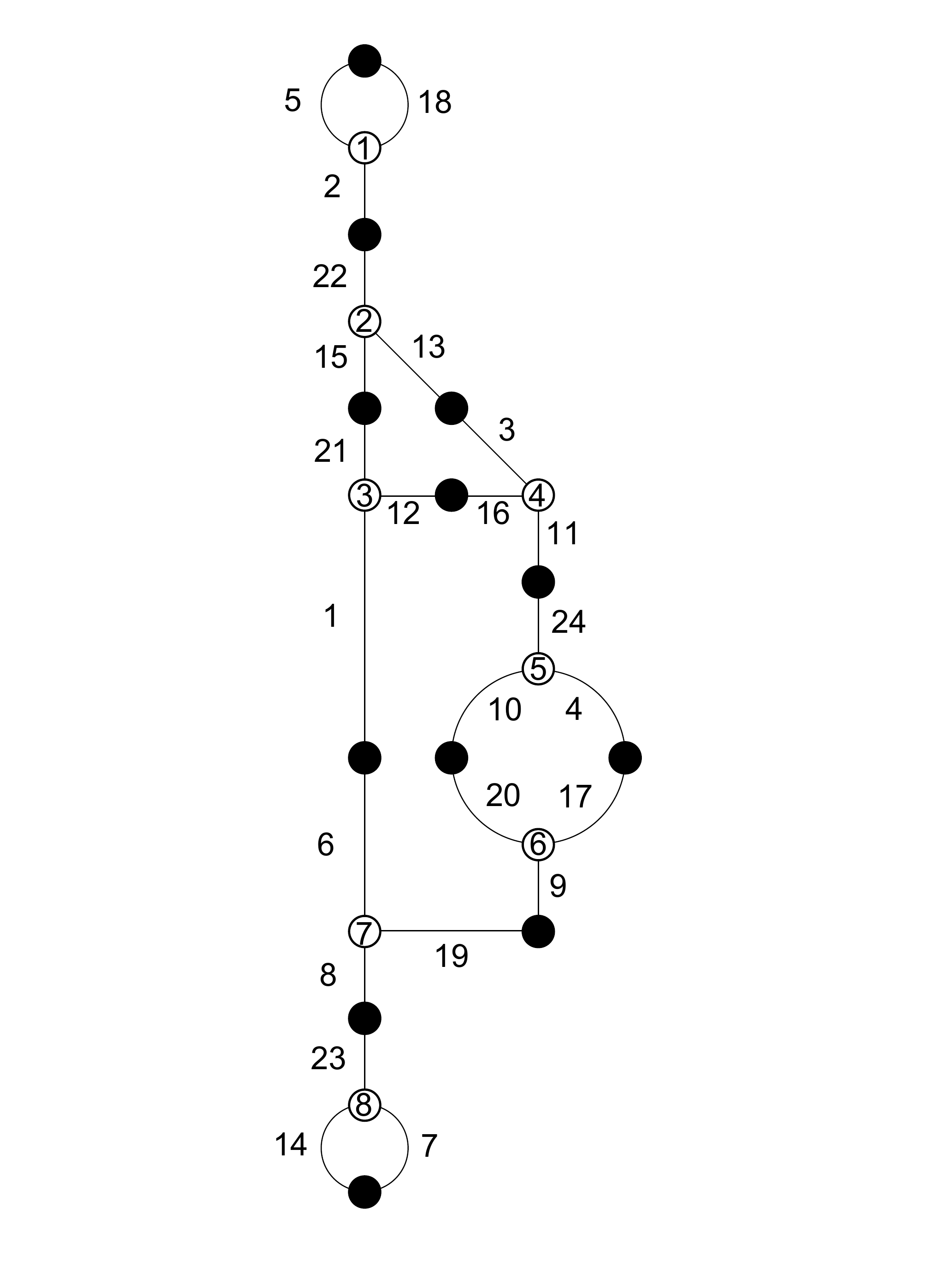}}}$
        $\vcenter{\hbox{\includegraphics[width=0.25\textwidth]{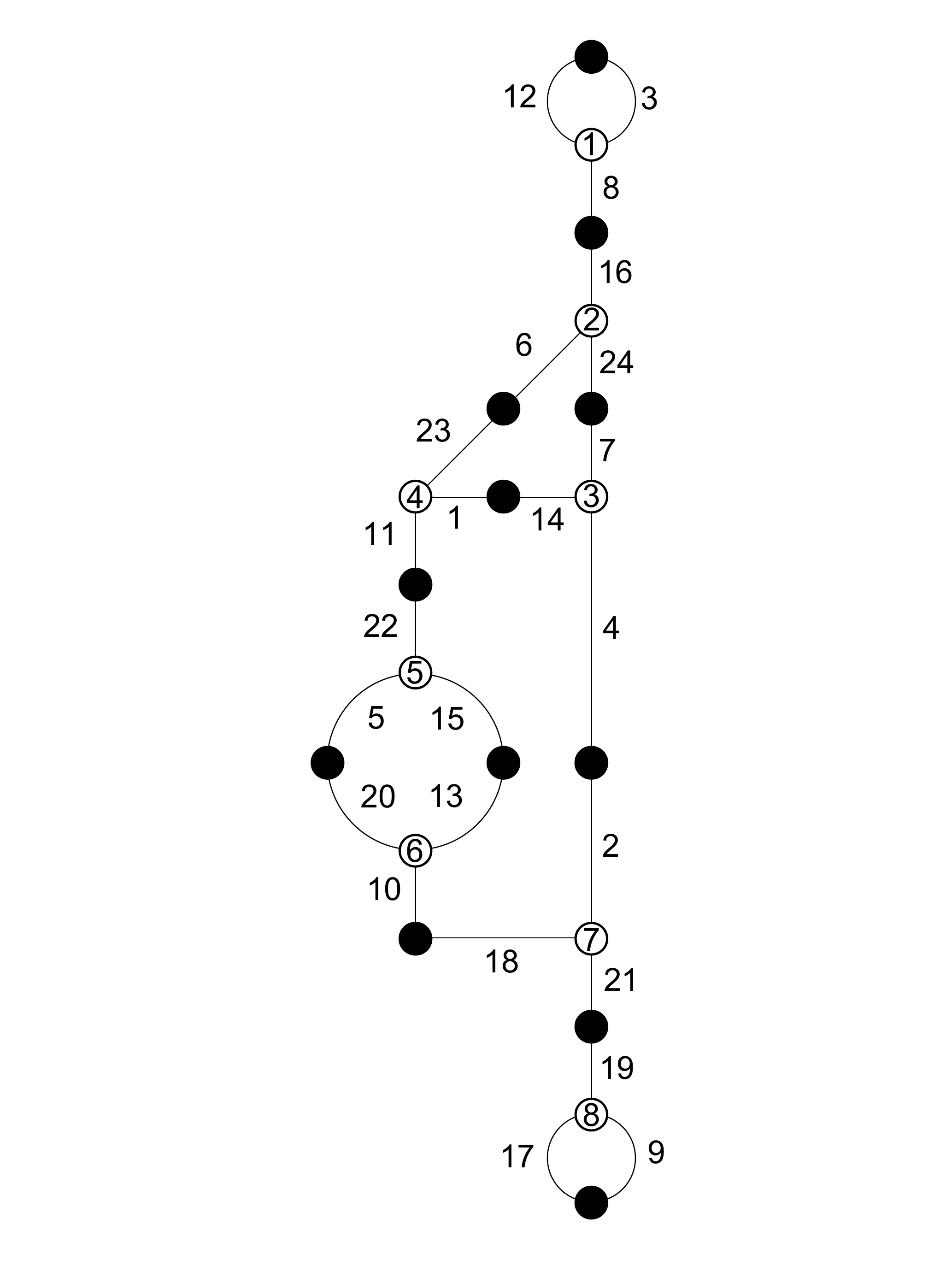}}}$
        \caption{ A: \{\{\{5,18,2\},\{22,13,15\},
        \{21,12,1\},\{3,11,16\},\{24,4,10\},
        \{17,9,20\},\{19,8,6\},\{23,7,14\}\}, \\ 
        \{\{14,7\},\{23,8\},\{19,9\},
        \{6,1\},\{20,10\},\{4,17\},
        \{24,11\},\{12,16\},\{21,15\},
        \{3,13\},\{22,2\},\{5,18\}\}\} \\
        B: \{\{\{12,3,8\},\{16,24,6\},
        \{7,4,14\},\{1,11,23\},\{22,15,5\},
        \{13,10,20\},\{2,21,18\},\{19,9,17\}\}, \\ 
        \{\{9,17\},\{19,21\},\{18,10\},
        \{13,15\},\{20,5\},\{22,11\},
        \{1,14\},\{23,6\},\{7,24\},
        \{16,8\},\{12,3\},\{2,4\}\}\}}
        \caption{12-5-3-2-1-1 A \& B (quartic)}
        \label{Dessin}
    \end{subfigure} \hfill
    \begin{subfigure}{0.5\textwidth}
        \centering \captionsetup{justification=centering}
        $\scalemath{0.75}{
        \displaystyle \begin{pmatrix}
            0 & 1 & 2 & 0 & 0 & 0 & 0 & 0\\ 
            1 & 2 & 0 & 0 & 0 & 0 & 0 & 0\\
            2 & 0 & 0 & 1 & 0 & 0 & 0 & 0\\
            0 & 0 & 1 & 0 & 1 & 1 & 0 & 0\\
            0 & 0 & 0 & 1 & 2 & 0 & 0 & 0\\
            0 & 0 & 0 & 1 & 0 & 0 & 1 & 1\\
            0 & 0 & 0 & 0 & 0 & 1 & 0 & 2\\
            0 & 0 & 0 & 0 & 0 & 1 & 2 & 0
        \end{pmatrix}}$
        $\vcenter{\hbox{\includegraphics[width=0.25\textwidth]{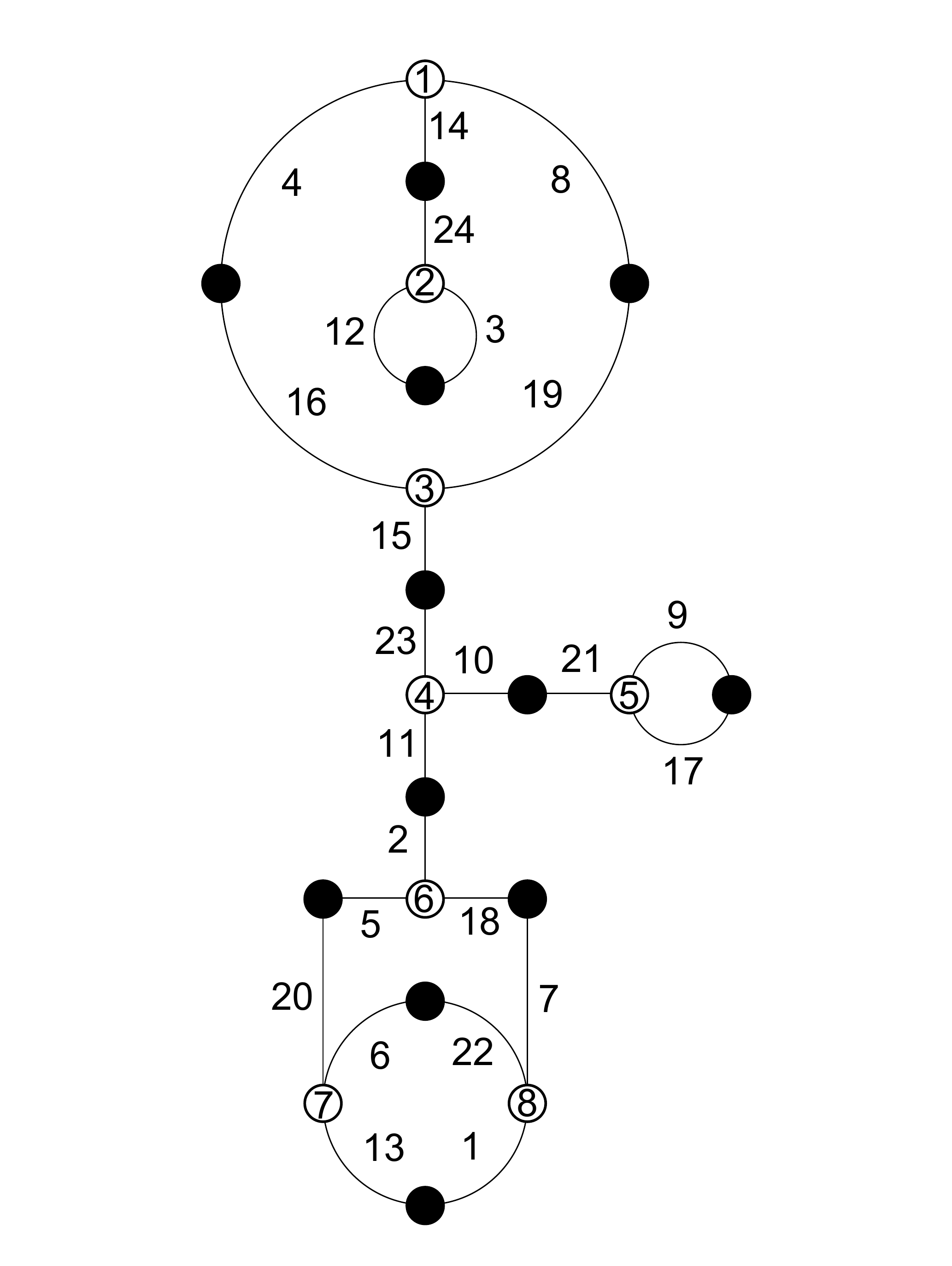}}}$
        $\vcenter{\hbox{\includegraphics[width=0.25\textwidth]{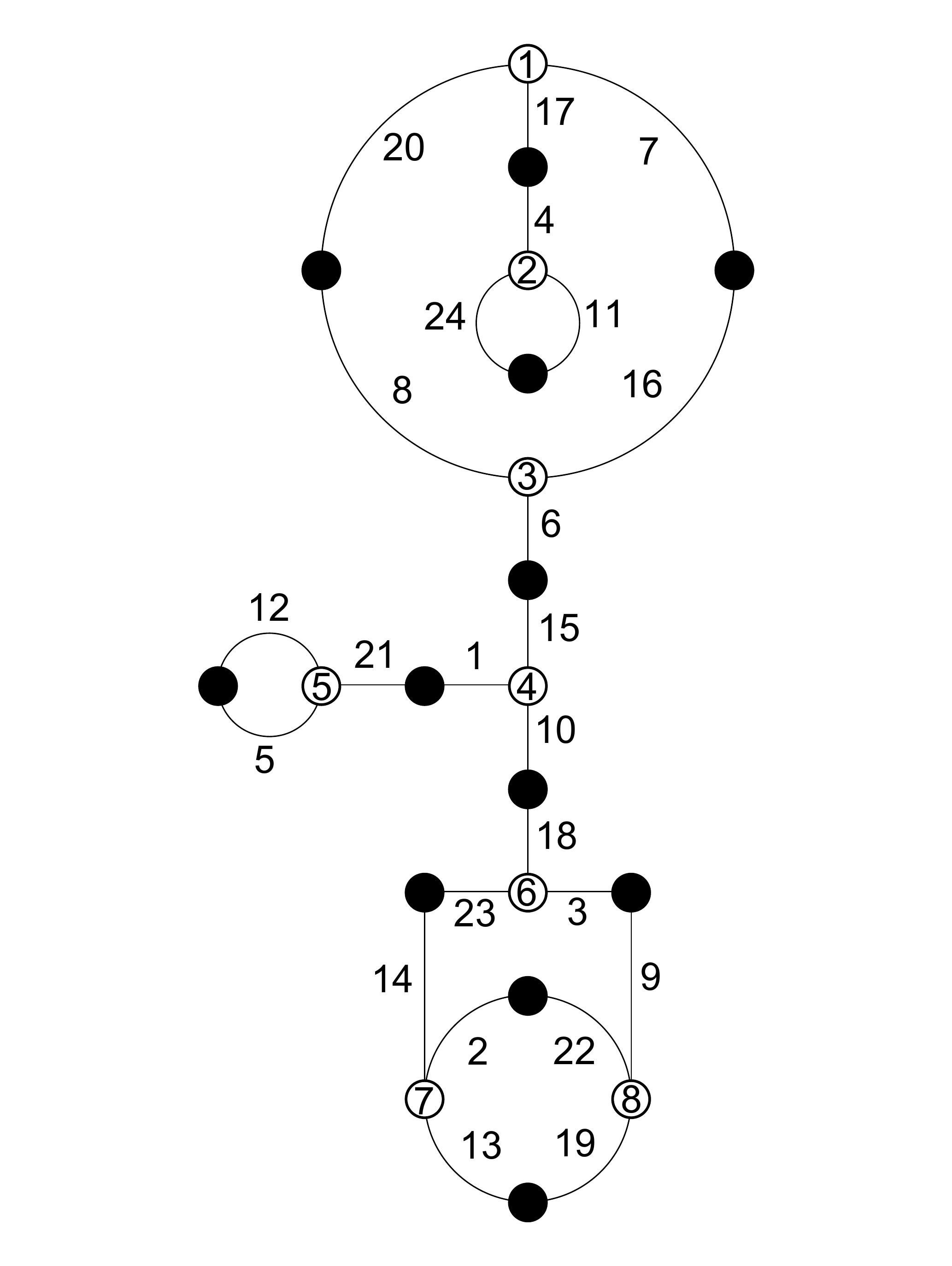}}}$
        \caption{ C: \{\{\{14,4,8\},\{24,3,12\},
        \{19,15,16\},\{23,10,11\},\{21,9,17\},
        \{2,18,5\},\{7,1,22\},\{6,13,20\}\}, \\
        \{\{13,1\},\{6,22\},\{18,7\},
        \{5,20\},\{2,11\},\{10,21\},
        \{9,17\},\{15,23\},\{19,8\},
        \{14,24\},\{4,16\},\{12,3\}\}\} \\
        D: \{\{\{20,7,17\},\{24,4,11\},
        \{8,16,6\},\{15,10,1\},\{21,5,12\},
        \{18,3,23\},\{9,19,22\},\{2,13,14\}\}, \\ 
        \{\{2,22\},\{13,19\},\{14,23\},
        \{3,9\},\{18,10\},\{1,21\},
        \{12,5\},\{15,6\},\{8,20\},
        \{7,16\},\{17,4\},\{24,11\}\}\}}
        \caption{12-5-3-2-1-1 C \& D (quartic)}
        \label{Dessin}
    \end{subfigure}\hfill
\end{figure}

\begin{figure}[H]
    \begin{subfigure}{0.5\textwidth}
        \centering \captionsetup{justification=centering}
        $\scalemath{0.75}{
        \displaystyle \begin{pmatrix}
            2 & 1 & 0 & 0 & 0 & 0 & 0 & 0\\ 
            1 & 0 & 1 & 1 & 0 & 0 & 0 & 0\\
            0 & 1 & 0 & 0 & 2 & 0 & 0 & 0\\
            0 & 1 & 0 & 0 & 0 & 2 & 0 & 0\\
            0 & 0 & 2 & 0 & 0 & 0 & 1 & 0\\
            0 & 0 & 0 & 2 & 0 & 0 & 1 & 0\\
            0 & 0 & 0 & 0 & 1 & 1 & 0 & 1\\
            0 & 0 & 0 & 0 & 0 & 0 & 1 & 2
        \end{pmatrix}}$
        $\vcenter{\hbox{\includegraphics[width=0.35\textwidth]{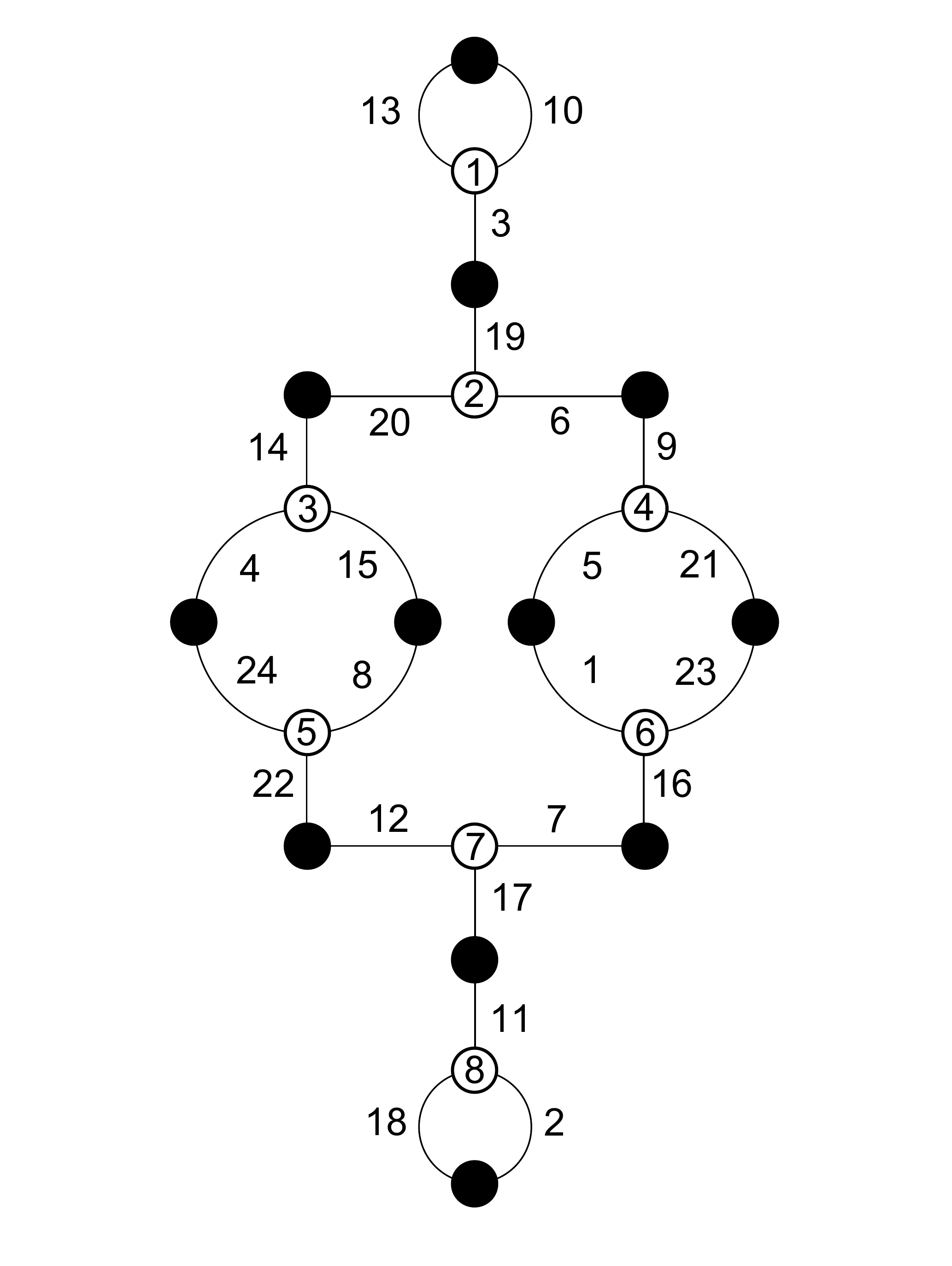}}}$
        \caption{ \{\{\{13,10,3\},\{19,6,20\},
        \{9,21,5\},\{14,15,4\},\{24,8,22\},
        \{1,23,16\},\{12,7,17\},\{11,2,18\}\}, \\ 
        \{\{2,18\},\{17,11\},\{7,16\},
        \{12,22\},\{8,15\},\{24,4\},
        \{1,5\},\{21,23\},\{9,6\},
        \{14,20\},\{19,3\},\{10,13\}\}\}}
        \caption{12-6-2-2-1-1 A $(\mathbb{Q})$}
        \label{Dessin}
    \end{subfigure} \hfill
    \begin{subfigure}{0.5\textwidth}
        \centering \captionsetup{justification=centering}
        $\scalemath{0.75}{
        \displaystyle \begin{pmatrix}
            0 & 1 & 0 & 2 & 0 & 0 & 0 & 0\\ 
            1 & 0 & 1 & 0 & 0 & 0 & 1 & 0\\
            0 & 1 & 2 & 0 & 0 & 0 & 0 & 0\\
            2 & 0 & 0 & 0 & 1 & 0 & 0 & 0\\
            0 & 0 & 0 & 1 & 0 & 2 & 0 & 0\\
            0 & 0 & 0 & 0 & 2 & 0 & 1 & 0\\
            0 & 1 & 0 & 0 & 0 & 1 & 0 & 1\\
            0 & 0 & 0 & 0 & 0 & 0 & 1 & 2
        \end{pmatrix}}$
        $\vcenter{\hbox{\includegraphics[width=0.35\textwidth]{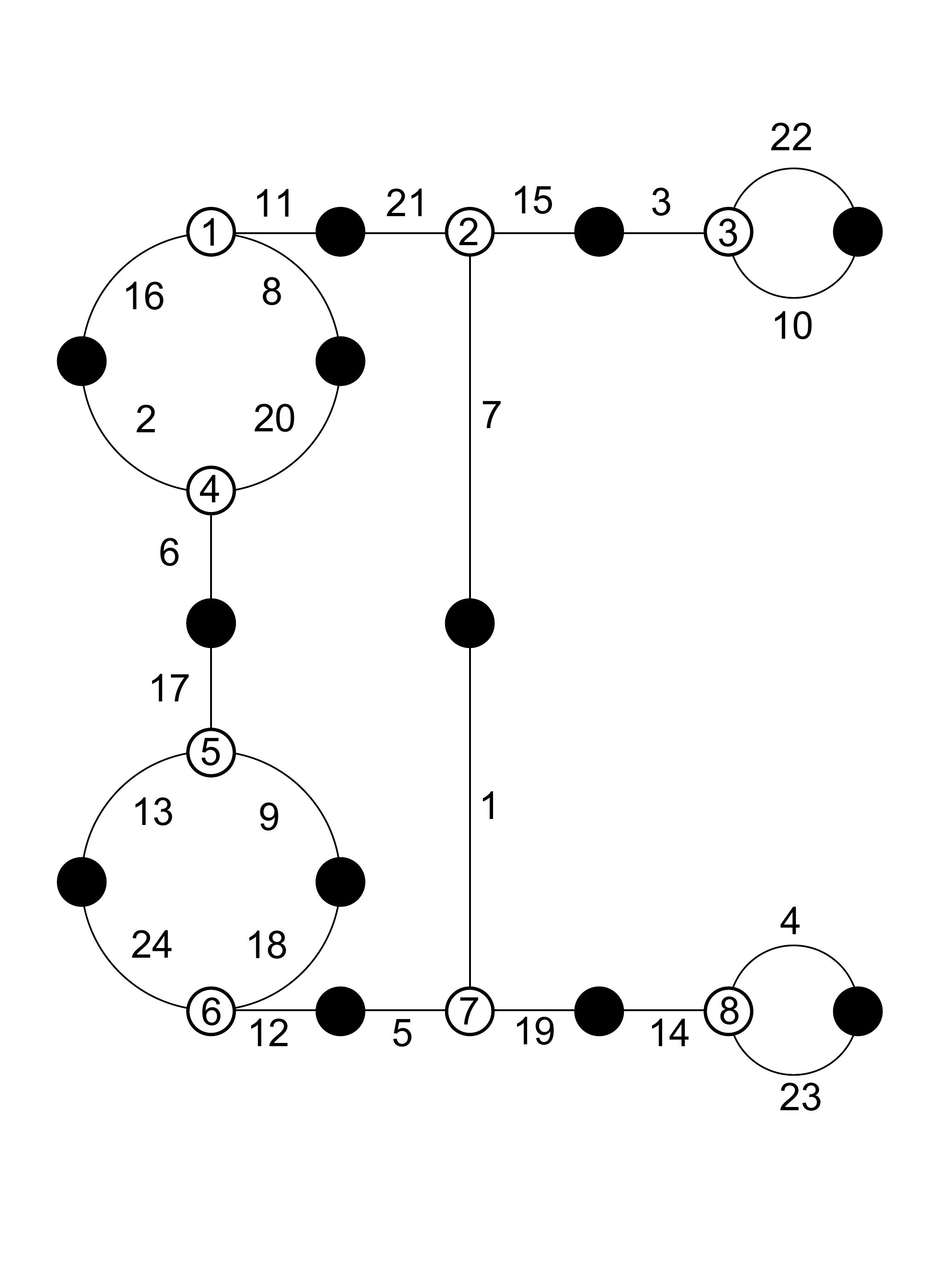}}}$
        \caption{ \{\{\{11,8,16\},\{20,6,2\},
        \{17,9,13\},\{24,18,12\},\{21,15,7\},
        \{3,22,10\},\{1,19,5\},\{14,4,23\}\}, \\ 
        \{\{22,10\},\{3,15\},\{21,11\},
        \{8,20\},\{16,2\},\{6,17\},
        \{13,24\},\{9,18\},\{12,5\},
        \{19,14\},\{4,23\},\{1,7\}\}\}}
        \caption{12-6-2-2-1-1 B $(\mathbb{Q})$}
        \label{Dessin}
    \end{subfigure}\hfill
\end{figure}

\begin{figure}[H]
    \begin{subfigure}{0.4\textwidth}
        \centering \captionsetup{justification=centering}
        $\scalemath{0.75}{
        \displaystyle \begin{pmatrix}
            2 & 1 & 0 & 0 & 0 & 0 & 0 & 0\\ 
            1 & 0 & 1 & 1 & 0 & 0 & 0 & 0\\
            0 & 1 & 2 & 0 & 0 & 0 & 0 & 0\\
            0 & 1 & 0 & 0 & 1 & 0 & 1 & 0\\
            0 & 0 & 0 & 1 & 0 & 0 & 1 & 1\\
            0 & 0 & 0 & 0 & 0 & 2 & 0 & 1\\
            0 & 0 & 0 & 1 & 1 & 0 & 0 & 1\\
            0 & 0 & 0 & 0 & 1 & 1 & 1 & 0
        \end{pmatrix}}$
        $\vcenter{\hbox{\includegraphics[width=0.35\textwidth]{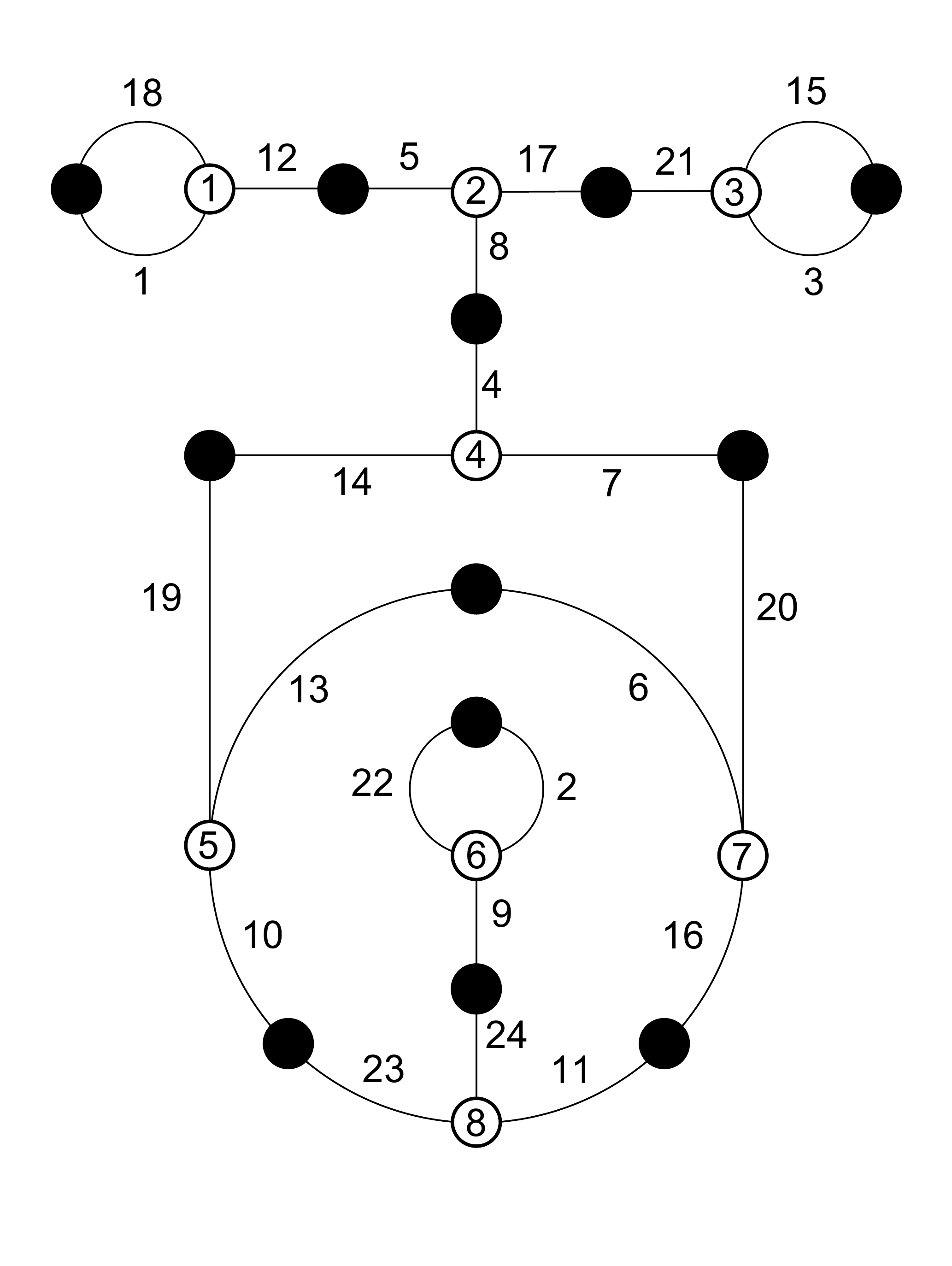}}}$
        \caption{ \{\{\{1,18,12\},\{5,17,8\},
        \{21,15,3\},\{4,7,14\},\{20,16,6\},
        \{13,10,19\},\{2,9,22\},\{24,11,23\}\}, \\ 
        \{\{9,24\},\{2,22\},\{23,10\},
        \{11,16\},\{20,7\},\{14,19\},
        \{6,13\},\{4,8\},\{17,21\},
        \{15,3\},\{5,12\},\{1,18\}\}\}}
        \caption{12-6-3-1-1-1 $(\mathbb{Q})$}
        \label{Dessin}
    \end{subfigure} \hfill
    \begin{subfigure}{0.6\textwidth}
        \centering \captionsetup{justification=centering}
        $\scalemath{0.75}{
        \displaystyle \begin{pmatrix}
            0 & 2 & 0 & 1 & 0 & 0 & 0 & 0\\ 
            2 & 0 & 0 & 0 & 1 & 0 & 0 & 0\\
            0 & 0 & 2 & 1 & 0 & 0 & 0 & 0\\
            1 & 0 & 1 & 0 & 1 & 0 & 0 & 0\\
            0 & 1 & 0 & 1 & 0 & 0 & 1 & 0\\
            0 & 0 & 0 & 0 & 0 & 2 & 1 & 0\\
            0 & 0 & 0 & 0 & 1 & 1 & 0 & 1\\
            0 & 0 & 0 & 0 & 0 & 0 & 1 & 2
        \end{pmatrix}}$
        $\vcenter{\hbox{\includegraphics[width=0.25\textwidth]{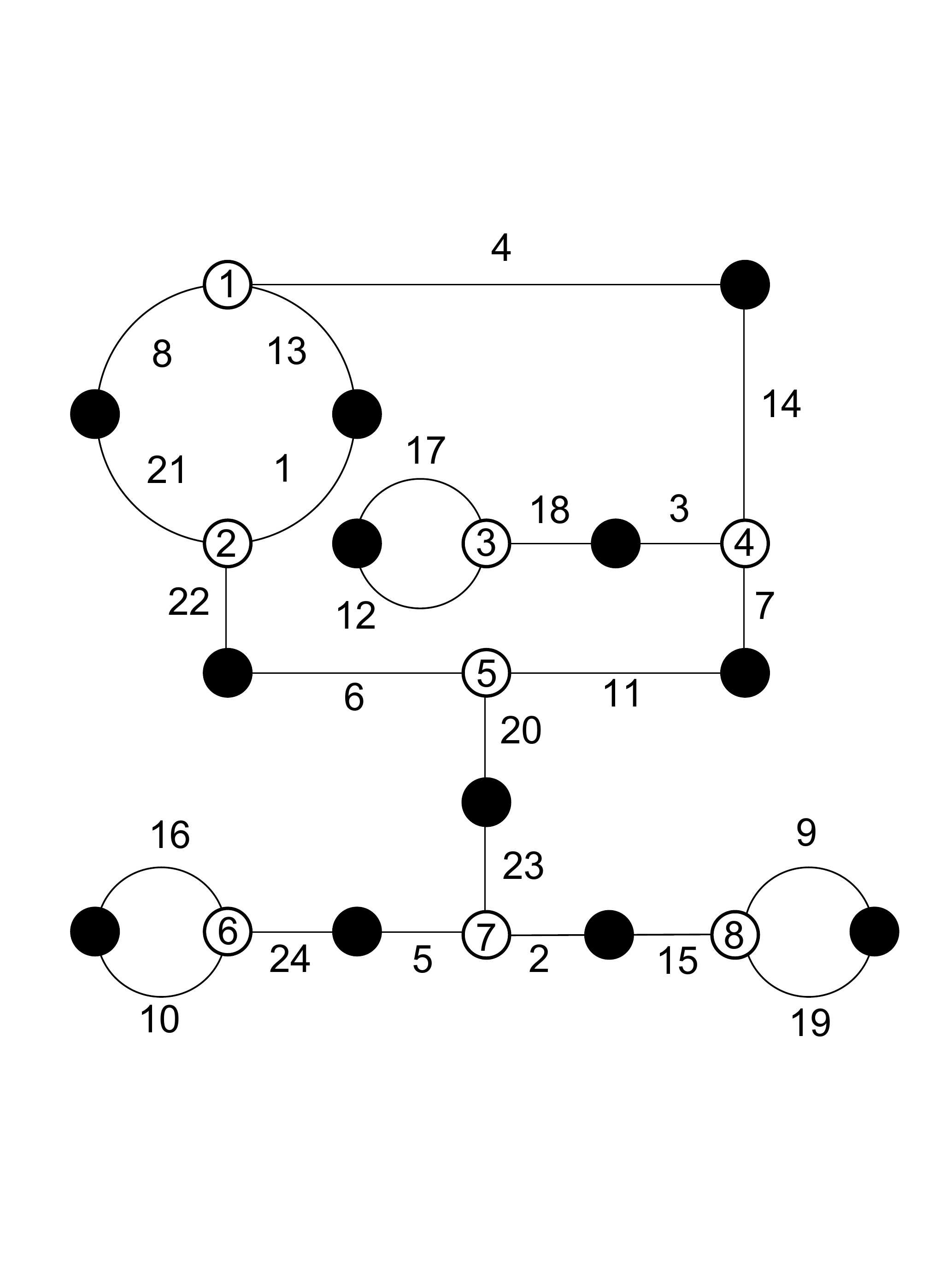}}}$
        $\vcenter{\hbox{\includegraphics[width=0.25\textwidth]{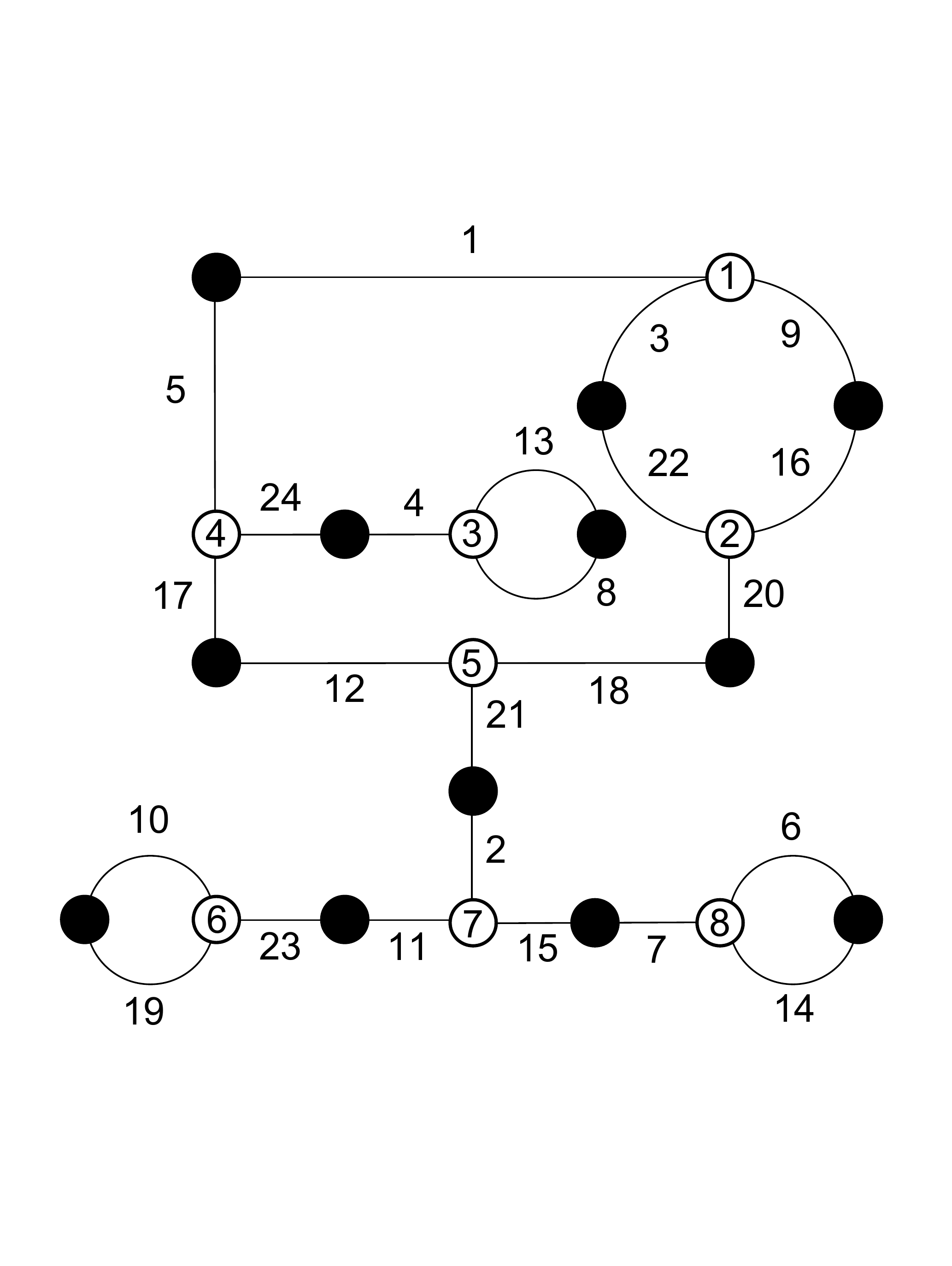}}}$
        \caption{ A: \{\{\{4,13,8\},\{14,7,3\},
        \{18,12,17\},\{1,22,21\},\{6,11,20\},
        \{23,2,5\},\{9,19,15\},\{24,10,16\}\}, \\ 
        \{\{10,16\},\{24,5\},\{2,15\},
        \{9,19\},\{23,20\},\{11,7\},
        \{6,22\},\{21,8\},\{1,13\},
        \{12,17\},\{4,14\},\{3,18\}\}\}\\
        B: \{\{\{1,9,3\},\{16,20,22\},
        \{13,8,4\},\{24,17,5\},\{12,18,21\},
        \{2,15,11\},\{6,14,7\},\{23,19,10\}\}, \\ 
        \{\{6,14\},\{7,15\},\{11,23\},
        \{19,10\},\{2,21\},\{18,20\},
        \{17,12\},\{24,4\},\{13,8\},
        \{5,1\},\{3,22\},\{9,16\}\}\}}
        \caption{12-7-2-1-1-1 A \& B $(\sqrt{-3})$}
        \label{Dessin}
    \end{subfigure}\hfill
\end{figure}

\begin{figure}[H]
    \begin{subfigure}{0.5\textwidth}
        \centering \captionsetup{justification=centering}
        $\scalemath{0.75}{
        \displaystyle \begin{pmatrix}
            0 & 2 & 1 & 0 & 0 & 0 & 0 & 0\\ 
            2 & 0 & 1 & 0 & 0 & 0 & 0 & 0\\
            1 & 1 & 0 & 0 & 1 & 0 & 0 & 0\\
            0 & 0 & 0 & 2 & 1 & 0 & 0 & 0\\
            0 & 0 & 1 & 1 & 0 & 1 & 0 & 0\\
            0 & 0 & 0 & 0 & 1 & 0 & 1 & 1\\
            0 & 0 & 0 & 0 & 0 & 1 & 0 & 2\\
            0 & 0 & 0 & 0 & 0 & 1 & 2 & 0
        \end{pmatrix}}$
        $\vcenter{\hbox{\includegraphics[width=0.35\textwidth]{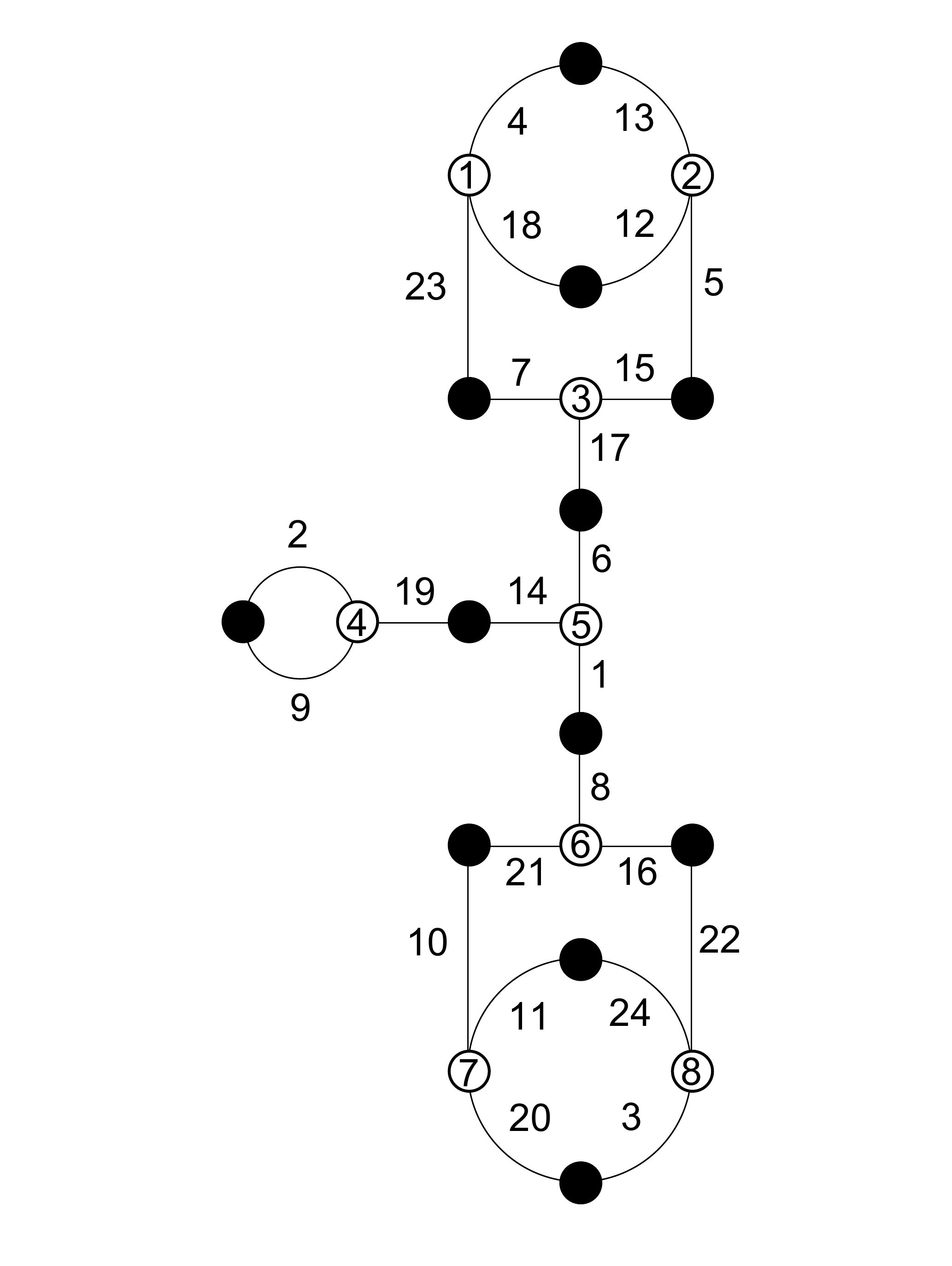}}}$
        \caption{ \{\{\{4,18,23\},\{13,5,12\},
        \{7,15,17\},\{6,1,14\},\{19,9,2\},
        \{8,16,21\},\{22,3,24\},\{11,20,10\}\}, \\ 
        \{\{20,3\},\{11,24\},\{22,16\},
        \{10,21\},\{8,1\},\{14,19\},
        \{2,9\},\{6,17\},\{7,23\},
        \{5,15\},\{18,12\},\{4,13\}\}\}}
        \caption{13-3-3-2-2-1 $(\mathbb{Q})$}
        \label{Dessin}
    \end{subfigure} \hfill
    \begin{subfigure}{0.5\textwidth}
        \centering \captionsetup{justification=centering}
        $\scalemath{0.75}{
        \displaystyle \begin{pmatrix}
            2 & 1 & 0 & 0 & 0 & 0 & 0 & 0\\ 
            1 & 0 & 1 & 1 & 0 & 0 & 0 & 0\\
            0 & 1 & 2 & 0 & 0 & 0 & 0 & 0\\
            0 & 1 & 0 & 0 & 1 & 1 & 0 & 0\\
            0 & 0 & 0 & 1 & 0 & 1 & 1 & 0\\
            0 & 0 & 0 & 1 & 1 & 0 & 0 & 1\\
            0 & 0 & 0 & 0 & 1 & 0 & 0 & 2\\
            0 & 0 & 0 & 0 & 0 & 1 & 2 & 0
        \end{pmatrix}}$
        $\vcenter{\hbox{\includegraphics[width=0.35\textwidth]{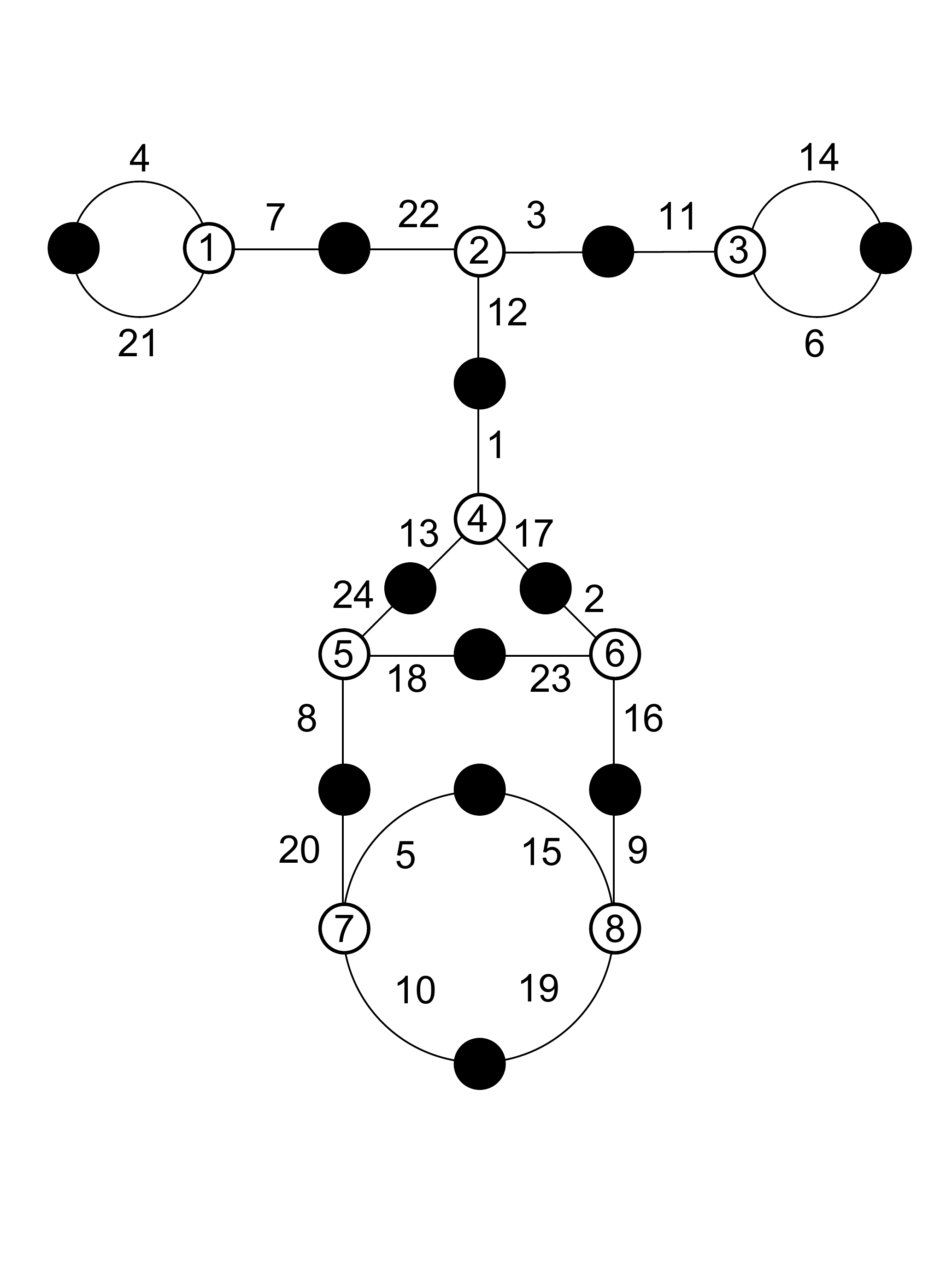}}}$
        \caption{ \{\{\{4,7,21\},\{22,3,12\},
        \{11,14,6\},\{1,17,13\},\{2,16,23\},
        \{24,18,8\},\{20,5,10\},\{15,9,19\}\}, \\ 
        \{\{10,19\},\{5,15\},\{9,16\},
        \{8,20\},\{18,23\},\{24,13\},
        \{2,17\},\{1,12\},\{3,11\},
        \{14,6\},\{22,7\},\{4,21\}\}\}}
        \caption{13-4-3-2-1-1 $(\mathbb{Q})$}
        \label{Dessin}
    \end{subfigure}\hfill
\end{figure}

\begin{figure}[H]
    \begin{subfigure}{0.5\textwidth}
        \centering \captionsetup{justification=centering}
        $\scalemath{0.75}{
        \displaystyle \begin{pmatrix}
            2 & 1 & 0 & 0 & 0 & 0 & 0 & 0\\ 
            1 & 0 & 1 & 1 & 0 & 0 & 0 & 0\\
            0 & 1 & 2 & 0 & 0 & 0 & 0 & 0\\
            0 & 1 & 0 & 0 & 1 & 1 & 0 & 0\\
            0 & 0 & 0 & 1 & 0 & 0 & 2 & 0\\
            0 & 0 & 0 & 1 & 0 & 0 & 0 & 2\\
            0 & 0 & 0 & 0 & 2 & 0 & 0 & 1\\
            0 & 0 & 0 & 0 & 0 & 2 & 1 & 0
        \end{pmatrix}}$
        $\vcenter{\hbox{\includegraphics[width=0.35\textwidth]{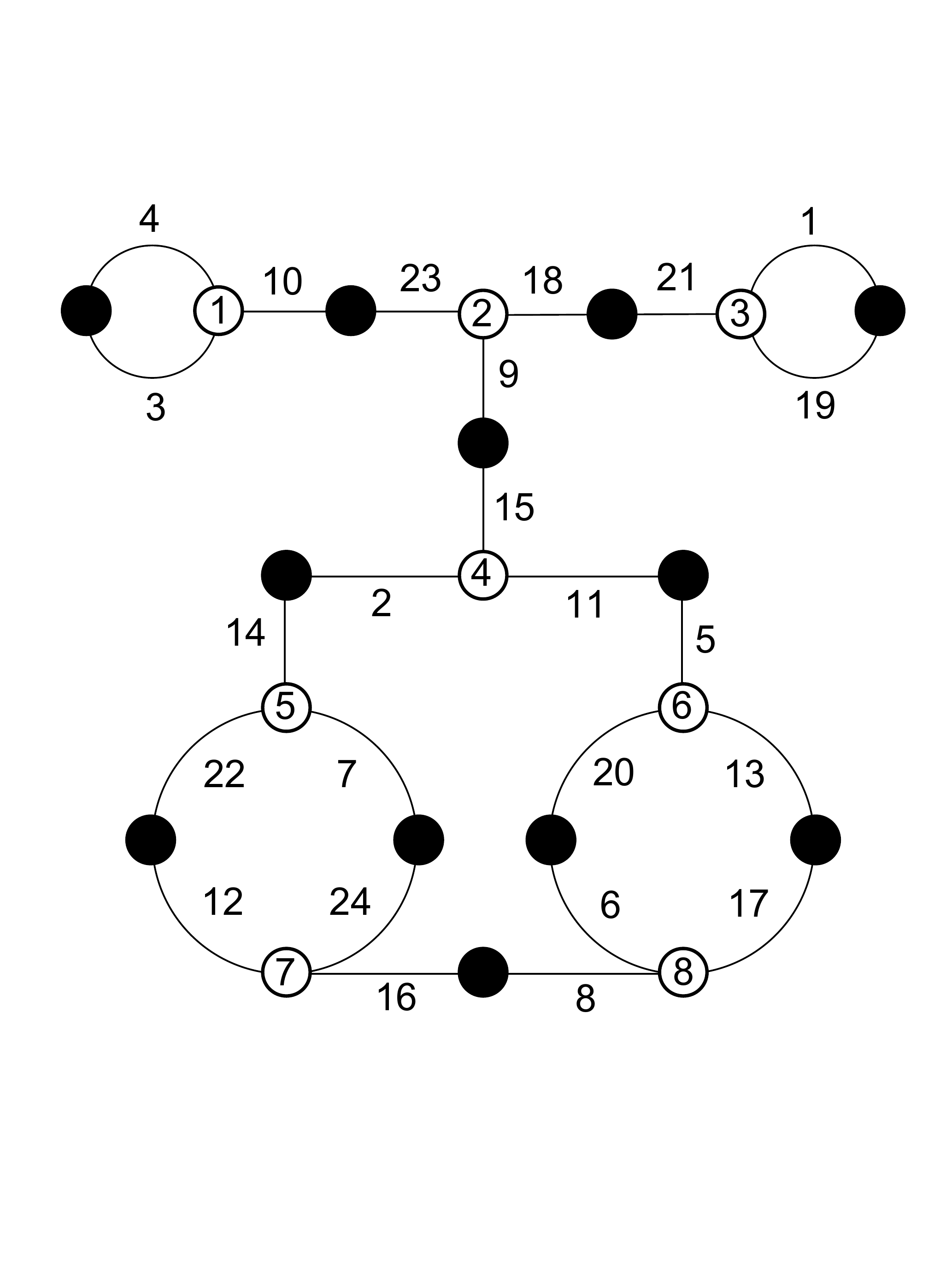}}}$
        \caption{ \{\{\{4,10,3\},\{23,18,9\},
        \{21,1,19\},\{15,11,2\},\{5,13,20\},
        \{17,8,6\},\{16,12,24\},\{22,14,7\}\}, \\ 
        \{\{16,8\},\{24,7\},\{22,12\},
        \{14,2\},\{11,5\},\{13,17\},
        \{6,20\},\{10,23\},\{9,15\},
        \{18,21\},\{1,19\},\{3,4\}\}\}}
        \caption{13-5-2-2-1-1 A $(\sqrt{65})$}
        \label{Dessin}
    \end{subfigure} \hfill
    \begin{subfigure}{0.5\textwidth}
        \centering \captionsetup{justification=centering}
        $\scalemath{0.75}{
        \displaystyle \begin{pmatrix}
            2 & 1 & 0 & 0 & 0 & 0 & 0 & 0\\ 
            1 & 0 & 2 & 0 & 0 & 0 & 0 & 0\\
            0 & 2 & 0 & 1 & 0 & 0 & 0 & 0\\
            0 & 0 & 1 & 0 & 2 & 0 & 0 & 0\\
            0 & 0 & 0 & 2 & 0 & 1 & 0 & 0\\
            0 & 0 & 0 & 0 & 1 & 0 & 0 & 2\\
            0 & 0 & 0 & 0 & 0 & 0 & 2 & 1\\
            0 & 0 & 0 & 0 & 0 & 2 & 1 & 0
        \end{pmatrix}}$
        $\vcenter{\hbox{\includegraphics[width=0.35\textwidth]{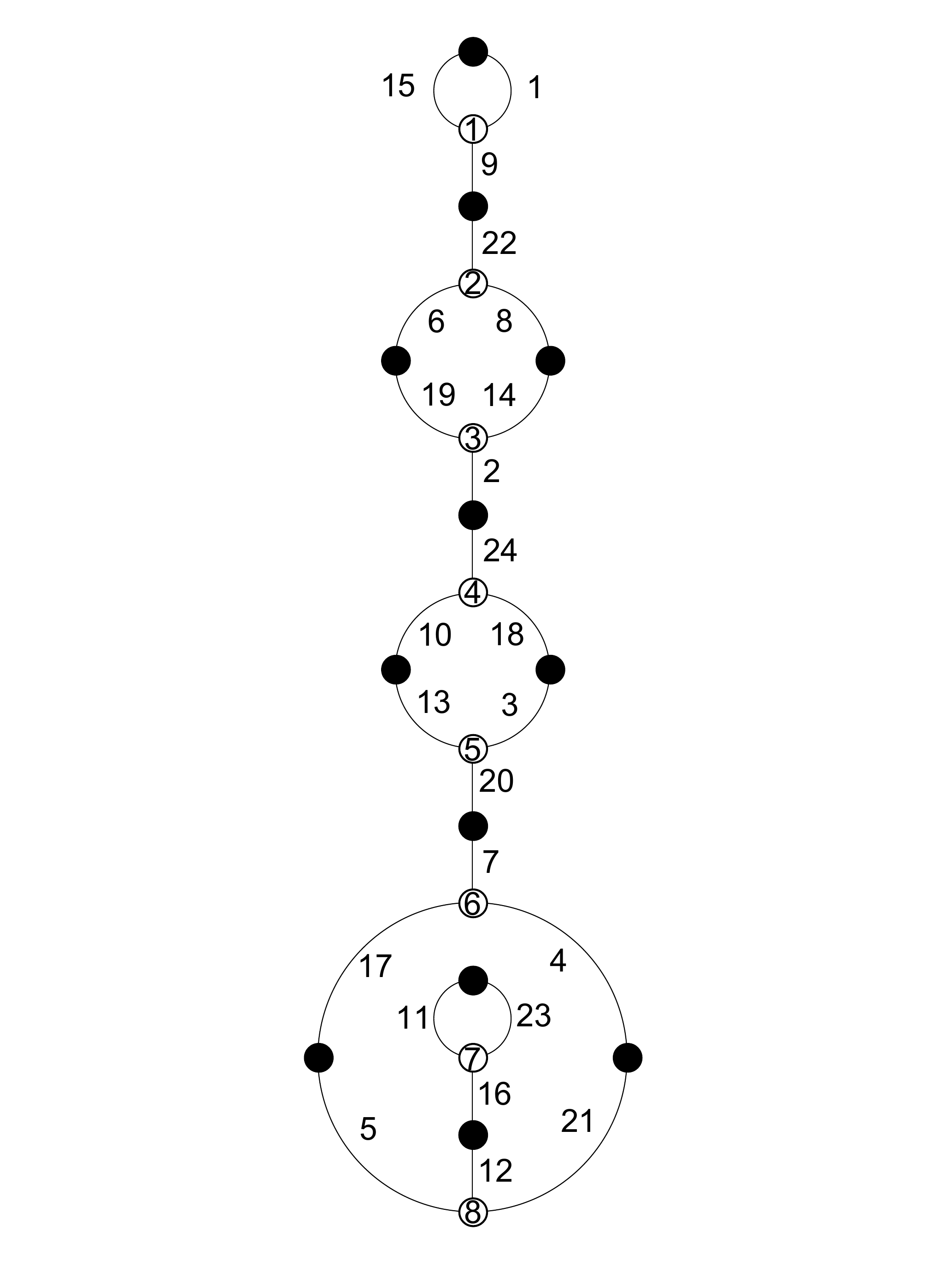}}}$
        \caption{ \{\{\{1,9,15\},\{22,8,6\},
        \{14,2,19\},\{24,18,10\},\{3,20,13\},
        \{7,4,17\},\{11,23,16\},\{12,21,5\}\}, \\ 
        \{\{12,16\},\{4,21\},\{17,5\},
        \{11,23\},\{7,20\},\{18,3\},
        \{10,13\},\{24,2\},\{19,6\},
        \{8,14\},\{9,22\},\{1,15\}\}\}}
        \caption{13-5-2-2-1-1 B $(\sqrt{65})$}
        \label{Dessin}
    \end{subfigure}\hfill
\end{figure}

\begin{figure}[H]
    \begin{subfigure}{0.4\textwidth}
        \centering \captionsetup{justification=centering}
        $\scalemath{0.75}{
        \displaystyle \begin{pmatrix}
            2 & 1 & 0 & 0 & 0 & 0 & 0 & 0\\ 
            1 & 0 & 1 & 0 & 1 & 0 & 0 & 0\\
            0 & 1 & 0 & 1 & 1 & 0 & 0 & 0\\
            0 & 0 & 1 & 2 & 0 & 0 & 0 & 0\\
            0 & 1 & 1 & 0 & 0 & 1 & 0 & 0\\
            0 & 0 & 0 & 0 & 1 & 0 & 0 & 2\\
            0 & 0 & 0 & 0 & 0 & 0 & 2 & 1\\
            0 & 0 & 0 & 0 & 0 & 2 & 1 & 0
        \end{pmatrix}}$
        $\vcenter{\hbox{\includegraphics[width=0.35\textwidth]{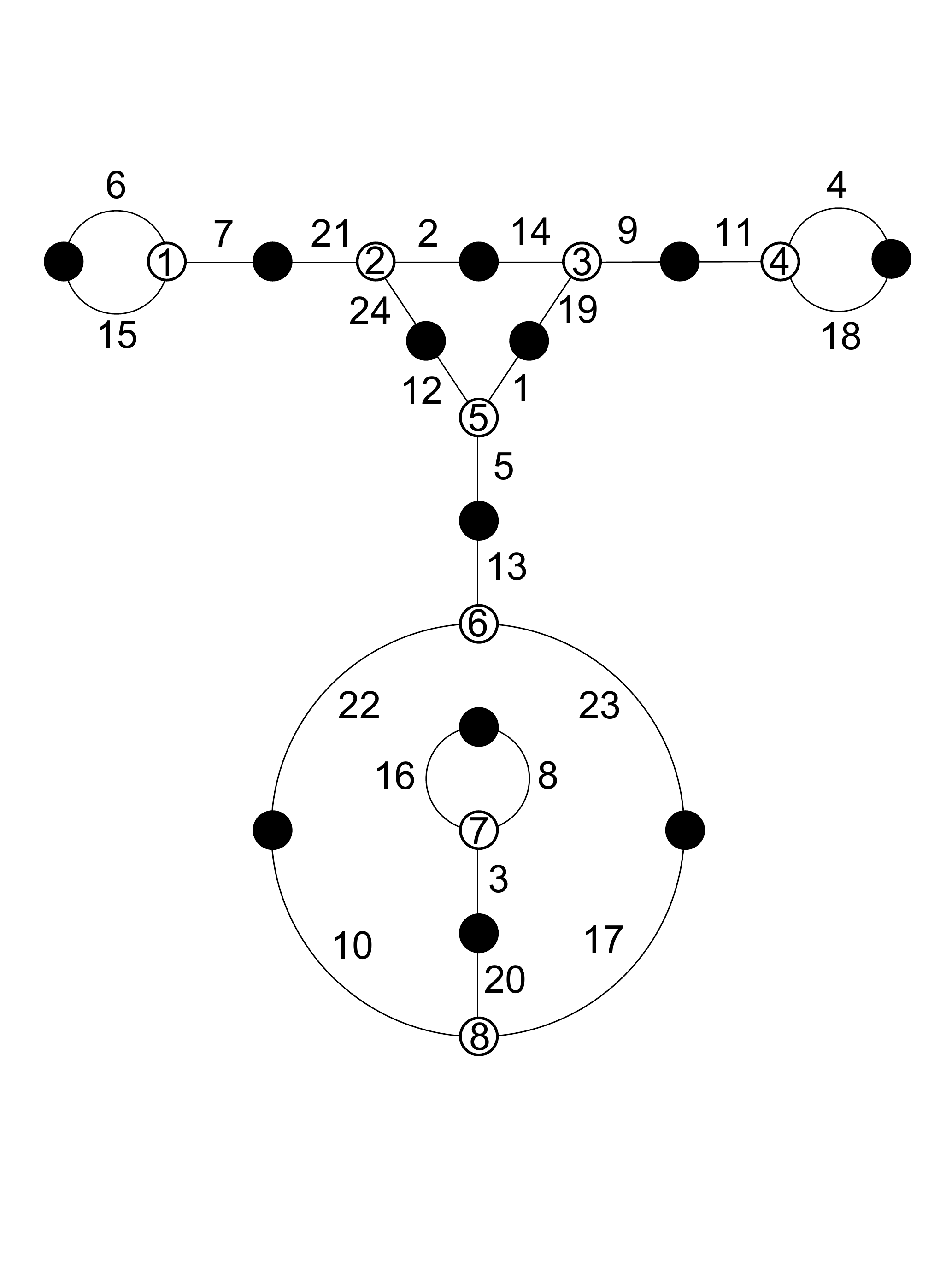}}}$
        \caption{ \{\{\{6,7,15\},\{2,24,21\},
        \{14,9,19\},\{4,18,11\},\{1,5,12\},
        \{13,23,22\},\{16,8,3\},\{20,17,10\}\}, \\ 
        \{\{3,20\},\{16,8\},\{17,23\},
        \{10,22\},\{5,13\},\{1,19\},
        \{12,24\},\{7,21\},\{6,15\},
        \{2,14\},\{9,11\},\{4,18\}\}\}}
        \caption{13-5-3-1-1-1 $(\mathbb{Q})$}
        \label{Dessin}
    \end{subfigure} \hfill
    \begin{subfigure}{0.6\textwidth}
        \centering \captionsetup{justification=centering}
        $\scalemath{0.75}{
        \displaystyle \begin{pmatrix}
            2 & 1 & 0 & 0 & 0 & 0 & 0 & 0\\ 
            1 & 0 & 2 & 0 & 0 & 0 & 0 & 0\\
            0 & 2 & 0 & 1 & 0 & 0 & 0 & 0\\
            0 & 0 & 1 & 0 & 0 & 1 & 1 & 0\\
            0 & 0 & 0 & 0 & 2 & 1 & 0 & 0\\
            0 & 0 & 0 & 1 & 1 & 0 & 1 & 0\\
            0 & 0 & 0 & 1 & 0 & 1 & 0 & 1\\
            0 & 0 & 0 & 0 & 0 & 0 & 1 & 2
        \end{pmatrix}}$
        $\vcenter{\hbox{\includegraphics[width=0.25\textwidth]{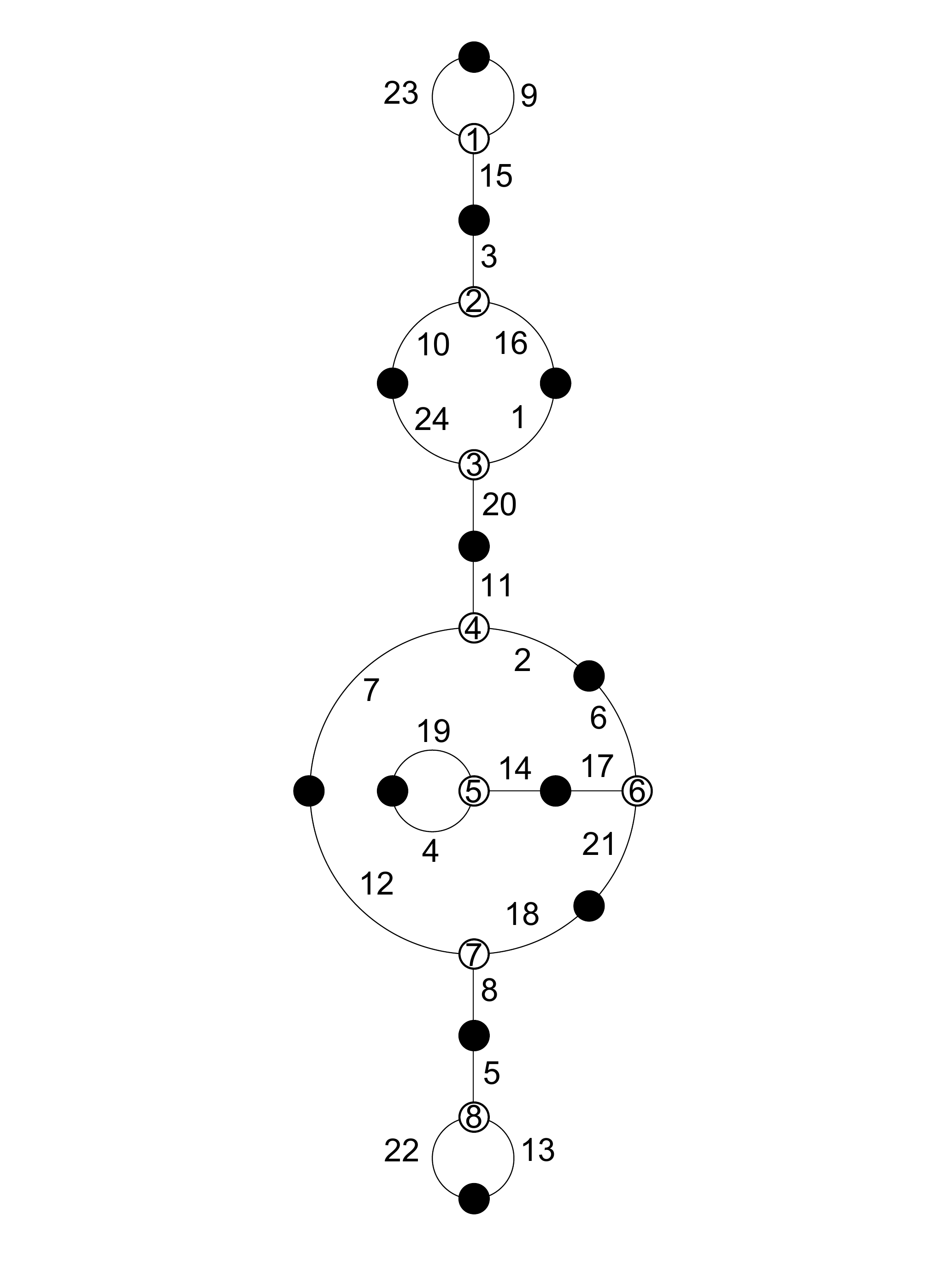}}}$
        $\vcenter{\hbox{\includegraphics[width=0.25\textwidth]{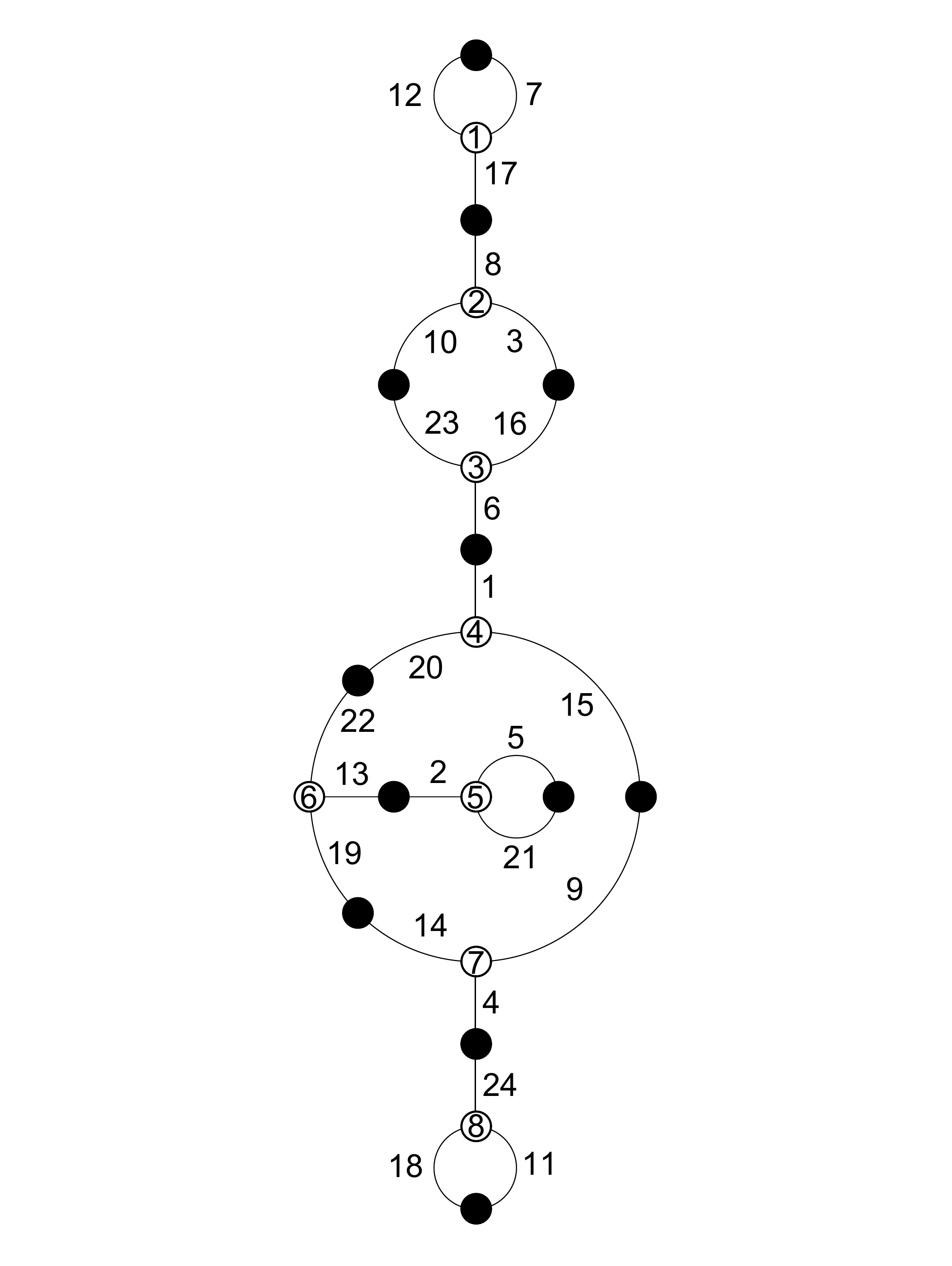}}}$
        \caption{ A: \{\{\{9,15,23\},\{3,16,10\},
        \{1,20,24\},\{11,2,7\},\{6,21,17\},
        \{14,4,19\},\{18,8,12\},\{5,13,22\}\}, \\ 
        \{\{22,13\},\{5,8\},\{18,21\},
        \{12,7\},\{2,6\},\{14,17\},
        \{4,19\},\{11,20\},\{1,16\},
        \{10,24\},\{3,15\},\{9,23\}\}\} \\
        B: \{\{\{17,12,7\},\{3,10,8\},
        \{23,16,6\},\{1,15,20\},\{22,13,19\},
        \{5,21,2\},\{9,4,14\},\{11,18,24\}\}, \\ 
        \{\{11,18\},\{4,24\},\{14,19\},
        \{9,15\},\{5,21\},\{13,2\},
        \{22,20\},\{1,6\},\{16,3\},
        \{23,10\},\{8,17\},\{7,12\}\}\}}
        \caption{13-6-2-1-1-1 A \& B $(\sqrt{-3})$}
        \label{Dessin}
    \end{subfigure}\hfill
\end{figure}

\begin{figure}[H]
    \begin{subfigure}{0.5\textwidth}
        \centering \captionsetup{justification=centering}
        $\scalemath{0.75}{
        \displaystyle \begin{pmatrix}
            2 & 1 & 0 & 0 & 0 & 0 & 0 & 0\\ 
            1 & 0 & 0 & 1 & 0 & 1 & 0 & 0\\
            0 & 0 & 2 & 1 & 0 & 0 & 0 & 0\\
            0 & 1 & 1 & 0 & 0 & 0 & 1 & 0\\
            0 & 0 & 0 & 0 & 2 & 1 & 0 & 0\\
            0 & 1 & 0 & 0 & 1 & 0 & 1 & 0\\
            0 & 0 & 0 & 1 & 0 & 1 & 0 & 1\\
            0 & 0 & 0 & 0 & 0 & 0 & 1 & 2
        \end{pmatrix}}$
        $\vcenter{\hbox{\includegraphics[width=0.35\textwidth]{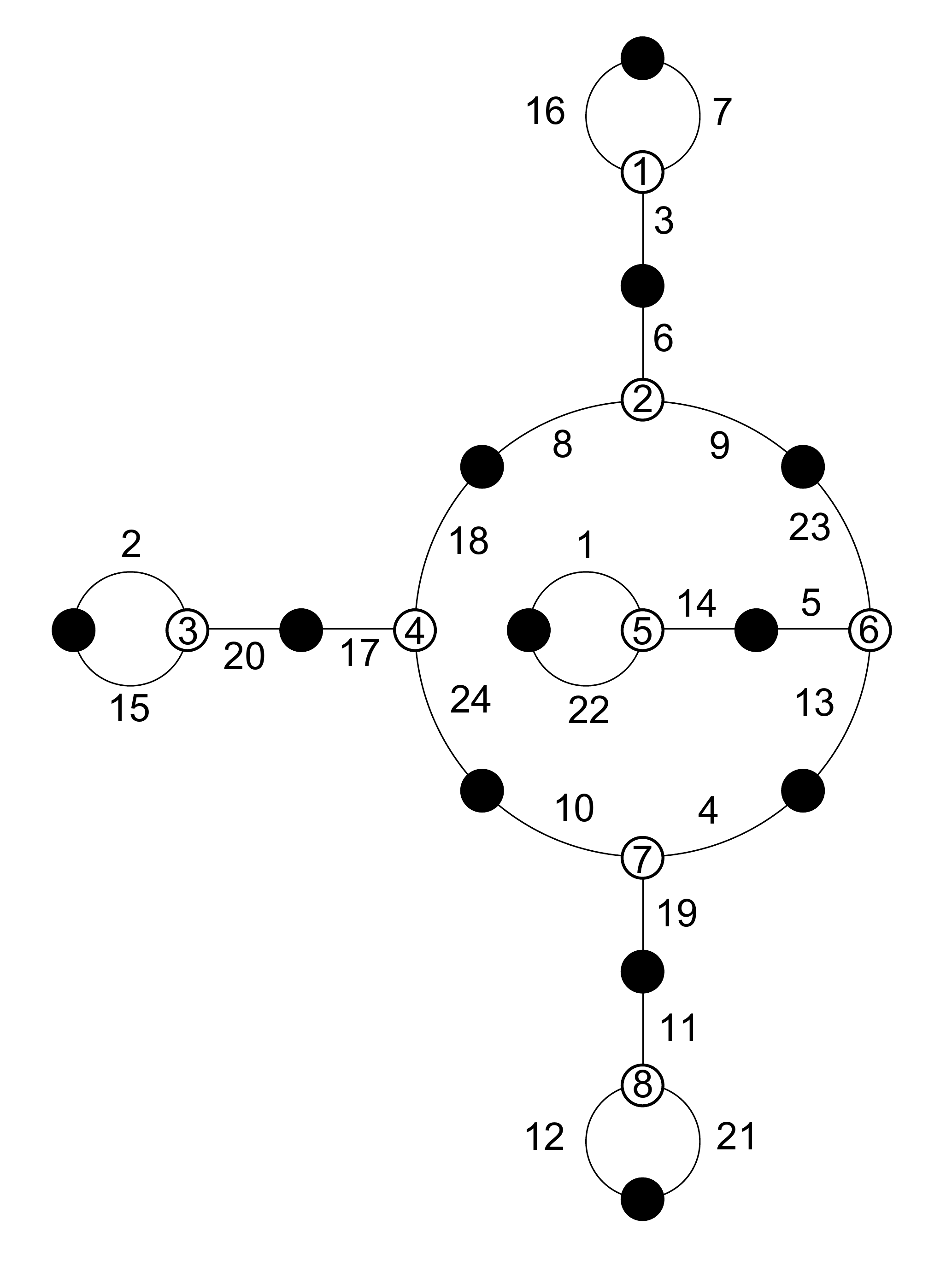}}}$
        \caption{ \{\{\{3,16,7\},\{6,9,8\},
        \{23,13,5\},\{1,14,22\},\{17,18,24\},
        \{2,20,15\},\{10,4,19\},\{11,21,12\}\}, \\ 
        \{\{12,21\},\{11,19\},\{4,13\},
        \{10,24\},\{20,17\},\{2,15\},
        \{18,8\},\{9,23\},\{5,14\},
        \{22,1\},\{6,3\},\{7,16\}\}\}}
        \caption{13-7-1-1-1-1 $(\mathbb{Q})$}
        \label{Dessin}
    \end{subfigure} \hfill
    \begin{subfigure}{0.5\textwidth}
        \centering \captionsetup{justification=centering}
        $\scalemath{0.75}{
        \displaystyle \begin{pmatrix}
            2 & 1 & 0 & 0 & 0 & 0 & 0 & 0\\ 
            1 & 0 & 2 & 0 & 0 & 0 & 0 & 0\\
            0 & 2 & 0 & 1 & 0 & 0 & 0 & 0\\
            0 & 0 & 1 & 0 & 2 & 0 & 0 & 0\\
            0 & 0 & 0 & 2 & 0 & 1 & 0 & 0\\
            0 & 0 & 0 & 0 & 1 & 0 & 1 & 1\\
            0 & 0 & 0 & 0 & 0 & 1 & 0 & 2\\
            0 & 0 & 0 & 0 & 0 & 1 & 2 & 0
        \end{pmatrix}}$
        $\vcenter{\hbox{\includegraphics[width=0.35\textwidth]{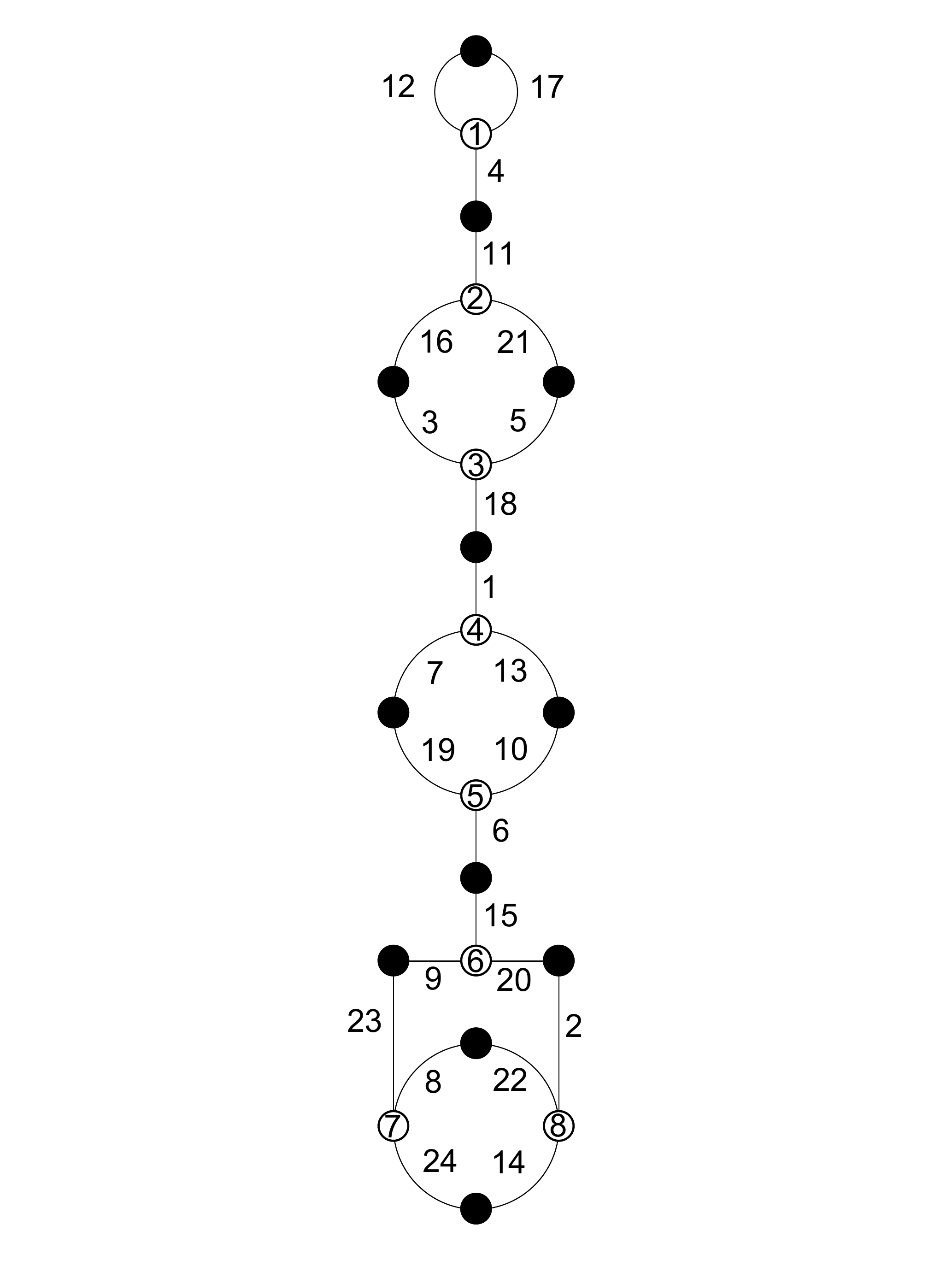}}}$
        \caption{ \{\{\{17,4,12\},\{11,21,16\},
        \{5,18,3\},\{1,13,7\},\{10,6,19\},
        \{15,20,9\},\{2,14,22\},\{8,24,23\}\}, \\ 
        \{\{24,14\},\{8,22\},\{2,20\},
        \{9,23\},\{15,6\},\{19,7\},
        \{13,10\},\{1,18\},\{3,16\},
        \{5,21\},\{11,4\},\{17,12\}\}\}}
        \caption{14-3-2-2-2-1 $(\mathbb{Q})$}
        \label{Dessin}
    \end{subfigure}\hfill
\end{figure}

\begin{figure}[H]
    \begin{subfigure}{0.5\textwidth}
        \centering \captionsetup{justification=centering}
        $\scalemath{0.75}{
        \displaystyle \begin{pmatrix}
            2 & 1 & 0 & 0 & 0 & 0 & 0 & 0\\ 
            1 & 0 & 1 & 0 & 1 & 0 & 0 & 0\\
            0 & 1 & 0 & 1 & 1 & 0 & 0 & 0\\
            0 & 0 & 1 & 2 & 0 & 0 & 0 & 0\\
            0 & 1 & 1 & 0 & 0 & 1 & 0 & 0\\
            0 & 0 & 0 & 0 & 1 & 0 & 1 & 1\\
            0 & 0 & 0 & 0 & 0 & 1 & 0 & 2\\
            0 & 0 & 0 & 0 & 0 & 1 & 2 & 0
        \end{pmatrix}}$
        $\vcenter{\hbox{\includegraphics[width=0.35\textwidth]{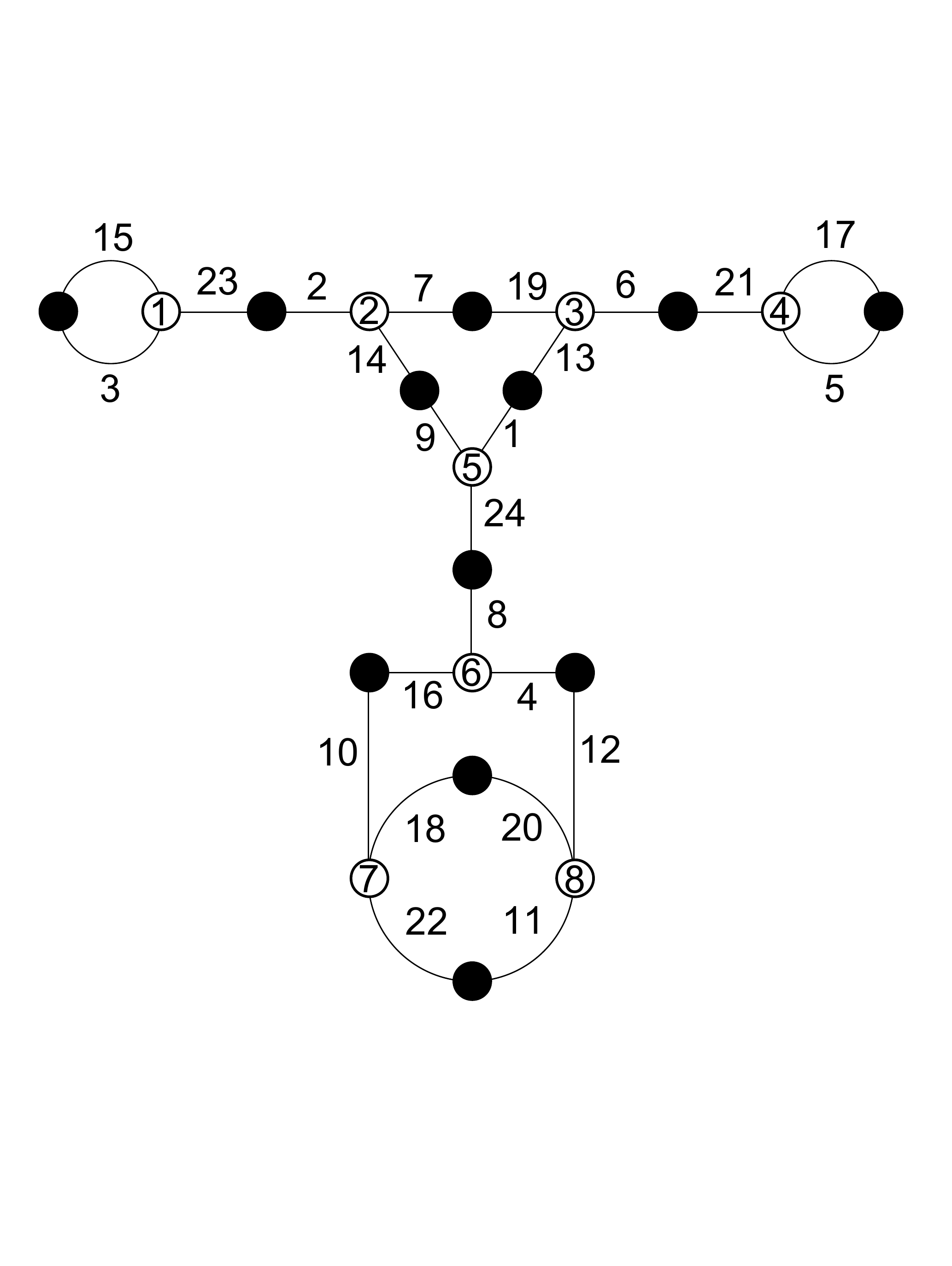}}}$
        \caption{ \{\{\{23,3,15\},\{2,7,14\},
        \{19,6,13\},\{21,17,5\},\{1,24,9\},
        \{8,4,16\},\{18,22,10\},\{20,12,11\}\}, \\ 
        \{\{22,11\},\{18,20\},\{10,16\},
        \{4,12\},\{8,24\},\{1,13\},
        \{9,14\},\{7,19\},\{6,21\},
        \{2,23\},\{3,15\},\{17,5\}\}\}}
        \caption{14-3-3-2-1-1 A $(\mathbb{Q})$}
        \label{Dessin}
    \end{subfigure} \hfill
    \begin{subfigure}{0.5\textwidth}
        \centering \captionsetup{justification=centering}
        $\scalemath{0.75}{
        \displaystyle \begin{pmatrix}
            2 & 1 & 0 & 0 & 0 & 0 & 0 & 0\\ 
            1 & 0 & 1 & 1 & 0 & 0 & 0 & 0\\
            0 & 1 & 0 & 1 & 1 & 0 & 0 & 0\\
            0 & 1 & 1 & 0 & 1 & 0 & 0 & 0\\
            0 & 0 & 1 & 1 & 0 & 1 & 0 & 0\\
            0 & 0 & 0 & 0 & 1 & 0 & 2 & 0\\
            0 & 0 & 0 & 0 & 0 & 2 & 0 & 1\\
            0 & 0 & 0 & 0 & 0 & 0 & 1 & 2
        \end{pmatrix}}$
        $\vcenter{\hbox{\includegraphics[width=0.35\textwidth]{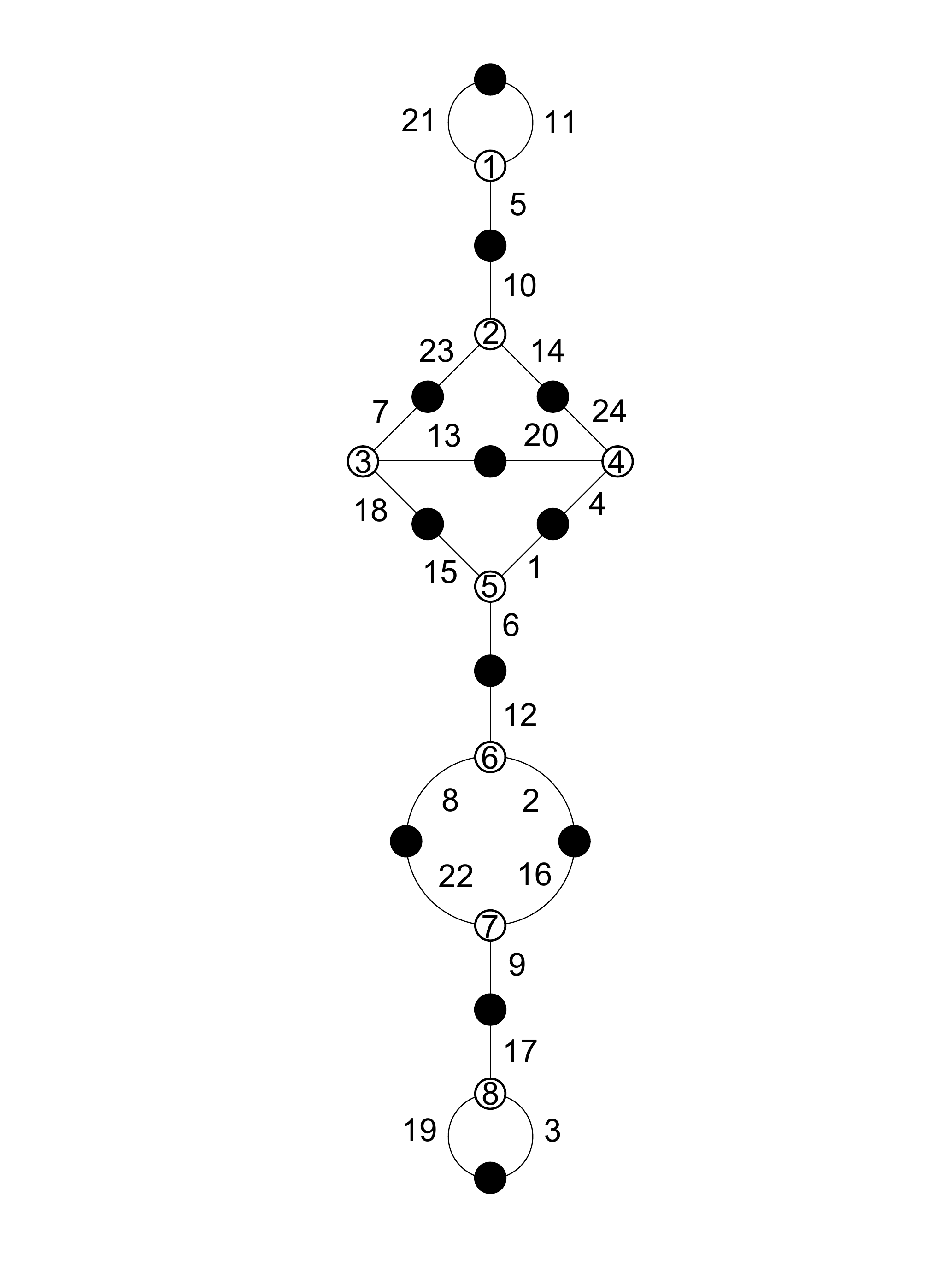}}}$
        \caption{ \{\{\{21,11,5\},\{10,14,23\},
        \{24,4,20\},\{13,18,7\},\{1,6,15\},
        \{12,2,8\},\{16,9,22\},\{17,3,19\}\}, \\ 
        \{\{19,3\},\{17,9\},\{2,16\},
        \{22,8\},\{12,6\},\{1,4\},
        \{15,18\},\{13,20\},\{7,23\},
        \{14,24\},\{10,5\},\{11,21\}\}\}}
        \caption{14-3-3-2-1-1 B $(\mathbb{Q})$}
        \label{Dessin}
    \end{subfigure}\hfill
\end{figure}

\begin{figure}[H]
    \begin{subfigure}{0.6\textwidth}
        \centering \captionsetup{justification=centering}
        $\scalemath{0.75}{
        \displaystyle \begin{pmatrix}
            2 & 1 & 0 & 0 & 0 & 0 & 0 & 0\\ 
            1 & 0 & 2 & 0 & 0 & 0 & 0 & 0\\
            0 & 2 & 0 & 1 & 0 & 0 & 0 & 0\\
            0 & 0 & 1 & 0 & 1 & 0 & 1 & 0\\
            0 & 0 & 0 & 1 & 0 & 2 & 0 & 0\\
            0 & 0 & 0 & 0 & 2 & 0 & 1 & 0\\
            0 & 0 & 0 & 1 & 0 & 1 & 0 & 1\\
            0 & 0 & 0 & 0 & 0 & 0 & 1 & 2
        \end{pmatrix}}$
        $\vcenter{\hbox{\includegraphics[width=0.25\textwidth]{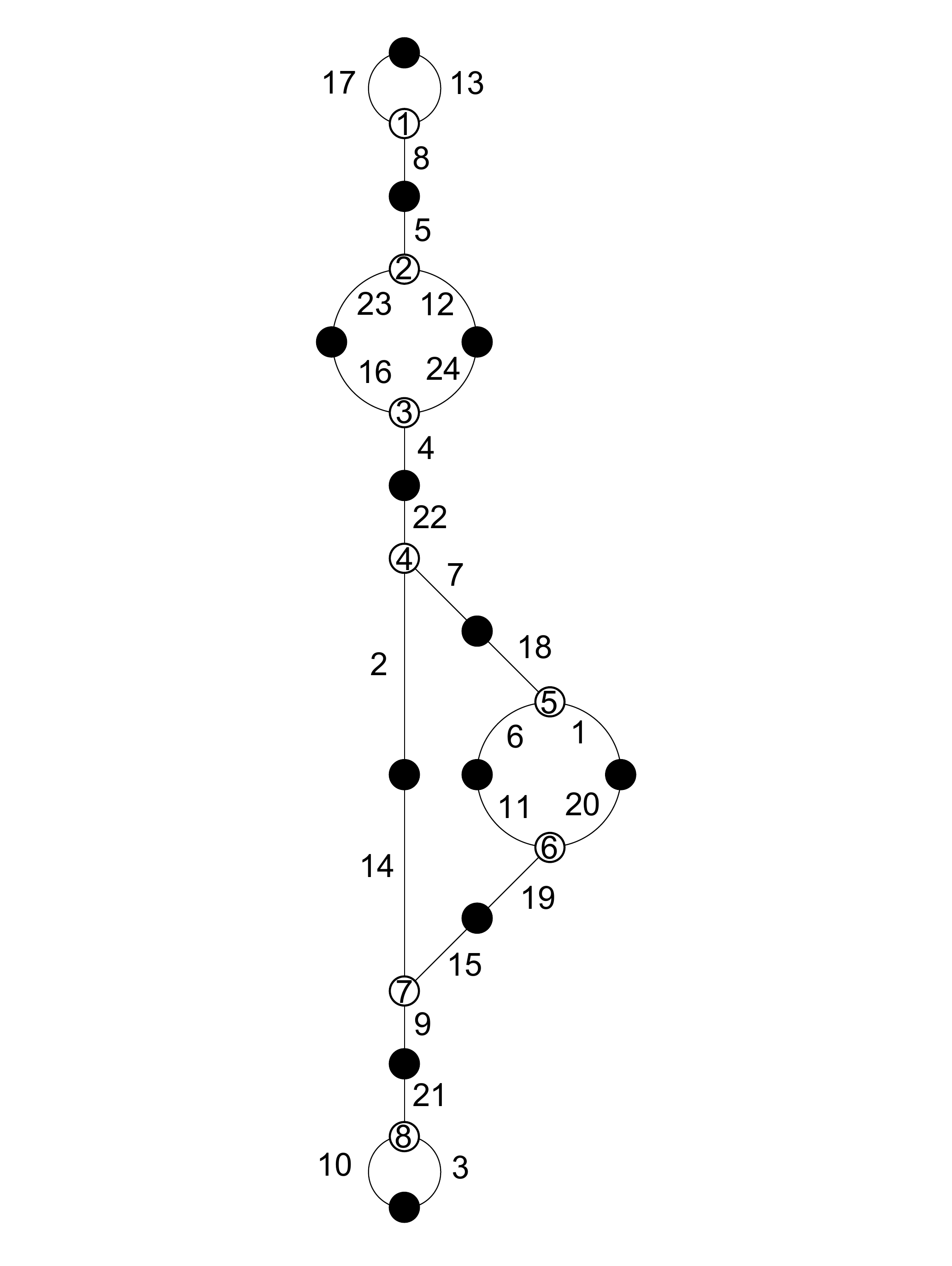}}}$
        $\vcenter{\hbox{\includegraphics[width=0.25\textwidth]{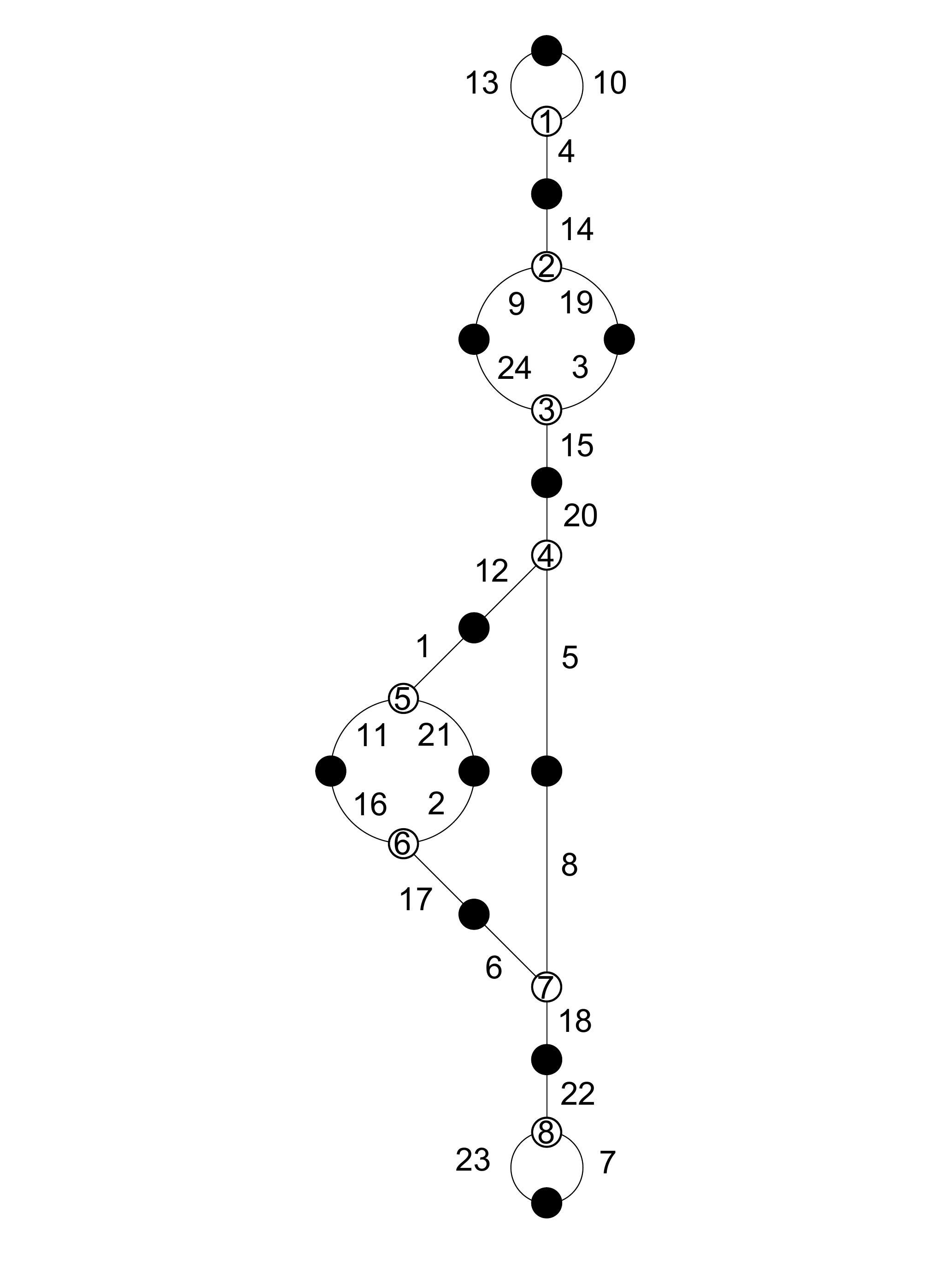}}}$
        \caption{ A: \{\{\{8,17,13\},\{5,12,23\},
        \{24,4,16\},\{22,7,2\},\{18,1,6\},
        \{20,19,11\},\{15,9,14\},\{21,3,10\}\}, \\ 
        \{\{3,10\},\{21,9\},\{15,19\},
        \{1,20\},\{6,11\},\{18,7\},
        \{2,14\},\{22,4\},\{24,12\},
        \{16,23\},\{5,8\},\{13,17\}\}\} \\
        B: \{\{\{13,10,4\},\{14,19,9\},
        \{3,15,24\},\{20,5,12\},\{1,21,11\},
        \{2,17,16\},\{8,18,6\},\{22,7,23\}\}, \\ 
        \{\{23,7\},\{18,22\},\{6,17\},
        \{5,8\},\{2,21\},\{11,16\},
        \{1,12\},\{20,15\},\{3,19\},
        \{9,24\},\{4,14\},\{10,13\}\}\}}
        \caption{14-4-2-2-1-1 A \& B $(\sqrt{-7})$}
        \label{Dessin}
    \end{subfigure} \hfill
    \begin{subfigure}{0.4\textwidth}
        \centering \captionsetup{justification=centering}
        $\scalemath{0.75}{
        \displaystyle \begin{pmatrix}
            2 & 1 & 0 & 0 & 0 & 0 & 0 & 0\\ 
            1 & 0 & 1 & 0 & 0 & 1 & 0 & 0\\
            0 & 1 & 0 & 1 & 0 & 0 & 1 & 0\\
            0 & 0 & 1 & 0 & 1 & 0 & 1 & 0\\
            0 & 0 & 0 & 1 & 2 & 0 & 0 & 0\\
            0 & 1 & 0 & 0 & 0 & 0 & 1 & 1\\
            0 & 0 & 1 & 1 & 0 & 1 & 0 & 0\\
            0 & 0 & 0 & 0 & 0 & 1 & 0 & 2
        \end{pmatrix}}$
        $\vcenter{\hbox{\includegraphics[width=0.35\textwidth]{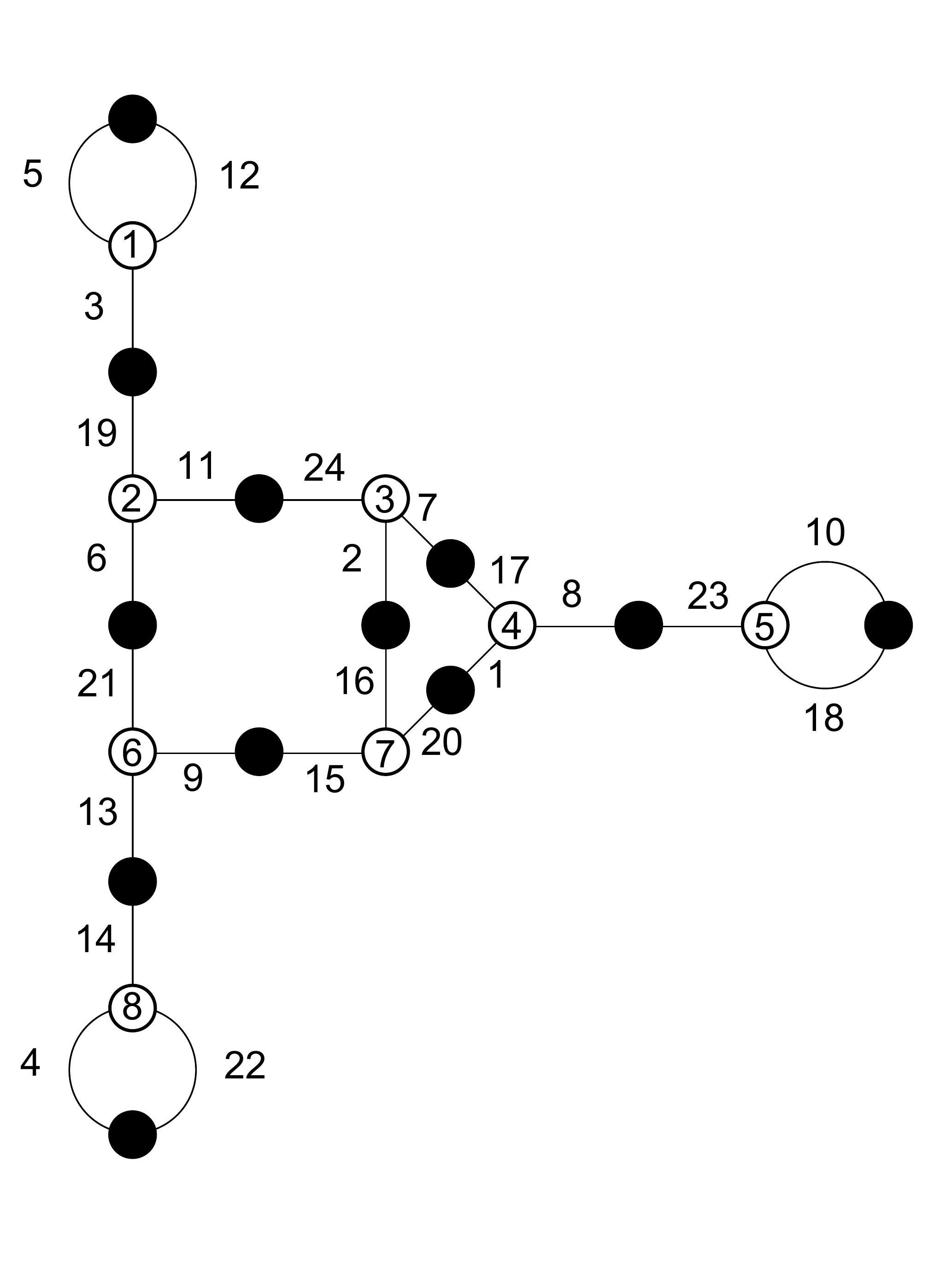}}}$
        \caption{ \{\{\{5,12,3\},\{19,11,6\},
        \{24,7,2\},\{17,8,1\},\{23,10,18\},
        \{20,15,16\},\{9,13,21\},\{14,22,4\}\}, \\ 
        \{\{22,4\},\{13,14\},\{9,15\},
        \{2,16\},\{7,17\},\{1,20\},
        \{8,23\},\{10,18\},\{11,24\},
        \{6,21\},\{3,19\},\{5,12\}\}\}}
        \caption{14-4-3-1-1-1 $(\mathbb{Q})$}
        \label{Dessin}
    \end{subfigure}\hfill
\end{figure}

\begin{figure}[H]
    \begin{subfigure}{0.5\textwidth}
        \centering \captionsetup{justification=centering}
        $\scalemath{0.75}{
        \displaystyle \begin{pmatrix}
            2 & 1 & 0 & 0 & 0 & 0 & 0 & 0\\ 
            1 & 0 & 1 & 1 & 0 & 0 & 0 & 0\\
            0 & 1 & 0 & 0 & 0 & 2 & 0 & 0\\
            0 & 1 & 0 & 0 & 1 & 0 & 1 & 0\\
            0 & 0 & 0 & 1 & 2 & 0 & 0 & 0\\
            0 & 0 & 2 & 0 & 0 & 0 & 1 & 0\\
            0 & 0 & 0 & 1 & 0 & 1 & 0 & 1\\
            0 & 0 & 0 & 0 & 0 & 0 & 1 & 2
        \end{pmatrix}}$
        $\vcenter{\hbox{\includegraphics[width=0.35\textwidth]{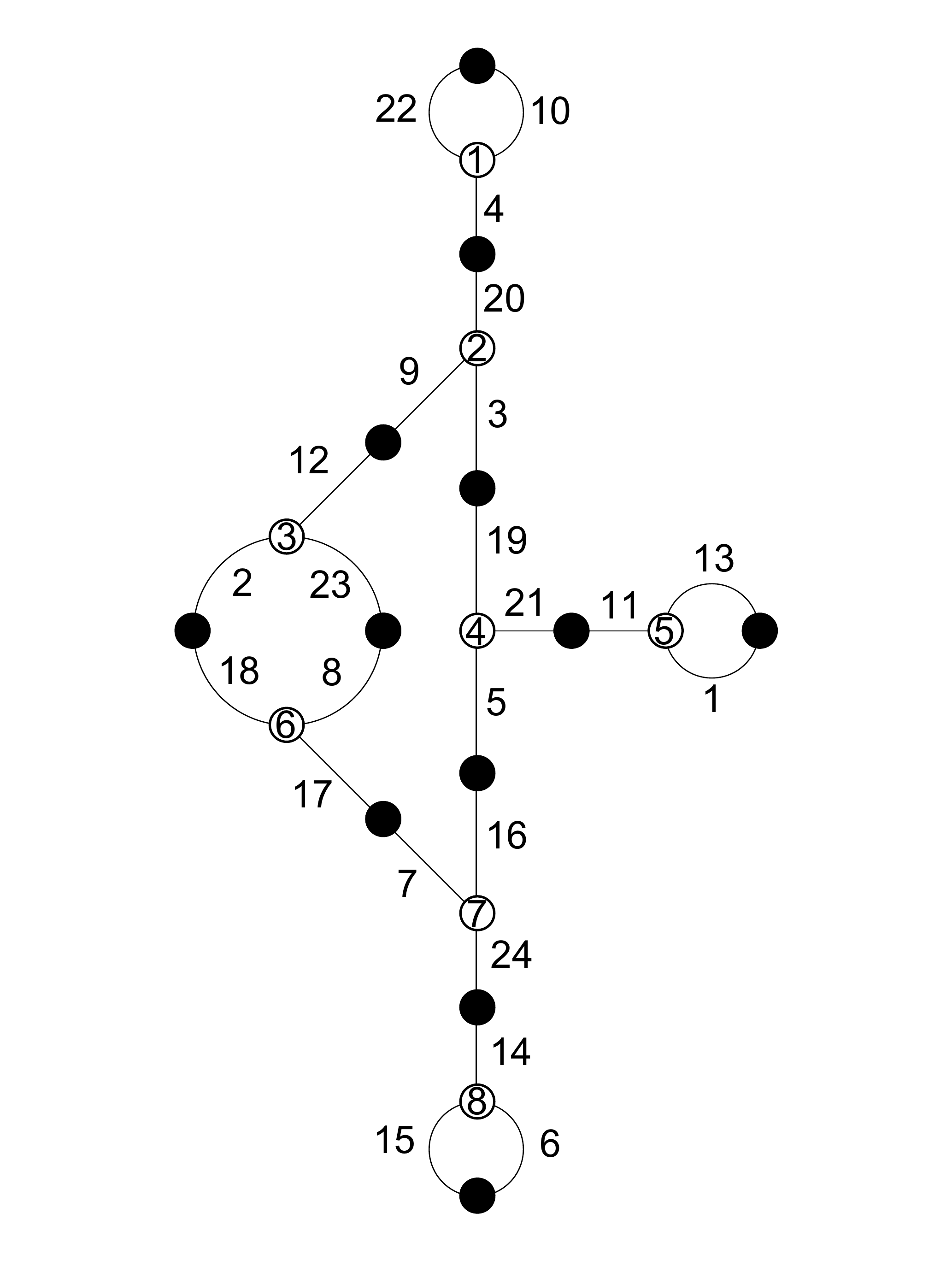}}}$
        \caption{ \{\{\{10,4,22\},\{20,3,9\},
        \{12,23,2\},\{19,21,5\},\{11,13,1\},
        \{16,24,7\},\{8,17,18\},\{14,6,15\}\}, \\ 
        \{\{15,6\},\{14,24\},\{5,16\},
        \{7,17\},\{8,23\},\{2,18\},
        \{9,12\},\{19,3\},\{21,11\},
        \{13,1\},\{20,4\},\{10,22\}\}\}}
        \caption{14-5-2-1-1-1 A $(\mathbb{Q})$}
        \label{Dessin}
    \end{subfigure} \hfill
    \begin{subfigure}{0.5\textwidth}
        \centering \captionsetup{justification=centering}
        $\scalemath{0.75}{
        \displaystyle \begin{pmatrix}
            2 & 1 & 0 & 0 & 0 & 0 & 0 & 0\\ 
            1 & 0 & 1 & 1 & 0 & 0 & 0 & 0\\
            0 & 1 & 2 & 0 & 0 & 0 & 0 & 0\\
            0 & 1 & 0 & 0 & 2 & 0 & 0 & 0\\
            0 & 0 & 0 & 2 & 0 & 1 & 0 & 0\\
            0 & 0 & 0 & 0 & 1 & 0 & 0 & 2\\
            0 & 0 & 0 & 0 & 0 & 0 & 2 & 1\\
            0 & 0 & 0 & 0 & 0 & 2 & 1 & 0
        \end{pmatrix}}$
        $\vcenter{\hbox{\includegraphics[width=0.35\textwidth]{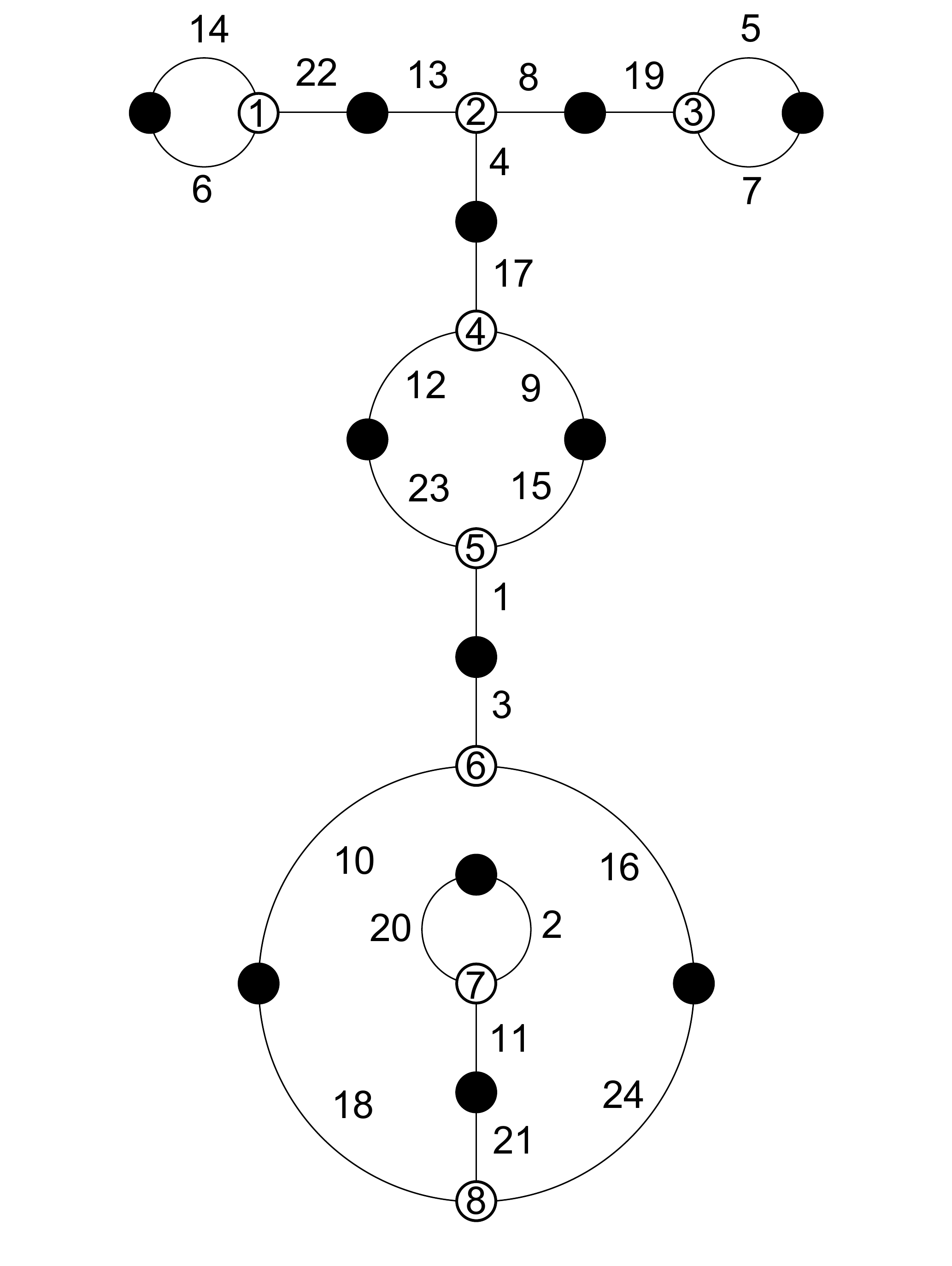}}}$
        \caption{ \{\{\{6,14,22\},\{13,8,4\},
        \{19,5,7\},\{9,12,17\},\{15,1,23\},
        \{3,16,10\},\{20,2,11\},\{21,24,18\}\}, \\ 
        \{\{20,2\},\{11,21\},\{24,16\},
        \{10,18\},\{1,3\},\{9,15\},
        \{12,23\},\{4,17\},\{8,19\},
        \{5,7\},\{13,22\},\{6,14\}\}\}}
        \caption{14-5-2-1-1-1 B (cubic)}
        \label{Dessin}
    \end{subfigure}\hfill
\end{figure}

\begin{figure}[H]
    \begin{subfigure}{0.5\textwidth}
        \centering \captionsetup{justification=centering}
        $\scalemath{0.75}{
        \displaystyle \begin{pmatrix}
            2 & 1 & 0 & 0 & 0 & 0 & 0 & 0\\ 
            1 & 0 & 2 & 0 & 0 & 0 & 0 & 0\\
            0 & 2 & 0 & 1 & 0 & 0 & 0 & 0\\
            0 & 0 & 1 & 0 & 1 & 1 & 0 & 0\\
            0 & 0 & 0 & 1 & 2 & 0 & 0 & 0\\
            0 & 0 & 0 & 1 & 0 & 0 & 0 & 2\\
            0 & 0 & 0 & 0 & 0 & 0 & 2 & 1\\
            0 & 0 & 0 & 0 & 0 & 2 & 1 & 0
        \end{pmatrix}}$
        $\vcenter{\hbox{\includegraphics[width=0.25\textwidth]{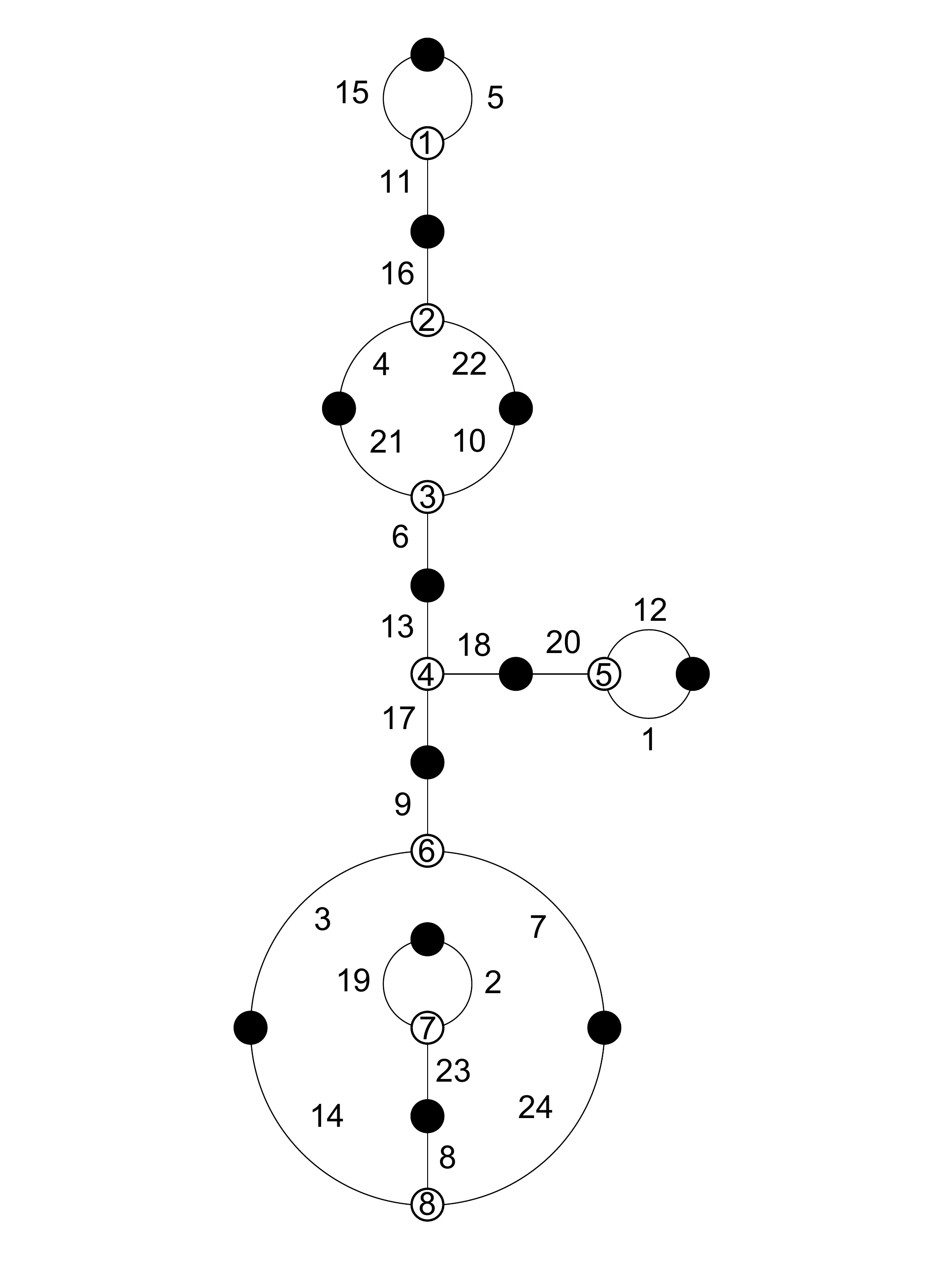}}}$
        $\vcenter{\hbox{\includegraphics[width=0.25\textwidth]{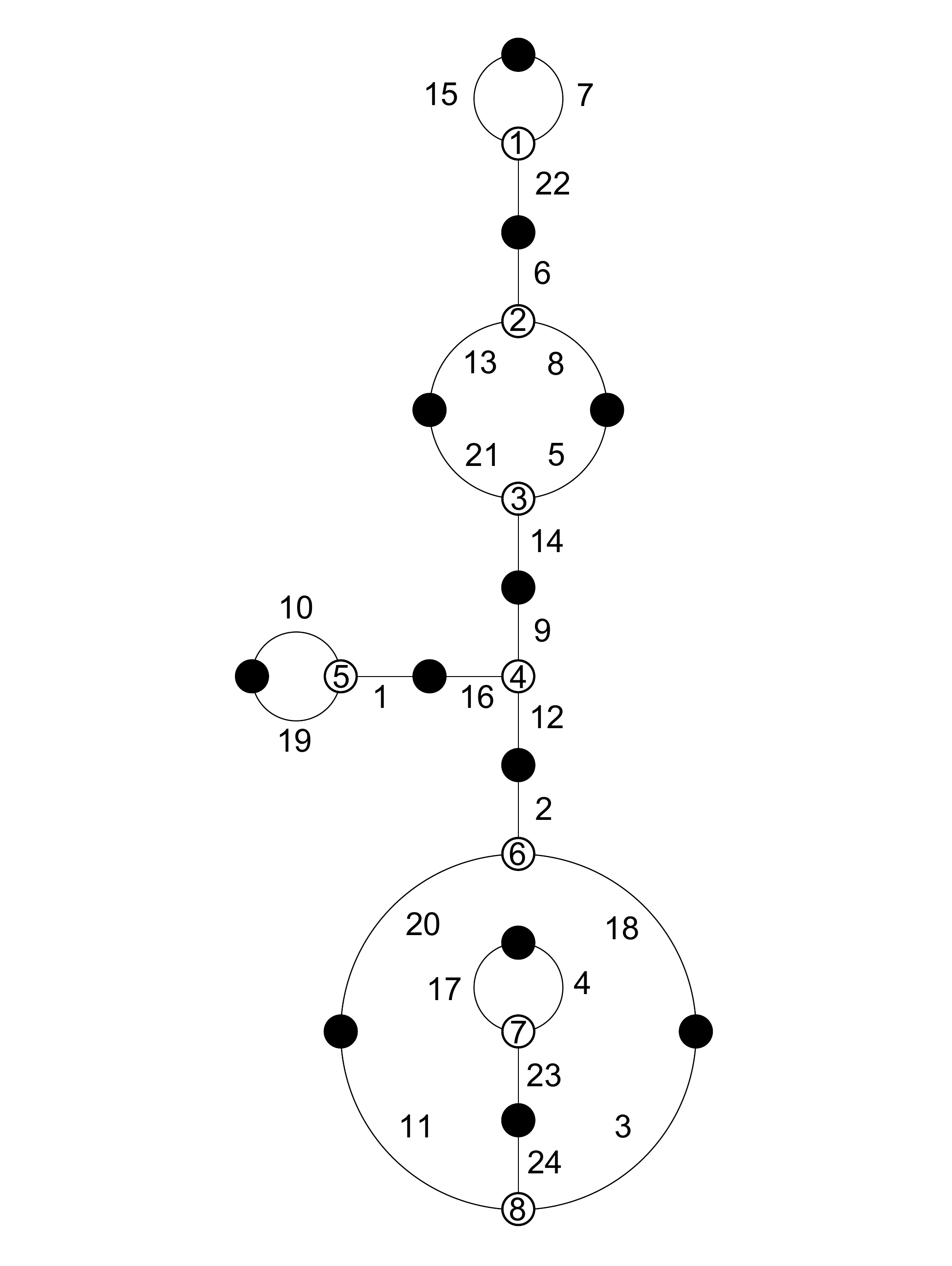}}}$
        \caption{ C: \{\{\{5,11,15\},\{4,16,22\},
        \{10,6,21\},\{13,18,17\},\{1,20,12\},
        \{9,7,3\},\{2,23,19\},\{8,24,14\}\}, \\
        \{\{19,2\},\{23,8\},\{7,24\},
        \{3,14\},\{9,17\},\{18,20\},
        \{1,12\},\{6,13\},\{10,22\},
        \{4,21\},\{11,16\},\{5,15\}\}\} \\
        D: \{\{\{22,15,7\},\{6,8,13\},
        \{5,14,21\},\{12,16,9\},\{10,1,19\},
        \{2,18,20\},\{4,23,17\},\{24,3,11\}\}, \\ 
        \{\{17,4\},\{23,24\},\{3,18\},
        \{11,20\},\{2,12\},\{9,14\},
        \{1,16\},\{10,19\},\{5,8\},
        \{21,13\},\{6,22\},\{7,15\}\}\}}
        \caption{14-5-2-1-1-1 C \& D (cubic)}
        \label{Dessin}
    \end{subfigure} \hfill
    \begin{subfigure}{0.5\textwidth}
        \centering \captionsetup{justification=centering}
        $\scalemath{0.75}{
        \displaystyle \begin{pmatrix}
            2 & 1 & 0 & 0 & 0 & 0 & 0 & 0\\ 
            1 & 0 & 1 & 0 & 1 & 0 & 0 & 0\\
            0 & 1 & 0 & 1 & 1 & 0 & 0 & 0\\
            0 & 0 & 1 & 2 & 0 & 0 & 0 & 0\\
            0 & 1 & 1 & 0 & 0 & 0 & 1 & 0\\
            0 & 0 & 0 & 0 & 0 & 2 & 1 & 0\\
            0 & 0 & 0 & 0 & 1 & 1 & 0 & 1\\
            0 & 0 & 0 & 0 & 0 & 0 & 1 & 2
        \end{pmatrix}}$
        $\vcenter{\hbox{\includegraphics[width=0.25\textwidth]{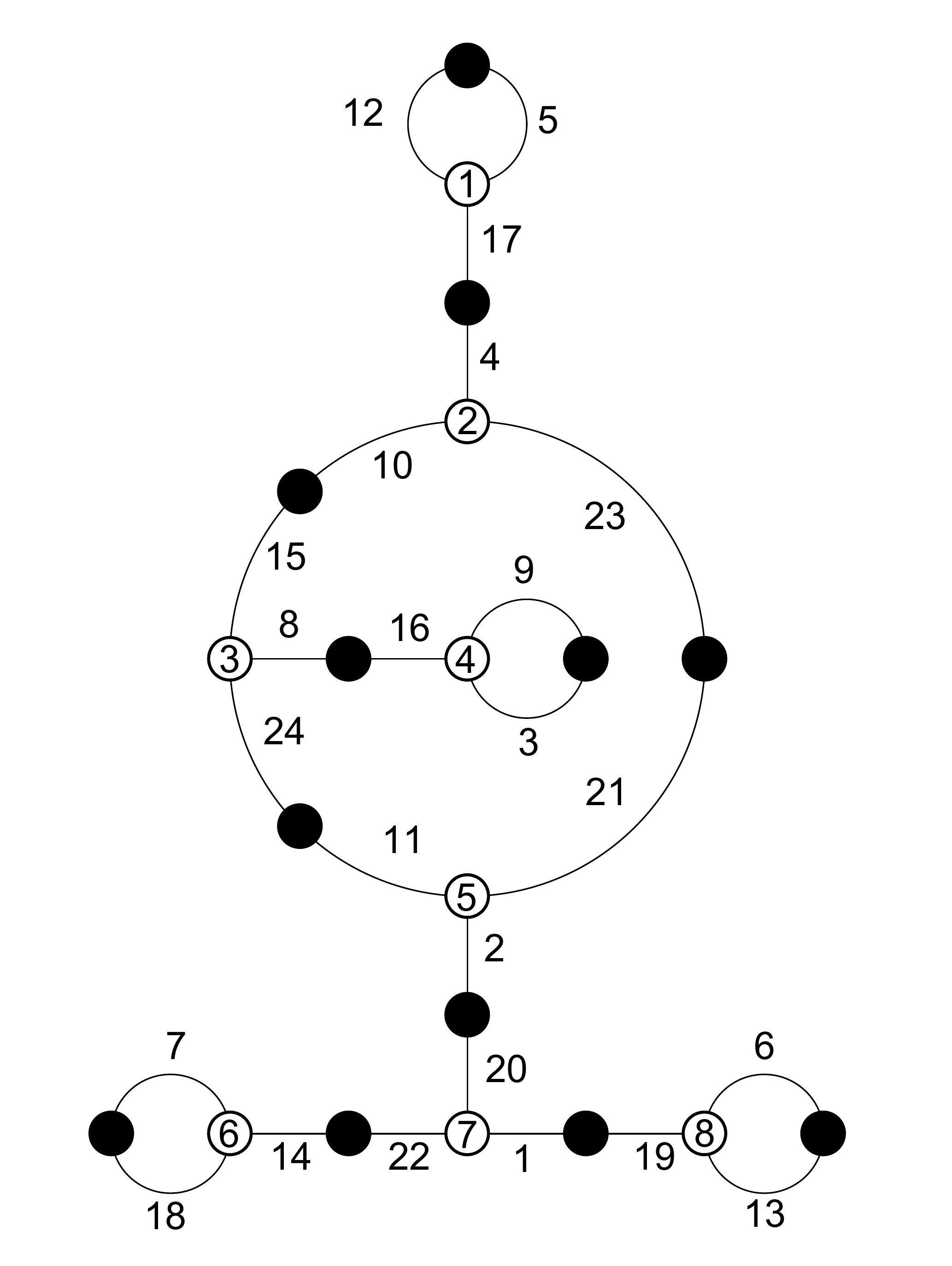}}}$
        $\vcenter{\hbox{\includegraphics[width=0.25\textwidth]{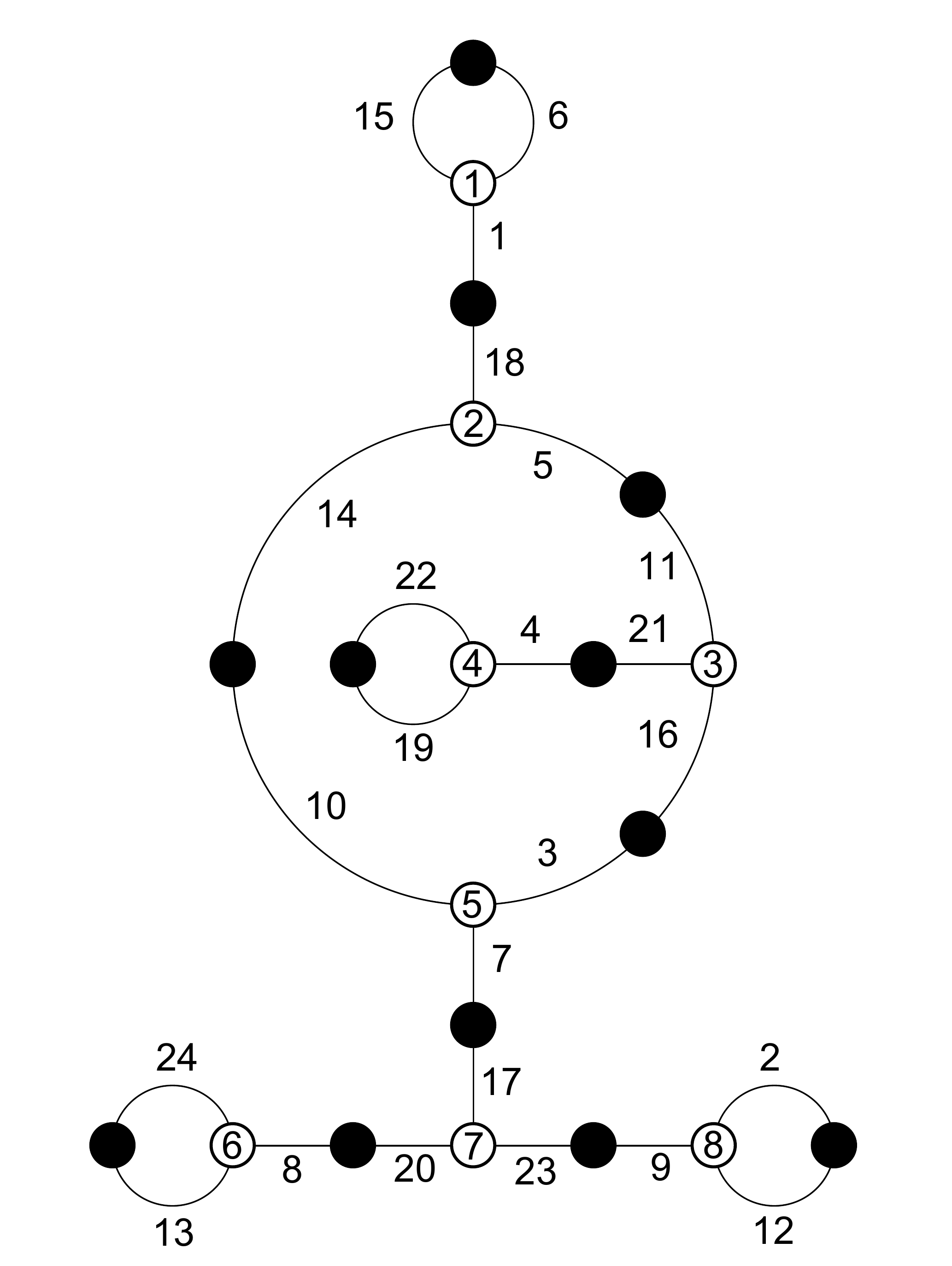}}}$
        \caption{ A: \{\{\{5,17,12\},\{4,23,10\},
        \{9,3,16\},\{8,24,15\},\{11,21,2\},
        \{20,1,22\},\{14,18,7\},\{6,13,19\}\}, \\ 
        \{\{6,13\},\{1,19\},\{20,2\},
        \{14,22\},\{7,18\},\{24,11\},
        \{8,16\},\{3,9\},\{21,23\},
        \{4,17\},\{10,15\},\{12,5\}\}\} \\
        B: \{\{\{1,15,6\},\{18,5,14\},
        \{11,16,21\},\{4,19,22\},\{3,7,10\},
        \{17,23,20\},\{9,2,12\},\{8,13,24\}\}, \\ 
        \{\{13,24\},\{8,20\},\{9,23\},
        \{2,12\},\{7,17\},\{3,16\},
        \{14,10\},\{19,22\},\{4,21\},
        \{5,11\},\{1,18\},\{6,15\}\}\}}
        \caption{14-6-1-1-1-1 A \& B $(\sqrt{-3})$}
        \label{Dessin}
    \end{subfigure}\hfill
\end{figure}

\begin{figure}[H]
    \begin{subfigure}{0.4\textwidth}
        \centering \captionsetup{justification=centering}
        $\scalemath{0.75}{
        \displaystyle \begin{pmatrix}
            2 & 1 & 0 & 0 & 0 & 0 & 0 & 0\\ 
            1 & 0 & 1 & 1 & 0 & 0 & 0 & 0\\
            0 & 1 & 2 & 0 & 0 & 0 & 0 & 0\\
            0 & 1 & 0 & 0 & 2 & 0 & 0 & 0\\
            0 & 0 & 0 & 2 & 0 & 1 & 0 & 0\\
            0 & 0 & 0 & 0 & 1 & 0 & 1 & 1\\
            0 & 0 & 0 & 0 & 0 & 1 & 0 & 2\\
            0 & 0 & 0 & 0 & 0 & 1 & 2 & 0
        \end{pmatrix}}$
        $\vcenter{\hbox{\includegraphics[width=0.35\textwidth]{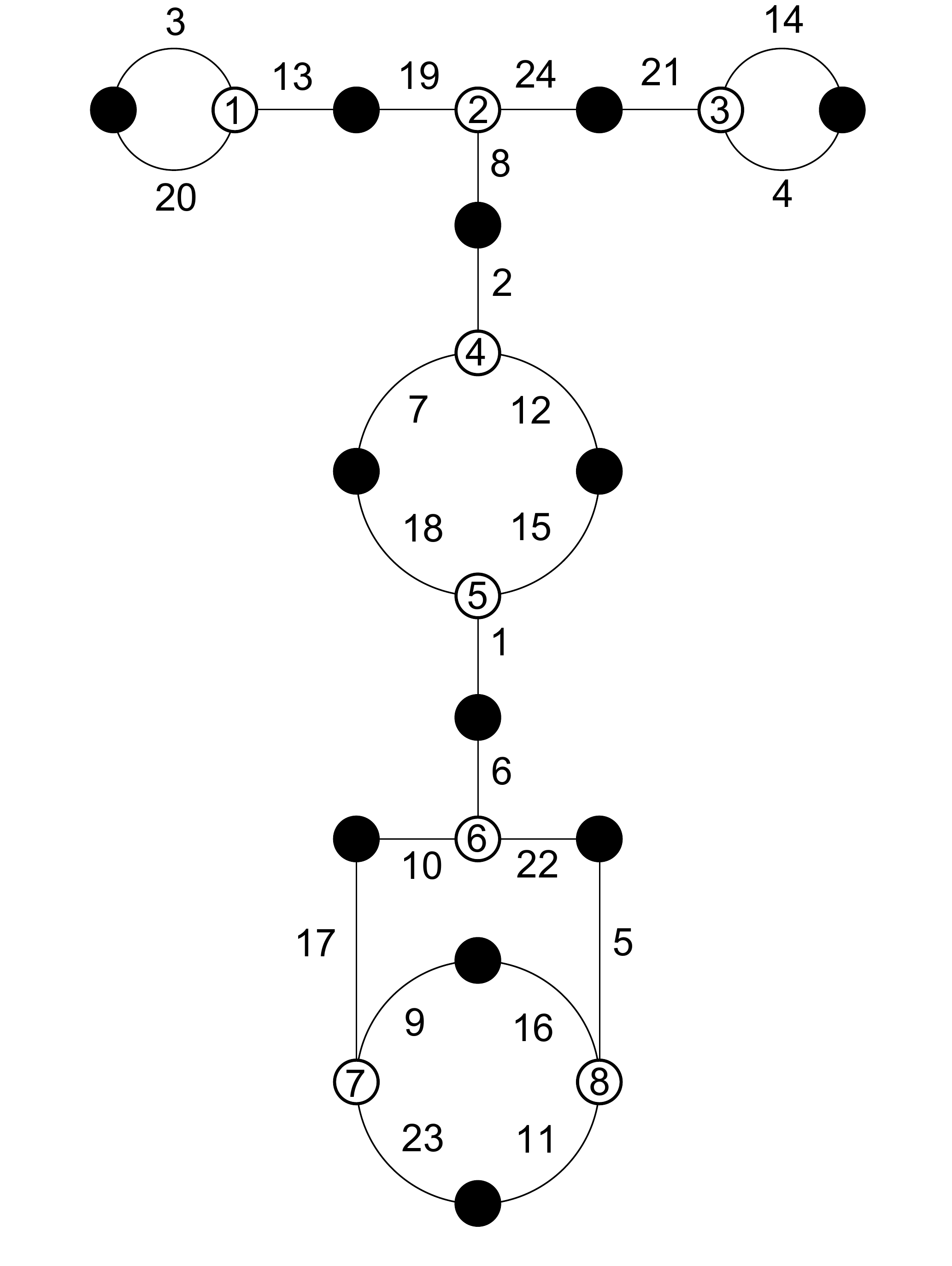}}}$
        \caption{ \{\{\{13,20,3\},\{19,24,8\},
        \{14,4,21\},\{2,12,7\},\{18,15,1\},
        \{6,22,10\},\{5,11,16\},\{23,17,9\}\}, \\ 
        \{\{9,16\},\{17,10\},\{11,23\},
        \{5,22\},\{1,6\},\{12,15\},
        \{7,18\},\{2,8\},\{19,13\},
        \{14,4\},\{3,20\},\{24,21\}\}\}}
        \caption{15-3-2-2-1-1 A $(\mathbb{Q})$}
        \label{Dessin}
    \end{subfigure} \hfill
    \begin{subfigure}{0.6\textwidth}
        \centering \captionsetup{justification=centering}
        $\scalemath{0.75}{
        \displaystyle \begin{pmatrix}
            2 & 1 & 0 & 0 & 0 & 0 & 0 & 0\\ 
            1 & 0 & 2 & 0 & 0 & 0 & 0 & 0\\
            0 & 2 & 0 & 1 & 0 & 0 & 0 & 0\\
            0 & 0 & 1 & 0 & 1 & 1 & 0 & 0\\
            0 & 0 & 0 & 1 & 2 & 0 & 0 & 0\\
            0 & 0 & 0 & 1 & 0 & 0 & 1 & 1\\
            0 & 0 & 0 & 0 & 0 & 1 & 0 & 2\\
            0 & 0 & 0 & 0 & 0 & 1 & 2 & 0
        \end{pmatrix}}$
        $\vcenter{\hbox{\includegraphics[width=0.25\textwidth]{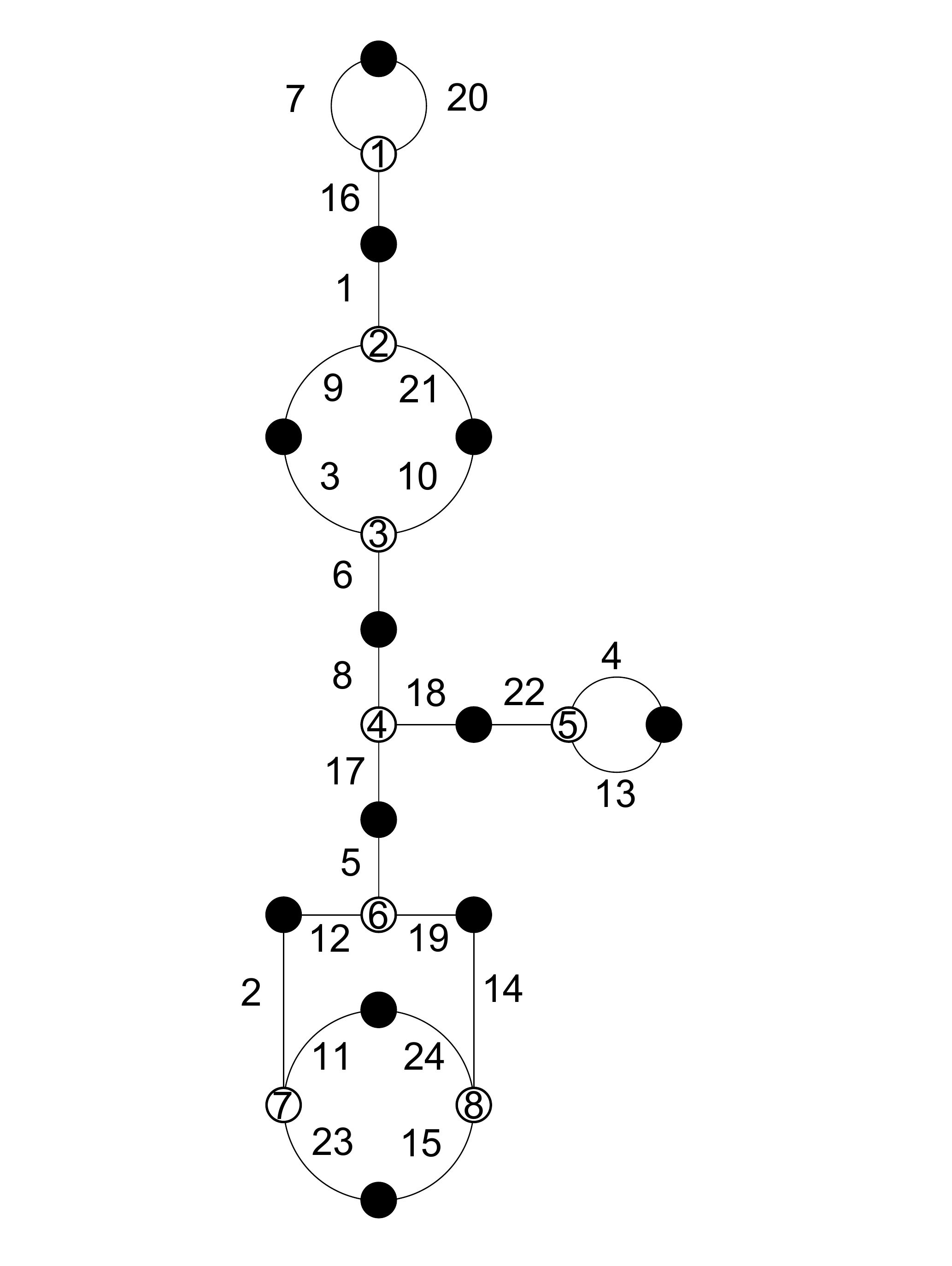}}}$
        $\vcenter{\hbox{\includegraphics[width=0.25\textwidth]{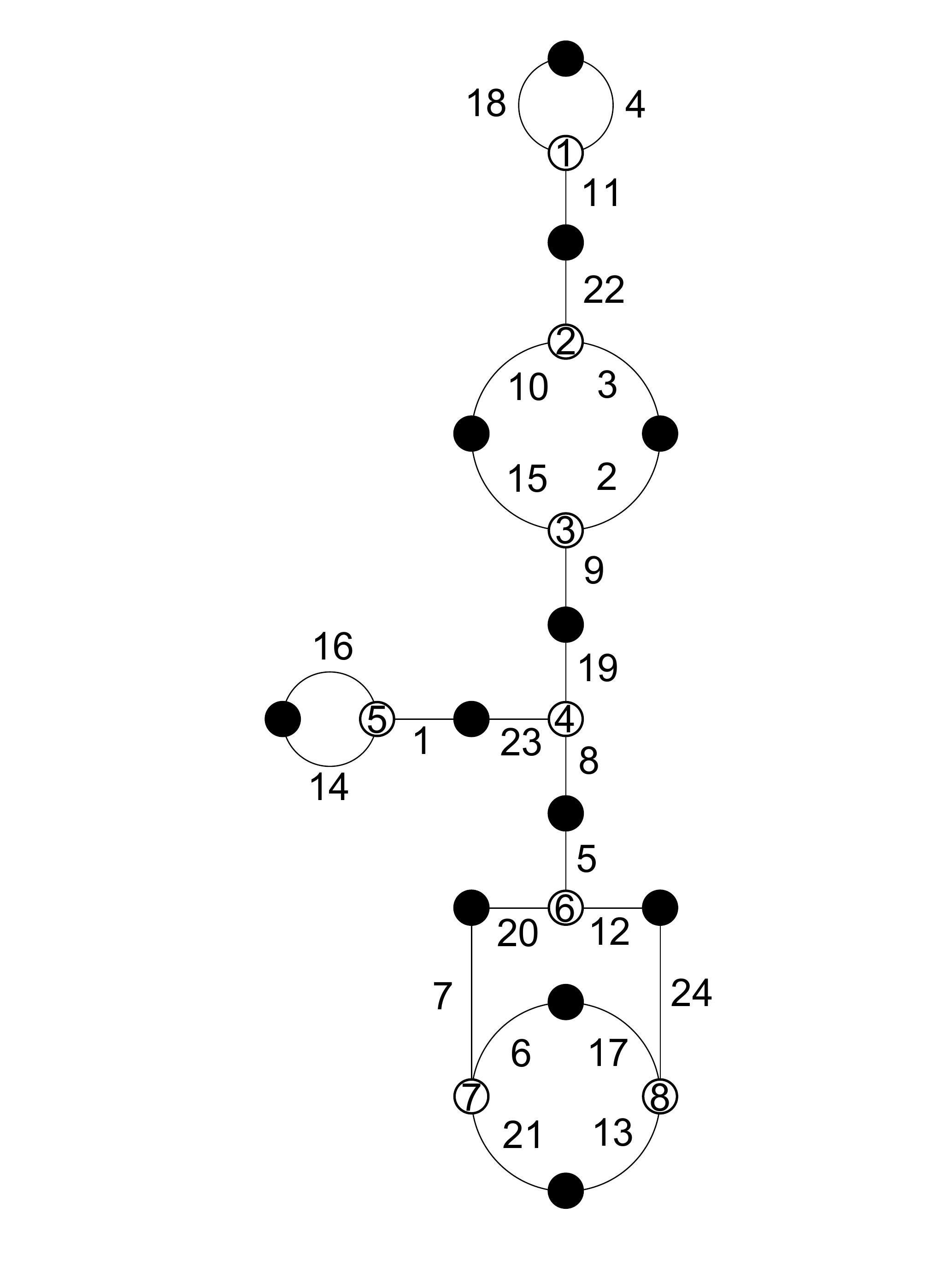}}}$
        \caption{ B: \{\{\{20,16,7\},\{1,21,9\},
        \{10,6,3\},\{18,17,8\},\{4,13,22\},
        \{5,19,12\},\{2,11,23\},\{24,14,15\}\}, \\ 
        \{\{11,24\},\{15,23\},\{14,19\},
        \{12,2\},\{5,17\},\{22,18\},
        \{13,4\},\{6,8\},\{10,21\},
        \{9,3\},\{1,16\},\{7,20\}\}\} \\
        C: \{\{\{18,4,11\},\{22,3,10\},
        \{15,2,9\},\{19,8,23\},\{1,14,16\},
        \{5,12,20\},\{24,13,17\},\{6,21,7\}\}, \\ 
        \{\{6,17\},\{21,13\},\{12,24\},
        \{7,20\},\{5,8\},\{1,23\},
        \{14,16\},\{19,9\},\{2,3\},
        \{10,15\},\{11,22\},\{4,18\}\}\}}
        \caption{15-3-2-2-1-1 B \& C $(\sqrt{-15})$}
        \label{Dessin}
    \end{subfigure}\hfill
\end{figure}

\begin{figure}[H]
    \begin{subfigure}{0.4\textwidth}
        \centering \captionsetup{justification=centering}
        $\scalemath{0.75}{
        \displaystyle \begin{pmatrix}
            2 & 1 & 0 & 0 & 0 & 0 & 0 & 0\\ 
            1 & 0 & 1 & 1 & 0 & 0 & 0 & 0\\
            0 & 1 & 0 & 1 & 1 & 0 & 0 & 0\\
            0 & 1 & 1 & 0 & 1 & 0 & 0 & 0\\
            0 & 0 & 1 & 1 & 0 & 0 & 1 & 0\\
            0 & 0 & 0 & 0 & 0 & 2 & 1 & 0\\
            0 & 0 & 0 & 0 & 1 & 1 & 0 & 1\\
            0 & 0 & 0 & 0 & 0 & 0 & 1 & 2
        \end{pmatrix}}$
        $\vcenter{\hbox{\includegraphics[width=0.35\textwidth]{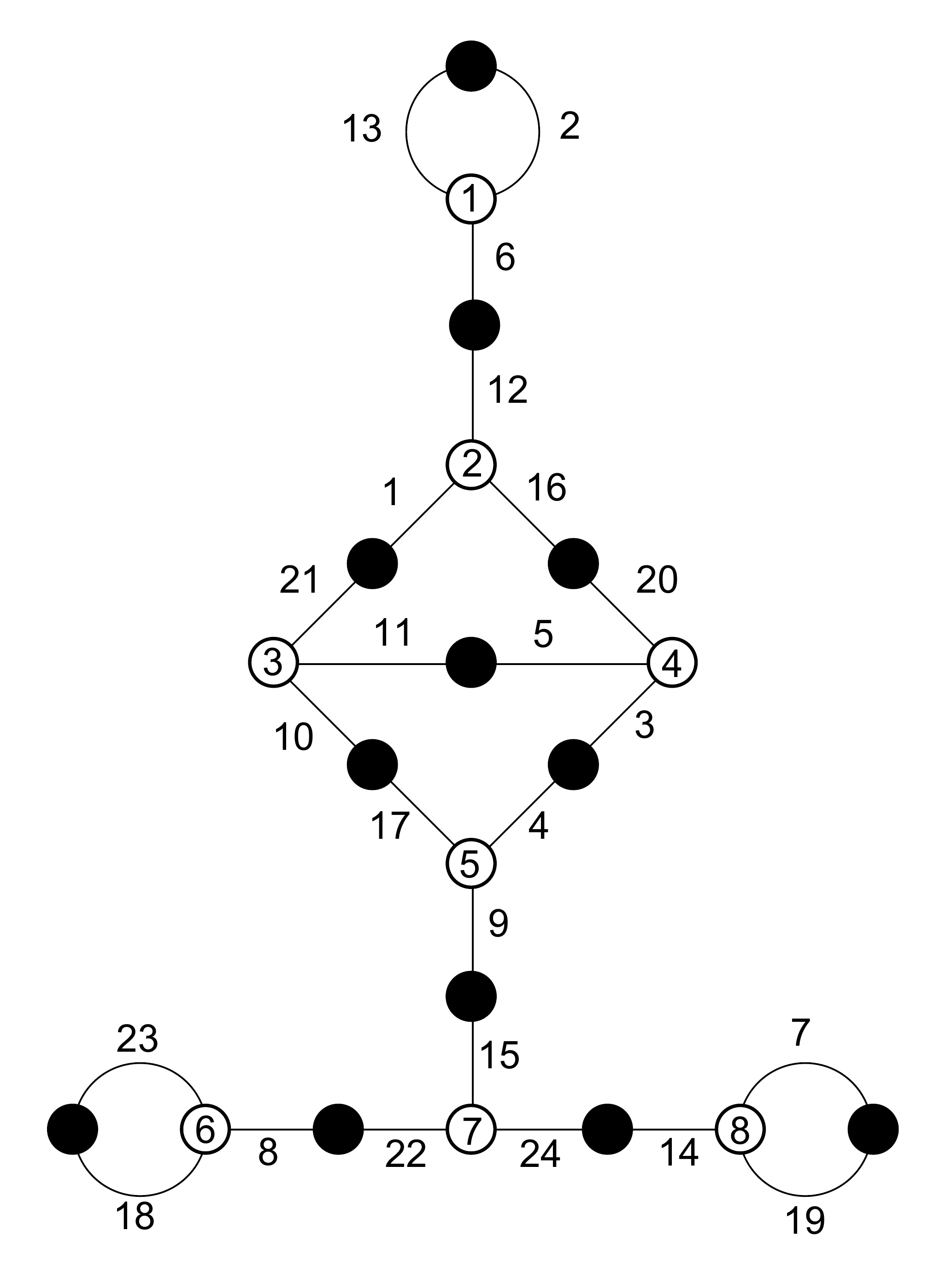}}}$
        \caption{ \{\{\{2,6,13\},\{1,12,16\},
        \{20,3,5\},\{11,10,21\},\{17,4,9\},
        \{15,24,22\},\{7,19,14\},\{8,18,23\}\}, \\ 
        \{\{18,23\},\{8,22\},\{24,14\},
        \{7,19\},\{9,15\},\{4,3\},
        \{10,17\},\{5,11\},\{16,20\},
        \{1,21\},\{6,12\},\{2,13\}\}\}}
        \caption{15-3-3-1-1-1 $(\mathbb{Q})$}
        \label{Dessin}
    \end{subfigure} \hfill
    \begin{subfigure}{0.6\textwidth}
        \centering \captionsetup{justification=centering}
        $\scalemath{0.75}{
        \displaystyle \begin{pmatrix}
            2 & 1 & 0 & 0 & 0 & 0 & 0 & 0\\ 
            1 & 0 & 1 & 0 & 1 & 0 & 0 & 0\\
            0 & 1 & 0 & 2 & 0 & 0 & 0 & 0\\
            0 & 0 & 2 & 0 & 1 & 0 & 0 & 0\\
            0 & 1 & 0 & 1 & 0 & 0 & 1 & 0\\
            0 & 0 & 0 & 0 & 0 & 2 & 1 & 0\\
            0 & 0 & 0 & 0 & 1 & 1 & 0 & 1\\
            0 & 0 & 0 & 0 & 0 & 0 & 1 & 2
        \end{pmatrix}}$
        $\vcenter{\hbox{\includegraphics[width=0.25\textwidth]{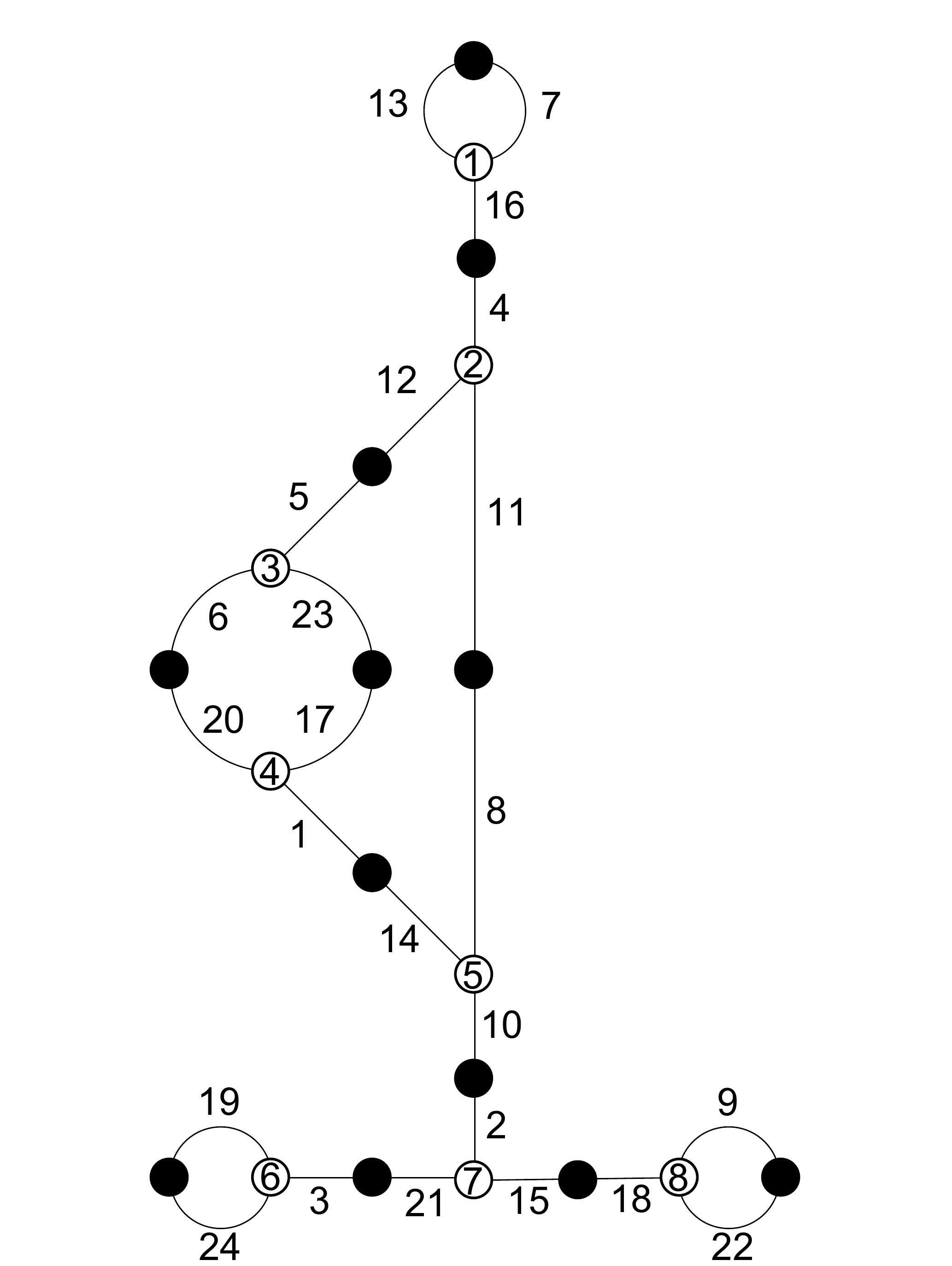}}}$
        $\vcenter{\hbox{\includegraphics[width=0.25\textwidth]{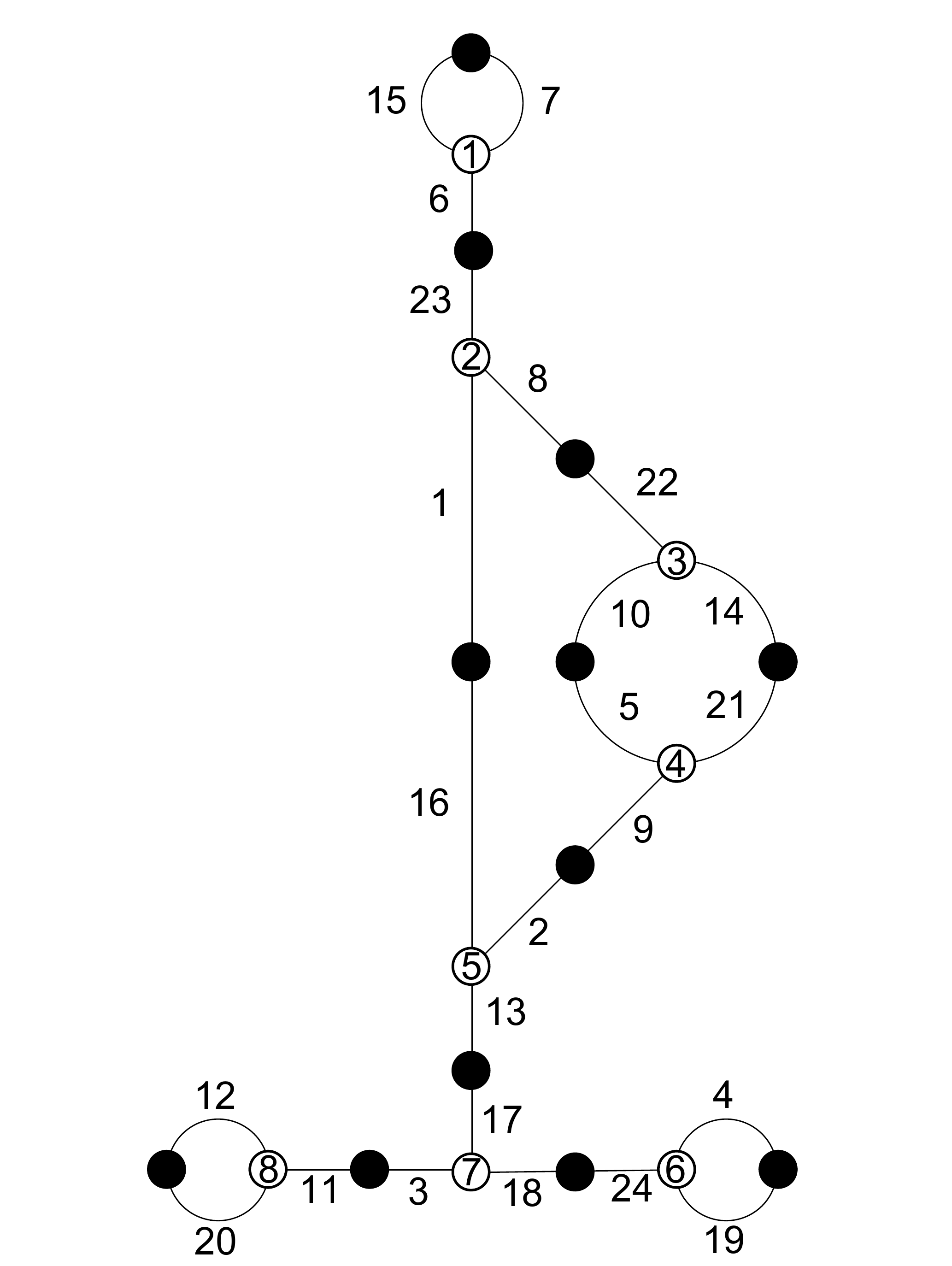}}}$
        \caption{ A: \{\{\{7,16,13\},\{4,11,12\},
        \{5,23,6\},\{20,17,1\},\{8,10,14\},
        \{2,15,21\},\{19,3,24\},\{9,22,18\}\}, \\ 
        \{\{9,22\},\{15,18\},\{2,10\},
        \{19,24\},\{3,21\},\{8,11\},
        \{14,1\},\{6,20\},\{23,17\},
        \{5,12\},\{16,4\},\{7,13\}\}\} \\
        B: \{\{\{7,6,15\},\{8,1,23\},
        \{22,14,10\},\{21,9,5\},\{2,13,16\},
        \{17,18,3\},\{4,19,24\},\{11,20,12\}\}, \\ 
        \{\{12,20\},\{3,11\},\{13,17\},
        \{18,24\},\{4,19\},\{2,9\},
        \{1,16\},\{8,22\},\{10,5\},
        \{14,21\},\{6,23\},\{7,15\}\}\}}
        \caption{15-4-2-1-1-1 A \& B $(\sqrt{-15})$}
        \label{Dessin}
    \end{subfigure}\hfill
\end{figure}

\begin{figure}[H]
    \begin{subfigure}{0.6\textwidth}
        \centering \captionsetup{justification=centering}
        $\scalemath{0.75}{
        \displaystyle \begin{pmatrix}
            2 & 1 & 0 & 0 & 0 & 0 & 0 & 0\\ 
            1 & 0 & 1 & 1 & 0 & 0 & 0 & 0\\
            0 & 1 & 2 & 0 & 0 & 0 & 0 & 0\\
            0 & 1 & 0 & 0 & 1 & 1 & 0 & 0\\
            0 & 0 & 0 & 1 & 2 & 0 & 0 & 0\\
            0 & 0 & 0 & 1 & 0 & 0 & 0 & 2\\
            0 & 0 & 0 & 0 & 0 & 0 & 2 & 1\\
            0 & 0 & 0 & 0 & 0 & 2 & 1 & 0
        \end{pmatrix}}$
        $\vcenter{\hbox{\includegraphics[width=0.25\textwidth]{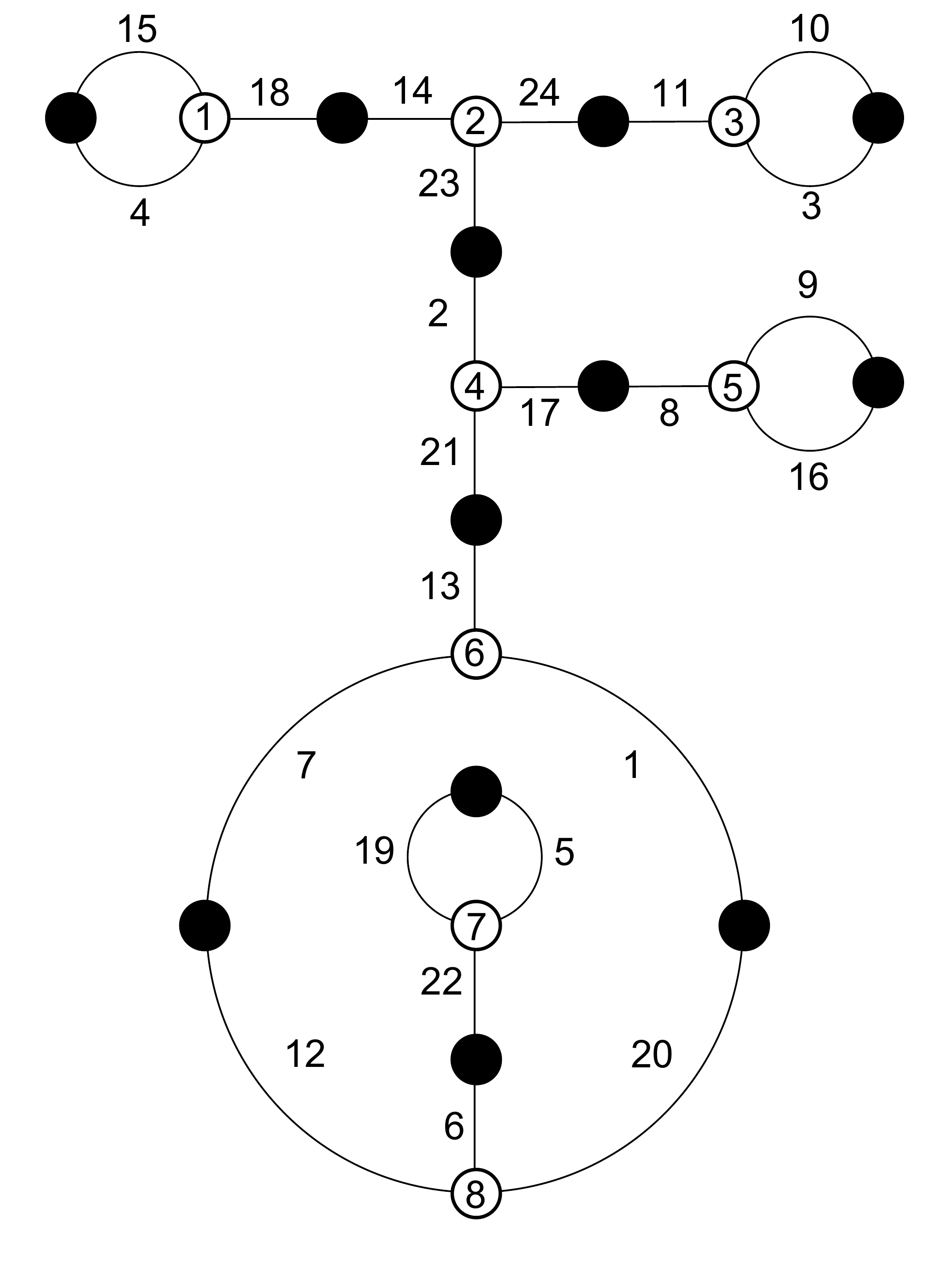}}}$
        $\vcenter{\hbox{\includegraphics[width=0.25\textwidth]{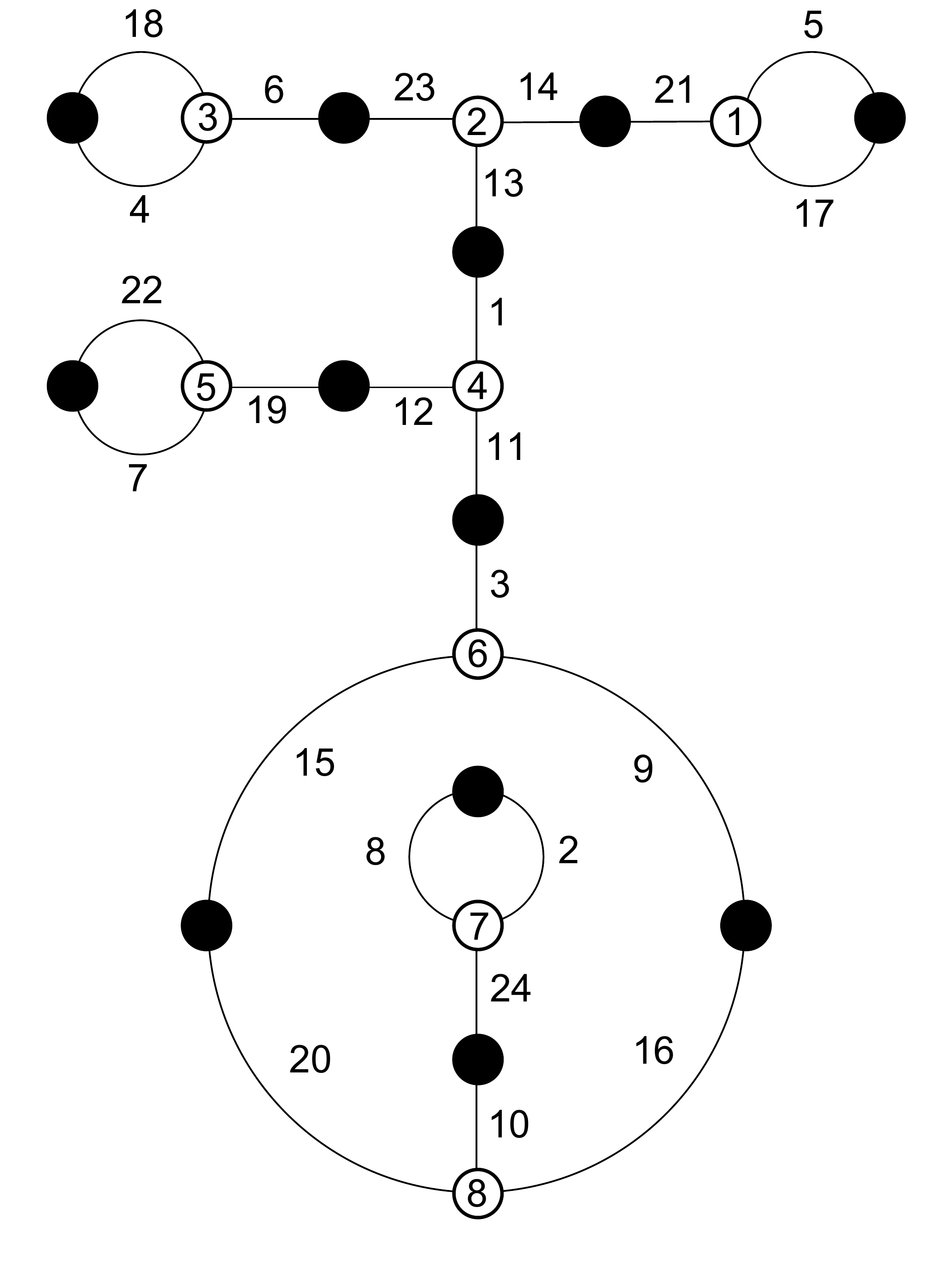}}}$
        \caption{ A: \{\{\{15,18,4\},\{14,24,23\},
        \{11,10,3\},\{9,16,8\},\{17,21,2\},
        \{13,1,7\},\{19,5,22\},\{6,20,12\}\}, \\ 
        \{\{19,5\},\{6,22\},\{1,20\},
        \{7,12\},\{13,21\},\{8,17\},
        \{9,16\},\{3,10\},\{11,24\},
        \{2,23\},\{14,18\},\{4,15\}\}\} \\
        B: \{\{\{4,18,6\},\{23,14,13\},
        \{5,17,21\},\{1,11,12\},\{7,22,19\},
        \{3,9,15\},\{2,24,8\},\{16,20,10\}\}, \\ 
        \{\{2,8\},\{24,10\},\{9,16\},
        \{15,20\},\{3,11\},\{12,19\},
        \{7,22\},\{1,13\},\{14,21\},
        \{5,17\},\{6,23\},\{4,18\}\}\}}
        \caption{15-5-1-1-1-1 A \& B $(\sqrt{-15})$}
        \label{Dessin}
    \end{subfigure} \hfill
    \begin{subfigure}{0.4\textwidth}
        \centering \captionsetup{justification=centering}
        $\scalemath{0.75}{
        \displaystyle \begin{pmatrix}
            2 & 1 & 0 & 0 & 0 & 0 & 0 & 0\\ 
            1 & 0 & 2 & 0 & 0 & 0 & 0 & 0\\
            0 & 2 & 0 & 1 & 0 & 0 & 0 & 0\\
            0 & 0 & 1 & 0 & 2 & 0 & 0 & 0\\
            0 & 0 & 0 & 2 & 0 & 1 & 0 & 0\\
            0 & 0 & 0 & 0 & 1 & 0 & 2 & 0\\
            0 & 0 & 0 & 0 & 0 & 2 & 0 & 1\\
            0 & 0 & 0 & 0 & 0 & 0 & 1 & 2
        \end{pmatrix}}$
        $\vcenter{\hbox{\includegraphics[width=0.35\textwidth]{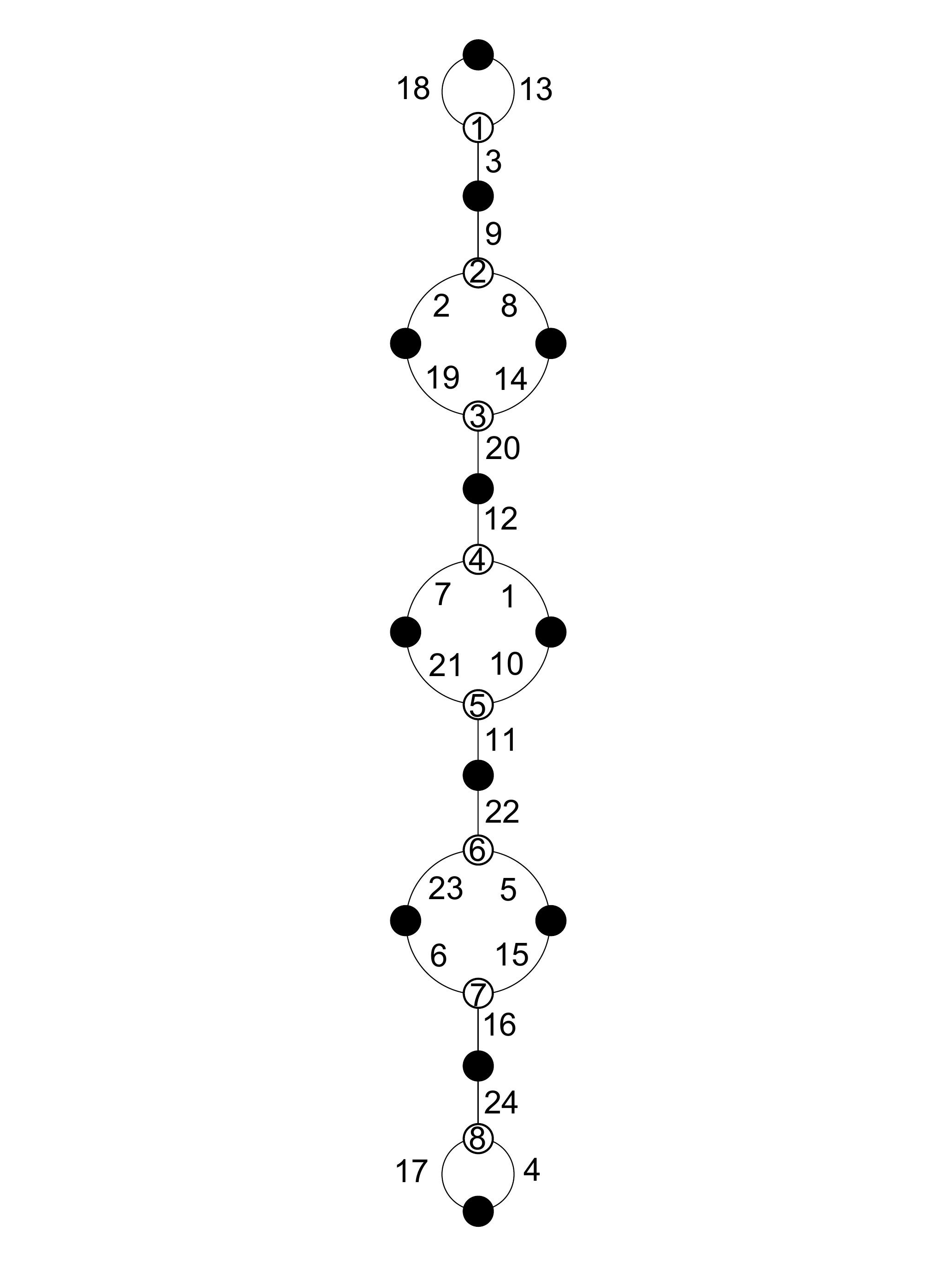}}}$
        \caption{ \{\{\{13,3,18\},\{9,8,2\},
        \{14,20,19\},\{12,1,7\},\{10,11,21\},
        \{22,5,23\},\{15,16,6\},\{24,4,17\}\}, \\ 
        \{\{17,4\},\{24,16\},\{6,23\},
        \{5,15\},\{22,11\},\{1,10\},
        \{7,21\},\{12,20\},\{8,14\},
        \{2,19\},\{3,9\},\{13,18\}\}\}}
        \caption{16-2-2-2-1-1 $(\mathbb{Q})$}
        \label{Dessin}
    \end{subfigure}\hfill
\end{figure}

\begin{figure}[H]
    \begin{subfigure}{0.4\textwidth}
        \centering \captionsetup{justification=centering}
        $\scalemath{0.75}{
        \displaystyle \begin{pmatrix}
            2 & 1 & 0 & 0 & 0 & 0 & 0 & 0\\ 
            1 & 0 & 2 & 0 & 0 & 0 & 0 & 0\\
            0 & 2 & 0 & 1 & 0 & 0 & 0 & 0\\
            0 & 0 & 1 & 0 & 1 & 1 & 0 & 0\\
            0 & 0 & 0 & 1 & 0 & 1 & 1 & 0\\
            0 & 0 & 0 & 1 & 1 & 0 & 0 & 1\\
            0 & 0 & 0 & 0 & 1 & 0 & 2 & 0\\
            0 & 0 & 0 & 0 & 0 & 1 & 0 & 2
        \end{pmatrix}}$
        $\vcenter{\hbox{\includegraphics[width=0.35\textwidth]{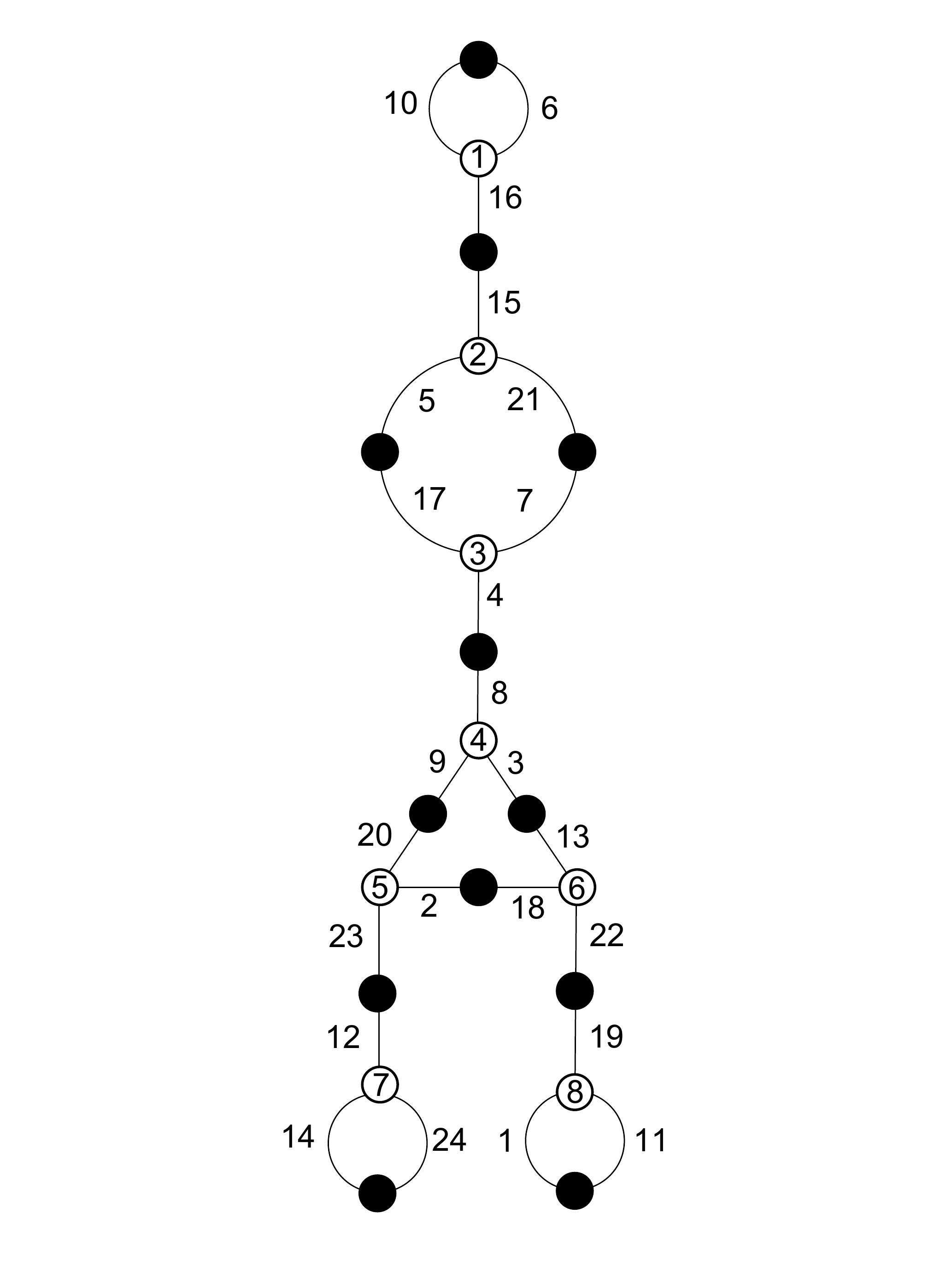}}}$
        \caption{ \{\{\{6,16,10\},\{15,21,5\},
        \{7,4,17\},\{8,3,9\},\{13,22,18\},
        \{20,2,23\},\{12,24,14\},\{19,11,1\}\}, \\ 
        \{\{1,11\},\{19,22\},\{23,12\},
        \{14,24\},\{2,18\},\{9,20\},
        \{3,13\},\{8,4\},\{7,21\},
        \{5,17\},\{15,16\},\{6,10\}\}\}}
        \caption{16-3-2-1-1-1 A $(\mathbb{Q})$}
        \label{Dessin}
    \end{subfigure} \hfill
    \begin{subfigure}{0.6\textwidth}
        \centering \captionsetup{justification=centering}
        $\scalemath{0.75}{
        \displaystyle \begin{pmatrix}
            2 & 1 & 0 & 0 & 0 & 0 & 0 & 0\\ 
            1 & 0 & 1 & 1 & 0 & 0 & 0 & 0\\
            0 & 1 & 2 & 0 & 0 & 0 & 0 & 0\\
            0 & 1 & 0 & 0 & 1 & 1 & 0 & 0\\
            0 & 0 & 0 & 1 & 2 & 0 & 0 & 0\\
            0 & 0 & 0 & 1 & 0 & 0 & 1 & 1\\
            0 & 0 & 0 & 0 & 0 & 1 & 0 & 2\\
            0 & 0 & 0 & 0 & 0 & 1 & 2 & 0
        \end{pmatrix}}$
        $\vcenter{\hbox{\includegraphics[width=0.25\textwidth]{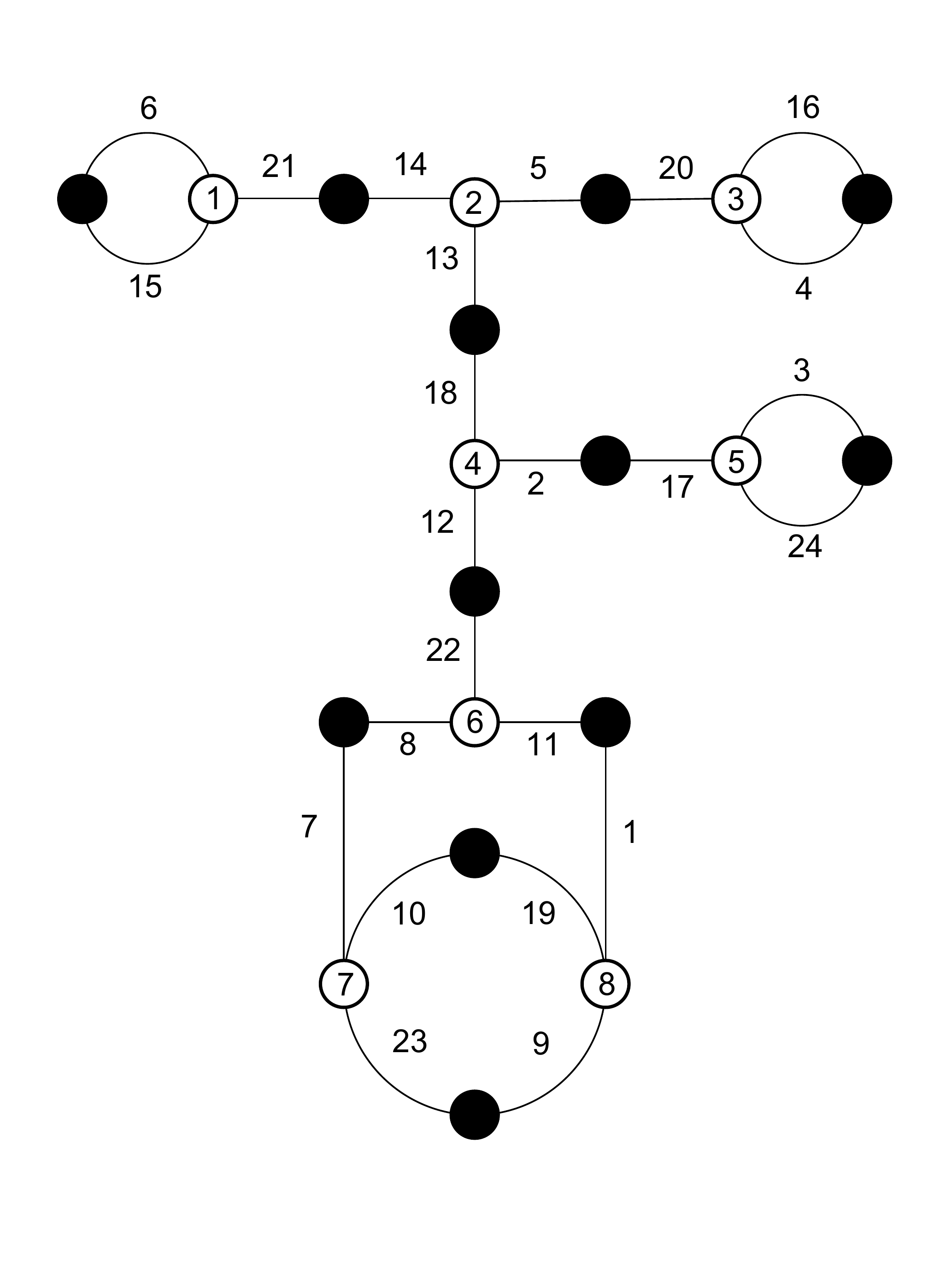}}}$
        $\vcenter{\hbox{\includegraphics[width=0.25\textwidth]{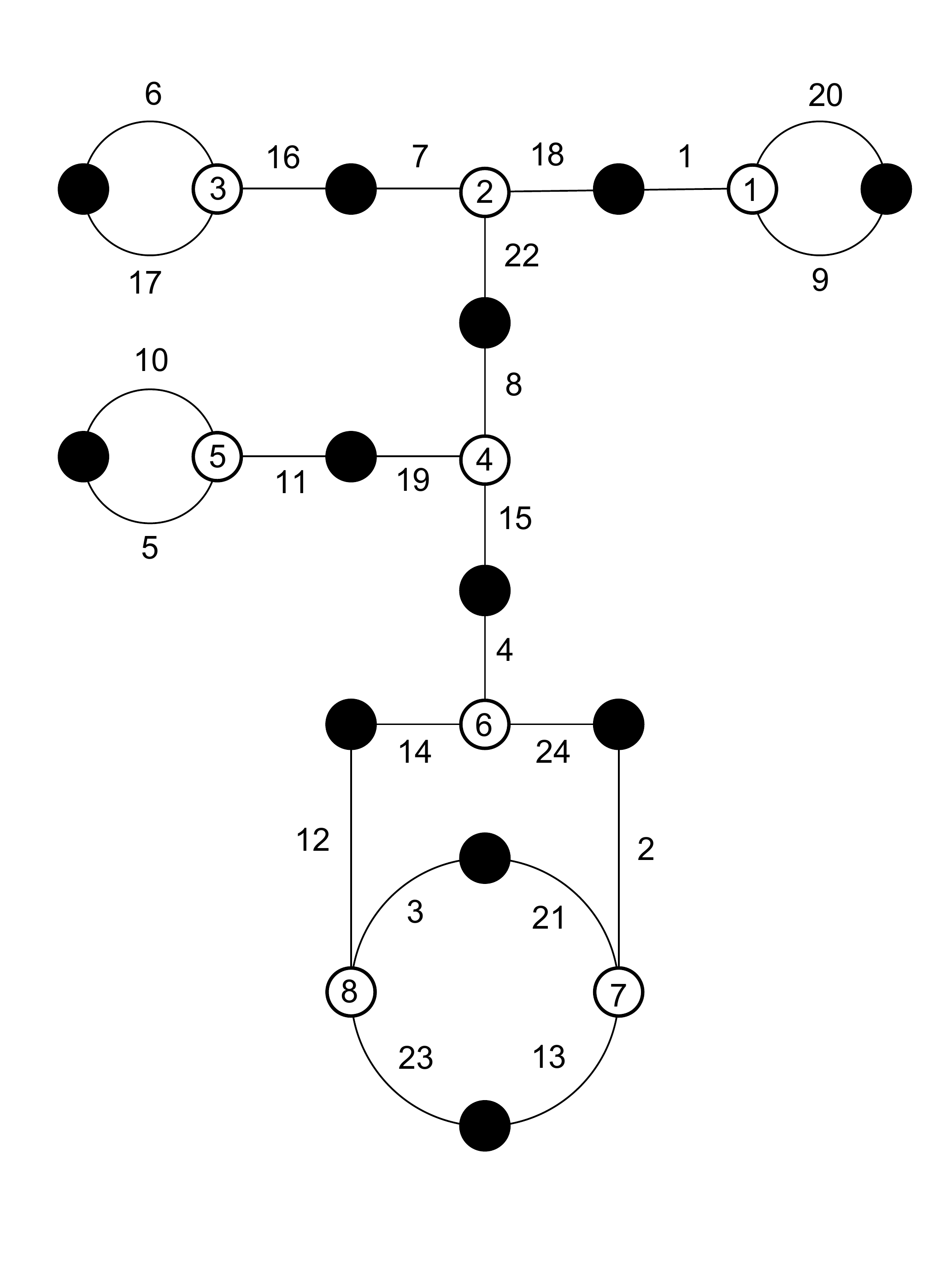}}}$
        \caption{ B: \{\{\{15,6,21\},\{14,5,13\},
        \{20,16,4\},\{2,12,18\},\{3,24,17\},
        \{22,11,8\},\{10,23,7\},\{1,9,19\}\}, \\ 
        \{\{9,23\},\{10,19\},\{1,11\},
        \{7,8\},\{22,12\},\{2,17\},
        \{3,24\},\{18,13\},\{5,20\},
        \{4,16\},\{21,14\},\{6,15\}\}\} \\
        C: \{\{\{3,23,12\},\{21,2,13\},
        \{24,14,4\},\{15,19,8\},\{11,5,10\},
        \{22,7,18\},\{20,9,1\},\{6,16,17\}\}, \\ 
        \{\{6,17\},\{16,7\},\{18,1\},
        \{20,9\},\{22,8\},\{19,11\},
        \{10,5\},\{4,15\},\{2,24\},
        \{12,14\},\{3,21\},\{23,13\}\}\}}
        \caption{16-3-2-1-1-1 B \& C $(\sqrt{-2})$}
        \label{Dessin}
    \end{subfigure}\hfill
\end{figure}

\begin{figure}[H]
    \begin{subfigure}{0.5\textwidth}
        \centering \captionsetup{justification=centering}
        $\scalemath{0.75}{
        \displaystyle \begin{pmatrix}
            2 & 0 & 1 & 0 & 0 & 0 & 0 & 0\\ 
            0 & 2 & 0 & 1 & 0 & 0 & 0 & 0\\
            1 & 0 & 0 & 1 & 1 & 0 & 0 & 0\\
            0 & 1 & 1 & 0 & 0 & 1 & 0 & 0\\
            0 & 0 & 1 & 0 & 0 & 1 & 1 & 0\\
            0 & 0 & 0 & 1 & 1 & 0 & 0 & 1\\
            0 & 0 & 0 & 0 & 1 & 0 & 2 & 0\\
            0 & 0 & 0 & 0 & 0 & 1 & 0 & 2
        \end{pmatrix}}$
        $\vcenter{\hbox{\includegraphics[width=0.35\textwidth]{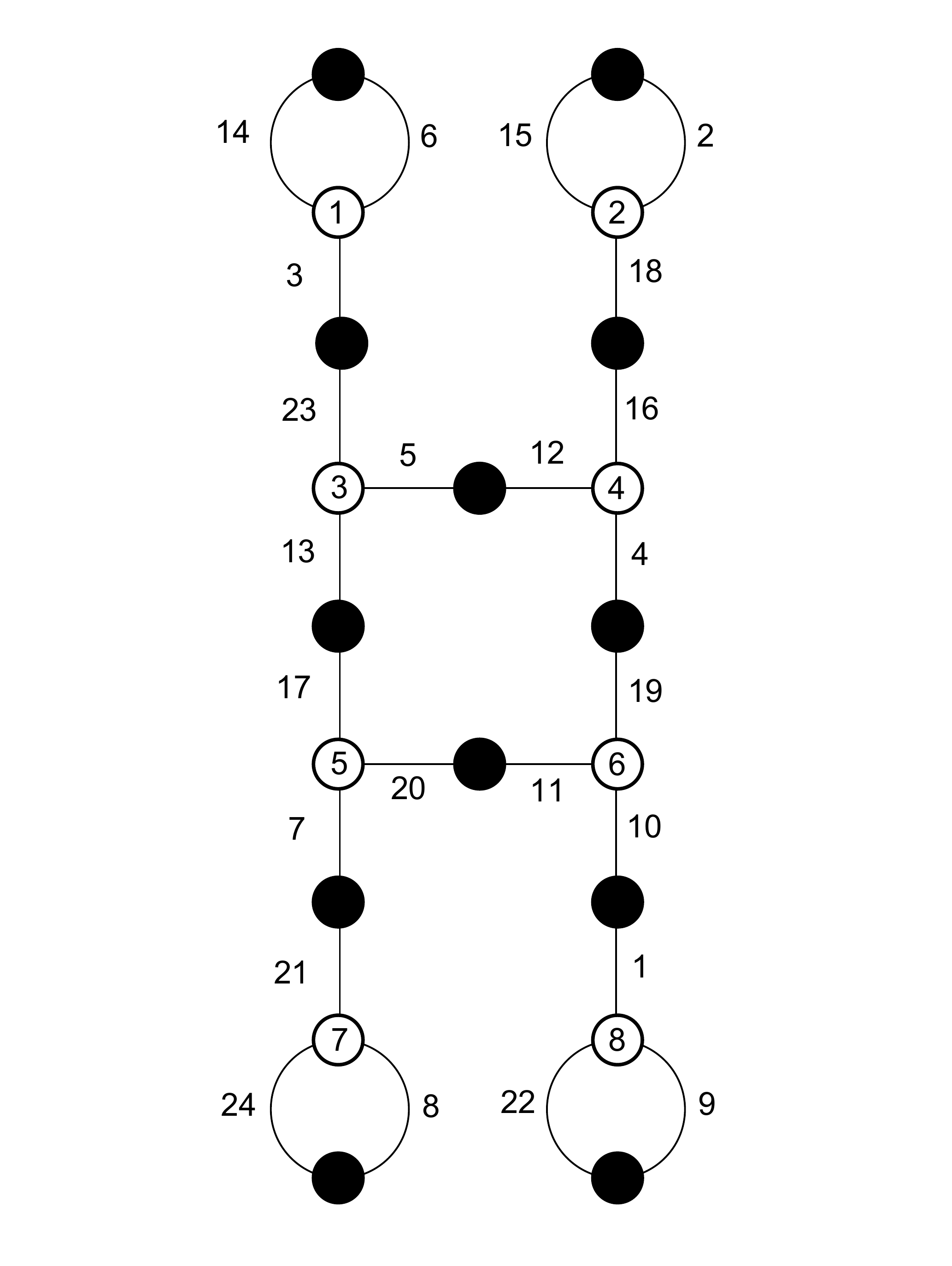}}}$
        \caption{ \{\{\{14,6,3\},\{23,5,13\},
        \{12,16,4\},\{2,18,15\},\{19,10,11\},
        \{17,20,7\},\{21,8,24\},\{1,9,22\}\}, \\ 
        \{\{8,24\},\{1,10\},\{9,22\},
        \{7,21\},\{20,11\},\{17,13\},
        \{4,19\},\{5,12\},\{16,18\},
        \{2,15\},\{6,14\},\{3,23\}\}\}}
        \caption{16-4-1-1-1-1 $(\mathbb{Q})$}
        \label{Dessin}
    \end{subfigure} \hfill
    \begin{subfigure}{0.5\textwidth}
        \centering \captionsetup{justification=centering}
        $\scalemath{0.75}{
        \displaystyle \begin{pmatrix}
            2 & 1 & 0 & 0 & 0 & 0 & 0 & 0\\ 
            1 & 0 & 2 & 0 & 0 & 0 & 0 & 0\\
            0 & 2 & 0 & 1 & 0 & 0 & 0 & 0\\
            0 & 0 & 1 & 0 & 2 & 0 & 0 & 0\\
            0 & 0 & 0 & 2 & 0 & 0 & 1 & 0\\
            0 & 0 & 0 & 0 & 0 & 2 & 1 & 0\\
            0 & 0 & 0 & 0 & 1 & 1 & 0 & 1\\
            0 & 0 & 0 & 0 & 0 & 0 & 1 & 2
        \end{pmatrix}}$
        $\vcenter{\hbox{\includegraphics[width=0.35\textwidth]{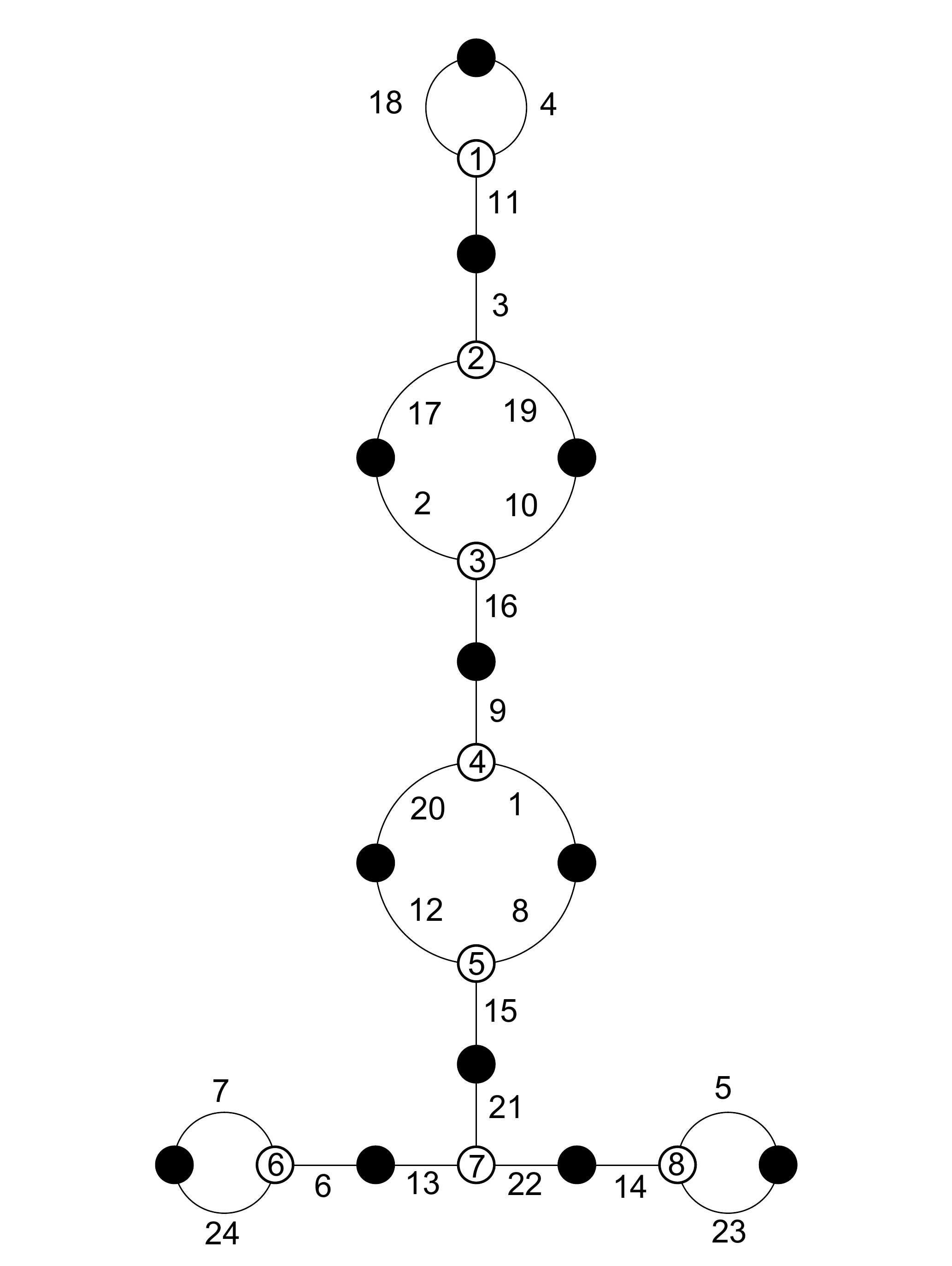}}}$
        \caption{ \{\{\{4,11,18\},\{3,19,17\},
        \{2,10,16\},\{9,1,20\},\{8,15,12\},
        \{21,22,13\},\{5,23,14\},\{6,24,7\}\}, \\ 
        \{\{7,24\},\{6,13\},\{22,14\},
        \{5,23\},\{21,15\},\{1,8\},
        \{20,12\},\{9,16\},\{10,19\},
        \{2,17\},\{3,11\},\{4,18\}\}\}}
        \caption{17-2-2-1-1-1 A $(\sqrt{17})$}
        \label{Dessin}
    \end{subfigure}\hfill
\end{figure}

\begin{figure}[H]
    \begin{subfigure}{0.5\textwidth}
        \centering \captionsetup{justification=centering}
        $\scalemath{0.75}{
        \displaystyle \begin{pmatrix}
            2 & 1 & 0 & 0 & 0 & 0 & 0 & 0\\ 
            1 & 0 & 2 & 0 & 0 & 0 & 0 & 0\\
            0 & 2 & 0 & 0 & 1 & 0 & 0 & 0\\
            0 & 0 & 0 & 2 & 1 & 0 & 0 & 0\\
            0 & 0 & 1 & 1 & 0 & 1 & 0 & 0\\
            0 & 0 & 0 & 0 & 1 & 0 & 2 & 0\\
            0 & 0 & 0 & 0 & 0 & 2 & 0 & 1\\
            0 & 0 & 0 & 0 & 0 & 0 & 1 & 2
        \end{pmatrix}}$
        $\vcenter{\hbox{\includegraphics[width=0.35\textwidth]{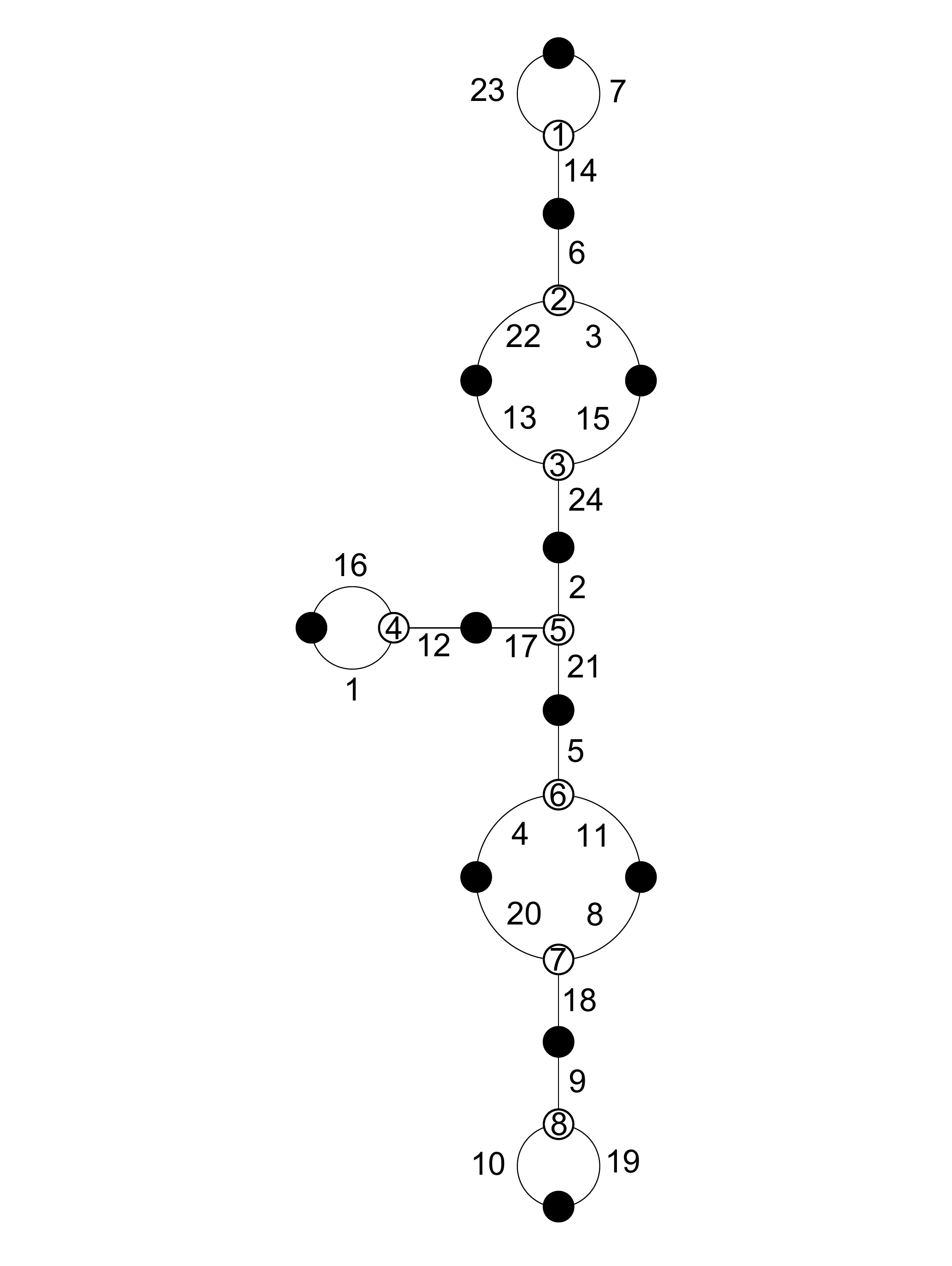}}}$
        \caption{ \{\{\{7,14,23\},\{6,3,22\},
        \{13,15,24\},\{2,21,17\},\{12,1,16\},
        \{5,11,4\},\{8,18,20\},\{9,19,10\}\}, \\ 
        \{\{10,19\},\{9,18\},\{4,20\},
        \{8,11\},\{5,21\},\{17,12\},
        \{1,16\},\{2,24\},\{13,22\},
        \{3,15\},\{6,14\},\{7,23\}\}\}}
        \caption{17-2-2-1-1-1 B $(\sqrt{17})$}
        \label{Dessin}
    \end{subfigure} \hfill
    \begin{subfigure}{0.5\textwidth}
        \centering \captionsetup{justification=centering}
        $\scalemath{0.75}{
        \displaystyle \begin{pmatrix}
            2 & 1 & 0 & 0 & 0 & 0 & 0 & 0\\ 
            1 & 0 & 1 & 1 & 0 & 0 & 0 & 0\\
            0 & 1 & 2 & 0 & 0 & 0 & 0 & 0\\
            0 & 1 & 0 & 0 & 1 & 1 & 0 & 0\\
            0 & 0 & 0 & 1 & 0 & 1 & 1 & 0\\
            0 & 0 & 0 & 1 & 1 & 0 & 0 & 1\\
            0 & 0 & 0 & 0 & 1 & 0 & 2 & 0\\
            0 & 0 & 0 & 0 & 0 & 1 & 0 & 2
        \end{pmatrix}}$
        $\vcenter{\hbox{\includegraphics[width=0.35\textwidth]{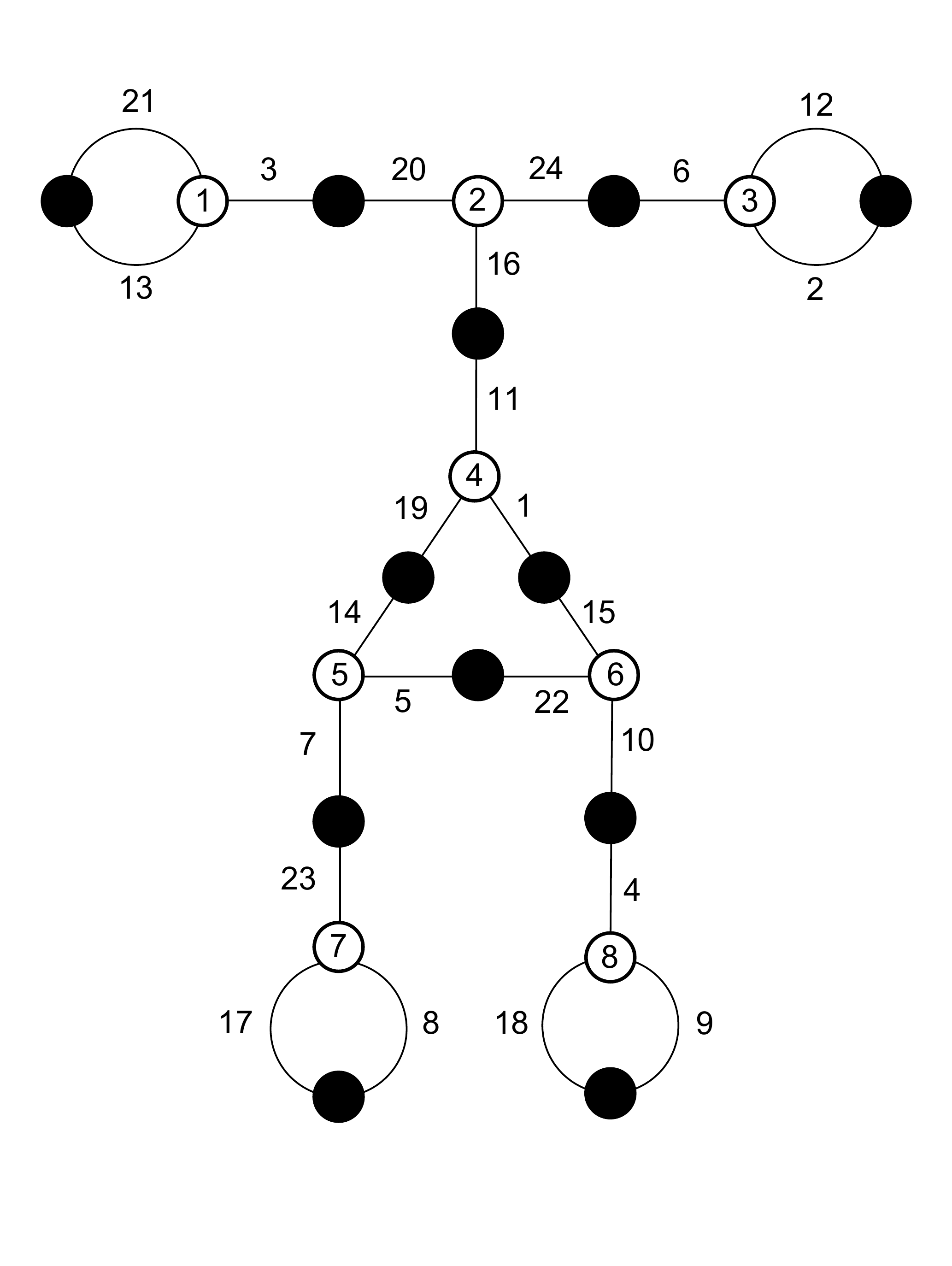}}}$
        \caption{ \{\{\{21,3,13\},\{20,24,16\},
        \{6,12,2\},\{11,1,19\},\{15,10,22\},
        \{14,5,7\},\{23,8,17\},\{4,9,18\}\}, \\
        \{\{13,21\},\{3,20\},\{6,24\},
        \{2,12\},\{11,16\},\{1,15\},
        \{19,14\},\{5,22\},\{7,23\},
        \{8,17\},\{9,18\},\{4,10\}\}\}}
        \caption{17-3-1-1-1-1 $(\mathbb{Q})$}
        \label{Dessin}
    \end{subfigure}\hfill
\end{figure}

\begin{figure}[H]
    \begin{subfigure}{0.4\textwidth}
        \centering \captionsetup{justification=centering}
        $\scalemath{0.75}{
        \displaystyle \begin{pmatrix}
            2 & 1 & 0 & 0 & 0 & 0 & 0 & 0\\ 
            1 & 0 & 1 & 1 & 0 & 0 & 0 & 0\\
            0 & 1 & 2 & 0 & 0 & 0 & 0 & 0\\
            0 & 1 & 0 & 0 & 2 & 0 & 0 & 0\\
            0 & 0 & 0 & 2 & 0 & 0 & 1 & 0\\
            0 & 0 & 0 & 0 & 0 & 2 & 1 & 0\\
            0 & 0 & 0 & 0 & 1 & 1 & 0 & 1\\
            0 & 0 & 0 & 0 & 0 & 0 & 1 & 2
        \end{pmatrix}}$
        $\vcenter{\hbox{\includegraphics[width=0.35\textwidth]{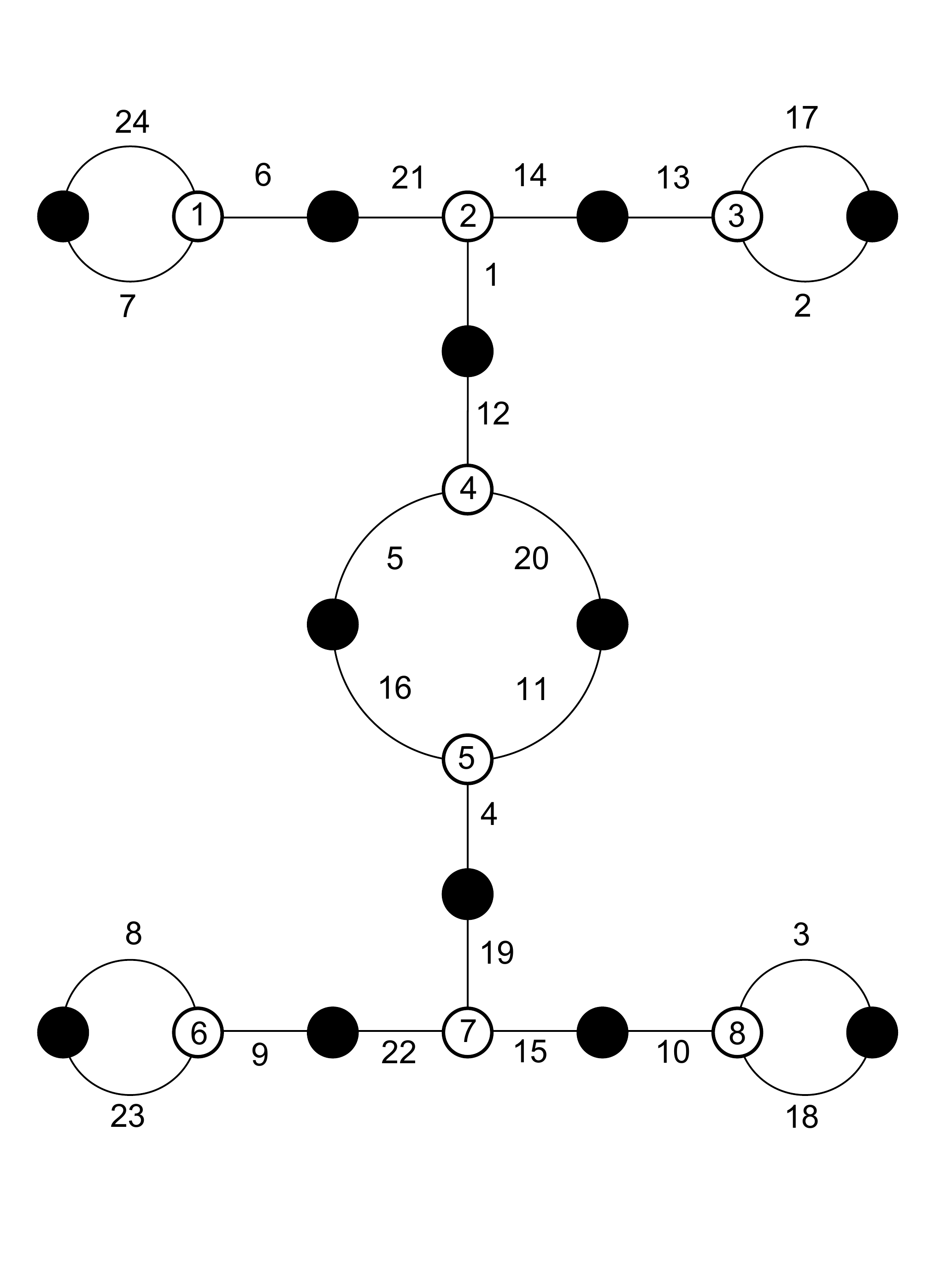}}}$
        \caption{ \{\{\{6,7,24\},\{21,14,1\},
        \{17,2,13\},\{12,20,5\},\{4,16,11\},
        \{19,15,22\},\{3,18,10\},\{9,23,8\}\}, \\ 
        \{\{8,23\},\{9,22\},\{15,10\},
        \{3,18\},\{4,19\},\{11,20\},
        \{5,16\},\{12,1\},\{24,7\},
        \{17,2\},\{13,14\},\{6,21\}\}\}}
        \caption{18-2-1-1-1-1 A $(\mathbb{Q})$}
        \label{Dessin}
    \end{subfigure} \hfill
    \begin{subfigure}{0.6\textwidth}
        \centering \captionsetup{justification=centering}
        $\scalemath{0.75}{
        \displaystyle \begin{pmatrix}
            2 & 1 & 0 & 0 & 0 & 0 & 0 & 0\\ 
            1 & 0 & 1 & 0 & 1 & 0 & 0 & 0\\
            0 & 1 & 2 & 0 & 0 & 0 & 0 & 0\\
            0 & 0 & 0 & 2 & 1 & 0 & 0 & 0\\
            0 & 1 & 0 & 1 & 0 & 1 & 0 & 0\\
            0 & 0 & 0 & 0 & 1 & 0 & 2 & 0\\
            0 & 0 & 0 & 0 & 0 & 2 & 0 & 1\\
            0 & 0 & 0 & 0 & 0 & 0 & 1 & 2
        \end{pmatrix}}$
        $\vcenter{\hbox{\includegraphics[width=0.25\textwidth]{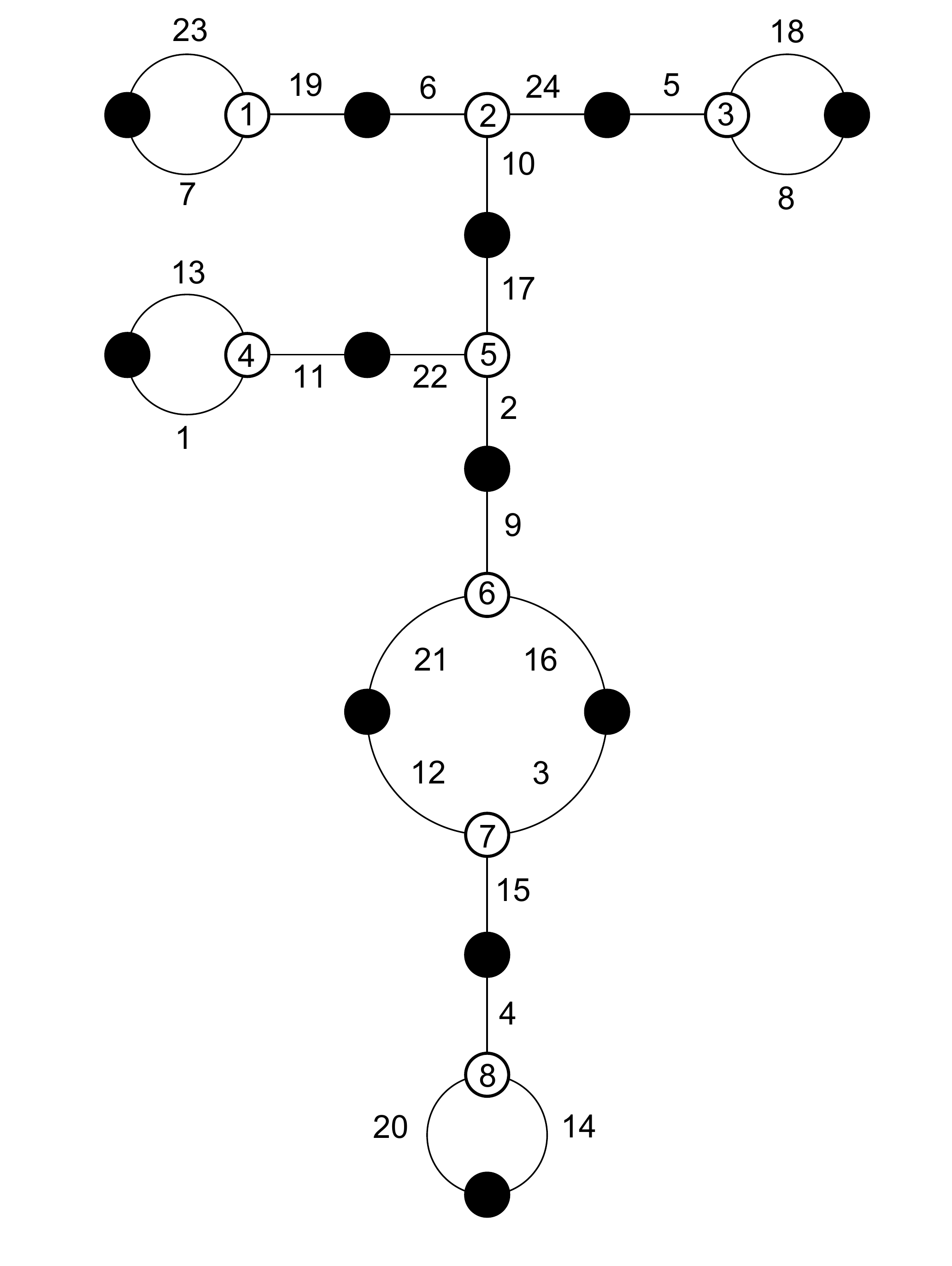}}}$
        $\vcenter{\hbox{\includegraphics[width=0.25\textwidth]{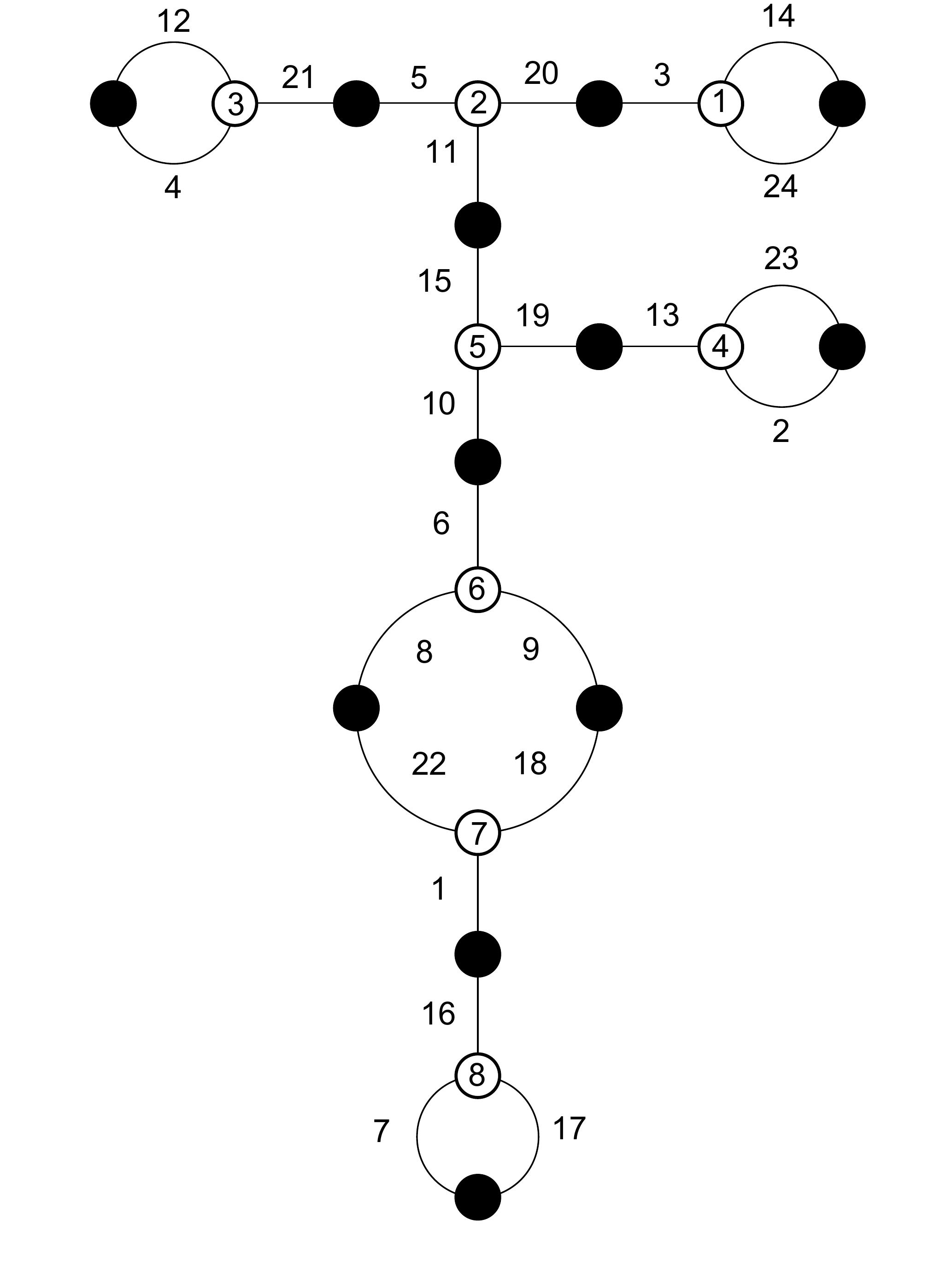}}}$
        \caption{ B: \{\{\{19,7,23\},\{24,10,6\},
        \{18,8,5\},\{17,2,22\},\{13,11,1\},
        \{9,16,21\},\{12,3,15\},\{4,14,20\}\}, \\ 
        \{\{14,20\},\{15,4\},\{3,16\},
        \{21,12\},\{9,2\},\{11,22\},
        \{1,13\},\{17,10\},\{19,6\},
        \{23,7\},\{24,5\},\{8,18\}\}\} \\
        C: \{\{\{14,24,3\},\{20,11,5\},
        \{12,21,4\},\{15,19,10\},\{2,13,23\},
        \{6,9,8\},\{18,1,22\},\{16,17,7\}\}, \\
        \{\{17,7\},\{1,16\},\{8,22\},
        \{9,18\},\{6,10\},\{13,19\},
        \{23,2\},\{11,15\},\{20,3\},
        \{12,4\},\{14,24\},\{5,21\}\}\}}
        \caption{18-2-1-1-1-1 B \& C $(\sqrt{-3})$}
        \label{Dessin}
    \end{subfigure}\hfill
\end{figure}

\begin{figure}[H]
    \begin{subfigure}{\textwidth}
        \centering \captionsetup{justification=centering}
        $\scalemath{0.75}{
        \displaystyle \begin{pmatrix}
            2 & 1 & 0 & 0 & 0 & 0 & 0 & 0\\ 
            1 & 0 & 1 & 0 & 1 & 0 & 0 & 0\\
            0 & 1 & 2 & 0 & 0 & 0 & 0 & 0\\
            0 & 0 & 0 & 2 & 1 & 0 & 0 & 0\\
            0 & 1 & 0 & 1 & 0 & 0 & 1 & 0\\
            0 & 0 & 0 & 0 & 0 & 2 & 1 & 0\\
            0 & 0 & 0 & 0 & 1 & 1 & 0 & 1\\
            0 & 0 & 0 & 0 & 0 & 0 & 1 & 2
        \end{pmatrix}}$
        $\vcenter{\hbox{\includegraphics[width=0.2\textwidth]{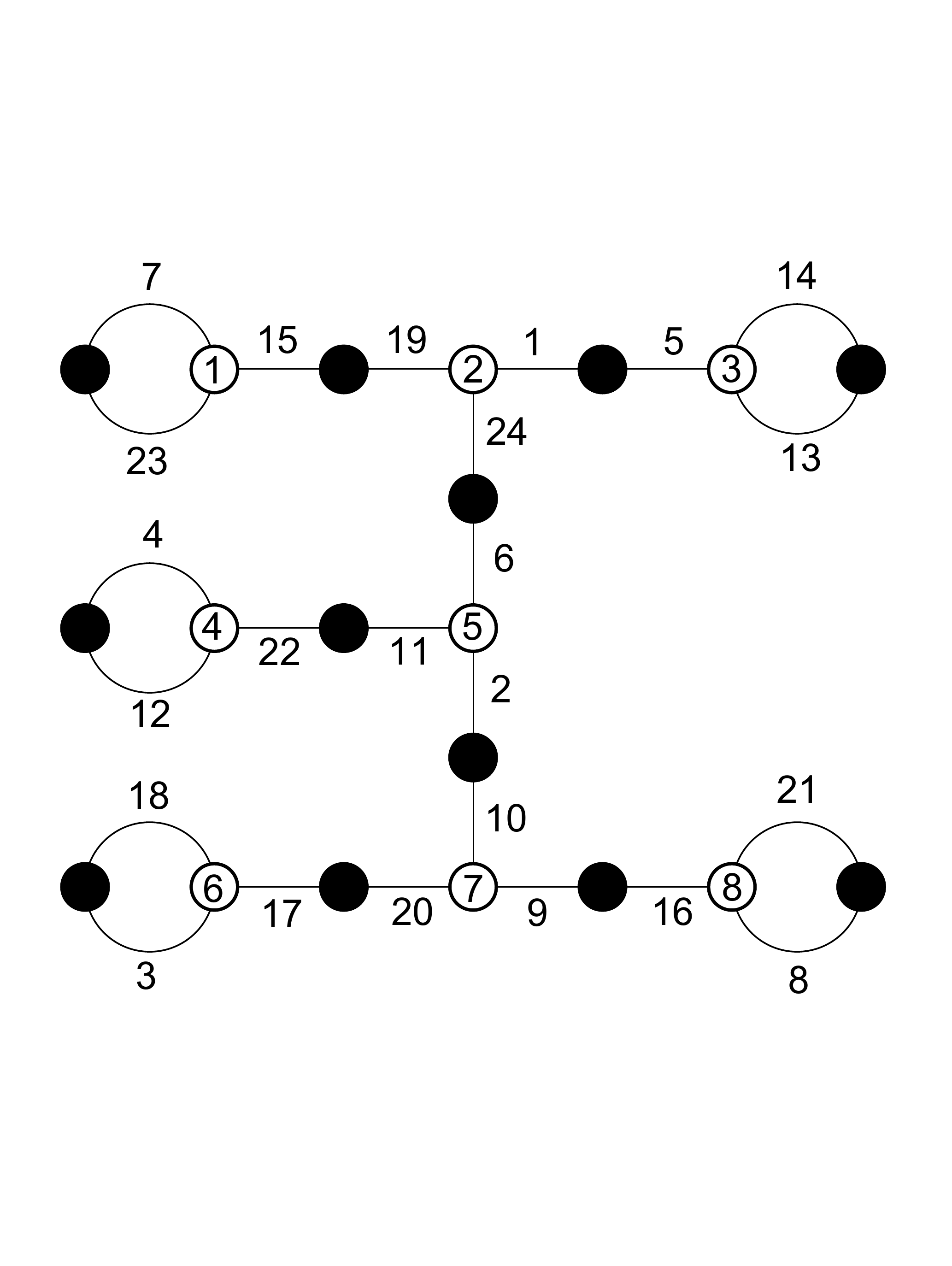}}}$
        \caption{ \{\{\{7,15,23\},\{6,2,11\},
        \{21,8,16\},\{17,3,18\},\{1,24,19\},
        \{14,13,5\},\{4,22,12\},\{10,9,20\}\}, \\ 
        \{\{8,21\},\{16,9\},\{10,2\},
        \{5,1\},\{14,13\},\{4,12\},
        \{3,18\},\{15,19\},\{24,6\},
        \{11,22\},\{17,20\},\{7,23\}\}\} }
        \caption{19-1-1-1-1-1 $(\mathbb{Q})$}
        \label{Dessin}
    \end{subfigure} \hfill
\end{figure}

}

\end{appendices}

\end{document}